\documentclass{optica-article}

\journal{opticajournal}

\articletype{Research Article}
\usepackage{lineno}
\usepackage[print-unity-mantissa=false]{siunitx}
\usepackage{braket}

\DeclareMathOperator{\sinc}{sinc}

\DeclareMathOperator{\sech}{sech}

\begin{document}


\title{Ultrafast second-order nonlinear photonics---from classical physics to non-Gaussian quantum dynamics}

\author{Marc~Jankowski\authormark{1,2,*,$\dagger$}, Ryotatsu~Yanagimoto\authormark{1,2,3,$\dagger$}, Edwin~Ng\authormark{1,2}, Ryan~Hamerly\authormark{2,4}, Timothy~P.~McKenna\authormark{1,2}, Hideo~Mabuchi\authormark{1}, M.~M.~Fejer\authormark{1}}

\address{\authormark{1}{Edward L. Ginzton Laboratory, Stanford University, Stanford, California, USA}\\
\authormark{2}NTT Research Inc. Physics and Informatics Labs, 940 Stewart Drive, Sunnyvale, California, USA\\
\authormark{3}{School of Applied and Engineering Physics, Cornell University, Ithaca, New York 14853, USA}\\
\authormark{4}{Research Laboratory of Electronics, MIT, 50 Vassar Street, Cambridge, MA 02139, USA}\\
\authormark{$\dagger$}{These authors contributed equally to this work}}

\email{\authormark{*}marc.jankowski@ntt-research.com} 


\begin{abstract}
    Photonic integrated circuits with second-order ($\chi^{(2)}$) nonlinearities are rapidly scaling to remarkably low powers. At this time, state-of-the-art devices achieve saturated nonlinear interactions with thousands of photons when driven by continuous-wave lasers, and further reductions in these energy requirements enabled by the use of ultrafast pulses may soon push nonlinear optics into the realm of single-photon nonlinearities. This tutorial reviews these recent developments in ultrafast nonlinear photonics, discusses design strategies for realizing few-photon nonlinear interactions, and presents a unified treatment of ultrafast quantum nonlinear optics using a framework that smoothly interpolates from classical behaviors to the few-photon scale. These emerging platforms for quantum optics fundamentally differ from typical realizations in cavity quantum electrodynamics due to the large number of coupled optical modes. Classically, multimode behaviors have been well studied in nonlinear optics, with famous examples including soliton formation and supercontinuum generation. In contrast, multimode quantum systems exhibit a far greater variety of behaviors, and yet closed-form solutions are even sparser than their classical counterparts. In developing a framework for ultrafast quantum optics, we will identify what behaviors carry over from classical to quantum devices, what intuition must be abandoned, and what new opportunities exist at the intersection of ultrafast and quantum nonlinear optics. While this article focuses on establishing connections between the classical and quantum behaviors of devices with $\chi^{(2)}$ nonlinearities, the frameworks developed here are general and are readily extended to the description of dynamical processes based on third-order $\chi^{(3)}$ nonlinearities.
\end{abstract}


\section{Introduction}

The development of tightly confining photonic integrated circuits with large second-order ($\chi^{(2)}$) nonlinearities is pushing nonlinear optics to radically new scales. The past few years have seen demonstrations of devices that achieve saturated nonlinear interactions with thousands of photons~\cite{lu2020toward,zhao2022ingap}, and recent theoretical proposals suggest that single-photon nonlinear interactions will soon be realized~\cite{Yanagimoto2022_temporal}. On the surface, the development of single-photon nonlinear photonics seems poised to revolutionize quantum optics for practical reasons; in contrast with well-established approaches based on circuit quantum electrodynamics (QED)~\cite{Fink2008,Wallraff2004} and cavity QED~\cite{Nogues1999, Brune1996quantum, Gleyzes2007, Thompson1992, Kimble1998}, nonlinear photonics operate at room temperature and atmospheric pressure, can be scaled to densely-integrated multi-functional platforms, and readily interface with telecom components. We emphasize, however, that nonlinear optical devices are fundamentally different than their traditional cavity QED counterparts. The large number of coupled optical modes encountered in broadband devices make possible an extraordinary diversity of classical behaviors, with well-known examples including solitons~\cite{Lugiato1987, werner1993simulton, Grelu2012, trillo1996bright}, mode-locking~\cite{mollenauer1984soliton, siegman1986lasers, Haus1975, Haus2000}, and supercontinuum generation~\cite{Dudley2006}. As these devices are scaled down to quantum limits, we arrive at a realization that is as exciting as it is unsettling: there likely exist devices sitting on optical tables that already exhibit novel quantum effects, yet given the massive Hilbert space associated with multimode systems no frameworks exist for rigorously analyzing these non-classical behaviors.

The purpose of this tutorial is three-fold. First, we provide an overview of recently developed theoretical tools needed to model the behavior of broadband nonlinear devices in the classical, semi-classical, and quantum limits. We note however, that this field is rapidly evolving, so we cannot provide a comprehensive overview of the many novel dynamical regimes that have already been explored in the literature. Instead, to build intuition we focus on a number of archetypal examples where closed form solutions exist. Particular focus will be given to behaviors shared across all of these regimes, and we make clear which classical intuition must be abandoned at the few-photon scale. The second goal of this tutorial is to provide a desk reference for quantum engineers who aim to build devices based on highly-nonlinear photonics. This tutorial will rely on the quantum-to-classical correspondence to establish rules for quickly calculating quantum parameters, such as the coupling rate in the quantum mechanical Hamiltonian, from empirically measured device behaviors, such as the normalized efficiency for second-harmonic generation. We emphasize here that these techniques are general, and can be readily extended to behaviors that are beyond the scope of this article such as third-order nonlinearities. With these techniques in mind we will review some recent experimental demonstrations, and wherever possible, provide a guide for comparing the performance of devices across different material systems. A recurring theme throughout this tutorial is that the relevant figures of merit and their scaling laws depend strongly on the particular process being studied, with different nonlinear processes (\textit{i.e.} pulsed quantum devices versus classical continuous-wave devices) favoring rather different material systems. The final goal of this tutorial is to provide a Rosetta stone that links the formalisms found in classical ultrafast optics with the frameworks of Gaussian quantum optics and cavity QED. In many cases, ideas that form the basis of commonly-held intuition in one context can appear rather non-intuitive in others. For this reason, we believe that having a common vocabulary to weave together insights from each of these communities will spur more rapid innovation in the field.

\subsection{State of the field}

\begin{figure}[t]
    \centering
    \includegraphics[width=\columnwidth]{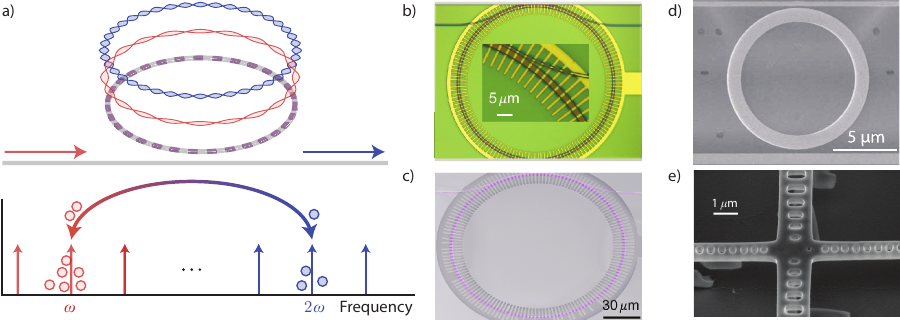}
    \caption{a) Continuous-wave (single-mode) second-order nonlinear interactions in resonators couple modes at fundamental ($\omega$) and second-harmonic ($2\omega$) frequencies. In the limit of strong coupling, cycles of up- and down-conversion manifest as Rabi oscillations. Ring resonators based on b,c) periodically-poled thin-film lithium niobate (TFLN)~\cite{lu2020toward} and d) InGaP~\cite{zhao2022ingap} have pushed towards the few-photon limit by balancing the trade-off between bending loss and nonlinear coupling. e) Alternative realizations based on GaAs and GaP photonic crystal cavities achieve the smallest mode volumes~\cite{rivoire2011multiply}, but to date have had prohibitively large losses. Figures (b) and (c) are adapted with permission from \cite{lu2020toward} Copyright 2020 Optica Publishing Group. Figure (d) is adapted with permission from~\cite{zhao2022ingap} Copyright 2022 Optica Publishing Group. Figure (e) is adapted with permission from~\cite{rivoire2011multiply} Copyright 2011 Optica Publishing Group.}
    \label{fig:1_ResonatorExamples}
\end{figure}

Rapid progress towards single-photon optical nonlinearities has been driven by the development of photonic integrated circuits that combine wavelength-scale confinement with large second-order ($\chi^{(2)}$) nonlinearities and low loss. At this time, the devices operating the closest to quantum scales, as shown in Fig.~\ref{fig:1_ResonatorExamples}, share a strong resemblance to many other integrated photonic platforms for cavity QED. In these realizations~\cite{luo2017chip,bruch2019opo,lu2019pplnring,lu2020toward,lu2021ultralow,McKenna2022} an optical microcavity confines two resonant waves, a fundamental with optical frequency $\omega$ and a second-harmonic with optical frequency $2\omega$, which are coupled by the second-order optical nonlinearity. Pairs of fundamental photons up-convert to second-harmonic photons through second-harmonic generation (SHG), and second-harmonic photons down-convert back to fundamental by optical parametric amplification (OPA). In the limit of strong coupling, where the field associated with a single intracavity photon is sufficiently strong to drive these processes into saturation, this system will execute Rabi oscillations between a biphoton of fundamental $\ket{20}$ and a single-photon Fock state of second harmonic $\ket{01}$.

At this time of writing, state-of-the-art devices based on continuous-wave SHG and OPA achieve saturation with thousands of photons~\cite{lu2020toward,zhao2022ingap}, and the push towards lower photon number has been limited by practical trade-offs. As an example, the ring resonators shown in Fig.~1 can achieve larger nonlinearities simply by fabricating tighter bends. However, with a decreasing radius of curvature comes increasing bending loss, and the trade-off between these two effects sets an optimal resonator geometry. Recent demonstrations using optimized ring geometries in indium phosphide~\cite{zhao2022ingap} and periodically-poled thin-film lithium niobate~\cite{lu2020toward} have led this effort, but are still two orders-of-magnitude (in nonlinearity) from the threshold of strong coupling. A similar set of limitations exist for photonic crystals. Point defect cavities can realize mode volumes on the order of a cubic wavelength, which result in the largest possible nonlinearities. However, in practice these devices have not yet been able to simultaneously resonate both harmonics with low loss, which cancels the benefits gained from tight confinement.

\begin{figure}[t]
    \centering
    \includegraphics[width=\columnwidth]{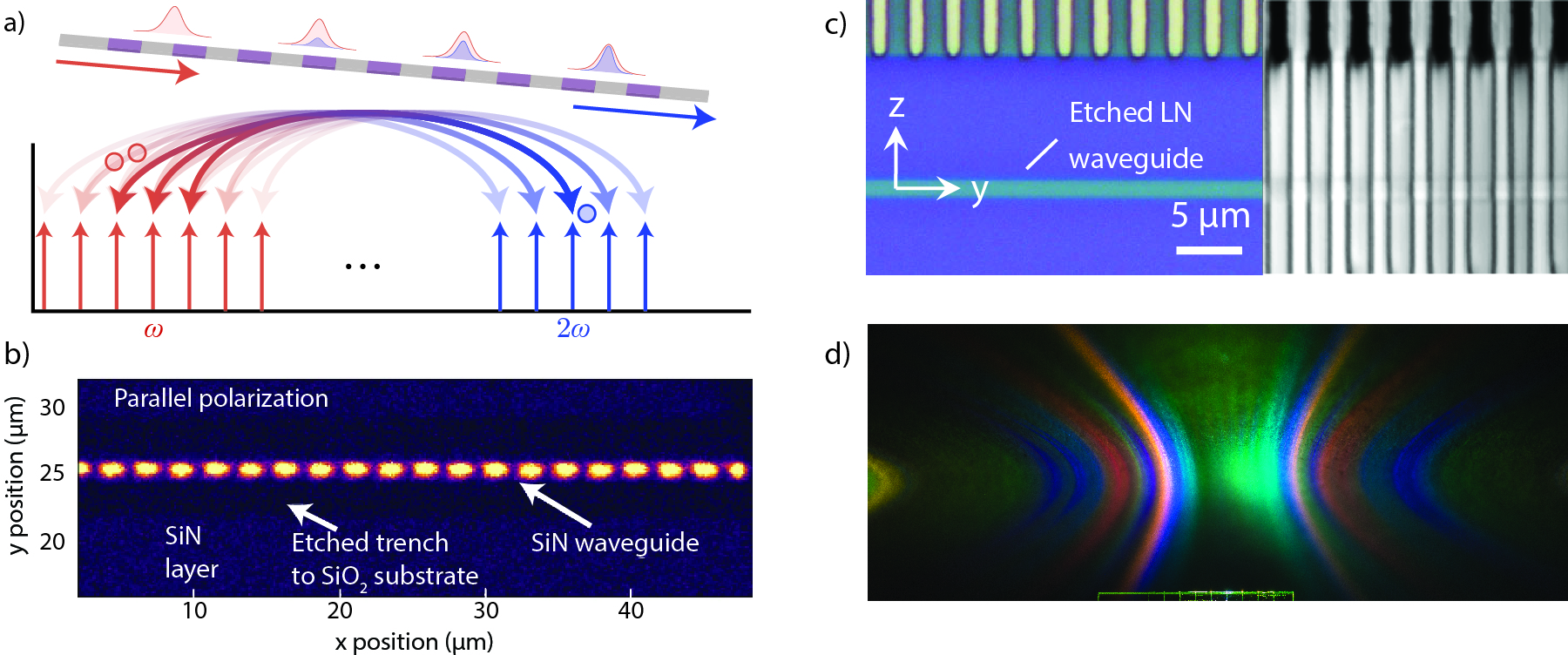}
    \caption{a) In nonlinear waveguides driven by femtosecond pulses, the pulse envelopes exchange photons during propagation. In these systems the pulse duration, combined with the transverse mode area, can play the role of the mode volume, and the interaction length plays the role of cavity lifetime. b,c) Examples of platforms used to realize ultrafast nonlinear optics: silicon nitride~\cite{Hickstein2019} and TFLN~\cite{zhao2020high}, respectively. d) Photo of a TFLN waveguide producing a broadband supercontinuum. In this system, the spectral broadening occurs due to distortions of the pulse envelopes that can occur in the highly saturated limit~\cite{Jankowski2020,Jankowski2021SCG}. Figure (b) is adapted with permission from~\cite{Hickstein2019}, Copyright 2019 Springer Nature. Figure (c) is adapted with permission from~\cite{zhao2020high}, Copyright 2020 Authors. Figure (d) is adapted with permission from~\cite{Jankowski2020}, Copyright 2020 Optica Publishing Group.}
    \label{fig:2_WaveguideExamples}
\end{figure}

A separate approach based on ultrafast interactions in dispersion-engineered $\chi^{(2)}$ waveguides has developed alongside efforts in resonant CW devices, as shown in Fig.~2. In these systems, two pulses exchange photons while propagating in a traveling-wave configuration. In an intuitive sense, which will be formalized later, space and time are interchanged in these devices when compared to resonators. Here longitudinal confinement comes from the short duration of the optical pulse, rather than the circumference of a ring resonator, and the interaction length is set by the distance over which the pulses overlap, rather than the lifetime of the resonator. In this picture, the interaction length is predominantly limited by the difference in group velocity between the two harmonics. We note, however, that pulsed interactions contain many modes, here represented by many frequencies, which lead to a far greater variety of behaviors. An example is shown in Fig.~\ref{fig:2_WaveguideExamples}(d). In this case, small phase drifts between the interacting pulses cause the envelopes to break up in the highly saturated limit, which leads to spectral broadening. Later, in the context of quantum nonlinear optics, we will see that care must be taken to harness these complicated multi-mode effects.

A key development in the field of $\chi^{(2)}$ nonlinear photonics has been the use of waveguide dispersion to eliminate conventional limitations to the interaction length such as temporal walk-off and dispersive pulse spreading~\cite{Hickstein2019,Singh2020,Jankowski2020}. This approach relies on the relationship between the shape of a waveguide and the dispersion relations of the guided modes. By engineering the geometry of the waveguide to eliminate these dispersive effects, the interaction length becomes limited only by the physical length of the waveguide, or in extreme limits, by the loss length of the waveguide. State-of-the-art devices that combine geometric dispersion engineering with ultrafast pump pulses have achieved saturated optical parametric fluorescence, commonly referred to as optical parametric generation (OPG), with millions of photons in single-pass millimeter-scale devices~\cite{Jankowski2022}. As these interaction lengths are scaled longer, the required photon number decreases quadratically, in principle enabling few-photon nonlinear interactions when the interaction length approaches the loss length.

The clearest route towards few-photon nonlinearities is to combine these two approaches. In the traveling-wave picture, embedding a femtosecond pulse inside a resonator enables many passes through the nonlinear section, effectively realizing an interaction length comparable to the loss length of the resonator. In the resonator picture, interference between many phase-coherent cavity modes creates a localized envelope with an effective mode volume determined by the spatial extent of this multimode field, which can be much smaller than that of any of the constituent single-frequency modes. A number of theoretical proposals~\cite{Yanagimoto2022_temporal,Jankowski2021-review} have suggested that this hybrid approach can combine the best of both worlds: effective mode volumes comparable to a cubic wavelength and lifetimes approaching limits set by the intrinsic loss of the material. To date no devices have yet been demonstrated based on this approach due numerous technical hurdles that must be resolved; efforts to address these issues are ongoing in several research groups.

\subsection{Themes of this article}

Below are brief summaries of four central themes covered in this tutorial. Each theme represents a common thread throughout the rest of the sections, and together, these summaries serve as a precis of the entire article.

\subsubsection{What does it mean for nonlinear optics to be quantum?}

In the classical description of light, each mode of the electromagnetic field (\textit{e.g.} each independent frequency, or wave confined to a resonator) is assumed to have a definite amplitude and is therefore fully described by a single complex number, $\alpha$. However in quantum mechanics, because quadrature observables do not commute, there must always be an uncertainty of each field mode amplitude around its (complex) expectation value. As a result, describing the quantum state of an optical mode requires not just a single complex number but rather a field operator, $\hat{a}$. In accordance with this intuition, we will find throughout this article that an insightful way to partition the optical field is
\begin{align}
    \hat{a}=\alpha+\delta\hat{a},
\end{align}
with no loss of generality. Here, the first term $\alpha$ is the classical mean-field amplitude, and the second term $\delta\hat{a}$ contains quantum corrections to this mean field, which can be interpreted as quantum fluctuations around the mean. The quantum state corresponding to a classical field (\textit{i.e.} the light output from a laser) is commonly taken to be a coherent state, $\ket{\alpha}$. In phase space, the coherent state is given by a symmetric Gaussian distribution centered on the mean amplitude $\alpha$ with half-a-photon's worth of quantum uncertainty in the amplitude and phase quadratures.

In classical nonlinear optics, the equations of motion, hereafter referred to as the coupled-wave equations, are derived from Maxwell’s equations to describe coupling between optical modes by a nonlinear polarization. A natural approach to derive the classical equations of motion from quantum mechanics is to assume that each interacting mode corresponds to a coherent state. Simply ignoring any role played by the fluctuations reduces the dynamics of the wavefunction to the evolution of mean-field amplitudes given by the coupled-wave equations. It turns out that this mean-field approach is an excellent approximation for many realistic experiments in optics due to a separation of scales argument. Until recently, conventional devices in nonlinear optics have typically operated with billions of photons. In contrast, the relatively weak quantum fluctuations rarely influence the dynamics of these mean fields, except in extreme limits involving saturated optical parametric fluorescence. Therefore, the influence of quantum fluctuations is rarely present in classical nonlinear devices unless care is taken to study the fluctuations themselves.

\begin{figure}
    \centering
    \includegraphics[width=0.6\textwidth]{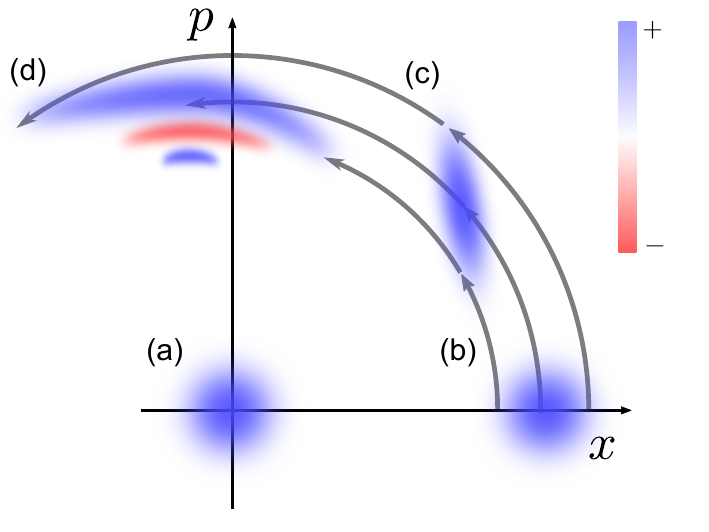}
    \caption{Illustration of the phase-space (Wigner function) evolution under the influence of a Kerr nonlinearity. (a) A vacuum state has symmetric quantum fluctuations around the origin. (b) A coherent state is generated by displacing a vacuum state in the phase space. (c) For weak nonlinearities, or equivalently a short interaction time, nonlinear deformation of the phase-space distribution can be approximated as linear squeezing and rotation, which keeps the state Gaussian. (d) After a long interaction time, the phase-space distribution becomes non-Gaussian and develops more exotic quantum features, such as Wigner function negativities. }
    \label{fig:introduction-phase-space}
\end{figure}

In reality, the same nonlinear interactions that cause the mean field to change, such as phase-sensitive amplification or self-phase modulation, similarly cause the quantum noise associated with each interacting mode to evolve. Figure~\ref{fig:introduction-phase-space} illustrates how nontrivial quantum fluctuations develop within natural nonlinear-optical dynamics. The action of the nonlinearity, in this case a Kerr nonlinearity, affects each volume element of phase space, here indicated by arrows encircling the origin. As a result, the distribution of quantum fluctuations around the mean field value $\alpha$ becomes distorted, and forms features that are rather distinct from that of classical coherent states. These behaviors are not unique to the simple case shown here; all coherent dynamical processes found in nonlinear optics can lead to a nontrivial evolution of the underlying quantum fluctuations. In this sense, quantum nonlinear optics is not a disjoint topic of study from classical nonlinear optics, but rather a natural extension~\cite{Armen2006} of the same behaviors to a treatment of the quantum fluctuations. These emergent quantum features become increasingly complicated with larger nonlinearity, and become significant as a consequence of scaling towards strong coupling and lower photon numbers.

In the regime of weak nonlinearity (or equivalently, short interaction time), the dynamics of quantum fluctuations can be approximately linearized around the mean-field dynamics to the lowest order, and their effects on the mean field and the quantum fluctuations can be ignored. In this linearized treatment, the dynamics of the phase-space distribution can be treated using operators that map one Gaussian distribution to another, namely, displacements, linear stretches (squeezing), and rotations. This limit, commonly referred to as Gaussian quantum optics~\cite{Olivares2012,Weedbrook2012,Quesada2022}, has been sufficient to capture the quantum behavior (\textit{e.g.}, squeezing) of almost all conventional nonlinear optical devices to date~\cite{Triginer2020,Guidry2022,Kashiwazaki2020,Vahlbruch2016, Bao2021}.

When quantum fluctuations grow sufficiently strong to drive both the mean field and the quantum fluctuations themselves (\textit{e.g.} in OPG, where vacuum fluctuations are amplified to macroscopic intensities), the linearized approximations used for Gaussian quantum optics break down. In principle, such dynamics can exhibit beyond-Gaussian quantum phenomena, \textit{e.g.}, non-Gaussian features in phase-space (such as Wigner-function negativities~\cite{Walschaers2021}). In practice, however, such features are difficult to observe experimentally since the large degree of anti-squeezing needed to deplete the pump makes them highly sensitive to experimental imperfections such as loss and phase noise. Thus, to resolve non-Gaussian quantum features, the nonlinearity of a photonic device must be strong enough that a \emph{mesoscopic} number (\textit{i.e.}, hundreds or dozens) of photons can cause saturated nonlinear dynamics~\cite{Ryo2023Mesoscopic}.  

These observations imply that the recent progress in nonlinear photonics towards ultra-low energy scale is pushing experimental quantum optics out of the scope of the conventional Gaussian formalism and into a much broader Hilbert space where far richer phenomena may occur. The states that arise in this mesoscopic regime may resemble the ones found in other contexts, such as a Schr\"{o}dinger cat-like state, but their dynamics can be qualitatively distinct. First, the non-Gaussian quantum features coexist with strong Gaussian quantum features, which critically modify and enhance the non-Gaussian quantum dynamics. Second, ultrashort pulses necessary to realize strong nonlinearity naturally involve millions of modes, leading to vastly more complicated physics than traditional few-mode quantum optics.

Finally, as nonlinear photonics reaches a scale where the field associated with only one or two photons suffices to cause saturated nonlinear dynamics, the distinction among mean-field, Gaussian, and non-Gaussian features collapses. In this deep-quantum regime photons behave like discrete particles~\cite{Cantu2020,Firstenberg2013,Chang2014}, \textit{e.g.}, as in cavity-QED experiments in the strong coupling regime~\cite{Nogues1999, Brune1996quantum, Gleyzes2007, Thompson1992, Kimble1998}. Even in this limit, however, the involvement of a large number of modes can lead to counterintuitive phenomena from the perspective of traditional few-mode quantum optics~\cite{Shapiro2006}, and careful engineering is essential to harness these multimode dynamics.

\subsubsection{From single to multimode interactions}
A core recurring theme of this article is how to extend the closed form solutions found for single-mode problems into an intuitive understanding of complicated multimode systems. In each section, from classical to quantum optics, we will consider a number of representative dynamical processes, and within each case we will follow a structure of single-mode, few-mode, and fully multimode treatments. As an example, one commonly used archetype for classical nonlinear optics is second harmonic generation (SHG), where an input wave of frequency $\omega$ generates a second-harmonic with frequency $2\omega$. The single-mode limit corresponds to continuous-wave SHG, where both the fundamental and second-harmonic each have one interacting mode. We then introduce the formalism needed for a few-mode treatment (\textit{e.g.} three-wave interactions), by allowing multiple fundamental modes to interact with one second-harmonic mode. Finally, we generalize to broadband interactions where each interacting harmonic contains many participating modes.

In classical nonlinear optics, closed-form solutions for broadband interactions rarely exist. This task only becomes more difficult in the context of quantum nonlinear optics. We can see this using a simple complexity argument. For a classical $M$-mode system, we require $\mathcal{O}(M)$ parameters to completely describe the interacting fields. On the other hand, in the presence of nonlinear interactions the quantum fluctuations associated with each mode can develop correlations. In the simplest (unsaturated) case, these correlations take the form of a covariance matrix, which is characterized by $\mathcal{O}(M^2)$ parameters. In the saturated regime, higher-order correlations become important, which require $\mathcal{O}(M^n)$ parameters to capture $n$th-order correlations. Eventually, as we enter the deep quantum limit, these systems can exhibit complicated dynamics that require exponentially many degrees of freedom with increasing mode number. Therefore, to understand the multimode quantum dynamics of photons, it is essential that we either limit the number of modes needed to analyse the nonlinear dynamics or that we limit the structure of the underlying correlations.

To treat complicated multimode physics, we will build intuition by placing restrictions on the equations of motion that simplify the dynamics sufficiently to admit closed form solutions. In almost every case, these simplified models will enable us to establish a link back to the few-mode or even single-mode treatment of the same phenomenon. The three tools used throughout this paper are (i) linearized (or ``undepleted'') treatments, where one of the interacting waves can be treated as a constant of motion, (ii) quasi-static heuristics, where linear optical behaviors such as dispersive pulse spreading are neglected, and (iii) model reduction, where we restrict our dynamics to a relevant subspace of possible behaviors. In the undepleted limit, the frequency-domain equations of motion typically simplify to a linear dynamical system with families of frequency modes coupled by the nonlinear polarization. These systems can be solved using standard approaches to ordinary differential equations. Conversely, quasi-static heuristics are useful for treating complicated multimode interactions in the saturated limit, provided that dispersion can be neglected. In this limit, the time-domain equations of motion reduce to an independent single-mode model for each point in time. These systems can therefore be fully solved using the solutions from the single-mode limit. Model reduction techniques rely on assumptions either about the structure of the mean field, \textit{e.g.} that the field can be described using a relatively small number of pulsed ``supermodes'', or about the structure of the correlations between modes. As an example, model reduction can build upon the linearized equations of motion by identifying a subset of generalized modes (often pulses, rather than independent frequencies), that dominate the underlying dynamics. Working within this mode basis renders the unsaturated dynamics trivial, and saturated behavior can be modeled as coupling between these modes. We emphasize here that all three of these approaches yield distinct insights into systems that otherwise have no closed-form solutions.


\subsubsection{A design workflow for the classical to quantum transition}

In the spirit of approaching the quantum physics from the classical side of nonlinear optics, in this article we offer illustrations of a ``workflow'' that allows us to (i) obtain experimentally meaningful parameters for quantum models and simulations, (ii) predict new mechanisms and behaviors arising from quantum effects, and (iii) formulate design rules (\textit{i.e.}, in terms of device parameters) for observing such new phenomena. 

{For the first point, we take a \emph{phenomenological} approach that leverages the classical-quantum correspondence to construct quantum models from classical ones, rather than the usual, more-involved approach of performing canonical second quantization from first principles~\cite{Hillery1984,Drummond1990,sipe2009photons,Drummond2014, Huttner1992,Raymer2020}. Assuming we start with wave equations rigorously derived via classical field theory to be consistent with Maxwell's equations, we show how a quantum model for the same dynamics can be obtained essentially by inspection, while still avoiding the common pitfalls involved in quantizing nonlinear optical systems~\cite{Hillery1984,sipe2009photons,Quesada2017,Quesada2022}. That is, once a correct classical field theory for a device is established, the form of the quantum Hamiltonian and the quantum equations of motion are fixed by the mean-field model. Crucially, this means that parameters of the quantum model, such as interaction rates, can be inferred directly from the calculated or measured parameters of the classical model. By taking this phenomenological approach, we can ensure that the quantum model recovers Maxwell's equation in the mean-field limit while greatly reducing the activation barrier to working with quantum models, at least for those practitioners who understand intimately and rigorously the classical behavior of a device. We note here that while this approach cannot not specify which fields have been quantized \emph{inside} the nonlinear medium, it does recover the statistics of the photons generated by the nonlinear dynamics.}

That said, the Heisenberg equations of motion (or the equivalent Schr\"{o}dinger equation) for the quantum model are often intractable to directly simulate, especially in the multimode setting. As a result, the crucial next step in this workflow is to apply model reduction methods to reformulate the quantum model into a more tractable form. This part is often the ``art'' in the engineering of such devices, and can, in principle, leverage any and all tools developed throughout the history of quantum mechanics. In this article, we focus on a few specific approaches that have been found to work particularly well for multimode bosonic systems (\textit{i.e.}, quantum pulse propagation), such as the aforementioned use of supermodes to truncate the mode basis, or the use of matrix product state to exploit the localized correlation structures among photons. In this context, the use of classical-quantum hybrid techniques has been especially fruitful in elucidating the physics in the classical-to-quantum transition.

Finally, it is worth emphasizing that while much of the new phenomenology found in the quantum regime can be very rich and complex, it is nevertheless possible to explicate concise \emph{design rules} for observing them, \textit{e.g.}, figures of merit and scaling laws for characteristic behavior as a function of device parameters. In many cases, similar expressions and figures of merit known for classical devices turn out to govern new quantum phenomena as well, perhaps with a slight change in scaling exponents. Thus, for the device engineer, there is a sense in which ``going quantum'' is not, figuratively speaking, a ``quantum leap'' conceptually; it merely requires utilizing a natural generalization of the classical model, some well-informed model reduction, and an eye for isolating and analyzing new engineerable mechanisms.




\subsubsection{Gaussian quantum noise is a gateway to quantum nonlinear optics}
Generally, a multimode Gaussian quantum state can be completely characterized by its mean and covariance, \textit{i.e.}, second-order correlations. As we go beyond Gaussian quantum optics and enter the mesoscopic regime higher-order correlations develop, forming non-Gaussian quantum features. Generally speaking, there are two approaches to understanding the onset of non-Gaussian quantum physics: (i) to build up the quantum state photon-by-photon to span a larger Fock space, or (ii) to build new frameworks on top of Gaussian quantum optics that smoothly interpolate down from classical and semiclassical regimes towards the few-photon scale. The former approach is difficult in broadband systems due to the exponential growth of Fock space with mode number. On the other hand, the latter approach is much more natural and often gives more insight into the behavior of these systems, including how the quantum-to-classical transition occurs and conditions for experimentally observing non-Gaussianity.

Building on the discussion above in which we partition the field operator into a mean field and its quantum fluctuations, we treat the formation of non-Gaussian features by further partitioning the physical features of the dynamics into mean-field, Gaussian, and non-Gaussian phenomena. More specifically, we show that we can factor out Gaussian dynamics in the form of an interaction frame, hereafter referred to as the Gaussian interaction frame~\cite{Yanagimoto2022-non-Gaussian,Tezak2017}. Using this frame, we can keep track of the residual non-Gaussian quantum features as a wavefunction ``riding'' on top of the mean-field and Gaussian features. The form of the effective Hamiltonian in the Gaussian interaction frame furthermore provides insights into how non-Gaussian features arise from semiclassical dynamics. In many cases the Gaussian evolution of the system, such as the formation of squeezing, comes with useful enhancements to the effective interaction rates of the residual non-Gaussian quantum state.

This scaffolding provided by Gaussian quantum optics is particularly useful when studying multimode nonlinear optical systems. Unlike typical cavity QED systems, where only a handful of modes are involved, it is not unusual to find nonlinear optical systems with millions of modes involved. To capture non-Gaussian quantum correlations among such an immense number of modes, we would na\"ively need an exponentially large number of parameters, which quickly becomes intractable. The Gaussian part of the hybrid state in the interaction frame can provide critical information on which specific (super)modes are evolving most rapidly and are most likely to generate non-Gaussian features. This predictive insight essentially allows us to restrict our Hilbert space to a small subset of relevant supermodes. While obviously indispensable for performing tractable numerical simulation, this analysis also deepens our understanding of how multimode phenomena take shape in quantum nonlinear optics.

Finally, in the deep-quantum regime of strong photon-photon interactions, the above hierarchy among mean-field, Gaussian, and non-Gaussian features collapses, and the most convenient approach returns to the usual cavity-QED technique of building up the Fock space photon-by-photon, which is more suitable in this regime of few-photon physics. In particular, we have found that modeling techniques from the field of many-body physics are effective in capturing the microscopic photon dynamics~\cite{Manzoni2017,Yanagimoto2021_mps,Lubasch2018}. In this way, we can obtain a direct chain of models and model reductions for numerical simulation and conceptual analysis that can fully span the classical, semiclassical, and quantum regimes of nonlinear optics.

\subsection{How to use this tutorial}
\begin{figure}[h]
    \centering
    \includegraphics[width=\textwidth]{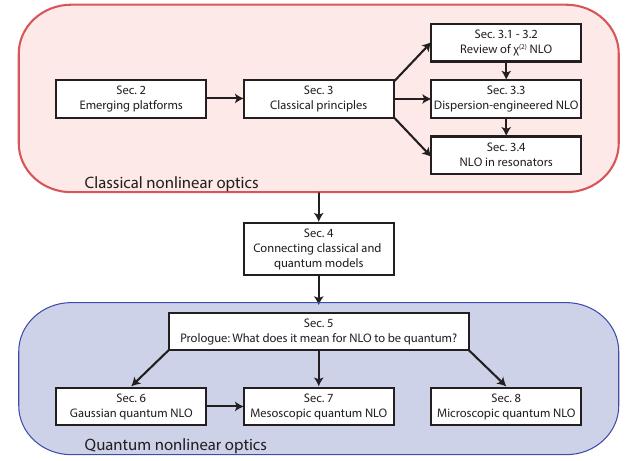}
    \caption{Structure of this tutorial. These headings are meant to provide a concise overview of the sections and are not necessarily the same as their actual titles. Some contents in the quantum NLO sections are adapted from Ref.~\cite{Yanagimoto2023-thesis}.}
    \label{fig:structure}
\end{figure}

This article is directed towards newcomers to the field who may have encountered nonlinear optics or quantum optics, but are not yet sure how these ideas fit together. Given the above themes, this tutorial proceeds with a scaffolded approach that builds up from classical to quantum nonlinear optics in seven sections (see Fig.~\ref{fig:structure}). While the intended audience will benefit the most from following this structure, readers more familiar with earlier topics, such as ultrafast optics or Gaussian quantum optics, should be able skip ahead to sections of interest where they may dive into new phenomena. We note, however, that some of the ideas used throughout this article are less commonly encountered in the literature. To facilitate readers interested in breaking this article up into separate modules, we briefly discuss how these sections fit together and what topics should be at least skimmed before diving into later parts of this article.


We begin with a brief overview of emerging platforms for nonlinear photonics in Sec.~\ref{sec:platforms}. This section is targeted towards experimentalists who are interested in understanding broadly what different approaches are being explored, and how to compare them. We discuss at a high level what considerations should go into selecting a platform for realizing few-photon nonlinear interactions, and compare recent progress across many promising material systems.

We then review the relevant aspects of classical nonlinear optics in Sec.~\ref{sec:classical_NLO}. This section introduces many of the tools used throughout the tutorial. These concepts include the coupled-wave equations, their closed-form solutions, and how to extract meaningful figures of merit from these solutions. This presentation differs from most textbook treatments by focusing on the bandwidths associated with nonlinear interactions, and their relationship to the energy requirements of devices driven by short pulses. Model reduction will first appear in the context of analysing unsaturated optical parametric amplification (OPA), and will be used throughout the subsection on fully-static ultrafast nonlinear optics. The end of this section introduces discrete maps, which are a useful framework for building up from the coupled-wave equations to nonlinear systems that comprise several parts, such as a resonator. The latter portion of this section (Sec.~\ref{sec:time-prop}-\ref{sec:prospects}) develops time-propagating equations of motion, which are useful for connecting the theory of classical nonlinear optics to the quantum theory. Our treatment of classical NLO concludes with Sec.~\ref{sec:prospects}, which discusses the prospects for realizing few-photon nonlinear interactions. Readers more familiar with ultrafast nonlinear optics who are curious about what new classical regimes exist in nonlinear photonics should consider reading later sections devoted to both quasi-static and fully-static nonlinear optics. These behaviors can only be accessed using the dispersion engineering available in tightly-confining nonlinear waveguides, and are less commonly encountered in the literature. Readers interested in developing few-photon nonlinear devices will greatly benefit from the material in Sec.~\ref{sec:resonators}.

We move from classical to quantum nonlinear optics in Sec.~\ref{sec:Rosetta_stone} by developing a set of rules for quickly converting the classical equations of motion into a quantum mechanical Hamiltonian. This section is crucial for anyone who wants to understand the behaviors of real devices. The treatment presented here builds on the presentation of Secs.~\ref{sec:time-prop}-\ref{sec:photon-normalized-units}, which establish classical time-propagating equations of motion in photon-number units. Based on the form of these classical equations of motion, we develop a phenomenological Hamiltonian that allows for a full quantum treatment of ultrafast nonlinear optics, and recovers the classical equations of motion in the mean-field limit. This approach gives the correct form of the Hamiltonian as calculated using canonical quantization, and enables straightforward calculations of quantum parameters by matching the parameters of the quantum model to the easily calculated (and measured) parameters of the classical model.

Having established the form of the quantum Hamiltonian for multimode $\chi^{(2)}$ devices, Sec.~\ref{sec:prologue} explains the general approach used throughout this tutorial to solve the equations of motion. The heart of this section is the introduction of the Gaussian interaction frame (GIF), a framework that separates out classical, semi-classical, and non-Gaussian quantum features in a hierarchical manner~\cite{Yanagimoto2022-non-Gaussian}. Through the lens of GIF, we can intuitively understand how the number of photons input to a nonlinear system makes a qualitative difference in how quantum features emerge, based on which we identify three regimes of quantum nonlinear optics: the macroscopic, mesoscopic, and microscopic regimes. The remaining sections will study these three regimes based on the formalism introduced here.


Gaussian quantum optics is reviewed in section~\ref{sec:semi-classical_NLO}. This section largely builds upon the linearized treatment of multimode OPA, and will be familiar to readers with a quantum optics background. For continuous-wave pumping, we will establish the link between OPA bandwidth and the correlation length of the downconverted photons. Similarly, in the pulsed case, a quasi-static treatment will establish a link between the generated power envelope and the correlation bandwidth of the generated spectrum. These localized correlation structures will later be used to motivate the use of matrix-product states to simulate broadband few-photon interactions. We close this section with a presentation of the Gaussian split-step Fourier (GSSF) method, which generalizes numerical methods commonly encountered in classical nonlinear optics to multimode Gaussian dynamics. By assuming that the underlying states remain Gaussian (\textit{i.e.} by neglecting any non-Gaussian contributions to the field correlations), this approach can efficiently simulate semi-classical dynamics (such as the formation of squeezing and Gaussian entanglement) with an arbitrary degree of saturation. The GSSF naturally elucidates the leading-order quantum behavior of otherwise intractable systems first encountered in classical nonlinear optics, such as supercontinuum generation and optical parametric generation.

The latter portion of this tutorial is focused on new phenomenology. The saturated nonlinear dynamics encountered at the end of section~\ref{sec:semi-classical_NLO} almost always coincide with the formation of non-Gaussian features, yet many saturated systems contain an extraordinarily large number of photons. In principle, behaviors such as those studied with the GSSF should contain some non-Gaussian features hidden under complicated Gaussian features. Section~\ref{sec:mesoscopic-quantum-nonlinear-optics} makes use of Gaussian interaction frames and model reduction to analyse the formation of such mesoscale features. This combination of techniques answers the question of how non-Gaussian features form in many-photon systems and gives an intuitive understanding of what conditions may soon lead to their observation. An example where the Gaussian interaction frame is particularly well suited is the case of saturated optical parametric fluorescence, where amplified quantum fluctuations become sufficiently bright to deplete the classical input pump. Revisiting these experiments through the lens of the Gaussian interaction frame reveals what one may intuitively expect, that pump depletion coincides with the formation of non-Gaussian features. A natural question, then, is why signatures of these behaviors have not yet been observed in experiments. The Gaussian interaction frame provides a simple, heuristic answer: Since it takes an extraordinary amount of parametric gain to trigger these non-Gaussian quantum dynamics, the resulting states become extremely squeezed, which makes them highly sensitive to any experimental imperfections such as loss and phase noise. As a result, the optical state devolves into an incoherent mixture of Gaussian quantum states.

Section~\ref{sec:quantum_NLO} is the final technical section of this tutorial and treats the few-photon regime. In contrast with textbook treatments of cavity QED, where single-mode dynamics are analysed in a Fock space containing a handful of photons, here we consider the limit where Fock numbers are limited to one or two, and the mode number is allowed to be arbitrarily large. We first contrast the case of OPA against the classical and semi-classical limits. This representative example provides readers with a preview of the qualitatively new dynamics one might expect to see in this limit (Rabi-like oscillations), but also illustrates that much of the classical intuition must be abandoned. Many common notions, such as the use of short pulses to enhance the strength of an interaction, simply do not carry over to this analysis. This section then proceeds to give an overview of model reduction techniques that can be used to analyze more general quantum systems in the few-photon limit: matrix product states, which facilitate efficient numerical simulations of multimode quantum systems by assuming a localized correlation structure, and a pulsed supermode basis. These reduced models are then applied to the case of quantum gate operation using pulsed two-photon SHG and reveal that pulsed enhancements to the coupling rate \emph{can} exist in quantum-nonlinear devices, but that multimode couplings often act as a decoherence channel that counteracts the benefits gained by these enhancements. This section concludes by revisiting fully-static nonlinear optics in the context of few-photon nonlinearities. In this case, the use of fully-static dynamics eliminates these multimode decoherence effects while retaining the enhancement of the coupling rate. Plugging in realistic experimental parameters suggests that these techniques can be used to realize optical CNOT gates.

\section{Emerging platforms for nonlinear photonics}\label{sec:platforms}

Many promising platforms for $\chi^{(2)}$ nonlinear photonics have proliferated in the last decade, each exhibiting a rather different set of features, and a natural question for newcomers to the field is how to select the appropriate platform for their desired application. In the context of quantum nonlinear optics, this question often boils down to how to select a platform that can realize few-photon nonlinearities. The answer turns out to depend quite strongly on the approach being taken, with resonant continuous-wave devices favoring rather different parameters than ultrafast devices. Similarly, constraining devices to operate at one wavelength, such as a 1560 nm for the fundamental harmonic, leads to a different set of considerations than leaving the wavelength a free parameter.  In this section we compare these emerging photonics platforms on two separate axes: (i) the maturity of these platforms at this time of writing, and (ii) relevant material properties, such as the nonlinear susceptibility. To better illustrate the trade-offs discussed above, we will compare the theoretical power requirements of traveling-wave SHG for these material systems. The design considerations for few-photon nonlinear devices will be revisited in Sec.~\ref{sec:prospects}, once the relevant figures of merit for single-mode and pulsed interactions have been defined.

In the context of second-order nonlinear photonics, we define a mature platform as having combined three features: i) wavelength-scale confinement, ii) low propagation loss, and iii) quasi-phasematched nonlinear interactions.  While the pursuit of low propagation loss and wavelength-scale confinement are common to nearly every emerging photonics platform, the development of quasi-phasematching (QPM) is equally important for realizing devices that take advantage of ultrafast pulses. Quasi-phasematched devices, where the $\chi^{(2)}$ coefficient is spatially patterned to correct for any phase-drifts between the interacting waves, have long been used to realize efficient nonlinear interactions~\cite{Armstrong1962,Franken1963,Fejer1992,Hum2007}. However, the recent development of quasi-phasematched nanophotonics~\cite{rao2016second, wang2018ultrahigh, rao2019actively, zhao2020high} have enabled an extraordinary degree of design freedom, in addition to much stronger nonlinear couplings. In the absence of QPM, phase-matched nonlinear interactions (\textit{e.g.} $n_{2\omega} = n_\omega$ for second harmonic generation) can be achieved in nonlinear photonics with a suitable choice of waveguide geometry. However, this approach, commonly referred to as modal phase-matching, often restricts the design to a small window of the parameter space and comes with reduced mode overlaps or the use of smaller off-diagonal $\chi^{(2)}$ elements, which results in weaker nonlinear couplings. In contrast, in ferroelectric or orientation-patterned materials the $\chi^{(2)}$ coefficient can almost always be patterned to quasi-phasematch any nonlinear process irrespective of waveguide geometry. This freedom allows the waveguide geometry to be used to engineer the group velocities and higher dispersion orders of the interacting waves, which is crucial for controlling the behavior of femtosecond pulses.

\begin{table}[h]
    \centering
    \begin{tabular}{c|c|c|c|c}
         Material & QPM Technique & QPM thin films & Low-loss photonics & Low loss + QPM\\
         \hline\hline
         LiNbO$_3$ & Periodic Poling & $\bullet$ & $\bullet$ & $\bullet$\\
         Silicon & EFISH & $\bullet$ & $\bullet$ & $\bullet$\\
         SiN$_x$ & EFISH & $\bullet$ & $\bullet$ & $\bullet$\\
         InGaP & Orientation patterning & $\bullet$ & $\bullet$ &  \\
         AlGaAs & Orientation patterning & $\bullet$ & $\bullet$ &  \\
         AlN & Orientation patterning & $\bullet$ & $\bullet$ &  \\
         GaN & Orientation patterning & $\bullet$ & $\bullet$ &  \\
         SiC & Orientation patterning & & $\bullet$ & \\
         LiTaO$_3$ & Periodic poling & & $\bullet$ & \\
         ZnSe & Orientation patterning & $\bullet$ & &  \\
         AlScN & Periodic poling & $\bullet$ & & \\
         AlBN & Periodic poling & $\bullet$ & & \\
         NbOCl$_2$ & Periodic poling & & & \\
    \end{tabular}
    \caption{Maturity of emerging photonics platforms, as defined by having realized wavelength-scale confinement with both low propagation loss and quasi-phasematched nonlinear interactions.}
    \label{tab:material_development}
\end{table}

With the above considerations in mind, Table~\ref{tab:material_development} summarizes the status of many promising platforms for nonlinear photonics. At this time, the most developed platforms are ferroelectrics, such as LiNbO$_3$, and materials with field-induced nonlinearities, such as silicon~\cite{Timurdogan2017,Singh2020,heydari2023degenerate} and silicon nitride~\cite{Billat2017,Porcel2017,Hickstein2019,XLu2020}, due to their ease of use. In these systems the $\chi^{(2)}$ can be patterned at any fabrication step; wafers can be diced into small chips and processed into linear or nonlinear photonics as desired, which allows for each component to be refined independently. In addition to these platforms, there has been substantial recent progress in the development of thin-film III-V semiconductors, including AlGaAs~\cite{chang2018heterogeneously,may2019second,chang2019low,chiles2019multifunctional,chang2019strong,stanton2020efficient,chang2020ultra,xie2020ultrahigh,mahmudlu2021algaas,may2021supercontinuum,castro2022expanding,wu2023algaas}, GaP~\cite{Rivoire2011second, Wilson2019, pantzas2022continuous}, and InGaP ternaries~\cite{ueno1997second,poulvellarie2021efficient,zhao2022ingap}. These materials exhibit broad transparency windows, large nonlinear susceptibilities ($100 - 300$ pm/V), and low propagation losses in wavelength-scale devices. However, quasi-phasematched interactions in these systems rely on orientation patterning~\cite{Skauli2002, yu2007growth,pantzas2022continuous}, where the $\chi^{(2)}$ is patterned during growth. To date, orientation-patterned waveguides have exhibited large propagation losses, since the different crystal orientations exhibit different etch rates, which results in strongly corrugated surfaces~\cite{yu2007growth, pantzas2022continuous}. The development of low-loss orientation patterned semiconductors will be a substantial step forward for the field. Similar developments have occurred with silicon carbide~\cite{Lukin2019,Song:19,Guidry:20,Lukin2020} and with III-V nitrides such as GaN~\cite{xiong2011integrated,hite2012development,stassen2019high,zheng2022integrated} and AlN~\cite{hickstein2017ultrabroadband, bruch2019chip,lu2020ultraviolet, liu2021aluminum, liu2023aluminum}, all of which have wider bandgaps than arsenides and phosphides, but smaller nonlinearities. Finally, we note that a number of exciting materials have been recently developed that have not yet been patterned into guided-wave devices. These include NbOX$_2$ ferroelectrics, ZnSe thin films, and ferroelectric III-V nitride semiconductors. NbOX$_2$ ferroelectrics, such as NbOCl$_2$~\cite{guo2023ultrathin} and NbOI$_2$~\cite{abdelwahab2022giant} have been shown to exhibit the largest nonlinear susceptibilities known to date at telecom wavelengths and, in principle, can be poled using ferroelectric domain inversion. Ferroelectric III-V nitrides, such as Al$_x$B$_{1-x}$N and Al$_{x}$Sc$_{1-x}$N ternaries, are a recent development~\cite{fichtner2019alscn, wang2021fully,zhu2021strongly,yoshioka2021strongly,suceava2023enhancement,yang2023domain}. These systems combine ferroelectric poling with extremely large bandgaps ($\sim 6$ eV), and intermediate-scale ($\sim 20 - 50$ pm/V) nonlinear susceptibilities, which may enable very low power nonlinear optics operating at visible or ultraviolet wavelengths. Orientation-patterned thin films of ZnSe~\cite{Tassev:19,Vangala:19,Schunemann2019} may enable the combination of a large ($\sim$ 100 pm/V) nonlinear susceptibility with an extremely wide transparency window.

The development of low-loss nonlinear photonics in each of these material systems is a rich topic, and a comprehensive discussion is beyond this scope of this tutorial. At a high level, there are an remarkable number of considerations that determine whether or not a material can be patterned into low-loss waveguides, including: whether a material can be chemically etched or must be physically etched, corrugations due to the differential etch rate between different crystal orientations (for QPM media), and how amenable a material is to surface passivation. Once losses from the surface of a material are eliminated, the bulk properties determine the remaining loss. These include: whether the material is epitaxially-grown or crucible-grown, and the resulting metal impurities, the presence of OH absorption overtones, and the transparency window of the medium as determined by the Urbach tail and multi-phonon absorption bands. Taking lithium niobate as an example, the bulk material loss limits, including the role of metal impurities, are discussed in~\cite{schwesyg2010light,Leidinger2015}, and further studies of the material absorption limits in thin films are given in~\cite{shams2022reduced}. Surface passivation is a relatively recent topic in TFLN, and is discussed in~\cite{li2023low, gruenke2023surface}. While surface corrugations are visible in scanning electron microscope images of PPLN waveguides~\cite{wang2018ultrahigh}, and light can be observed to scatter from these defects~\cite{Zhao2023unveiling}, there have been no quantitative studies of the loss imparted by surface corrugations in thin-film lithium niobate. At this time, propagation losses in TFLN are dominated by line-edge roughness of the waveguide surface, and state-of-the-art devices still exhibit more than an order of magnitude more loss than the limit set by bulk material absorption~\cite{zhang2017monolithic,shams2022reduced}.

\begin{figure}[t]
    \centering
    \includegraphics[width=\columnwidth]{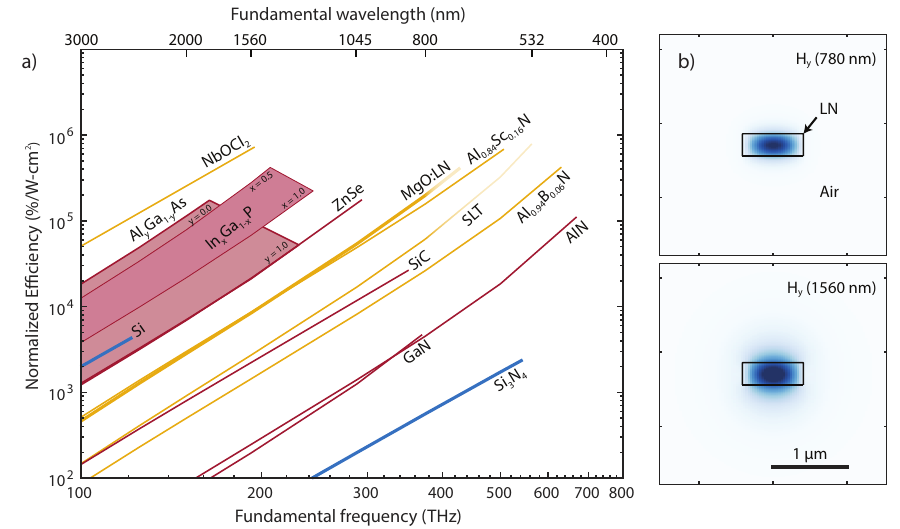}
    \caption{a) Comparison of the theoretical normalized efficiency for SHG, $\eta_0$, as a function of fundamental wavelength for many emerging platforms for nonlinear photonics. Each material is assumed to be a suspended air-clad ridge, with the height and width of the waveguide optimized at each wavelength to achieve the largest possible normalized efficiency. Smart-cut thin films, such as SLT and MgO:LN are shaded for wavelengths where extrinsic absorption of the second harmonic can be significant. In principle, the impurities that cause this absorption may be removed by using epitaxially grown thin films, rather than bonding crucible-grown materials. b) Example optimized waveguide geometry for doubling a 1560-nm fundamental in a suspended lithium niobate ridge.}
    \label{fig:eta0 comparison}
\end{figure}

When comparing material properties to assess the potential of realizing low-power nonlinear interactions, there is a strong tendency to favor semiconductors with large nonlinear susceptibilities and large refractive indices. This intuition is informed by the idea that waveguides with large refractive indices achieve much tighter spatial confinement, and therefore materials with both a large $\chi^{(2)}$ and a large refractive index will naturally require the least power. In practice, the nonlinear susceptibility alone is poor predictor of the power requirements of different nonlinear media when wavelength is also taken as a free parameter. To illustrate these trade-offs, we consider the normalized efficiency for second-harmonic generation,
\begin{equation}
	\eta_0 = \frac{2 Z_0 \omega^2 d_\mathrm{eff}^2}{c^2 n_\omega^2 	n_{2\omega}A_{\mathrm{eff}}},\label{eqn:eta0}
\end{equation}
where $A_\text{eff}$ is the effective interaction area of a pair of waveguide modes, $Z_0$ is the impedance of free space, $\omega$ is the frequency of the fundamental, $n_\omega$ is the effective refractive index of the relevant waveguide mode at frequency $\omega$, and $d_\text{eff}$ is the effective nonlinear susceptibility in m/V ($2d_{ijk} = \chi^{(2)}_{ijk}$ for SHG and $d_\text{eff} = 2\max_{ijk}|d_{ijk}|/\pi$ for QPM by a square-wave grating with 50\% duty cycle). Equation~\ref{eqn:eta0}, typically quoted in $\%/$W-cm$^2$ is a commonly used figure of merit that determines the power required to achieve saturated second-harmonic generation. We note here that Eqn.~\ref{eqn:eta0} is derived in Appendix~\ref{sec:CWEs} and will be discussed in more detail in the following sections. Devices with a large $\eta_0$ achieve saturated SHG with low optical power, or alternatively, with a smaller footprint and therefore $\eta_0$ is a realistic proxy for the relevant figures of merit for quantum nonlinear optics, such as a coupling rate (in Hertz), that will be introduced in later sections. 

The normalized efficiency, $\eta_0$, grows rapidly with the frequency of the interacting waves and scales inversely with both the refractive index of the modes and the effective area of the nonlinear interaction, $A_\text{eff}$. Noting that the effective area for a wavelength-scale device is $A_\text{eff}\approx (\lambda/n_\omega)^2$ (assuming an air-clad device), we find that the reduction of $\eta_0$ due to the explicit $n_\omega^2 n_{2\omega}$ scaling of the denominator largely cancels the decrease in $A_\text{eff}$ due to having a high-index core. Furthermore, given the $\omega^{-2}$ scaling of $A_\text{eff}$ due to the scale-invariance of Maxwell's equations, we should expect $\eta_0$ to scale as $\omega^4$. This suggests that when the fundamental wavelength is taken as a free parameter, the bandgap of the nonlinear medium strongly influences the largest normalized efficiency that can be achieved in a given medium. Figure~\ref{fig:eta0 comparison} is a more rigorous comparison of the maximum normalized efficiency attainable in many emerging platforms for nonlinear photonics as a function of fundamental wavelength. For each platform, we have assumed a quasi-phasematched interaction with $d_\text{eff} = 2\max_{ijk}|d_{ijk}|/\pi$, and an air-clad ridge waveguide with the waveguide dimensions optimized to maximize $\eta_0$ at each wavelength. For each material, we restrict the range of fundamental wavelengths ($2\hbar\omega < 0.9 E_g$) to avoid absorption of the second-harmonic by the Urbach tail of the nonlinear medium. A key insight from this analysis is that many materials can obtain the same normalized efficiency when $2\hbar\omega = 0.9 E_g$, despite the nonlinear susceptibility of these materials varying by an order of magnitude, since materials with large nonlinear susceptibilities tend to have small bandgaps. In addition, the dispersion of $\chi^{(2)}$ itself, here approximated using Miller's delta scaling, contributes a slight increase in the power law for each material system, with a typical scaling of $\omega^{4.2} - \omega^{4.5}$. As a result of this strong scaling, we find that in general, the platforms with the largest normalized efficiencies have i) a large nonlinear susceptibility, ii) a large bandgap, and iii) a low refractive index.



\begin{figure}
    \centering
    \includegraphics[width=\textwidth]{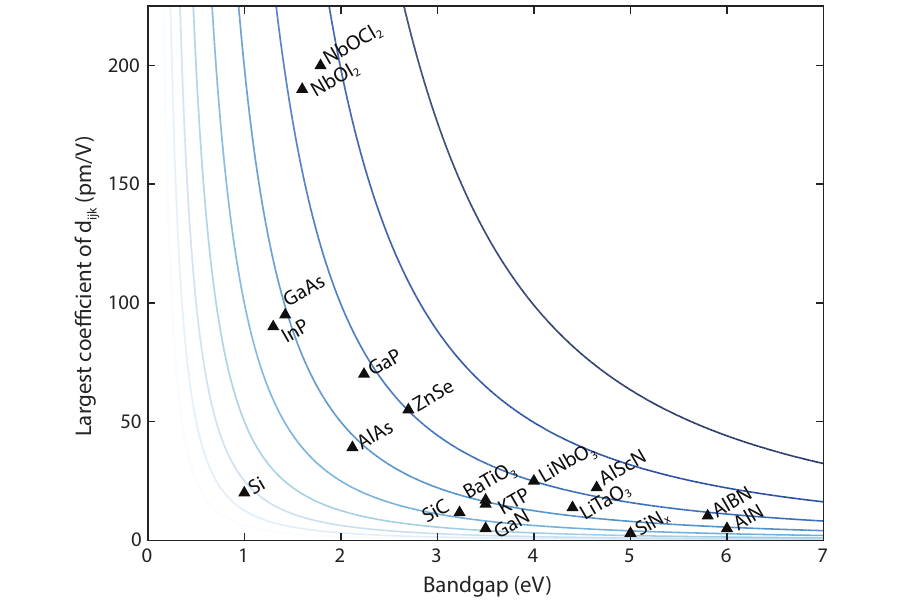}
    \caption{Comparison of the bandgap $E_g$ and the largest component of the nonlinear susceptibility, $d_\text{ijk}$, for many emerging nonlinear platforms. The contour lines correspond to 3-dB steps of the heuristic figure of merit defined here, $d_\mathrm{eff}E_g^2$. Reported values for silicon and silicon nitride correspond to the nonlinear susceptibility measured for field-induced nonlinearities.}
    \label{fig:materials_eg_dijk}
\end{figure}

The above scalings suggests that for realizing single-photon nonlinearities $d_\text{eff}$ and bandgap (more specifically $E_g^2$) are on equal footing. In later sections we will further refine these comparisons once the relevant figures of merit for quantum nonlinear interactions have been defined; the relative importance of the bandgap and the nonlinear susceptibility will depend on whether continuous-wave or pulsed interactions are being used. For now, as a means of comparison, we define a simplified heuristic figure of merit, $d_\mathrm{eff}E_g^2$, to compare the above material systems. This figure of merit, as a function of bandgap and nonlinearity, is plotted for in Fig.~\ref{fig:materials_eg_dijk}. By this metric many less-explored materials appear quite promising: AlN and LiTaO$_3$ can be made to realize comparable nonlinearities to GaAs and InP by operating at shorter wavelengths, and materials such as ZnSe and GaP are comparable to LiNbO$_3$. Emerging ferroelectric materials such as III-V ferroelectric nitrides and NbOCl$_2$ appear particularly promising, since they combine large bandgaps with large nonlinearities.

\section{Principles of classical nonlinear photonics}\label{sec:classical_NLO}

\subsection{Nonlinear interactions between waveguide modes}\label{sec:NLO_modes}

Optical waveguides have been crucial in the development of nonlinear optics, since efficient nonlinear interactions typically require both long interaction lengths and tight field confinement. Without waveguides, these two prescriptions are typically in conflict; focusing fields more tightly in a nonlinear crystal causes them to diffract more rapidly, which limits the length scales over which the waves can interact. Any treatment of nonlinear photonics therefore relies on the well-developed field of guided-wave optics. We briefly summarize the relevant aspects of optical waveguide theory here.

\begin{figure}
    \centering
    \includegraphics[width=\columnwidth]{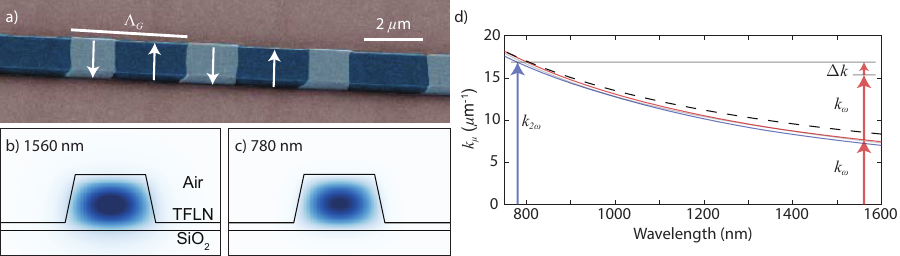}
    \caption{a) Example of a nonlinear waveguide in thin-film lithium niobate designed to generate 780-nm light from a fundamental at 1560 nm~\cite{wang2018ultrahigh}. Here, ferroelectric domains (false color: light and dark blue) are periodically inverted to enable QPM for efficient nonlinear interactions. b,c) The transverse mode profiles, $H_y(x,y,\omega)$, for the fundamental and second harmonic, respectively. d) Typical dispersion relations for a waveguide (Red: fundamental ($k_\mu$), blue: second harmonic ($k_\nu$), dashed black: bulk lithium niobate). In the absence of periodic poling, the small phase-mismatch ($\Delta k$) prevents efficient conversion from occurring. Figure (a) is adapted from Wang \textit{et al.}, Optica {\bf 5}, 11 (2018). Copyright 2018 Optica Publishing Group.}
    \label{fig:2_nlo_setup}
\end{figure}

In the absence of a nonlinear polarization, the electric and magnetic fields in a waveguide can be decomposed into independent bound modes, or eigenfunctions. More formally, for each frequency $\omega$, we have a number of discrete transverse modes, $\mathbf{E}_\mu(x,y,\omega)$ and $\mathbf{H}_\mu(x,y,\omega)$, each of which have an associated propagation constant, $k_\mu$. Figure~\ref{fig:2_nlo_setup}(a) shows a typical waveguide used in nonlinear optics, here made out of periodically-poled thin-film lithium niobate, with the associated transverse TE$_{00}$ modes $H_y(x,y,\omega)$ and $H_y(x,y,2\omega)$ shown in Figs.~\ref{fig:2_nlo_setup}(b-c). The dispersion relations of a typical waveguide are shown in Fig.~\ref{fig:2_nlo_setup}(d). Propagation of an electric field along $z$ is then given by the superposition of these bound modes,
\begin{align}
\mathbf{E}(\mathbf{r},t) &= \int_{-\infty}^{\infty}\left(\sum_{\mu}a_{\mu}(\omega)\mathbf{E}_{\mu}(x,y,\omega)\exp(-\mathrm{i}\omega t+i k_\mu(\omega) z)\right)\frac{d\omega}{2\pi},\label{eqn:waveguide_mode}
\end{align}
with corresponding expressions for $\mathbf{H}(r,t)$ and $\mathbf{D}(r,t)$. It can be shown using Poynting's theorem that the power spectral density associated with mode $\mu$ is given by $P|a_\mu(\omega)|^2$, where $P = 1$ Watt is chosen to normalize the mode profiles, $\frac12\text{Re}\left(\int \mathbf{E}_\mu\times\mathbf{H}_\mu^* dA\right)=P$.

In the presence of a nonlinear polarization all of these modes become coupled together, which is modeled by allowing the Fourier components associated with each mode to evolve during propagation, $a(\omega)\mapsto a(z,\omega)$. In general, these coupled-wave equations are given by
\begin{align}
    \partial_z a_{\mu}(z,\omega) = \frac{-i\omega}{4\mathrm{P}}\exp(i k_{\mu}(\omega)z)\int \mathbf{E}_{\mu}^*(x,y,\omega) \cdot \mathbf{P}_\mathrm{NL}(x,y,\omega) dx dy, \label{eqn:CWE_general}
\end{align}
where $\mathbf{P}_\mathrm{NL}(x,y,\omega)$ is a complex phasor representing the nonlinear polarization induced at frequency $\omega$. A full derivation of the coupled-wave equations from Maxwell's equations can be found in Appendix~\ref{sec:CWEs}. To make this formalism more concrete, we restrict our focus for now to the particular case of SHG. The evolution of the second harmonic is given by
\begin{equation*}
    \partial_z A_{2\omega}(z) = -i\kappa A_\omega^2\exp(i\Delta k z),
\end{equation*}
where $A_{2\omega} = \sqrt{P}a_\nu(z,2\omega)$ and $A_{\omega} = \sqrt{P}a_\mu(z,\omega)$ are the complex field amplitudes in power-normalized units ($|A|^2$ has units of Watts) for the relevant transverse modes ($\mu$ and $\nu$) associated with the fundamental and second harmonic, respectively. The coupling coefficient $\kappa$ contains an overlap integral that captures the strength of the interaction between some relevant subset of modes and will be discussed in more detail in the following sections. The phase-mismatch is given by $\Delta k = k_{2\omega} - 2k_\omega$. For quasi-phasematched devices, this phase-mismatch is shifted by the k-vector of the grating, $\Delta k = k_{2\omega} - 2k_\omega-k_G$, where $k_G = 2\pi/\Lambda_G$ is the angular wavenumber of the QPM grating and $\Lambda_G$ is the period of the grating, as shown in Fig.~\ref{fig:2_nlo_setup}.

From an engineer's perspective, the heart of nonlinear photonics is figuring out how to control these parameters to realize some desirable functionality. In the context of simple CW interactions, the typical goal is to make $\kappa$ large and $\Delta k = 0$ in order to achieve the greatest possible conversion efficiency with the least amount of power. For broadband interactions involving many frequencies, many more behaviors are possible. In these systems, engineering the dispersion relations $k_\mu(\omega)$ and the input pulse envelopes (\textit{e.g.} $A_\omega(z=0,\omega)$) can control how all of the interacting pulses change their shape during propagation. In the simplest of cases, we may simply choose to engineer these systems to take advantage of the large intensity associated with the peak of a pulse to realize efficient interactions with very low average power. However, in many contexts, the goals of nonlinear optics broaden beyond efficient frequency conversion. Here the nonlinear dynamics can be used to control the bandwidth generated by a nonlinear process, or the duration of a generated pulse, or to selectively amplify one pulse shape from a sea of pulses. We will see that often the dispersion relations $k_\mu(\omega)$ are among the most powerful parameters for controlling these operating regimes. In a similar vein, these ideas will generalize in the context of quantum nonlinear optics to how we can control these dynamics to generate a particular quantum state. The purpose of this section is to develop these engineering principles.

\subsection{The undepleted limit}\label{sec:NLO_review}

A natural starting point for analyzing nonlinear devices is the undepleted limit, where nonlinear interactions can often be reduced to a linear system of ordinary differential equations (ODEs). The simple solutions found by this linearized treatment are used to guide almost all physical intuition regarding the behavior of nonlinear devices. Furthermore, the experimental characterization of devices in the undepleted limit is used both to extract device parameters, such as $\kappa$, and as a diagnostic to characterize nonidealities. Our approach throughout this section will be to build up from single mode to multimode behavior for two key nonlinear processes, namely second-harmonic generation (SHG) and optical parametric amplification (OPA). We begin with second-harmonic generation since this is the simplest (and most common) process in nonlinear optics, and since SHG captures most of the essential dynamics of other nonlinear processes. We then generalize this treatment to a few modes by considering three-wave interactions, namely sum- and difference-frequency generation (SFG and DFG, respectively). Having established the continuous-wave treatment of SHG, SFG, and DFG, we have all of the theoretical tools needed to treat pulsed SHG in an intuitive way. We then repeat this scaffolded approach, from single-mode, to two-mode, to pulsed and multimode interactions, for optical parametric amplification.

\subsubsection{Continuous-wave SHG}

The solutions for continuous-wave (CW) SHG in the undepleted limit establish two crucial figures of merit for nonlinear devices, i) the normalized efficiency, which is the typical measure of nonlinearity in a waveguide, and ii) the SHG transfer function~\cite{Fejer1992,Imeshev2000a,Imeshev2000b}. The normalized efficiency, determined by measuring the generated second-harmonic power as a function of input fundamental power, characterizes the power required to achieve saturation given an interaction length $L$. The SHG transfer function, determined by measuring the generated second-harmonic power as a function of input wavelength, provides a quantitative measurement of the dispersion relations $k_\mu(\omega)$, as well as a measure of device quality. Imperfections in a fabricated device, such as spatial variations of $\kappa$ and $\Delta k$ manifest as broadening and skewing of the SHG transfer function. We will see in Sec.~\ref{sec:pulsed_SHG} that these transfer functions are fundamentally related to the dynamics of ultrafast pulses.

The coupled-wave equations (CWEs) for continuous-wave SHG are given in power-normalized units by 
\begin{subequations}
\begin{align}
\partial_z A_\omega(z) &= -i\kappa A_{2\omega}(z)A_\omega^*(z) \exp(-i\Delta k z)\label{eqn:CWE01}\\
\partial_z A_{2\omega}(z) &= -i\kappa A_\omega^2(z) \exp(i\Delta k z),\label{eqn:CWE02}    
\end{align}
\end{subequations}
where $A_\omega$ has units of W$^{-1/2}$. This choice of normalization is convenient for experiments since the power contained in the waveguide mode at frequency $\omega$ is $|A_\omega|^2=P_\omega$. In addition, this choice of normalization renders the nonlinear coupling, $\kappa$, the same in each equation, which naturally follows from power conservation,
\begin{equation}
    \partial_z P_{\omega}(z) = -\kappa |A_{2\omega}(z)||A_\omega(z)|^2 \sin(\Delta \theta(z)) = -\partial_z P_{2\omega}(z),
\end{equation}
where $-\Delta \theta = \phi_{2\omega}-2\phi_{\omega}-\Delta k z$. For quasi-phasematched waveguides, the phase-mismatch between the interacting harmonics is given by $\Delta k = k_{2\omega} - 2k_\omega - k_G$. 

The nonlinear coupling is derived using the complex reciprocity relations in Appendix~\ref{sec:kappa}, and is given by
\begin{equation}
\kappa = \frac{\sqrt{2 Z_0} \omega d_\mathrm{eff}}{c n_\omega \sqrt{n_{2\omega}A_{\mathrm{eff}}}},\label{eqn:kappa}
\end{equation}
where $d_\mathrm{eff} = 2\max_{ijk}|d_{ijk}|/\pi$ is the effective nonlinear coefficient. The factor of $2/\pi$ is introduced by assuming that the first-order Fourier component of a $50\%$ duty cycle square-wave is used to realize quasi-phasematching. $Z_0=377$ Ohms is the impedance of free space, and $n_\omega \equiv k_\mu(\omega) c/\omega$ is the effective refractive index of the relevant mode at frequency $\omega$. The effective area, $A_\mathrm{eff}$, provides a measure of the relative strength of the nonlinear interaction due to both tight confinement and the overlap of the mode with the nonlinear medium. The effective area (Eqn.~\ref{eqn:Aeff}) is more rigorously defined in Sec.~\ref{sec:kappa}, and typical values of $A_\text{eff}$ in wavelength-scale devices are on the order of $A_\mathrm{eff}\sim$1 $\mu$m$^2$ for a fundamental of 1560 nanometers. Typical effective areas in diffused waveguides and machined waveguides are on the order of $A_\mathrm{eff}\sim$20 $\mu$m$^2$. In bulk nonlinear devices driven by Gaussian beams the effective area is set by matching the confocal length to the crystal length, and is on the order of hundreds or thousands of $\mu$m$^2$ in centimeter-scale devices. As discussed in Sec.~\ref{sec:platforms} rescaling guided wave devices to shorter wavelengths (by rescaling each of the waveguide dimensions) results in a scaling of the effective area by $\lambda^{-2}$, leading to a quadratic scaling of $\kappa$ as a given device is scaled to operate at shorter wavelengths.

In the undepleted limit, Eqns~\ref{eqn:CWE01}-\ref{eqn:CWE02} are solved by assusming the input fundamental power to be constant, $A_\omega(z)\approx A_\omega(0)$. In the absence of phase-mismatch, $\Delta k = 0$, the generated second harmonic found by integrating Eqn.~\ref{eqn:CWE02} is given by
\begin{equation}
    A_{2\omega}(z) = -i\kappa A_\omega^2(0)z,\label{eqn:SHG_example_CW}
\end{equation}
or in terms of the power input to the nonlinear section at $z=0$ and output at $z=L$,
\begin{equation}
    P_{2\omega}(L) = \kappa^2 P_\omega^2(0)L^2.
\end{equation}
The conversion efficiency $\eta(z) = P_{2\omega}(z)/P_\omega(0) = \kappa^2 P_\omega(0) z^2$ grows quadratically with the propagation length $z$, and linearly with the input power $P_\omega(0)$. Since the conversion efficiency can be made arbitrarily large by using a more intense input pump or a longer waveguide, the conventional figure of merit for guided-wave SHG is the normalized efficiency,
\begin{equation}
    \eta_0 \equiv \frac{\eta(L)}{P_\omega(0)L^2} = \kappa^2.
\end{equation}
$\eta_0$ is typically quoted in units of $\%/\left(\text{W}\cdot\text{cm}^{2}\right)$, and is commonly determined by measuring the generated second-harmonic power as a function of the pump power input to a nonlinear device of known length, $L$. We note here that $z_\mathrm{sat}^{-2}=\eta_0 P_\omega(0)$ sets the characteristic length scale for efficient conversion in a nonlinear device, and similarly, $P_\mathrm{sat}^{-1}=\eta_0 L^2$ sets the characteristic power scale.

\begin{figure}[t]
    \centering
    \includegraphics[width=\columnwidth]{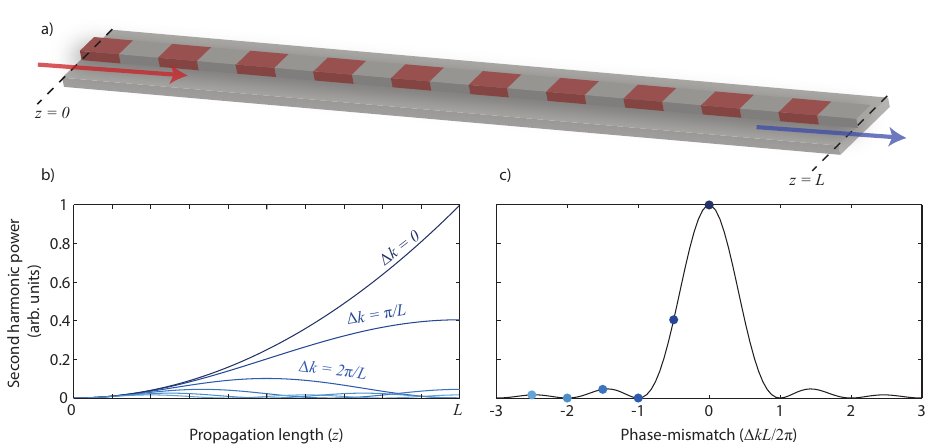}
    \caption{a) SHG in a nonlinear waveguide. Here, an input fundamental (red) is upconverted to the second harmonic (blue) during propagation from $z=0$ to $z=L$. b) Generated second-harmonic power $P_{2\omega}(z)$ as a function of propagation length for select values of phase-mismatch, assuming an undepleted fundemental. In the case of phase-matched SHG, the second-harmonic power grows as $z^2$, whereas phase-mismatched solutions exhibit oscillatory behavior. c) The SHG transfer function, given by $|H_\text{SHG}|^2(L)$ (Blue dots: $|H_\text{SHG}|^2(L)$ for the values of $\Delta k$ plotted in (b)).}
    \label{fig:SHG_basics}
\end{figure}

When the generated second harmonic is phase-mismatched with respect to the input fundamental, we can integrate the CWEs to find
\begin{equation}
    A_{2\omega}(z) = -i\kappa A_\omega^2(0)\int_0^z\exp\left(i\Delta k z\right) = -\frac{\kappa A_\omega^2(0)}{\Delta k}\left(\exp\left(i\Delta k z\right)-1\right).
\end{equation}
We may gain further insight by recasting this equation in two more forms
\begin{subequations}
\begin{align}
    A_{2\omega}(z) = \frac{-2i\kappa A_\omega^2(0)}{\Delta k}\exp\left(i\Delta k z/2\right)\sin\left(\frac{\Delta k z}{2}\right),\label{eqn:phase-mismatched_SHG_1}\\
    = -i\kappa A_\omega^2(0)z\exp\left(i\Delta k z/2\right)\sinc\left(\frac{\Delta k z}{2}\right).\label{eqn:phase-mismatched_SHG_2}
\end{align}    
\end{subequations}
Equation~\ref{eqn:phase-mismatched_SHG_1} is useful to visualize the evolution of the generated second-harmonic power during propagation (Fig.~\ref{fig:SHG_basics}(a)). The generated SH power undergoes sinusoidal oscillations during propagation with a peak power (located at odd multiples of the coherence length $L_\mathrm{coh} = \pi/\Delta k$) given by $P_\mathrm{2\omega,\mathrm{max}}=4\eta_0 P_\omega^2(0)/\Delta k ^2$. The conversion period is given by $L_\mathrm{conv} = 2 L_\text{coh} = 2\pi/|\Delta k|$. This behavior is due to periodic phase drifts between the generated second harmonic and the nonlinear polarization driving the field. For $z\in\left[0, L_\mathrm{coh}\right)$ the propagating second-harmonic field and the driving polarization have the same sign, and therefore SH photons radiated by the nonlinear polarization constructively interfere with the SH photons propagating in the waveguide. For $z\in\left[L_\mathrm{coh}, 2L_\mathrm{coh}\right)$ the photons generated by the nonlinear polarization destructively interfere with the propagating SH photons, thereby causing back-conversion to the fundamental.

Equation~\ref{eqn:phase-mismatched_SHG_2} is useful for evaluating the generated second-harmonic output from a device as a function of the phase-mismatch,
\begin{subequations}
\begin{align}
    P_{2\omega}(L) = \eta_0 P_\omega^2(0)L^2|H_\text{SHG}(\Delta k L/2)|^2,\\
    H_\text{SHG}(\Delta k L/2) = \exp\left(\frac{i\Delta k L}{2}\right)\sinc\left(\frac{\Delta k L}{2}\right),
    \label{eqn:SHG_TF}
\end{align}    
\end{subequations}
where $|H_\text{SHG}|^2$ is commonly referred to as the SHG transfer function and is shown in Fig.~\ref{fig:SHG_basics}(b). This transfer function reflects the response of the generated SH to the driving nonlinear polarization in the undepleted limit, and is a useful diagnostic tool for nonlinear devices. The deviation of a measured SHG transfer function from the theoretical $\sinc^2$ provides a qualitative measurement of device nonuniformity, and the width between the zeros of the transfer function as measured by detuning the frequency of the input laser provides an indirect measurement of the waveguide dispersion. We note here that more sophisticated techniques can be used to extract the complex transfer function, $H_\text{SHG}$, which enable quantitative measurements of device inhomogeneities~\cite{Chang2014}. Similar transfer functions will be used in the analysis of three-wave mixing and parametric amplifiers, and have fundamental implications for ultrafast pulsed interactions.

\begin{figure}
    \centering
    \includegraphics[width=\columnwidth]{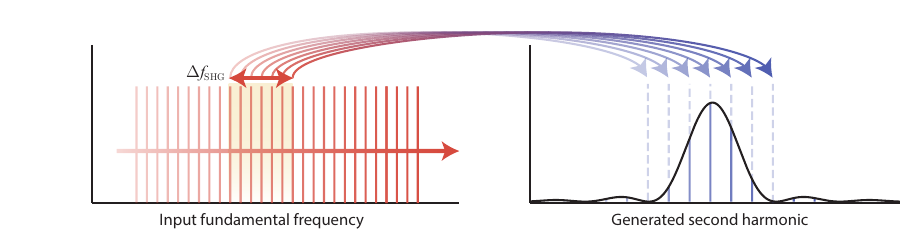}
    \caption{When driven by a swept CW laser (represented by discrete red lines) the second harmonic generated by a nonlinear waveguide is filtered by a transfer function (solid black). The bandwidth between the first zeros of the transfer function, $\Delta f_\text{SHG}$, is determined by the group velocity mismatch between the interacting waves.}
    \label{fig:SHG_TF}
\end{figure}

In most cases, the SHG transfer function is measured by sweeping the frequency of the fundamental input to the waveguide. The variation of phase-mismatch with respect to a frequency detuning  $\Omega/(2\pi)$ around a nominal operating point, $\omega/(2\pi)$, is given by
\begin{subequations}
    \begin{align}
        \Delta k(\Omega) &= k(2\omega + 2\Omega) - 2k(\omega + \Omega) - k_G\\
        &= \Delta k_0 + 2\Delta k'\Omega + \left(2k_{2\omega}''-k_\omega''\right)\Omega^2 + \mathcal{O}(\Omega^3),\label{eqn:dk_SHG}
    \end{align}
\end{subequations}
where the latter form follows from expanding $k(\omega+\Omega)$ as a Taylor series in $\Omega$. Here $\Delta k_0 = \Delta k(\Omega=0)$ is the phase-mismatch at $\Omega=0$, $\Delta k' = k'_{2\omega} - k'_\omega = v_{g,2\omega}^{-1}-v_{g,\omega}^{-1}$ is the group-velocity-mismatch (GVM) between the interacting waves, and $k_\omega''$ represents the group velocity dispersion (GVD) at frequency $\omega$. The first zeros of the SHG transfer function occur when $\Delta k(\Omega_\mathrm{\pm})L = \pm2\pi$. For a phase-matched interaction at $\Omega=0$ (and ignoring the GVD terms for now), we have
\begin{equation}
\pm 2\pi = 2\Delta k'L\Omega_\mathrm{\pm}.
\end{equation}
Figure~\ref{fig:SHG_TF} shows the generated second harmonic power (blue lines) as a function of the frequency input to the waveguide (red lines). The full frequency span around $\omega$ between the first two zeros of the SHG transfer function, $2\pi\Delta f_\mathrm{SHG} = |\Omega_+-\Omega_-|$, is given by 
\begin{equation}
\Delta f_\mathrm{SHG}^{-1} = \Delta k' L.
\end{equation}
The frequency bandwidth of the SHG transfer function decreases with both $\Delta k'$ and $L$, and provides a direct measurement of the GVM for a device with known nonlinear length $L$. In our later analysis of pulsed interactions (Sec.~\ref{sec:pulsed_SHG}) we will see that the GVM, and therefore the SHG bandwidth, is a critical parameter that determines how easily a device may be driven into saturation. In anticipation of this discussion, we link the frequency-domain bandwidth observed here to a time domain description in terms of temporal walk-off; noting that the accumulated group delay between the fundamental and second harmonic due to temporal walk-off is $\tau_\mathrm{walk-off} = \Delta k' L$, then the full width between the zeros of the SHG transfer function is given simply by
\begin{equation}
\Delta f_\mathrm{SHG}^{-1} = \tau_\mathrm{walk-off}.
\end{equation}

\subsubsection{A comment about alternative normalizations and phase conventions}

Thus far we have considered the evolution of slowly-varying envelopes $A$ in power-normalized units. While this choice of both normalization and phase reference (co-moving with the carrier frequency of each envelope) are common, there exist a number of alternative choices that are more convenient in other contexts, especially in quantum optics. We briefly address these conventions here, and will use them when convenient in later sections.

First, we note that the use of flux-normalized, rather than power-normalized units are extremely common, both in three-wave interactions and in quantum nonlinear optics. If we define the flux amplitude as $a_\omega = A_\omega/\sqrt{\hbar \omega}$ for the fundamental, and $a_{2\omega} = A_{2\omega}/\sqrt{\hbar 2\omega}$, then $|a_{\omega}|^2$ and $|a_{2\omega}|^2$ now contain the flux rate (in photons/s) of the fundamental and second harmonic respectively. At this point, we have not invoked quantization in any way. These quantities are still c-number fields, corresponding to the mean flux rate of each mode. Substituting these definitions into Eqn.~\ref{eqn:CWE01}-\ref{eqn:CWE02} yields the flux-normalized coupled-wave equations,
\begin{subequations}
\begin{align}
\partial_z a_\omega(z) &= -i\epsilon a_{2\omega}(z)a_\omega^*(z) \exp(-i\Delta k z),\label{eqn:CWE_flux_01}\\
\partial_z a_{2\omega}(z) &= -i\frac\epsilon2 a_\omega^2(z) \exp(i\Delta k z),\label{eqn:CWE_flux02}    
\end{align}
\end{subequations}
where $\epsilon=\kappa\sqrt{2\hbar \omega}$. Here, the physical origin of the factor of $\frac12$ for the second-harmonic comes from there being two generated fundamental photons from every down-converted second-harmonic photon, and one generated second-harmonic photon from each pair of up-converted fundamental photons, $a_\omega^*\partial_z a_\omega = -\frac12 a_{2\omega}\partial_z a_{2\omega}^*$.

In addition to our choice of normalization, there are a variety of useful phase references that can provide greater insights in different contexts. Both in the context of optical parametric amplification, and in the following sections on quantum optics, we will work in a rotating frame that renders the right-hand side of Eqns.~\ref{eqn:CWE01}-\ref{eqn:CWE02} translation invariant. In general, such shifts of the phase reference can be obtained by defining $\tilde{A}_{\omega} = A_{\omega}\exp(-i k_1 z)$ and $\tilde{A}_{2\omega} = A_{2\omega}\exp(-i k_2 z)$. As an example, in the context of OPA we will use $k_2 = 0$, $k_1 = \Delta k/2$, which generates the following translation-invariant CWEs,
\begin{align*}
\partial_z \tilde{A}_\omega(z) &= -i\frac{\Delta k}{2}-i\kappa \tilde{A}_{2\omega}(z)\tilde{A}_\omega^*(z),\\
\partial_z \tilde{A}_{2\omega}(z) &= -i\kappa \tilde{A}_\omega^2(z).
\end{align*}

\subsubsection{Three-wave mixing}

Having reviewed the key aspects of undepleted SHG, we now generalize this treatment to three-wave mixing (TWM) in the limit of an undepleted input wave, where the coupled-wave equations reduce to a system of coupled linear ODEs with simple closed-form solutions. We begin by considering the case where one of the input fields is zero, and the other two fields are constant, which yields nearly identical solutions to those found above for the case of SHG. In this case, TWM is also characterized by a normalized efficiency, $\eta_0$, and a $\sinc$ transfer function. However, in contrast with SHG, two of the three interacting waves can be detuned, which gives these transfer functions more complicated behavior. We also consider the special case of SFG near degeneracy, where each pair of long wavelength photons at angular frequencies $\omega_1$ and $\omega_2$ sums to the same wavelength, $\omega_3 = \omega_1 + \omega_2$. In this case, we introduce the concept of an SFG transfer function~\cite{Imeshev2000a,Imeshev2000b}, which measures how much bandwidth near degeneracy can be summed to a single frequency around the second harmonic. The SHG transfer function introduced previously and SFG transfer function introduced here will both be useful tools in the analysis of pulsed interactions. We close this section by generalizing this treatment to the case where only one of the long waves is sufficiently bright that we may assume a constant field intensity during propagation. The insights provided by this more general solution will be useful in Sec.~\ref{sec:OPA}, where we consider optical parametric amplification.

For three-wave interactions, the coupled-wave equations are given by
\begin{subequations}
\begin{align}
\partial_z A_1(z) &= -i\kappa_1 A_{3}(z)A_{2}^*(z) \exp(-i\Delta k z)\label{eqn:CWE_TWM_01},\\
\partial_z A_2(z) &= -i\kappa_2 A_{3}(z)A_{1}^*(z) \exp(-i\Delta k z)\label{eqn:CWE_TWM_02},\\
\partial_z A_{3}(z) &= -i\kappa_3 A_{1}(z)A_{2}(z) \exp(i\Delta k z),\label{eqn:CWE_TWM_03} 
\end{align}
\end{subequations}
where each envelope $A_j$ corresponds to the field amplitude at frequency $\omega_j$, with $\omega_3 = \omega_1 + \omega_2$ and $\omega_3>\omega_2>\omega_1$. The phase mismatch is now given by $\Delta k = k(\omega_3) - k(\omega_2) - k(\omega_1) - k_G$, and the nonlinear coupling is given by
\begin{equation}
\kappa_j = \frac{\sqrt{2 Z_0} \omega_j d_\mathrm{eff}}{c \sqrt{n_1 n_2 n_3 A_{\mathrm{eff}}}},\label{eqn:kappa_TWM}
\end{equation}
where $n_j$ is the refractive index of the relevant waveguide mode at $\omega_j$. The coupled-wave equations for TWM exhibit slightly different conservation laws than SHG. Here, power conservation is given by $\partial_z P_3  = -\partial_z (P_1 + P_2)$, and number conservation occurs between the short ($\omega_3$) wave and both long waves separately, $\partial_z P_3/(\hbar\omega_3) = -\partial_z P_1/(\hbar\omega_1) = -\partial_z P_2/(\hbar\omega_2)$. Stated more simply, each down-converted photon from $\omega_3$ generates both a signal photon ($\omega_2$) and an idler ($\omega_1$) photon, and pairs of long-wavelength photons are up-converted to form $\omega_3$ photons. Together, these conservation laws give the Manley-Rowe relations,
\begin{equation}
    \partial_z \frac{P_3(z)}{\hbar\omega_3} = -\partial_z \frac{P_2(z)}{\hbar\omega_2} = -\partial_z \frac{P_1(z)}{\hbar\omega_1}.\label{eqn:Manley-Rowe}
\end{equation}
These relations suggest that closed-form solutions for three-wave mixing are more naturally found using flux-normalized units, where the coupled-wave equations take the form
\begin{subequations}
\begin{align}
\partial_z a_1(z) &= -i\epsilon a_{3}(z)a_{2}^*(z) \exp(-i\Delta k z)\label{eqn:CWE_TWM_flux_01},\\
\partial_z a_2(z) &= -i\epsilon a_{3}(z)a_{1}^*(z) \exp(-i\Delta k z)\label{eqn:CWE_TWM_flux_02},\\
\partial_z a_{3}(z) &= -i\epsilon a_{1}(z)a_{2}(z) \exp(i\Delta k z),\label{eqn:CWE_TWM_flux_03} 
\end{align}
\end{subequations}
with a nonlinear coupling given by
\begin{equation}
\epsilon = \frac{\sqrt{2 Z_0 \hbar \omega_1 \omega_2 \omega_3}  d_\mathrm{eff}}{c \sqrt{n_1 n_2 n_3 A_{\mathrm{eff}}}}.\label{eqn:epsilon_TWM}
\end{equation}
The flux-normalized coupled-wave equations are useful both when considering depletion and optical parametric amplification.

The simplest solutions for TWM can be found by assuming one of the input waves to have zero power, and the other two fields to be sufficiently bright that they remain undepleted during propagation. In the case of sum-frequency generation, we have $A_3(0)=0$ and the other two waves are undepleted ($A_2(z) = A_2(0)$ and $A_1(z) = A_1(0)$). Integrating Eqn.~\ref{eqn:CWE_TWM_03}, we find
\begin{equation}
    A_3(z) = -i\kappa_3 z A_1(0)A_2(0)\exp\left(\frac{i \Delta k z}{2}\right)\sinc\left(\frac{\Delta k z}{2}\right),
\end{equation}
or in terms of the powers input and output from a device with a nonlinear section of length $L$,
\begin{equation}
    P_3(L) = \eta_{0,3} z P_1(0)P_2(0)\sinc^2\left(\frac{\Delta k L}{2}\right).\label{eqn:TWM_undepleted}
\end{equation}
The functional form of Eqn.~\ref{eqn:TWM_undepleted} is identical to the undepleted behavior encountered for SHG, with a normalized efficiency given by $\eta_{0,3}\equiv\kappa_3^2$. A similar expression may be obtained for difference-frequency generation where $A_1(0) = 0$ and the remaining waves are undepleted ($A_3(z)=A_3(0)$, $A_2(z) = A_2(0)$), with normalized efficiency $\eta_{0,1}\equiv\kappa_1^2$. Solutions for the case where $A_2(z) = 0$ may be obtained by using the symmetry of Eqns~\ref{eqn:CWE_TWM_01}-\ref{eqn:CWE_TWM_02} with respect to an interchange of indices ($1 \leftrightarrow 2$); the resulting solutions are identical.

The phase-mismatch can now be varied by detuning the frequency of any of the three waves. Denoting the frequency detuning from $\omega_1$ and $\omega_2$ by $\Omega_1$ and $\Omega_2$, respectively, the phase-mismatch is given by
\begin{subequations}
    \begin{align}
        \Delta k(\Omega_1,\Omega_2) &= k(\omega_3 + \Omega_1 + \Omega_2) - k(\omega_2 + \Omega_2) - k(\omega_1 + \Omega_1) - k_G\\
        &= \Delta k_0 + \Delta k_{3-1}'\Omega_1 + \Delta k_{3-2}'\Omega_2 + \mathcal{O}(\Omega^2),
    \end{align}
\end{subequations}
where $\Delta k_{3-1}' = k'(\omega_3)-k'(\omega_1)=v_{g,\omega_3}^{-1} - v_{g,\omega_1}^{-1}$. We may establish a correspondence to the SHG case studied above by instead parameterizing $\Delta k$ using a symmetric detuning, $\Omega$, and anti-symmetric detuning, $\Omega'$, defined by the relations $\Omega_1 = \Omega + \Omega'$ and $\Omega_2 = \Omega - \Omega'$. In this case, the phase-mismatch is given by
\begin{subequations}
    \begin{align}
        \Delta k(\Omega,\Omega') &= k(\omega_3 + 2\Omega) - k(\omega_2 + \Omega + \Omega') - k(\omega_1 + \Omega - \Omega') - k_G\\
        &= \Delta k_0 + (\Delta k_{3-1}'+\Delta k_{3-2}')\Omega + (\Delta k_{2-1}')\Omega' + \mathcal{O}(\Omega^2).\label{eqn:dk_TWM}
    \end{align}
\end{subequations}
Comparing Eqn.~\ref{eqn:dk_TWM} to Eqn.~\ref{eqn:dk_SHG}, we see that the symmetric detuning, $\Omega$, plays an identical role to the frequency detuning for SHG, with the generated sum-frequency being detuned by $2\Omega$ and the rate change of the phase-mismatch now being determined by the mean GVM of the three waves rather than the GVM between the fundamental and second harmonic. The anti-symmetric detuning, $\Omega'$, determines the range of frequency pairs that can be efficiently summed to $\omega_3$, or the range of frequencies that can be generated by DFG of a tunable signal with frequency $\omega+\Omega'$ against a short-wavelength pump with frequency $\omega_3 = 2\omega$.

A special case of TWM is SFG and DFG near degeneracy ($\omega_1 = \omega_2=\omega, \omega_3 = 2\omega$), which will be useful in understanding the behavior of pulsed nonlinear devices. In this case, the phase-mismatch is given by
\begin{equation}
    \Delta k(\Omega,\Omega') = \Delta k_0 + 2\Delta k'\Omega + (2k_{2\omega}''-k_\omega'')\Omega^2 + \frac12 k_\omega''\left(\Omega'\right)^2 + \mathcal{O}(\Omega^3).\label{eqn:dk_SFG}    
\end{equation}
Our choice of definition for $\Omega$ and $\Omega'$ renders Eqn.~\ref{eqn:dk_SFG} identical to Eqn.~\ref{eqn:dk_SHG}, up to second order in $\Omega$. We note here that if we set $\Omega=0$ Eqn.~\ref{eqn:dk_SFG} only contains even orders of $\Omega'$. This behavior suggests that the range of frequencies that can be generated around $2\omega$ by SFG is determined by the SHG transfer function, and that the range of frequency pairs around $\omega$ that can contribute to SFG at frequency $2\omega$ is determined by even derivatives of $k(\omega)$ at frequency $\omega$.

\begin{figure}
    \centering
    \includegraphics[width=\columnwidth]{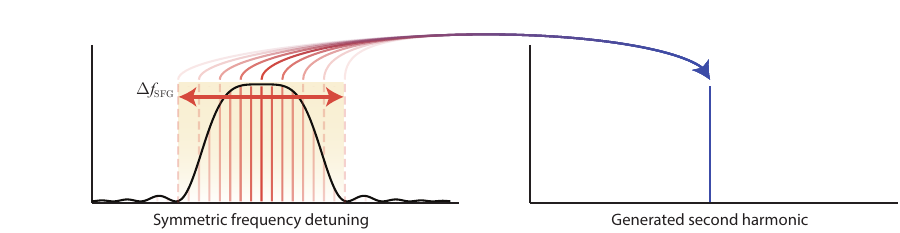}
    \caption{The total bandwidth of frequency pairs (denoted by pairs of discrete lines) around $\omega$ that may be summed to frequency $2\omega$ is determined by the SFG transfer function (solid black). The characteristic bandwidth between the zeros of the transfer function, $\Delta f_\text{SFG}$ is determined by the group delay dispersion of the fundamental, $k_\omega''L$.}
    \label{fig:SFG_TF}
\end{figure}

We further formalize this notion by defining an SFG transfer function, in analogy to the previously defined SHG transfer function, as
$$H_\mathrm{SFG}(\Omega')=\sinc(\Delta k(\Omega=0,\Omega') L/2).$$
Assuming $\Delta k(\Omega=0,\Omega')$ is dominated by $k_\omega''$ the first zeros of the SFG transfer function occur at
\begin{equation}
    4 \pi = |k_\omega''| L\left(\Omega_\pm'\right)^2,
\end{equation}
where $\Omega_+'$ and $\Omega_-'$ are defined to be the positively detuned and negatively detuned zero, respectively. The frequency span between the first zeros of the transfer function, $2\pi\Delta f_\mathrm{SFG} = \Omega_+' - \Omega_-'$, is given by
\begin{equation}
    \Delta f_\mathrm{SFG}^{-1} = \frac12 \sqrt{\pi k_\omega'' L}.
\end{equation}
Figure~\ref{fig:SFG_TF} shows a typical SFG transfer function for summing contributions from many frequencies around $\omega$ to a single frequency $2\omega$. We note here that the spacing between successive zeros of the SFG transfer function shrinks with increasing detuning since the argument is quadratic in $\Omega'$. We may again link the frequency-domain bandwidth observed here to an intuitive time domain description by noting that the group delay dispersion accumulated by a pulse propagating at frequency $\omega$ is given by $\phi''=k_\omega'' L$, resulting in dispersive spreading of the width of a typical pulse by twice its transform-limited duration.

\subsubsection{Pulsed SHG}\label{sec:pulsed_SHG}

Thus far in our consideration of traveling-wave interactions the conversion efficiency for a given input power has been determined entirely by the normalized efficiency, $\eta_0$, and the interaction length $L$. As second-order nonlinear photonics approach wavelength scale confinement, the effective areas that determine the strength of $\eta_0$ reach their theoretical limits. Noting that the conversion efficiency scales as $P_\text{in}$, a common approach to realizing low-average-power nonlinearities is the use of short pulses, where now the instantaneous peak power located around the peak of the pulse can drive the nonlinear interaction. We now consider pulsed SHG, where the dispersion relations of the nonlinear waveguide become critical in determining the efficiency and shape of the generated second harmonic. We present closed-form solutions in the undepleted limit, introduce scaling laws for pulsed interactions, and emphasize here that these time-domain dynamics can be analyzed intuitively using the CW transfer functions introduced in the previous sections~\cite{Imeshev2000a,Imeshev2000b}. We note here that in the presence of dispersion few closed-form solution exists for pulsed SHG in the saturated limit. Closed-form solutions that accurately describe saturated behaviors in dispersion-engineered devices will be presented in Sec.~\ref{sec:QSNLO}.

For broadband pulses, the coupled-wave equations for $A_\mu(\omega_1)$ and $A_\mu(\omega_3)$ now contain contributions from all possible frequency pairs that sum to $\omega$,
\begin{align}
    \partial_z A_\mu(z,\omega_1) &= -i\int_{-\infty}^{\infty}\kappa_{\mu\nu\mu}(\omega_1,\omega_3)A_{\nu}(z,\omega_3)A_{\mu}^*(z,\omega_3-\omega_1)\exp\left(-i\Delta k(\omega_1,\omega_3)z\right)\frac{d\omega_2}{2\pi},\label{eqn:broadband_SHG1}\\
    \partial_z A_\nu(z,\omega_3) &= -i\int_{-\infty}^{\infty}\kappa_{\nu\mu\mu}(\omega_3,\omega_1)A_{\mu}(z,\omega_1)A_{\mu}(z,\omega_3-\omega_1)\exp\left(i\Delta k(\omega_1,\omega_3)z\right)\frac{d\omega_1}{2\pi},\label{eqn:broadband_SHG2}
\end{align}
where $\mu$ and $\nu$ refer to the relevant mode for the fundamental and second harmonic, respectively. $\kappa_{\mu\nu\mu}(\omega_1,\omega_3)$ takes the form of the nonlinear coupling for a three-wave interaction,
\begin{align}
    \kappa_{\nu\mu\mu}(\omega_1,\omega_3) = \kappa_{\mu\nu\mu}(\omega_1,\omega_3) &= \frac{\sqrt{2 Z_0} \omega_1 d_\mathrm{eff}}{c \sqrt{A_{\mathrm{eff}}(\omega_1,\omega_3)n_{\mu}(\omega_1)n_{\nu}(\omega_3)n_{\mu}(\omega_3-\omega_1)}}.
\end{align}
The effective area, $A_\text{eff}$, now depends on the confinement and overlap of the modes at each of the three interacting frequencies. The phase-mismatch is given by $\Delta k(\omega_1,\omega_3)=k_\nu(\omega_3)-k_\mu(\omega_1) - k_\mu(\omega_3-\omega_1) - k_G$. A more detailed derivation of the coupled-wave equations and the expressions for the effective area are given in Appendix~\ref{sec:CWEs}. In many cases, the dispersion of $A_\text{eff}$ and $n$ can be neglected in calculating these integrals, which allows $\kappa(\omega_1,\omega_3)$ to be factored out of the integral. For treating extremely broadband behaviors, alternative normalizations have been developed that further suppress the dispersion of $\kappa$~\cite{phillips2012broadband}.

Our goal is to convert these equations of motion into a system of coupled differential equations that describe the instantaneous power of a pulse envelope centered around frequencies $\omega$ and $2\omega$. We first note that even when the transverse modes are the same ($\mu=\nu$) the envelopes tend to be localized around discrete carrier frequencies, which lets us continue to use Eqns.~\ref{eqn:broadband_SHG1}-\ref{eqn:broadband_SHG2}, provided that the two envelopes refer to bandwidths localized around carrier frequencies $\omega$ and $2\omega$. We therefore break up the total optical signal in each transverse mode into discrete envelopes centered around these carriers, $A_\omega(z,\Omega_1) = A_\mu(z,\omega + \Omega_1)$ and $A_{2\omega}(z,\Omega_2) = A_\nu(z,2\omega + \Omega_2)$. Since typical devices only have phase-matching between one relevant spatial mode of fundamental and one of second harmonic, we have dropped the subscripts $\mu$ and $\nu$ to simply let each of these envelopes refer to the appropriate pair of modes in a given context. In quasi-phasematched devices, the fundamental and second-harmonic are typically contained in the same transverse mode, such as TE$_{00}$ for x-cut lithium niobate, and TM$_{00}$ for z-cut. In waveguides relying on modal phase-matching the fundamental and second harmonic are often orthogonally polarized, \textit{e.g.} TE$_{00}$ and TM$_{20}$, respectively.

Having defined these envelopes, we now move into a rotating frame that removes the fast spatial variations of the envelope.
\begin{equation*}
\tilde{A}_\omega(\Omega_1)=A_\omega(\Omega_1)\exp(-i k_\omega(\Omega_1)z + i k_\omega(0) z),
\end{equation*}
This substitution converts the equations of motion into a convolution integral,
\begin{align}
    \partial_z \tilde{A}_\omega(z,\Omega_1) = &-i\left(k_\omega(\Omega_1)-k_\omega(0)\right)\tilde{A}_\omega(z,\Omega_1)\label{eqn:broadband_SHG3}\\
    &-i\kappa\int_{-\infty}^{\infty}\tilde{A}_{2\omega}(z,\Omega_2)\tilde{A}_{\omega}^*(z,\Omega_2-\Omega_1)\exp\left(-i\Delta k z\right)\frac{d\Omega_2}{2\pi},\nonumber\\
    \partial_z \tilde{A}_{2\omega}(z,\Omega_2) = &-i\left(k_{2\omega}(\Omega_2)-k_{2\omega}(0)\right)\tilde{A}_{2\omega}(z,\Omega_2)\label{eqn:broadband_SHG4}\\
    &-i\kappa\int_{-\infty}^{\infty}\tilde{A}_{\omega}(z,\Omega_1)\tilde{A}_{\omega}(z,\Omega_2-\Omega_1)\exp\left(i\Delta k z\right)\frac{d\Omega_1}{2\pi},\nonumber
\end{align}
where $\Delta k = k_{2\omega}(0) - 2 k_\omega(0) - k_G$ now represents the phase-mismatch between the carrier frequencies. These convolution integrals are more naturally evaluated in the time domain, where they take the form of a product between the two pulse envelopes. To inverse-Fourier transform these equations, we first series expand $k_\omega(\Omega_1)$ and $k_\omega(\Omega_2)$. For compactness, we write these equations in terms of dispersion operators,
\begin{align}
	k_\omega(\Omega_1) - k_\omega(0) & = k_\omega'\Omega_1 + \frac12 k_\omega''\Omega_1^2 + \frac16 k_\omega'''\Omega_1^3 + ...  = k_\omega'\Omega_1 + D_{\text{int},\omega}(i\Omega_1),\\
	k_{2\omega}(\Omega_2)  - k_{2\omega}(0)& = k_{2\omega}'\Omega_2 + \frac12 k_{2\omega}''\Omega_2^2 + \frac16 k_{2\omega}'''\Omega_2^3 + ... = k_{2\omega}'\Omega_2 + D_{\text{int},2\omega}(i\Omega_2),
\end{align}
where $D_{\text{int},\omega}(i\Omega_1) = \sum_{m=2}^{\infty} \frac{(-i)^m}{m!}
(i\Omega_1)^m k_\omega^{(m)}$ is the integrated dispersion for the fundamental at $\omega$ and $k^{(m)} = \partial_{\Omega} k$ is the $m$th derivative of $k(\Omega)$, evaluated at $\Omega = 0$.

We now inverse Fourier transform these equations of motion using Fourier's rule for derivatives, $i\Omega\leftrightarrow\partial_t$, and the convolution theorem to find the time-domain coupled-wave equations
\begin{align}
    \partial_z A_\omega(z,t) = &-k_\omega'\partial_t A_\omega(z,t) - iD_{\text{int},\omega}(\partial_t)A_\omega(z,t)\label{eqn:broadband_SHG5}\\
    &-i\kappa A_{2\omega}(z,t)A_{\omega}^*(z,t)\exp\left(-i\Delta k z\right)\nonumber\\
    \partial_z A_{2\omega}(z,t) = &-k_{2\omega}'\partial_t A_{2\omega}(z,t) -iD_{\text{int},2\omega}(\partial_t)A_{2\omega}(z,t)\label{eqn:broadband_SHG6}\\
    &-i\kappa A_{\omega}^2(z,t)\exp\left(i\Delta k z\right).\nonumber
\end{align}
Equations~\ref{eqn:broadband_SHG5}-\ref{eqn:broadband_SHG6} often provide the most intuitive insights into the nonlinear dynamics of pulses since the nonlinearity is completely localized: $\kappa$ only couples together $A_\omega$ and $A_{2\omega}$ at the same point of space and time. We will see that this property can often greatly simplify the solutions to the equations of motion. The clearest way to interpret $A_\omega(z,t)$ is to consider a discrete point in the waveguide, $z_0$. Here, $|A_\omega(z_0,t)|^2$ is the power envelope that will pass through $z_0$, and therefore the signal that would appear when detected by a receiver with infinite bandwidth located at $z_0$. Propagating along $z$ from the input of the waveguide to the output then produces the time-domain waveform output from the waveguide. We note here that in defining our rotating waves, we have freedom in how the linear terms that go as $k' \Omega$ are absorbed into the envelopes. We can, for example, define a rotating frame where $\tilde{A}_\omega$ contains no linear terms, and $\tilde{A}_{2\omega}$ contains terms that go as $(k'_{2\omega} - k'_\omega)\Omega_2'$. This eliminates $k_\omega'$ from Eqn.~\ref{eqn:broadband_SHG5}, and modifies Eqn.~\ref{eqn:broadband_SHG6} to now have a temporal walk-off term $(k_{2\omega}'-k_\omega')\partial_t A_{2\omega}$. For every problem in the following sections, we will modify Eqns.~\ref{eqn:broadband_SHG5}-\ref{eqn:broadband_SHG6} to be co-moving with whichever wave is most convenient.

Having established the coupled-wave equations for ultrafast pulses, we now consider the case of undepleted SHG with a non-dispersing fundamental. In this case, we again assume the fundamental envelope to be undepleted $A_\omega(z,t) = A_\omega(0,t)$, and following the above discussion, choose a reference group velocity co-moving with the fundamental. In this case, the equation of motion for the second harmonic becomes
\begin{align*}
    \partial_z A_{2\omega}(z,t) = &-\Delta k '\partial_t A_{2\omega}(z,t) -iD_{\text{int},2\omega}(\partial_t)A_{2\omega}(z,t)-i\kappa A_{\omega}^2(0,t)\exp\left(i\Delta k z\right),
\end{align*}
where $\Delta k' = \left(k_{2\omega}'-k_{\omega}'\right)$ is the group-velocity mismatch between the interacting waves. For a phase-matched interaction, and ignoring higher order dispersion $D_{\text{int},2\omega} \approx 0$, the second-harmonic envelope is given by integrating along the characteristic $t' = t + \Delta k'z$,
\begin{align*}
	A_{2\omega}(z,t) &= \kappa\int_0^{z} A_\omega^2(0, t+\Delta k'(z-z')) dz'\\
	&= \frac{\kappa}{\Delta k'}\int_{t-\Delta k' z}^{t} A_\omega^2(0,t')dt'.
\end{align*}
The qualitative features of this solution are a weak function of the shape of $A_\omega(0,t)$. We take as an example $A_\omega(0,t) = A_0\sech(t/\tau)$, in which case the integral evaluates to 
\begin{align}
	A_{2\omega}(z,t) &= A_0^2\frac{\kappa \tau}{\Delta k'}\left(\tanh\left(\frac{t}{\tau}\right)-\tanh\left(\frac{t-\Delta k' z}{\tau}\right)\right).\label{eqn:SHG_example}
\end{align}

\begin{figure}
    \centering
    \includegraphics[width=\linewidth]{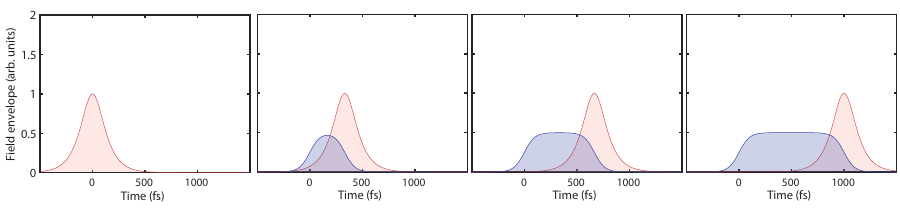}
    \caption{In the presence of temporal walk-off, the fundamental (red) leaves behind a tail of generated second-harmonic (blue), as determined by Eqn.~\ref{eqn:SHG_example} for $\tau = 100$ fs and $\Delta k' L = 1$ ps. Here the four sub-figures are rendered for $z = 0, L/3, 2L/3,$ and $L$. This process effectively limits the interaction length of the two waves to $L_\text{eff}=\tau/\Delta k'$.}
    \label{fig:SHG_pulsed_example}
\end{figure}

Eqn.~\ref{eqn:SHG_example} is shown in Fig.~\ref{fig:SHG_pulsed_example}. Here a second-harmonic pulse builds up until $\Delta k' z > \tau$, after which the generated second harmonic becomes a top-hat pulse that grows wider with increasing propagation length. Comparing Eqn.~\ref{eqn:SHG_example} with Eqn.~\ref{eqn:SHG_example_CW}, we can identify several key differences between the pulsed and CW case. Here, as expected, the peak power rather than the average power, determines the conversion efficiency. However, the interaction length is now effectively limited to a walk-off length $L_\text{eff} = \tau/\Delta k'$. Noting that $A_0^2 = \text{U}_0/(2\tau)$ for a sech pulse, the increase in conversion efficiency due to the increased field intensity (by reducing $\tau$) is canceled by a corresponding reduction in interaction length. Similarly, the increase in conversion efficiency due to increasing device length is only due to having generated a second-harmonic pulse with a longer duration, rather than a greater peak intensity. For a fixed $\Delta k'$ the route towards efficient, or low power, ultrafast interactions is by a simultaneous rescaling of $\tau$ and $L$. If we consider a device of length $L = \tau/\Delta k'$, a simultaneous increase of $L$ and $\tau$ by a factor $s_\tau$ will rescale \emph{both} the duration and the peak of the generated second-harmonic pulse by $s_\tau$, thereby recovering the quadratic growth of the generated second-harmonic energy with increasing device length.

There is an intuitive link between a frequency domain picture enabled by the SHG transfer function and the time domain envelope of the generated second harmonic. Returning to Eqn.~\ref{eqn:broadband_SHG2} in the undepleted limit, and only retaining the group-velocity mismatch in $\Delta k(\omega_1,\omega_2)$, the equations of motion can be evaluated by first integrating over the length of the waveguide,
\begin{align}
A_{2\omega}(z,\Omega_2) &= -i\kappa z H_\text{SHG}\left(\frac{\Delta k'\Omega_2 z}{2}\right)A_{\text{NL},2\omega}(0,\Omega_2),\label{eqn:broadband_SHG_filter}
\end{align}
where $A_{\text{NL},2\omega}(0,\Omega_2) = \int_{-\infty}^{\infty}A_{\omega}(z,\Omega_1)A_{\omega}(z,\Omega_2-\Omega_1)d\Omega_1/(2\pi)$ is the Fourier transform of $A_\omega^2(0,t)$. Therefore, the generated second harmonic is given simply by filtering the bandwidth of $A_\omega^2(0,t)$ by the SHG transfer function. Since the functional form of this transfer function is given by $\sinc(\Delta k'\Omega_2 z/2)$, the time-domain behavior (up to an overall delay) corresponds to convolving $A_\omega^2(0,t)$ with a top-hat of temporal width $\Delta k' z$. We note here that for a transform-limited input pulse, Eqn.~\ref{eqn:broadband_SHG_filter} can be used to obtain $|H_\text{SHG}|^2$ from the power spectral density of the second harmonic.

\begin{figure}
    \centering
    \includegraphics[width=\columnwidth]{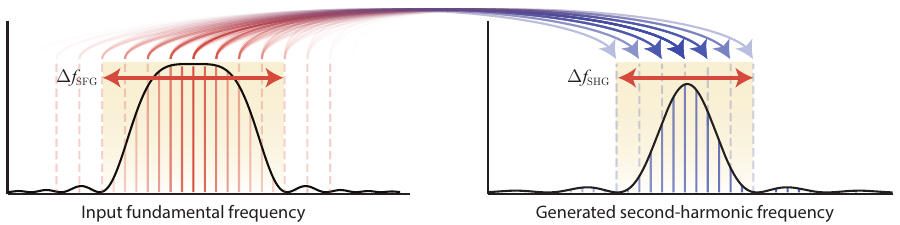}
    \caption{Frequency-domain representation of pulsed SHG. The response of the second harmonic to the nonlinear polarization is filtered by the SHG transfer function, with the width of the $\sinc^2$ transfer function in frequency determined by the temporal walk-off in the time-domain picture. When $\Delta k(\Omega, \Omega')$ can be separated into terms that depend only on $\Omega$ and $\Omega'$, the range of frequencies around the fundamental that can contribute to any one frequency of the second harmonic is determined by the SFG transfer function.}
    \label{fig:shg_filter}
\end{figure}

We may further develop the link between the dynamics of ultrafast pulses and the transfer functions for continuous-wave interactions by repeating this analysis with $\Delta k(\Omega_1,\Omega_2)$ series expanded to second order. Re-introducing the symmetric and anti-symmertric detunings, $\Omega$ and $\Omega'$, Eqn.~\ref{eqn:broadband_SHG_filter} becomes 
\begin{align}
A_{2\omega}(z,2\Omega) &= -i\kappa z H_\text{SHG}\left(\frac{\Delta k(\Omega) z}{2}\right)A_{\text{NL},2\omega}(0,2\Omega),\label{eqn:broadband_SHG_filter2}
\end{align}
where $\Delta k(\Omega) = 2\Delta k'\Omega + (2k_{2\omega}''-k_\omega'')\Omega^2$, and $A_{\text{NL},2\omega}(0,2\Omega)$ is now filtered by the SFG transfer function, 
\begin{equation}
    A_{\text{NL},2\omega}(0,2\Omega) = \int_{-\infty}^{\infty}A_{\omega}(z,\Omega + \Omega')A_{\omega}(z,\Omega-\Omega')H_\text{SFG}(\Omega')d\Omega_1/(2\pi).\label{eqn:broadband_SHG_filter3}
\end{equation}
The SFG transfer function is determined, as before, by the GVD of the fundamental,
\begin{equation}
    H_\text{SFG}(\Omega') = \sinc\left(\frac12 k_\omega''\left(\Omega'\right)^2 z\right).
\end{equation}
Equations~\ref{eqn:broadband_SHG_filter2}-\ref{eqn:broadband_SHG_filter3} allow pulsed SHG to be treated as a two step process, as shown in Fig.~\ref{fig:shg_filter}. First, the fundamental generates an intermediate field, $A_{\text{NL},2\omega}(0,2\Omega)$, which plays a role similar to the nonlinear polarization. Here the bandwidth of the input fundamental that may contribute to $A_{\text{NL},2\omega}(0,2\Omega)$ is determined by the SFG transfer function, which effectively determines how much bandwidth around the fundamental can contribute to any one frequency of the second harmonic. Then, the response of the second-harmonic to the intermediate field is filtered by the SHG transfer function, thereby limiting the generated bandwidth around $2\omega$. This analysis can be extended to any case where $\Delta k(\Omega, \Omega')$ can be separated into terms that depend only on $\Omega$ and $\Omega'$.   Later, when we consider dispersion-engineered interactions, the leading-order terms in the series expansion of $\Delta k(\Omega, \Omega')$, such as $\Delta k'$ and $k_\omega''$, will be made small. This naturally draws into question the validity of separating $H_\text{SHG}$ and $H_\text{SFG}$, since the series expansion of $\Delta k(\Omega, \Omega')$ contains many cross-terms once third- and fourth-order dispersion dominate the phase-mismatch. In practice, this approach can still work well for realistic ($\sim 50$ fs) pulse durations since $2 k_\omega''$ and $k_\omega''$ are often strongly mismatched, in which case $\Delta k(\Omega,\Omega')\approx (2 k_{2\omega}''-k_\omega'')\Omega^2$. For the special case where $2 k_{2\omega}''\approx k_\omega''$ and $\Delta k'=0$, the response of the second harmonic to the input fundamental bandwidth is better evaluated using the full integral in Eqn.~\ref{eqn:broadband_SHG2}.

We close this section by emphasizing that the techniques used here will be extended to many more contexts in subsequent sections. In general, frequency-domain analysis in the undepleted limit is well suited for identifying the dominant dispersion orders for a nonlinear process, how these dispersion orders limit interaction lengths, and how generated bandwidths scale in the presence of a particular dominant order. We will later revisit these behaviors in the context of dispersion-engineered devices in Sec.~\ref{sec:QSNLO}. The most salient advantage of dispersion-engineered waveguides is that the bandwidths and interaction lengths of a nonlinear process can be greatly enhanced by reducing or eliminating $\Delta k'$ and $k_\omega''$. These limits will coincide with substantial reductions in the required power to achieve saturated behavior. In addition, in many cases dispersion-engineered devices admit closed-form solutions for saturated behavior in the time domain. These solutions will provide much more insight into the behavior of highly-nonlinear waveguides.

\subsubsection{Continuous-wave optical parametric amplification}\label{sec:OPA}

While most three-wave interactions do not deviate meaningfully from the previous analysis for SHG, optical parametric amplification (OPA) warrants a separate treatment. Parametric gain is a crucial resource in many nonlinear systems, where it can be used to generate coherent broadband light at arbitrary wavelengths and as a source of nonclassical light at convenient wavelengths for quantum optics. As with SHG, a natural starting point for OPA is a continuous-wave analysis, where closed-form solutions are easily attainable. In analogy to the SFG bandwidth for TWM around degeneracy, we establish a link between the gain bandwidth and the group-velocity dispersion (GVD) of the fundamental.

Optical parametric amplification occurs when a bright pump at $\omega_3$ provides gain to signal and idler pairs at $\omega_2$ and $\omega_1$, respectively. Since each photon of pump generates a pair of signal and idler photons (Eqn.~\ref{eqn:Manley-Rowe}), the CWEs are more easily solved in flux-normalized units, $|a_j(z)|^2=|A_j(z)|^2/(\hbar\omega_j)$. Assuming an undepleted pump, Eqns.~\ref{eqn:CWE_TWM_01}-\ref{eqn:CWE_TWM_02} become
\begin{subequations}
\begin{align}
\partial_z a_1(z) &= \gamma_0 a_{2}^*(z) \exp(-i\Delta k z)\label{eqn:CWE_OPA_01},\\
\partial_z a_2^*(z) &= \gamma_0 a_{1}(z) \exp(i\Delta k z)\label{eqn:CWE_OPA_02},
\end{align}
\end{subequations}
where $\gamma_0 = -i\sqrt{\kappa_1\kappa_2}A_3(0) = \sqrt{\kappa_1\kappa_2 P_3(0)}$ is chosen to be a positive real number without loss of generality. Eqns.~\ref{eqn:CWE_OPA_01}-\ref{eqn:CWE_OPA_02} are useful in the weak gain limit ($a_1(z)\approx a_1(0)$, $a_2(z)\approx a_2(0)$), where the equations of motion can be directly integrated. In the more general case, we can define rotating envelopes, $\tilde{a}_j=a_j\exp(-i\Delta k z/2)$, to convert Eqns~\ref{eqn:CWE_OPA_01}-\ref{eqn:CWE_OPA_02} into a translation-invariant linear system of ODEs,
\begin{equation}
    \partial_z
    \begin{pmatrix}
    \tilde{a}_1(z)\\
    \tilde{a}_{2}^*(z)
    \end{pmatrix}
    =
    \underbrace{\begin{pmatrix}
    \frac{-i\Delta k}{2} & \gamma_0\\
    \gamma_0 & \frac{i\Delta k}{2}
    \end{pmatrix}}_{=M}
    \begin{pmatrix}
    \tilde{a}_1(z)\\
    \tilde{a}_{2}^*(z)
    \end{pmatrix}\label{eqn:CWE_OPA_system_two-mode}.
\end{equation}
Formally, these equations of motion can be solved using a matrix exponential, or equivalently, using an eigenvalue decomposition,
\begin{equation}
    \begin{pmatrix}
    \tilde{a}_1(z)\\
    \tilde{a}_{2}^*(z)
    \end{pmatrix}
    =
    \exp(M z)
    \begin{pmatrix}
    \tilde{a}_1(0)\\
    \tilde{a}_{2}^*(0)
    \end{pmatrix}
    =
    V\exp(\Lambda z)V^{-1}
    \begin{pmatrix}
    \tilde{a}_1(0)\\
    \tilde{a}_{2}^*(0)
    \end{pmatrix}\label{eqn:CWE_OPA_solution_two-mode},
\end{equation}
where $\Lambda = diag(\lambda_1, \lambda_2)$ is a diagonal matrix of eigenvalues and $V = (v_1, v_2)$ is a matrix with columns corresponding to the eigenvectors of $M$. Some care must be taken to interpret these solutions since $M$ is not Hermitian. As a consequence, the eigenvalues $\lambda_j$ may be complex, and the corresponding eigenvectors are not orthonormal (\textit{i.e.} $V$ is not unitary). There is, however, some sense in which the dynamics can be thought of as decomposing the input $\left(\tilde{a}_1(0),\tilde{a}_2^*(0)\right)^\intercal$ into eigenvectors, propagating them, and reconstituting the output. Noting that $V^{-1}=\left(w_1, w_2\right)^\dagger$ is a matrix with rows given by the left eigenvectors associated with each $\lambda_j$, that is $M^\dagger w_j = \lambda_j^* w_j$. Eqn.~\ref{eqn:CWE_OPA_solution_two-mode} can be interpreted as evaluating the component of the input along the \emph{left} eigenvectors $c_j = w_j^\dagger\left(\tilde{a}_1(0),\tilde{a}_2^*(0)\right)^\intercal$, evolving each component using $\exp(\lambda_j z)$, and reconstituting the output in terms of eigenvectors $v_j$. The propagator $\exp(M z)$ is sometimes referred to as a Green's function $G(z,z'=0)$ since it describes the response of each frequency generated at $z$ to a single frequency input at $z'=0$.

While Eqn.~\ref{eqn:CWE_OPA_solution_two-mode} gives the general solution for CW-pumped OPA, more intuition can be gained by taking a closer look at these solutions in limits of practical interest. In analogy to SHG, we first consider the degenerate case, where $\omega_2 = \omega_1$. In this case, Eqn.~\ref{eqn:CWE_OPA_system_two-mode} becomes a two-by-two system that describes the coupling between $\tilde{a}_1$, and $\tilde{a}_1^*$. In the absence of phase-mismatch, the signal is given by
\begin{equation}
    \tilde{a}_1(z) = \tilde{a}_+(0)\exp(\gamma_0 z) - i\tilde{a}_-(0)\exp(-\gamma_0 z),\label{eqn:DOPA_example}
\end{equation}
where $\tilde{a}_\pm = (\tilde{a}_1 \pm \tilde{a}_1^*)/2$. In contrast with SHG and SFG, where the generated field grows linearly with $z$, the solutions of OPA are characterized by exponential growth for the in-phase component, $\tilde{a}_+$, and deamplification of the quadrature component, $\tilde{a}_-$. In the case of non-degenerate OPA, the evolution of the signal and idler are given by
\begin{equation}
    \left(\begin{array}{c}
         \tilde{a}_2(z)  \\
         \tilde{a}_1^*(z) 
    \end{array}\right)=
    \left(\begin{array}{cc}
         \cosh(\gamma_0 z) & \sinh(\gamma_0 z) \\
         \sinh(\gamma_0 z) & \cosh(\gamma_0 z)
    \end{array}\right)
    \left(\begin{array}{c}
         \tilde{a}_2(0)  \\
         \tilde{a}_1^*(0) 
    \end{array}\right),\label{eqn:NDOPA_example}    
\end{equation}
where $\gamma_0$ can again be identified as the field gain coefficient by considering the the limit of a strong pump, which simplifies $\cosh(\gamma_0 z) \approx \sinh(\gamma_0 z) \approx \exp(\gamma z)/2$. Under typical experimental conditions only one of the waves is seeded, in which case $\cosh(\gamma_0 z)$ describes the growth of the seeded wave and $\sinh(\gamma_0 z)$ captures the growth of the generated wave. In all further discussion, we assume that the signal ($a_2$) is seeded without loss of generality. These solutions exhibit qualitatively different behavior in the low gain ($\gamma_0 z \ll 1$) and high gain ($\gamma_0 z\gg 1$) limits. For small parametric gains, the generated fields exhibit polynomial growth in $z$, \textit{e.g.} $\cosh(\gamma_0 z)\approx 1 + (\gamma_0 z)^2/2$ and $\sinh(\gamma_0 z) = \gamma_0 z$. The seeded signal wave is essentially constant for small $z$, since the generated signal photons represent a small contribution relative to the input photon flux. The generated idler wave is entirely composed of downconverted photons due to DFG between the undepleted pump and seed, and therefore grows linearly with $z$. We note here that in the context of CW OPA, the solutions for degenerate OPA (Eqn.~\ref{eqn:DOPA_example}) can be recovered from Eqn.~\ref{eqn:NDOPA_example} simply by replacing $\left(\tilde{a}_2(z), \tilde{a}_1^*(z)\right)^\intercal$ with $\left(\tilde{a}_1(z), \tilde{a}_1^*(z)\right)^\intercal$.

\begin{figure}[t]
    \centering
    \includegraphics[width=\columnwidth]{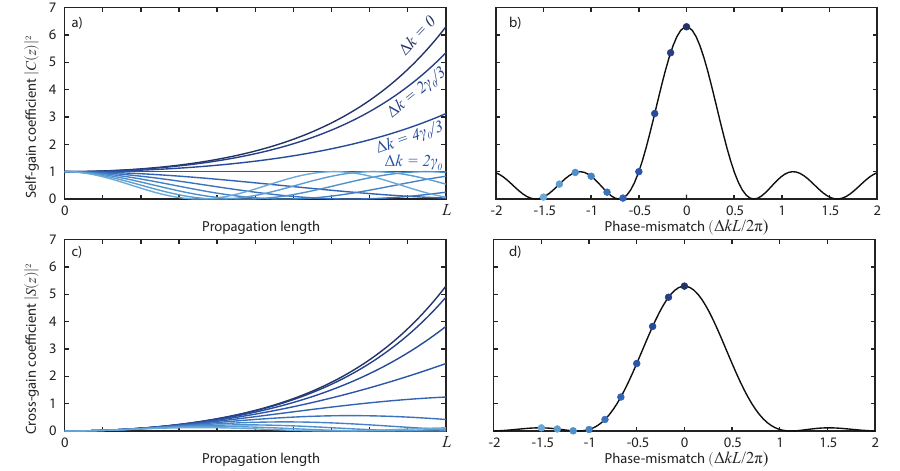}
    \caption{a,c) The self- and cross-gain coefficients $|C(z)|^2$ and $|S(z)|^2$, respectively, undergo transitions from exponential to oscillatory solutions with increasing $\Delta k$. b,d) The OPA gain spectrum, given by $|C(L)|^2$ and $|S(L)|^2$, respectively. Blue dots correspond to the select values of $\Delta k$ plotted in (a) and (c). Here, for simplicity, we take $L = 1$ and $\gamma_0 L = \pi/2$ to set the crossover from exponential to oscillatory solutions at $\Delta k L = \pi$.}
    \label{fig:OPA_basics}
\end{figure}

In the general case ($\Delta k \neq 0$), the eigenvalues of $M$ are given by $\gamma_\pm = \pm\sqrt{\gamma_0^2 - (\Delta k/2)^2}$. For $|\Delta k| < 2\gamma_0$, the main role played by phase-mismatch is to reduce the gain coefficient, $\gamma_\pm$, and to impart a small shift in the propagation constants of the interacting waves. Conversely, for large phase-mismatch ($|\Delta k| > 2\gamma_0$), the eigenvalues become purely imaginary and the associated field evolution undergoes a transition from exponentially growing to oscillatory solutions. We emphasize here that when $\gamma_\pm$ are real, there is always an amplified mode ($\gamma_+>0$) and a corresponding deamplified mode ($\gamma_- = - \gamma_+$). Whenever the eigenvalues are imaginary, they are complex conjugates $\gamma_+ = \gamma_-^*$. These features will carry over to the multimode case, and can guide intuition about the underlying dynamics of rather complicated systems. The eigenvectors of $M$ associated with $\gamma_\pm$ are \emph{not} orthogonal and are given by $v_\pm = \left(\gamma_0, i\Delta k/2 + \gamma_\pm\right)^\intercal/\sqrt{N}$, where the arbitrary normalization constant $N$ can be chosen such that $v_\pm$ has unity magnitude. With $\gamma_\pm$ and $v_\pm$, we can evaluate the Green's function, which results in
\begin{subequations}
\begin{align}
    \left(\begin{array}{c}
         \tilde{a}_2(z)  \\
         \tilde{a}_1^*(z) 
    \end{array}\right)&=
    \left(\begin{array}{cc}
         C(z) & S(z) \\
         S^*(z) & C^*(z)
    \end{array}\right)
    \left(\begin{array}{c}
         \tilde{a}_2(0)  \\
         \tilde{a}_1^*(0) 
    \end{array}\right)\label{eqn:undepleted_OPA01},\\
    C(z) &= \mathrm{cosh}(\gamma z)+\frac{i\Delta k z}{2\gamma z}\mathrm{sinh}(\gamma z),\label{eqn:undepleted_OPA02}\\
    S(z) &= \frac{\gamma_0 z}{\gamma z}\mathrm{sinh}(\gamma z).\label{eqn:undepleted_OPA03}
\end{align}
\end{subequations}
Here, as before, the eigenvalue $\gamma=\gamma_0\sqrt{1-(\Delta k)^2/(2\gamma_0)^2}$ can be identified as the field gain coefficient by considering the the limit of a strong pump ($\gamma_0\gg\Delta k$). Eqns.~\ref{eqn:undepleted_OPA01}-\ref{eqn:undepleted_OPA03} admit a wide variety of behaviors, depending on the relative magnitudes of $\gamma_0 z$ and $\gamma_0/\Delta k$ that arise in rather different experimental contexts. These behaviors are shown in Fig.~\ref{fig:OPA_basics}. As discussed previously for $\Delta k = 0$, the flux amplitudes undergo a transition from polynomial to exponential growth around $\gamma z \approx 1$. Similarly, for constant $\gamma_0$ and varying $\Delta k$, the flux amplitudes undergo a transition from exponential growth to oscillatory solutions at $|\Delta k| = 2\gamma_0$. In the limit of small $\gamma_0$, $S(z)\mapsto\gamma_0 z \sinc(\Delta k z/2)$ recovers the typical transfer function encountered for weakly-depleted three-wave interactions. Classically, this low gain limit recovers the behavior of phase-matched DFG, and the high-gain limit is relevant for OPA. In the case of semi-classical interactions (Sec.~\ref{sec:semi-classical_NLO}), where the the input fields correspond to a vacuum state, the low-gain limit will be used to describe spontaneous parametric down-conversion (SPDC), and the high-gain limit will coincide with vacuum squeezing. We emphasize here that since the eigenvalue $\gamma = \gamma_0\sqrt{1 - (\Delta k)^2/(2\gamma_0)^2}$ converges to $\gamma_0$ in the limit of high gain, the gain coefficient $\gamma_0 = \sqrt{\kappa_1\kappa_2 P_3(0)}$ can be measured experimentally by fitting the exponential growth of the unseeded wave with respect to the in-coupled pump power, which is free of any background. This diagnostic is an essential tool for experimentally characterizing the behavior of nonlinear devices.

\begin{figure}
    \centering
    \includegraphics[width=\columnwidth]{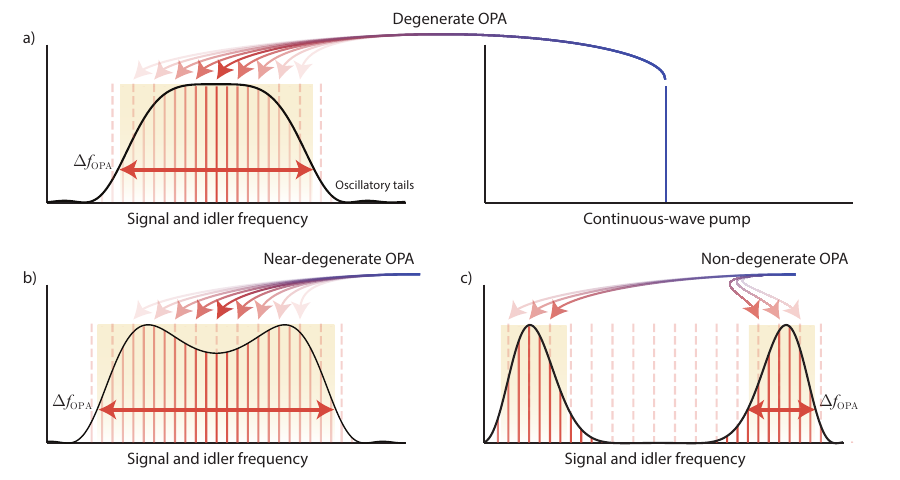}
    \caption{The OPA gain spectrum $|S(L)|^2$ generated by a CW pump (blue line) for the operating regimes described in the main text. a) Degenerate OPA ($\Delta k_0 k_\omega'' \geq 0$), exhibits a single gain peak at $\omega_3/2$. b) Near-degenerate OPA ($\Delta k_0 k_\omega'' <0$, $2\gamma_0>\Delta k_0$), exhibits two local maxima in the gain spectrum, and one continuous band of exponentially growing solutions (shaded yellow). c) Non-degenerate OPA ($\Delta k_0 k_\omega'' <0$, $2\gamma_0<\Delta k_0$) exhibits two separate bands corresponding to amplified signal and idler frequencies, respectively.}
    \label{fig:OPA_TF}
\end{figure}

The bandwidth generated by a CW-pumped OPA can be determined by the behavior of $\lambda_+$ for fixed $\gamma_0$ and varying $\Delta k$. Since the gain coefficient $\gamma$ undergoes a transition from purely real to purely imaginary when $|\Delta k| > 2\gamma_0$, and the associated solutions generated by $C(z)$ and $S(z)$ undergo a corresponding transition from exponentially growing to oscillatory (sinusoidal), we define the OPA bandwidth $\Delta f_\text{OPA}$ as the range of frequencies that satisfy $|\Delta k| \leq \gamma_0$. For frequency pairs detuned by $\pm\Omega'$ around degeneracy, the phase-mismatch is given to second order in $\Omega'$ by
\begin{equation}
\Delta k(\Omega=0, \Omega') = \Delta k_0 + \frac{k_\omega''}{2}(\Omega')^2.\label{eqn:dk_OPA}
\end{equation}
Figure~\ref{fig:OPA_TF}(a-c) compares $|S(L)|^2$ for three qualitatively distinct regimes, referred to as degenerate, near-degenerate, and non-degenerate OPA, respectively. Degenerate operation occurs when $\Delta k_0 k_\omega'' \geq 0$. In this case, the gain spectrum exhibits a local maximum at the degenerate point, $\Omega'=0$, with a maximum gain coefficient given by $\gamma_\text{max} = \sqrt{\gamma_0^2 - (\Delta k_0/2)^2}$. The OPA bandwidth around degeneracy is given by
\begin{equation}
\Delta f_\text{OPA, degenerate} = \frac{1}{\pi}\sqrt{\frac{4\gamma_0 - 2\Delta k_0}{|k_\omega''|}}.
\end{equation}
The special case of near-degenerate operation (Fig.~\ref{fig:OPA_TF}(b)) occurs for a narrow range of $\Delta k_0$ that simultaneously satisfy $\Delta k_0 k_\omega'' < 0$ and $\Delta k_0 < 2\gamma_0$. Here, the gain spectrum exhibits two local maxima at the phase-matched frequencies $\Omega_{\text{max},\pm} = \pm \sqrt{2\Delta k_0/k_\omega''}$, with gain coefficients given by $\gamma_{\text{max},\pm} = \gamma_0$. In this case, the OPA bandwidth is
\begin{equation}
\Delta f_\text{OPA, near-degenerate} = \frac{1}{\pi}\sqrt{\frac{4\gamma_0 + 2\Delta k_0}{|k_\omega''|}}.
\end{equation}
Near-degenerate operation has previously been used to extend the gain bandwidth of OPAs, at the cost of reduced parametric gain at the degenerate point~\cite{Crouch1988}. Nondegenerate operation occurs when $\Delta k_0$ simultaneously satisfies $\Delta k_0 k_\omega'' < 0$ and $\Delta k_0 > 2\gamma_0$. In this limit, the gain spectrum splits into two distinct bands with oscillatory solutions in between, as shown in Fig.~\ref{fig:OPA_TF}(c). The gain bandwidth within each of these bands is determined by the GVM between the signal and idler waves centered around $\omega \pm \Omega_{\text{max},\pm}$,
\begin{equation}
\Delta f_\text{OPA, non-degenerate} = \frac{2\gamma_0}{\pi\Delta k_{2-1}'},
\end{equation}
where $\Delta k_{2-1}' = k'(\omega+\Omega_{\text{max},+})-k'(\omega-\Omega_{\text{max},-})$. We note here that in contrast with SHG, where the relevant bandwidths are determined by the total temporal walk-off ($\Delta k' L$) or group delay dispersion ($k_\omega''L$) accumulated over the length of the nonlinear section, the OPA bandwidths are determined by the dispersion accumulated per gain length, \textit{i.e.} $k_\omega''/\gamma_0$.

In our later treatment of quantum nonlinear optics, we will see that OPA is one of the simplest processes that enables the observation of uniquely quantum effects. For this reason, the above analysis will be revisited in several sections, including our discussions of Gaussian, mesoscopic, and single-photon quantum nonlinear optics. In the context of Gaussian quantum optics, these solutions will generalize to allow for spontaneous parametric fluorescence, where photons down-convert from the pump in the absence of any coherent seed. In this case, the fluorescence bandwidth and rate is determined entirely by the cross-gain coefficient, $|S(L)|^2$. In addition, we will find that the correlations between these down-converted photons are determined by the OPA bandwidth. Later, in the context of few-photon dynamics, we will again revisit OPA as a coupling between a single-photon of pump and a continuum of down-converted signal and idler modes. Here we will see that the choice of degenerate versus non-degenerate operation will define two qualitatively distinct regimes, termed ``dispersive'' or ``dissipative'' coupling, respectively. In the former regime, the $2\omega$ pump weakly excites a localized bi-photon of fundamental with a conversion efficiency set by the phase-mismatch between the two harmonics. In the dissipative coupling regime, the pump down-converts to a broadband signal and idler, exhibiting Rabi-like oscillations with a period determined by the phase-mismatch. In these later sections, space and time will be interchanged, now with interaction times playing the role of interaction lengths, spatial-frequency bandwidth playing the role of optical frequency bandwidths, and frequency detunings playing the role of phase-mismatch.

\subsubsection{Pulsed optical parametric amplification}

As with second-harmonic generation, the parametric gain generated by OPA can be greatly enhanced by using ultrafast pulses. The treatment in this section mirrors that of pulsed second-harmonic generation, albeit with rather different behaviors. We first consider the solutions to the coupled-wave equations in the absence of dispersion, phase-mismatch, and pump depletion. This simple example will define the interaction lengths for pulsed devices, and start to build intuition for how pulsed OPA differs from the continuous-wave case. We then discuss more general solutions to the coupled-wave equations using a frequency domain analysis.


In the absence of higher order dispersion and pump depletion, the coupled-wave equation for OPA of a signal centered around a fundamental frequency $\omega$ is given by
\begin{subequations}
\begin{align}
        \partial_z A_\omega(z,t) &=-i\kappa A_{2\omega}(z,t)A_\omega^*(z,t),
        \label{eqn:TD_OPA_degen_1}
\end{align}
\end{subequations}
with an undepleted pump given by $A_{2\omega}(z,t) = A_{2\omega}(0,t-\Delta k' z)$. We assume the signal is a real function and that the pump has the appropriate relative phase to amplify the signal. With these assumptions, the equation for the signal can be solved simply by integration,
\begin{subequations}
\begin{align}
        A_\omega(z,t) &=\exp\left(\int_0^z \kappa A_{2\omega}(0,t-\Delta k' z')dz'\right)A_\omega(0,t).\label{eqn:OPA_gain_walkoff}
\end{align}
\end{subequations}
This solution is readily generalized for signals with a phase offset relative to the case considered above by noting that one quadrature is amplified and the other is de-amplified. The field gain experienced by the signal is $g(z,t) = \exp(\gamma_0(z,t) z)$, where $\gamma_0(z,t) = \frac1z\int_0^z \kappa A_{2\omega}(0,t-\Delta k' z')dz'$. The gain coefficient $\gamma_0(z,t)$ now has a temporal envelope that evolves during propagation, which makes the overall gain experienced by the signal pulse a function of the relative timing between the input signal and pump.

\begin{figure}
    \centering
    \includegraphics[width=\columnwidth]{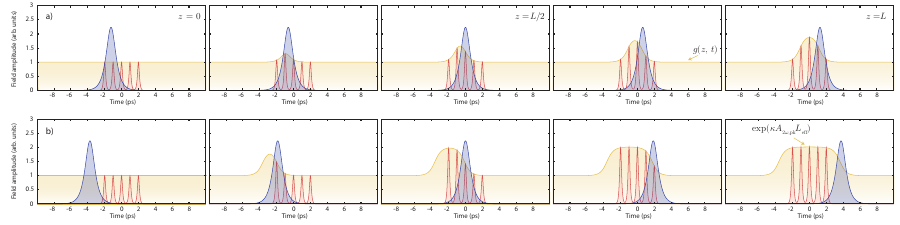}
    \caption{a) Amplification of a closely-spaced train of signal pulses (red) by a 500-fs pump pulse (blue). For a total temporal walk-off comparable to the duration of the pump pulse, most of the gain occurs around $t=0$, here chosen to achieve symmetric walk-off of the pump pulse from $-\Delta k' L/2$ to $\Delta k'L/2$. b) For a temporal walk-off much longer than the pulse duration, the field gain $g(z,t)$ (yellow) forms a flat-top and all of the pulses within this window become equally amplified. The length where this flat-top behavior begins to occur is the effective interaction length of the OPA.}
    \label{fig:gain_clipping}
\end{figure}

Figure~\ref{fig:gain_clipping} plots the amplification of a closely spaced train of pulses as a function of $z$ for two device lengths. The shaded yellow region corresponds to $g(z,t)$. We first consider the case where the total temporal walk-off $\Delta k'L$ is the same order of magnitude as the pump pulse duration, $\Delta k'L = 5\tau$, in Fig.~\ref{fig:gain_clipping}(a). Here, the signal pulse located around $t=0$ that experiences symmetric walk-off relative to the pump exhibits the most gain, with adjacent pulses seeing less gain due to a poorer temporal overlap with the peak of $g(z,t)$. Figure~\ref{fig:gain_clipping}(b) shows the case where the temporal walk-off is much longer than the pump pulse duration $\Delta k' L = 15\tau$. In this case, the field gain forms a flat top in time, and all of the closely spaced pulses are equally amplified. Intuitively, each point in time can only experience a field gain given by the duration of the pump pulses, the nonlinear coupling, and the temporal walk-off. To better understand these behaviors, we consider a pump pulse envelope of the form $A_{2\omega}(0,t) = \sech(t/\tau)$. The gain coefficient can be evaluated as
\begin{align*}
	\gamma_0(z,t) = \kappa A_{2\omega,\text{pk}}\frac{2\tau}{\Delta k' z}\left[\tan^{-1}\left(\exp\left(\frac{t}{\tau}\right)\right)-\tan^{-1}\left(\exp\left(\frac{t-\Delta k'z}{\tau}\right)\right) \right].
\end{align*}
For $\Delta k' z\gg\tau$, the terms inside of the brackets evaluate to $\pi$, in which case we have $\gamma_\text{max}(z,t)z = \kappa A_{2\omega,\text{pk}}L_\text{eff}$ where $L_\text{eff} = 2\pi\tau/\Delta k'$. As with SHG, the interaction length is determined by $\tau/\Delta k'$.

In many respects the behaviors of degenerate OPA resemble the time-reverse of SHG. Here, the interaction length set by $\tau/\Delta k'$ determines the maximum gain experienced by the fundamental, or equivalently how much second-harmonic bandwidth can contribute to gain. In our previous analysis of pulsed SHG, the interaction length $\tau/\Delta k'$ determined the local conversion efficiency from fundamental to second harmonic, or equivalently, the bandwidth of the generated second harmonic. Similarly, the continuous-wave analysis of OPA showed that the amount of bandwidth that can be generated around the fundamental is determined by $k_\omega''$, whereas for the CW analysis of SFG around degeneracy the GVD of the fundamental determined how much bandwidth can contribute to any one second-harmonic frequency.

Having established the role played by group-velocity mismatch in determining the effective interaction length over which gain can be accumulated, we now generalize this treatment to allow for dispersion to arbitrary order. This presentation is similar to the frequency domain approach used for SHG. However, to better establish a correspondence with the few-mode treatment of OPA, we will work with discrete Fourier modes, rather than continuous Fourier integrals. The behavior of the OPA can then be understood in terms of Green's functions and eigenvectors. This discrete analysis is valid for waveguides driven by mode-locked lasers, which contain discrete frequencies, and is well-suited for numerical analysis.

We begin by considering degenerate OPA and choose our phase reference to be co-moving with the pump. In this case, the time-domain coupled-wave equation for the signal becomes
\begin{align}
        \partial_z A_\omega(z,t) &=-i\kappa A_{2\omega}(z,t)A_\omega^*(z,t)\exp(-i\Delta k z) + \Delta k'\partial_t A_\omega(z,t) -iD_{\text{int},\omega}(\partial_t)A_\omega(z,t).
        \label{eqn:TD_OPA_degen_2}
\end{align}
The coupled-wave equations for discrete temporal modes are found by expanding each envelope in terms of a Fourier series,
$$A(z,t) = \sum_m A_{m}(z)\exp(2\pi i m t/T),$$
where $T$ can be chosen to be the repetition period of the pump, or the width of an arbitrary window sufficiently large to contain both pulse envelopes. For the pump envelope, we extract the peak amplitude, $A_{2\omega}(z,t) = A_\text{pk}f(t)$, and series expand the pulse shape only, $f(t) = \sum_m c_{m}\exp(2\pi i m t/T)$. Following our treatment of CW OPA, we define flux amplitudes $a_m = A_m/\sqrt{\hbar\omega_m}$, and move into a rotating frame given by $\tilde{a}_m = a_m\exp(-i\Delta k_0 z/2)$. With these definitions, Eqn.~\ref{eqn:TD_OPA_degen_2} becomes
\begin{align}
	\partial_z \tilde{a}_m(z) =  -i\Delta k_m \tilde{a}_m(z) + \gamma_\text{pk}\sum_n c_{m+n} \tilde{a}_n^*(z),\label{eqn:Fourier_CWE}
\end{align}
where $\Delta k_m = \frac12 \Delta k_0 + \Delta k'(2\pi m/T) + \frac12 k''(2\pi m/T)^2 + ...$ contains all contributions to the phase mismatch from mode $m$, and the peak gain coefficient $\gamma_\text{pk} = -i\kappa A_\text{pk}$ is assumed to be real. In the absence of pump depletion, Eqn.~\ref{eqn:Fourier_CWE} is a linear system of equations for the flux amplitudes $\tilde{a}_m$. Following the treatment of Sec.~\ref{sec:OPA}, it is more insightful to write the coupled-wave equations as a matrix for $\tilde{a}_m$ and $\tilde{a}_n^*$,
\begin{equation}
    \partial_z
    \left(\begin{array}{c}
    \vdots\\
    \tilde{a}_{m}(z)\\
    \vdots\\
    \hline
    \vdots\\
    \tilde{a}_{n}^*(z)\\
    \vdots\\
    \end{array}\right)
    =
    \left(
    \begin{array}{c c c | c c c}
    \ddots & & & & & \\
    & -i\Delta k_m & & & \gamma_{mn} &\\
    & & \ddots & & & \\
    \hline
	& & & \ddots & & \\
    & \gamma_{nm} & & & i\Delta k_n &\\
 	& & & & & \ddots\\
    \end{array}\right)
    \left(\begin{array}{c}
    \vdots\\
    \tilde{a}_{m}(z)\\
    \vdots\\
    \hline
    \vdots\\
    \tilde{a}_{n}^*(z)\\
    \vdots\\
    \end{array}\right)
    \label{eqn:CWE_OPA_system_multimode},
\end{equation}
where $\gamma_{mn} = \gamma_\text{pk}c_{m+n}$ are matrix entries containing the coupling between each pair of signal frequencies due to parametric gain. Using our notation from Sec.~\ref{sec:OPA}, we can write Eqn.~\ref{eqn:CWE_OPA_system_multimode} as
\begin{equation}
    \partial_z
    \begin{pmatrix}
    \tilde{\mathbf{a}}(z)\\
    \tilde{\mathbf{a}}^*(z)\\
    \end{pmatrix}
    =
    M
    \begin{pmatrix}
    \tilde{\mathbf{a}}(z)\\
    \tilde{\mathbf{a}}^*(z)\\
    \end{pmatrix},
    \label{eqn:CWE_OPA_system_multimode_simple}
\end{equation}
where $\tilde{\mathbf{a}} = (\tilde{a}_1, \tilde{a}_2, ...)^\intercal$ are vectors containing the flux amplitudes.

In general, the solution to Eqn.~\ref{eqn:CWE_OPA_system_multimode_simple} is given by
\begin{equation}
	\begin{pmatrix}
    \tilde{\mathbf{a}}(z)\\
    \tilde{\mathbf{a}}^*(z)\\
    \end{pmatrix}
    =
    \exp(M z)
    \begin{pmatrix}
    \tilde{\mathbf{a}}(0)\\
    \tilde{\mathbf{a}}^*(0)\\
    \end{pmatrix},
\end{equation}
where the matrix exponential $\exp(M z)$ is the Green's function for multimode OPA. As with continuous-wave OPA, Eqns.~\ref{eqn:CWE_OPA_system_multimode} are more easily obtained using an eigenvalue decomposition, $M = V \Lambda V^{-1}$, with corresponding Green's function given by $V\exp(\Lambda z)V^{-1}$. Closed-form solutions for the multimode coupled-wave equations are rare in practice~\cite{Sukhorukov1971,Danielius1993,Manzoni2016,Charbonneau_Lefort_2010}, but we may obtain some qualitative insights about the behavior of OPA by studying the eigenvalues. First, we note that for $\gamma_{mn}=0$, $M$ is a diagonal matrix. The phase-mismatch manifests as conjugate pairs of purely imaginary eigenvalues, with associated eigenvectors corresponding to oscillatory solutions. As the gain increases the trace of $M$ remains zero, and pairs of eigenvalues both move inward towards the real axis and outwards away from the imaginary axis, which coincides with the formation of eigenvectors that extract gain during propagation. These complex eigenvalues come in sets of four: $\lambda$, $\lambda^*$, $-\lambda$, and $-\lambda^*$. Therefore, for each amplified eigenvector with eigenvalue $\lambda$, there exists a de-amplified eigenvector with eigenvalue $-\lambda$. Furthermore, for any oscillatory solution $\lambda$, there exists a conjugate eigenvalue $
\lambda^*$ that interferes to form sinusoidal and cosinusoidal terms. In practice, the most common approach for understanding this behavior is to study the singular value decomposition of the Green's function, which enables a description in terms of a handful of dominant modes that extract the most gain. This technique will be discussed in more detail in Sec.~\ref{sec:Gaussian-pulse-pumped-OPA}.

\subsection{Dispersion-engineered nonlinear interactions}\label{sec:QSNLO}

In our treatment of pulsed nonlinear interactions, we found that GVM and higher-order dispersion limit the effective interaction lengths, and therefore partially mitigate the benefits gained from the large peak intensities associated with short pulses. In addition, we restricted our treatment of pulsed interactions to the undepleted limit, since general closed-form solutions have not been found for dispersive propagation. We now consider dispersion-engineered nonlinear devices, which overcome both of these limitations. The approach taken here is motivated by the scale invariance of Eqns.~\ref{eqn:broadband_SHG1}-\ref{eqn:broadband_SHG2}, copied here for convenience,
\begin{align*}
    \partial_z A_\omega(z,t) = &-k_\omega'\partial_t A_\omega(z,t) - iD_{\text{int},\omega}(\partial_t)A_\omega(z,t)\\
    &-i\kappa A_{2\omega}(z,t)A_{\omega}^*(z,t)\exp\left(-i\Delta k z\right)\nonumber\\
    \partial_z A_{2\omega}(z,t) = &-k_{2\omega}'\partial_t A_{2\omega}(z,t) -iD_{\text{int},2\omega}(\partial_t)A_{2\omega}(z,t)\\
    &-i\kappa A_{\omega}^2(z,t)\exp\left(i\Delta k z\right).\nonumber
\end{align*}
The limitations to interaction length by dispersion of \emph{arbitrary} order can be found by finding the transformations that leave Eqns.~\ref{eqn:broadband_SHG1}-\ref{eqn:broadband_SHG2} unchanged. On inspection, we find that the coupled-wave equations are scale invariant with respect to
\begin{align}
    z \mapsto s z\qquad \Delta k \mapsto \frac{\Delta k}{s} \qquad \Delta k' \mapsto \frac{\Delta k'}{s} \qquad \hat{D}\mapsto\frac{\hat{D}}{s} \qquad A(z,t)\mapsto \frac{\hat{A}(z,t)}{s},\label{eqn:scale_invariance}
\end{align}
where we have dropped the subscripts to denote that these scalings are applied simultaneously to the fundamental and second harmonic. Equation~\ref{eqn:scale_invariance} shows that a simultaneous reduction of phase-mismatch, temporal walk-off, and the dispersion operators by a factor $s$ enables an increase of the interaction length by $s$, thereby realizing a \emph{quadratic} reduction of the power requirements for second-order nonlinear waveguides.

In reality, the simultaneous rescalings in Eqn.~\ref{eqn:scale_invariance} cannot be realized for realistic waveguides. Instead, we consider eliminating the dominant dispersion orders for a given nonlinear process and ignoring the effects of less impactful dispersion orders. In all further discussion, this operating regime will be referred to as quasi-static operation, since the length scales over which dispersion causes the pulse envelopes to distort are far greater than a typical device length. Quasi-static devices can be realized in practice by first identifying the relative importance of each dispersion order for a given nonlinear process, and then using geometric dispersion engineering~\cite{Jankowski2020,Singh2020,Hickstein2019} to eliminate the dominant terms. There are two key benefits to this approach. Quasi-static operation enables arbitrarily long interaction lengths, and therefore extremely low power requirements, and quasi-static equations of motion admit closed-form solutions that often capture the observed behaviors of realistic devices. In the following sections, we review these closed-form solutions for saturated SHG~\cite{Armstrong1962,Eckardt1984,Jankowski2020} in both the phase-matched and phase-mismatched cases. In the case of phase-mismatched SHG, saturated interactions cause the pulse envelopes to distort and bifurcate, which limits the total conversion efficiency and can generate octaves of bandwidth. This limit provides the most rigorous test of quasi-static heuristics. We then discuss the role played by higher-order dispersion, and show that these behaviors place practical limitations on the efficiency of quasi-static devices. Finally, we treat fully-static nonlinear optics, where a temporal waveguide is used to define the pulse envelopes. In this limit, the pulse envelopes no longer change shape during propagation, which allows all of the benefits of quasi-static operation to be retained without any limitations to conversion efficiency.

\subsubsection{Quasi-static SHG}
\label{sec:quasi-static-shg}

In the case of both SHG and degenerate OPA the dominant dispersion orders are given by $\Delta k'$ and $k_\omega''$. Quasi-static interactions are realized by engineering these terms to be zero,
\begin{subequations}
    \begin{align}
        \partial_z A_\omega &=-i\kappa A_{2\omega}A_\omega^* \exp(-i\Delta k z) + \underbrace{\frac{i}{2} k_\omega''\partial_t^2 A_{\omega}}_{= 0} + \underbrace{\mathcal{O}(\partial_t^3)}_{\text{ignored}},\label{eqn:TD-CWE1}\\
        \partial_z A_{2\omega} &= -i\kappa A_\omega^2 \exp(i\Delta k z)-\underbrace{\Delta k'\partial_t A_{2\omega}}_{=0}+ \underbrace{D_{\text{int},2\omega}A_{2\omega}}_{\text{ignored}}.\label{eqn:TD-CWE2}
    \end{align}
\end{subequations}
Having eliminated the leading-order terms (\textit{e.g.} with geometric dispersion engineering), we ignore the remaining higher order terms to obtain heuristic solutions to the coupled-wave equations. Within these approximations, we recover the CW coupled-wave equations, with each time bin acting as an independent wave,
\begin{subequations}
    \begin{align}
        \partial_z A_\omega(z,t) &=-i\kappa A_{2\omega}(z,t)A_\omega^*(z,t) \exp(-i\Delta k z),\label{eqn:QS-CWE1}\\
        \partial_z A_{2\omega}(z,t) &= -i\kappa A_\omega^2(z,t)(z,t) \exp(i\Delta k z).\label{eqn:QS-CWE2}
    \end{align}
\end{subequations}
We solve for the dynamics of each time-slice separately, treating them effectively as independent CW modes, following the treatment of~\cite{Armstrong1962,Eckardt1984, Jankowski2021SCG}. A self-contained analysis is given in Appendix~\ref{sec:QS_theory}. The heart of this technique is to make use of the fact that each time bin locally conserves power, which allows us to normalize the fields into the notation used by Bloembergen, $\sqrt{P(t)}u(z,t) = \rho_\omega(z,t)$, $\sqrt{P(t)}v(z,t) = \rho_{2\omega}(z,t)$. Here $P(t) = |A_\omega(z,t)|^2 + |A_{2\omega}(z,t)|^2$, and $\rho_\omega = |A_\omega|$. With the fields written in terms of $u$ and $v$, the evolution of the field amplitudes depends only on $u$, $v$, and the relative phase of the two harmonics given by $\theta$. Once solutions are obtained for $u$ and $v$, they can be substituted back into the coupled-wave equations to determine the phase evolution of each harmonic separately. 

For the case of phase-matched SHG, the fundamental and second harmonic are given by
\begin{subequations}
\begin{align}
A_{2\omega}(z,t) &= -i A_\omega(0,t)\tanh(\kappa A_\omega(0,t) z),\label{eqn:saturated_SHG_1}\\
A_{\omega}(z,t) &= A_\omega(0,t)\mathrm{sech}(\kappa A_\omega(0,t) z).\label{eqn:saturated_SHG_2}
\end{align}
\end{subequations}
As expected, each time bin separately evolves from undepleted SHG ($A_{2\omega}(z,t) \approx -i\kappa A_\omega^2(0,t)z$) to full conversion ($A_{2\omega} = -iA_\omega(0,t)$). We emphasize here that the relevant length scale for conversion is now given by the peak intensity of the pulse, rather than the average power. However, we note here that for femtosecond pulses, the small residual dispersion in Eqns.~\ref{eqn:TD-CWE1}-\ref{eqn:TD-CWE2} limits the degree of saturation that can occur in practice. This behavior is analysed in Sec.~\ref{sec:Gouy-phase}. We note here that the solutions for phase-matched OPA can be recovered from~\ref{eqn:saturated_SHG_1}-\ref{eqn:saturated_SHG_2} by shifting the origin to a negative value of $z$ until the initial conditions match the fields at the boundary of the waveguide.

In the presence of phase-mismatch, we may similarly solve the equations of motion using the CW solutions corresponding to the instantaneous field intensity. In this case, the solutions take the form of a Jacobi elliptic sine, $\text{sn}(u|m)$. Defining the instantaneous field conversion efficiency as $|v(z,t)| = |A_{2\omega}(z,t)/A_\omega(0,t)|$, the field envelopes of the fundamental and second harmonic are 
\begin{subequations}
\begin{align}
A_\omega(z,t) = &\sqrt{1-v^2(z,t)}|A_\omega(0,t)|\exp(i\phi_\omega(z,t)),\label{eqn:je1}\\
A_{2\omega}(z,t) &= v(z,t)|A_\omega(0,t)|\exp(i\phi_{2\omega}(z,t))\label{eqn:je2},
\end{align}
\end{subequations}
where $v(z,t)=v_b(t) \mathrm{sn}\left(\kappa A_\omega(0,t)v_b^{-1}(t) z | v_b^4(t)\right)$. Here $\mathrm{sn}(u|m)$ is the Jacobi elliptic sine and $v_b(t)=-|\Delta k/(4 \kappa A_\omega(0,t))|+\sqrt{1+|\Delta k/(4 \kappa A_\omega(0,t))|^2}$. We note here that $v_b^2(t)$ is the maximum pump depletion attainable for a given $\Delta k$ as a function of the local field amplitude input to the waveguide $A_\omega(0,t)$. The Jacobi-elliptic sine smoothly interpolates between the conventional undepleted solutions for phase-mismatched SHG and the saturated solutions for phase-matched SHG. Noting that $\text{sn}(u|0)=\sin(u)$, we find that in the limit of large $\Delta k$ the elliptic sine becomes $\sin(\Delta k z/2)$ and we recover the conventional solutions for undepleted SHG. Similarly, noting that $\text{sn}(u|1)=\tanh(u)$, in the limit as $\Delta k\rightarrow 0$ we recover the $\tanh$ solutions for phase-matched SHG.

\begin{figure*}[t]
\centering
\includegraphics[width=\textwidth]{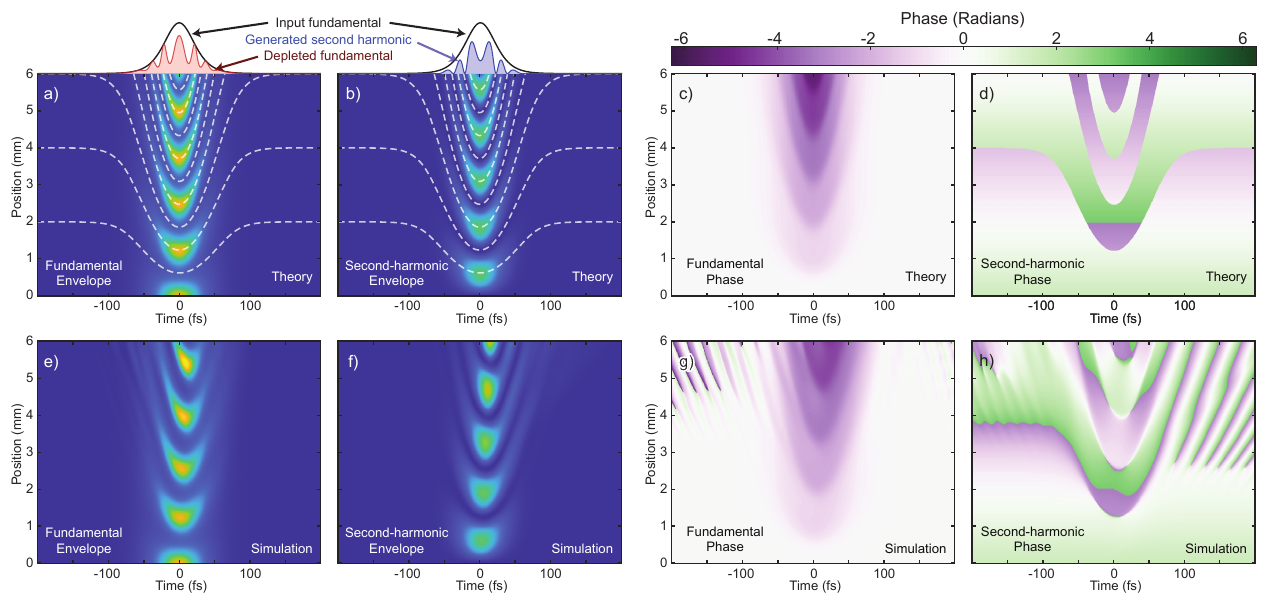}
\caption{\label{fig:jetheory} a,b) Theoretical evolution of $|A_\omega(t)|^2$ and $|A_{2\omega}(t)|^2$ based on Eqns.~\ref{eqn:je1}-\ref{eqn:je2}. Dashed white lines: conversion half-periods during propagation, solid black lines: in-coupled fundamental pulse. c,d) The phase of the fundamental and second harmonic calculated using Eqns.~\ref{eqn:phase1}-\ref{eqn:phase2}. Both harmonics form plateaus of constant phase, and therefore spectral broadening is predominantly due to femtosecond-scale oscillations of the pump depletion. e-h) Comparison of quasi-static theory with a split-step Fourier simulation, including dispersion to third order and self-steepening. The structure of each harmonic is largely unchanged by waveguide dispersion, so that the essential behavior is accurately captured by the simple quasi-static heuristic. Figure adapted with permission from~\cite{Jankowski2021SCG}, Copyright 2023 Authors.}
\end{figure*}

For saturated SHG, the conversion period now varies across the pulse envelope due to the variation of the instantaneous field intensity. This conversion period is given by
\begin{equation}
L_\mathrm{conv}(t)=\frac{2 v_b(t) K(v_b^4(t))}{\kappa A_\omega(0,t)},
\end{equation}
where $K$ is the complete elliptic integral of the first kind. The contribution to $L_\mathrm{conv}$ from $K(v_b^4(t))$ is almost always a weak functuion of $\nu_b$, since $K(v_b^4(t))$ varies slowly for most physically encountered values of $v_b(t)$. In the absence of any depletion, we have $K(v_b^4=0)=\pi/2$. For a peak depletion of 90\%, $K(v_b^4=0.81)\approx 1.45 \pi/2$. Therefore the variation of $L_\mathrm{conv}(t)$ is dominated by $v_b(t)/(\kappa A_\omega(0,t))$. Figure~\ref{fig:jetheory}(a-b) shows an example of the theoretical evolution of a 50-fs-wide (3 dB) sech$^2$ pulse in a 6-mm-long dispersion-engineered TFLN waveguide given by Eqns.~\ref{eqn:je1}-\ref{eqn:je2}. The dashed white lines correspond to the $m^\mathrm{th}$ half-period, $L_m(t)= mL_\mathrm{conv}(t)/2$, where even $m$ coincide with the local maxima and minima of the fundamental and second harmonic, respectively. Near the peak of the pulse the conversion period is the shortest and both harmonics undergo $\sim 5$ conversion periods as the field propagates through the waveguide. The oscillations of the power in the tails of the pulse asymptotically approach the conversion period associated with undepleted SHG (equal to twice the conventional coherence length in this limit), $L_\mathrm{conv}(\infty) = 2\pi/|\Delta k|$. The power at the peak oscillates three times faster than in the tails of the pulse, which gives rise to a pulse shape with rapid temporal amplitude oscillations as each portion of the pulse cycles through a different number of conversion periods.

Having solved for the fields, these Jacobi elliptic solutions can be used to predict phase envelopes for the fundamental and second harmonic,
\begin{subequations}
\begin{align}
\phi_\omega(z,t) - \phi_\omega(0,t) &= -\frac{\Delta k}{2}\int_0^z \frac{v^2(0,t)-v^2(z',t)}{1-v^2(z',t)}dz'\label{eqn:phase1}\\
&=\frac{1}{2}\sin^{-1}\left(\frac{\Delta k |A_{2\omega}(z,t)|}{2\kappa |A_{\omega}(z,t)|^2}\right)-\frac{\Delta k z}{4},\nonumber\\
\phi_{2\omega}(z,t) - 2\phi_\omega(0,t) &= - \pi/2 + \frac{\Delta k z}{2},\label{eqn:phase2}
\end{align}
\end{subequations}
respectively. Equations~\ref{eqn:phase1}-\ref{eqn:phase2} are plotted in Fig.~\ref{fig:jetheory}(c,d) for an unchirped fundamental input to the waveguide, $\phi_\omega(0,t)=0$. We note here that the rate of phase accumulation has a fixed sign determined by $\Delta k$, and therefore in this context $\sin^{-1}(\sin(x))=x$ is defined to be a monotonic function. The fundamental accumulates phase most rapidly around values of $z$ and $t$ that correspond to local maxima of $v(z,t)$, where input fundamental is most strongly depleted. This behavior causes the fundamental to accumulate phase in sharp jumps, with the total accumulated phase plateauing across large time bins (Fig.~\ref{fig:jetheory}(c)). The phase envelope of the second harmonic is independent of time, $\phi_{2\omega}(z,t)=\phi_{2\omega}(z,0)$, up to the sign changes of $\text{sn}(u,m)$. The second-harmonic phase shown in Fig.~\ref{fig:jetheory}(d), $\angle A_{2\omega}(z,t)$, contains contributions from both $\phi_{2\omega}(z,t)$ and sign changes of $\nu(z,t)$, and therefore exhibits phase discontinuities of $\pm \pi$ every $L_\mathrm{conv}(t)$.

This heuristic model for pulse propagation can be verified in three ways. First, comparing Eqn.~\ref{eqn:je1}-\ref{eqn:je2} against a split-step Fourier simulation of the quasi-static equations of motion, we find no difference between theory and simulation, which verifies that this heuristic model captures all of the physics associated with saturation and phase-mismatch. We repeat this comparison, with the split-step Fourier methods now accounting for both self-steepening and dispersion to third order to determine whether or not the quasi-static equations of motion are a reasonable approximation for realistic devices. In this case, we use the parameters obtained from a waveguide simulation of the dispersion-engineered waveguides studied in~\cite{Jankowski2021SCG}: a temporal walk-off of $\Delta k'=$5 fs/mm, group velocity dispersion for the fundamental and second harmonic of $k_\omega''=$9.5 fs$^2$/mm and $k_{2\omega}''=$70 fs$^2$/mm, respectively, and third-order dispersion given by $k_\omega'''=$-1100 fs$^3$/mm and $k_{2\omega}'''=$1200 fs$^3$/mm. The time-domain instantaneous power associated with each envelope, $|A_\omega|^2$ and $|A_{2\omega}|^2$, is shown in Fig. \ref{fig:jetheory}(e-f), respectively. The phase associated with each envelope is shown in Fig. \ref{fig:jetheory}(g-h). We note here that while the fundamental phase envelope is unwrapped to better visualize the phase accumulated during propagation, a similar procedure cannot be applied to the second harmonic due to the phase discontinuities accumulated around $L_\mathrm{conv}(t)$. To facilitate comparisons between theory and simulation we have left the second-harmonic phase wrapped. While the simulated pulse envelopes exhibit some distortion due to second- and third-order dispersion, the key aspects of the heuristic Jacobi elliptic approach, such as the field oscillations, are largely preserved, which suggests that quasi-static devices may be well described by the heuristic models developed here.

\begin{figure*}[t]
\centering
\includegraphics[width=\textwidth]{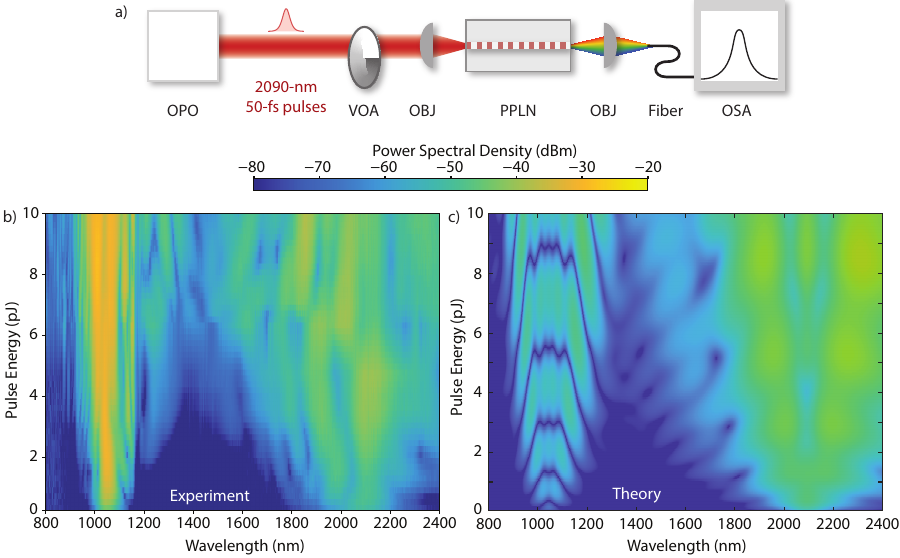}
\caption{\label{fig:jacobi_elliptic_data} a) Experimental setup for characterizing spectral broadening generated by quasi-static SHG. VOA: variable optical attenuator, OBJ: reflective objective, OSA: optical spectrum analyzer. Figure adapted with permission from~\cite{Jankowski2021SCG}, Copyright 2023 Authors.}
\end{figure*}

The most rigorous test of quasi-static behaviors is the experimental study given in~\cite{Jankowski2021SCG}, which determines whether any additional dynamics are neglected by the quasi-static model. Figure~\ref{fig:jacobi_elliptic_data} shows a side-by-side comparison between the heuristic theory presented here and an experimentally measured power spectral density generated by the waveguide as a function of in-coupled pump pulse energy. In this case, we observe reasonable agreement between the theoretical and experimental power spectral density as a function of the input pulse energy. In both experiment and in theory the fundamental and second harmonic are observed to broaden for input pulse energies in excess of 100 femtojoules, with the two harmonics merging at the -40 dB level for pulse energies as low as 4-pJ. In addition, there are a number of qualitative similarities between the spectra observed in theory and experiment. For pulse energies between one to five picojoules, the power spectrum of the fundamental exhibits a local minimum around the carrier wavelength of 2090 nanometers. For pulse energies greater than five picojoules, this local minimum splits into two minima centered symmetrically around the carrier frequency, with a local maximum at 2090 nanometers. Similar patterns occur in the tails of the spectra. The spectrum of the fundamental forms successive local minima and maxima in the band between 1600 and 1800 nanometers with increasing pulse energy, and the second harmonic exhibits oscillatory tails between 1200 and 1400 nanometers. Given the sensitivity of these fine-scale spectral fringes to the time-domain phase envelope of the pulses, one might not expect these features to survive in the presence of realistic waveguide dispersion.

We close this section by emphasizing that quasi-static devices are one of the most accessible routes towards extremely low-power devices, and that the dynamical regimes of quasi-static devices are still a topic of active study. Early demonstrations of quasi-static devices in TFLN achieved saturated SHG with tens of femtojoules of input fundamental~\cite{Jankowski2020}, with similar demonstrations in silicon~\cite{Singh2020} and silicon nitride~\cite{Hickstein2019}. Soon after, phase-sensitive amplification was demonstrated in a quasi-static OPA~\cite{Ledezma2022}, as well as saturated OPG with record-low pulse energies~\cite{Jankowski2022}. The mechanisms for spectral broadening by quasi-static SHG and their scaling laws are studied in~\cite{Jankowski2021SCG}. More recent work has demonstrated ultrabroadband squeezing using quasi-static OPA~\cite{nehra2022few}. We discuss some technical limitations of quasi-static devices, and their potential resolution, in the following sections.

\subsubsection{The role of higher-order dispersion}\label{sec:Gouy-phase}

The heuristic solutions to the quasi-static equations of motion rely on the assumption that dispersion is negligible to arbitrary order. A natural question to ask is what effects higher-order dispersion can have in quasi-static devices. After all, these terms are not zero, they are simply too weak to reshape the envelopes. In this section, we show that higher-order dispersion can be treated as an effective shift of the propagation constant associated with each field envelope. In a simplified limit, where the only contribution to higher-order dispersion is group velocity dispersion ($k''$), this process has a direct correspondence to the Gouy phase shift that occurs for focused Gaussian beams~\cite{gouy1890phase} via Akhmanov's space-time analogy~\cite{akhmanov1969nonstationary}. Here we obtain closed form solutions that describe this Gouy phase for arbitrary pulse envelopes in the presence of any dispersion relation, and discuss limits on the conversion efficiency imparted by this process. These propagation constants undergo dynamical changes as the evolving envelopes are distorted by saturation. As a result, the envelopes effectively become phase-mismatched, which causes them to undergo Jacobi-elliptic oscillations in the highly depleted limit.

The approach taken here is motivated by the Gouy phase observed for a Gaussian beam going through a focus. Near the focus, the Gaussian envelope acquires an additional phase shift, $\phi_g = \tan^{-1}(z/z_r)$, where $z_r$ is the Rayleigh length of the Gaussian beam. This effect contributes a $\pi$ phase shift as the Gaussian beam goes from a positive to negative radius of curvature. An analogous effect occurs for a Gaussian pulse $\exp((t/\tau_0)^2)$ in the presence of second-order dispersion. As the pulse goes from positively to negatively chirped during propagation an additional phase shift is accumulated, given by $\phi_p = -\frac12 \tan^{-1}(z/z_d)$. Here $z_d=\tau_0^2/(2k'')$ is the dispersion length of the pulse in the presence of second-order dispersion. Our goal is to generalize this effect to arbitrary envelopes in the presence of arbitrary dispersion.

To treat this generalized Gouy phase, we assume that the higher-order dispersion is sufficiently weak that the pulse envelope does not change shape. Under these assumptions, we can use an ansatz of the form $A(z,t) = A(0,t)\exp(i \phi(z))$. In the frequency domain, the field envelope evolves in a co-moving frame as
\begin{equation}
	A(z,\Omega) = \exp(-iD_{\text{int}}(\Omega)z)A(0,\Omega),
\end{equation}
which leads to the propagation equation,
$$\partial_z A(z,\Omega) = -iD_{\text{int}}(\Omega)A(z,\Omega),$$
with our usual definition of the dispersion operator,
$$D_{\text{int}}(\Omega) = \left(\frac{k_\omega''}{2}\Omega^2 + \frac{k_\omega'''}{3!}\Omega^3+...\right)$$.
The time-domain field envelope is given by
$$A(z,t) = \int_{-\infty}^{\infty}\exp(i\Omega t)A(z,\Omega)\frac{d\Omega}{2\pi}.$$
Noting that our choice of phase reference has the peak of the pulse located at $t=0$, and that the only contribution of the phase to $t=0$ is the Gouy phase, we restrict our attention to the peak of the envelope given by
$$A(z,0) = \int_{-\infty}^{\infty}A(z,\Omega)\frac{d\Omega}{2\pi},$$
and define the Gouy phase as $\phi(z,0)$. The evolution of the peak field is given by
$$
\partial_z A(z,0) = \int_{-\infty}^{\infty} -iD_{\text{int}}(\Omega)A(z,\Omega)\frac{d\Omega}{2\pi}.
$$
Putting $A(z,0)=r\exp(i\phi)$ in phase-amplitude form,
$$
\left(\frac{\partial_z r}{r}+i\partial_z\phi \right)A(z,0) = \int_{-\infty}^{\infty} -iD_{\text{int}}(\Omega)A(z,\Omega)\frac{d\Omega}{2\pi},
$$
we find the rate change of the Gouy phase, or the local propagation constant, entirely in terms of a frequency-domain integral of the field envelopes,
\begin{equation}
\partial_z\phi  = \mathrm{Im}\left[\frac{\int_{-\infty}^{\infty} -iD_{\text{int}}(\Omega)A(z,\Omega)\frac{d\Omega}{2\pi}}{\int_{-\infty}^{\infty}A(z,\Omega)\frac{d\Omega}{2\pi}}\right].\label{eqn:Gouy}
\end{equation}
Eqn.~\ref{eqn:Gouy} is the main result of this analysis, which allows us to solve for the Gouy-phase of an arbitrary pulse shape in the presence of arbitrary waveguide dispersion.

We can verify Eqn.~\ref{eqn:Gouy} by considering a Gaussian pulse, in which case we have
\begin{equation*}
    \partial_z \phi_\text{Gaussian} = -\frac{1}{2}\frac{z_d}{z^2+z_d^2}.
\end{equation*}
This agrees exactly with our previous expression, $\phi_p = -\frac12 \tan^{-1}(z/z_d)$. Further verification can be obtained by comparing  the numerically calculated Gouy phase for $\sech$ and $\sech^2$ envelopes with Eqn.~\ref{eqn:Gouy}, and we observe strong agreement in every case. We note here that while Eqn.~\ref{eqn:Gouy} rarely admits closed-form integrals for most pulse envelopes, we can often find closed-form answers for the propagation constant of an unchirped pulse,
\begin{equation}
\partial_z\phi  = \mathrm{Im}\left[\frac{\int_{-\infty}^{\infty} -iD_{\text{int}}(\Omega)A(0,\Omega)\frac{d\Omega}{2\pi}}{\int_{-\infty}^{\infty}A(0,\Omega)\frac{d\Omega}{2\pi}}\right].\label{eqn:Gouy_k}
\end{equation}
For quasi-static nonlinear interactions with unchirped pulse envelopes, Eqn.~\ref{eqn:Gouy_k} is often sufficient for modeling the dynamical phase-shifts that occur during pulse propagation. For the simple case where $D_{\text{int}}$ is approximated by group velocity dispersion, we find that pulse envelopes of the form $\sech(t/\tau)$ acquire a shift of their propagation constant given by $- k''/(2 \tau^2)$. Similarly, a pulse envelope of the form $\sech^2(t/\tau)$ acquires a shift given by $- k''/(6\tau^2)$.

These behaviors suggest that the dominant effect played by higher order dispersion is simply to change the propagation constants of the envelopes, which can be compensated by an overall change in the grating period used to phase-match the harmonics~\cite{Major2008}. In practice, the field envelopes become distorted in the saturated limit, which causes the phase-mismatch between these envelopes to dynamically evolve during propagation. As a simple example, we consider second-harmonic generation of a sech-pulse of fundamental in the presence of a small group-velocity dispersion for the second harmonic, $k_{2\omega}''$. For weak amounts of depletion, the second harmonic envelope is given by $\sech^2(t/\tau)$, which has a shift of the propagation constant given by $- k_{2\omega}''/(6\tau^2)$. In the highly saturated limit, the second-harmonic envelope will become $\sech(t/\tau)$, with a corresponding shift of the propagation constant given by $- k_{2\omega}''/(2\tau^2)$. For typical numbers ($k_{2\omega}'' = 100$ fs$^2$/mm and $\tau = 100$ fs), the phase-mismatch between the undepleted and depleted cases can change by 3300 m$^{-1}$, or a full $\pi$ of phase error in a 1-mm-long waveguide. These large phase-shifts will cause the envelopes to undergo Jacobi-elliptic oscillations, which prevents SHG from being driven far into saturation. While these Gouy phase-shifts can, in principle, be compensated for by solving the quasi-static equations of motion and evaluating Eqn.~\ref{eqn:Gouy_k} with the full dispersion relations of the waveguide, this technique only compensates for the accumulated Gouy phase for a given input pulse energy. In the presence of any small phase error, such as those occurring due to fabrication errors, the fields will undergo back-conversion in a highly-saturated limit. As a result, this interplay between the evolution of the saturated pulse envelopes and higher-order dispersion will ultimately limit the degree of saturation that can be achieved in any real device.

\subsubsection{Fully-static nonlinear optics}\label{sec:FSNLO}

In the previous sections we saw that saturated nonlinearities can greatly distort the interacting envelopes due to the presence of a small phase-mismatch, even in the absence of dispersion. Furthermore, even when the fundamental and second harmonic are phase-mismatched (including a Gouy phase), these pulse distortions can couple to higher-order dispersion to create a dynamical phase-mismatch, which ultimately limits the degree of conversion attainable in pulsed devices. In this section we discuss recent theoretical proposals aimed at eliminating these effects by a process known as temporal trapping~\cite{Yanagimoto2022_temporal,Babushkin2022_temporal}. In this approach, the pulses are guided in time, in addition to space, by cross-phase modulation (XPM) between a bright ``trapping'' pulse and the two interacting harmonics. We will later see that temporal trapping can greatly simplify the behavior of broadband quantum systems by eliminating multimode behaviors. When properly designed, temporal traps can recover single-mode behavior where the interacting modes are now pulses, rather than Fourier modes.

The dynamics of temporally-trapped waves are captured by the coupled-wave equations with an additional contribution from XPM. We assume that all three waves are group-velocity matched, and include dispersion up to second order. We also assume that the dispersion length of the trap pulse is much longer than any relevant nonlinear length, and that the field associated with the trap pulse is sufficiently large that it is unperturbed by any interactions with the trapped waves. In this limit, the coupled-wave equations become
\begin{align}
    \partial_z A_\omega(z,t) = &\frac{i}{2} k_{\omega}'' \partial_t^2 A_\omega(z,t)-i\kappa A_{2\omega}(z,t)A_{\omega}^*(z,t)-i\gamma_\text{xpm,1}|A_\text{trap}|^2A_{\omega},\label{eqn:CWE_trap_fh}\\
    \partial_z A_{2\omega}(z,t) = &\frac{i}{2} k_{2\omega}'' \partial_t^2 A_{2\omega}(z,t)-i\kappa A_{\omega}^2(z,t)-i\gamma_\text{xpm,2}|A_\text{trap}|^2A_{2\omega}.\label{eqn:CWE_trap_sh}
\end{align}
Unlike quasi-static devices, here we consider the limit where cross-phase modulation and group-velocity dispersion are sufficiently strong that the second-order nonlinearity is a weak perturbation to these interactions. In this case, the pulse envelope of each harmonic is determined by the solutions for a one dimensional waveguide, where now the refractive index variation is in time (due to XPM) instead of space. Without loss of generality we consider the fundamental,
\begin{align}
    \partial_z A_\omega(z,t) = &\frac{i}{2} k_{\omega}'' \partial_t^2 A_\omega(z,t)-i\gamma_\text{xpm,1}|A_\text{trap}(t)|^2A_{\omega}(z,t).\label{eqn:wg_trap_fh}
\end{align}
Equation~\ref{eqn:wg_trap_fh} admits eigenfunctions of the form $u_\omega(t)\exp(-i k_{\text{NL},\omega} z)$, with associated eigenvalue $-ik_{\text{NL},\omega}$. For the special case of $A_\text{trap}(t) = A_0\sech(t/\tau)$, where $\gamma_\text{xpm,1}A_0^2 = -k_\omega''\tau^{-2}$, the lowest-order eigenmode for the fundamental is given by $\sech(t/\tau)$, and the associated eigenvalue given by $k_{\text{NL},\omega} = -\frac12 k_\omega''\tau^{-2}$. We note here that the eigenvalue of this mode is equivalent to the Gouy phase of an unchirped $\sech$ pulse in the presence of second-order dispersion.

In general, Eqn.~\ref{eqn:wg_trap_fh} admits a complete orthonormal basis of eigenfunctions, with a similar expression holding for the second harmonic. We can therefore expand both harmonics in their respective bases,
\begin{align*}
	A_\omega(z,t) = \sum_m a_{\omega,m}(z) u_{\omega,m}(t)\exp(-ik_{\text{NL},\omega,m}z)\\
	A_{2\omega}(z,t) = \sum_m a_{2\omega,m}(z) u_{2\omega,m}(t)\exp(-ik_{\text{NL},2\omega,m}z).
\end{align*}
Noting that $u_m$ has units of s$^{-1/2}$ (\textit{i.e.} $\int u_m^*(t)u_n dt = \delta_{mn}$, where $\delta_{mn}$ is the Kronecker delta function), $|a_m|^2$ is the energy contained in mode $m$. In the presence of the second-order nonlinear coupling, the coefficients $a_{\omega,m}$ and $a_{2\omega,m}$ can now evolve during propagation. Their equations of motion can be extracted from Eqns.~\ref{eqn:CWE_trap_fh}-\ref{eqn:CWE_trap_sh} using orthogonality,
\begin{align*}
	\partial_z a_{\omega,\ell}(z) = -i\kappa \sum_{m,n} a_{2\omega,n}(z) a_{\omega,m}^*(z) \exp(-ik_n z+ik_m z+ik_\ell z)\int_{-\infty}^{\infty}u_{2\omega,n}(t)u_{\omega,m}^*(t)u_{\omega,\ell}^*(t)dt,\\
	\partial_z a_{2\omega,n}(z) = -i\kappa \sum_{\ell,m} a_{\omega,\ell}(z) a_{\omega,m}(z) \exp(ik_n z-ik_m z-ik_\ell z)\int_{-\infty}^{\infty}u_{2\omega,n}^*(t)u_{\omega,m}(t)u_{\omega,\ell}(t)dt.
\end{align*}
Finally, noting that higher-order eigenfunctions have larger eigenvalues, we can eliminate interactions between all but one set of phase-matched eigenfunctions by considering the limit where both $k_\omega''$ and $k_{2\omega}''$ are large. This corresponds to having larger splittings between adjacent eigenvalues, which strongly phase-mismatches coupling into these higher order modes. With this approximation, we recover the CW coupled-wave equations where now the two interacting modes are pulses,
\begin{align*}
	\partial_z a_{\omega,\text{trap}}(z) = -i\kappa a_{2\omega}(z) a_{\omega,\text{trap}}^*(z) \int_{-\infty}^{\infty}u_{2\omega}(t)u_{\omega}^*(t)u_{\omega}^*(t)dt,\\
	\partial_z a_{2\omega,\text{trap}}(z) = -i\kappa a_{\omega,\text{trap}}^2(z) \int_{-\infty}^{\infty}u_{2\omega,n}^*(t)u_{\omega,m}(t)u_{\omega,\ell}(t)dt.
\end{align*}

Taking the $\sech(t/\tau)$ soliton envelope as an example, these equations evaluate to
\begin{align*}
	\partial_z a_{\omega,\text{trap}}(z) = -i\kappa a_{2\omega,\text{trap}}(z) a_{\omega,\text{trap}}^*(z)\frac{\pi}{4\sqrt{2\tau}},\\
	\partial_z a_{2\omega,\text{trap}}(z) = -i\kappa a_{\omega,\text{trap}}^2(z) \frac{\pi}{4\sqrt{2\tau}}.
\end{align*}
Finally, to compare these equations to a continuous-wave interaction, we convert the field coefficients in Joules$^{1/2}$ back to Watts$^{1/2}$ by a factor of pulse repetition frequency, $T_r$. Defining the amplitude in mode $a_{\text{trap}}$ as $A_{\text{trap}}\sqrt{T_r} = a_{\text{trap}}$, we have
\begin{align*}
	\partial_z A_{\omega,\text{trap}}(z) = -i\kappa A_{2\omega,\text{trap}}(z) A_{\omega,\text{trap}}^*(z)\frac{\pi}{4}\sqrt{\frac{T_r}{2\tau}},\\
	\partial_z A_{2\omega,\text{trap}}(z) = -i\kappa A_{\omega,\text{trap}}^2(z)\frac{\pi}{4}\sqrt{\frac{T_r}{2\tau}}.
\end{align*}
We see that the power requirements of trapped pulses are reduced relative to a continuous-wave interaction by the duty cycle,
\begin{equation}
	\kappa_\text{trap} = \kappa \frac{\pi}{4}\sqrt{\frac{T_r}{2\tau}}.
\end{equation}
While temporal trapping comes with significant experimental overhead, we will later see in the context of quantum nonlinear optics that the benefits seem well worth the difficulty. Using this technique, the pulse envelopes no longer change shape during propagation. As a result, we can completely recover the behaviors of single-mode interactions, while retaining the power reductions associated with using short pulses. We note here that higher order dispersion manifests as additional couplings between the pulse envelopes. These effects can be further suppressed by increasing $k''$, which increases the splitting between the eigenvalues, thereby increasing the phase-mismatch between all of the envelopes and suppressing this loss channel. Therefore, the benefits of temporal trapping extend to more general cases containing higher order dispersion, and may enable arbitrarily large degrees of saturation. In the few-photon limit, we will see that temporal trapping can be used to realize Rabi-oscillations between a biphoton of fundmental and a single photon of second harmonic.

\subsubsection{Approaches to dispersion engineering}

Throughout our discussion of pulsed interactions we have focused on the roles played by the dominant dispersion orders of the interacting waveguide modes. At this time, all demonstrations of dispersion-engineered $\chi^{(2)}$ interactions have relied on relatively simple approaches to dispersion engineering, in which the geometry of a ridge waveguide is chosen to realize a desirable group velocity mismatch, group velocity dispersion, or both. However, as more complicated multi-wave (or short wavelength) interactions become of interest a natural progression for $\chi^{(2)}$ nonlinear photonics is the development of more sophisticated approaches to dispersion engineering that enable greater control over each of the interacting waves. We briefly review these recently developed approaches here.

\begin{figure}
    \centering
    \includegraphics[width=\textwidth]{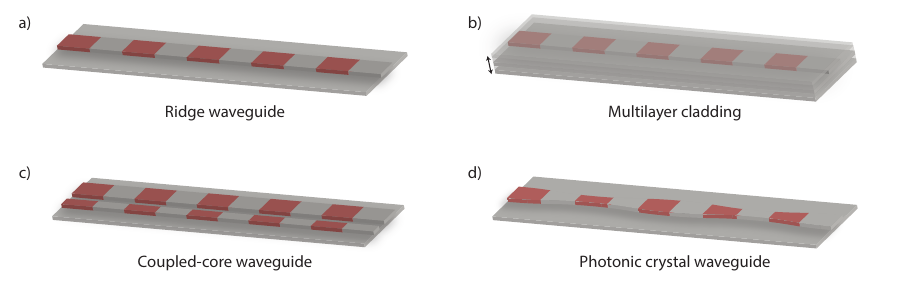}
    \caption{In addition to geometric dispersion engineering enabled by tightly confining ridge waveguides shown in (a), emerging approaches that enable greater design flexibility include b) multilayer claddings, c) coupled-core waveguides, and d) photonic crystal waveguides. We note here that these approaches are readily extended to dispersion-engineered nonlinear resonators.}
    \label{fig:new_disp_eng}
\end{figure}

Figure~\ref{fig:new_disp_eng} summarizes several emerging approaches to dispersion engineering in nonlinear nanophotonics. The simplest extension of the ridge waveguide is the use of a multilayer cladding, in analogy to the multilayer claddings used to realize dispersion-shifted fibers. For both ridge waveguides and multilayer structures, the shift of the group velocity induced by the structure of the waveguide is best intuited using Eqn.~\ref{eqn:disp2}, copied below for convenience,
\begin{align*}
k'_\mu(\omega) &= \frac{\frac{1}{4}\int_{A_\infty}\mu_0\mathbf{H}_\mu\cdot\mathbf{H}_\mu^*+\mathbf{E}_\mu^*\cdot\partial_\omega(\omega\epsilon(x,y,\omega))\cdot\mathbf{E}_\mu dxdy}{\frac{1}{2}\int_{A_\infty}\mathrm{Re}\left(\mathbf{E}_\mu(x,y,\omega)\times\mathbf{H}^*_\mu(x,y,\omega_2)\right)\cdot\hat{\mathbf{z}}dxdy}.
\end{align*}
Here we see that the inverse group velocity of each mode is given by the overlap of the mode fields with the underlying materials. This facilitates a simple picture of dispersion engineering; to modify the group-velocity mismatch of the interacting waves one can introduce materials that overlap with the evanescent tails of the long-wavelength fields (\textit{e.g.} the signal, or the trapping pulse), which modifies the group velocity of these waves relative to the more tightly confined short-wavelengths. Since  multilayer structures can be made of many different materials, and each layer may be deposited with a different thickness, in principle this simple approach may introduce sufficiently many degrees of freedom to enable much finer control over each wave.

A number of more sophisticated approaches have been recently studied in $\chi^{(3)}$-nonlinear media, include the development of coupled-core waveguides~\cite{Guo2020,lukashchuk2019advanced}, and photonic crystal resonators~\cite{Lucas2023,moille2023fourier,Vercruysse2020}. All of these approaches rely on hybridization between pairs of modes to control the propagation constant of the emergent supermode. In the case of coupled-core waveguides, the supermodes are formed by linear combinations of evanescently-coupled waveguide modes, where the gap between the two waveguides controls the coupling strength, and the geometry of the two waveguides determines the dispersion relations of the two uncoupled modes. This approach has been used to control the dynamics of Kerr solitons and $\chi^{(3)}$ supercontinua. In the case of photonic crystal resonators, each waveguide mode is coupled to a backwards-propagating mode by the Fourier components of the surrounding dielectric perturbations. In~\cite{Vercruysse2020} the dispersion relations of a photonic crystal were engineered to realize slow light by inverse-designing the hole pattern cut around the waveguide. In~\cite{Lucas2023,moille2023fourier}, inverse-designed surface corrugations were used to realize custom-tailored dispersion relations with unprecedented control over the operating regimes of short pulses evolving under the influence of $\chi^{(3)}$ nonlinearities. These approaches are particularly flexible since there is a one-to-one map between the k-vector of the surface corrugation and the pair of coupled forward and backward going modes. This one-to-one mapping allows the couplings of each mode pair to be engineered independently using the amplitudes of each Fourier component of the surface corrugation. 

At this time, all of these approaches are relatively new and have yet to be applied to devices with $\chi^{(2)}$ nonlinearities. Further development of these techniques and their application to realizing both low-power operation and qualitatively new dynamical regimes is an extremely ripe area of study. We note here that ultimately the approaches most likely to become commonly adopted will enable greater design flexibility while operating with realistic fabrication tolerances and without significant additional loss. Thus far, the studies of \cite{Lucas2023,moille2023fourier} suggest that these approaches may be realized with minimal degradation to typical waveguide losses, with demonstrated quality factors of 1-2 million.

\subsection{Nonlinear interactions in resonators}~\label{sec:resonators}


Until now, we have considered traveling-wave interactions in nonlinear waveguides. At this time, the lowest power nonlinear devices all use resonators to enhance the interaction length of the nonlinear interaction. In this section, we will connect the behaviors of traveling-wave devices to those of resonators by using discrete maps. This approach is commonly used to treat resonators that comprise many discrete components~\cite{Haus1975,Ikeda1979,Lugiato1987,Haus2000,Hamerly2016}, where each component is modeled independently and the output from one component serves as the input into the next. In the context of nonlinear optics, the use of discrete maps for modeling resonators is sometimes colloquially referred to as the ``Ikeda map''~\cite{Hamerly2016} since Ikeda first applied discrete maps to develop models of resonators with $\chi^{(3)}$ nonlinearities~\cite{Ikeda1979}.

In the high-finesse limit, where the fields experience small changes on each round trip of the cavity, these discrete maps can converted to a differential equation that describes how the fields evolve over many round trips of the cavity~\cite{Haus1975,Haus2000,Hamerly2016}. We will see that these equations are identical to the traveling-wave coupled wave equations with a coarse time-scale, $T$, taking the role of the propagation coordinate $z$, and the addition of driving and loss terms. A key difference, is that the propagation coordinate is now the number of round trips in the resonator, which effectively increases the nonlinear length of a waveguide into a loss length.

In anticipation of our transition to quantum nonlinear optics, where the mean fields are related to intra-cavity photon number, $n$, or the number density, we will explicate concise rules for converting between the instantaneous power envelope $|A(z,t)|^2$ and the energy density, $|u(z,t)|^2$. This analysis is restricted to the case of classical fields, and will later be used to establish a correspondence between the parameters in quantum equations of motion generated by a phenomenological Hamiltonian and the known parameters of a classical device. We emphasize here that discrete maps are a general technique that can be extended to contexts beyond the presentation contained in this tutorial. As an example of how discrete maps can be applied to quantum states, the evolution of a Gaussian state undergoing both nonlinear propagation and measurement feedback is presented in~\cite{Ng2022sampling}.

\begin{figure}
    \centering
    \includegraphics[width=\linewidth]{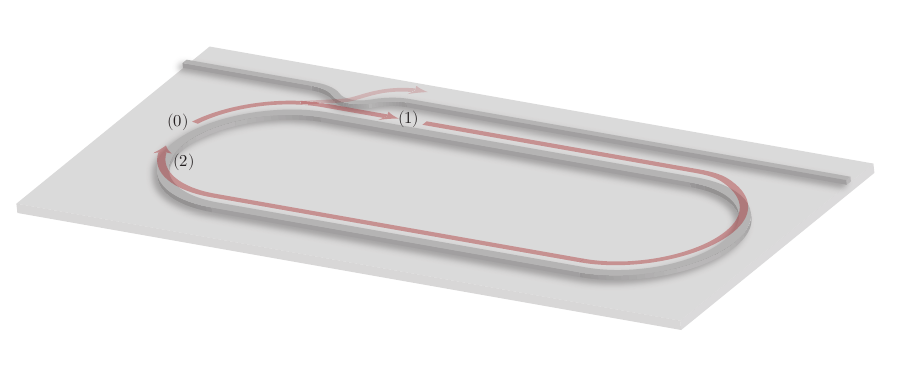}
    \caption{Propagation in a linear resonator can be broken up into a sequence of discrete steps. For this example, propagation from (0) to (1) is modeled using a 2x2 beamsplitter, which partially outcouples the intracavity field to a bus waveguide. The closed loop from (1) to (2) contains both the propagation loss of the resonator and the phase accumulated during linear propagation. Taken together, (0) through (2) represent one single pass around the resonator; a full model of resonator dynamics can then be obtained by iterating this process over many round trips.}
    \label{fig:2.4.Ikeda.1}
\end{figure}

\subsubsection{Linear propagation in a resonator}
 
To best illustrate the process of applying discrete maps and converting them to differential equations in the high-finesse limit, we begin by considering the simple case of linear propagation. We start by instantiating a pulse $A_{\omega}(t)$ at a reference position $z_0$ in the cavity. Propagation around the cavity is broken into a sequence of steps, as shown in Fig.~\ref{fig:2.4.Ikeda.1}, between discrete positions, $z_n$, in the cavity. Starting with a reference position $z_0$ located before the directional coupler (labeled $(0)$ in Fig.~\ref{fig:2.4.Ikeda.1}), the field at each successive point $A_\omega(z_n,t)$ will be denoted with the short-hand $A_\omega(z_n) \equiv A_\omega^{(n)}(t)$.

The first step in this example of a linear cavity is the directional coupler, here assumed to be lossless, which out-couples the intracavity field to a bus waveguide, $A_{\omega}^{\text{oc}} = i t_{\text{oc},\omega}A_\omega^{(0)}$. Here $it_{\text{oc},\omega}$ is often referred to as the transmission coefficient or the scattering coefficient for the ``cross port''. The remaining intracavity field is given by $A_{\omega}^{(1)} = r_{\text{oc},\omega} A_{\omega}^{(0)}$, where $r_{\text{oc},\omega}$ is colloquially referred to as the reflection coefficient, or the scattering coefficient for the ``bar port''. In power-normalized units, the scattering coefficients satisfy $|r_{\text{oc},\omega}|^2+|t_{\text{oc},\omega}|^2 \equiv R_{\text{oc},\omega} + T_{\text{oc},\omega}= 1$ for a lossless directional coupler. For simplicity, we have ignored any dispersion in $r_{\text{oc},\omega}$ and $t_{\text{oc},\omega}$; the more general case is easily treated by Fourier-transforming $A_{\omega}(t)$ and applying $r_{\text{oc},\omega}(\Omega)$ and $t_{\text{oc},\omega}(\Omega)$ to each frequency $\omega + \Omega$. When the resonator is driven by a pump field, $A_\omega^{(p)}(t)$, this field is in-coupled to the resonator through the ``cross port'',
\begin{equation}
    A_{\omega}^{(1)}(t) = r_{\text{oc},\omega} A_{\omega}^{(0)}(t) + it_{\text{oc},\omega} A_{\omega}^{(p)}(t).
\end{equation}

The second step is dispersive propagation around the resonator. Here we can either propagate each frequency independently, or equivalently, we can use the propagation rule for time-domain envelopes (Eqn.~\ref{eqn:broadband_SHG5}, as derived in Sec.~\ref{sec:pulse_prop}), 
\begin{equation}
\partial_z A_\omega(z,t) = -k_\omega'\partial_t A_\omega(z,t) + v_{g,\text{ref}}^{-1}\partial_t A_\omega(z,t)  - iD_{\text{int},\omega}(\partial_t)A_\omega(z,t),\label{eqn:resonator_pulse_prop}\\
\end{equation}
where $v_{g,\text{ref}}$ is our arbitrary choice of reference group velocity. We can additionally add in propagation loss, $\partial_z A_\omega = -\frac12 \alpha_{\ell,\omega}A_\omega$, where $\alpha_{\ell,\omega}$ is the power loss coefficient. After linear propagation, the pulse envelope arriving at $z_2$ is attenuated by $\exp(-\alpha_{\ell,\omega}L_\text{cav}/2)$, delayed by a group delay $T_{\text{rt},\omega} = k_\omega' L_\text{cav}$, and deformed by higher-order dispersion,
\begin{equation}
    A_{\omega}^{(2)}(t) = \exp\left(-\frac{\alpha_{\ell,\omega}}{2}L_\text{cav}-i D_{\text{int},\omega}(\partial_t)L_\text{cav})A_{\omega}^{(1)}(t-T_{\text{rt},\omega}-v_{g,\text{ref}}^{-1}L_\text{cav}\right).
\end{equation}
Here, to prevent confusion between the round-trip time of the cavity and the power transmission coefficient of the outcoupler, we have labeled them $T_\text{rt}$ and $T_\text{oc}$, respectively. For now, we consider the case where $v_{g,\text{ref}}^{-1} = 0$.

Noting that $A_{\omega}^{(2)}$ is the field envelope after one round-trip of the cavity, we \emph{overload} the variable $A_{\omega}$ to have two arguments, $A_\omega(m,t)$, where $m$ is an integer describing the number of round trips in the cavity, and $t$ describes the time-domain waveform that passes through the reference position $z_0$ on the $m^{\text{th}}$ round trip of the cavity. The discrete map, $\mathcal{M}$, iterates the round trip number $m$ by applying the sequence of steps used to model the cavity, $A_\omega(m+1,t) = \mathcal{M}A_\omega(m,t)$. Successive round trips are given by $A_\omega(m+n,t) = \mathcal{M}^nA_\omega(m,t)$. The full time-history of the intracavity electric field can be constructed using
\begin{equation*}
E(z_0,t) = \sqrt{\frac{2Z_0}{n_\omega A_{\mathrm{mode},\omega}}}\exp(i\omega t)\sum_n \big(\exp(-ik_\omega L_\text{cav})\mathcal{M}\big)^n A_\omega(m=0,t),
\end{equation*}
For linear propagation, the phase factor $\exp(-ik_\omega L_\text{cav})$ commutes with the discrete map $\mathcal{M}$. Later, in the context of nonlinear optics, this term will no longer commute and the overall phase accumulated by the fields on successive round trips can determine how power flows between the interacting waves. Noting that the field coupled out into the bus waveguide is given by $it_{\text{oc},\omega}A_\omega^{(0)}(t)$, the full time-history of the pulse train emerging in the bus waveguide is simply given by $E_{\text{oc}}(t) = -it_{\text{oc},\omega}E(z_0,t) + r_{\text{oc},\omega} E^{(p)}(t).$

Having established that the field envelope on successive passes $A_\omega(m,t)$ is sufficient to synthesize the full time-history of the intracavity field in the resonator, for convenience we now choose $v_{g,\text{ref}}^{-1}=k_\omega'=v_{g,\omega}^{-1}$. The time-history of the intracavity electric field is now obtained by time-delaying each of the successive envelopes in the co-moving frame by $T_{\text{rt},\omega}$,
\begin{equation}
E(z_0,t) = \sqrt{\frac{2Z_0}{n_\mu A_{\mathrm{mode},\mu}}}\exp(i\omega t)\sum_n \big(\exp(-ik_\omega L_\text{cav})\mathcal{M}\big)^n A_\omega(0,t-T_{\text{rt},\omega}).\label{eqn:Ikeda_field_synthesis}
\end{equation}
In the co-moving frame, the discrete map for the cavity is given by
\begin{equation}
A_\omega(m+1,t) = r_{\text{oc},\omega}\exp(-\frac{\alpha_{\ell,\omega}}{2}L_\text{cav}-i D_{\text{int},\omega}(\partial_t)L_\text{cav})A_{\omega}(m,t) + i\tilde{t}_{\text{oc},\omega} A_{\omega}^{(p)}(t),\label{eqn:Ikeda_linear_comoving}
\end{equation}
where $\tilde{t}_{\text{oc},\omega} = t_{\text{oc},\omega}\exp(-\frac{\alpha_{\ell,\omega}}{2}L_\text{cav}-i D_{\text{int},\omega}(\partial_t)L_\text{cav})$ has absorbed the effects of propagating the pump pulse into the definition of the transmission coefficient (note that $|\tilde{t}_{\text{oc},\omega}|^2 + |r_{\text{oc},\omega}|^2 \neq 1$).

While Eqn.~\ref{eqn:Ikeda_linear_comoving} is sufficient for treating pulse propagation in \emph{any} linear cavity, we are often concerned with the high-finesse limit, where the change on each round trip is small. In this case we rewrite $r_{\text{oc},\omega}=\exp(\ln(r_{\text{oc},\omega}))$ and, assuming that the change in the field envelope due to dispersive pulse propagation is small, we series expand the exponential,
\begin{align}
A_\omega(m+1,t) = &A_\omega(m,t) + i\tilde{t}_{\text{oc},\omega} A_{\omega}^{(p)}(t).\label{eqn:Ikeda_linear_highfinesse}
\\&+(-\ln(r_{\text{oc},\omega})-\frac{\alpha_{\ell,\omega}}{2}L_\text{cav}-i D_{\text{int},\omega}(\partial_t)L_\text{cav})A_{\omega}(m,t)\nonumber
\end{align}
Subtracting $A_\omega(m,t)$ from both sides of Eqn.~\ref{eqn:Ikeda_linear_highfinesse} and dividing by $T_{\text{rt},\omega}$, we convert Eqn.~\ref{eqn:Ikeda_linear_highfinesse} into a differential equation for the field envelope,
\begin{equation}
\partial_T A_\omega(T,t) = (-\kappa_{\text{ex},\omega}-\kappa_{\text{in},\omega}-i D_{\text{int},\omega}(\partial_t)v_{g,\omega})A_{\omega}(m,t) + \frac{i\tilde{t}_{\text{oc},\omega}}{T_{\text{rt},\omega}} A_{\omega}^{(p)}(t).\label{eqn:master_eqn_linear_highfinesse}
\end{equation}
Here $\kappa_{\text{ex},\omega}=-\ln(r_{\text{oc},\omega}) T_{\text{rt},\omega}^{-1}$ is the extrinsic field loss rate due to coupling from resonator to the bus waveguide, and $\kappa_{\text{in},\omega}=\frac12 \alpha_{\ell,\omega}L_\text{cav}T_{\text{rt},\omega}^{-1}=\frac12 \alpha_{\ell,\omega}v_{g,\omega}$ is the intrinsic field loss rate due to propagation in the cavity. The propagation coordinate  of the envelope, $T$, is referred to as the ``slow'' or ``coarse'' time, $T = m T_{\text{rt},\omega}$, and the envelope coordinate $t$ is referred to as the ``fast'' time since $T$ describes how the pulse envelope $A_\omega(t)$ evolves over long timescales. The round-trip time of the cavity is commonly expressed using the free-spectral range around frequency $\omega$, $T_{\text{rt},\omega}^{-1} = \Delta f_{\text{fsr},\omega}$. We note here that when calculating $\kappa_{\text{ex},\omega}$, it is common practice to expand the feedback coefficient $-\ln(r_{\text{oc},\omega})$ as a Taylor series in the high-finesse limit, $-\ln(r_{\text{oc},\omega})=-\frac12\ln(1-T_{\text{oc},\omega}) \approx \frac12T_{\text{oc},\omega}$. However, this small error in the coefficient of an exponent often leads to meaningful differences in the predicted field after a few round trips in the cavity~\cite{siegman1986lasers}.

\subsubsection{A comment about phase references}\label{sec:resonator_frames}

As with traveling-wave nonlinear optics and our later treatment of quantum nonlinear optics, it is common to find a variety of phase references in use throughout the literature. Noting that the electric field synthesized by the summation in Eqn.~\ref{eqn:Ikeda_field_synthesis} contains terms that go as $(\exp(-ik_\omega L_\text{cav})\mathcal{M})^n$, a common and useful choice of convention is the rotating frame $\tilde{A}_\omega(z,t) = \exp(-i k_\omega z)A_\omega(z,t)$, in which case the overall phase factor is absorbed into the discrete map,
\begin{equation*}
\tilde{\mathcal{M}}^n = \big(\exp(-ik_\omega L_\text{cav})\mathcal{M}\big)^n.
\end{equation*} 
With this choice of phase reference, the full time history of the intracavity electric field is given by
\begin{equation}
E(z_0,t) = \sqrt{\frac{2Z_0}{n_\mu A_{\mathrm{mode},\mu}}}\exp(i\omega t)\sum_n \big(\tilde{\mathcal{M}}\big)^n A_\omega(m=0,t).\label{eqn:Ikeda_field_synthesis_rotating}
\end{equation}
While this choice of phase reference does not modify the results obtained by the discrete map, the differential equation obtained in the high-finesse limit for the complex field amplitude is now generalized to be able to treat off-resonant behavior. Assuming $\phi_\omega = k_\omega L_\text{cav} ~\text{mod}~2\pi$ is small, the field evolution is now given by
\begin{equation}
\partial_T \tilde{A}_\omega(T,t) = (-i\delta_\omega-\kappa_{\text{ex},\omega}-\kappa_{\text{in},\omega}-i D_{\text{int},\omega}(\partial_t)v_{g,\omega})\tilde{A}_{\omega}(m,t) + \frac{i\tilde{t}_{\text{oc},\omega}}{T_{\text{rt},\omega}} A_{\omega}^{(p)}(t),\label{eqn:master_eqn_detuned_highfinesse}
\end{equation}
where $\delta_\omega = \phi_\omega T_{\text{rt},\omega}^{-1}$ represents the frequency detuning between $\omega$ and the cavity resonance. We note here that the same equation of motion can be obtained by moving into a rotating frame where the phase reference is given by the nearest cavity mode, that is $\tilde{A}_\omega(z,t) = \exp(-i (k_\omega - k_\text{ref}) z)A_\omega(z,t)$, where $k_\text{ref} = k_\omega + v_{g,\omega}^{-1}\delta_\omega$ satisfies $k_\text{ref} L = 0~\text{mod}~2\pi$. We will see in the following sections that this latter choice of phase reference is useful for nonlinear optics, since the phase associated with each field envelope evolves slowly in the rotating frame, and the interplay between detuning and nonlinearity is now treated natively by the dynamics generated by the equations of motion for $\tilde{A}_\omega$. In each section, we will briefly mention how our choice of reference frame is related to the typical traveling-wave coupled-wave equations, before dropping the overhead tildes from the envelopes for compactness. In the following subsections we will repeat this analysis for SHG in the presence of $\chi^{(2)}$ nonlinearities. We note here that the more familiar extension of Eqn.~\ref{eqn:master_eqn_detuned_highfinesse} to $\chi^{(3)}$-nonlinear resonators is commonly referred to as the Lugiato–Lefever equation (LLE)~\cite{Lugiato1987}.

\subsubsection{Canonical examples of linear resonator behaviors}

Before we analyze nonlinear dynamics with discrete maps, we briefly consider some illustrative examples of linear propagation in resonators, and compare the solutions generated by the discrete map $\mathcal{M}$ to those generated by the equations for the field evolution in a high-finesse resonator. While the connection between these two models is seemingly straightforward, it is instructive to see where these formalisms disagree. 

\begin{figure}
\centering
\includegraphics[width=\linewidth]{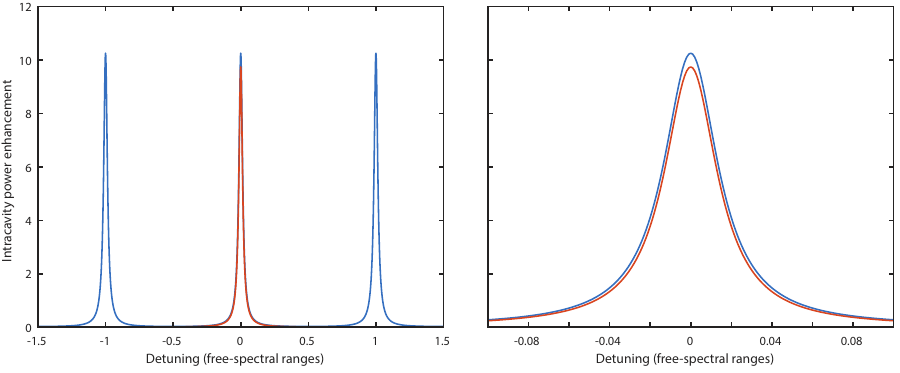}
\caption{\label{fig:Ikeda_resonance_comparison}Comparison of resonator lineshapes predicted by discrete maps (blue) and the continuous-time model (red) derived in the high-finesse limit. The continuous-time model predicts a Lorentzian lineshape for a single resonance centered around $\delta_\omega=0$, whereas the discrete map yields an Airy function (Eqn.~\ref{eqn:airy_func}). Here we have assumed $r = 0.95$ and $\exp(-\alpha_{\ell,\omega}L_\text{cav}/2)=0.95$. The slight differences of field enhancement and lineshape are due to our choice of $r = 0.95$; in the limit as $r\rightarrow 1$ and $\alpha_{\ell,\omega}\rightarrow 0$ the two approaches yield the same lineshape.}
\end{figure}

We begin by considering the textbook case of resonance in a cavity. Here we assume a driving field of the form, $A_\omega^{(p)}(t) = \sqrt{P_\text{in}}$, and solve for the steady-state intracavity field. In the discrete map approach, the field envelope within the rotating frame defined in Sec.~\ref{sec:resonator_frames} after one round trip of the cavity is given by
\begin{equation*}
A_\omega(m+1) = r_{\text{oc},\omega} \exp(-i\phi_\omega - \alpha_{\ell,\omega}L_\text{cav}/2)A_\omega(m)+ it_{\text{oc},\omega}\sqrt{P_\text{in}}.
\end{equation*}
While the full electric field can be synthesized by taking the infinite series in Eqn.~\ref{eqn:Ikeda_field_synthesis_rotating}, a simpler approach for finding the steady-state intracavity field is to assert that $A_\omega(m+1) = A_\omega(m)$. In this case we find that the intracavity power is given by Airy functions, as is well known for the analysis of Fabry-Perot resonators,
\begin{equation}
P_\omega = \frac{T_{\text{oc},\omega}P_\text{in}}{\left(
1-r\exp\left(-\frac{\alpha_{\ell,\omega}L_\text{cav}}{2}\right)\right)^2 + 4 r\exp\left(-\frac{\alpha_{\ell,\omega}L_\text{cav}}{2}\right)\sin^2\left(\frac{\phi_\omega}{2}\right)}.\label{eqn:airy_func}
\end{equation}
In the high-finesse model, the equations of motion are given by
\begin{equation*}
\partial_T A_\omega(T) = (-i\delta_\omega - \kappa_{\text{ex},\omega} - \kappa_{\text{in},\omega})A_\omega(t) + \frac{i t_{\text{oc},\omega}}{T_\text{cav}}\sqrt{P_\text{in}},
\end{equation*}
which yield a steady-state intra-cavity power given by a Lorentzian  distribution,
\begin{equation*}
P_\omega = \frac{2 \kappa_{\text{ex},\omega} P_\text{in} \Delta f_\text{fsr}}{\left(\kappa_{\text{ex},\omega} + \kappa_{\text{in},\omega}\right)^2 + \delta_\omega^2}.
\end{equation*}
Figure~\ref{fig:Ikeda_resonance_comparison} compares the Lorentzian lineshape predicted by the high-finesse model with the Airy function given by the discrete maps for $r = 0.95$ and an intracavity propagation loss given by $\exp(-\alpha_{\ell,\omega}L_\text{cav}/2)=0.95$. For the small out-coupling chosen here, the two approaches largely agree about both the shape of the resonance and the intracavity field enhancement. Both approaches also predict the same ``critical coupling'' condition for maximizing the intracavity field, $\kappa_{\text{in},\omega} = \kappa_{\text{ex},\omega}$. We note, however, that the continuous model can only be used to model isolated resonances and that discrete maps must be used to model low-finesse cavities. The intracavity field enhancement predicted by the two models disagrees by more than 10\% when $r<0.9$, and the continuous-time model produces unphysical results for $r < \exp(-0.5)$.

\begin{figure}
\centering
\includegraphics[width=\linewidth]{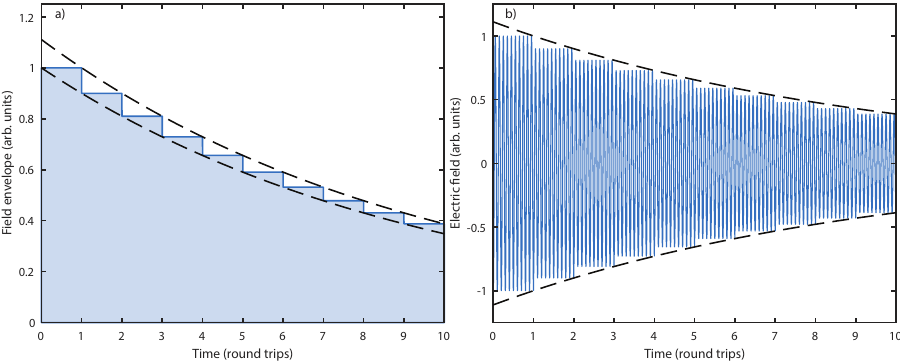}
\caption{\label{fig:Ikeda_ringdown}Simulation of a) the intracavity field envelope, and b) the intracavity electric field during cavity ringdown using discrete maps. Dashed black lines: exponential decay predicted from a continuous-time model.}
\end{figure}

We now consider the case of cavity ringdown spectroscopy, where we assume a cavity is driven on resonance and into steady state. Then the pump laser is turned off at $t=0$. In this case, the discrete map exhibits some peculiarities that disagree with the more well-known continuous model. We discuss how these disagreements can be resolved, and note that such differences do not appear in most practical examples. We begin with the familiar case of the continuous model, in which case $\partial_T A_\omega(T) = -\kappa_{\text{ex},\omega}A_\omega(T)$ is solved simply by $A_\omega(T) = A_\omega(0)\exp(-\kappa_{\text{ex},\omega} T)$. For the discrete model, we assume a field $A_\omega(m=0,t)$ is instantiated throughout the cavity from $t = 0$ to $t = T_{\text{rt},\omega}$. Ignoring propagation loss, the field on the $m^\text{th}$ round trip is given simply by $A_\omega(m,t) = r_{\text{oc},\omega}^m A_\omega(0,t)$. Fig.~\ref{fig:Ikeda_ringdown} shows the full time history of $A_\omega(m,t)$ and the resulting electric field generated by successive applications of the discrete map using Eqn.~\ref{eqn:Ikeda_field_synthesis_rotating} for $r = 0.9$. As opposed to the exponential decay predicted by the continuous-time model (Fig.~\ref{fig:Ikeda_ringdown}, dashed black lines), the discrete model predicts cavity ringdown to occur as a staircase with an exponentially decaying envelope. The origin of these differences is in how we have constructed the two envelopes. In the discrete case, we have instantiated the cavity with a uniform field distribution with the driving field turned off at $t=0$. The formation of a staircase is due to the finite time taken by the hard edge of the pump to propagate around the cavity. In the continuous-time model, we have associated a single quantity, $A_\omega(T)$, with the entire intracavity field, which assumes the field to be uniform by construction. This effectively distributes the loss throughout the cavity, resulting in a continuous exponential decay. We can recover the staircase solution in a continuous-time model by instead using the fast and slow time, $A_\omega(T,t)$, with $A_\omega(0,t)$ having a uniform distribution throughout the cavity. In this case, the solution $A_\omega(T,t) = A_\omega(0,t)\exp(-\kappa_{\text{ex},\omega}T)$ still exhibits a continuous decay with slow-time $T$. However, to reconstruct the time history of the electric field we must sample $A_\omega(T,t)$ at discrete points $T = m T_{\text{rt},\omega}$. The resulting envelopes are given by $A_\omega(m,t)\equiv A_\omega(m T_{\text{rt},\omega},t)$, where the left-hand side of the equality refers to the discrete envelopes and the right-hand side refers the the continuous-time envelopes. Using these envelopes with Eqn.~\ref{eqn:Ikeda_field_synthesis_rotating} yields the same solution as the discrete map.

\begin{figure}
\centering
\includegraphics[width=\linewidth]{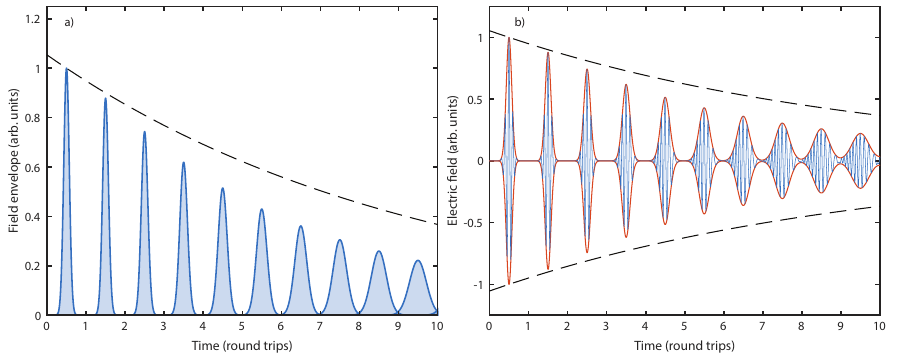}
\caption{\label{fig:Ikeda_pulse_ringdown} Simulation of a) the intracavity field envelope, and b) the intracavity electric field during cavity ringdown of an ultrafast pulse using discrete maps. Dashed black lines: exponential decay predicted from a continuous-time model. Here, the reduction of the field amplitude in excess of the continuous-time model is due to dispersive pulse spreading. }
\end{figure}

We close this section by considering a more general case to illustrate how the full time-history of the electric field can be constructed from $A_\omega(m,t)$. Here we instantiate a transform-limited Gaussian pulse in the cavity and include both outcoupling ($r=0.9$) and group velocity dispersion. Fig.~\ref{fig:Ikeda_pulse_ringdown}(a, shaded red) shows the field envelopes calculated by iterating the discrete map. The dashed black line shows the exponential decay of the peak value of the envelopes due to linear loss, with the peak amplitude of each envelope decaying more rapidly due to dispersive pulse spreading. Fig.~\ref{fig:Ikeda_pulse_ringdown}(b) shows the full time history of the electric field obtained from the coherent sum of each independent envelope (Eqn~\ref{eqn:Ikeda_field_synthesis_rotating}). We note here that when the pulses spread beyond a cavity round-trip time, the interference between the two envelopes produces the correct field in a linear cavity. In a nonlinear cavity, these effects are better simulated using the time evolution of a spatial envelope, rather than a temporal envelope, as developed later in this section. We note, however, that in most experimental settings the pulse envelopes rarely extend beyond a cavity free-spectral range, and therefore these considerations usually do not have a meaningful impact in any real system.

\subsubsection{Nonlinear propagation}

\begin{figure}
    \centering
    \includegraphics[width=\linewidth]{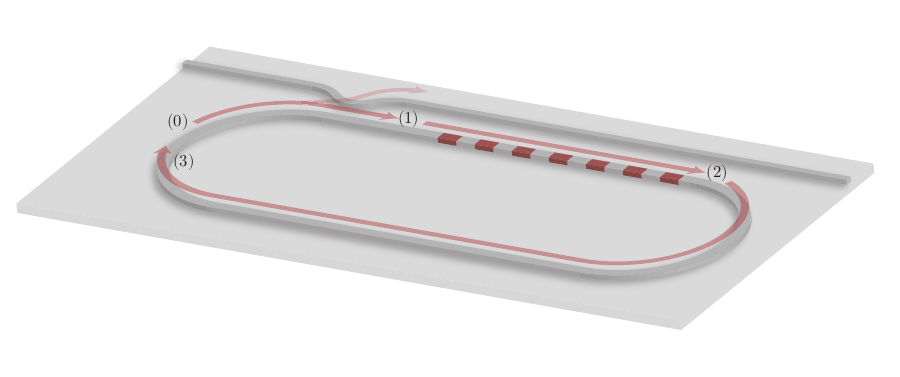}
    \caption{Propagation in a nonlinear resonator can be broken up into a sequence of three discrete steps. As before, propagation from (0) to (1) is modeled using a 2x2 beamsplitter and the closed loop from (2) to (3) contains both the propagation loss of the resonator and the phase accumulated during linear propagation. Propagation from (1) to (2) is now modeled by the coupled-wave equations.}
    \label{fig:2.4.Ikeda.2}
\end{figure}

We now consider a nonlinear resonator, as shown in Fig.~\ref{fig:2.4.Ikeda.2}. We proceed as before, but now track the evolution of both a fundamental and second harmonic envelope in the resonator. As discussed in Sec.~\ref{sec:resonator_frames}, we work in a rotating frame that is not only co-moving with the fundamental, but also has a phase reference subtracted from each envelope, $\tilde{A}_\omega(z,t) = \exp(-i (k_\omega - k_{\text{ref},\omega}) z)A_\omega(z,t)$, and $\tilde{A}_{2\omega}(z,t) = \exp(-i (k_{2\omega} - 2 k_{\text{ref},\omega}) z)A_{2\omega}(z,t)$, where the subscript $\omega$ in $k_{\text{ref},\omega}$ denotes that we have chosen the nearest resonance to $\omega$ as a reference. We reiterate here that this choice is arbitrary; for example, when studying intra-cavity OPA in subsequent sections both the reference velocity and $k_{\text{ref}}$ will be chosen using the $2\omega$ wave. In all further discussion in this sub-section, we will assume to be in a reference frame determined by $\omega$, and drop the overhead tildes from each envelope.

For the case considered here, the resonator is broken into three steps: i) a directional coupler, ii) a nonlinear section, and iii) a linear feedback loop. Each field evolves independently in each of the linear components. Starting with the directional coupler, we have
\begin{subequations}
\begin{align}
    A_{\omega}^{(1)}(t) &= r_{\text{oc},\omega} A_{\omega}^{(0)}(t) + it_{\text{oc},\omega} A_{\omega}^{(p)}(t),\label{eqn:Ikeda_NL_start_1}\\
    A_{2\omega}^{(1)}(t) &= r_{\text{oc},2\omega} A_{2\omega}^{(0)}(t) + it_{\text{oc},2\omega} A_{2\omega}^{(p)}(t).\label{eqn:Ikeda_NL_start_2}
\end{align}
\end{subequations}
Next, in a nonlinear section that extends from $z_1$ to $z_1 + L_\text{qpm}$, the fields evolve according to the coupled-wave equations,
\begin{subequations}\
\begin{align}
    A_{\omega}^{(2)}(t) - A_{\omega}^{(1)}(t) = \int_{z_1}^{z_1 + L_\text{qpm}}\partial_{z'} A_{\omega}(z',t) dz'\equiv \Delta A_{\omega,\text{NL}},\\
    A_{2\omega}^{(2)}(t) - A_{2\omega}^{(1)}(t) = \int_{z_1}^{z_1 + L_\text{qpm}}\partial_{z'} A_{2\omega}(z',t) dz'\equiv  \Delta A_{2\omega,\text{NL}},
\end{align}
\end{subequations}
where the field evolution occurs in a reference frame co-propagating with the signal, and the phase evolution of the envelopes determined by our choice of rotating frame,
\begin{align}
    \partial_z A_\omega(z,t) = & \left(-i v_{g,\omega}^{-1}\delta_\omega - iD_{\text{int},\omega}(\partial_t)\right)A_\omega(z,t)\label{eqn:CWE_Ikeda_1}\\
    &-i\kappa A_{2\omega}(z,t)A_{\omega}^*(z,t)\nonumber\\
    \partial_z A_{2\omega}(z,t)+\Delta k'\partial_t A_{2\omega}(z,t) = & \left(-i v_{g,\omega}^{-1}\delta_{2\omega}-iD_{\text{int},2\omega}(\partial_t)\right)A_{2\omega}(z,t)\label{eqn:CWE_Ikeda_2}\\
    &-i\kappa A_{\omega}^2(z,t).\nonumber
\end{align}
Here $v_{g,\omega}^{-1}\delta_{2\omega} = k_{2\omega} - 2k_{\text{ref},\omega}$, and as before, $\Delta k' = k_{2\omega}'-k_\omega'$ describes the temporal walk-off between the two envelopes. We note here that in addition to exchanging energy between the two fields, integrating the coupled-wave equations contributes a small group delay to the second harmonic due to temporal walk-off, and distorts the two envelopes due to the higher-order dispersion of the waveguide. We have ignored propagation loss in Eqn.~\ref{eqn:CWE_Ikeda_1}-\ref{eqn:CWE_Ikeda_2} since the total accumulated loss in the nonlinear section is often small compared to the total loss accumulated in the resonator. In the high-finesse and weakly-dispersive limit, these linear contributions, namely walk-off, dispersion, and propagation loss, decouple from nonlinear propagation and can be incorporated into the linear feedback section. Finally, propagating through the linear feedback section of length $L_\text{lin}$, where $L_\text{cav} = L_\text{lin} + L_\text{qpm}$, we have
\begin{subequations}
\begin{align}
    A_{\omega}^{(3)}(t) = \exp(-i v_{g,\omega}^{-1}\delta_{\omega}L_\text{lin}-\frac{\alpha_{\ell,\omega}}{2}L_\text{lin}-i D_{\text{int},\omega}(\partial_t)L_\text{lin})A_{\omega}^{(2)}(t),\label{eqn:Ikeda_NL_stop_1}\\
    A_{2\omega}^{(3)}(t) = \exp(-i v_{g,\omega}^{-1}\delta_{2\omega}L_\text{lin}-\frac{\alpha_{\ell,2\omega}}{2}L_\text{lin}-i D_{\text{int},2\omega}(\partial_t)L_\text{lin})A_{2\omega}^{(2)}(t).\label{eqn:Ikeda_NL_stop_2}
\end{align}
\end{subequations}
Eqns.~\ref{eqn:Ikeda_NL_start_1}-\ref{eqn:Ikeda_NL_stop_2} comprise the Ikeda map $\mathcal{\tilde{M}}$ for a $\chi^{(2)}$-nonlinear resonator driven by ultrafast pulses. At this stage, no approximations have been made, and the field evolution generated by successive applications of $\mathcal{\tilde{M}}$ is valid for any amount of loss, dispersion, or detuning.

\subsubsection{Nonlinear propagation in the high-finesse limit}

As with linear propagation, when the fields undergo small changes on each round trip, the discrete map $\tilde{\mathcal{M}}$ can be converted to an LLE-like equation. To treat the nonlinear interaction we rewrite $A_{\omega}^{(2)}(t)$ as
\begin{equation*}
	A_{\omega}^{(2)}(t) = A_{\omega}^{(1)}(t) + \Delta A_{\omega,\text{NL}} =  A_{\omega}^{(1)}(t)\left(1 + \frac{\Delta A_{\omega,\text{NL}}}{A_{\omega}^{(1)}(t)}\right)\approx A_{\omega}^{(1)}(t)\exp\left(\frac{\Delta A_{\omega,\text{NL}}}{A_{\omega}^{(1)}(t)}\right),\\
\end{equation*}
where the latter form assumes that the change $\Delta A_{\omega,\text{NL}}$ is small compared to $A_\omega^{(1)}(t)$. An identical expression holds for $A_{2\omega}^{(1)}(t)$ and $\Delta A_{2\omega,\text{NL}}$. The small change to each field is evaluated by assuming that each field is constant within a single pass of the resonator and integrating the coupled-wave equations,
\begin{align}
    \Delta A_{\omega,\text{NL}}(t) = & \left(-i v_{g,\omega}^{-1}\delta_\omega - iD_{\text{int},\omega}(\partial_t)\right)L_\text{qpm}A_\omega(z_1,t)\label{eqn:Integrated_CWE_Ikeda_1}\\
    &-i\kappa L_\text{qpm} A_{2\omega}(z_1,t)A_{\omega}^*(z_1,t)\nonumber\\
    \Delta A_{2\omega,\text{NL}}(t) = & \left(-i v_{g,\omega}^{-1}\delta_{2\omega}-\Delta k'\partial_t-iD_{\text{int},2\omega}(\partial_t)\right)L_\text{qpm}A_{2\omega}(z_1,t)\label{eqn:Integrated_CWE_Ikeda_2}\\
    &-i\kappa L_\text{qpm} A_{\omega}^2(z_1,t).\nonumber
\end{align}
Putting these steps together, we obtain an pair of coupled LLE-like equations,
\begin{subequations}
\begin{align}
\partial_T A_\omega(T,t) = & \frac{i\tilde{t}_{\text{oc},\omega}}{T_{\text{rt},\omega}} A_{\omega}^{(p)}(t) -i\kappa \frac{L_\text{qpm}}{T_{\text{rt},\omega}} A_{2\omega}(T,t)A_{\omega}^*(T,t)\label{eqn:master_eqn_nonlinear_highfinesse_1}\\
&(-i\delta_\omega-\kappa_{\text{ex},\omega}-\kappa_{\text{in},\omega}-i D_{\text{int},\omega}(\partial_t)v_{g,\omega})A_{\omega}(T,t),\nonumber\\
\partial_T A_{2\omega}(T,t) = & \frac{i\tilde{t}_{\text{oc},2\omega}}{T_{\text{rt},\omega}} A_{2\omega}^{(p)}(t) -i\kappa \frac{L_\text{qpm}}{T_{\text{rt},\omega}} A_{\omega}^2(T,t)-\frac{\Delta T_\text{rt}}{T_{\text{rt},\omega}}\partial_t A_{2\omega}(T,t)\label{eqn:master_eqn_nonlinear_highfinesse_2}\\
&(-i\delta_{2\omega}-\kappa_{\text{ex},2\omega}-\kappa_{\text{in},2\omega}-i D_{\text{int},2\omega}(\partial_t)v_{g,\omega})A_{2\omega}(T,t),\nonumber
\end{align}
\end{subequations}
where $\Delta T_{\text{rt}} = (v_{g,2\omega}^{-1}-v_{g,\omega}^{-1})L_\text{cav}$ is the timing mismatch between the two waves. The appearance of $v_{g,\omega}$ and $T_{\text{rt},\omega}$ rather than $v_{g,2\omega}$ and $T_{\text{rt},2\omega}$ in Eqn.~\ref{eqn:master_eqn_nonlinear_highfinesse_2} stems from our choice of $v_{g,\omega}$ as the reference velocity. Eqns.~\ref{eqn:master_eqn_nonlinear_highfinesse_1}-\ref{eqn:master_eqn_nonlinear_highfinesse_2} are better interpreted in their discrete map form; $A_\omega(m,t)$ and $A_{2\omega}(m,t)$ are the time-domain waveform passing through the reference position $z_0$ after the fundamental $\omega$ has executed $m$ round trips in the cavity. This asymmetry stems from the use of two time coordinates, a ``fast time'' $t$ and a ``slow time'' $T$, with the latter determined by the round-trip time of the reference wave, and will be eliminated later in this section when we move to time-propagating spatial envelopes.

\subsubsection{Examples: Resonant SHG and OPO}\label{sec:resonator_examples}

To illustrate how the above techniques are applied, and to develop intuition about nonlinear resonators, we briefly discuss a few textbook examples, namely resonant SHG and OPO. We first consider the case of a resonant fundamental and a traveling-wave second harmonic, which is naturally treated using discrete maps since the second-harmonic is not resonant. We then consider the case of a high-finesse cavity with all waves resonant, which is more naturally treated using the LLE-like formalism. We note here that even though these two configurations only contain a fundamental and second harmonic, they are sometimes colloquially referred to as doubly- and triply-resonant since two photons of fundamental correspond to resonant modes and one photon of pump corresponds to a resonant mode. This naming convention is tied to the observed resonance behaviors of degenerate OPOs, which only oscillate at discrete cavity lengths, similar to non-degenerate OPOs with a separately resonant signal and idler. 

For simplicity, we will assume continuous-wave interactions on resonance throughout this section. We note, however, that while this operating condition is a useful theoretical tool for building intuition, a more detailed study of the interplay between detuning and dispersion shows that this operating point coincides with the boundary between degenerate and non-degenerate operation. A comprehensive theoretical study of the operating regimes of near-degenerate doubly-resonant OPOs, including the extension to pulsed operation is given in~\cite{Hamerly2016}. The transition from single- to multi-mode dynamics in triply-resonant OPOs is studied in detail in~\cite{McKenna2022}. As a result of these behaviors, real devices are always operated with a finite detuning from perfect resonance.

We begin by considering the simple case of resonant SHG to clarify the role played by a resonator in enhancing the strength of a nonlinear interaction. Here we take $r_{\text{oc},2\omega}=0$, $t_{\text{oc},2\omega}=1$, and $A_{2\omega}^{(p)}=0$. Starting from position $(1)$ in the cavity, we have
\begin{align*}
    A_{\omega}^{(1)} &= r_{\text{oc},\omega} A_{\omega}^{(0)} + it_{\text{oc},\omega} A_{\omega}^{(p)},\\
    A_{2\omega}^{(1)} &= 0.
\end{align*}
Assuming a high-finesse cavity for the fundamental, $A_\omega^{(1)}$ is negligibly depleted during SHG. Therefore, the the fields at $(2)$ are given by 
\begin{align*}
    A_{\omega}^{(2)} &\approx A_{\omega}^{(1)},\\
    A_{2\omega}^{(2)} &= -i\kappa L_\text{qpm}\left(A_\omega^{(1)}\right)^2,
\end{align*}
Finally, we assume the fundamental incurs a small propagation loss after propagating around the cavity, $A_\omega^{(3)} = A_\omega^{(2)}\exp(-\alpha_{\ell,\omega}L_\text{cav}/2)$, where we have assumed that any loss accumulated during propagating in the QPM section de-couples from nonlinear propagation and can therefore be incorporated in the linear feedback section.

Since we assume the fundamental is negligibly depleted by SHG, the intracavity field has the well-known solution for a linear resonator. Writing $r_{\text{oc},\omega}$ and $\alpha_{\ell,\omega}$ in terms of the intrinsic and extrinsic loss rates, we have
\begin{equation*}
	A_{\omega}^{(1)} = \frac{i\sqrt{1 - \exp\left(-2\kappa_{\text{ex},\omega}T_{\text{rt},\omega}\right)}}{1-\exp\left(-(\kappa_{\text{ex},\omega}+\kappa_{\text{in},\omega})T_{\text{rt},\omega}\right)}A_{\omega}^{(p)},
\end{equation*}
which yields a generated second harmonic field given by
\begin{equation*}
	A_{2\omega}^{(2)} = -i\kappa L_\text{qpm}\left(\frac{\sqrt{1 - \exp\left(-2\kappa_{\text{ex},\omega}T_{\text{rt},\omega}\right)}}{1-\exp\left(-(\kappa_{\text{ex},\omega}+\kappa_{\text{in},\omega})T_{\text{rt},\omega}\right)}\right)^2\left(i A_{\omega}^{(p)}\right)^2.
\end{equation*}
Up to an overall phase, $A_{2\omega}^{(2)}$ is equivalent to the solution found in traveling-wave SHG, now with an effective interaction length given by
\begin{equation*}
L_\text{eff} = L_\text{qpm}\left(\frac{\sqrt{1 - \exp\left(-2\kappa_{\text{ex},\omega}T_{\text{rt},\omega}\right)}}{1-\exp\left(-(\kappa_{\text{ex},\omega}+\kappa_{\text{in},\omega})T_{\text{rt},\omega}\right)}\right)^2.
\end{equation*}
The effective interaction length is maximized when the cavity is critically coupled at the fundamental ($\kappa_{\text{ex},\omega}=\kappa_{\text{in},\omega}$), in which case
\begin{equation}
L_\text{eff, critical} = \frac{L_\text{qpm}}{1-\exp\left(-2\kappa_{\text{ex},\omega}T_{\text{rt},\omega}\right)} = \frac{L_\text{qpm}}{T_{\text{oc},\omega}}\approx \frac{L_\text{qpm}}{(\kappa_{\text{ex},\omega}+\kappa_{\text{in},\omega})T_{\text{rt},\omega}}.\label{eqn:resonant_SHG}
\end{equation}
The latter form of Eqn.~\ref{eqn:resonant_SHG} provides the clearest intuition; the interaction length for resonant SHG is effectively enhanced by the lifetime of the cavity. Put simply, the use of resonant devices enables interaction lengths approaching the loss-length of the nonlinear waveguide in a compact footprint. Typical devices have $(\kappa_{\text{ex},\omega}+\kappa_{\text{in},\omega}) T_{\text{rt},\omega}\approx 0.1$~\cite{McKenna2022}, with state-of-the-art devices achieving $(\kappa_{\text{ex},\omega}+\kappa_{\text{in},\omega}) T_{\text{rt},\omega}\approx 0.01$~\cite{zhang2017monolithic}.

In practice, to better describe resonant SHG devices with large conversion efficiencies Eqn.~\ref{eqn:resonant_SHG} can be corrected to account for the single-pass depletion of the fundamental. Noting that the steady-state second-harmonic field inside the nonlinear section is given by $A_{2\omega} = -i\kappa A_\omega^2 z$ in the weakly-depleted limit, a first-order correction accounting for pump depletion is given by
\begin{subequations}
\begin{align}
    A_{\omega}^{(2)} &\approx A_{\omega}^{(1)}\left(1 - \frac{\eta_0 L_\text{qpm}^2}{2} |A_\omega^{(1)}|^2\right),\label{eqn:resonant_SHG_depleted_FH}\\
    A_{2\omega}^{(2)} &= -i\kappa L_\text{qpm}\left(A_\omega^{(1)}\right)^2,\label{eqn:resonant_SHG_depleted_SH}
\end{align}
\end{subequations}
where the second term in Eqn.~\ref{eqn:resonant_SHG_depleted_FH} is an effective nonlinear loss due to second-harmonic generation. The self-consistent intracavity field may be found by solving
\begin{equation}
	it_{\text{oc},\omega} A_{\omega}^{(p)} + r_{\text{oc},\omega}\exp(-\alpha L_\text{cav}/2)\left(1 - \frac{\eta_0 L_\text{qpm}^2}{2} |A_\omega^{(1)}|^2\right)A_\omega^{(1)} = A_\omega^{(1)},\label{eqn:self_consistency_resonant_SHG}
\end{equation}
which requires finding the roots of a cubic polynomial. We may simplify this solution by noting that efficient operation requires an impedance matched resonator, where now $r_{\text{oc},\omega}$ is matched to both the linear and nonlinear loss accumulated by the fundamental in a single pass, $r_{\text{oc},\omega} = \exp(-\alpha L_\text{cav}/2)\left(1 - \frac12 \eta_0 L_\text{qpm}^2 |A_\omega^{(1)}|^2\right)$. In this case, the self-consistent intracavity power can be found by solving a quadratic equation, $P_\text{in} \approx P_\omega^{(1)}\left(\alpha L_\text{cav} + \eta_0 P_\omega^{(1)} L_\text{qpm}^2\right)$. In other words, when designing efficient SHG devices with a known power budget, intracavity propagation loss, and normalized efficiency, we may first solve for the circulating intracavity power, $P_\omega^{(1)}$, and then solve for the outcoupling $t_{\text{oc},\omega} = \sqrt{1 - r_{\text{oc},\omega}^2}$ needed to achieve a matched resonator at the desired operating power.

We now consider the case of an optical parametric oscillator (OPO), where an input pump $A_{2\omega}^{(p)}=0$ generates a resonant wave of fundamental. Degenerate OPOs are a rich topic in both classical and quantum optics, and we only provide a brief introduction here. These systems exhibit many dynamical regimes, including both continuous-wave and pulsed operation, and have rather different coherence properties than non-degenerate OPOs. Above threshold, these devices are commonly used as an efficient source of coherent light. Below threshold, degenerate OPOs are commonly used to generate squeezed light. In the context of this tutorial, we will see that OPOs offer a realistic path towards few-photon operation, and we will use the classical behavior of OPOs to better intuit the typical figures of merit used in quantum nonlinear optics.

Proceeding as before we start from position $(1)$, noting that $t_{\text{oc},2\omega} = 1$,
\begin{align*}
    A_{\omega}^{(1)} &= r_{\text{oc},\omega} A_{\omega}^{(0)},\\
    A_{2\omega}^{(1)} &= iA_{2\omega}^{(p)}.
\end{align*}
The signal out-coupled from the directional coupler is given by $A_{\omega}^{(\text{out})} = it_{\text{oc},\omega} A_{\omega}^{(1)}$, and therefore to describe the total external conversion efficiency of the OPO it suffices to solve for the steady-state intracavity field at $A_{\omega}^{(1)}$. The fields at position $(2)$ will be calculated by integrating the coupled-wave equations again assuming a negligibly-depleted signal, and the signal at $(3)$ is given by $A_{\omega}^{(3)} = A_{\omega}^{(2)}\exp(-\kappa_{\text{ex},\omega}T_{\text{rt},\omega})=A_{\omega}^{(1)}$. To solve for the self-consistent intracavity field, we make use of the fact that the coupled-wave equations for propagating from $(1)$ to $(2)$ conserve power, 
\begin{equation}
	\left|A_{\omega}^{(2)}\right|^2 - \left|A_{\omega}^{(1)}\right|^2 = \left|A_{2\omega}^{(1)}\right|^2 - \left|A_{2\omega}^{(2)}\right|^2.\label{eqn:self_consistency}
\end{equation}

Noting that the linear loss accumulated by propagating from the output of the OPA section to its input, $A_{\omega}^{(2)}\exp(-(\kappa_{\text{ex},\omega}+\kappa_{\text{in},\omega})T_{\text{rt},\omega})=A_{\omega}^{(1)}$, must in steady-state be compensated by the gain of the OPA, we express the fractional power gained by the signal in each pass through the saturated OPA as $\ell_\omega = \exp\left(2(\kappa_{\text{ex},\omega}+\kappa_{\text{in},\omega})T_{\text{rt},\omega}\right)-1 = (P_\omega^{(2)} - P_\omega^{(1)})/P_\omega^{(1)}$, which balances the fractional power lost on each round trip. In the high-finesse limit, the fractional power loss is typically expressed as $\ell_\omega \approx 2(\kappa_{\text{ex},\omega}+\kappa_{\text{in},\omega})T_{\text{rt},\omega}$. The pump power depleted during saturated OPA may be obtained by integrating the coupled-wave equations, $A_{2\omega}^{(2)} = A_{2\omega}^{(1)} + \Delta A_{2\omega}$, where $\Delta A_{2\omega} \approx -i\kappa L_\text{qpm} \left(A_{\omega}^{(1)}\right)^2$ for a high-finesse cavity. Equation.~\ref{eqn:self_consistency} can be solved (assuming $A_\omega^{(1)}$ is real) to find
\begin{equation}
	P_\omega^{(1)} = 2 \frac{P_{\text{th,DRO}}}{\ell_\omega}\left(\sqrt{\frac{P_{\text{in},2\omega}}{P_\text{th,DRO}}}-1\right),\label{eqn:OPO_steady_state}
\end{equation}
where $P_\text{th,DRO}=\ell_\omega^2/(\kappa^2 L_\text{qpm}^2)$, is the threshold power above which $P_\omega^{(1)} > 0$. The signal power generated by the OPO is maximized when $P_{\text{in},2\omega} = 4 P_{\text{th,DRO}} \equiv P_\text{sat}$, which corresponds to $100\%$ conversion from $P_{\text{in},2\omega}$ to $P_{\text{out},\omega}=T_{\text{oc},\omega}P_{\omega}^{(1)}$ in the absence of intrinsic loss. Looking ahead to later portions of this tutorial, where we connect the classical treatment of resonator behaviors to the quantum theory, we note that the photon number $N = P_{2\omega} T_{\text{rt},2\omega}/(2\hbar\omega)$ at saturation is often used as a measure of how nonclassical the behaviors of an OPO are. $N_\text{sat} \leq 1$ is conventionally chosen as the threshold below which highly nonclassical dynamics occur. A theoretical example is given in~\cite{Onodera2022}, which studied the formation of features similar to cat states in pulsed OPOs.

We now repeat the analysis of continuous-wave SHG and OPO with all waves resonant (\textit{i.e.} triply-resonant), which is more naturally treated using the LLE-like form of the coupled-wave equations. Starting with the case of resonant SHG, we have (from Eqns.~\ref{eqn:master_eqn_nonlinear_highfinesse_1}-\ref{eqn:master_eqn_nonlinear_highfinesse_2})
\begin{subequations}
\begin{align}
\partial_T A_\omega(T) = & \frac{i\tilde{t}_{\text{oc},\omega}}{T_{\text{rt},\omega}} A_{\omega}^{(p)} -i\kappa \frac{L_\text{qpm}}{T_{\text{rt},\omega}} A_{2\omega}(T)A_{\omega}^*(T)-(\kappa_{\text{ex},\omega}+\kappa_{\text{in},\omega})A_{\omega}(T)\label{eqn:master_eqn_SHG_highfinesse_1}\\
\partial_T A_{2\omega}(T) = & -i\kappa \frac{L_\text{qpm}}{T_{\text{rt},\omega}} A_{\omega}^2(T)-(\kappa_{\text{ex},2\omega}+\kappa_{\text{in},2\omega})A_{2\omega}(T).\label{eqn:master_eqn_SHG_highfinesse_2}
\end{align}
\end{subequations}
Solving Eqn.~\ref{eqn:master_eqn_SHG_highfinesse_2} for the steady-state second-harmonic field, we find that the intracavity second harmonic is given by the field generated in a single pass of the nonlinear stage, with an enhancement given by the ratio of the cavity lifetime and round-trip time,
\begin{equation}
A_{2\omega}=\frac{-i\kappa L_\text{qpm}}{T_{\text{rt},\omega}(\kappa_{\text{ex},2\omega}+\kappa_{\text{in},2\omega})} A_{\omega}^2.
\end{equation}
The intracavity fundamental found by solving Eqn.~\ref{eqn:master_eqn_SHG_highfinesse_1} contains a resonantly-enhanced pump, in addition to a depletion term due to conversion from fundamental to second harmonic,
\begin{equation}
A_{\omega}=\underbrace{\frac{i\tilde{t}_{\text{oc},\omega}}{T_{\text{rt},\omega}(\kappa_{\text{ex},\omega}+\kappa_{\text{in},\omega})} A_{\omega}^{(p)}}_{\text{pump}}-\underbrace{\frac{\kappa^2 L_\text{qpm}^2 |A_\omega|^2 A_\omega}{(\kappa_{\text{ex},2\omega}+\kappa_{\text{in},2\omega})(\kappa_{\text{ex},\omega}+\kappa_{\text{in},\omega})T_{\text{rt},\omega}^2}}_{\text{nonlinear loss}}.\label{eqn:resonant_FH_highfinesse}
\end{equation}
As with Eqn.~\ref{eqn:self_consistency_resonant_SHG}, Eqn.~\ref{eqn:resonant_FH_highfinesse} can be solved by finding the roots of a cubic polynomial. There are two approaches to obtaining simple, intuitive solutions. First, noting that Eqn.~\ref{eqn:resonant_FH_highfinesse} has the same form as Eqn.~\ref{eqn:self_consistency_resonant_SHG}, we may find solutions corresponding to efficient operation by imposing the impedance matching condition, $\kappa_{\text{ex},\omega} = \kappa_{\text{in},\omega}+\kappa_{\text{NL},\omega}$, where $\kappa_{\text{NL},\omega}$ is the loss rate due to SHG of the fundamental. This approach again reduces the equations for the self-consistent intracavity power to a quadratic equation. The second approach to obtaining simple closed-form solutions is to incorporate pump depletion perturbatively. Noting that in the absence of pump depletion $A_{\omega,0} \approx i\tilde{t}_{\text{oc},\omega}(T_{\text{rt},\omega}(\kappa_{\text{ex},\omega}+\kappa_{\text{in},\omega}))^{-1} A_{\omega}^{(p)}$ we may incorporate depletion as $A_{\omega} \approx A_{\omega,0}(1 - \Delta)$, where
\begin{align*}
	&\Delta = \frac{\delta A_{\omega,0}^2}{2 + 6 \delta A_{\omega,0}^2},\\
	\delta &= \frac{\kappa^2 L_\text{qpm}^2}{(\kappa_{\text{ex},2\omega}+\kappa_{\text{in},2\omega})(\kappa_{\text{ex},\omega}+\kappa_{\text{in},\omega})T_{\text{rt},\omega}^2}.
\end{align*}

At this time, resonant SHG is typically characterized in the undepleted limit, in which case the intracavity second-harmonic given by
\begin{equation}
A_{2\omega}=\frac{-i\kappa L_\text{qpm}}{T_{\text{rt},\omega}(\kappa_{\text{ex},2\omega}+\kappa_{\text{in},2\omega})} \left(\frac{i\tilde{t}_{\text{oc},\omega}}{T_{\text{rt},\omega}(\kappa_{\text{ex},\omega}+\kappa_{\text{in},\omega})} A_{\omega}^{(p)}\right)^2
\end{equation}
has the same form found in traveling-wave SHG, with an interaction length enhanced by the cavity lifetime. It is common practice in the literature to quote a normalized SHG efficiency $\eta = P_{2\omega}^{\text{out}}/P_{\omega}^{\text{in}} = T_{\text{oc},\omega}|A_{2\omega}|^2/|A_{\omega}^{(p)}|^2$ in units of \%/Watt. We note, however, that care must be taken when interpreting this figure of merit. It is common to find demonstrations of resonators with large normalized SHG efficiencies where the low power requirements are obtained by minimizing $\kappa_{\text{ex}}$ for each wave. In practice these resonators exhibit large normalized efficiencies, but cannot achieve large external conversion efficiencies, since most of the intracavity fundamental and second harmonic are absorbed rather than out-coupled. Furthermore, the intracavity two-photon loss imparted on the fundamental by pump depletion effectively introduces an impedance mismatch between the pump and the intracavity resonance, which further reduces the overall end-to-end conversion efficiency in the depleted limit. For continuous-wave inputs, the clearest approach towards achieving both low-power operation and efficient operation is to realize a large single-pass nonlinear coupling $\kappa$ in a resonator designed to satisfy the nonlinear impedance matching condition.

Finally, we consider the case of triply-resonant OPO. Here, the coupled-wave equations are given by
\begin{subequations}
\begin{align}
\partial_T A_\omega(T) = & -i\kappa \frac{L_\text{qpm}}{T_{\text{rt},\omega}} A_{2\omega}(T)A_{\omega}^*(T)-(\kappa_{\text{ex},\omega}+\kappa_{\text{in},\omega})A_{\omega}(T)\label{eqn:master_eqn_OPO_highfinesse_1}\\
\partial_T A_{2\omega}(T) = & \frac{i\tilde{t}_{\text{oc},2\omega}}{T_{\text{rt},\omega}} A_{2\omega}^{(p)}-i\kappa \frac{L_\text{qpm}}{T_{\text{rt},\omega}} A_{\omega}^2(T)-(\kappa_{\text{ex},2\omega}+\kappa_{\text{in},2\omega})A_{2\omega}(T)\label{eqn:master_eqn_OPO_highfinesse_2}
\end{align}
\end{subequations}
Solving Eqn.~\ref{eqn:master_eqn_OPO_highfinesse_1} in steady state, assuming $A_\omega(T)$ to be real without loss of generality, we find an intracavity pump given by
\begin{equation}
	A_{2\omega} = \frac{(\kappa_{\text{ex},\omega}+\kappa_{\text{in},\omega})T_{\text{rt},\omega}}{-i\kappa L_\text{qpm}}.\label{eqn:OPO_steadystate_2}.
\end{equation}
Equation~\ref{eqn:master_eqn_OPO_highfinesse_2} can be solved for the steady-state intracavity signal field with Eqn.~\ref{eqn:OPO_steadystate_2} for the intracavity pump,
\begin{equation*}
	P_\omega = \frac{4P_\text{th,TRO}t_{\text{oc},2\omega}^2}{\ell_\omega\ell_{2\omega}}\left(\sqrt{\frac{P_\text{in}}{P_\text{th,TRO}}}-1\right),
\end{equation*}
where the threshold power is given by
\begin{equation}
P_\text{th,TRO} = \left(\frac{\ell_\omega\ell_{2\omega}}{4\kappa L_\text{qpm}t_{\text{oc},2\omega}}\right)^2 = P_\text{th,DRO}\frac{\ell_{2\omega}^2}{4 t_{\text{oc},2\omega}^2}.
\end{equation}
Here, for simplicity, we have made use of the relations $\ell_{\omega} = 2(\kappa_{\text{ex},\omega}+\kappa_{\text{in},\omega})T_{\text{rt},\omega}$ and $\ell_{2\omega} = 2(\kappa_{\text{ex},2\omega}+\kappa_{\text{in},2\omega})T_{\text{rt},\omega}$ in the high-finesse limit.

On paper, the behaviors of triply-resonant OPOs do not deviate meaningfully from those of doubly-resonant OPOs, aside from having a substantially reduced threshold power. In reality, the behaviors of triply-resonant OPOs depend strongly on the detunings, $\delta_\omega$ and $\delta_{2\omega}$, of each wave from resonance. The condition for stable single-mode operation are derived for doubly- and triply-resonant OPOs in~\cite{Hamerly2016} and~\cite{McKenna2022}, respectively.

\subsubsection{Towards quantum nonlinear optics: time-evolving fields}\label{sec:time-prop}

Thus far, our discussion has centered on the evolution of the complex field envelope $A_\omega(z,t)$ in the cavity. In this approach, the envelopes describe the instantaneous power, $P_\omega(z,t)=|A_\omega(z,t)|^2$, centered around frequency $\omega$ flowing through a point $z$. Evolution occurs in $z$, that is, integrating the equations of motion from $z_1$ to $z_2$ yields the instantaneous power envelope flowing through $z_2$. In cavity QED, the object of study is more commonly the photon number contained in the normal modes of a resonator, such as the standing wave modes of a Fabry-Perot cavity or the azimuthal modes of a ring resonator. In the mean-field limit, these models describe the time evolution of the energy density throughout the cavity. To better establish the connection between classical nonlinear optics and the mean-field behavior of cavity QED, we first reformulate classical nonlinear optics in terms of the time-evolving envelopes, $u_\omega(z,t)$, that describe the spatial distribution of the energy density within a nonlinear medium.

\begin{figure}
\centering
\includegraphics[width=\linewidth]{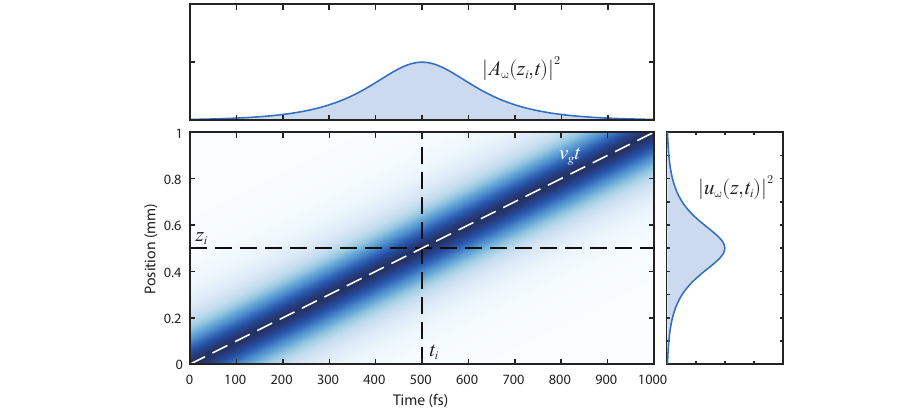}
\caption{The field envelopes $A_\omega(z,t)$ and $u_\omega(z,t)$ obtained by horizontal and vertical slices of the field distribution in the $t-z$ plane are equivalent descriptions of the field evolution. The equations of motion for $A_\omega(z,t)$ propagate a time-varying waveform from $z=0$ to $z=z_i$ (horizontal dashed line), whereas the equations of motion for $u_\omega(z,t)$ propagate a spatial distribution from $t=0$ to $t=t_i$ (vertical dashed line).\label{fig:space-time-plot}}
\end{figure}

The connection between $A_\omega(z,t)$ and $u_\omega(z,t)$ is best established by first considering linear propagation in the absence of higher-order dispersion in a non-moving frame ($v_{g,\text{ref}}=0$). For simplicity we first consider an isolated pulse in a waveguide, where the fields may extend over an arbitrarily large domain. In this case, both the fundamental and second harmonic walk off from the origin at their respective group velocities, as shown in Fig.~\ref{fig:space-time-plot}. For any position $z_i$, the field envelope $A_\omega(z_i,t)$ is given by a horizontal slice in the $t-z$ plane, with the power flowing through $z_i$ as a function of time given by $|A_\omega(z_i,t)|^2 dt$. The spatial envelope describing the energy density at a time $t_i$ is given by a vertical slice in the $t-z$ plane, $A_\omega(z, t_i) = u_\omega(z, t_i)\sqrt{v_{g,\omega}}$, where $t_i v_{g,\omega} = z_i$. As an example, if we consider a sech-pulse, we find
\begin{equation*}
A_\omega(z,t) = \frac{1}{\sqrt{2\tau}}\sech\left(\frac{t-v_{g,\omega}^{-1}z}{\tau}\right)
\leftrightarrow
u_\omega(z,t) = \frac{1}{\sqrt{2\tau v_{g,\omega}}}\sech\left(\frac{z-v_{g,\omega}t}{v_{g,\omega}\tau}\right).
\end{equation*}
The field envelopes $A_\omega(z,t)$ and $u_\omega(z,t)$ are clearly equivalent descriptions of how the fields are distributed within the medium, and either can easily be used to calculate the total energy, $U_\omega$, stored within a pulse envelope 
\begin{equation*}
\int_{-\infty}^{\infty} |A_\omega(z_i,t)|^2 dt = \int_{-\infty}^{\infty} |u_\omega(z,t_i)|^2 dz = U_\omega.
\end{equation*}
Thus far, with only temporal walk-off the equations of motion are given by
\begin{subequations}
\begin{align}
\partial_t u_\omega(z,t) = -v_{g,\omega}\partial_z u_\omega(z,t),\label{eqn:time-walkoff_1}\\
\partial_t u_{2\omega}(z,t) = -v_{g,2\omega}\partial_z u_{2\omega}(z,t).\label{eqn:time-walkoff_2}
\end{align}
\end{subequations}
We note here that the conversion between spatial and temporal envelopes can be nontrivial in the presence of higher-order dispersion. We will treat this more general case at the end of this section.

While the correspondence between $A_\omega(z,t)$ and $u_\omega(z,t)$ undergoing linear propagation is clear in the non-moving frame used above, the transformation to a co-moving frame commonly used to model nonlinear interactions contains subtleties that are worth further discussion. The definition of a co-moving frame in this context is \emph{not} a Lorentz transformation of the underlying coordinates, but rather a choice of phase-reference in the Fourier domain $A(z,\Omega)\rightarrow A(z,\Omega)\exp(-i v_{g,\text{ref}}^{-1}\Omega z)$. Defining a co-moving frame for $A_\omega(z,t)$ does not distort or Doppler-shift either $A_\omega$ or $A_{2\omega}$. Instead, the centroid of $A_\omega(z,t)$ at each point $z$ is translated to $t=0$, which deforms the dashed white line in Fig.~\ref{fig:space-time-plot} to a vertical line in the $t-z$ plane. This transformation skews $u_\omega(z,t)$; while the horizontal cuts in the $t-z$ plane used to extract $A_\omega(z,t)$ are not distorted by a horizontal shift of the time-axis for each point $z$, the vertical cut used to extract $u_\omega(z,t)$ are deformed into lines running parallel to $z=-v_{g,\omega}t$. While the coupled-wave equations are more easily solved in a co-moving frame, the simplest way to convert between these two descriptions of the field is to move back into a non-moving frame. Therefore, the approach we advocate in converting between these descriptions is as follows: i) solve the equations of motion in a convenient co-moving frame within one description of the field evolution, ii) convert to a non-moving frame, iii) calculate $A_\omega(z,t)$ or $u_\omega(z,t)$ using either horizontal or vertical line cuts of the field envelopes in the non-moving frame. With this in mind, we note that a similar co-moving frame can be defined for the spatial envelopes by Fourier transforming  $u_\omega(z,t)$ and applying a phase shift of $\exp(-i v_{g,\text{ref}} \delta k t)$ to translate the centroid of $u_\omega(z,t)$ down to $z=0$. Here $\delta k$ is the detuning of the wavenumber from the mean angular wavenumber of the envelope, in analogy to $\Omega$ being the angular frequency detuning from the carrier frequency of the envelope. These co-moving frames for the spatial envelopes will be used throughout the following sections on ultrafast quantum nonlinear optics.

Having established basic rules for converting between temporal and spatial envelopes, we now consider nonlinear propagation. Taking SHG with temporal walk-off as a canonical example to establish the connection between spatial and temporal propagation, we begin with the coupled-wave equations in the non-moving frame,
\begin{align*}
	\partial_z A_{\omega}(z,t) = -v_{g,\omega}^{-1}\partial_t A_{\omega}(z,t) - i \kappa A_{2\omega}(z,t)A_{\omega}^*(z,t),\\
	\partial_z A_{2\omega}(z,t) = -v_{g,2\omega}^{-1}\partial_t A_{2\omega}(z,t) - i \kappa A_{\omega}^2(z,t).
\end{align*}
Multiplying each side by the respective group velocity and converting from temporal to spatial envelopes using $A_\omega(z, t) = u_\omega(z, t)\sqrt{v_{g,\omega}}$, we find the time-propagating form of the coupled-wave equations
\begin{subequations}
\begin{align}
	\partial_t u_{\omega}(z,t) = -v_{g,\omega}\partial_z u_{\omega}(z,t) - i \sigma u_{2\omega}(z,t) u_{\omega}^*(z,t),\label{eqn:cwe_time_basic_1}\\
	\partial_t u_{2\omega}(z,t) = -v_{g,2\omega}\partial_z u_{2\omega}(z,t) - i \sigma u_{\omega}^2(z,t),\label{eqn:cwe_time_basic_2}
\end{align}
\end{subequations}
where $\sigma = \kappa\sqrt{v_{g,\omega}^2 v_{g,2\omega}}$ is the coupling coefficient for the field densities, $u_{\omega}$ and $u_{2\omega}$, here defined as the square-root of the energy density. We note here that in the absence of temporal walk-off, Eqn.~\ref{eqn:cwe_time_basic_1}-\ref{eqn:cwe_time_basic_2} conserve the local energy density, and in general equations of this form will conserve the total energy $\int |u_\omega(z,t)|^2 + |u_{2\omega}(z,t)|^2 dz$. These equations of motion can be converted into the more familiar co-moving form by shifting our choice of phase reference in the Fourier domain as described above,
\begin{subequations}
\begin{align}
	\partial_t u_{\omega}(z,t) = - i \sigma u_{2\omega}(z,t)u_{\omega}^*(z,t),\label{eqn:cwe_time_comoving_1}\\
	\partial_t u_{2\omega}(z,t) = -\Delta v_{g}\partial_z u_{2\omega}(z,t) - i \sigma u_{\omega}^2(z,t).\label{eqn:cwe_time_comoving_2}
\end{align}
\end{subequations}

At first glance, there is no meaningful difference between Eqns.~\ref{eqn:cwe_time_comoving_1}-\ref{eqn:cwe_time_comoving_2} and the more commonly encountered CWEs discussed throughout this tutorial. After all, Eqns.~\ref{eqn:cwe_time_comoving_1}-\ref{eqn:cwe_time_comoving_2} were obtained with a few algebraic manipulations, and have the same functional form as the space-propagating CWEs with space and time interchanged. There are, however, some subtle differences with how these equations are applied to describe physical phenomena. Take, for example, the single-mode case given by
\begin{subequations}
\begin{align}
	\partial_t u_{\omega}(t) = - i \sigma u_{2\omega}(t)u_{\omega}^*(t),\label{eqn:cwe_time_singlemode_1}\\
	\partial_t u_{2\omega}(t) = - i \sigma u_{\omega}^2(t).\label{eqn:cwe_time_singlemode_2}
\end{align}
\end{subequations}
Eqns.~\ref{eqn:cwe_time_singlemode_1}-\ref{eqn:cwe_time_singlemode_2} do \emph{not} describe traveling-wave SHG in a nonlinear waveguide. While the fundamental is a uniform distribution of both power and energy in traveling-wave SHG, the second harmonic has a quadratic spatial variation. In contrast, in Eqns.~\ref{eqn:cwe_time_singlemode_1}-\ref{eqn:cwe_time_singlemode_2} we are considering a nonlinear medium that either extends to infinity or has periodic boundary conditions extending from $z=0$ to $z=L$. This medium is excited by a uniform field $u_{\omega}(t)$ instantiated throughout the entire nonlinear medium at $t=0$. In this case, in the undepleted limit the second-harmonic field builds up with a quadratic \emph{time} dependence and remains spatially uniform for all time. While these solutions do not describe traveling-wave SHG, they do correspond well to the behavior of extremely high-finesse resonators, where the field throughout the nonlinear medium is well approximated by a uniform distribution. The usual solutions for traveling-wave SHG can be recovered from the time-propagating model by working in the non-moving frame (Eqns.~\ref{eqn:cwe_time_basic_1}-\ref{eqn:cwe_time_basic_2}) and solving for the envelopes $u_\omega(z,t)$ and $u_{2\omega}(z,t)$ in steady-state ($\partial_t \mapsto 0$) everywhere inside the waveguide. The time-propagating equations of motion are not commonly used in traveling-wave nonlinear optics, since the boundary conditions for $u_\omega(z,t)$ involve adding a time-varying driving term to Maxwell's equations.

\subsubsection{Normal modes}

Thus far, our treatment has considered the evolution of an isolated pulse in an arbitrarily long waveguide. In reality, all nonlinear devices have a finite spatial extent. In the case of traveling-wave devices or Fabry-Perot resonators, the spatial extent of each is constrained to $z\in[0,L]$. Similarly, for ring resonators and microtoroids, the fields have periodic bounds, $z\in[0,L]$, where $z$ is now the azimuthal coordinate and $L$ is the circumference of the resonator. In all of these cases, the intracavity field can be written in terms of a Fourier series over spatial modes, $u_\omega(z,t) = \sum_m u_{\omega,m}\exp(2\pi i m z/L)$. Throughout this section we will use $k_m = 2\pi m/L$ to denote the angular wavenumber of the $m^\text{th}$ Fourier mode. For the case of a finite-domain signal, such as a straight waveguide, $u_\omega(z,t)$ can be obtained by windowing the periodic signal with a rectangle function. A more detailed discussion of normal modes is given in Appendix~\ref{sec:time-propagating-spatial-modes}.

When the fields are constrained to a finite domain, the solutions to Maxwell's equations are given by
\begin{align}
\mathbf{E}(\mathbf{r},t) &= \frac12\left(\sum_{\mu,m}a_{\mu,m}\mathbf{E}_{\mu}(x,y,\omega_\mu(k_m))\exp(-\mathrm{i}\omega_\mu(k_m)t+i k_m z) + c.c.\right),\label{eqn:normal_mode_E}
\end{align}
with identical expressions holding for $\mathbf{H}$ and $\mathbf{D}$. Here $k_m = 2\pi m /L$ is the angular wavenumber of each mode, and $a_{\mu, m}$ is a dimensionless number that describes the fraction of the overall energy contained in the $m^\text{th}$ Fourier component of mode $\mu$. We will see below that the physical interpretation of $a_{\mu,m}$ will depend on our choice of mode normalization. Equation~\ref{eqn:normal_mode_E} is identical to Eqn.~\ref{eqn:waveguide_mode} for waveguide modes, now with the Fourier series expansion taken over spatial modes rather than frequencies. For the case of an isolated pulse, as above, the summation over $m$ can be replaced with a Fourier integral. The normal modes
\begin{align*}
	\mathbf{E}_{\mu,m} &= \exp(-i\omega_\mu(k_m)t+ik_m z)\mathbf{E}_{\mu}(x,y,\omega_\mu(k_m)),\\
	\mathbf{H}_{\mu,m} &= \exp(-i\omega_\mu(k_m)t+ik_m z)\mathbf{H}_{\mu}(x,y,\omega_\mu(k_m)),\\
	\mathbf{D}_{\mu,m} &= \exp(-i\omega_\mu(k_m)t+ik_m z)\epsilon(x,y,\omega_\mu(k_m))\mathbf{E}_{\mu}(x,y,\omega_\mu(k_m)),
\end{align*}
each independently satisfy Maxwell's equations, however $\omega_\mu(k_m)$ is now interpreted as the eigenvalue associated with propagation constant $k_m$, with corresponding eigenfunctions given by $\mathbf{E}_\mu$ and $\mathbf{H}_\mu$. The normal modes are each normalized to a characteristic energy, $\text{U}_0$, using the expression for the energy contained in a linear dispersive dielectric~\cite{Raymer2020,haus1984waves}
\begin{equation}
    \frac14\int \mu_0\mathbf{H}_\mu\cdot\mathbf{H}_\mu^*+\mathbf{E}_\mu^*\cdot\left(\epsilon(x,y,\omega)+\omega\partial_\omega\epsilon(x,y,\omega)\right)\cdot \mathbf{E}_\mu dV=\text{U}_0.\label{eqn:spatialmode_energy}
\end{equation}
$\text{U}_0$ will be taken to be 1 Joule for classical nonlinear optics, or will later be chosen independently for each mode $U_{\mu,m} = \hbar \omega_\mu(k_m)$. In the former case, $\text{U}_0|a_{\mu,m}|^2$ corresponds to the energy (in Joules) contained in a mode. Noting that in the latter case $\hbar \omega_\mu(k_m)|a_{\mu,m}|^2$ is the energy contained in mode $\mu$, we can identify $|a_{\mu,m}|^2$ as the number of photons contained in a mode.

As with pulse propagation in traveling-wave devices, we can define a spatial envelope associated with transverse mode $\mu$ given by,
\begin{align}
u_\mu(z,t) = &\sqrt{\frac{\text{U}_0}{L}}\sum_m a_{\mu,m}\exp\left(-i(\omega_\mu(k_m)-\omega_\text{ref})t\right)\label{eqn:energy_envelope}\\
&\times \exp\left(i(k_m - k_\text{ref}) z + i (k_m - k_\text{ref}) v_{g,\text{ref}} t\right),\nonumber
\end{align}
where $k_\text{ref}$ is the mean spatial frequency of the envelope, $\omega_\text{ref}$ is an arbitrary reference frequency ($\omega_\text{ref}$ need not be $\omega_\mu(k_\text{ref})$), and $v_{g,\text{ref}}$ is an arbitrary reference velocity. As with space-propagating envelopes, the envelope $u_\mu(z,t)$ describes the continuous field distribution obtained from the normal modes by Fourier synthesis. We may derive a propagation rule for $u_\mu(z,t)$ that describes the displacement and spreading of the field distribution due to dispersion by taking the derivative of Eqn.~\ref{eqn:energy_envelope} with respect to time. Series expanding $\omega_\mu(k_m)$ around $k_\text{ref}$, we find
\begin{align}
\partial_t u_\mu(z,t) = & -i(\omega_\mu(k_\text{ref})-\omega_\text{ref})u_\mu(z,t)-\left(v_{g,\mu}-v_{g,\text{ref}}\right)\partial_z u_\mu(z,t)\label{eqn:normal_mode_linear_prop}\\ &-iD_{\text{int},\mu}(\partial_z)u_\mu(z,t),\nonumber
\end{align}
where $D_{\text{int},\mu}(\partial_z) = \sum_{m=2}^{\infty} \frac{(-i)^m}{m!}
(\partial_z)^m \omega_\mu^{(m)}$ is the integrated dispersion for mode $\mu$, and $\omega_\mu^{(m)} = \partial_k \omega_\mu(k)|_{k_\text{ref}}$ is the $m$th derivative of $\omega_\mu(k)$, evaluated at $k_\text{ref}$. We reiterate here that Eqn.~\ref{eqn:normal_mode_linear_prop} is the Fourier dual of the more commonly encountered propagation rule $\partial_t u_{\mu,m}(t) = -i\omega_\mu(k_m) u_{\mu,m}(t)$ in a rotating frame defined by $\omega_\text{ref}$ and $v_{g,\text{ref}}$.

We can connect $u_\mu$ back to the spatial envelope derived in Sec.~\ref{sec:time-prop} by working in a non-moving frame ($v_{g,\text{ref}}=0$) and setting $\omega_\text{ref} = \omega_\mu(k_\text{ref})$. In this case, the propagation rule is equivalent to Eqn.~\ref{eqn:time-walkoff_1}-\ref{eqn:time-walkoff_2}, now generalized to include higher order dispersion. In the time-propagating picture, the integrated dispersion contains dispersion orders generated by series expanding $\omega$ with respect to $k$, rather than $k$ with respect to $\omega$, and contains spatial derivatives $\partial_z$. The integrated dispersion $D_{\text{int},\mu}(\partial_t)$ previously encountered for temporal envelopes is not converted to $D_{\text{int},\mu}(\partial_z)$ by substituting $v_g\partial_z = \partial_t$. Instead, each dispersion order contains chain factors, as tabulated in Table~\ref{tab:dispersion}. There are two limiting cases where the approximate dispersion orders given by $\omega^{(n)}(k)\approx k^{(n)}(\omega)v_g^{n+1}$ can be used to estimate the dispersion relations of $\omega(k)$: when the dispersion relations can be truncated at second-order, or when $k''=0$. In the latter case, the chain factors for both third and fourth-order dispersion become zero.

\begin{table}[]
    \centering
    \begin{tabular}{c|c}
    	\multicolumn{2}{c}{Dispersion orders of $\omega(k)$, as determined by derivatives of $k(\omega)$}\\
    	\hline
    	\hline
         $\omega'(k)$ & $\frac{1}{k'(\omega)} = v_g$ \\
         $\omega''(k)$ & $-k''(\omega)v_g^3$\\
         $\omega'''(k)$ & $-k'''(\omega)v_g^4 + 3(k''(\omega))^2v_g^5$\\
         $\omega''''(k)$ & $-k''''(\omega)v_g^5 + 10k''(\omega)k'''(\omega)v_g^6 - 15\left(k''(\omega)\right)^3v_g^7$
    \end{tabular}
    \caption{Relation between the $n$th dispersion order of $\omega(k)$ and the dispersion orders of $k(\omega)$. Here, the right-hand side is evaluated using the chain rule. In many practical applications the \emph{approximate} rule $\omega^{(n)}(k)\approx k^{(n)}(\omega)v_g^{n+1}$ can be used to estimate the dispersion relations of $\omega(k)$, as described in the main text.}
    \label{tab:dispersion}
\end{table}

A word of caution is in order when converting between these two formalisms. In working with highly broadband pulses, the time-propagating and space-propagating pictures can produce different results depending on how the Taylor series of $\omega$ or $k$ is truncated. Care must be taken to determine whether these differences have a physical or numerical origin. For example, in a dispersion-engineered waveguide with a large GVD and $k'''=0$, the behavior of space-propagating models is well understood. However, for a time-propagating model the non-zero third order dispersion $\omega''' = 3(k'')^2 v_g^5$ will cause the envelope $u(z,t)$ to skew. This distortion is not a numerical artifact, and can be verified by using space-time diagrams such as Fig.~\ref{fig:space-time-plot}. If we now consider a Kerr microresonator with the same dispersion relations, one model will predict a Kerr soliton that radiates dispersive waves, while the other model will not. This behavior is an artifact of having only truncated the two models at third-order dispersion, and can be reconciled by using the full dispersion relations. Before considering nonlinear interactions between normal modes, we close this section by briefly discussing loss. In contrast with discrete maps, where the loss rate was set by our choice of reference wave, here $\kappa_\text{ex}$ and $\kappa_\text{in}$ for $\omega$ and $2\omega$ will depend on the group velocity of each wave, $v_{g,\omega}$ and $v_{g,2\omega}$, respectively. We limit our treatment to non-dispersive losses, which enables a simple time-domain analysis.

We begin with propagation loss. Noting that the distance traveled by the envelope $u_\omega(z,t)$ in a given time is $z = v_{g,\omega}t$, the field decays to $u_\omega(z-v_{g,\omega}t,0)\exp(-\alpha_{\ell,\omega}z/2) = u_\omega(z,0)\exp(-\kappa_{\text{in},\omega}t)$. Similarly, for the second harmonic $u_{2\omega}(z-v_{g,2\omega}t,0)\exp(-\alpha_{\ell,2\omega}z/2) = u_{2\omega}(z,0)\exp(-\kappa_{\text{in},2\omega}t)$. We may extend this analysis to extrinsic losses by considering a discrete out-coupler with transmission and reflection coefficients $it_{\text{oc},\omega}$ and $r_{\text{oc},\omega}$, respectively. Noting that the extrinsic loss rate for a round-trip time of $T_\text{rt}$ is given by $\kappa_\text{ex}=-\ln(r_\text{oc})/T_\text{rt}$, the extrinsic loss rate for each wave is given by $\kappa_{\text{ex},\omega}=-\ln(r_{\text{oc},\omega})/T_{\text{rt},\omega}$ and $\kappa_{\text{ex},2\omega}=-\ln(r_{\text{oc},2\omega})/T_{\text{rt},2\omega}$.

A natural question is whether or not these altering definitions of loss rate between the space- and time-propagating models contribute any measurable difference in the predicted behaviors of nonlinear resonators. The apparent contradiction between these two approaches can be resolved in the high-finesse limit by noting that in either case the only physically meaningful parameters are the total loss accumulated per round trip. This can be seen by noting that in all of the examples considered throughout Sec.~\ref{sec:resonator_examples}, the loss rates $\kappa_{\text{ex}}$ and $\kappa_\text{in}$ can always be grouped with the round-trip time of the reference wave $T_{\text{rt},\omega}$, which eliminates the reference velocity for every expression for the loss. Similar behaviors occur for the dispersion and phase. 

\subsubsection{Nonlinear interactions between normal modes}

We can incorporate nonlinearity into Eqn.~\ref{eqn:normal_mode_linear_prop} by assuming that $\chi^{(2)}$ nonlinearities act locally in space and do not couple to higher-order dispersion. In this case, comparing Eqn.~\ref{eqn:normal_mode_linear_prop} to Eqns.~\ref{eqn:cwe_time_basic_1}-\ref{eqn:cwe_time_basic_2} suggests equations of the form
\begin{subequations}
\begin{align}
\partial_t u_\omega(z,t) = & -\left(v_{g,\omega}-v_{g,\text{ref}}\right)\partial_z u_\omega(z,t)\label{eqn:normal_mode_nonlinear_prop_1}\\ &-iD_{\text{int},\omega}(\partial_z)u_\omega(z,t)- i \sigma u_{2\omega}(z,t)u_{\omega}^*(z,t),\nonumber\\
\partial_t u_{2\omega}(z,t) = & -\left(v_{g,2\omega}-v_{g,\text{ref}}\right)\partial_z u_{2\omega}(z,t)\label{eqn:normal_mode_nonlinear_prop_2}\\ &-iD_{\text{int},2\omega}(\partial_z)u_{2\omega}(z,t)- i \sigma u_{\omega}^2(z,t),\nonumber
\end{align}
\end{subequations}
for a phase-matched interaction. Here, we have dropped the subscript $\mu$ in favor of $\omega$ and $2\omega$ to denote the relevant spatial mode for the fundamental and second harmonic. Phase-mismatch can be incorporated by working in a non-rotating frame,
\begin{subequations}
\begin{align}
\partial_t u_{\omega}(z,t) = & -i\omega(k_\text{ref})u_{\omega}(z,t)-\left(v_{g,\omega}-v_{g,\text{ref}}\right)\partial_z u_\omega(z,t)\label{eqn:normal_mode_nonlinear_general_1}\\ &-iD_{\text{int},\omega}(\partial_z)u_\omega(z,t)- i \sigma u_{2\omega}(z,t)u_{\omega}^*(z,t),\nonumber\\
\partial_t u_{2\omega}(z,t) = & -i\omega(2k_\text{ref})u_{2\omega}(z,t)-\left(v_{g,2\omega}-v_{g,\text{ref}}\right)\partial_z u_{2\omega}(z,t)\label{eqn:normal_mode_nonlinear_general_2}\\ &-iD_{\text{int},2\omega}(\partial_z)u_{2\omega}(z,t)- i \sigma u_{\omega}^2(z,t),\nonumber
\end{align}
\end{subequations}
where the choice of arguments $k_\text{ref}$ and $2k_\text{ref}$ will be clarified below. As with traveling-wave NLO, we now have many choices of rotating frame. Choosing a reference frequency of $\omega(k_\text{ref})$ for the fundamental and $\omega(2k_\text{ref})$ for the second harmonic recovers a more familiar form of the coupled-wave equations,
\begin{subequations}
\begin{align}
\partial_t u_{\omega}(z,t) = & -\left(v_{g,\omega}-v_{g,\text{ref}}\right)\partial_z u_\omega(z,t)-iD_{\text{int},\omega}(\partial_z)u_\omega(z,t)\label{eqn:normal_mode_nonlinear_familiar_1}\\ &- i \sigma u_{2\omega}(z,t)u_{\omega}^*(z,t)\exp(-i\Delta \omega t),\nonumber\\
\partial_t u_{2\omega}(z,t) = & -\left(v_{g,2\omega}-v_{g,\text{ref}}\right)\partial_z u_{2\omega}(z,t)-iD_{\text{int},2\omega}(\partial_z)u_{2\omega}(z,t)\label{eqn:normal_mode_nonlinear_familiar_2}\\ &- i \sigma u_{\omega}^2(z,t)\exp(i\Delta \omega t),\nonumber
\end{align}
\end{subequations}
where $\Delta \omega = \omega(2k_\text{ref}) - 2\omega(k_\text{ref})$. Another common choice of reference frequency is $\omega(2k_\text{ref})$ for the second harmonic, and $\omega(2k_\text{ref})/2$ for the fundamental, which yields
\begin{subequations}
\begin{align}
\partial_t u_{\omega}(z,t) = & \frac{-i\Delta \omega}{2} u_{\omega}(z,t)-\left(v_{g,\omega}-v_{g,\text{ref}}\right)\partial_z u_\omega(z,t)\label{eqn:normal_mode_nonlinear_opa_1}\\ &-iD_{\text{int},\omega}(\partial_z)u_\omega(z,t)- i \sigma u_{2\omega}(z,t)u_{\omega}^*(z,t),\nonumber\\
\partial_t u_{2\omega}(z,t) = & -\left(v_{g,2\omega}-v_{g,\text{ref}}\right)\partial_z u_{2\omega}(z,t)\label{eqn:normal_mode_nonlinear_opa_2}\\ &-iD_{\text{int},2\omega}(\partial_z)u_{2\omega}(z,t)- i \sigma u_{\omega}^2(z,t).\nonumber
\end{align}
\end{subequations}
This latter choice of rotating frame was used throughout Sec.~\ref{sec:OPA}, and will be a common choice throughout the following sections on quantum nonlinear optics.

The origin of $\Delta \omega$, rather than $\Delta k$, as the phase-mismatch for normal modes can be clarified by revisiting the Fourier domain now with nonlinear coupling terms. As before, using the periodic bounds of the resonator to express the envelopes as $u_{\omega}(z,t) = \sum_m u_{\omega,m}(t)\exp(2\pi i m z/L - i k_\text{ref}z)$ for the fundamental and $u_{2\omega}(z,t) = \sum_n u_{2\omega,n}(t)\exp(2\pi i n z/L - 2 i k_\text{ref}z)$ for the second harmonic, Eqns.~\ref{eqn:normal_mode_nonlinear_general_1}-\ref{eqn:normal_mode_nonlinear_general_2} become 
\begin{subequations}
\begin{align}
\partial_t u_{\omega,m}(t) = & -i\omega(k_\text{ref})u_{\omega,m}(t)-\left(v_{g,\omega}-v_{g,\text{ref}}\right)(i\delta k_m) u_{\omega,m}(t)\label{eqn:normal_mode_nonlinear_fourier_1}\\ &-iD_{\text{int},\omega}(i\delta k_m)u_{\omega,m}(t)- i \sigma \sum_n u_{2\omega,n}(t)u_{\omega, n-m}^*(t),\nonumber\\
\partial_t u_{2\omega,n}(t) = & -i\omega(2k_\text{ref})u_{2\omega,n}(t)-\left(v_{g,2\omega}-v_{g,\text{ref}}\right)(i\delta k_n) u_{2\omega,n}(t)\label{eqn:normal_mode_nonlinear_fourier_2}\\ &-iD_{\text{int},2\omega}(i\delta k_n)u_{2\omega,n}(t)- i \sigma \sum_m u_{\omega,m}(t)u_{\omega,n-m}(t),\nonumber
\end{align}
\end{subequations}
where $\delta k_m = 2\pi m/L - k_\text{ref}$. Written in this form, we see that interactions between normal modes must conserve momentum due to Fourier's rule. We note here that the conditions for quasi-phasematching are unchanged from the traveling-wave picture. If $\sigma\mapsto\sigma(z)$ is a periodic function, then we may also Fourier series expand $\sigma(z)$ and repeat the above derivation of the Fourier-domain coupled-wave equations. In this case, we find that the Fourier components of $\sigma(z)$ shift the pairs of Fourier components coupled together by the nonlinearity, which effectively shifts the reference k-vector of one of the harmonics, \textit{e.g.} $k_{\text{ref},2\omega} = 2k_{\text{ref},\omega} - k_G$. This shifts the frequency of the second-harmonic envelope, thereby reducing $\Delta \omega$.

\subsubsection{Photon-normalized units}\label{sec:photon-normalized-units}

As with the use of flux-normalized units in classical nonlinear optics, the use of photon-normalized units if often convenient for time-propagating models. This is particularly useful for connecting mean-field behavior of the quantum theory to the behaviors of the classical model. Rather than choosing our normalization energy $\text{U}_0 = 1$ Joule, we normalize each resonator mode to contain an energy $U_m = \hbar \omega(k_m)$. With this choice of normalization, the Fourier components $\alpha_m(t)$ are related to the number of photons in each longitudinal mode, $|\alpha_m|^2 = N_m$. In later sections on quantum nonlinear optics, $\alpha_m$ will be promoted from a c-number to an operator. In this case, the spatial envelope associated with the intracavity photons must be obtained using a unitary Fourier series, $\alpha(z,t) = \sum_m \alpha_m(t)\exp(2\pi i m z/L)/\sqrt{L}$, where $\int_0^L |\alpha(z,t)|^2 dz = \sum_m |\alpha_m|^2 = \sum_m N_m$ is the total number of intracavity photons contained in the envelope. We adopt this Fourier convention here, since it will be used throughout our treatment of quantum nonlinear optics, and refer to $\alpha(z,t)$ as the complex amplitude density, since $|\alpha(z,t)|^2$ is the number density. Given this choice of Fourier convention, the connection between the energy-normalized Fourier modes and number-normalized Fourier modes is given by $\alpha_m(t) = u_m(t)\sqrt{L/\hbar \omega_m}$. The connection between spatial envelopes is given by $\alpha(z,t) \approx u(z,t)/\sqrt{\hbar\omega}$ for envelopes with a narrow spectral bandwidth compared to $\omega$. We note here that the factor of $\sqrt{L}$ appears when converting Fourier modes, but not spatial envelopes, due to this change in Fourier convention.

In photon-number units, Eqns.~\ref{eqn:normal_mode_nonlinear_fourier_1}-\ref{eqn:normal_mode_nonlinear_fourier_2} become
\begin{subequations}
\begin{align}
\partial_t \alpha_{\omega,m}(t) = & -i\omega(k_\text{ref})\alpha_{\omega,m}(t)-\left(v_{g,\omega}-v_{g,\text{ref}}\right)(i\delta k_m) \alpha_{\omega,m}(t)\label{eqn:cwes_photon_number_1}\\ &-iD_{\text{int},\omega}(i\delta k_m)\alpha_{\omega,m}(t)- i g \sum_n \alpha_{2\omega,n}(t)\alpha_{\omega, n-m}^*(t),\nonumber\\
\partial_t \alpha_{2\omega,n}(t) = & -i\omega(2k_\text{ref})\alpha_{2\omega,n}(t)-\left(v_{g,2\omega}-v_{g,\text{ref}}\right)(i\delta k_n) \alpha_{2\omega,n}(t)\label{eqn:cwes_photon_number_2}\\ &-iD_{\text{int},2\omega}(i\delta k_n)\alpha_{2\omega,n}(t)- i \frac{g}{2} \sum_m \alpha_{\omega,m}(t)\alpha_{\omega,n-m}(t),\nonumber
\end{align}
\end{subequations}
where $g = \sqrt{2\hbar\omega L^{-1}}\sigma$ is the single-mode coupling rate and $|\alpha_{2\omega,n}|^2$ is the number of photons contained in mode $n$. Here the factor of $\sqrt{L^{-1}}$ in the coupling rate comes from working with the total photon number, rather than the number density. In the non-rotating frame, we can sum the series expansion contained in $D_{\text{int},\omega}$ and $D_{\text{int},2\omega}$ to write the equations of motion more succinctly as
\begin{subequations}
\begin{align}
\partial_t \alpha_{\omega,m}(t) = & -i\omega_m\alpha_{\omega,m}(t)- i g \sum_n \alpha_{2\omega,n}(t)\alpha_{\omega, n-m}^*(t),\label{eqn:cwe_photon_conventional_1}\\
\partial_t \alpha_{2\omega,n}(t) = & -i\omega_n\alpha_{2\omega,n}(t)- i \frac{g}{2} \sum_m \alpha_{\omega,m}(t)\alpha_{\omega,n-m}(t).\label{eqn:cwe_photon_conventional_2}
\end{align}
\end{subequations}
Equations~\ref{eqn:cwe_photon_conventional_1}-\ref{eqn:cwe_photon_conventional_2} are one of the main results of this section, which will enable us to connect the classical equations of motion and the coupling coefficient $g$ to the quantum equations of motion.

Similarly, working in real space rather than the Fourier domain, Eqns.~\ref{eqn:normal_mode_nonlinear_general_1}-\ref{eqn:normal_mode_nonlinear_general_2} become
\begin{subequations}
\begin{align}
\partial_t \alpha_{\omega}(z,t) = & -i\omega(k_\text{ref})\alpha_{\omega}(z,t)-\left(v_{g,\omega}-v_{g,\text{ref}}\right)\partial_z \alpha_\omega(z,t)\label{eqn:cwe_quantum_envelope_1}\\ &-iD_{\text{int},\omega}(\partial_z)\alpha_\omega(z,t)- i r \alpha_{2\omega}(z,t)\alpha_{\omega}^*(z,t),\nonumber\\
\partial_t \alpha_{2\omega}(z,t) = & -i\omega(2k_\text{ref})\alpha_{2\omega}(z,t)-\left(v_{g,2\omega}-v_{g,\text{ref}}\right)\partial_z \alpha_{2\omega}(z,t)\label{eqn:cwe_quantum_envelope_2}\\ &-iD_{\text{int},2\omega}(\partial_z)\alpha_{2\omega}(z,t)- i \frac{r}{2} \alpha_{\omega}^2(z,t),\nonumber
\end{align}
\end{subequations}
where the coupling rate is given by
\begin{equation}
r = \sqrt{2 \hbar \omega}\sigma = \sqrt{2\hbar\omega \eta_0 v_{g,\omega}^2 v_{g,2\omega}}=g\sqrt{L}.\label{eqn:r_QNLO}
\end{equation}
Here $|\alpha_\omega(z,t)|^2$ is the photon number density (in m$^{-1}$). Note that for the choice of Fourier convention used for $\alpha_\omega(z,t)$ and $\alpha_{\omega,m}(t)$, Parseval's theorem takes the form $\int_0^L |\alpha_\omega(z,t)|^2 dz = \sum_m |\alpha_{\omega,m}(t)|^2$. These classical equations of motion will be revisited in Sec.~\ref{sec:Rosetta_stone}, where they will be shown to correspond to the mean-field limit of the quantum equations of motion.







\subsubsection{The meaning of $g/\kappa$}

Throughout literature on few-photon nonlinear optics, a common figure of merit is the ratio of the single-mode coupling rate, $g$, to a characteristic loss rate, $\kappa$, not to be confused with the traveling-wave nonlinear coupling coefficient. Having established the classical equations of motion for nonlinear resonators in photon-number-normalized units, we are equipped now to discuss the physical meaning of $g/\kappa$ in more detail. As a canonical example, we consider a single-mode resonator, accounting only for loss and nonlinear coupling
\begin{subequations}
\begin{align}
\partial_t \alpha_{\omega}(t) = & -(\kappa_{\text{ex},\omega}+\kappa_{\text{in},\omega})\alpha_{\omega}(t)- i g \alpha_{2\omega}(t)\alpha_{\omega,}^*(t),\label{eqn:cwe_photon_singlemode_1}\\
\partial_t \alpha_{2\omega}(t) = & -(\kappa_{\text{ex},2\omega}+\kappa_{\text{in},2\omega})\alpha_{2\omega}(t)-i \frac{g}{2} \alpha_{\omega}^2(t).\label{eqn:cwe_photon_singlemode_2}
\end{align}
\end{subequations}

To build intuition, we consider two cases: SHG and OPA. First, we consider instantiating two-photons-worth of field in the fundamental, $\alpha_\omega^2(0) = 2$, and solve Eqns.~\ref{eqn:cwe_photon_singlemode_1}-\ref{eqn:cwe_photon_singlemode_2} in the absence of loss. The second-harmonic field is given by the solution for saturated SHG, $\alpha_{2\omega}(t) = -i\tanh(g t)$. The nonlinear coupling, $g$, sets the timescale over which a pair of photons from the fundamental up-converts to a single photon of second harmonic, with saturation occurring when $gt\approx 1$. The ratio $g/(\kappa_{\text{ex},2\omega}+\kappa_{\text{in},2\omega})$ measures the degree to which few-photon signals can achieve saturated SHG before the second harmonic decays from the cavity. Similarly, for the case of OPA, we instantiate the cavity with a single-photon-worth of field at each harmonic, $\alpha_\omega(0) = 1$ and $\alpha_{2\omega}(0) = 1$. In this case, $g$ sets the time-scale over which the second harmonic down-converts to fundamental by saturated OPA, and the ratio $g/(\kappa_{\text{ex},\omega}+\kappa_{\text{in},\omega})$ measures the degree to which the small-signal gain imparted by the pump photon for small $t$ exceeds the loss of the fundamental.

In both of the above cases, the coupling rate $g$ is being compared to a characteristic loss rate $\kappa$, and the condition $g>\kappa$ denotes that few-photon nonlinear behaviors become important. We note, however, that at this time there is no uniform choice for the characteristic loss rate $\kappa$, or the figure of merit. The value $g/(\kappa_{\text{ex},\omega}+\kappa_{\text{in},\omega})$ is commonly reported as the relevant figure of merit in triply-resonant devices,  however this neglects decoherence due to loss of the second harmonic. $g/\sqrt{(\kappa_{\text{ex},\omega}+\kappa_{\text{in},\omega})(\kappa_{\text{ex},2\omega}+\kappa_{\text{in},2\omega})}$ was proposed in~\cite{Yanagimoto2022_temporal} to give equal weight to the fundamental and second harmonic, and other arrangements such as $g/\left((\kappa_{\text{ex},\omega}+\kappa_{\text{in},\omega})^2(\kappa_{\text{ex},2\omega}+\kappa_{\text{in},2\omega})\right)^{1/3}$ could be reasonably argued by comparing the factors of group velocity that enter into the numerator and denominator. Similarly, $g/\text{max}(\kappa_{\text{ex},\omega}+\kappa_{\text{in},\omega},\kappa_{\text{ex},\omega}+\kappa_{\text{in},\omega})$ could be argued as a more conservative measure. In addition to these differing conventions for $g/\kappa$, there also exist differing conventions for the underlying parameters. In the above treatment of few-photon OPA, $g$ and $\kappa$ correspond to the \emph{field} gain and loss rates, respectively, but differing conventions exist where $2g$ and $\kappa/2$ are the field gain and loss rates, respectively. Put simply, care must be taken when comparing reported numbers for $g/\kappa$ for triply-resonant devices in the literature, and apples-to-apples comparisons are only meaningful when the same figure of merit is used.

To better interpret these differing conventions, it can be helpful to recast $g/\kappa$ in terms of physical behaviors, such as the number of intracavity pump photons at threshold, or at saturation. For doubly-resonant OPOs, there is no ambiguity regarding the loss rate, and $g/(\kappa_{\text{ex},\omega}+\kappa_{\text{in},\omega})$ is commonly taken as the relevant figure of merit. Here, we can build further intuition by rewriting Eqn.~\ref{eqn:OPO_steady_state} in photon-number-normalized units. Noting that the intracavity pump photon number is given by $N_{2\omega} = P_{2\omega} T_{\text{rt},2\omega}/(2\hbar\omega)$, the threshold photon number is $N_\text{th} = (2(\kappa_{\text{ex},\omega}+\kappa_{\text{in},\omega})/g)^2$. For the conventions chosen here, $g/\kappa = 1$ corresponds to $N_\text{th} = 4$. If instead we used $\kappa$ to denote the \emph{power} loss rate, then $g/\kappa = 1$ corresponds to $N_\text{th} = 1$. Using alternative conventions for $g$ ($\rightarrow 2g$), also yields $N_\text{th} = 1$ when $g/\kappa = 1$, and adopting \emph{both} conventions yields $N_\text{th} = 1/4$. Repeating this conversion to photon-normalized units for a triply-resonant OPO, we have
\begin{equation*}
    P_\text{th,TRO} = \left(\frac{\kappa_{\text{ex},\omega}+\kappa_{\text{in},\omega}}{g}\right)^2\left(\frac{\left(\kappa_{\text{ex},2\omega} + \kappa_{\text{in},2\omega}\right)^2 T_{\text{rt},2\omega}}{2\kappa_{\text{ex},2\omega}}\right)\left(\frac{2\hbar\omega}{T_{\text{rt},2\omega}}\right),
\end{equation*}
where we have used $2\kappa_{\text{ex},2\omega}\approx T_{\text{oc},2\omega} T_{\text{rt},2\omega}^{-1}$. Noting that the intracavity pump power is 
\begin{equation*}
    P_\text{th,intracavity} = P_\text{th,TRO}\left(\frac{2\kappa_{\text{ex},2\omega}}{\left(\kappa_{\text{ex},2\omega} + \kappa_{\text{in},2\omega}\right)^2 T_{\text{rt},2\omega}}\right),
\end{equation*}
and that intracavity photon number is
\begin{equation}
    N_\text{th,intracavity} = P_\text{th,intracavity}\left(\frac{T_{\text{rt},2\omega}}{2\hbar\omega}\right),
\end{equation}
we find that $g/(\kappa_{\text{ex},\omega}+\kappa_{\text{in},\omega}) = 1$ corresponds to the condition that the gain imparted by a single intracavity pump photon exceeds the signal loss. We note here that employing physical comparisons, such as the intracavity photon number at threshold, is an extremely useful practice in addition to comparing the definitions of $g$ and $\kappa$ in the equations of motion when interpreting results found in the literature. More examples of the connection between physical phenomena and $g/\kappa$ are given in~\cite{zhao2022ingap}.


\subsubsection{Prospects for observing few-photon quantum nonlinearities}~\label{sec:prospects}

Having established $g/\kappa > 1$ as the condition for observing few-photon nonlinearities, we briefly discuss experimental routes towards realizing such strongly-coupled nonlinear resonators. We note here that the practical lower bound of $\kappa$ is set by the bulk absorption properties of the underlying waveguide materials, and that the upper bound for $g$ set by the material nonlinear susceptibility, bandgap, and refractive index. Therefore, for any choice of material system, there is a material-limited upper bound for $g/\kappa$. We first consider the trade-offs between $g$ and $\kappa$ in continuous-wave resonators, before discussing proposals for enhancing the coupling rate using femtosecond pulses. In principle, this latter approach circumvents the limitations of continuous-wave devices by enabling small effective mode volumes irrespective of the physical size of the resonator. This flexibility allows for the coupling rate and loss rate to be optimized independently. We note here that a comprehensive discussion of the loss mechanisms in each emerging material system are beyond the scope of this tutorial. To familiarize readers with the current state of the field, we will discuss loss numbers in thin-film lithium niobate, which achieves similar loss numbers to other state-of-the-art platforms for nonlinear photonics.

\begin{figure}
     \centering
     \includegraphics[width=\columnwidth]{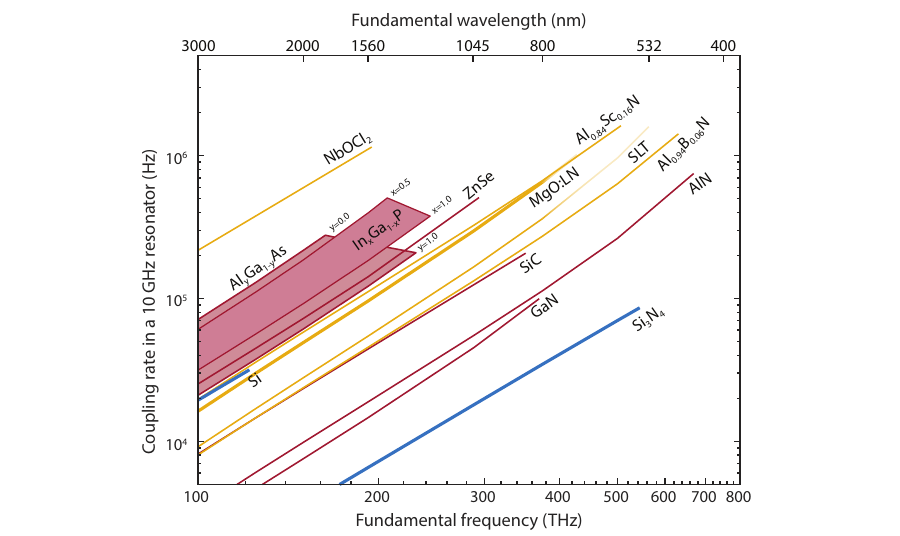}
     \caption{\label{fig:g_comparison}Comparison of the continuous-wave coupling rate $g$ for many emerging platforms for nonlinear photonics, here assuming a cavity free-spectral range of $\Delta f_{\text{fsr},\omega} = v_{g,\omega}/L = 10$~GHz and the optimized waveguide geometries used to generate Fig.~\ref{fig:eta0 comparison}. Here the nonlinear coupling exhibits an $\omega^{2.5}-\omega^{2.75}$ scaling, where the $\omega^{2.5}$ comes from $g \propto \sqrt{\omega \eta_0} \propto \sqrt{d_\text{eff}^2 \omega^5}$, and the additional power-law scaling of 0 - 0.25 comes from the dispersion of $d_\text{ijk}$.}
\end{figure}

We begin by discussing limitations to the coupling rate, $g$. Noting that $g = \sqrt{2\hbar\omega v_{g,\omega}^2 v_{g,2\omega}\eta_0/L}$, we identify three components necessary for realizing strong coupling, namely, a large operating frequency $\omega$, a large normalized efficiency $\eta_0$ (\textit{e.g.} by optimizing the transverse confinement as in Fig.~\ref{fig:eta0 comparison}), and a small cavity length, $L$. We note, however, that there often exist practical limitations that set a minimum cavity length $L$. For example, ring resonators are limited to a minimum bend radius $R$, with the circumference $2\pi R$ playing the role of cavity length, due to an increase in bending loss and a decrease in nonlinear overlap that occurs with increasing $R$. In principle, point-defect photonic crystal nanocavities can be used to confine the fundamental and second harmonic to a wavelength-scale mode volume, however such small mode volumes are typically accompanied by large losses at at least one harmonic, as well as additional decoherence due to thermorefractive noise~\cite{panuski2020fundamental}. For these reasons, we compare the largest single-mode $g$ attainable in each emerging material system as a function of wavelength assuming a constant free-spectral range, $\Delta f_{\text{FSR},\omega} = v_{g,\omega}/L = 10$~GHz, in Fig.~\ref{fig:g_comparison}. In contrast with our previous comparison of the normalized efficiency, $\eta_0$, which scales as $d_\text{eff}^2\omega^4$, $g$ exhibits a $d_\text{eff} \omega^{2.5}$ scaling since $g \propto \sqrt{\omega \eta_0} \propto d_\text{eff} \omega^{2.5}$. Here the additional factor of $\sqrt{\omega}$ is due to a factor of $\sqrt{\omega}$ in the field-per-photon. In reality, $g$ exhibits an additional scaling exponent of $0-0.25$ depending on the material, due to the dispersion of $d_\text{eff}$. Given this strong frequency dependence, we find that materials with large bandgaps tend to exhibit the largest nonlinear couplings.

There are many practical considerations in designing resonators with low loss rates. These include the intrinsic material absorption, the ability to pattern a material with low surface and sidewall roughness, and the geometry of the resonator itself. For example, a ring resonator with a near-lossless directional coupler is predominantly limited by the propagation loss accumulated over a single round trip of the ring. Therefore, as the resonator dimensions are scaled smaller, both the round-trip propagation loss and the round-trip time are rescaled by the same factor, which renders the loss rate invariant with respect to resonator size (until bending loss becomes significant). Conversely, a Fabry-Perot resonator with discrete mirror losses exhibits a loss rate that increases linearly as the length of the resonator is decreased, since the round-trip time is decreased and the single-pass loss is unchanged. In other words, the nature of the dominant loss mechanism (discrete versus distributed) determines the extent to which the $g\propto L^{-1/2}$ scaling of the nonlinear coupling can be utilized to increase $g/\kappa$. Noting that both discrete and distributed losses are present in any real resonator, in reality there is always a turning point in $L$ at which $g/\kappa$ is optimized for a particular design. 

As an example of realistic loss numbers in nonlinear photonics, we consider the case of thin-film lithium niobate. At this time, typical propagation losses at a wavelength of 1560-nm at 30 dB/m in poled waveguides~\cite{McKenna2022}, 3 dB/m in state-of-the-art ring resonators~\cite{zhang2017monolithic}, and 0.3 dB/m at the bulk material limit set by various defects, such as in-diffused metals and OH absorption overtones~\cite{schwesyg2010light,Leidinger2015,shams2022reduced}. For a typical group index of 2.3, these values correspond to an intrinsic loss rate $\kappa_{\text{in},\omega}/(2\pi)$ of 76~MHz, 7.6~MHz, and 760~KHz, respectively. Comparing these values to the coupling rate for optimized TFLN ridge waveguides (see Fig.~\ref{fig:eta0 comparison}), we find an optimized coupling rate of $g/(2\pi)=$100 kHz in a resonator with a free-spectral range of 10 GHz, and $g/(2\pi)\sim$10 MHz in a wavelength-scale (100 THz free-spectral range) resonator. In other words, the best-case scenario, namely a wavelength-scale resonator operating at the material-loss limit, may achieve $g/\kappa \approx 10$, when operating at a fundamental wavelength of 1560-nm. In reality, given the practical tradeoffs between $g$ and $\kappa$ discussed above, more realistic resonators operate at $g/\kappa \approx 0.001 - 0.1$, with current state-of-the-art devices operating at $g/\kappa \approx 0.01$~\cite{lu2020toward}.

\begin{figure}
     \centering
     \includegraphics[width=\columnwidth]{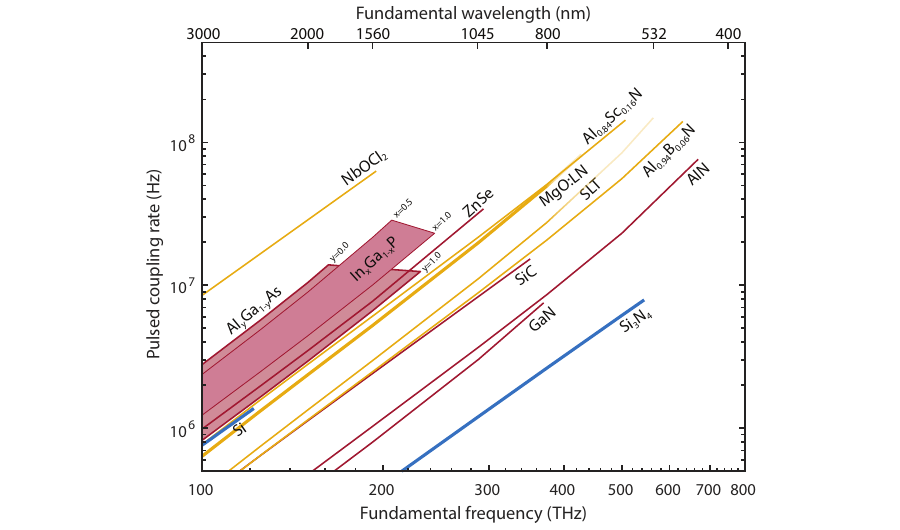}
     \caption{Comparison of the effective pulsed coupling rate, $g_\text{pulsed}$, for many emerging platforms for nonlinear photonics using the optimized geometries from Fig.~\ref{fig:eta0 comparison}. Each material system exhibits a power law scaling of $\omega^3-\omega^{3.25}$, where the excess exponent $>3$ is due to the dispersion of $d_{ijk}$ estimated by Miller's delta scaling.}
     \label{fig:pulsed_g_comparison}
\end{figure}

An alternative route is the development of resonators driven by ultrafast pulses, where enhancements to the field per photon enabled by multimode interference can be used to realize small effective mode volumes in resonators of arbitrary size. The full quantum theory of pulsed nonlinear interactions will be developed in the following sections. For the present discussion, we note that the clearest point of comparison between continuous-wave interactions and pulsed interactions is the temporally-trapped dynamics studied in Sec.~\ref{sec:FSNLO}. In these systems, the trapped pulses behave identically to CW fields, with an effective coupling rate given by
\begin{equation}
g_\text{pulsed} = g\frac{\pi}{4}\sqrt{\frac{T_\text{rt}}{2\tau}}=\frac{\pi}{4}\sqrt{\frac{2\hbar\omega v_{g}^2\eta_0}{2\tau}},
\end{equation}
where $\tau$ is the duration of the trapped pulses, here assuming a sech$^2$ envelope. We note here that temporal trapping requires the two interacting waves to be group-velocity-matched, and therefore $T_\text{rt}=T_{\text{rt},\omega}=T_{\text{rt},2\omega}$ is the cavity round trip time. For interactions between trapped pulses, the pulse duration $\tau$ (more specifically $\tau v_g$) now plays the role of the longitudinal confinement $L$. For resonators driven by few-cycle pulses, the pulsed coupling rate becomes comparable to that of a wavelength-scale resonator, irrespective of the cavity free-spectral range set by $L$. This feature allows for the resonator geometry to be chosen to minimize loss, \textit{e.g.} by fabricating rings with a large radius of curvature. A comparison of optimized pulsed coupling rates for several emerging nonlinear photonics platforms is given in Fig.~\ref{fig:pulsed_g_comparison}, here assuming a pulse duration of two cycles. We note here that since $\tau = m \omega$, where $m$ is the number of cycles, $g_\text{pulsed}$ exhibits an additional $\omega^{1/2}$ scaling relative to the single-mode coupling rate in a continuous-wave resonator. As a result, pulsed devices favor short wavelength operation: in thin-film lithium niobate, $g_\text{pulsed}\approx 5-6$~MHz at a fundamental wavelength of 1560-nm, whereas coupling rates approaching 100~MHz become possible at much shorter wavelengths. Since ring resonators with lifetimes approaching 10~MHz have already been demonstrated, and the bulk material loss supports loss rates as low as 1~MHz, pulsed devices appear to be a promising route towards realizing few-photon nonlinearities with $g/\kappa > 10$.


\subsection{Closing remarks}

The past few sections have broadly addressed the routes towards realizing extremely low power nonlinear devices. For continuous-wave interactions low power is enabled simply by the combination of tight transverse confinement, and long interaction lengths. In a resonator, the long interaction length is instead replaced by a long lifetime, which enables much greater conversion efficiencies in physically compact devices. Pulsed interactions can greatly reduce the energy required to achieve saturation, and in principle may approach comparable energy requirements to wavelength-scale resonators by using few-cycle pulses. In these systems, the interaction length is limited by the dispersion of the waveguide, and in principle, the use of dispersion-engineered waveguides can enable saturated behaviors to occur with relatively low photon number. This raises the following questions: How extendable are these benefits to quantum devices? Does the field enhancement seen in classical devices always improve the performance of quantum devices? Can we still produce simple models in the quantum limit? The latter half of this tutorial develops a more careful treatment of the quantum behaviors of ultrafast pulses. This formalism will enable us to address these questions, and to identify new opportunities that exist at the boundary of ultrafast and quantum nonlinear optics.

\section{From classical to quantum NLO}\label{sec:Rosetta_stone}


\subsection{Introduction}

At first glance, the coupled-wave equations found in classical nonlinear optics are rather different than the formalism of quantum optics. In the former, the evolution of a complex field envelope $\alpha(z,t)$ generally takes the form
\begin{align}
    \label{eq:rosetta-coupled-wave-generic}\partial_z\alpha(z,t)=f_\mathrm{c}(\alpha(z,t),\alpha^*(z,t)),
\end{align}
where the subscript ``c'' stands for ``classical''. In this picture, we associate a complex number with each point, $t$, in the pulse envelope, and the fields evolve with respect to propagation along $z$ in a nonlinear waveguide. In contrast, the quantum formalism assigns to each mode an operator
\begin{align}
    \hat{a}=\alpha+\delta\hat{a},
\end{align}
where $\alpha=\langle\hat{a}\rangle$ is the mean-field amplitude and $\delta\hat{a}$ is a quantized operator that captures the quantum fluctuations around this mean. These modes may be Fourier modes in a resonator, or modes localized around points $z$ within a waveguide. In the Schr\"odinger picture, the quantum state $\ket{\psi}$ evolves under the influence of a Hamiltonian according to the Schr\"{o}dinger equation,
\begin{align}
\label{eq:intro-schroedinger}
i\hbar\partial_t\ket{\psi}=\hat{H}\ket{\psi}.
\end{align}
Given how disparate these formalisms are, a natural question is how to establish a quantum-classical correspondence between the coupled-wave equations and the Schr\"odinger equation. To do this, we need to resolve several discrepancies: (i) The former concerns optical field amplitudes, described by complex numbers, while the latter describes the dynamics of a wavefunction $\ket{\psi}$. (ii) The former is usually a system of nonlinear ODEs, while the latter is always a linear ODE. (iii) The former propagates in the spatial coordinate $z$, while the latter propagates in the time coordinate $t$. The main purpose of this section is to bridge these gaps and reconcile the classical description of nonlinear optics with the quantum mechanical pictures. This prepares readers for the quantum treatments discussed in the rest of the tutorial by linking many of the observed behaviors back to the dynamics of classical devices, and establishes the notation used throughout the quantum section. Furthermore, by using the quantum-classical correspondence, we find the form of the quantum Hamiltonian in such a way that its parameters are directly tied to the usual parameters found in the classical theory.

To close the gaps (i) and (ii), we first discuss a general procedure to derive the evolution of optical field amplitudes for a given quantum model. For this purpose, it is more insightful to work in the Heisenberg picture rather than the Schr\"odinger picture described by Eqn.~\ref{eq:intro-schroedinger}. In this picture the evolution of the field operators (rather than the state) as determined by a Hamiltonian, $\hat{H}=H(\hat{a},\hat{a}^\dagger)$, is given by
\begin{align}
    \label{eq:rosetta-generic-heisenberg}
    \partial_t\hat{a}=F_\mathrm{q}(\hat{a},\hat{a}^\dagger),
\end{align}
where $F_\mathrm{q}(\hat{a},\hat{a}^\dagger)=i[\hat{H},\hat{a}]/\hbar$. Here the subscript ``q'' denotes that these equations of motion are determined by the quantum Hamiltonian. For a given Hamiltonian, the Heisenberg equations of motion can be evaluated using a commutator, and conversely, given the Heisenberg equations of motion and the known commutators between $\hat{a}$ and $\hat{a}^\dagger$, the equations of motion can be inverted to determine the quantum Hamiltonian, up to a constant offset. Starting from \eqref{eq:rosetta-generic-heisenberg}, we can derive the evolution of the \emph{classical} optical fields, \textit{i.e.}, the expectation-values of the field operators, by performing the substitutions $\hat{a}=\alpha+\delta\hat{a}$ and ignoring terms of order $\delta\hat{a}$ (\textit{i.e.} $\alpha\delta\hat{a}=0$, $\delta\hat{a}\alpha = 0$, and $\delta\hat{a}\delta\hat{a}=  0$). With these approximations, Eqn.~\ref{eq:rosetta-generic-heisenberg} for the field operators will only contain the mean fields,
\begin{align}
    \label{eq:rosetta-generic-heisenberg-alpha}
    \partial_t{\alpha}=F_\mathrm{q}(\alpha,\alpha^*),
\end{align}
which resolves the first discrepancy (i). Intuitively, we can understand this step as ignoring any fluctuations of the field due to $\delta\hat{a}$, thereby keeping only the evolution of the mean field. Note that $F_\mathrm{q}$ is often a nonlinear function of $\alpha$ and $\alpha^*$, meaning that \eqref{eq:rosetta-generic-heisenberg-alpha} is a nonlinear ODE. This provides a resolution to (ii) as well; nonlinear ODEs for field amplitudes arise from the Heisenberg equations of motion, rather than the Schr\"odinger equation.

Finally, to resolve the discrepancy between the space- and time-propagating pictures (iii) we must recast the classical equations of motion in terms of spatial field envelopes that evolve in time rather than temporal envelopes that evolve in space. A heuristic derivation of the time-propagating equations of motion was presented in Sec.~\ref{sec:time-prop}, and a more formal derivation based on waveguide and resonator normal modes is given in Appendix~\ref{sec:time-propagating-spatial-modes}. In either of these cases, the classical coupled-wave equations take the form
\begin{align}
    \label{eq:rosetta-coupled-wave-generic-t}\partial_t\alpha=F_\mathrm{c}(\alpha,\alpha^*),
\end{align}
where the complex scalar $\alpha$ is normalized in photon-number units. For normal modes, $|\alpha(t)|^2$ corresponds to the expected photon number contained in each mode. We therefore identify Fourier-domain coupled-wave equations that account for the full time evolution of the normal modes, namely Eqns.~\ref{eqn:cwe_photon_conventional_1}-\ref{eqn:cwe_photon_conventional_2}, as the point of connection between the classical and quantum theories.

Having resolved the above discrepancies between the classical and quantum models, we can determine the quantum Hamiltonian by reversing these steps. That is, starting from the classical time-evolving equations of motion, $F_\mathrm{c}$, we promote $\alpha$ and $\alpha^*$ to quantum operators, $\hat{a}$ and $\hat{a}^\dagger$, and enforce commutation relations $\left[\hat{a},\hat{a}^\dagger\right] = 1$. Since the classical and quantum equations of motion describe the same field amplitudes in the mean-field limit, we assert that the quantum equations $F_\mathrm{q}(\hat{a},\hat{a}^\dagger)$ must have the same form as $F_\mathrm{c}(\alpha,\alpha^*)$ with $\alpha$ and $\alpha^*$ replaced by $\hat{a}$ and $\hat{a}^\dagger$. Finally, with $F_\mathrm{q}(\hat{a},\hat{a}^\dagger)$, we can infer the form of the quantum Hamiltonian $\hat{H}$ by inspection.

A benefit of this top-down approach is that the resultant quantum model is ensured to agree with classical predictions regarding the mean-field dynamics. In addition, since the classical equations of motion are derived directly from Maxwell’s equations, the resultant Hamiltonian, and thus our quantum model, always recovers Maxwell's equations by construction. A drawback is that we do not immediately have an answer to the question ``what exactly is the field quantized here?'' While such questions can, of course, be answered by deriving the Hamiltonian using canonical field quantization~\cite{Hillery1984,Drummond1990,sipe2009photons,Quesada2017,Quesada2022,Raymer2020,Drummond2014}, this bottom-up approach to quantization is often theoretically involved, and a failure to properly choose the fundamental fields to be quantized can lead to systematic errors in the resulting quantum model~\cite{Quesada2017, Quesada2022,Hillery1984}. While lacking the rigor of canonical field quantization, this phenomenological approach facilitates simpler calculations for experimentalists, whose interests are simply in obtaining a form of Hamiltonian that predicts their experimental outcomes correctly.

{The rest of the section is structured as follows. In Sec.~\ref{sec:rosetta-derive-coupled-wave}, we introduce a more detailed derivation of mean-field equation from a quantum Hamiltonian, using a $\chi^{(2)}$ system as an example. The purpose of this subsection is to provide intuition on how the functional form of a Hamiltonian and the corresponding mean-field equations are related to each other. In Sec.~\ref{sec:rosetta-waveguide-hamiltonian}, we write down a quantum Hamiltonian for a $\chi^{(2)}$ nonlinear resonator so that the Heisenberg equation of motion agrees with the classical CWEs in the mean-field limit. Finally, in Sec.~\ref{sec:small-large-limits}, we consider the small and large size limit of $\chi^{(2)}$ nonlinear resonators, which recover the behaviors of single-mode continuous-wave resonators and traveling-wave pulse propagation, respectively. We reiterate here that the Hamiltonians presented in these subsequent sections are equivalent to those derived more formally using canonical quantization~\cite{Drummond2014, Quesada2017}.}


\subsection{Obtaining coupled-wave equations from the Heisenberg equations of motion}
\label{sec:rosetta-derive-coupled-wave}

To recover the classical equations of motion we assume that the quantum state associated with each mode corresponds to a coherent state for all time,
\begin{align}
\label{eq:intro-ket-classical}
\ket{\psi_\mathrm{classical}}=\hat{D}(\alpha)\ket{0}
\end{align}
where $\ket{0}$ is a vacuum, and $\alpha$ characterizes a complex field amplitude. The displacement operator $\hat{D}(\alpha)$ moves a phase-space distribution of a vacuum from the origin, to excite a coherent state with mean amplitude $\alpha$. In making this assumption, we effectively confine the equations of motion onto a low-dimensional manifold spanned by the coherent states. This mean-field approximation is obtained by taking expectation values of the Heisenberg equations of motions with respect to $\ket{\psi_\mathrm{classical}}$ as
\begin{align}
\label{eq:rosetta-generic-heisenberg2}
\partial_t\hat{a}=-\frac{\mathrm{i}}{\hbar}\partial_ {\hat{a}^\dagger} H(\hat{a},\hat{a}^\dagger)\mapsto \partial_t\alpha=-\frac{\mathrm{i}}{\hbar}\partial_{\alpha^*}H(\alpha,\alpha^*),
\end{align}
where $\hat{H}=H(\hat{a},\hat{a}^\dagger)$ is the Hamiltonian of the system. The mean-field equation \eqref{eq:rosetta-generic-heisenberg2} can be interpreted as a time-propagating coupled-wave equation. We can easily extend the equation to multimode dynamics involving a vector of field operators $\boldsymbol{\hat{a}}$ as
\begin{align}
\label{eq:rosetta-generic-heisenberg-multimode}
\partial_t\boldsymbol{\hat{a}}=-\frac{\mathrm{i}}{\hbar}\partial_ {\boldsymbol{\hat{a}}^\dagger} H(\boldsymbol{\hat{a}},\boldsymbol{\hat{a}}^\dagger)\mapsto \partial_t\boldsymbol{\alpha}=-\frac{\mathrm{i}}{\hbar}\partial_{\boldsymbol{\alpha}^*}H(\boldsymbol{\alpha},\boldsymbol{\alpha}^*),
\end{align}
where $\boldsymbol{\alpha}$ is a vector containing complex mean-field values and $\partial_{\alpha^*}$ is a  gradient operator.

For instance, consider a Hamiltonian of the form
\begin{align}
    \hat{H}/\hbar=\frac{g}{2}(\hat{a}^{\dagger2}\hat{b}+\hat{a}^2\hat{b}^\dagger),
\end{align}
which is usually used to describe $\chi^{(2)}$ nonlinear interactions between the FH and SH modes with annihilation operators $\hat{a}$ and $\hat{b}$, respectively. The Heisenberg equations of motion are
\begin{align}
    &\partial_t\hat{a}=-\mathrm{i}g\hat{a}^\dagger\hat{b} &\partial_t\hat{b}=-\frac{\mathrm{i}g}{2}\hat{a}^{2}.
\end{align}
These equations can be put to the form \eqref{eq:rosetta-generic-heisenberg-multimode} by summarizing the operators in a vector form $\boldsymbol{\hat{a}}=(\hat{a},\hat{b})^\intercal$. Taking expectation values of both sides with respect to the product state $\ket{\psi_\mathrm{classical}}=\ket{\alpha}\ket{\beta}$, we obtain coupled-wave equations describing the evolution of the coherent state amplitudes
\begin{align}
    &\partial_t\alpha=-\mathrm{i}g\alpha^*\beta &\partial_t\beta=-\frac{\mathrm{i}g}{2}\alpha^{2}.\label{eqn:cwe_meanfield_example}
\end{align}

Here, it is worth asking what are the features that are excluded from the mean-field approximation. To see this, we rewrite the full wavefunction $\ket{\psi}$ in a form
\begin{align}
    \ket{\psi}=\hat{D}\ket{\varphi_\mathrm{D}},
\end{align}
where classical dynamics are factored out in the form of a displacement operator. The wavefunction $\ket{\varphi_\mathrm{D}}$ is the state in the displaced frame, which evolves under the Hamiltonian $\hat{H}_\mathrm{D}(t)=\hat{D}^\dagger\hat{H}\hat{D}-\mathrm{i}\hat{D}^\dagger\partial_t\hat{D}$. Thus, we find mean-field approximation to be equivalent to assuming $\hat{H}_\mathrm{D}\approx0$ and ignoring any deviation of quantum fluctuations in $\ket{\varphi_\mathrm{D}}$ from a vacuum.

\subsection{A phenomenological quantum model for a $\chi^{(2)}$ nonlinear waveguide}
\label{sec:rosetta-waveguide-hamiltonian}
In this subsection, we introduce the Hamiltonian for a $\chi^{(2)}$ nonlinear waveguide. Following the same approach as was used in the classical section, we first consider a finite spatial window $-L/2\leq z\leq L/2$ over which we assume a periodic boundary condition. Such a finite spatial window leads to discrete wavespace modes, which we can quantize through the standard approach to the electric field quantizations. Starting from a model with finite $L$, we can take various limits to derive models depending on the goal. If our aim is to model the propagation of an isolated pulse, we can take the limit of $L\rightarrow\infty$ to obtain a model for an infinite waveguide. In this picture, $L$ is an artificial parameter introduced solely for the purpose of quantization, and any finite-size effects that depends on $L$ are artifacts. On the other hand, $L$ can be set to model a physical resonator with size $L$, in which case finite-size effects are real. Below, we first establish a correspondence between quantum and classical theories using a model with finite $L$. We then consider the limit of small and large $L$, which correspond to single-mode resonator and free propagation of an optical pulse, respectively.

We denote the carrier angular wavenumber for the FH as $k_{a}=\frac{2\pi}{L}m_a$, which sets the carrier wavenumber of the SH to $k_{b}=2k_a$. We label wavenumber modes with $k_{u,m}=k_u+\delta k_m$ for $u\in\{a,b\}$, where $\delta k_m=2\pi m/L$ is the deviation of the mode from the carrier wavenumber $k_u$, so that $k_{u,m}=\frac{2\pi}{L}(m_u+m)$. The corresponding modes are characterized by the annihilation operators $\hat{a}_m$ and $\hat{b}_m$ for FH and SH, respectively, and they fulfill the canonical commutation relations $[\hat{a}_m,\hat{a}_{m'}^\dagger]=[\hat{b}_m,\hat{b}_{m'}^\dagger]=\delta_{m,m'}$. 

With these wavespace operators, we can also denote field operators in the real space via Fourier summations
\begin{align}
    &\hat{a}_z=\frac{1}{\sqrt{L}}\sum_{m=-\infty}^\infty e^{-\mathrm{i}\delta k_m z}\hat{a}_m, &\hat{b}_z=\frac{1}{\sqrt{L}}\sum_{m=-\infty}^\infty e^{-\mathrm{i}\delta k_m z}\hat{b}_m.
\end{align}
Intuitively, $\hat{a}_z^\dagger$ ($\hat{b}_z^\dagger$) creates a localized excitation at position $z$ in the FH (SH) mode. The definition of the field operators ensures that the commutation relations $[\hat{a}_z,\hat{a}_{z'}^\dagger]=[\hat{b}_z,\hat{b}_{z'}^\dagger]=\delta(z-z')$ are fulfilled. We note here that the delta function $\delta(z-z')$ has a unit of $[\mathrm{length}^{-1}]$, the field operator $\hat{a}_z$ has a unit of $[\mathrm{length}^{-1/2}]$.

Below, we write down the functional form of the Hamiltonian by inspection from the classical CWEs, which, in general, takes the form
\begin{subequations}
\label{eq:multimode-hamiltonian-finite-L}
\begin{align}
    \hat{H}=\hat{H}_\text{L}+\hat{H}_\text{NL}.
\end{align}
The linear term $\hat{H}_\text{L}$ represents the dispersion of the waveguide in the absence of any nonlinearity, and $\hat{H}_\text{NL}$ describes nonlinear interactions. Nominally, the dispersion of the waveguide can be denoted as $\omega(k)$, which is the frequency (\textit{i.e.}, energy) of a mode $\omega$ as a function of the wavenumber of the mode $k$. Here, we expand the dispersion function around the carrier wavenumbers $k_u$ for FH ($u=a$) and SH ($u=b$), defining $\omega_v(\delta k)=\omega(k_u+\delta k)$. With this notation, the linear Hamiltonian is expressed as
\begin{align}
\begin{split}
    \hat{H}_\text{L}/\hbar&=\sum_{u\in\{a,b\}}\int\mathrm{d}z\,\hat{u}_z^\dagger \omega_u(-\mathrm{i}\partial_z)\hat{u}_z=\sum_{u\in\{a,b\}}\sum_m\,\hat{u}_m^\dagger \omega_u(\delta k_m)\hat{u}_m.
\end{split}
\end{align}

We then consider the nonlinear terms. Classically, the nonlinear polarization generates an interaction of the form given by Eqn.~\ref{eqn:cwe_meanfield_example} at each point in space, which suggests a localized interaction in which two fundamental photons may ``collide'' to produce a second-harmonic photon (and vice versa). These observations motivate us to denote the nonlinear Hamiltonian as
\begin{align}
\begin{split}
    \hat{H}_\text{NL}/\hbar&=\frac{r}{2}\int\mathrm{d}z\,\left(\hat{a}_z^2\hat{b}_z^\dagger+\hat{a}_z^{\dagger 2}\hat{b}_z\right)=\frac{r}{2\sqrt{L}}\sum_{m_1,m_2}\left(\hat{a}_{m_1}\hat{a}_{m_2}\hat{b}_{m_1+m_2}^\dagger+\hat{a}_{m_1}^\dagger\hat{a}_{m_2}^\dagger\hat{b}_{m_1+m_2}\right),
\end{split}
\end{align}
\end{subequations}
where $r$ is a phenomenological nonlinear coefficient. The structure of the nonlinear Hamiltonian as written in the wavespace, $\hat{H}_\text{NL}\propto(\hat{a}_{m_1}\hat{a}_{m_2}\hat{b}_{m_1+m_2}^\dagger+\hat{a}_{m_1}^\dagger\hat{a}_{m_2}^\dagger\hat{b}_{m_1+m_2})$, implies that the photon-photon interaction conserves the total momentum (\textit{i.e.}, $(m_1+m_2) - m_1 - m_2 = 0$).

With this phenomenological Hamiltonian, we now are ready to derive mean-field equations that recover the dynamics of the classical coupled-wave equations. The evolution of the field operators, as described by the Heisenberg equations of motion, are given by
\begin{align}
    &\partial_t\hat{a}_z=-\mathrm{i}\omega_a(-\mathrm{i}\partial_z)\hat{a}_z-\mathrm{i}r\hat{a}_z^\dagger\hat{b}_z, &\partial_t\hat{b}_z=-\mathrm{i}\omega_b(-\mathrm{i}\partial_z)\hat{b}_z-\mathrm{i}\frac{r}{2}\hat{a}_z^{2}.
\end{align}
Taking an expectation value of both sides with respect to a coherent ansatz state, we obtain
\begin{align}
    &\partial_t\alpha_z=-\mathrm{i}\omega_a(-\mathrm{i}\partial_z)\alpha_z-\mathrm{i}r\alpha_z^*\beta_z, &\partial_t\beta_z=-\mathrm{i}\omega_b(-\mathrm{i}\partial_z)\beta_z-\mathrm{i}\frac{r}{2}\alpha_z^{2},\label{eqn:mean_field_limit}
\end{align}
where we have denoted $\alpha_z=\langle\hat{a}_z\rangle$ and $\beta_z=\langle\hat{b}_z\rangle$.\footnote{The expectation values of products of operators can be decomposed to product of expectation values for coherent states. How this classical treatment breaks down through the classical-quantum transitions is discussed in Ref.~\cite{Armen2006}, for instance.}  Eqns.~\ref{eqn:mean_field_limit} are identical to the classical Eqns.~\ref{eqn:cwe_quantum_envelope_1}-\ref{eqn:cwe_quantum_envelope_2}, up to an overall choice of the rotating frame. Comparing the coefficients in front of the nonlinear terms, we find
\begin{align}
\label{eq:nonlinear-r-expression}
    r =  \sqrt{2\hbar\omega_av_{g,a}^2v_{g,b}\eta_0},
\end{align} 
where we have denoted the group velocity of the FH and SH modes as $v_{g,a}$ and $v_{g,b}$, respectively. The SHG normalized efficiency $\eta_0$ has a unit of $[\mathrm{power}^{-1}\cdot \mathrm{length}^{-2}]$, and therefore the overall nonlinear coefficient $r$ has a unit of $[\mathrm{length}^{1/2}\cdot\mathrm{time}^{-1}]$.

In principle, solving the equations of motion generated by the Hamiltonian \eqref{eq:multimode-hamiltonian-finite-L} yields the full multi-mode quantum state generated by the nonlinear waveguide. However, as with the classical coupled-wave equations, it is often more convenient to consider the Hamiltonian in a rotating frame where the overall phase and group delay accumulated by the envelopes due to linear propagation is factored out. More specifically, we consider a co-propagating frame with a reference velocity $v_\mathrm{r}$ and phase $\omega_\mathrm{r}$ via the unitary transformation
\begin{align}
    \hat{U}(t)=\exp\left\{-\mathrm{i}\sum_m\left[\left(\omega_\mathrm{r}+v_\mathrm{r}\delta k_m\right)\hat{a}_m^\dagger \hat{a}_m +\left(2\omega_\mathrm{r}+v_\mathrm{r}\delta k_m\right)\hat{b}_s^\dagger \hat{b}_s \right]t\right\},
\end{align}
which induces operator transformations
\begin{align}
    &\hat{U}^\dagger\hat{a}_m\hat{U}=e^{-\mathrm{i}(\omega_r+\mu_r\delta k_m)t}\hat{a}_m, &\hat{U}^\dagger\hat{b}_m\hat{U}=e^{-\mathrm{i}(2\omega_r+\mu_r\delta k_m)t}\hat{b}_m
\end{align}
These transformation leaves $\hat{H}_\text{NL}$ invariant, while the linear Hamiltonian is transformed as 
\begin{align}
\begin{split}
    \hat{H}_\text{L}/\hbar=\sum_{u\in\{a,b\}}\int\mathrm{d}{\bar{z}}\,\hat{u}_{\bar{z}}^\dagger \delta\omega_u(-\mathrm{i}\partial_{\bar{z}})\hat{u}_{\bar{z}}=\sum_{u\in\{a,b\}}\sum_m\hat{u}_m^\dagger \delta\omega_u(\delta k_m)\hat{u}_m,
\end{split}
\end{align}
where ${\bar{z}}=z-v_\mathrm{r}t$ is the co-moving coordinate, and 
\begin{align}
    &\delta\omega_a(\delta k)=\omega_a(\delta k)-\omega_\mathrm{r}-v_\mathrm{r}\delta k ,&\delta\omega_b(\delta k)=\omega_b(\delta k)-2\omega_\mathrm{r}-v_\mathrm{r}\delta k 
\end{align}
contain the residual phase accumulation within this rotating frame. 

We note here that this transformation is formally equivalent to our definition of envelope quantities in the context of classical nonlinear optics, here generalized for field operators rather than c-number quantities, with the residual dispersion terms $\delta \omega_\mu$ now playing the role of the integrated dispersion $D_{\text{int},\mu}$ encountered in $z$-propagating models. While the choice of $\omega_\mathrm{r}$ and $v_\mathrm{r}$ is arbitrary, in practice, it is useful to choose them such that the trivial phase and spatial dynamics of the system are eliminated. This is particularly important for numerical evaluations; typically, the nonlinear coupling rate is on the order of few megahertz, while $\omega_\mathrm{r}$ is on the order of hundreds of terahertz, and simultaneously keeping track of these vastly different timescales is computationally expensive. To eliminate this overhead, as with our previous treatment of ultrafast behaviors, we will define rotating frames that eliminate the explicit linear phase evolution of the carrier frequency, as well as the group delay accumulated by the envelopes. As an example, taking $\omega_\mathrm{r}$ and $v_\mathrm{r}$ to be the phase-velocity and the group-velocity of the FH carrier, we can eliminate the $0$th- and $1$st-order dependence of $\delta\omega_a$ on $\delta k$. For $\delta\omega_b$, the residual $0$th order term $\delta\omega_b(0)$ and $1$st order term $\delta\omega_b'(0)$ physically represent phase-mismatch and group-velocity-mismatch, respectively. For notational simplicity, we use the $z$ subscript to denote co-moving coordinate for the rest of this tutorial. Moving into the rotating frame, the overall Hamiltonian takes the form
\begin{align}
\label{eq:Hamiltonian-multimode-finite-L}
    \hat{H}/\hbar&=\sum_{u\in\{a,b\}}\int\mathrm{d}z\,\hat{u}_z^\dagger \omega_u(-\mathrm{i}\partial_z)\hat{u}_z+\frac{r}{2}\int\mathrm{d}z\,\left(\hat{a}_z^2\hat{b}_z^\dagger+\hat{a}_z^{\dagger 2}\hat{b}_z\right)\nonumber\\
    &=\sum_{u\in\{a,b\}}\sum_m\hat{u}_m^\dagger \delta\omega_u(\delta k_m)\hat{u}_m+\frac{r}{2\sqrt{L}}\sum_{m_1,m_2}\left(\hat{a}_{m_1}\hat{a}_{m_2}\hat{b}_{m_1+m_2}^\dagger+\hat{a}_{m_1}^\dagger\hat{a}_{m_2}^\dagger\hat{b}_{m_1+m_2}\right)
\end{align}

\subsection{Small and large $L$ limits}
\label{sec:small-large-limits}
Here, we start from the quantum model we derived in the previous section to consider two extreme cases of $L$, \textit{i.e.}, small and large $L$ limits, from which we obtain a single-mode and a continuous multimode Hamiltonian, respectively. The Hamiltonians and notations we introduce in this subsection will be the basis for the discussions in the following sections.

\paragraph{Small $L$ limit} First, we consider the small $L$ limit $L\rightarrow0$, which corresponds to the physics of a small resonator. In this limit, the gap between the wavenumbers of neighboring modes grows to infinity, \textit{i.e.}, $\delta k_{m+1}-\delta k_{m}=2\pi/L\rightarrow\infty$. Consequently, all the modes get isolated from each other, and nonlinear interactions can mediate coupling among only few modes that are phase-matched. For instance, when only the carrier modes can interact strongly, the Hamiltonian effectively reduces to a single-mode model
\begin{align}
\label{eq:single-mode-small-L-Hamiltonian}
    \hat{H}/\hbar=\frac{g}{2}(\hat{a}_0^2\hat{b}_{0}^\dagger+\hat{a}_0^{\dagger2}\hat{b}_{0})+ \delta\omega_a(0)\hat{a}_0^\dagger\hat{a}_0+\delta\omega_b(0)\hat{b}_0^\dagger\hat{b}_0.
\end{align}
The $L$ dependence of the CW coupling strength
\begin{align}
    g=\frac{r}{\sqrt{L}}
\end{align}
indicates that smaller resonators can produce stronger nonlinear interactions, reflecting the enhancement of electric field amplitude per photon inside the resonator. This is the same expression as we obtained in Sec.~\ref{sec:photon-normalized-units}.

Note that \eqref{eq:single-mode-small-L-Hamiltonian} can be further simplified via an appropriate choice of reference frequency $\omega_\mathrm{r}$; We can eliminate $\delta\omega_a(0)$ and $\delta\omega_b(0)$ by setting $\omega_\mathrm{r}=\omega_a(0)$ and $\omega_\mathrm{r}=\omega_b(0)/2$, respectively.

\paragraph{Large $L$ limit} The other limit, $L\rightarrow\infty$, corresponds to an infinitely long waveguide. In this limit, the spacings among the wavenumber modes vanishes, \textit{i.e.}, $\delta k_{m+1}-\delta k_{m}=2\pi/L\rightarrow0$, forming a continuum. We model such continuous feature by defining field operators for continuous wavespace modes
\begin{align}
    &\hat{a}_s=\int\mathrm{d}z\,e^{-2\pi\mathrm{i}s z}\hat{a}_z &\hat{b}_s=\int\mathrm{d}z\,e^{-2\pi\mathrm{i}s z}\hat{b}_z.
\end{align}
Here and in the following of this tutorial, we use $s$ to denote non-angular wavenumber, which is related to the nominal angular wavenumber $k$ via $k=2\pi s$. Such non-angular wavenumber may seem slightly unconventional but is beneficial from notational perspective as it eliminates extra prefactors in front of the integrations.\footnote{If we use angular wavenumber $k$, the transform pairs $\hat{a}_k=\frac{1}{\sqrt{2\pi}}\int\mathrm{d}z\,e^{-ikz}\hat{a}_z$ and $\hat{a}_z=\frac{1}{\sqrt{2\pi}}\int\mathrm{d}z\,e^{ ikz}\hat{a}_k$ must be used in order to ensure commutation relationships $[\hat{a}_k,\hat{a}_{k'}^\dagger]=\delta(k-k')$, which are accompanied with an extra prefactor $\frac{1}{\sqrt{2\pi}}$ in front of the integrals.}  Intuitively, $\hat{a}_s^\dagger$ ($\hat{b}_s^\dagger$) creates a delta-function-like FH (SH) excitation at wavenumber $s$, and the operators fulfill the commutation relations $[\hat{a}_s,\hat{a}_{s'}^\dagger]=[\hat{b}_s,\hat{b}_{s'}^\dagger]=\delta(s-s')$. Since $\delta(s-s')$ has a unit of $[\mathrm{length}]$, the field operator in the wavespace $\hat{a}_s$ has a unit of $[\mathrm{length}^{1/2}]$. The overall Hamiltonian is written as
\begin{align}
\label{eq:broadband-chi2-hamiltonian}
   \hat{H}/\hbar=\sum_{u\in\{a,b\}}\int\mathrm{d}s\,\hat{u}_s^\dagger \delta\omega_u(2\pi s)\hat{u}_s+\frac{r}{2}\iint\mathrm{d}s_1\mathrm{d}s_2\,\left(\hat{a}_{s_1}\hat{a}_{s_2}\hat{b}_{s_1+s_2}^\dagger+\hat{a}_{s_1}^\dagger\hat{a}_{s_2}^\dagger\hat{b}_{s_1+s_2}\right),
\end{align}
which is the continuous wavenumber limit of \eqref{eq:Hamiltonian-multimode-finite-L}. In Sec.~\ref{sec:microscopic-monochromatic}, we will study the dynamics generated by these Hamiltonians in the small $L$, large $L$, and intermediate limits to better understand how the number of interacting modes qualitatively impacts device behaviors.

\section{Prologue for quantum nonlinear optics beyond mean field}
\label{sec:prologue}


In Section \ref{sec:Rosetta_stone}, we established a procedure for constructing a quantum Hamiltonian for a given nonlinear-optical device. We used a phenomenological quantization approach, determining the Hamiltonian parameters so that the mean-field-approximated quantum dynamics match the classical dynamics studied in Section \ref{sec:classical_NLO}. With this Hamiltonian, we can explore the quantum physics of nonlinear optics beyond the mean-field approximation, which is the goal of the subsequent sections. In this context, this section serves as a prologue to the following sections and introduces the reader to quantum physics in nonlinear optics. The heart of this section is the introduction of the Gaussian interaction frame (GIF), a framework that captures various features of light, from classical to exotic quantum ones, in a hierarchical manner. Through the lens of GIF, we can intuitively understand how the number of photons input to a nonlinear system makes a qualitative difference in how quantum features emerge, based on which we identify three regimes of nonlinear optics: the macroscopic, mesoscopic, and microscopic regimes.

\subsection{Gaussian quantum optics}
\label{sec:prologue-gaussian-quantum}
In classical nonlinear optics, the evolution of the quantum state is determined entirely by that of the mean-field, $\alpha(t)$, leading to a quantum-mechanical description strictly in terms of coherent states
\begin{align}
\label{eq:beyond-mean-classical-ket}
\ket{\psi_\mathrm{classical}(t)}=\hat{D}\bigl(\alpha(t)\bigr)\ket{0},
\end{align}
where $\hat{D}(\alpha)$ is the displacement operator that generates a coherent state of amplitude $\alpha$ from vacuum. Formally, what we covered in Section~\ref{sec:classical_NLO} is only quantum-mechanically correct in so far as the true quantum state $\ket{\psi(t)}$ is close to $\ket{\psi_\text{classical}(t)}$. However, under strong nonlinearities, or equivalently, long interaction time in the absence of dissipation, the distribution of quantum fluctuations can become deformed due to the nonlinear phase-space flow, forming features that do not resemble those of coherent states (see Fig.~\ref{fig:introduction-phase-space} for illustration).

\begin{figure}[h]
    \centering
    \includegraphics[width=0.9\textwidth]{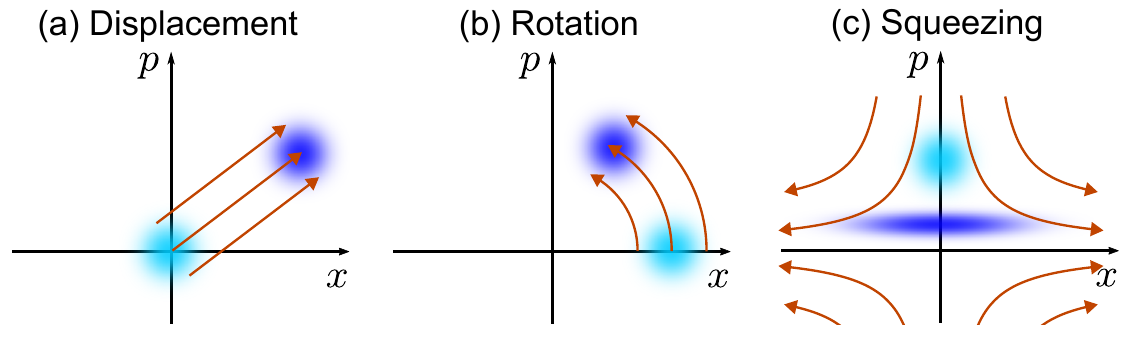}
    \caption{Single-mode Gaussian unitary operations. Light blue and dark blue distributions represent the phase-space portraits before and after each operation, respectively. (a) Displacement operation shifts the overall phase-space distribution. (b) The rotation operation rotates the distribution around the origin. (c) Squeezing operation extends the distribution linearly along one axis while shrinking it along the other axis.}
    \label{fig:disp-rot-sq}
\end{figure}

Nevertheless, to lowest order, such distortions can be well approximated as only stretches and rotations, as opposed to shears, etc. In this case, while the isotropic distribution of a coherent state around its mean is no longer accurate, the distorted distribution is still a Gaussian, just stretched and rotated. These transformations that preserve Gaussianity are referred \emph{Gaussian transformations} in general, and the subset of operations beyond displacement, such as rotationsand squeezing, are symplectic transformations. See Fig.~\ref{fig:disp-rot-sq} for illustrations of these Gaussian transformations in the phase space. If we want to bring into our model quantum states that feature these symplectic distortions then we can introduce a unitary $\hat{G}$, generating stretches and rotations, leading to generic states of the form
\begin{align}
   \label{eq:intro-semi-classical} \ket{\psi_\mathrm{Gaussian}(t)}=\hat{D}\bigl(\alpha(t)\bigr) \hat{G}\bigl(C(t), S(t)\bigr) \ket{0}.
\end{align}
The study of optical states of the form \eqref{eq:intro-semi-classical} is the basis of the field of Gaussian quantum optics.
In this description, the quantum state can still be concisely represented by three sets of c-numbers, $\alpha(t)$ for $\hat{D}$, and $C(t)$ and $S(t)$ for $\hat{G}$. Formally, these Gaussian unitaries are defined by the transformations
\begin{subequations}
\begin{align}
    \hat{D}^\dagger(\alpha)\hat{a}\hat{D}(\alpha)&=\hat{a}+\alpha \\\hat{G}^\dagger(C,S)\hat{a}\hat{G}(C,S)&=C\hat{a}+S\hat{a}^\dagger, \label{eq:beyond-mean-field-Gaussian-trans}
\end{align}
\end{subequations}
so that $C$ and $S$ together comprise a propagator, or Green's function~\cite{Wasilewski2006,Braunstein2005,Olivares2012}, for the operators in the Heisenberg picture, consistent with their behavior in Section~\ref{sec:classical_NLO}. Thus, specifying the evolution of $\alpha$, $C$, and $S$ in Gaussian quantum optics is a full quantum description of the system dynamics, provided the true quantum state $\ket{\psi}$ is close to $\ket{\psi_\text{Gaussian}}$.

However, it remains to be specified how $C(t)$ and $S(t)$ should be determined. The most widely used (but not the only) approach is to employ a \emph{linearized approximation}, as follows. First, as already described, we use a mean-field approximation to derive classical mean-field dynamics for $\alpha(t)$. Then, to determine what remains after taking these mean-field dynamics into account, we move to a displaced frame given by $\hat{D}(\alpha)$, which induces a transformation on the system Hamiltonian according to $\hat{H}_\mathrm{D}=\hat{D}^\dagger\hat{H}\hat{D}-\mathrm{i}\hat{D}^\dagger\partial_t\hat{D}$.
Crucially, we can partition the terms of $\hat H_\text{D}$ according to
\begin{align}
\hat H_\text{D} =\hat{H}_\mathrm{G}+\hat{H}_\mathrm{NG},
\end{align}
where, by construction, $\hat{H}_\mathrm{G}$ contains all and only the
quadratic terms, such as $\hat a^\dagger \hat a$ and $\hat a \hat a$, that generate symplectic operations. In this case, $\hat{H}_{\mathrm{NG}}$ contains the remaining terms; when $\hat{D}$ is chosen properly, the dominant linear terms (first-order in operators) are eliminated from $\hat{H}_\mathrm{D}$, only leaving higher-order nonlinear (non-Gaussian) terms in $\hat{H}_{\mathrm{NG}}$. In conventional experimental regimes of nonlinear optics, the magnitude of $\hat{H}_\mathrm{G}$ is much greater than $\hat{H}_\mathrm{NG}$, which justifies our approximation $\hat{H}_\mathrm{D}\approx \hat{H}_\mathrm{G}$. As a result, the leading-order corrections to the quantum dynamics of $\ket{\psi_\text{classical}}$ are captured by the Gaussian unitary
\begin{align}
\label{eq:gaussian-unitary-define}
    \hat{G}(t)=\mathcal{T}\exp\left(-\frac{\mathrm{i}}{\hbar}\int^t_0\mathrm{d}t'\,\hat{H}_\mathrm{G}(t')\right)
\end{align}
for $\ket{\psi_\text{Gaussian}}$. Here, $\mathcal{T}$ denotes the time-ordering operator. Then, $C(t)$ and $S(t)$ are obtained \emph{from} $\hat G(t)$ so defined: If we take a time derivative of \eqref{eq:beyond-mean-field-Gaussian-trans} and use the equality $\mathrm{i}\hbar\partial_t\hat{G}=\hat{H}_\mathrm{G}\hat{G}$, it follows that
\begin{align}
\partial_tC\hat{a}+\partial_tS\hat{a}^\dagger=\frac{\mathrm{i}}{\hbar}\hat{G}^\dagger[\hat{H}_\mathrm{G},\hat{a}]\hat{G}.
\end{align}
That is, we obtain a differential equation for $C$ and $S$, which can in principle be solved to generate their evolution.
Going one step further, we can also compare the coefficients of $\hat{a}$ and $\hat{a}^\dagger$ on the two sides of this linear equation (remembering that $\hat{H}_\mathrm{G}$ is a quadratic function of operators to simplify the commutator). This tells us that
\begin{align}
\label{eq:abstract-CS-evolution}
    &\partial_tC=\frac{\mathrm{i}}{\hbar}[\hat{G}^\dagger[\hat{H}_\mathrm{G},\hat{a}]\hat{G},\hat{a}^\dagger] &\partial_tS=-\frac{\mathrm{i}}{\hbar}[\hat{G}^\dagger[\hat{H}_\mathrm{G},\hat{a}]\hat{G},a].
\end{align}
These treatments can be easily extended to multimode scenario by treating $\alpha$ as a vector and $C$ and $S$ as matrices~\cite{Olivares2012}.

The linearized approximation $\ket{\psi_\text{Gaussian}}$ specified by $\alpha(t)$, $C(t)$ and $S(t)$ is surprisingly effective in capturing realistic experiments, and almost all experimental studies in quantum nonlinear optics have been based on this approximation~\cite{Triginer2020,Guidry2022,Kashiwazaki2020,Vahlbruch2016, Bao2021}. As discussed before, the success of such an approximation can be understood in terms of a separation of energy scales: so long as the energy contained in the mean-field excitation and the quantum fluctuations are significantly different, the quantum fluctuations see the effect of the mean field ($\hat G$ is determined by $\alpha$), but not the other way around. In Section~\ref{sec:semi-classical_NLO}, we apply this linearized approximation to OPAs to elucidate their Gaussian quantum dynamics. 

\subsection{Gaussian interaction frame}
\label{sec:gif-intro}
Conversely, the breakdown of the linearized approximation implies that quantum fluctuations start to affect the evolution of both the mean field and the quantum fluctuations themselves. For instance, in optical parametric generation (OPG)~\cite{Jankowski2022, Harris1967}, where a coherent pump field induces strong vacuum squeezing of the signal field, the exponential growth of the down-converted signal photons eventually grows to a comparable intensity to the pump. At this point, the signal quantum fluctuations begin to deplete the mean field of the pump~\cite{Xing2023}. Such dynamics can only be explained by retaining terms in the Hamiltonian that are ignored in the linearized approximation $\hat H \approx \hat H_\text{G}$; these term can in principle produce \emph{non-Gaussian} features in the quantum state.

To capture these non-Gaussian quantum features, we present a framework that builds on Gaussian quantum optics instead of completely abandoning it. This begins by noting that we can write the true quantum state in the form
\begin{align}
    \label{eq:intro-GIF-wavefunction}
    \ket{\psi}=\hat{D}\hat{G}\ket{\varphi_\mathrm{I}}
\end{align}
with no loss of generality. Here, an insightful way to interpret \eqref{eq:intro-GIF-wavefunction} is to see $\hat{U}=\hat{D}\hat{G}$ as a unitary defining a Gaussian interaction frame (GIF), in which classical mean-field and Gaussian quantum dynamics have been factored out. In contrast to the approach in Gaussian quantum optics, the residual non-Gaussian quantum fluctuations are captured by the interaction-frame state $\ket{\varphi_\mathrm{I}}$, which evolves under the Hamiltonian in the interaction frame
\begin{align}
\label{eq:beyond-mean-field-interaction-frame}
   \hat{H}_\mathrm{I}=\hat{U}^\dagger\hat{H}\hat{U}-\mathrm{i}\hat{U}^\dagger\partial_t\hat{U}.
\end{align}

\begin{figure}[tb]
    \centering
    \includegraphics[width=0.6\textwidth]{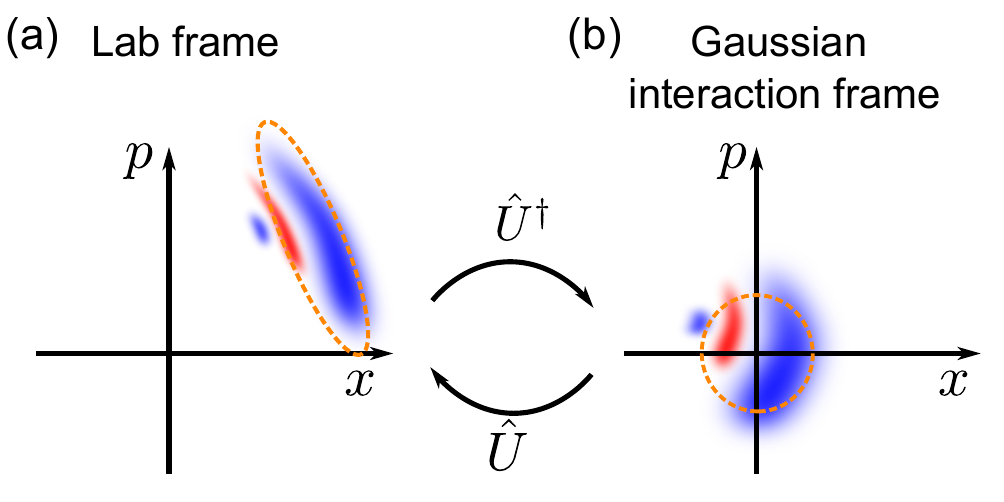}
    \caption{Illustration of how a compressed quantum-state representation is realized using the GIF. (a) In the lab frame, the phase space distribution can be far away from the origin, which requires a large Fock space for an accurate state representation. (b) By factoring out the Gaussian dynamics in the form of a Gaussian unitary $\hat{U}$, the state description in the GIF can be more efficient. The orange ellipses represent the Gaussian approximation of the true state in each frame. Figure is adapted from Ref.~\cite{Yanagimoto2023-thesis}.}
    \label{fig:gif}
\end{figure}

The GIF offers both practical and conceptual benefits. First, the GIF enables a more efficient numerical model than a lab-frame model. To capture a non-Gaussian quantum state with $n$ photons, one would na\"ively need $n$-dimensional Fock space to numerically represent the wavefunction, which is a daunting task considering that it is not unusual to find millions of photons in nonlinear optics. The situation gets only worse when multiple modes are involved, where the Hilbert space dimension scales as $n^M$ for the number of modes $M$. In practice, most of these photon excitations correspond to mean-field and Gaussian quantum features, such as the displacement and anti-squeezing of the quantum state. Therefore, as shown in Fig.~\ref{fig:gif}, we can factor out these trivial features in the form of a Gaussian unitary $\hat{U}$ to minimize the excitation in the GIF, thereby enabling a concise representation of the quantum state.

Second, using the GIF provides crucial insights into how non-Gaussian quantum interactions occur, which are harder to see in the lab frame. As we can see in Eqn.~\eqref{eq:beyond-mean-field-interaction-frame}, the form of the non-Gaussian interaction depends on the structure of the Gaussian unitary $\hat{U}$. Since Gaussian quantum features typically have a much larger magnitude than the higher-order non-Gaussian features, the Gaussian part of the dynamics provides crucial information on where and how non-Gaussian quantum features emerge. For instance, as shown in Ref.~\cite{Yanagimoto2023-qnd, Qin2022}, a dominant Gaussian Hamiltonian $\hat{H}_\mathrm{G}$ can confine non-Gaussian dynamics to its eigenbasis, realizing emergent functionalities, \textit{e.g.}, quantum-nondemolition measurements. In Section \ref{sec:mesoscale_sqeezing}, we use the knowledge of $\hat{U}$ to identify a handful of modes that predominantly experience non-Gaussianity, helping to inform a tractable model for multimode non-Gaussian dynamics.

\subsection{Macroscopic, mesoscopic, and microscopic quantum nonlinear optics} 
\label{sec:intro-macro-meso-micro}
\begin{figure}[h]
    \centering
    \includegraphics[width=\textwidth]{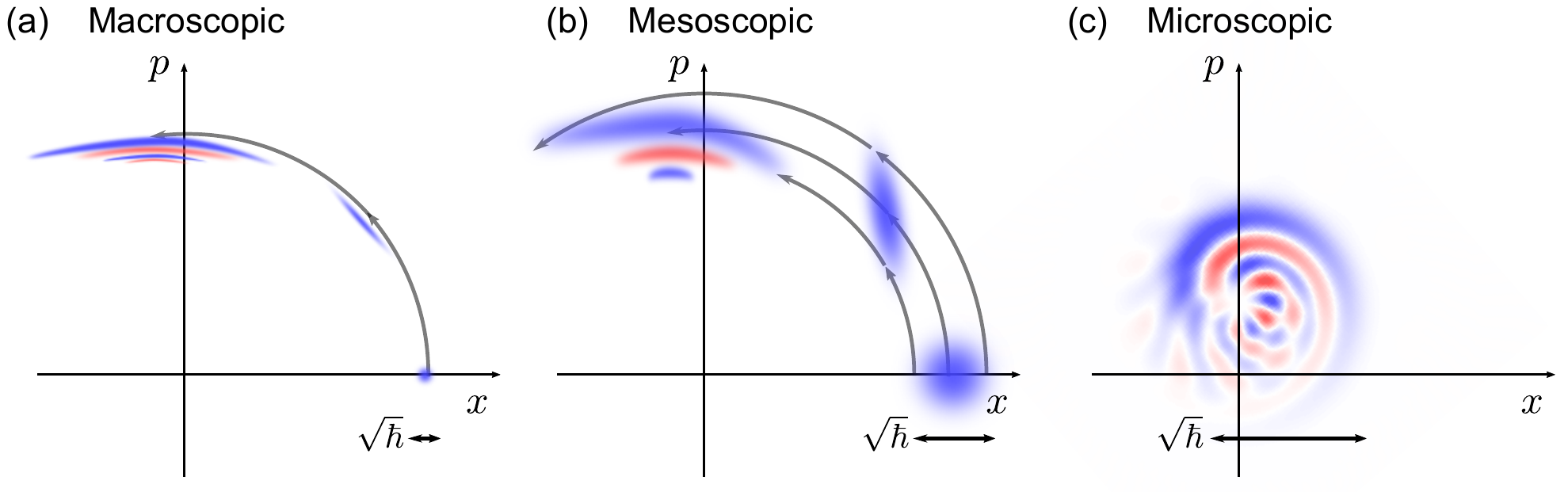}
    \caption{Schematics for the phase-space dynamics under Kerr nonlinearity. The black arrows depict the scale of vacuum-level quantum fluctuations. (a) When a macroscopic number of photons are involved, non-Gaussian quantum features appear on top of strong squeezing, making these quantum features highly prone to decoherence. (b) In the mesoscopic regime, non-Gaussian quantum features appear on top of a moderate magnitude (\textit{i.e.}, hundreds to dozens of photons) of Gaussian quantum features. (c) In the microscopic regime, quantum fluctuations are so large that the hierarchies among the mean field, Gaussian quantum features, and non-Gaussian quantum features collapse.}
    \label{fig:macro-meso-micro}
\end{figure}

Typically, the onset of non-Gaussian quantum physics coincides with the point where classical dynamics exhibit strong saturation. Since the nonlinear dynamical rate typically increases as the number of pump photons, it is, in principle, possible to trigger saturation on a given nonlinear device by simply driving it hard enough, no matter how weak the nonlinearity is. Then, what has prevented us from using such operations to realize, \textit{e.g.}, non-Gaussian quantum light sources? Here, a key to answering this question is the number of photons required to trigger saturated nonlinear dynamics. Before significant non-Gaussian quantum features are generated, Gaussian quantum fluctuations first have to grow to the magnitude that can affect the mean field. As a result, when a strongly-driven weakly-nonlinear device reaches saturation, huge Gaussian quantum features (\textit{e.g.}, squeezing and mean field) should be present, and non-Gaussian quantum features can only be present on top of them. These Gaussian features make the exotic quantum features that ride on top of them extremely prone to any experimental imperfections (\textit{e.g.}, loss and phase noise), making it impossible to observe them in realistic experiments, and one can only detect their semi-classical remnants~\cite{Florez2020} (see Fig.~\ref{fig:macro-meso-micro}(a)). To resolve non-Gaussian quantum features, it is thus essential to have strong enough nonlinearity to induce saturation with a non-macroscopic number of pump photons. We discuss how the losses affect quantum features in more detail in Sec.~\ref{sec:role-of-loss}. 

These considerations motivate us to single out an intermediate \emph{mesoscopic} regime of nonlinear quantum optics. In this regime, only hundreds to dozens of photons suffice to induce saturated nonlinear dynamics and squeezing that is comparable to the intensity of the mean field, and all of the mean field, Gaussian quantum, and non-Gaussian quantum features crucially contribute to the dynamics (see Fig.~\ref{fig:macro-meso-micro}(b)). Because of the moderate level of Gaussian features, non-Gaussian quantum features could reliably be generated and sustained. In Section \ref{sec:mesoscopic-quantum-nonlinear-optics}, we unravel non-Gaussian dynamics of OPAs in the mesoscopic regime as a case study, showing the essential roles the GIF plays in the analysis. 

Beyond the mesoscopic regime, as the nonlinearity and interaction length increase further, the threshold pump photon number to cause saturation eventually reaches the order of unity. In this regime of \emph{microsopic} nonlinear optics, the magnitude of quantum fluctuations is equivalent to that of the mean field, and thus, their distinction becomes meaningless. In this regime, it is more appropriate to perceive photons as particles interacting with each other, like a photon gas, for which a discrete-variable (DV) formulation becomes natural. Such DV pictures naturally find analogies in atomic physics, exhibiting behaviors reminiscent of atomic physics, \textit{e.g.}, Rabi-like oscillations in photon conversion dynamics~\cite{Langford2011}. Many theoretical analyses in this regime played significant roles in the initial conception of photonic quantum information~\cite{Chuang1995, Milburn1989, Nielsen2000}. In Section \ref{sec:quantum_NLO}, we provide a case study of optical parametric interactions among a microsopic number of photons, showing that many classical intuitions break down in this regime. 

\paragraph{Further reading} Readers can refer to Ref.~\cite{Ryo2023Mesoscopic} for a review on both theoretical and experimental efforts towards understanding and engineering emergent non-Gaussian quantum physics in the mesoscopic regime.

\subsection{The role of loss in quantum nonlinear optics}
\label{sec:role-of-loss}
For the experimental observation of exotic quantum features in nonlinear quantum optics, photon loss is expected to be a major roadblock, and in this subsection, we explain when and how photon loss diminishes quantum features. Here, it is our intention to keep the discussions to the most basic and intuitive level. While the physics of loss and quantum decoherence constitutes a fertile field of open quantum systems, even unitary (\textit{i.e.}, lossless) dynamics of quantum nonlinear optics are rich enough that we cannot cover beyond unitary physics in this tutorial. We thus lead interested readers to further references in open quantum systems at the end of this subsection. 

\begin{figure}[tb]
    \centering
    \includegraphics[width=1\textwidth]{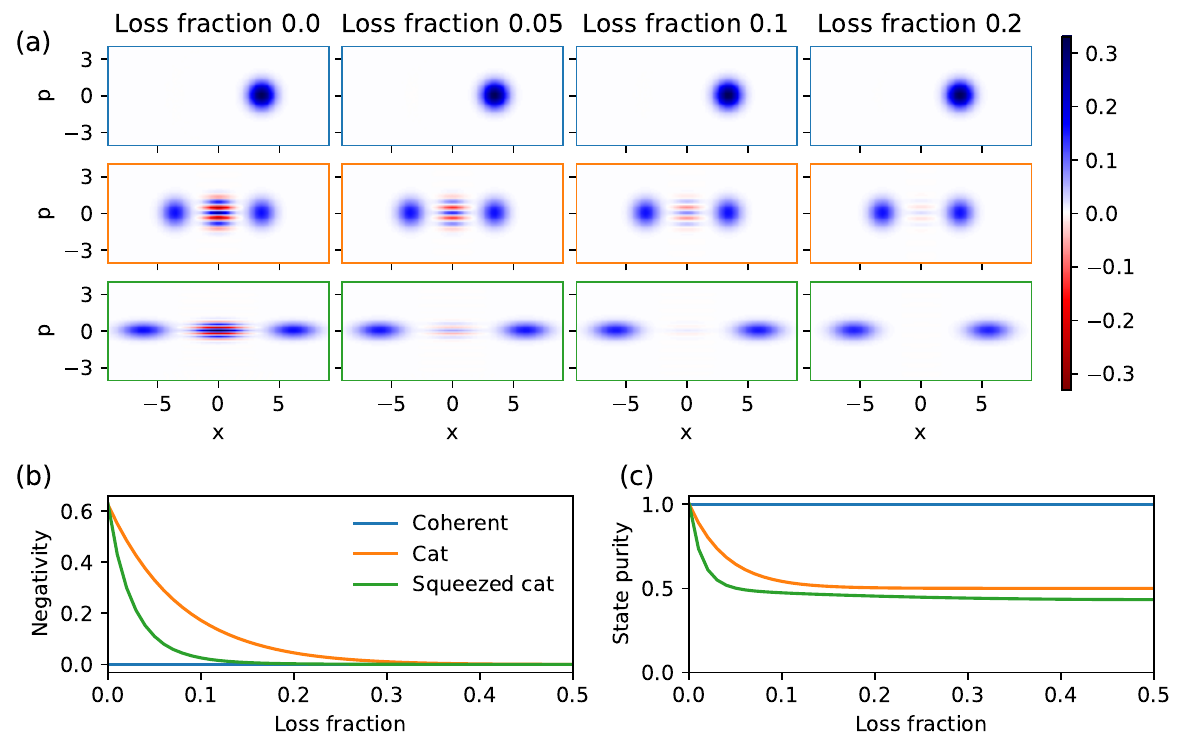}
    \caption{(a) Wigner functions of a coherent state (top row, blue frames), a Schr\"odinger's cat state (middle row, orange frames), and a squeezed cat state (bottom row, green frames) under various ratios of photon loss. The coherent state has a displacement of $\alpha=2.5$. The cat state is an even superposition of coherent states with $\alpha=\pm 2.5$. The squeezed cat state is obtained by applying a squeezing operation with power gain of $3$ to the cat state, where the $x$ quadrature gets amplified. (b) Volume of Wigner function negativity~\cite{Kenfack2004} and (c) purity of the states for various losses.}
    \label{fig:state-decoherence}
\end{figure}

It is worth noting that the impact of loss is usually much more severe for quantum states than classical ones. Intuitively, Wigner functions of a state after loss are given as a convolution between the original Wigner function (with appropriate scaling) and a Gaussian distribution, which ``blurs'' features in the phase space. Consequently, more exotic quantum states with finer phase-space features lose their coherence more quickly. To see this more visually, we show Wigner functions of a coherent state, a Schr\"odinger's cat state (\textit{i.e.}, a coherent superposition of two coherent states), and a squeezed Schr\"odinger's cat state under various losses in Fig.~\ref{fig:state-decoherence}(a). While the coherent state does not exhibit any qualitative change in its phase space portrait under loss except for its amplitude, the loss critically affects the (squeezed) cat states, washing away the nonclassical interference patterns. These variations can be seen more clearly in Fig.~\ref{fig:state-decoherence}(b) and (c), where we show the Wigner function negativity~\cite{Kenfack2004} and state purity~\cite{Chuang1995}, respectively. The negativity of the Wigner function is used to quantify the nonclassicality, which we observe is quickly lost under photon loss for the (squeezed) cat states. On the other hand, a coherent state does not have any negativity to begin with, due to its classical nature. The state purity measures the coherence of a quantum state, and any impurity implies that quantum information is lost due to the photon loss. Here, while the (squeezed) cat states become impure quickly, the coherent state remains pure under any amount of loss, reflecting the robustness of classical features to decoherence. 

It is worth noting that the squeezed cat state loses quantum features much more quickly than the normal cat state, despite they having the same amount of negativities to being with. This is because additional squeezing increases the magnitude of quantum fluctuations, increasing the effective rate at which photons are lost. Such increased magnitude of quantum fluctuations is also correlated with the horizontal stretch of the interference patterns, making them more prone to the blurring effect that the loss induces. These observations imply the fragility of quantum features in the macroscopic regime: Due to the strong Gaussian dynamics in this regime, non-Gaussian quantum features get strongly stretched in the phase space, getting hit more severely by photon loss.

\paragraph{Further reading} 
For introductory contents for general physics of open quantum systems, readers can refer to Refs.~\cite{Wiseman2010,Breuer2002}. For nonlinear-optical systems, Refs.~\cite{Roberts2020,Rivera2023,Seibold2022,Onodera2022} discuss physics that arises due to the unique interplay among quantum nonlinearity, linear loss, and nonlinear loss. Sometimes, decoherence occurs in a non-Markovian manner, which is reviewed in Ref.~\cite{deVega2017}.

\section{Gaussian quantum physics of nonlinear optics}\label{sec:semi-classical_NLO}

\subsection{Introduction}
In the previous section on classical nonlinear optics (Sec.~\ref{sec:classical_NLO}), we have seen that coupled-wave equations fully capture nonlinear interactions among classical lightwaves. As the classical mean fields evolve under the coupled-wave equations, the phase-space distribution of quantum fluctuations also evolves under the same dynamics, experiencing deformations. To the lowest order, such deformation can be well-approximated as linear displacement, rotation, and squeezing, which map a Gaussian state to another, and thus is referred to as \emph{Gaussian} unitary (see Fig.~\ref{fig:disp-rot-sq} for illustration). Here, displacement and rotation map an initial coherent state only to another coherent state, implying that they are essentially classical operations that can be physically realized only using linear optics (\textit{i.e.}, delay line, coherent light source, and beam splitters). On the other hand, the squeezing operations decrease or increase the variance of quantum fluctuations and can produce a nonclassical state of light, \textit{i.e.}, squeezed light~\cite{Walls1983,Wu1986}, from an initial coherent state. Such squeezing operations inherently require an active gain medium~\cite{Braunstein2005} and have filled a role for quantum science and technology that nonlinear optics is uniquely suited to providing~\cite{LIGO2013,Furusawa1998,Takeda2019}.

This section introduces how to model and understand Gaussian quantum features that that appear naturally in nonlinear-optical dynamics. A central tool we employ is a linearized approximation, where we assume that the quantum fluctuations do not affect the dynamics of the mean field and the quantum fluctuations. Under this approximation, the photon dynamics can be modeled as Gaussian unitary. We apply these theoretical tools to unravel the physics of an OPA as a representative example. Before directly going into the contents of this section, we encourage readers to visit a short prologue, Sec.~\ref{sec:prologue}, to get familiarized with the terminologies and theory tools essential for the quantum formulation of nonlinear optics.

The structure of this section follows the theme of the overall tutorial, \textit{i.e.}, single-mode to multimode. In Sec.~\ref{sec:Gaussian-single-mode-opa}, we introduce a most elementary case of single-mode OPA, a basic building block for more complicated setups. In Sec.~\ref{sec:cw-pumped-broadband}, we consider a broadband OPA on a nonlinear waveguide pumped with CW light, showing that seemingly multimode dynamics of the broadband field can be decomposed as independently squeezed pairs of modes in the frequency domain. Finally, in Sec.~\ref{sec:Gaussian-pulse-pumped-OPA}, we consider the most complicated scenario of pulse-pumped OPA on a nonlinear waveguide. Even in this case, it is shown that the system dynamics can be decomposed to independent squeezing of \emph{supermodes}, a linear combination of canonical frequency modes. 

\subsection{Single-mode OPA}
\label{sec:Gaussian-single-mode-opa}
In this subsection, we introduce a most basic model for Gaussian quantum optics, \textit{i.e.}, single-mode OPA. Based on the model, we show how to use the undepleted pump approximation to derive the Gaussian quantum approximation for the system dynamics. As a critical theory tool to concisely capture such Gaussian quantum dynamics, we use the formalism of propagators, which characterizes the linear evolution of the optical fields. We specifically highlight how the competition between phase mismatch and parametric gain determines the qualitative behavior of an OPA. While natural OPA dynamics in a nonlinear waveguide are often inherently multimode (and special engineering efforts are required to strictly enforce single-mode interactions), the single-mode physics we discuss in this section provides valuable insights into more general multimode OPA dynamics discussed in latter subsections.

The following discussions follow the pipeline presented in Sec. \ref{sec:prologue-gaussian-quantum}. The Hamiltonian for a single-mode OPA takes a form
\begin{align}
\label{eq:gaussian-single-mode-hamiltonian}
    \hat{H}/\hbar=\frac{g}{2}(\hat{a}^2\hat{b}^\dagger+\hat{a}^{\dagger2}\hat{b})+\delta\hat{a}^\dagger\hat{a},
\end{align}
where $\hat{a}$ and $\hat{b}$ are the annihilation operators for the signal and pump modes, respectively, $g$ is the $\chi^{(2)}$ nonlinear coupling constant, and $\delta$ is the phase-mismatch. These parameters can be connected to classical experimental parameters as discussed in Sec.~\ref{sec:small-large-limits}. Using the mean-field approximation, we obtain classical coupled-wave equations
\begin{align}
&\mathrm{i}\partial_t\alpha=g\alpha^*\beta+\delta\alpha&\mathrm{i}\partial_t\beta=\frac{g}{2}\alpha^{2},
\end{align}
where $\alpha$ and $\beta$ are signal and pump mean-field amplitudes. To solve the mean-field evolution analytically, we perform an undepleted pump approximation, asserting that the signal field amplitude is small enough $\alpha\approx 0$ that it does not deplete the pump field. As a result of this approximation, we get a time-independent pump amplitude
\begin{align}
    \beta(t)=\mathrm{i}\beta_0,
\end{align}
where we have taken $\mathrm{i}\beta_0$ to be the initial amplitude. Then, we linearize the dynamics of quantum fluctuations to derive the Gaussian unitary $\hat{G}$, which is generated by the Gaussian Hamiltonian terms
\begin{align}
\label{eq:single-mode-constant}
    \hat{H}_\mathrm{G}/\hbar=-\frac{\mathrm{i}u}{2}(\hat{a}^2-\hat{a}^{\dagger2})+\delta\hat{a}^\dagger\hat{a},
\end{align}
where we have defined a parametric gain $u=g\beta_0$. The Gaussian unitary $\hat{G}$ can be completely characterized by the propagators, whose equations of motion follow from \eqref{eq:abstract-CS-evolution} as
\begin{align}
&\partial_tC(t)=uS^*(t)-\mathrm{i}\delta C(t), &\partial_tS(t)=uC^*(t)-\mathrm{i}\delta S(t),
\end{align}
which can be summarized in a vector form 
\begin{align}
    \partial_t\begin{pmatrix}
        C\\S^*
    \end{pmatrix}=M\begin{pmatrix}
        C\\S^*
    \end{pmatrix}
\end{align}
with
\begin{align}
\label{eq:M-matrix-single-mode}
    M=\begin{pmatrix}
        -\mathrm{i}\delta&u\\
        u&\mathrm{i}\delta
    \end{pmatrix}.
\end{align}
Note that this equation of motion takes an equivalent form to the one we obtained in classical formalism, with the propagation coordinate changed from space to time. With the boundary conditions $C(t=0) = 1$ and $S(t=0) = 0$, the propagator dynamics are solved analytically as \footnote{For $u^2-\delta^2<0$, the arguments inside the sinh and cosh functions become imaginary. The functional expressions of the propagators are well defined in this regime as well via $\cosh(ix)=\cos(x)$ and $\sinh(ix)=i\sin(x)$.}
\begin{subequations}
\label{eq:undepleted-single-cs}
\begin{align}
\label{eq:undepleted-single-c}
    C(t)&=\cosh\left(\sqrt{u^2-\delta^2}\,t\right)-\frac{\mathrm{i}\delta}{\sqrt{u^2-\delta^2}}\sinh\left(\sqrt{u^2-\delta^2}\,t\right),\\
\label{eq:undepleted-single-s}
S^\mathrm{*}(t)&=\frac{u}{\sqrt{u^2-\delta^2}}\sinh\left(\sqrt{u^2-\delta^2}\,t\right).
\end{align}
\end{subequations}

\begin{figure}[bt]
    \centering
    \includegraphics[width=\textwidth]{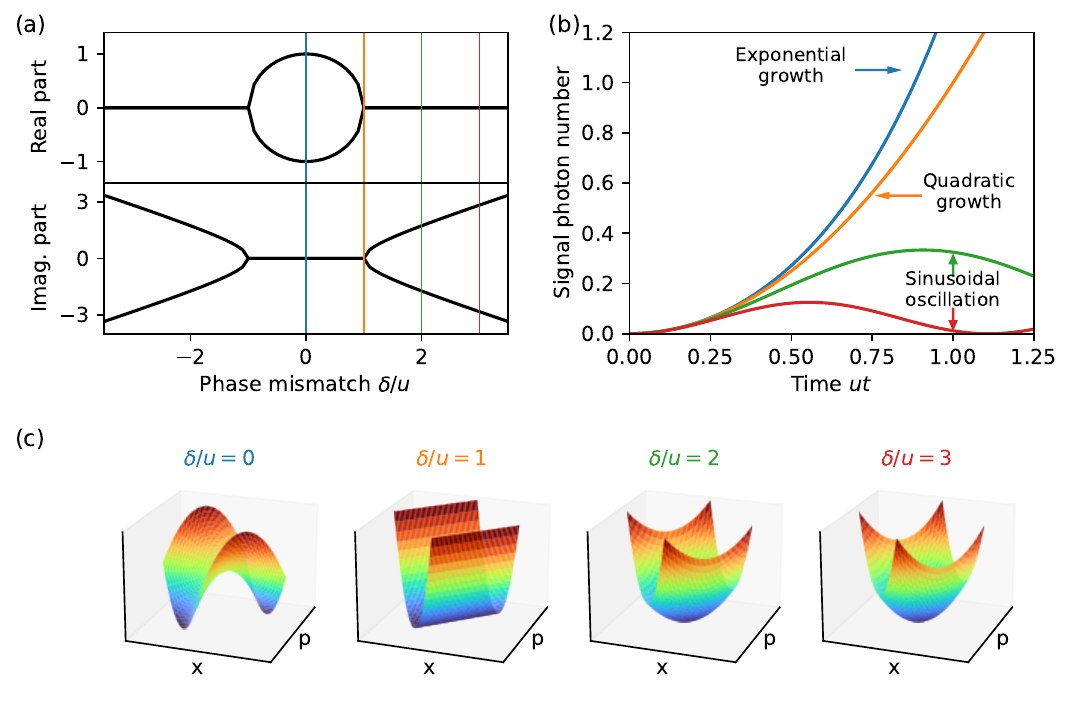}
    \caption{(a) Real and imaginary parts of the eigenvalues of $M$ shown as functions of normalized phase mismatch $\delta/u$. At the exceptional points $\delta/u=\pm1$, the eigenspectrum changes from purely real to purely imaginary. (b) Dynamics of the signal photon number (\textit{i.e.}, power of parametric fluorescence) for various values of phase mismatch (the values of $\delta/u$ correspond with the colors shown in the labels in (c)). (c) Potential function $\hat{H}_\mathrm{G}=u(\hat{x}\hat{p}+\hat{p}\hat{x})+\delta(\hat{x}^2+\hat{p}^2-1/2)$ plotted in the quadrature space for various values of the phase mismatch. At the exceptional points $|\delta/u|=1$, potential shapes change from quadratic to hyperbolic. Figure is adapted from Ref.~\cite{Yanagimoto2023-thesis}.}
    \label{fig:single-mode-opg}
\end{figure}

To intuitively understand the action of the Gaussian unitary $\hat{G}$, we rewrite the propagators as
\begin{align}
\label{eq:undepleted-svg-single-mode}
&C(t)=A^\mathrm{out*}\cosh\lambda\,A^\mathrm{in}&S(t)=A^\mathrm{out*}\sinh\lambda\,A^\mathrm{in*},
\end{align}
where $\lambda\geq0$ is the squeezing parameter that gives the field gain as $e^\lambda$. The complex amplitudes $A^\mathrm{in/out}$ are normalized as $|A^\mathrm{in/out}|^2=1$ and characterize the input/output modes that experience quadrature squeezing via $\hat{a}^\mathrm{in/out}=A^\mathrm{in/out*}\hat{a}$. The motivation behind the choice of $A^\mathrm{in/out}$ will become clear shortly.

We then have
\begin{align}
\hat{G}^\dagger\hat{a}^\mathrm{out}\hat{G}=\cosh \lambda\,\hat{a}^\mathrm{in}+\sinh \lambda\,\hat{a}^\mathrm{in\dagger},
\end{align}
with which we obtain
\begin{align}
\label{eq:single-mode-quadrature-map}
&\hat{G}^\dagger\hat{x}^\mathrm{out}\hat{G}=e^\lambda \hat{x}^\mathrm{in}, &\hat{G}^\dagger\hat{p}^\mathrm{out}\hat{G}=e^{-\lambda}\hat{p}^\mathrm{in}
\end{align}
for $\hat{x}=(\hat{a}+\hat{a}^\mathrm{\dagger})/2$ and $\hat{p}=(\hat{a}-\hat{a}^\mathrm{\dagger})/2\mathrm{i}$. The quadrature operator transformations \eqref{eq:single-mode-quadrature-map} show that $\hat{G}$ plays the role of a phase-sensitive amplifier with gain $e^\lambda$, where the quadrature of the input mode $\hat{x}^\mathrm{in}$ ($\hat{p}^\mathrm{in}$) is amplified (deamplified) to be mapped to the quadrature of the output mode $\hat{x}^\mathrm{out}$ ($\hat{p}^\mathrm{out}$). Note that the parametrization \eqref{eq:undepleted-svg-single-mode} was made such that the $x$ ($p$) quadrature of the output mode is proportional to the $x$ ($p$) quadrature of the input mode. 

Specifically, for the case of vacuum squeezing, the final state becomes a squeezed vacuum state whose $\hat{p}^\mathrm{out}$ quadrature is squeezed by the field gain of $e^\lambda$, whereas its $\hat{x}^\mathrm{out}$ is anti-squeezed by the same gain. This can also be seen from the quadrature variances
\begin{align}
&\langle(\hat{x}^{\mathrm{out}})^2\rangle=\frac{e^{2\lambda}}{4} &\langle(\hat{p}^{\mathrm{out}})^2\rangle=\frac{e^{-2\lambda}}{4},
\end{align}
where $1/4$ is the vacuum noise level, so the $\hat{p}^{\mathrm{out}}$ quadrature exhibits sub-vacuum quantum noise.

A useful quantity to characterize the system behavior is the mean photon number
\begin{align}
\label{eq:undepleted-single-power}
\langle\hat{a}^\dagger\hat{a}\rangle=|S(t)|^2=\frac{u^2}{u^2-\delta^2}\sinh^2\left(\sqrt{u^2-\delta^2}\,t\right).
\end{align}
The value of $\langle\hat{a}^\dagger\hat{a}\rangle$ can intuitively imply the signal state we have; When $\langle\hat{a}^\dagger\hat{a}\rangle\ll 1$ holds, the signal state is approximately a superposition of a vacuum state and a small amplitude of biphoton state, \textit{i.e.}, $\ket{0}+\epsilon\ket{2}$ with $\epsilon\ll 1$. This low-gain regime is conventionally called the limit of spontaneous parametric downconversion (SPDC) because the stimulated downconversion process can be ignored in this regime~\cite{Conteau2018}. In the other limit $\langle\hat{a}^\dagger\hat{a}\rangle\gg 1$, a large population of signal photons induces stimulated downconversion, forming a strongly squeezed vacuum state. 

Below, based on the signal photon population, we discuss the behavior of vacuum squeezing for various system parameters, which we find is strongly tied to the structure of the generator of the dynamics $M$ \eqref{eq:M-matrix-single-mode}. More specifically, we will find the eigenstructure of $M$ plays a crucial role. The eigenvectors of $M$ are $v^\pm=(\delta\pm\sqrt{\delta^2-u^2},\mathrm{i}u)^\intercal$ with
eigenvalues $\pm\mathrm{i}\sqrt{\delta^2-u^2}$, and their qualitative features change at the boundary value $|\delta/u|=1$, at which point the eigenspectrum of $M$ changes from purely imaginary values to purely real values (see Fig.~\ref{fig:single-mode-opg}(a)). In the language of non-Hermitian physics, this phase-transition point is referred to as an exceptional point (EP)~\cite{El-Ganainy2018}. 

For $|\delta/u|>1$, corresponding to a regime with a large phase mismatch, the eigenvalues of $M$ are purely imaginary. In this regime, the hyperbolic functions in \eqref{eq:undepleted-single-power} become trigonometric functions with oscillation frequency $\sqrt{\delta^2-u^2}$, which leads to sinusoidal oscillation of the power of parametric fluorescence as shown in Fig.~\ref{fig:single-mode-opg}(b). As we approach the EP $|\delta/u|\mapsto 1$, the amplitude of the oscillation in signal power \eqref{eq:undepleted-single-power} diverges, \textit{i.e.}, $u^2/\sqrt{\delta^2-u^2}\rightarrow\infty$, while the oscillation frequency converges to zero $\sqrt{\delta^2-u^2}\rightarrow0$, resulting in a purely quadratic growth of the parametric fluorescence $\langle\hat{a}^\dagger\hat{a}\rangle=u^2t^2$. For $|\delta/u|<1$, corresponding to the regime with a small phase mismatch, the eigenvalues become purely real. In this phase-matched regime, the hyperbolic functions in \eqref{eq:undepleted-single-power} lead to an exponential growth of the signal parametric fluorescence.

To intuitively see how the qualitative behavior of parametric interaction changes at the boundary value $|\delta/u|=1$, it is instructive to rewrite the Hamiltonian in the quadrature variables
\begin{align}
\hat{H}_\mathrm{G}/\hbar=u(\hat{x}\hat{p}+\hat{p}\hat{x})+\delta(\hat{x}^2+\hat{p}^2-1/2).
\end{align}
The shape of the potential in the $x$-$p$ space is depicted in Fig.~\ref{fig:single-mode-opg}(c) for various $|\delta/u|$. We here note that classical trajectories of the quadrature variables follow equipotential lines of the Hamiltonian. For $|\delta/u|>1$, the shape of the potential is a 2D quadratic function with bounded equipotential orbits in the phase space, resulting in oscillatory behavior. For $|\delta/u|<1$, on the other hand, the potential shape turns hyperbolic, and equipotential lines are unbounded. As a result, quantum fluctuations are amplified indefinitely to induce exponential growth of parametric fluorescence.

\subsection{CW-pumped traveling-wave OPA}
\label{sec:cw-pumped-broadband}
In this subsection, we study a most basic Gaussian quantum dynamical system in a nonlinear waveguide, \textit{i.e.}, CW-pumped OPA. Using the undepleted pump approximation, we show that the CW-pumped OPA can be decomposed into independent single-mode OPAs in the wavespace. As a result, we can directly adopt the theoretical analysis of single-mode OPA from the previous section to analytically solve the multimode quantum behavior of the system.

Based on these analytic results, we study the phenomenology of CW-pumped vacuum squeezing. Specifically, we show that the power of parametric fluorescence grows polynomially at the beginning (SPDC-like regime) as a function of propagation distance, which turns into exponential growth at longer propagation distance (vacuum-squeezing-like regime). We show that the cross-over point between these regimes coincides with the distance where the number of photons in the characteristic photon-photon correlation length exceeds an order of unity, corresponding to a transition from spontaneous downconversion to stimulated downconversion.

\begin{figure}[bt]
\centering
\includegraphics[width=\textwidth]{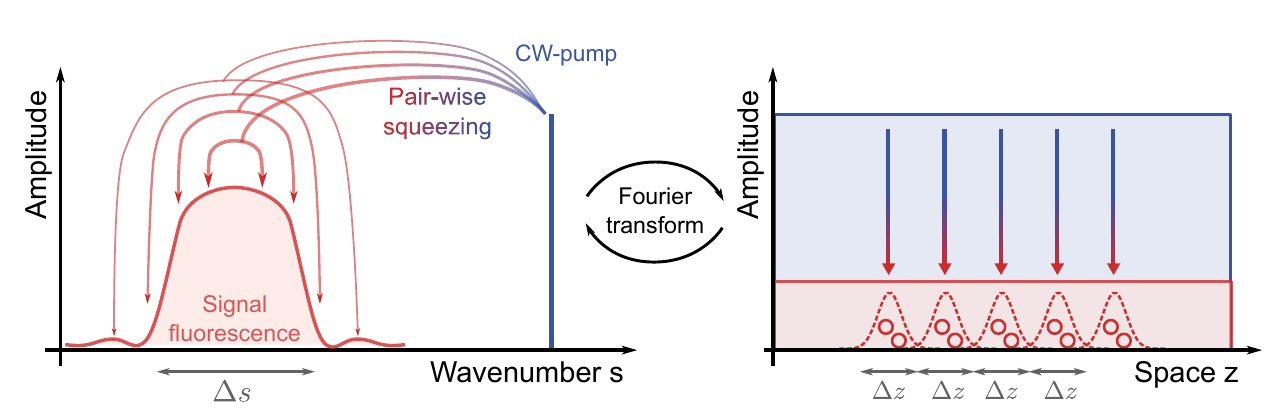}
\caption{Illustration of the CW-pumped broadband vacuum squeezing. Left figure: In the wavespace, CW-pump field induces independent pair-wise squeezing in of the signal field. The signal excitation is significant only within the phase-matching window with size $\sim \Delta s$. Right figure: In the spatial domain, finite squeezing window translates to finite spatial correlations among photons with characteristic correlation length of $\Delta z=1/\Delta s$. Figure is adapted from Ref.~\cite{Yanagimoto2023-thesis}.}
\label{fig:schematic-cw-opg}
\end{figure}

For a given initial coherent pump profile in the wavespace $\langle\hat{b}_s\rangle=\beta_s(0)$, under the undepleted pump approximation, the mean-field dynamics at a given interaction time $t$ (which is related to the propagation distance via the group-velocity of light) can be solved as $\beta_s(t)=e^{-\mathrm{i}\delta\omega_b(2\pi s)t}\beta_s(0)$. As we consider a CW pump field at the carrier wavenumber (\textit{i.e.}, $s=0$), the initial pump field is a delta-function-like excitation in the wavespace $\beta_s(0)=\mathrm{i}\beta_0\delta(s)$. By an appropriate choice of the reference phase-velocity, we can always choose a frame in which $\delta\omega_b(0)=0$ (see Section \ref{sec:rosetta-waveguide-hamiltonian} for more details), which leads to a time-independent pump field
\begin{align}
    \beta_s(t)=\mathrm{i}\beta_0\delta(s).
\end{align}
Physically, $|\beta_0|^2$ corresponds to the spatial pump photon number density. 
Using the linearized approximation we introduced in Sec.~\ref{sec:prologue-gaussian-quantum}, we obtain a Gaussian Hamiltonian
\begin{align}
\label{eq:Gaussian-chi2-Hamiltonian}
    \hat{H}_\mathrm{G}/\hbar=-\frac{\mathrm{i}u}{2}\int\mathrm{d}s\,\left(\hat{a}_{s}\hat{a}_{-s}-\hat{a}_{s}^\dagger\hat{a}_{-s}^\dagger\right)+\int\mathrm{d}s\,\hat{a}_s^\dagger \delta\omega_a(2\pi s)\hat{a}_s
\end{align}
with a parametric gain $u=\beta_0r$. We emphasize here that in contrast with Sec.~\ref{sec:Gaussian-single-mode-opa}, where $|\beta_0|^2$ represented the total number of photons in a given spatial mode, $|\beta_0|^2$ now represents the spatial photon number density (with units of $[\mathrm{length}^{-1}]$). As in Sec.~\ref{sec:photon-normalized-units} we may connect $\beta_0$ to the pump power using $P_b=\hbar\omega_bv_\mathrm{g,b}|\beta_0|^2$, and the nonlinear coefficient $r$ (Eqn.~\ref{eqn:r_QNLO}) has units of $[\mathrm{length}^{1/2}\cdot\mathrm{time}^{-1}]$. Therefore, the parametric gain parameter $u$ still has units of frequency as in Sec.~\ref{sec:Gaussian-single-mode-opa}. The Heisenberg equations of motions follow
\begin{align}
    \partial_t\hat{a}_s=\frac{u}{2}\hat{a}_{-s}^\dagger-\mathrm{i}\delta\omega_a(2\pi s)\hat{a}_s.
\end{align}
Here, by moving to a rotating frame given by the odd part of the dispersion, \textit{i.e.}, $\frac{1}{2}(\delta\omega_a(2\pi s)-\delta\omega_a(-2\pi s))$, we can eliminate their contributions from \eqref{eq:Gaussian-chi2-Hamiltonian} without changing its form. Based on this observation, we are always allowed to assume $\delta\omega_a(2\pi s)$ is an even function of $s$, which lets us further simplify the Hamiltonian as
\begin{align}
\hat{H}_\mathrm{G}/\hbar=\int_0^\infty\mathrm{d}s\,(\hat{H}_{\mathrm{G},s}^++\hat{H}_{\mathrm{G},s}^-)
\end{align}
with
\begin{align}
    \label{eq:multimode-constant}
    \hat{H}_{\mathrm{G},s}^\pm/\hbar=\mp\frac{\mathrm{i}u}{2}(\hat{a}_s^{\pm2}-\hat{a}_s^{\pm\dagger2})+\delta_s\hat{a}_s^{\pm\dagger}\hat{a}_s^\pm,
\end{align}
where we have introduced 
\begin{align}
\delta_s\equiv\delta\omega_a(2\pi s)
\end{align}
as a short-hand notation to represent the phase-mismatch as a function of wavenumber $s$. $\delta_s$ is an even function of $s$, and
\begin{align}
\hat{a}^\pm_s=\frac{\hat{a}_s\pm\hat{a}_{-s}}{\sqrt{2}}
\end{align}
is a mode composed of symmetric/anti-symmetric pairs of wavespace modes. Note that \eqref{eq:multimode-constant} takes the same form as its single-mode $\chi^{(2)}$ Hamiltonian \eqref{eq:single-mode-constant}, indicating that the CW-pumped broadband optical parametric amplification can be decomposed to independent single-mode squeezings of paired-wavespace modes $\hat{a}_s^\pm$, for which we can directly adopt the analysis performed in the previous Section \ref{sec:Gaussian-single-mode-opa}. The pair-wise squeezing structure in the wavespace is depicted in Fig.~\ref{fig:schematic-cw-opg}. Note that in this picture, the parametric amplification is broadband in the wavespace, which, when converted to the space-propagating picture, corresponds to broadband in the frequency space.

Because a paired-wavespace mode $\hat{a}_s^\pm$ experiences single-mode squeezing, its evolution under the Gaussian unitary $\hat{G}$ should be given as a single-mode Bogoliubov transformation
\begin{align}
    \hat{G}^\dagger\hat{a}^\pm_s\hat{G}=C_s^\pm\hat{a}^\pm_s+S_s^\pm\hat{a}^{\pm\dagger}_s.
\end{align}
The propagators for the paired-wavespace modes can be related to the ones for the canonical wavespace modes, which are defined via
\begin{align}
\hat{G}^\dagger\hat{a}_s\hat{G}=\int\mathrm{d}p\,(C_{sp}\hat{a}_p+S_{sp}\hat{a}_p^\dagger),
\end{align}
as
\begin{subequations}
\label{eq:undepleted-broadband-pair-correlation}
\begin{align}
    C_{sp}&=\frac{1}{2}(C_s^++C_{-s}^-)\delta(s-p)+\frac{1}{2}(C_s^+-C_{-s}^-)\delta(s+p)\\
    S_{sp}&=\frac{1}{2}(S_s^++S_{-s}^-)\delta(s-p)+\frac{1}{2}(S_s^+-S_{-s}^-)\delta(s+p).
\end{align}
\end{subequations}
Below, we determine the values of the propagators $C_s^\pm$ and $S_s^\pm$ using the results already obtained for the single-mode OPA. Similarly to the single-mode case, we can summarize the equations of motion of the propagators in a vector form as
\begin{align}
    \partial_t\begin{pmatrix}
        C_s^\pm\\S_s^{\pm*}
    \end{pmatrix}=M_s^\pm\begin{pmatrix}
        C_s^\pm\\S_s^{\pm*}
    \end{pmatrix},
\end{align}
where the time evolution is generated by
\begin{align}
    M^\pm_s=\begin{pmatrix}
        -\mathrm{i}\delta_s&\pm u\\
        \pm u&\mathrm{i}\delta_s
    \end{pmatrix}.
\end{align}
By directly adopting the results in the single-mode section, we obtain
\begin{subequations}
\label{eq:undepleted-broadband-cs}
\begin{align}
    C_s^\pm(t)&=\cosh\left(\sqrt{u^2-\delta^2_s}\,t\right)-\frac{\mathrm{i}\delta_s}{\sqrt{u^2-\delta^2_s}}\sinh\left(\sqrt{u^2-\delta^2_s}\,t\right)
\end{align}
and
\begin{align}
S^{\pm\mathrm{*}}_s(t)&=\frac{\pm u}{\sqrt{u^2-\delta^2_s}}\sinh\left(\sqrt{u^2-\delta^2_s}\,t\right).
\end{align}
\end{subequations}
Notice that \eqref{eq:undepleted-broadband-cs} takes exactly the same form as \eqref{eq:undepleted-single-cs} with the only difference in the overall sign. We note that the symmetries $C_s^+=C_s^-$ and $S_s^+=-S_s^-$ allow us to explicitly simplify \eqref{eq:undepleted-broadband-pair-correlation} to
\begin{align}
    &C_{sp}(t)=C_s^+(t)\delta(s-p) &S_{sp}(t)=S_s^+(t)\delta(s+p).
\end{align}
As far as the undepleted pump approximation holds, \eqref{eq:undepleted-broadband-cs} provides the analytic solution to the quantum dynamics of optical parametric amplification.

Specifically, in the following, we analyze the case of vacuum input, where the OPA outputs a broadband squeezed vacuum state. To this goal, we first calculate the two-photon correlation functions in the wave space (\textit{i.e.}, spectral correlation functions)
\begin{align}
\Sigma_{ss'}&=\langle\hat{a}_s^\dagger\hat{a}_{s'}\rangle=\int\mathrm{d}p\,S^*_{sp}S_{ps'}=|S_s^+|^2\delta(s-s'),\\
\Pi_{ss'}&=\langle\hat{a}_s\hat{a}_{s'}\rangle=\int\mathrm{d}p\,S^*_{sp}C_{ps'}=S_s^{+*}C_s^+\delta(s+s').
\end{align}
The photon number spectral density of the parametric fluorescence $P_s$ is given by the diagonal elements of $\Sigma_{ss'}$ (\textit{i.e.}, the self-correlation terms),
\begin{align}
\label{eq:Ps}
P_s&=\frac{u^2}{u^2-\delta^2_s}\sinh^2\left(\sqrt{u^2-\delta^2_s}\,t\right).
\end{align}

\begin{figure}[tb]
    \centering
    \includegraphics[width=\textwidth]{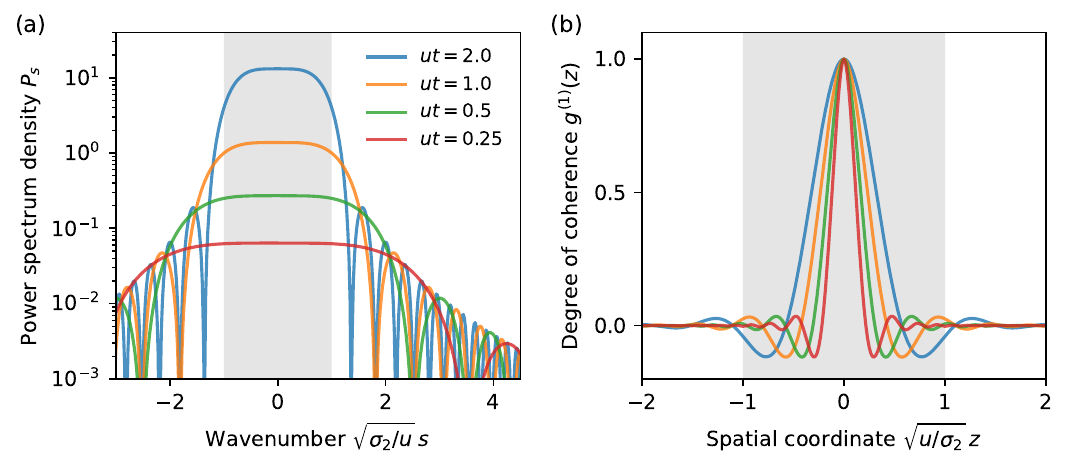}
    \caption{(a) Power spectral density $P_s$ and (b) first-order correlation function $g^{(1)}(z)$ for CW-pumped vacuum squeezing. The shaded area in (a) represents the OPA bandwidth, $|s|\leq \Delta s$, with associated correlation length given by $|z|\leq \Delta z=1/\Delta s$ shown in (b). As the interaction time is increased, only frequencies $s$ within the OPA bandwidth grow exponentially, which causes the zeros of the correlation function to shift asymptotically to $\Delta z=1/\Delta s$. For the quadratic dispersion assumed here, $\delta_s=\sigma_2s^2$, the phase-matching bandwidth is $\Delta s=(u/\sigma_2)^{1/2}$. We note that the power spectral density shown in (a) is determined by the cross-gain coefficient for traveling-wave OPA shown in Fig.~\ref{fig:OPA_basics}, here with space and time interchanged as in Sec.~\ref{sec:time-prop}. This figure is adapted from Ref.~\cite{Yanagimoto2023-thesis}.}
    \label{fig:broadband-fluorescence}
\end{figure}
In a waveguide, the phase-mismatch $\delta_s$ varies as a function of wavenumber $s$ due to dispersion, which changes the qualitative behavior of the parametric amplification for different wavenumber components. For the wavenumber modes with $|\delta_s/u|>1$, the parametric interaction is phase-mismatched, and the parametric fluorescence $P_s$ exhibits sinusoidal oscillation with an amplitude $2u^2/(\delta_s^2-u^2)$. As a result, no strong parametric amplification can occur for these modes. For modes pairs that fulfill $|\delta_s/u|<1$, the quantum fluctuations of the signal grow exponentially, exhibiting strong squeezing of the field. At the boundary $|\delta_s/u|=1$, which corresponds to an exceptional point of $M_s^\pm$, the signal power exhibits quadratic growth as a function of the propagation time, $P_s=2u^2t^2$, marking the transition between exponential growth and sinusoidal oscillation. Based on this physical understanding, we define the phase-matching window as the range of wavenumbers that fulfills $|\delta_s/u|<1$, within which parametric gain grows exponentially as a function of propagation distance. This window corresponds to the phase-matching bandwidth analyzed in the classical sections. In Fig.~\ref{fig:broadband-fluorescence}(a), we show the power spectral density for the signal fluorescence, where we observe these features reproduced numerically.

So far, our discussions have focused on the structure of the signal correlations in the wavenumber space (corresponding to frequency space). Here, we switch our attention to the spatial domain to see how the spectral features translate to nontrivial spatial structures. As a measure of photon-photon correlation, we consider the first order spatial coherence function
\begin{align}
g^{(1)}(z,z')=\frac{\langle\hat{a}^\dagger_{z}\hat{a}_{z'}\rangle}{\sqrt{\langle\hat{a}^\dagger_{z}\hat{a}_{z}\rangle\langle\hat{a}^\dagger_{z'}\hat{a}_{z'}\rangle}}.
\end{align}
Since the power envelopes of the generated signal are translationally invariant (\textit{i.e.}, $\langle\hat{a}^\dagger_{z}\hat{a}_{z}\rangle=\langle\hat{a}^\dagger_{0}\hat{a}_{0}\rangle$), we can set the reference point to $z'=0$ without loss of generality, leading to
\begin{align}
\label{eq:degree-of-coherence}
g^{(1)}(z)=\frac{\langle\hat{a}^\dagger_{z}\hat{a}_{0}\rangle}{\langle\hat{a}^\dagger_{0}\hat{a}_{0}\rangle}=\mathcal{\rho}^{-1}\int\mathrm{d}s\,e^{2\pi\mathrm{i}sz}P_s,
\end{align}
where we have used $\langle\hat{a}^\dagger_{z}\hat{a}_{0}\rangle=\int\mathrm{d}s\,e^{2\pi\mathrm{i}sz}P_s$, and 
\begin{align}
\label{eq:photon-deisnity-space}
\mathcal{\rho}(t)=\langle\hat{a}^\dagger_{0}\hat{a}_{0}\rangle=\int\mathrm{d}s\,P_s
\end{align}
is the spatial photon-number density. Physically, \eqref{eq:degree-of-coherence} implies that the spatial correlation length of the generated signal photons is given by the Fourier transform of the photon number spectral density (see Fig.~\ref{fig:schematic-cw-opg} for illustration). Here, it is worth reminding that the photon number spectral density takes significant amplitudes only within the phase-matching window. As a result, for a given window size $\Delta s$, the spatial correlation should have a width $\Delta z\sim 1/\Delta s$, \textit{i.e.}, a broader phase-matching window leads to narrower spatial correlation length. Fig.~\ref{fig:broadband-fluorescence}(b) shows numerically calculated $g^{(1)}(z)$, where we observe localized but nontrivial spatial photon-photon correlation structures.

Here, it might seem possible to argue that one can ignore the spatial photon-photon correlation structure when the signal dispersion is small enough, because the large phase-matching bandwidth would lead to vanishing spatial correlation length $\Delta z$. This is a tempting simplification especially because the dynamics then become local in time (\textit{i.e.}, space), which would significantly simplify the physical picture. We would like to emphasize, however, that we \emph{cannot} ignore the signal dispersion for optical parametric amplification, and thus, finite temporal correlation structure is intrinsic to the broadband squeezing. To see this more concretely, assume that signal dispersion does not exist, in which case, $\delta_s$ becomes independent of $s$. Then, the power spectrum density $P_s$ also becomes independent of $s$, resulting in an infinite power flux density $\mathcal{\rho}(t)=\int\mathrm{d}s\,P_s(t)\rightarrow\infty$, which is unphysical. In reality, finite dispersion prevents the signal fluorescence from exhibiting infinite magnitude.

Typically, the role of dispersion is most crucial in the SPDC limit (\textit{i.e.}, during the early time $ut\ll 1$), where the signal field is mostly in the vacuum state. Because SPDC is a process stimulated by vacuum fluctuations, which have an infinite bandwidth, a pump photon could downconvert to an infinite bandwidth of signal fields in the absence of dispersion. In reality, competition between the broadband parametric gain and the signal dispersion determines the fluorescence bandwidth in a nontrivial manner. Below, we derive analytic expressions for the dynamics of vacuum squeeizng in this SPDC limit to unravel the role of dispersion.

For this purpose, we first assume a purely quadratic dispersion $\delta_s=\sigma_2 s^2$, where $\sigma_2$ is proportional to the group-velocity dispersion at the signal carrier frequency.\footnote{Expanding signal dispersion up to second order and using the conversion formulas in Table~\ref{tab:dispersion}, we write the series expansion of the phase-mismatch as $\delta\omega_a(k)=\frac{1}{2}\delta\omega_a''(0)k^2=-\frac{1}{2}k''_av_{g,a}^3k^2$. Thus, $\delta_s=-4\pi^2k_a''v_{g,a}^3s^2$, meaning that the value of $\sigma_2$ is related to classical experimental parameters via $\sigma_2=-4\pi^2k_a''v_{g,a}^3$.} 
We note that the value of the photon number spectral density $P_s=u^2t^2\mathrm{sinc}^2\left(\sqrt{\delta_s^2-u^2}\,t\right)$ as given in \eqref{eq:Ps} is relatively flat up to the point $\sqrt{\delta_s^2-u^2}\sim t^{-1}$, implying that we can approximate $\sqrt{\delta_s^2-u^2}\approx \delta_s$, which lets us convert \eqref{eq:photon-deisnity-space} to
\begin{align}
\label{eq:broadband-flux-quadratic}
    \rho(t)\approx u^2t^2\int_{-\infty}^\infty\mathrm{d}s\,\mathrm{sinc}^2\left(\sigma_2 s^2 t\right)=\frac{4\sqrt{\pi}}{3}\Delta s\,(ut)^{3/2}
\end{align}
with the phase-matching bandwidth
\begin{align}
   \Delta s=(u/\sigma_2)^{1/2}.
\end{align}
The expression \eqref{eq:broadband-flux-quadratic} shows that the PDC rate is directly proportional to the signal phase-matching bandwidth $\Delta s$, and the result becomes unphysical when zero dispersion is assumed (\textit{i.e.}, $\sigma_2\rightarrow0$ implies $\Delta s\rightarrow\infty$). 

It is worth noting that parametric fluorescence shows a fractional time scaling $\rho(t)\sim t^{3/2}$, instead of the quadratic scaling $\sim t^2$ that one would na\"ively expect from single-mode model. Such fractional scaling is caused by the narrowing of the sinc function in \eqref{eq:broadband-flux-quadratic} in the wavespace as a function of time, leading to a partial cancellation of the overall prefactor with $\sim t^2$ scaling. 

Generally, the exact time scaling is determined by the dominant dispersion order in the OPA bandwidth; When $m$th order dispersion dominates, the signal power should scale as $\sim t^{2-\frac{1}{m}}$. For instance, with a purely quartic dispersion $\delta_s=\sigma_4s^4$, we obtain
\begin{align}
\label{eq:broadband-flux-quartic}
    \rho(t)\approx u^2t^2\int_{-\infty}^\infty\mathrm{d}s\,\mathrm{sinc}^2\left(\sigma_4 s^4 t\right)=\frac{8\sqrt{1+\sqrt{2}}\,\Gamma(1/4)}{21}\Delta s\,(ut)^{7/4},
\end{align}
where $\Gamma(x)$ is the Gamma function, and the phase-matching bandwidth is now given as
\begin{align}
    \Delta s=(u/\sigma_4)^{1/4}.
\end{align}
Notice that $\rho(t)$ is again proportional to the phase-matching bandwidth, and we observe an expected polynomial time scaling $\rho(t)\sim t^{7/4}=t^{2-\frac{1}{4}}$.

\begin{figure}[bt]
\centering
\includegraphics[width=0.7\textwidth]{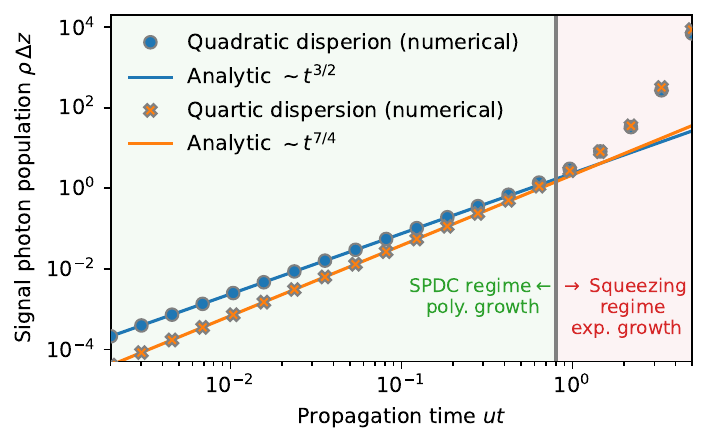}
    \caption{Dynamics of the signal photon population within a characteristic spatial correlation length $\rho(t)\Delta z$ for CW-pumped broadband vacuum squeezing, where we consider purely quadratic and quartic signal dispersion. The grey vertical line (marking the time around which $\rho(t)\Delta z\sim 1$ holds) represents the cross-over point between SPDC-like and vacuum-squeezing-like regimes of optical parametric interactions. We note here that the saturated limit, where anti-squeezed vacuum depletes the pump, is commonly referred to as optical parametric generation (OPG) in the literature. Figure is adapted from Ref.~\cite{Yanagimoto2023-thesis}.}
    \label{fig:fluorescence-scaling}
\end{figure}

Fig.~\ref{fig:fluorescence-scaling} compare signal photon population, obtained by solving \eqref{eq:photon-deisnity-space} numerically, to the analytic expressions \eqref{eq:broadband-flux-quadratic} and \eqref{eq:broadband-flux-quartic}. Specifically, we show the dynamics of $\rho(t)\Delta z$, which is the expected number of signal photons within a characteristic correlation length $\Delta z=1/\Delta s$. Intuitively, $\Delta z$ is a distance over which signal photons can see each other, and thus, $\rho(t)\Delta z$ can be interpreted as the number of photons inside a spatial ``bin'' within which signal photons can interact. When $\rho(t)\Delta z\ll1$ holds, each ``bin'' is almost in a vacuum state, and the pump photons downconvert only spontaneously, marking the SPDC-like regime of optical parametric interaction. In this regime, the signal photon population grows polynomially as a function of the interaction time. As the magnitude of $\rho(t)\Delta z$ approaches the order of unity, a finite population of signal photons within each spatial ``bin'' stimulates further downconversion of pump photons, leading to an exponential growth of signal photon population, which marks the transition to a vacuum-squeezing-like regime. Such cross-over from polynomial scaling (SPDC-like regime) to exponential scaling (vacuum-squeezing-like regime) around $\rho(t)\Delta z\sim 1$ can be clearly identified in Fig.~\ref{fig:fluorescence-scaling}.

Finally, we would like to provide an alternative interpretation of the physics in the SPDC limit to relate it to a later section \ref{sec:PDC}. It turns out that we can rewrite \eqref{eq:broadband-flux-quadratic} as
\begin{align}
\label{eq:broadband-flux-gpdc}
    \rho(t) = \frac{4\sqrt{\pi}}{3}\beta_0^2\,(g_\mathrm{pdc}t)^{3/2}
\end{align}
with a characteristic rate
\begin{align}
    g_\mathrm{pdc}=(r^4/\sigma_2)^{1/3}.
\end{align}
As shown later in Sec.~\ref{sec:PDC}, $g_\mathrm{pdc}$ is a rate at which a single pump photon downconverts to a signal photon pair, which explains \eqref{eq:broadband-flux-gpdc} as a summation of signal fluorescence independently produced by each pump photon undergoing PDC process (notice that $\rho(t)$ is proportional to the pump flux density $\beta_0^2$). This is a reflection of the fact that stimulated parametric downconversion is negligible in the SPDC limit. The same discussion applies to the case of purely quartic dispersion, where the single-photon PDC rate is to be modified to $g_\mathrm{pdc}=(r^8/\sigma_4)^{1/7}$.

\paragraph{Further reading} Readers can refer to Refs.\cite{Pysher2009,Mondain2019,Kashiwazaki2020} for recent experimental works that have demonstrated CW-pumped broadband vacuum squeezing using $\chi^{(2)}$ nonlinear waveguides. Though not covered in this work, an OPA can also be realized in a $\chi^{(3)}$ nonlinear waveguide/fiber using dual-pump configuration~\cite{Quesada2022}, and readers can refer to Refs.~\cite{Riemensberger2022,Hansryd2001} for experimental $\chi^{(3)}$-based OPAs.

\subsection{Pulse-pumped traveling-wave OPA}
\label{sec:Gaussian-pulse-pumped-OPA}
In this section, we study the most generic case of traveling-wave optical parametric amplification, \textit{i.e.}, pulse-pumped broadband optical parametric amplification~\cite{nehra2022few}. Due to the involvement of multiple frequency components of the pump pulse, we can no longer decompose the system dynamics into pair-wise independent squeezing of wavenumber components as was done for the CW-pumped case, and no general analytic solution exists. However, it turns out that we can still identify discrete ``supermodes'' as independent entities that experience independent single-mode squeezing, providing a concise physical picture for the multimode quantum dynamics. The formalism we introduce in this section to capture generic multimode Gaussian quantum dynamics provides a crucial scaffolding in Section \ref{sec:mesoscopic-quantum-nonlinear-optics}, where we include non-Gaussian quantum features on top of the Gaussian quantum dynamics.

\begin{figure}[bt]
    \centering
    \includegraphics[width=\textwidth]{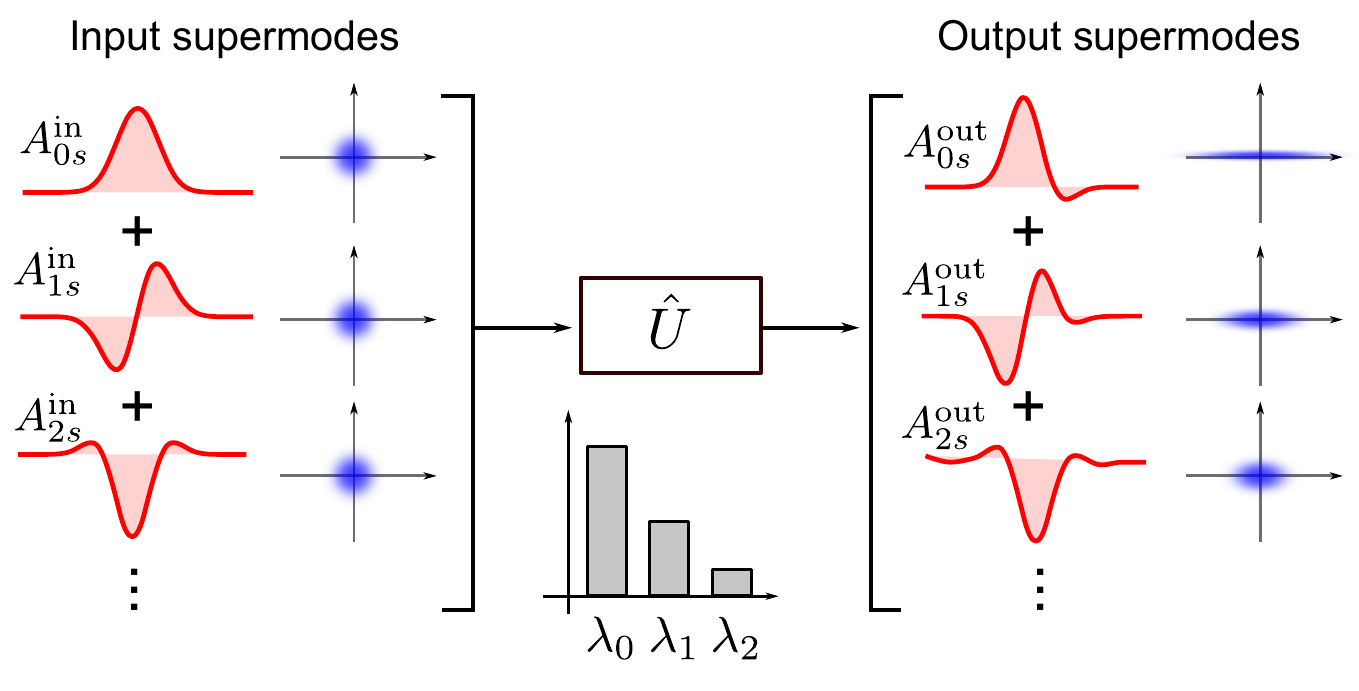}
    \caption{Illustration for multimode squeezing induced by pulse-pumped optical parametric amplification, which can be decomposed into independent squeezing of squeezing supermodes. Each of input supermode with waveform $\hat{A}_{ms}^{\mathrm{in}}$ experiences single-mode squeezing with field gain $e^{\lambda_m}$, and the output is encoded to the output waveform $\hat{A}_{ms}^{\mathrm{out}}$. Figure is adapted from Ref.~\cite{Yanagimoto2023-thesis}.}
    \label{fig:supermode}
\end{figure}

For a generic initial pump field amplitude $\beta_s(t=0)$, we employ the undepleted pump approximation to calculate the mean-field dynamics for following times with \eqref{eq:rosetta-generic-heisenberg-multimode} as
\begin{align}
    \beta_s(t)=e^{-\mathrm{i}\delta\omega_b(2\pi s)t}\beta_s(0).
\end{align}

Due to the finite dispersion of the pump pulse, the Gaussian Hamiltonian given in Sec.~\ref{sec:prologue-gaussian-quantum} generally depends on time as
\begin{align}
\label{eq:undepleted-chi2-waveguide}
    \hat{H}_\mathrm{G}(t)/\hbar&=\frac{r}{2}\iint\mathrm{d}s_1\mathrm{d}s_2\,\left(\hat{a}_{s_1}\hat{a}_{s_2}\beta_{s_1+s_2}^*(t)+\hat{a}_{s_1}^\dagger\hat{a}_{s_2}^\dagger\beta_{s_1+s_2}\right)+\int\mathrm{d}s\,\hat{a}_s^\dagger \delta\omega_a(2\pi s)\hat{a}_s.
\end{align}
The Gaussian unitary $\hat{G}$ as defined in \eqref{eq:gaussian-unitary-define} is completely characterized via the multimode Bogoliubov transformation using propagators as
\begin{align}
\hat{G}^\dagger\hat{a}_s\hat{G}=\int\mathrm{d}p\,(C_{sp}\hat{a}_p+S_{sp}\hat{a}_p^\dagger).
\end{align}
The values of the propagators for a given time can be obtained by solving their equations of motion
\begin{subequations}
\label{eq:undepleted-greens-broadband-eom}
\begin{align}
    \partial_tC_{sp}(t)&=-\mathrm{i}r\int\mathrm{d}q\,\beta_{s+q}(t)S^*_{qp}(t)-\mathrm{i}\delta\omega_a(2\pi s)C_{sp}(t)\\
    \partial_tS_{sp}(t)&=-\mathrm{i}r\int\mathrm{d}q\,\beta_{s+q}(t)C^*_{qp}(t)-\mathrm{i}\delta\omega_a(2\pi s)S_{sp}(t),
\end{align}
\end{subequations}
which are obtained from obtained from \eqref{eq:abstract-CS-evolution}, with boundary conditions $C_{sp}(0)=\delta(s-p)$ and $S_{sp}(0)=0$. Because a generic pump field $\beta_s(t)$ contains multiple frequency components, \eqref{eq:undepleted-greens-broadband-eom} cannot be decomposed into separate components as we did in Section \ref{sec:cw-pumped-broadband} for CW-pumped OPA. Thus, for a generic pump pulse shape and gain, we need to calculate the propagators numerically.

Notably, for a given $C_{sp}$ and $S_{sp}$, it is always possible to perform singular value decomposition (SVD) to obtain
\begin{align}
    &C_{sp}(t)=\sum_{m=0}^\infty A^{\mathrm{out}}_{ms}\cosh\lambda_m\,A_{mp}^\mathrm{in*} &S_{sp}(t)=\sum_{m=0}^\infty A^{\mathrm{out}}_{ms}\sinh\lambda_m\,A_{mp}^{\mathrm{in}},
\end{align}
which is to be seen as a multimode extension of \eqref{eq:undepleted-svg-single-mode}. Here, $\sinh\lambda_m\geq 0$ and $\cosh\lambda_m\geq 0$ are the $m$th singular values of $S_{sp}(t)$ and $C_{sp}(t)$ sorted in an ascending order, respectively. The singular vectors $A^{\mathrm{in/out}}_{ms}$ are normalized as
\begin{align}
&\int\mathrm{d}s\,A^{\mathrm{in/out}*}_{ms}A^{\mathrm{in/out}}_{m's}=\delta_{mm'} &\sum_{m=0}^\infty A^{\mathrm{in/out}*}_{ms}A^{\mathrm{in/out}}_{ms'}=\delta(s-s')
\end{align}
With these expansions, we can rewrite the operator transformation under the Gaussian unitary $\hat{G}^\dagger\hat{a}_s\hat{G}$ as
\begin{align}
\label{eq:supermode-inout-output}
\hat{G}^\dagger\hat{a}_m^\mathrm{out}\hat{G}=\cosh\lambda_m\,\hat{a}_m^\mathrm{in}+\sinh\lambda_m\,\hat{a}_m^{\mathrm{in}\dagger}
\end{align}
where the ``input'' and ``output'' supermodes are defined as
\begin{align}
    &\hat{a}_m^\mathrm{in}=\int\mathrm{d}s\,A^{\mathrm{in}*}_{ms}\hat{a}_s &\hat{a}_m^\mathrm{out}=\int\mathrm{d}s\,A^{\mathrm{out}*}_{ms}\hat{a}_s.
\end{align}
To see the meaning of ``input'' and ``output'' more clearly, we rewrite \eqref{eq:supermode-inout-output} using their quadratures as
\begin{align}
\label{eq:supermode-quadrature-map}
&\hat{G}^\dagger\hat{x}_m^\mathrm{out}\hat{G}=e^{\lambda_m}\,\hat{x}_m^\mathrm{in}&\hat{G}^\dagger\hat{p}_m^\mathrm{out}\hat{G}=e^{-\lambda_m}\,\hat{p}_m^\mathrm{in}.
\end{align}
These equations imply that $\hat{G}(t)$ acts independently on input pulse supermode $\hat{a}_m^\mathrm{in}$, applies single-mode squeezing with field gain $e^{\lambda_m}$, and encodes the result to the output pulse supermode $\hat{a}_m^\mathrm{out}$. Such elegant decomposition of broadband pulsed squeezing into independent squeezing of supermodes was established in Ref.~\cite{Wasilewski2006} (see Fig.~\ref{fig:supermode} for illustration). 

It is worth emphasizing that the waveforms of input/output supermodes are different in general (\textit{i.e.}, $A^{\mathrm{in}}_{ms}\neq A^{\mathrm{out}}_{ms}$), meaning that two physical processes are involved in $\hat{U}$, \textit{i.e.}, squeezing and pulse-waveform distortions. Also, due to the time dependence of the propagators, their singular vectors $A^{\mathrm{in/out}}$ also depend nontrivially on time, implying that the structure of squeezing supermodes is not static and ``morphs'' in time~\cite{Gouzien2020}.

For a vacuum input, pulse-pumped OPA outputs broadband vacuum-squeezed light. Using \eqref{eq:supermode-quadrature-map}, quadrature variances for the final state can be written as
\begin{align}
&\langle(\hat{x}_m^{\mathrm{out}})^2\rangle=\frac{e^{2\lambda_m}}{4} &\langle(\hat{p}_m^{\mathrm{out}})^2\rangle=\frac{e^{-2\lambda_m}}{4},
\end{align}
which exhibits sub-vacuum quadrature fluctuations for $\lambda_m>0$. Notice that higher-order supermodes (\textit{i.e.}, modes with $m\geq 1$) generally contain finite excitation, and thus, the resultant state is \emph{not} a single-mode squeezed state in general, unless $\lambda_{m\geq 1}=0$ holds. Quantitatively, the multimode-ness of the pulsed squeezed state is measured by the Schmidt number~\cite{Brecht2015}
\begin{align}
    K=\frac{(\sum_m\sinh^2\lambda_m)^2}{\sum_m\sinh^4\lambda_m},
\end{align}
where $K=1$ is fulfilled only when the state is single-mode squeezed, and $K>1$ indicates that the state is multimode-squeezed.

For many applications in quantum information science, it is essential to suppress such multimode squeezing effects; For instance, heralded photon subtraction is a powerful and widely-used technique to generate non-Gaussian quantum states, \textit{e.g.}, single-photon state and Schr\"odinger's cat state. When higher-order supermodes are not in vacuum state, however, we cannot tell whether a subtracted photon has come subtracted from the primary supermode or the higher-order supermodes, which critically limits the purity of the generated state. In the SPDC limit of pulsed squeezing, dispersion engineering enables one to realize a single-mode weakly squeezed state. In the high-gain regime, however, suppressing multimode effects is more challenging, and realization of strong single-mode squeezed light source is an active area of research.

\paragraph{Further reading} Strong demand for high-quality heralded single-photon sources has motivated the study of pulsed squeezing in the SPDC limit. In this limit, one can ignore the time-ordering of the operators and focus on the biphoton wavefunction of the generated state, which enables in-depth analytic studies~\cite{Grice1997,Law2000,Uren2005,Keller1997}. Readers can refer to Ref.~\cite{Ansari2018} for a comprehensive review. Ref.~\cite{Quesada2014} provide a thorough analysis on the effects of the operator time-ordering.




\section{Mesoscopic quantum nonlinear optics}
\label{sec:mesoscopic-quantum-nonlinear-optics}

\subsection{Introduction}
In a typical setup of nonlinear optics, the intensity of quantum fluctuations is much smaller than the mean field. In this regime, the photon dynamics can be approximated as Gaussian unitary, which has been the topic of the previous section ~\ref{sec:semi-classical_NLO}. As the nonlinearity or the interaction time increases, however, the growth of Gaussian quantum fluctuations can eventually reach a point where they affect the dynamics of the mean field, and non-Gaussian quantum features can emerge. The onset of this regime often coincides with the point where classical dynamics exhibit strong saturation. 

The magnitude of Gaussian quantum features required to affect the mean field strongly correlates with the number of photons driving the nonlinear device, intuitively, large quantum fluctuations are required to affect a strong classical mean field. Such strong Gaussian quantum features (\textit{i.e.}, squeezing) make non-Gaussian quantum features that may appear on top highly fragile and sensitive to experimental imperfections, making the observation of the latter infeasible, as was discussed in Sec.~\ref{sec:intro-macro-meso-micro}. These considerations motivate us to explore mesoscopic nonlinear optics, where only a few hundred photons can lead to saturated dynamics. This regime offers a promising opportunity to investigate non-Gaussian quantum physics and develop new concepts in the field.

This section provides a case study of optical parametric generation (OPG) operated in the mesoscopic regime, discussing the unique interplay among mean-field, Gaussian quantum, and non-Gaussian quantum features. Since we assume the knowledge of the Gaussian interaction frame (GIF), a theoretical framework to capture non-Gaussian quantum optical dynamics concisely, we encourage the readers to visit the short prologue Sec.~\ref{sec:prologue} before this section. The construction of a GIF involves solving Gaussian-approximated dynamics of the system, for which we cite the contents of Sec.~\ref{sec:semi-classical_NLO}, which concerns Gaussian quantum physics in nonlinear optics. 

The rest of the section is structured as follows. In Sec. \ref{subsec:GIF}, we introduce the framework of the Gaussian interaction frame to capture the onset of non-Gaussian quantum physics concisely. Instead of completely disposing of Gaussian quantum optics, this framework allows us to account for non-Gaussian quantum features on top of a scaffold provided by a Gaussian model, for which we can directly adopt the analysis from the previous Sec. \ref{sec:semi-classical_NLO}. We also introduce a nonlinear Gaussian model that can account for pump depletion effects, which allows us to develop a further refined Gaussian interaction frame. Finally, in Sec. \ref{sec:mesoscale_sqeezing}, we apply the developed formalism to study broadband squeezing pumped with a mesoscopic number of photons. The Gaussian interaction frame provides essential information on where and how non-Gaussian quantum features emerge, which we can leverage to construct a concise numerical approach for analyzing this otherwise intractable system.

\subsection{Gaussian interaction frames for single-mode OPA}
\label{subsec:GIF}
In this subsection, we apply the framework of GIF to unravel the non-Gaussian quantum dynamics of a single-mode OPG efficiently. The Hamiltonian takes the form
\begin{align} \label{eq:single-mode-opg}
    \hat{H}/\hbar=\frac{g}{2}(\hat{a}^{\dagger2}\hat{b}+\hat{a}^2\hat{b}^\dagger),
\end{align}
where $\hat{a}$ and $\hat{b}$ are annihilation operators for the signal and pump modes, respectively, and $g$ is the $\chi^{(2)}$ nonlinear coupling rate. See Sec.~\ref{sec:rosetta-waveguide-hamiltonian} for the derivations of the Hamiltonian. For OPG, we start from an initial state $\ket{\psi(t=0)}=\ket{0}_a\ket{\mathrm{i}\beta_0}_b$, where $\ket{0}_a$ and $\ket{\mathrm{i}\beta_0}_b$ are the vacuum signal state and coherent pump state with amplitude $\mathrm{i}\beta_0$, respectively.

A conventional approach to performing full-quantum simulation is to numerically expand the state and the Hamiltonian in the Fock space as
\begin{align}
    &\ket{\psi(t)}=\sum_{k=0}^{N_a}\sum_{\ell=0}^{N_b}\psi_{k,\ell}\ket{k}_a\ket{\ell}_b,&\hat{H}=\sum_{k_1,k_2}^{N_a}\sum_{\ell_1,\ell_2}^{N_b}H_{k_1,\ell_1,k_2,\ell_2}\ket{k_1}_a\ket{\ell_1}_b\bra{k_2}_a\bra{\ell_2}_b,
\end{align}
where $N_a$ and $N_b$ represent the truncation of the photon-number basis. Because Schr\"odinger's equation involves multiplication of the Hamiltonian by the wavefunction, the numerical cost to simulate the Schr\"odinger's equation is equivalent to the matrix-vector multiplication with dimension $N_aN_b$, which is $\mathcal{O}(N_a^2N_b^2)$. It is worth noting that the pump amplitude can easily be a macroscopic quantity. For instance, even $\SI{10}{fJ}$ of pump pulse at $\SI{750}{nm}$ wavelength contains $\sim 40000$ photons. If we are interested in, say, $\SI{20}{dB}$ of squeezing induced by the nonlinear optical dynamics, this translates to $\sim 100$ signal photons, leading to a total of $N_aN_b\sim 4\times 10^6$ dimensional Hilbert space. Clearly, this approach is not scalable, especially in considering an extension to broadband pulse dynamics, where thousands of modes can be involved. 

To circumvent the challenge, we factor out the trivial Gaussian quantum features in the form of a  Gaussian unitary $\hat{U}=\hat{D}\hat{G}$, which defines a GIF (see Fig.~\ref{fig:gif} for illustration and Sec.~\ref{sec:gif-intro} for more detailed introduction on GIF). A compressed state can now capture the residual non-Gaussian quantum fluctuations in the GIF $\ket{\varphi_\mathrm{I}}$. Below, we show a concrete procedure to construct a GIF for the OPG dynamics. 

First, we construct $\hat{D}$ so that it optimally accounts for the mean-field dynamics. Due to the parity symmetry of the signal mode $\hat{a}\mapsto -\hat{a}$, we expect no mean-field displacement on the signal. As a result, $\hat{D}$ is only parameterized by the pump mean-field estimate $\beta(t)$. A straightforward approach to set $\beta(t)$ is by means of a linearized approximation, where we directly use the solution for a classical model (see Sec.~\ref{sec:prologue} for an introduction). While this approach provides an intuitive construction of a GIF, it turns out that we can further refine the estimate for the pump mean-field by incorporating finite pump depletion effects. In anticipation that we will later incorporate these two approaches to derive clearer insights, we leave $\beta(t)$ undetermined at this stage.\footnote{It is, in principle, possible to construct a physically valid GIF based on any choice of $\hat{D}$ and $\hat{G}$, but the resultant representation of $\ket{\varphi_\mathrm{I}}$ becomes inefficient. For an efficient and concise representation of the quantum state, it is essential that we find a good estimate of the Gaussian part of the dynamics based on semi-classical intuition.} 

Following the discussion in Sec.~\ref{sec:Gaussian-single-mode-opa}, for a given $\beta(t)$, we can derive the Gaussian Hamiltonian
\begin{align}
    H_\mathrm{G}(t)/\hbar=\frac{g}{2}\left(\beta(t)\hat{a}^{\dagger2}+\beta^*(t)\hat{a}^2\right),
\end{align}
which induces symplectic transformation $\hat{G}$. The action of the unitary $\hat{G}$ can be completely characterized by the propagators $C$ and $S$, whose values are determined by their equations of motion (see \eqref{eq:abstract-CS-evolution})
\begin{align}
\label{eq:mesoscopic-propagators-eom}
    &\partial_tC=-\mathrm{i}g\beta S^*
    &\partial_tS=-\mathrm{i}g\beta C^*
\end{align}
with boundary conditions $C(0)=1$ and $S(0)=0$. 

For a given $\beta(t)$, we can determine $C(t)$ and $S(t)$ using \eqref{eq:mesoscopic-propagators-eom}, which completes the characterization of a GIF unitary $\hat{U}=\hat{D}\hat{G}$, where $\hat{D}=\exp(\beta\hat{b}^\dagger-\beta^*\hat{b})$ is a displacement operator. The Hamiltonian in the GIF takes a form
\begin{align}
\label{eq:H-interaction}
\begin{split}
    \hat{H}_\mathrm{I}&=\hat{H}_\mathrm{C}+\hat{H}_\mathrm{L},
\end{split}
\end{align}
where 
\begin{align}
\label{eq:h-nl}
    \hat{H}_\mathrm{C}/\hbar=\frac{g}{2}\left(C^*\hat{a}^\dagger+S^*\hat{a}\right)^2\hat{b}+\mathrm{h.c.}
\end{align}
is the cubic nonlinear term, and
\begin{align}
    \hat{H}_\mathrm{L}/\hbar=-\mathrm{i}\hat{D}^\dagger\partial_t\hat{D}=-\mathrm{i}\partial_t\beta\hat{b}^\dagger+\mathrm{i}\partial_t\beta^*\hat{b}
\end{align}
is the linear term. Now, our remaining task is to identify an optimal time dependence of $\beta(t)$ so that $\hat{U}$ maximally factors out the Gaussian quantum features in $\ket{\psi}$.

\subsubsection{Linearized approximation}
\begin{figure}[tb]
    \centering
    \includegraphics[width=\textwidth=1.0]{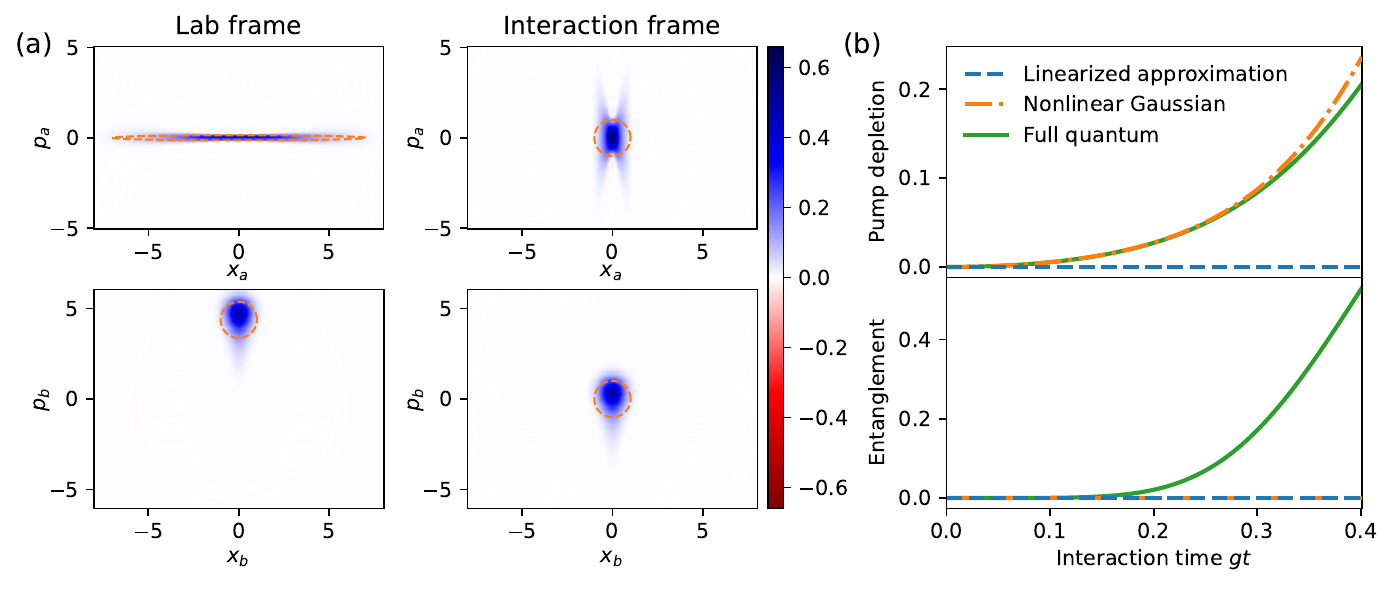}
    \caption{(a) Phase-space representation of the signal (upper row) and the pump state (bottom row) shown in the lab frame (left column) and the Gaussian interaction frame given by the nonlinear Gaussian model (right column). Orange ellipses represent $1/e^2$-width of the variance predicted by the Gaussian-approximated state $\ket{\psi_\mathrm{nlin}'}$. (b) Pump depletion ratio $R(t)$ and signal-pump entanglement shown as functions of time for linearized-approximated state $\ket{\psi_\mathrm{lin}'}$, Gaussian-approximated state $\ket{\psi_\mathrm{nlin}'}$ with the nonlinear Gaussian model, and full-quantum state $\ket{\psi}$. Entanglement is measured by the entanglement entropy. For all the simulations, we use initial pump amplitude $\beta_0=\sqrt{50}$, and the states in (a) are for time $t=0.6t_\mathrm{dep}\approx 0.4g^{-1}$. Figure is adapted from Ref.~\cite{Yanagimoto2023-thesis}.}
    \label{fig:gif-combined}
\end{figure}

One way to minimize the dynamics induced by $\hat{H}_\mathrm{I}$ shown in \eqref{eq:H-interaction} in the GIF would be first to ignore the higher-order term $\hat{H}_\mathrm{C}$ and eliminate the remaining linear contribution $\hat{H}_\mathrm{L}$ via an appropriate choice of  $\beta(t)=\beta_\mathrm{lin}(t)$ (the meaning for the subscripts will become clear shortly). This can be done by simply taking
\begin{subequations}
\label{eq:full-eq-undepleted}
\begin{align}
\label{eq:beta-eq-undepleted}
    \partial_t\beta_\mathrm{lin}=0,
\end{align}
under which the linear term vanishes $\hat{H}_\mathrm{L}=0$. The solution of \eqref{eq:beta-eq-undepleted} is a trivial constant solution $\beta_\mathrm{lin}(t)=\mathrm{i}\beta_0$, where $\mathrm{i}\beta_0=\langle\psi(0)\vert\hat{b}\vert\psi(0)\rangle$ with $\beta_0\in\mathbb{R}$ is the initial pump amplitude. For this choice of $\beta$, we can solve for the propagator dynamics analytically as
\begin{align}
    &C_\mathrm{lin}(t)=\cosh(\beta_0gt)  &S_\mathrm{lin}(t)=\sinh(\beta_0gt),
\end{align}
\end{subequations}
which defines a Gaussian unitary $\hat{U}_\mathrm{lin}$ and a Gaussian-approximated state $\ket{\psi'_\mathrm{lin}}=\hat{U}_\mathrm{lin}\ket{0}$. The mean-field estimate $\beta_\mathrm{lin}$ (and consequently the unitary $\hat{U}_\mathrm{lin}$) obtained by this procedure is the same as what we would obtain by the linearized approximation (thus the subscript ``lin''), which gives another interpretation to the linearized approximation that it is a treatment of ignoring $\hat{H}_\mathrm{C}$ in an interaction frame $\hat{U}_\mathrm{lin}$ defined by \eqref{eq:full-eq-undepleted}.

The suboptimality of the linearized approximation is seen most clearly in the pump depletion ratio
\begin{align}
    R(t)=1-\frac{N_\mathrm{b}(t)}{N_\mathrm{b}(0)}
\end{align}
where $N_\mathrm{b}(t)$ is the number of pump photons. For the linearized-approximated state $\ket{\psi'_\mathrm{lin}}$, we have constant pump photon number $N_\mathrm{b}(t)=N_\mathrm{b}(0)$, resulting in $R(t)=0$. However, in reality, the pump photon number should decrease (\textit{i.e.}, deplete) as the photons downconvert to the signal mode. As shown in Fig.~\ref{fig:gif-combined}, the full-quantum simulation predicts quadratic growth of $R(t)$ as a function of interaction time, which the linearized approximation fails to capture. Such misprediction of the pump field amplitude generally results in inefficient state representation in the interaction frame, leading to unnecessarily large requirements for the Fock space truncation when we desire to capture non-Gaussian quantum dynamics.

The consequence of failing to capture pump depletion can also be seen in the Manley-Rowe invariant
\begin{align}
   N_\mathrm{MR}(t)=N_\mathrm{b}(t)+\frac{N_\mathrm{a}(t)}{2},
\end{align}
which is a constant of motion under optical parametric interactions. However, the linearized-approximated state $\ket{\psi_\mathrm{lin}'}$ leads to
\begin{align}
    N_\mathrm{MR}(t)=|\beta_\mathrm{lin}(t)|^2+\frac{|S_\mathrm{lin}(t)|^2}{2}=\beta_0^2+\sinh^2(\beta_0gt).
\end{align}
which increases as a function of time. 

\subsubsection{Nonlinear Gaussian model} \label{sec:nonlinear-gaussian-model}
We have seen that linearized approximation, which directly uses the classical undepleted solution as the estimate for pump mean field $\beta(t)$, leads to a suboptimal GIF. It turns out that we can find a more sophisticated choice of $\beta(t)$ so that it accounts for pump depletion and further compresses the quantum fluctuations in the GIF. 

Our starting point is to notice that the nonlinear Hamiltonian \eqref{eq:h-nl} still contains linear terms that induce trivial phase-space displacement. For instance, terms of the form $\propto\hat{b}^\dagger\hat{a}\hat{a}^\dagger$ can induce displacement upon acting on a vacuum (which is the initial state in $\ket{\varphi_\mathrm{I}}$) since $\hat{b}^\dagger\hat{a}\hat{a}^\dagger=\hat{b}^\dagger+\hat{b}^\dagger\hat{a}^\dagger\hat{a}$. To see this more clearly, we order operators in $\hat{H}_\mathrm{C}$ \eqref{eq:h-nl} in the normal order to obtain
\begin{align}
    \hat{H}_\mathrm{C}/\hbar=\underbrace{\frac{g}{2}\left\{\left(S^{*2}\hat{a}^2+2S^*C^*\hat{a}^\dagger\hat{a}+C^{*2}\hat{a}^{\dagger 2}\right)\hat{b}+\mathrm{h.c.}\right\}}_{\hat{H}_\mathrm{C}'}+\underbrace{\frac{g}{2}\left(S^*C^*\hat{b}+SC\hat{b}^\dagger\right)}_{\delta\hat{H}_\mathrm{L}},
\end{align}
where $\hat{H}_\mathrm{C}'$ is a normally-ordered nonlinear term, and the residual linear terms are separated as $\delta\hat{H}_\mathrm{L}$. This partition establishes another way of expressing the interaction-frame Hamiltonian
\begin{align}
    \hat{H}_\mathrm{I}=\hat{H}_\mathrm{C}+\hat{H}_\mathrm{L}=\hat{H}_\mathrm{C}'+\hat{H}_\mathrm{L}'
\end{align}
with a new linear term
\begin{align}
    \hat{H}_\mathrm{L}'/\hbar=(\hat{H}_\mathrm{L}+\delta\hat{H}_\mathrm{L})/\hbar=\left(\frac{g}{2}SC-\mathrm{i}\partial_t\beta\right)\hat{b}^\dagger+\left(\frac{g}{2}S^*C^*+\mathrm{i}\partial_t\beta^*\right)\hat{b}.
\end{align}
To maximally factor out the linear dynamics, we choose the time-dependence of $\beta$ so that $\hat{H}_\mathrm{L}'$ (instead of $\hat{H}_\mathrm{L}$) is canceled, which is possible by setting
\begin{align}
\label{eq:gif-beta}
    \partial_t\beta_\mathrm{nlin}=-\frac{\mathrm{i}g}{2}C_\mathrm{nlin}S_\mathrm{nlin},
\end{align}
and we refer to the unitary it defines as the nonlinear Gaussian model $\hat{U}_\mathrm{nlin}(t)$. Because of the right-hand side of \eqref{eq:gif-beta}, $\beta_\mathrm{nlin}$ varies as a function of time, accounting for pump-depletion dynamics. 

It is worth mentioning that the nonlinear Gaussian model has two explicitly conserved quantities
\begin{align}
    &|C_\mathrm{nlin}(t)|^2-|S_\mathrm{nlin}(t)|^2=1 &|\beta_\mathrm{nlin}(t)|^2+\frac{|S_\mathrm{nlin}(t)|^2}{2}=\beta_0^2,
\end{align}
where the second equality ensures that the Manley-Rowe invariant remains constant for the Gaussian-approximated state $\ket{\psi_\mathrm{nlin}'}=\hat{U}_\mathrm{nlin}\ket{0}$. By leveraging these conserved quantities, we can analytically solve the nonlinear differential equations \eqref{eq:mesoscopic-propagators-eom} and \eqref{eq:gif-beta} using Jacobi elliptic functions as
\begin{subequations}
\label{eq:full-eq-nonlinear-depleted}
\begin{align}
    \beta_\mathrm{nlin}&=\mathrm{i}\beta_0\,\mathrm{sn}(\tau+K(m),m)\\
    C_\mathrm{nlin}&=-\sqrt{2(\beta_0^2+1/2)}\,\mathrm{dn}(\tau+K(m),m)\\
    S_\mathrm{nlin}&=-\sqrt{2}\beta_0\,\mathrm{cn}(\tau+K(m),m)
\end{align}
\end{subequations}
where
\begin{align}
    &\tau=\sqrt{\beta_0^2+1/2}\,gt
\end{align}
is the scaled time, $\mathrm{sn}$ and $\mathrm{cn}$ are the elliptic sine and cosine functions, respectively, and $\mathrm{dn}$ is the delta amplitude function. They exhibit oscillation with a period $4K(m)$ in $\tau$ with parameter
\begin{align}
m=\frac{\beta_0^2}{\beta_0^2+1/2}
\end{align}
and the complete elliptic integral of the first kind
\begin{align}
    K(m)=\int_0^1\frac{\mathrm{d}z}{(1-z^2)(1-mz^2)}.
\end{align}
The Gaussian-approximated state $\ket{\psi_\mathrm{nlin}'(t)}$ predicts perfect pump depletion at a characteristic depletion time $t_\mathrm{dep}$ which we define such that $\beta_\mathrm{nlin}(t_\mathrm{dep})=0$. The value of $t_\mathrm{dep}$ as a function of initial pump field amplitude is given as
\begin{align}
    t_\mathrm{dep}=g^{-1}\frac{K(m)}{\beta_0^2+1/2}\approx \frac{1}{2g\beta_0}\log(32\beta_0^2).
\end{align}
The analytic expression for $t_\mathrm{dep}$ gives us an approximate scaling for the time over which we expect to see non-Gaussian quantum physics for a given pump field amplitude $\beta_0$.

In Fig.~\ref{fig:gif-combined}(a), we show the phase-space portraits of the full-quantum simulation shown in the lab frame and the GIF defined by the nonlinear Gaussian model, where we can see that the use of the GIF compresses quantum fluctuations around the origin. In Fig.~\ref{fig:gif-combined}(b), we compare the time-evolution of the pump depletion ratio for the linearized approximated state $\ket{\psi'_\mathrm{lin}(t)}$, Gaussian-approximated state with the nonlinear Gaussian model $\ket{\psi'_\mathrm{nlin}(t)}$, and the full-quantum state $\ket{\psi(t)}$. The Gaussian approximation under the linearized and nonlinear Gaussian models are characterized by \eqref{eq:full-eq-undepleted} and \eqref{eq:full-eq-nonlinear-depleted}, respectively. The full-quantum state $\ket{\psi(t)}$ is calculated using a standard quantum simulation package~\cite{Kraemer2018}, where the quantum dynamics are expanded in the photon-number basis. While the linearized approximation overestimates the pump field amplitude, the nonlinear Gaussian model predicts the amount of pump depletion with high accuracy. This allows us to use minimal Fock space size for the pump mode for the simulation in the interaction frame.

Finally, we point out that even a nonlinear Gaussian model cannot capture the full quantum state. Nonlinear terms in the interaction frame induce non-Gaussian quantum features that cannot be captured by Gaussian-approximated states, which are clearly seen in the phase-space portraits of the states shown in the figures. Furthermore, as shown in Fig.~\ref{fig:gif-combined}, such nonlinear quantum dynamics induce non-Gaussian entanglement between signal and pump modes, which cannot be captured by a Gaussian unitary. When the pump mode is traced out, signal-pump entanglement results in excess signal noise, limiting the level of attainable squeezing. To capture this emergent non-Gaussian quantum physics, we include the contribution from the interaction-frame state $\ket{\varphi_\mathrm{I}}$, which can be performed efficiently using the GIF.  

\subsection{Mesoscopic pulsed squeezing}\label{sec:mesoscale_sqeezing}

In the previous Sec. \ref{subsec:GIF}, we studied single-mode OPG and saw that a GIF can reduce requirements for the size of Fock-space, realizing an efficient numerical model. For the case of multimode systems, however, the numerical complexity can still be exponential even after an application of a GIF. Assume, for instance, that each of $M$ modes can be populated with either $0$ or $1$ photons, which is the smallest nontrivial Fock space size. Even for such a scenario, we would na\"ively need $2^M$ parameters to describe the system state. Therefore, to analyze multimode non-Gaussian quantum dynamics, we need an additional model-reduction technique to realize a tractable quantum model. To this goal, the structure of the GIF provides essential information on where and how non-Gaussian quantum features emerge, with which we can truncate a mode basis to realize a polynomial-size numerical model.

In this subsection, as a case study on multimode non-Gaussian quantum dynamics, we consider pulse-pumped broadband OPG. By inspecting the structure of the Hamiltonian in the GIF, we show that principal squeezing supermodes predominantly accumulate non-Gaussian quantum features during the dynamics. By approximating higher-order squeezing supermodes as being populated only with Gaussian states, we realize a concise model for these otherwise intractable system dynamics.

\subsubsection{Construction of a GIF}
We construct a GIF using the nonlinear Gaussian model for pulse-pumped OPG. Because the pump field can constitute a broadband excitation, the pump mean-field estimate takes a form $\beta_s(t)$. Below, we use the results presented in Sec.~\ref{sec:Gaussian-pulse-pumped-OPA} to derive the Gaussian dynamics of the system. For given $\beta_s(t)$, the Gaussian Hamiltonian takes the form 
\begin{align}
\hat{H}_\mathrm{G}/\hbar=\frac{r}{2}\iint\mathrm{d}s_1\mathrm{d}s_2\,\left(\beta_{s_1+s_2}^*\hat{a}_{s_1}\hat{a}_{s_2}+\beta_{s_1+s_2}\hat{a}_{s_1}^\dagger\hat{a}_{s_2}^\dagger\right)+\sum_{u\in\{a,b\}}\int\mathrm{d}s\,\hat{u}_s^\dagger \delta\omega_u(2\pi s)\hat{u}_s,
\end{align}
which induces a Gaussian unitary $\hat{G}$. The transformations of the field operators are given as multimode Bogoliubov transformations
\begin{subequations}
\begin{align}
&\hat{U}^\dagger\hat{a}_s\hat{U}=\int\mathrm{d}p\,\left(C_{sp}\hat{a}_p+S_{sp}\hat{a}_p^\dagger\right)\\ &\hat{U}^\dagger\hat{b}_s\hat{U}=e^{-\mathrm{i}\delta\omega_b(2\pi s)t}\hat{b}+\beta_s,
\end{align}
\end{subequations}
where the values of the propagators $C_{sp}$ and $S_{sp}$ are to be determined by solving their equations of motions
\begin{subequations}
\label{eq:mesoscopic-multimode-eom}
\begin{align}
    &\partial_tC_{sp}=-\mathrm{i}r\int\mathrm{d}q\,\beta_{s+q}S^*_{qp}-\mathrm{i}\delta\omega_a(2\pi s)C_{sp}\\
    &\partial_tS_{sp}=-\mathrm{i}r\int\mathrm{d}q\,\beta_{s+q}C^*_{qp}-\mathrm{i}\delta\omega_a(2\pi s)S_{sp}.
\end{align}
\end{subequations}

The remaining step to construct a GIF is to choose the dynamics of $\beta_s(t)$ so that mean-field dynamics are optimally accounted for. As in the single-mode case, the interaction-frame Hamiltonian $\hat{H}_\mathrm{I}$ can be decomposed as
\begin{align}
\hat{H}_\mathrm{I}=\hat{H}_\mathrm{C}'+\hat{H}_\mathrm{L}'
\end{align}
with
\begin{align}
    \hat{H}_\mathrm{L}'/\hbar=&\int\mathrm{d}s\,e^{\mathrm{i}\delta\omega_b(2\pi s)t}\left(-\mathrm{i}\partial_t\beta_s+\delta\omega_b(2\pi s)\beta_s+\frac{r}{2}\int\mathrm{d}p\mathrm{d}q\,C_{pq}S_{s-p,q}\right)\hat{b}_s^\dagger+\mathrm{h.c.}
\end{align}
The linear term $\hat{H}_\mathrm{L}'$ can be eliminated by setting the time dependence of $\beta_s$ as
\begin{align}
\label{eq:beta-multimode}
    \partial_t\beta_s=-\mathrm{i}\delta\omega_b(2\pi s)\beta_s-\frac{\mathrm{i}r}{2}\int\mathrm{d}p\mathrm{d}q\,C_{pq}S_{s-p,q},
\end{align}
which determines the value of $\beta_s$ (and consequently the entire $\hat{U}$). The Hamiltonian in the GIF becomes
\begin{align}
\label{eq:gif-hamiltonian-multimode}
\begin{split}
    \hat{H}_\mathrm{I}/\hbar=&\frac{r}{2}\int\mathrm{d}s\mathrm{d}s'\mathrm{d}p\mathrm{d}p'e^{\mathrm{i}\delta\omega_b(2\pi (s+s'))t}\hat{b}_{s+s'}^\dagger\\
    &\times\left(S_{sp}S_{s'p'}\hat{a}_{p}^\dagger\hat{a}_{p'}^\dagger+2S_{sp}C_{s'p'}\hat{a}_{p}^\dagger\hat{a}_{p'}+C_{sp}C_{s'p'}\hat{a}_{p}\hat{a}_{p'}\right)+\mathrm{h.c.}.
\end{split}
\end{align}

\subsubsection{Supermode expansion}
\begin{figure}[h]
    \centering
    \includegraphics[width=0.85\textwidth]{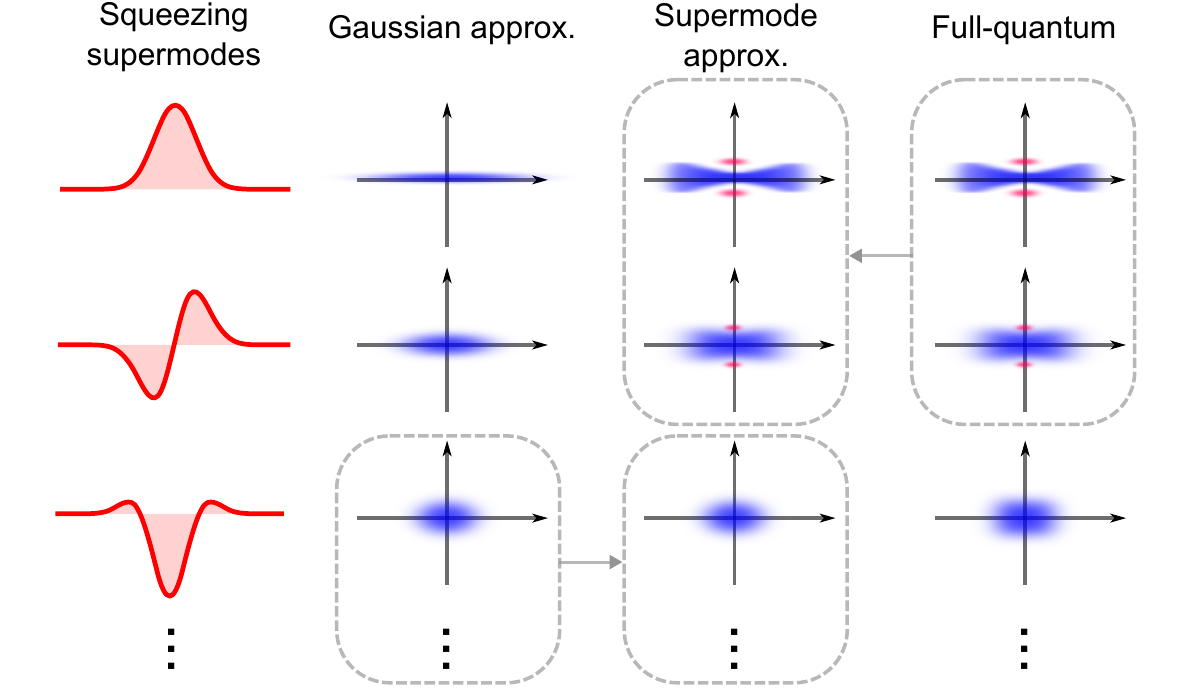}
    \caption{For an OPG pumped by a mesoscopic number of pump photons, non-Gaussian quantum features predominantly develop on canonical squeezing supermodes (and pump modes coupled to them) obtained by Gaussian approximations. In a supermode approximation, we include non-Gaussian quantum dynamics only on these principal supermodes to realize an efficient numerical model.}
    \label{fig:non-Gaussian-supermode}
\end{figure}

The GIF can factor out trivial quantum features to create a compressed state description. However, the Hamiltonian \eqref{eq:gif-hamiltonian-multimode} still involves nonlinear couplings among multiple modes and can na\"ively cause exponentially complicated dynamics. To realize a tractable quantum model, we identify a few principal supermodes, which are waveforms formed by combining frequency modes, that predominantly exhibit non-Gaussian quantum features, and the GIF is essential in providing information about these supermodes. In the case of pulsed OPG, these principal non-Gaussian supermodes coincide with canonical squeezing supermodes (and pump modes coupled to them), as shown in Fig.~\ref{fig:non-Gaussian-supermode}. Based on this knowledge, we can truncate the supermode basis to include only the primary non-Gaussian quantum features, enabling an efficient quantum model.

A generic supermode basis is characterized by field operators
\begin{align}
    &\hat{a}_m(t)=\int\mathrm{d}s\,A^*_{ms}(t)\hat{a}_s &\hat{b}_m(t)=\int\mathrm{d}s\,B^*_{ms}(t)\hat{b}_s.
\end{align}
The waveforms of the $m$th signal and pump supermodes are given as $A_{ms}$ and $B_{ms}$, respectively, as functions of the wavenumber $s$. These waveforms form respective complete bases and are normalized as $\int\mathrm{d}s\,C_{ms}^*C_{m's}=\delta_{m,m'}$ and $\sum_{m=0}^\infty C^*_{ms}C^*_{ms'}=\delta(s-s')$ with $C\in\{A,B\}$. Using this basis, the Hamiltonian in the GIF can be rewritten as
\begin{align}
    \hat{H}_\mathrm{I}/\hbar=\frac{r}{2}\sum_{\ell mn}^\infty\left(\hat{b}_\ell^\dagger \left(\mu_{\ell mn}\hat{a}_m^\dagger\hat{a}_n^\dagger+\nu_{\ell mn}\hat{a}_m^\dagger\hat{a}_n+\xi_{\ell mn}\hat{a}_m\hat{a}_n\right)\right)+\mathrm{h.c.},
\end{align}
where the nonlinear coupling tensors are given as
\begin{subequations}
\begin{align}
\label{eq:mesoscopic-coupling-mu}
    \mu_{\ell mn}&=\int\mathrm{d}s\mathrm{d}s'\mathrm{d}p\mathrm{d}p'\,e^{\mathrm{i}\delta \omega_b(2\pi (s+s'))t}S_{sp}S_{s'p'}B^*_{\ell,(s+s')}A^*_{mp}A^*_{np'}\\
    \nu_{\ell mn}&=\int\mathrm{d}s\mathrm{d}s'\mathrm{d}p\mathrm{d}p'\,e^{\mathrm{i}\delta \omega_b(2\pi (s+s'))t}S_{sp}C_{s'p'}B^*_{\ell,(s+s')}A^*_{mp}A_{np'}\\
    \xi_{\ell mn}&=\int\mathrm{d}s\mathrm{d}s'\mathrm{d}p\mathrm{d}p'\,e^{\mathrm{i}\delta \omega_b(2\pi (s+s'))t}C_{sp}C_{s'p'}B^*_{\ell,(s+s')}A_{mp}A_{np'}.
\end{align}
\end{subequations}
Since the supermode basis itself depends on time, the dynamics in this new basis must include the non-inertia term in the Hamiltonian as
\begin{align}
\label{eq:mesoscopic-effective-h}
\hat{H}_\mathrm{eff}=\hat{H}_\mathrm{I}+\hat{H}_\mathrm{non{\text -}inertia},
\end{align}
with
\begin{align}
    \hat{H}_\mathrm{non{\text -}inertia}/\hbar=\sum_{C\in\{A,B\}}\sum_{m,n}\int\mathrm{d}s\,(\mathrm{i}C_{ns}\partial_tC^*_{ms})\hat{c}_m^\dagger\hat{c}_n.
\end{align}

To identify principal supermodes that predominantly experience non-Gussian quantum dynamics, we note that any term in $\hat{H}_\mathrm{I}$ that contains an annihilation operator does not affect the state to the lowest order because the initial state in the GIF is a vacuum. As a result, the terms of the form $\propto\hat{B}_\ell^\dagger\hat{A}_m^\dagger\hat{A}_n^\dagger$ crucially determine the onset of non-Gaussian quantum physics, and we inspect their coefficients $\mu_{\ell mn}$ to unravel their structures.

As shown in \eqref{eq:mesoscopic-coupling-mu}, the coupling tensor $\mu_{\ell mn}$ comprises propagator $S_{sp}$, which we can be decomposed as
\begin{align}
\label{eq:svd-S}
S_{sp}=\sum_{m=0}^\infty A^{\mathrm{out}}_{ms}\sinh\lambda_m\,A_{mp}^{\mathrm{in}}
\end{align}
(see Sec.~\ref{sec:Gaussian-pulse-pumped-OPA} for full discussions). Intuitively, $A_{mp}^\mathrm{in}$ represents the signal spectral waveform that is coupled to the pump mode via the integration $\int\mathrm{d}q\,S_{sp}A_{mp}^*$ with an amplitude $\sinh \lambda_m$. Therefore, a reasonable way to choose principal signal supermodes is to set $A_{mp}=A_{mp}^\mathrm{in}$. With this choice of signal supermodes, the coupling tensor takes the form
\begin{align}
    \mu_{\ell mn}=\int\mathrm{d}s\,\mathcal{I}_{mn,s}B^*_{\ell,s}
\end{align}
with 
\begin{align}
\label{eq:I_mn}
\mathcal{I}_{mn,s}=\sinh\lambda_m\sinh\lambda_n \int\mathrm{d}q\,e^{\mathrm{i}\delta \omega_b(2\pi s)t}A_{ms}^\mathrm{out}A_{n,(q-s)}^\mathrm{out},
\end{align}
implying that $m$th and $n$th signal supermodes are coupled to pump waveform $\propto \mathcal{I}_{mn,s}$.

Based on these discussions, our prescription for choosing $m_a$ signal principal supermodes and corresponding pump principal supermodes is shown below.
\begin{enumerate}
    \item We choose signal principal supermodes as the input squeezing supermodes of the GIF, according to \eqref{eq:svd-S}, \textit{i.e.}, $A_{ms}=A^\mathrm{in}_{ms}$, for a given mode number truncation $m\leq m_a$.
    \item We calculate pump waveforms $\mathcal{I}_\mathrm{mn,s}$, according to \eqref{eq:I_mn}, for $m,n\leq m_a$. Note that some of $\mathcal{I}_\mathrm{mn,s}$ can vanish due to the symmetry of signal supermodes. As a result, we have at most $m_a(m_a+1)/2$ independent non-zero waveforms.
    \item We perform orthonormalization on the non-zero waveforms to obtain principal pump supermodes.
\end{enumerate}
For numerical simulations, we simply ignore terms consisting of higher-order supermodes in \eqref{eq:mesoscopic-effective-h} to calculate an approximate GIF state $\ket{\varphi_\mathrm{I}'}$, and the lab-frame state becomes $\hat{D}\hat{G}\ket{\varphi_\mathrm{I}'}$. Results of numerical simulations can be found, \textit{e.g.}, in Ref.~\cite{Yanagimoto2022-non-Gaussian}.

\subsection{Dynamical moment expansion and Gaussian split-step Fourier methods} \label{sec:GSSF}

As we have seen, the GIF framework is useful for obtaining optimal Gaussian approximations of nonlinear dynamics, as well as for model reduction in mesoscopic systems.
Given this general utility of the method, we would like to be able to derive the equivalent of Eqns.~\eqref{eq:gif-beta} and \eqref{eq:mesoscopic-propagators-eom}, but for any arbitrary nonlinear optical system.
The method presented in Sec.~\ref{sec:nonlinear-gaussian-model} involves finding exact cancellations of particular low-order terms in the interaction-frame Hamiltonian, which often requires heavy operator algebra.
While the approach is general in nature, it can become rather involved for more complicated scenarios, such as seeded OPA, saturated SHG, or even multimode four-wave-mixing (\textit{e.g.}, under a Kerr effect or an optical cascade).

In these situations, it can be helpful to turn to other means for obtaining Gaussian coordinates for the GIF (namely, the equivalent of $\beta$, $C$, and $S$).
Obviously, the mean-field limit and the linearized approach are both examples of simpler Gaussian approximations that do not require operator manipulation, but as we have seen in this section and the last, they are suboptimal in that they neglect certain Gaussian phenomena, thus defining a GIF that still has residual Gaussian dynamics in the interaction-frame Hamiltonian.

Here, we briefly review an alternative approach, which we call \emph{dynamical moment expansion}, that can be used to generate equations of motion for the Gaussian coordinates that can be used to define a high-quality (in many cases, optimal) Gaussian interaction frame.

The method is based on the idea that associated with each value of $(\alpha,C,S)$ in a GIF is a pure Gaussian state that is given by applying the propagator $(C,S)$ to the vacuum and displacing the result by $\alpha$.
Thus, we can instead simply work on generating a good Gaussian-state approximation to the dynamics.
For an $M$-mode state, a Gaussian state $\ket{G}$ is fully specified by a vector of $M$ mean fields $\alpha_i = \braket{G|\hat a_i|G}$ and two $M \times M$ covariance matrices $\Sigma_{ij} = \braket{G | \delta\hat a_i \delta\hat a_j | G}$ and $\Pi_{ij} = \braket{G | \delta\hat a_i^\dagger \delta\hat a_j | G}$.
Conventionally, in Gaussian quantum optics, we use one covariance matrix of size $2M \times 2M$ written in terms of the quadrature operators, but converting between the conventions is straightforward, and this form will prove more convenient for us. Furthermore, for a pure Gaussian state, the relation between $\Sigma$ and $\Pi$ with $C$ and $S$ are also easy to derive.

The moment expansion technique attempts to calculate the time evolution of $\alpha_i$, $\Sigma_{ij}$, and $\Pi_{ij}$ starting from the quantized coupled-wave equations, while making the sole assumption that \emph{the quantum state is Gaussian at all times}~\cite{Schack1990}.
Note that this does not mean we lose access to all non-Gaussian effects from this point onward; this assumption is merely useful for deriving the GIF, under which non-Gaussian physics can be analyzed later.

We illustrate the dynamical moment expansion technique by an example, from which it should become clear the technique works for any system for which we have quantized coupled-wave equations.
Consider the single-mode OPG Hamiltonian in Eq.~\eqref{eq:single-mode-opg}, which produces the coupled-wave equations
\begin{align}
    \partial_t \hat a &= -\mathrm{i} g \hat a^\dagger \hat b &
    \partial_t \hat b &= -\frac{\mathrm{i}g}{2} \hat a^2.
\end{align}
We start by computing
\begin{subequations} \label{eq:moment-expansion-0}
\begin{align}
    \partial_t \braket{\hat a} &= -\mathrm{i} g \left(
    \braket{\hat a^\dagger}\braket{\hat b} + \braket{\delta \hat a^\dagger \delta \hat b}
    \right) \\
    \partial_t \braket{\hat b} &= -\frac{\mathrm{i}g}{2} \left(
    \braket{\hat a}^2 + \braket{\delta\hat a^2}
    \right),
\end{align}
\end{subequations}
where we have used the fact that $\braket{\delta\hat a} = \braket{\delta\hat b} = 0$ in general.
Hence we see that the mean fields are driven by the covariances.

To obtain the evolution of the covariances, we first calculate
\begin{subequations} \label{eq:moment-expansion-1}
\begin{align}
    \partial_t \delta\hat a &= -\mathrm{i} g \left[
    \left(\delta\hat a^\dagger \delta\hat b - \braket{\delta\hat a^\dagger \delta\hat b}\right) + \braket{\hat b} \delta\hat a^\dagger + \braket{\hat a^\dagger} \delta\hat b
    \right] \\
    \partial_t \delta\hat b &= -\frac{\mathrm{i}g}{2} \left[ \left(\delta\hat a^2 - \braket{\delta\hat a^2} \right) + 2 \braket{\hat a} \delta\hat a \right].
\end{align}
\end{subequations}
These expressions allow us to calculate the following equations of motion for the covariances by employing (i) the product rule $\partial_t(\hat z_1 \hat z_2) = \hat z_1 (\partial_t\hat z_2) + (\partial_t \hat z_1) \hat z_2$, and (ii) the fact that \emph{for Gaussian states}, third-order central moments such as $\braket{\delta\hat a^\dagger \delta\hat a \delta\hat b} = 0$.
(As an algebra hint, note that in doing so, we can effectively ignore the first inner-bracketed terms of Eqn.~\eqref{eq:moment-expansion-1}, since they both end up contributing odd-order central moments.)
\begin{subequations} \label{eq:moment-expansion-2}
\begin{align}
    \partial_t \braket{\delta\hat a^2} &= -\mathrm{i} g \left[
    \braket{\hat b} \left(
    \braket{\delta\hat a \delta\hat a^\dagger} + \braket{\delta\hat a^\dagger \delta\hat a}
    \right) + 2\braket{\hat a^\dagger} \braket{\delta\hat a \delta\hat b}
    \right] \\
    \partial_t \braket{\delta\hat a^\dagger\delta\hat a} &= -\mathrm{i} g \left(
    \braket{\hat b^\dagger} \braket{\delta\hat a^2} + \braket{\hat a} \braket{\delta\hat a \delta\hat b^\dagger}
    \right) + \text{H.c.} \\
    \partial_t \braket{\delta\hat a \delta\hat b} &= -\mathrm{i}g \left[
    \braket{\hat a} \braket{\delta\hat a^2} + \braket{\hat a^\dagger} \braket{\delta\hat b^2} + \braket{\hat b} \braket{\delta\hat a^\dagger \delta\hat b}
    \right] \\
    \partial_t \braket{\delta\hat a^\dagger \delta\hat b} &= \mathrm{i} g \left[ 
    \braket{\hat b^\dagger} \braket{\delta\hat a \delta\hat b} + \braket{\hat a} \braket{\delta\hat b^\dagger \delta\hat b} - \braket{\hat a} \braket{\delta\hat a^\dagger \delta\hat a} \right] \\
    \partial_t \braket{\delta\hat b^2} &= -2\mathrm{i} g \braket{\hat a} \braket{\delta\hat a \delta\hat b} \\
    \partial_t \braket{\delta\hat b^\dagger \delta\hat b} &= \mathrm{i} g \braket{\hat a^\dagger} \braket{\delta\hat a^\dagger \delta\hat b} + \text{H.c.}
\end{align}
\end{subequations}
The result is that we now have a set of \emph{Gaussian coupled-wave equations}, \textit{i.e.}, Eqns.~\eqref{eq:moment-expansion-0} and \eqref{eq:moment-expansion-2}, where instead of just the equations of motion for two mean fields, we also have six additional equations for the covariances; all eight are in general coupled dynamically.

While the above might also seem like a good deal of algebraic manipulation, it is worth noting the remarkable generality of the above equations: Whereas the calculations done earlier in this section applied only to the case of vacuum squeezing, these equations describe generic, degenerate three-wave-mixing between single-mode pump and signal, which includes for example saturated SHG.
In fact, we can easily obtain the special case considered previously by noting that for vacuum squeezing, we have $\braket{\hat a} = \braket{\delta\hat a \delta\hat b} = \braket{\delta\hat a^\dagger \delta\hat b} = \braket{\delta\hat b^2} = \braket{\delta\hat b^\dagger \delta\hat b} = 0$ for all time.
In this case, we have the much simpler equations
\begin{align}
    \partial_t \beta &= -\frac{\mathrm{i}g}{2} \Sigma_{aa}, &
    \partial_t \Sigma_{aa} &= -\mathrm{i}g \beta \left( 2\Pi_{aa} + 1 \right), &
    \partial_t \Pi_{aa} &= -\mathrm{i} g \beta^* \Sigma_{aa} + \text{H.c.},
\end{align}
where we recall $\beta = \braket{\hat b}$, $\Sigma_{aa} = \braket{\delta\hat a^2}$ and $\Pi_{aa} = \braket{\delta\hat a^\dagger \delta\hat a}$.
It can be shown that these equations are equivalent to Eqns.~\eqref{eq:gif-beta} and \eqref{eq:mesoscopic-propagators-eom} and they result in the same Gaussian interaction frame.

Because this dynamical moment expansion results in equations of motion that are functionally nothing more than generalized coupled-wave equations, it follows that the method is manifestly compatible with split-step Fourier (SSF) methods for treating multimode pulse propagation.
Specifically, we need only derive the moment expansion equations for the nonlinear step of the SSF method, which follows exactly the same procedure as above, but for an $M$-mode envelope, we now have covariance matrices that are $M \times M$, whose elements are all coupled to one another (and to the mean vectors).
Numerically, one then needs to Fourier transform these matrices to the frequency domain using 2D FFTs (keeping in mind that $\braket{\delta\hat a_i^\dagger \delta\hat a_j}$ transforms differently across its dimensions, \textit{e.g.}, forward FFT in the columns and backward FFT in the rows, or vice versa).
The linear step in the frequency domain is then the simple application of a scalar multiplication (representing phase shift and loss), as in the usual SSF method.
In total, the time needed to compute a step of this \emph{Gaussian SSF (GSSF) method} goes as $O(M^2\log M)$, compared to the cost of $O(M\log M)$ for classical SSF, which is in a sense optimal given that we are explicitly tracking the quantum correlations among $M$ modes up to second order.

\begin{figure}[h]
    \centering
    \includegraphics[width=\textwidth]{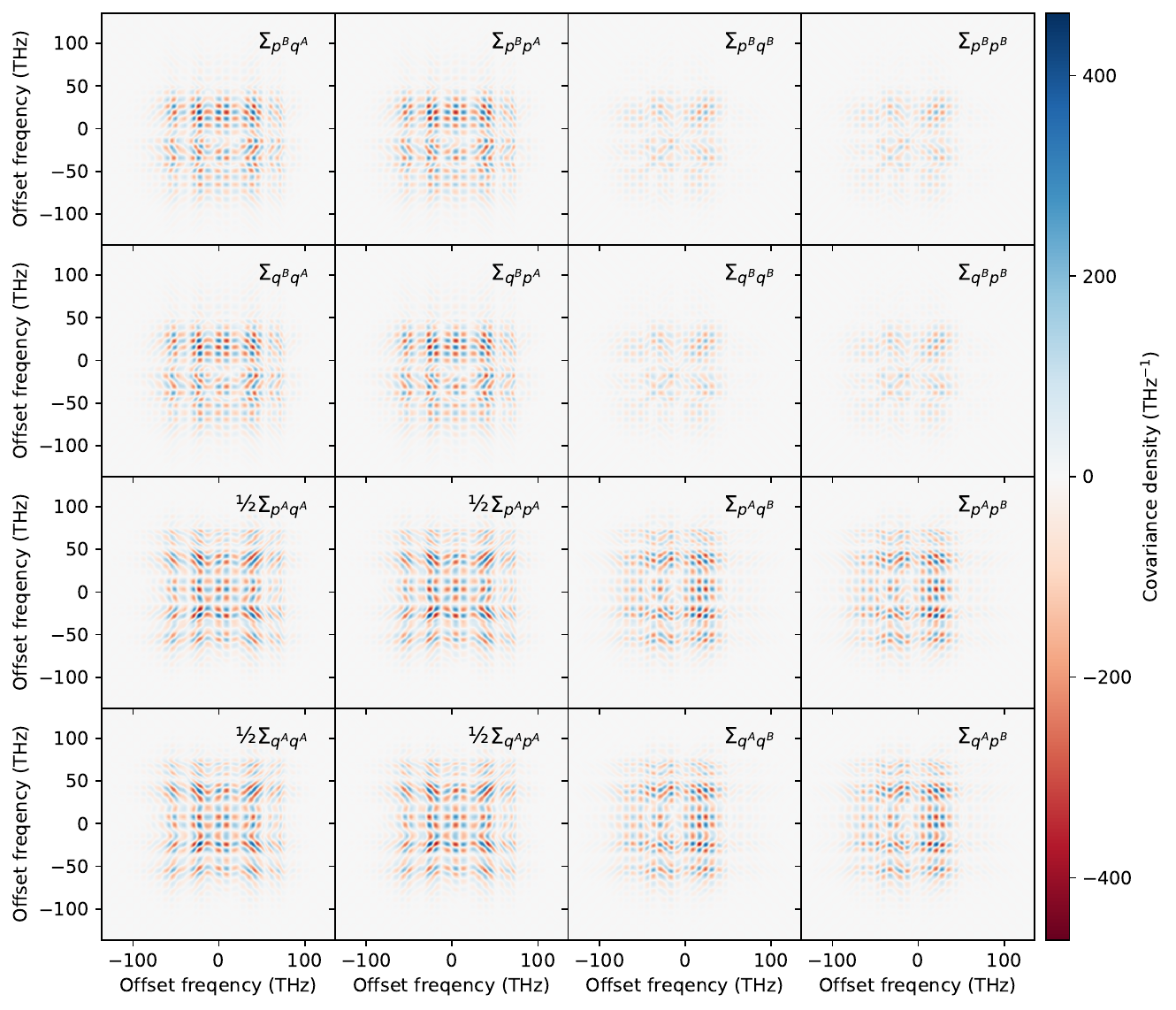}
    \caption{Gaussian-state approximation to the semiclassical dynamics of supercontinuum generation based on quasi-static SHG (see Sec.~\ref{sec:quasi-static-shg}). The full multimode Gaussian state is represented here by its covariance matrix in the frequency domain, expressed in terms of the quadratures $\hat q^A$ and $\hat p^A$ of the FH and $\hat q^B$ and $\hat p^B$ of the SH. The constant $\frac12 \delta$-function diagonal of the covariance matrix due to the contribution from vacuum is not shown, but units are chosen such that the integral over frequency of the diagonal (minus that of the vacuum) equals the total number of fluorescence (\textit{i.e.}, non-mean-field or $\braket{\delta\hat a_i^\dagger \delta\hat a_i}$) photons. Note that some blocks of the covariance matrix are halved for better contrast.}
    \label{fig:scg-covariance}
\end{figure}

Reference~\cite{Ng2023} provides a detailed derivation of the GSSF method with explicit expressions for the multimode nonlinear steps, in both $\chi^{(2)}$ and $\chi^{(3)}$ systems. (For the latter, it is particularly difficult to derive optimal GIFs using the operator approach due to the higher order of the $\chi^{(3)}$ Hamiltonian.)
Using the GSSF method, Ref.~\cite{Ng2023} studies multimode quantum noise dynamics in Kerr solitons, saturated degenerate OPG, and SHG-based supercontinuum generation.
As an illustration of the method here, Fig.~\ref{fig:scg-covariance} shows the full covariance matrices of the $\chi^{(2)}$ SCG discussed in Sec.~\ref{sec:quasi-static-shg}.
It is worth noting that even with a relatively na\"ive implementation RK4IP~\cite{Hult2007} SSF implementation on an Ampere A100 GPU, this simulation requires $<5$ seconds of compute using $2^{10}$ points per envelope.
A software package implementing GSSF in Julia for GPU deployment (via CUDA.jl~\cite{CUDAjl}) is available as GaussianSSF.jl~\cite{GaussianSSGjl}.

Finally, we note that moment expansion can be generalized to higher-order correlations, realizing systematic approximations of non-Gaussian quantum physics~\cite{Huang2022,Plankensteiner2021}. For instance, in Ref.~\cite{Xing2023}, they show higher-order correlations can capture non-Gaussian signal-pump entanglement in pump-depleted dynamics of an OPA. However, while such higher-order expansion techniques have been effective in studying systems with a handful modes, application to highly multimode systems can become progressively difficult as the order of expansion increases. Generally, $\mathcal{O}(M^n)$ parameters are required to describe $n$th order correlation among $M$ modes. Thus, to capture next-order (\textit{i.e.}, cubic) correlations in the SCG shown in Fig.~\ref{fig:scg-covariance}, we need to keep track of $(2\times 2^{10})^3\sim10^{10}$ parameters, which is possible but numerically demanding. Thus, to efficiently capture multimode, higher-order correlations, one should also pursueadditional model-reduction steps, \textit{e.g.}, supermode truncation.

\section{The deep-quantum regime}\label{sec:quantum_NLO}

\subsection{Introduction}
In the limit of strong nonlinearity, we enter the regime of deep-quantum nonlinear optics, where even a microscopic (i.e., an order of unity) number of photons can trigger saturated optical dynamics. In this limit, the hierarchy among mean-field, Gaussian quantum, and non-Gaussian quantum features collapse, and the photon-number (Fock) basis becomes the natural representation of the light. Highly quantum photon dynamics in this regime can defy classical intuition, and experimental hallmarks of such physics include photon blockade~\cite{Birnbaum2005}, quantum Rabi oscillation~\cite{Brune1996quantum}, and formation of photon bound states~\cite{Darrick2014, Firstenberg2013, Cantu2020}. We will see throughout this section that compared to traditional cavity QED systems, nonlinear optical systems in the deeply quantum limit can host an immense number of modes alongside strong photon-photon coupling, leading to rich but complicated phenomenology. A recurring theme is that multimode coupling can act as a decoherence channel and that quasi-static devices tend to operate in a regime of large nonlinearity and high decoherence. 

A representative case study is given in optical parametric interactions in the deep-quantum limit, \textit{i.e.}, single-photon-pumped parametric downconversion (PDC). We can obtain a number of crucial insights from this analysis: i) contrary to the classical and semiclassical limits, the pulse shape of the pump does not enhance the gain or fluorescence rate of the fundamental. Instead, only the fluorescence \emph{bandwidth} contributes to enhancing the effective coupling rate. ii) The scaling laws for this effective coupling rate differs radically from the usual coupling rates for single-mode dynamics. In particular, we find that the effective coupling rate for multimode parametric fluorescence is invariant with respect to the size of a physical resonator. In contrast, the coupling rate of a single-mode resonator grows monotonically with the cavity-free spectral range. iii) We observe Rabi-like oscillations, including \SI{100}{\percent} downconversion from a single-photon pump to a biphoton of fundamental. The behavior of these oscillation dynamics is fundamentally different from that of single-mode vacuum Rabi oscillations, and the contrast of these oscillations decays during propagation without any loss. This decaying fringe contrast arises from the multimode nature of the nonlinear interaction, and is the first instance in this section where the multimode dynamics can be seen to act as a decoherence channel.

The rest of the section is structured as follows. In Sec.~\ref{sec:PDC}, we introduce the physics of single-photon-pumped PDC. We start from a basic single-mode case and extend the analysis to a more complicated broadband case. The contrast between the two instances highlights the unique physics arising from multimode interactions. Analytic solutions for the broadband dynamics are obtained by establishing an analogy between single-photon PDC and atomic autoionization~\cite{Fano1961,Ryo2020Fano}. In Sec.~\ref{sec:MPS}, we introduce a model reduction technique based on matrix product state (MPS) as a tool to realize efficient numerical simulations of general photon dynamics in this deep-quantum limit. We show how the involvement of multiple pump photons in PDC leads to deviation from the case of a single-photon pump, highlighting the utility of the MPS to interpolate between the microscopic and mesoscopic regimes. In Sec.~\ref{sec:decoherence}, we take a deep dive into the phenomenology of multimode decoherence, for which we use nonlinear-optical quantum gate operation as a touchstone. Finally, in Sec.~\ref{sec:temporal_trapping}, we introduce temporal trapping to circumvent multimode decoherence effects, enabling high-fidelity quantum gate operations using nonlinear optics.


\subsection{Single-photon-pumped PDC}
\label{sec:PDC}

\subsubsection{Introduction}
\begin{figure}[tbh]
    \centering
\includegraphics[width=\textwidth]{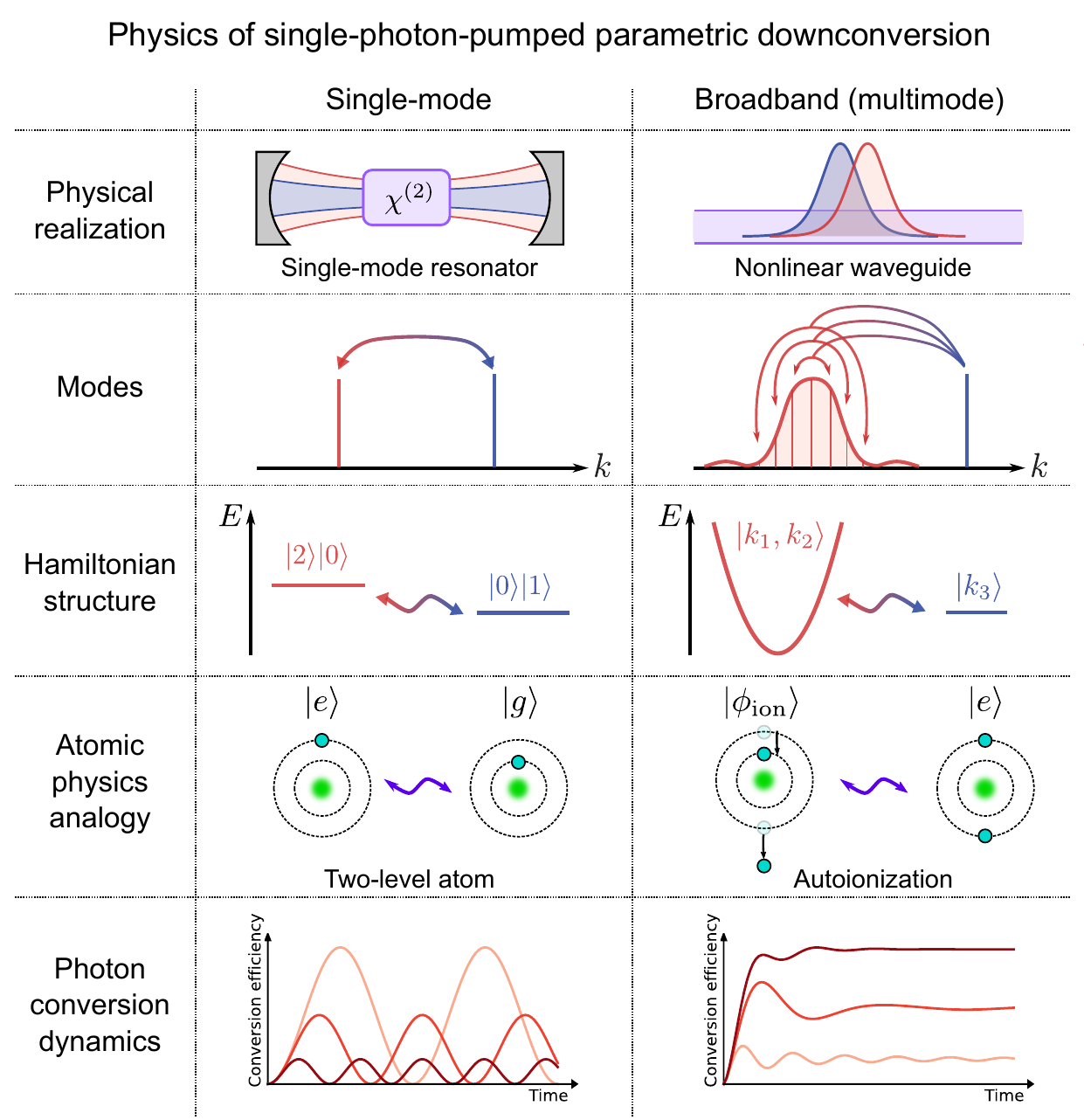}
    \caption{Illustrations of the physics of single-photon-pumped PDC in a single-mode resonator and broadband waveguide. Figure is adapted from Ref.~\cite{Yanagimoto2023-thesis}.}
    \label{fig:fano-schematics}
\end{figure}

In the previous sections, we have studied the physics of parametric interactions among macroscopic (Sec.~\ref{sec:semi-classical_NLO}) and mesoscopic (Sec.~\ref{sec:mesoscopic-quantum-nonlinear-optics}) numbers of photons. As the nonlinear coupling strength increases further, even fewer pump photons suffice to trigger a strong nonlinear optical response. In this limit of the deep-quantum regime, even a \emph{single} pump photon can downconvert to a pair of signal photons with high efficiency, which we refer to as single-photon-pumped PDC. The phenomenology of single-photon-pumped PDC serves as an insightful case study providing essential insights into what it means for nonlinear optics to be ``quantum.''

To understand the physics of broadband parametric downconversion, it is insightful to sketch out the comparison to single-mode PDC as illustrated in Fig.~\ref{fig:fano-schematics}. For single-mode PDC, \textit{e.g.}, inside a resonator containing mode spacings comparable to the optical frequencies, the large energy gaps between modes allow one to isolate a pair of signal and pump modes, which interact via the $\chi^{(2)}$ nonlinear interaction. The Hamiltonian that couples an initial single-photon pump state $\ket{0\,1}$ to a two-photon signal state $\ket{2\,0}$ has the same structure as that of a driven two-level atom. As a result, the photon-conversion dynamics between the signal and pump exhibit clear Rabi oscillations.

On the other hand, for broadband PDC, \textit{e.g.}, in a dispersion-engineered waveguide or a highly multimode resonator, a pump photon with momentum $k_3$ can downconvert to a signal photon pair with momentum $k_1$ and $k_2$ as long as momentum conservation is fulfilled $k_1+k_2=k_3$. This extra degree of freedom, where a single-mode pump can couple to many distinct signal and idler pairs, leads to a characteristic structure of the Hamiltonian where a discrete pump state is coupled to a continuum of signal states. Such discrete-continuum interactions can be found in atomic autoionization, where an atomic excited state is coupled to a continuum of ionized states. By coupling to a continuum of signal and idler, the photon conversion dynamics no longer show clear Rabi oscillation. Instead, the dephasing induced by the continuum leads to characteristic damped oscillations. This unique behavior of multimode single-photon-pumped PDC was originally discovered in Ref.~\cite{Antonosyan2014}, followed by an analytic study~\cite{Ryo2020Fano}, and a conception of unit-efficiency photon-pair generation~\cite{Solntsev2022}.

The physics of broadband PDC in the deep-quantum limit cannot be recast in terms of independently squeezed supermodes (see Sec.~\ref{sec:semi-classical_NLO}), which is in stark contrast to the semiclassical parametric interactions (\textit{e.g.}, PDC). Intuitively, this complication is due to the creation of a single signal photon pair in single-photon-pumped PDC completely depleting the pump photon, which suppresses further downconversion from happening. As a result, all the signal modes effectively get coupled to each other via the pump depletion, and we observe inherently multimode physics with no single-mode analog. In the rest of this subsection, we discuss the physics of single-photon-pumped PDC in more detail. Since the mathematical treatments are highly involved, we keep the discussions high-level in the main and provide the full derivations in Appendix~\ref{sec:single-photon-pumped-PDC-appendix}.

\subsubsection{Single-mode PDC}

\begin{figure}
    \centering
    \includegraphics[width=\textwidth]{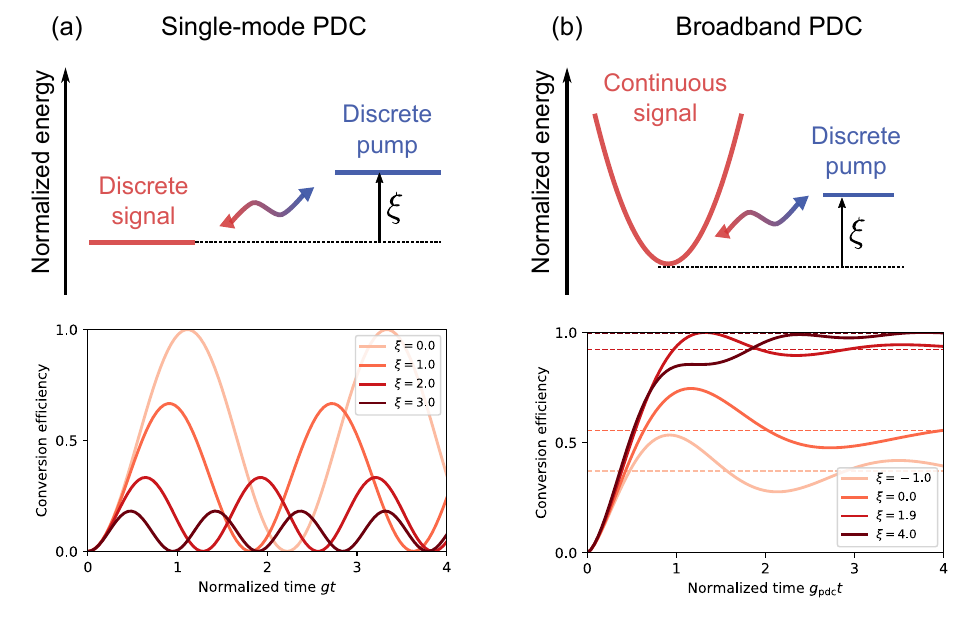}
    \caption{The single-photon-pumped PDC physics in a (a) single-mode system and (b) broadband system. Top row: Illustration for the energies of the states involved in the single-photon-pumped PDC dynamics. Bottom row: Conversion efficiencies of PDC for various dimensionless normalized phase-mismatch $\xi$ as defined in \eqref{eq:xi-definition-single-mode} and \eqref{eq:definition-xi} for single-mode and broadband PDC, respectively. The bottom figures are adapted from Ref.~\cite{Yanagimoto2023-thesis}.}
    \label{fig:fano-summary}
\end{figure}

In a single-mode PDC, the single-photon pump state is coupled to the two-photon signal state with the coupling strength $g$, which leads to a Rabi oscillation between these two states. A full derivations of these behaviors is provided in Appendix~\ref{sec:single-mode-pdc-appendix}. As is well known from the physics of two-level systems, the overall behavior of the Rabi oscillation is characterized by the ratio between the coupling and the energy difference between the energy levels involved, which for the case of the PDC is the dimensionless normalized phase mismatch $\xi$ (as defined in Eqn.\eqref{eq:xi-definition-single-mode}; see Fig.~\ref{fig:fano-summary}(a) for illustration). Intuitively, PDC is most efficient when the phase-matching condition is met, \textit{i.e.}, $\xi=0$, while large $\xi$ indicates the interaction is phase-mismatched, and the conversion efficiency is limited. We can derive closed-form solutions for the eigenstates, with which the PDC conversion efficiency can be derived as (see Eqn.\eqref{eq:conversion-single-mode-PDC})
\begin{align}
\label{eq:pdc-conversion-efficiency-single-mode}
\mathcal{E}(\tau)=\frac{2}{\xi^2+2}\sin^2\left(\frac{1}{2}\sqrt{\xi^2+2}\,\tau\right),
\end{align}
where $\tau=gt$ is the normalized time set by the nonlinear coupling rate $g$. In Fig.~\ref{fig:fano-summary}(a), we show $\mathcal{E}$ for various values of normalized phase-mismatch $\xi$. As expected, the conversion efficiency $\mathcal{E}$ exhibits sinusoidal oscillations that reach unit efficiency for $\xi=0$.

\subsubsection{Single-photon-pumped broadband PDC}
In this subsection, we consider broadband PDC pumped by a single pump photon, and we provide the full details in Appendices \ref{sec:Hamiltonian-structure-appendix}, \ref{sec:fano-broadband-eigen}, \ref{sec:fano-pump-photon-dynamics} and \ref{sec:fano-signal-photon-dynamics}. Via the $\chi^{(2)}$ nonlinear interactions, the single-photon pump state with wavenumber $k_3$ is coupled to all the two-photon signal states that fulfill momentum conservation, \textit{i.e.}, $k_1+k_2=k_3$, where $k_1$ and $k_2$ are wavenumbers of the signal photons. Consequently, the Hamiltonian takes a characteristic structure, where the discrete pump state is coupled to the continuum of signal states (copied from Appendix.~\ref{sec:Hamiltonian-structure-appendix}),
\begin{align*}
   \hat{H}/\hbar=\sum_{u\in\{a,b\}}\int\mathrm{d}s\,\hat{u}_s^\dagger \delta\omega_u(2\pi s)\hat{u}_s+\frac{r}{2}\iint\mathrm{d}s_1\mathrm{d}s_2\,\left(\hat{a}_{s_1}\hat{a}_{s_2}\hat{b}_{s_1+s_2}^\dagger+\hat{a}_{s_1}^\dagger\hat{a}_{s_2}^\dagger\hat{b}_{s_1+s_2}\right).
\end{align*}
When the dispersion $\delta\omega_a$ is truncated to second order, we can normalize the interaction time to a characteristic coupling rate $g_\text{pdc}$, and the spatial extend of the envelope to a characteristic correlation length set by the OPA bandwidth. Once the equations of motion are nondimensionalized, the normalized phase-mismatch $\xi$ determines the qualitative feature of the PDC dynamics. We define a normalized time $\tau=g_\text{pdc}t$ using the characteristic PDC rate
\begin{equation}
    g_\mathrm{pdc}=(r^4/4\pi^2\delta\omega_a''(0))^{1/3}.
\end{equation}
Physically, $\xi$ represents the normalized energy offset between the pump state and the bottom of the signal continuum (see Fig.~\ref{fig:fano-summary}(b)). While the Hamiltonian involves multimode interactions, we can derive analytic expressions for the eigenstates using Fano's theory for discrete-continuum interaction~\cite{Fano1961}, where we find one discrete photon-bound-state solution and a continuum of eigenstates (see Appendix~\ref{sec:fano-broadband-eigen} for the full derivations). This discrete bound state was first identified in Ref.~\cite{Drummond1997}, where it was referred to as an optical optical meson. With these solutions, an analytic expression for the PDC conversion efficiency is derived as
\begin{align}
\label{eq:pdc-conversion-efficiency-broadband}
    \mathcal{E}(\tau)=1-\Biggl\vert \underbrace{\left(1+\frac{\pi}{4\bar{\lambda}^{3/2}_\xi}\right)^{-1}}_{\mathrm{optical~meson~contribution}}+\int_0^\infty\mathrm{d}\lambda\,\underbrace{\frac{2\sqrt{\lambda}\,e^{-\mathrm{i}(\lambda+\bar{\lambda}_\xi)\tau}}{\pi^2+4\lambda(\lambda-\xi)^2}}_{\mathrm{continuum~contribution}}\Biggr\vert^2,
\end{align}
where we indicate contributions from the optical meson and the continuum of eigenstates. Here, $\bar{\lambda}_\xi$ is the binding energy of the optical meson and is given as the solution of the equation
\begin{align}
\bar{\lambda}_\xi=-\xi+\frac{\pi}{2\sqrt{\bar{\lambda}_\xi}}.
\end{align}

As is apparent from the expression \eqref{eq:pdc-conversion-efficiency-broadband}, the broadband photon conversion dynamics are much more complicated than the sinusoidal oscillations we observe in the single-mode case \eqref{eq:pdc-conversion-efficiency-single-mode}. In Fig.~\ref{fig:fano-summary}(b), we show the time evolution of the conversion efficiency $\mathcal{E}$ for various $\xi$, where we observe Rabi-like oscillations with decaying amplitudes. Interestingly, the asymptotic conversion efficiency $\mathcal{E}(\tau\rightarrow\infty)$ gets higher as the value of phase-mismatch $\xi$ \emph{increases}, which is in a stark contrast to the single-mode case, where $\xi=0$ leads to the highest efficiency. How can we make sense of such a counterintuitive behavior? 

\begin{figure}[bt]
    \centering
\includegraphics[width=0.9\textwidth]{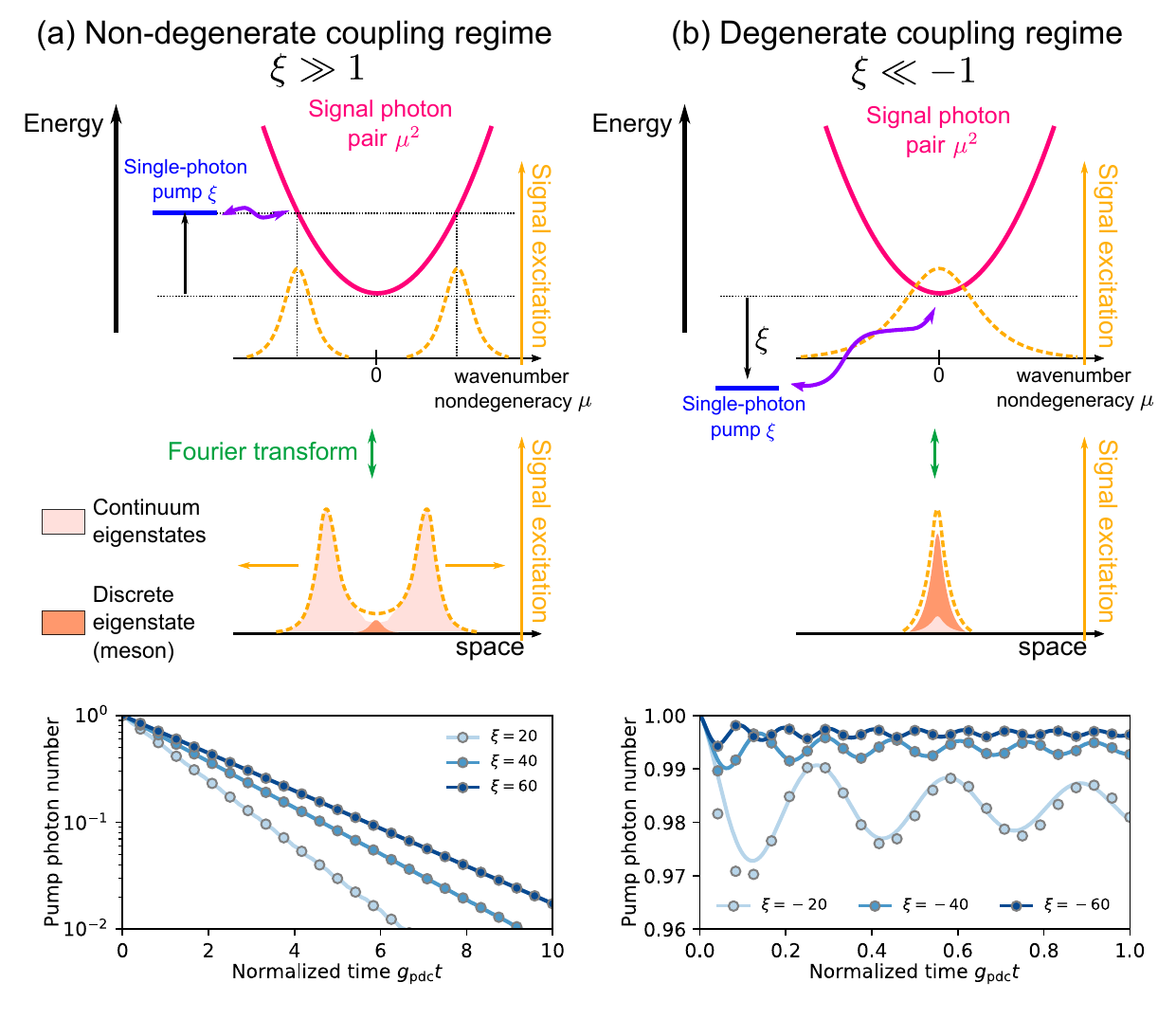}
    \caption{Illustrations for the phenomenology of (a) the non-degenerate and (b) the degenerate coupling regimes of single-photon-pumped broadband PDC. The shaded regions in the spatial distribution represent contributions from the continuum (light orange) and the discrete (dark orange) eigenstates. The bottom figures show the dynamics of the pump photon population in each regime, where the solid lines and circles represent numerical results and approximate expressions. The approximate expressions for the non-degenerate and degenerate cases are shown in \eqref{eq:approximate-dissipative-E} and \eqref{eq:approximate-dispersive-E}, respectively. Figure is adapted from Ref.~\cite{Yanagimoto2023-thesis} with modifications. We note that for the choice of rotating frame used here the energy-wavenumber diagrams in Figures (a) and (b) are equivalent to the traveling-wave phase mismatch $\Delta k(\Omega')$, with space and time interchanged as in Sec.~\ref{sec:time-prop}. The transition from degenerate to non-degenerate operation is determined by the same conditions for as in Fig.~\ref{fig:OPA_TF}.}
    \label{fig:fano-dissipative-dispersive}
\end{figure}

A key to answer this question is to look at the dynamics of the signal photons in the spatial domain. As shown in Fig.~\ref{fig:fano-dissipative-dispersive}(a), with $\xi\gg 1$, the energy of the discrete pump state lies in the middle of the signal energy band. Thus, the signal photon pairs excited by the PDC have opposite offset of wavenumber from the center wavenumber of the parabolic dispersion (signal dispersion included up to second order)
and thus have opposite group velocity in the co-moving frame. As a result, in this regime of non-degenerate coupling, downconverted signal photons spatially move away from each other, which suppresses backconversion, thereby leading to high conversion efficiency. The decrease of the pump photon population approximately follows an exponential function, leading to a conversion efficiency of
\begin{align}
\label{eq:approximate-dissipative-E}
    \mathcal{E}(\tau)\approx 1-e^{-\pi \mathrm{\tau}/\sqrt{\xi}}.
\end{align}

In the other limit of $\xi\ll -1$, the pump energy lies well below the bottom of the signal energy band. Here, PDC only excites a narrow band of signal states at the bottom of the dispersion curve, which regime we refer to as the degenerate coupling regime (see Fig.~\ref{fig:fano-dissipative-dispersive}(b)). In this regime, a large portion of the signal excitation comprises the discrete optical meson, in which the down-converted signal photons exhibit local spatial correlations due to an interplay between nonlinearity and dispersion. The interference between the signal components of the meson and the continuum of eigenstates leads to oscillations in the photon conversion dynamics, where the oscillation amplitude decays due to the spatial dispersion of the continuum contributions. The conversion efficiency in this degenerate coupling regime can be approximated as
\begin{align}
\label{eq:approximate-dispersive-E}
    \mathcal{E}(\tau)\approx 1-\left\vert 1-\frac{\pi}{4(-\xi^{3/2})}+\frac{\sqrt{\pi}}{2\xi^2\sqrt{\tau}}e^{\mathrm{i}(\xi\tau-\pi/4)}\right\vert,
\end{align}
which exhibits sinusoidal oscillations with sub-polynomial amplitude decay with scaling $\sim\tau^{-1/2}$. 

The behavior of the PDC for an intermediate phase-mismatch can be understood as an interpolation between the non-degenerate and degenerate coupling regimes. Note that due to the involvement of continuum contribution, $\xi=0$ does not necessarily coincide with the largest conversion efficiency. Rather, we numerically find that $\xi\approx 1.9$ leads to a unit efficiency at a transient time $\tau=1.32$, due to the complete destructive interference of the pump amplitudes (see Ref.~\cite{Ryo2020Fano} for full discussions).

\subsubsection{Broadband PDC: pulsed pump}
\label{sec:fano-pulse-pump}

\begin{figure}[bt]
    \centering
    \includegraphics[width=\textwidth]{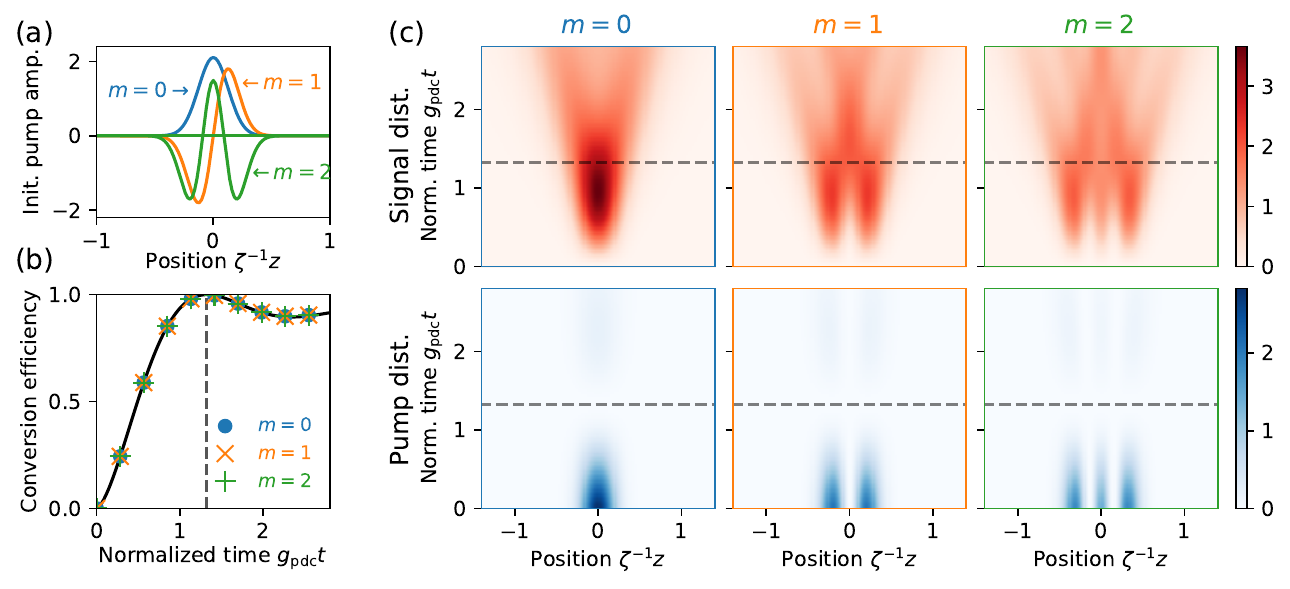}
    \caption{Numerically simulated dynamics of single-photon-pumped broadband PDC with a pulsed pump. (a) Initial pump waveforms in the spatial domain $h_z\propto \exp(-z^2/2\sigma^2) H_m(z/\sigma)$ with Hermite polynomial $H_m$ and pulse width $\sigma=0.2\zeta^{-1}_\mathrm{pdc}$. The characteristic photon-correlation length $\zeta_\mathrm{pdc}$ is defined in Appendix.~\ref{sec:fano-broadband-eigen}. (b) The overall conversion efficiency of the single-photon PDC. The black solid line represents the theoretical prediction \eqref{eq:pdc-conversion-efficiency-broadband}. (c) Spatial distributions of the signal and pump photons $\zeta^{-1}_\mathrm{pdc}\langle\hat{u}_z^\dagger\hat{u}_z\rangle~(u\in\{a,b\})$ as functions of time. Grey dashed lines represent $\tau=g_\mathrm{pdc}t\approx 1.32$, at which time, a complete conversion is achieved for the assumed phase-mismatch $\xi\approx 1.9$. Figure is adapted from Ref.~\cite{Yanagimoto2023-thesis}.}
    \label{fig:fano-pulse-pumped}
\end{figure}

When PDC is pumped by a pulse, classical intuition tell us that the rate of PDC should go up. However, in the deeply quantum limit where the pump pulse comprises a single photon, it turns out that the PDC rate is almost independent of the shape of the pump pulse. This can be clearly seen in the full-quantum simulation shown in Fig.~\ref{fig:fano-pulse-pumped}, where we show the dynamics of the single-photon pulse-pumped PDC with various pump-pulse shapes (see (a)). The overall conversion efficiency (see (b)) takes an identical trajectory regardless of the shape of the pump pulse, despite the fact that they exhibit seemingly distinctive propagation dynamics in the spatial domain (see (c)).

To see how this a counter-intuitive result arises, it is insightful to revisit how the PDC rate gets enhanced in semi-classical regime. Here, for small $t$, the fluorescence rate is dominated by the OPA bandwidth and therefore the rate at which PDC occurs is set entirely by $g_\text{pdc}$ (see, \textit{e.g.} Eqn.~\ref{eq:broadband-flux-gpdc}). When PDC is pumped with a strong coherent-state pulse, these spontaneously generated signal photons can stimulate further PDC without depleting the pump. The pulsed enhancement occurs because these down-converted photons are localized around the peak of the pump pulse; the large pump field co-located with the signal photons provides a large parametric gain. When PDC is pumped by a single-photon pump pulse, however, pump depletion occurs within the timescale set by $g_\text{pdc}$. In other words, since the spontaneously generated photon pair (with generation rate set by only the OPA bandwidth) depletes the pump, no further stimulated emission occurs.

\begin{figure}
    \centering
    \includegraphics[width=0.8\textwidth]{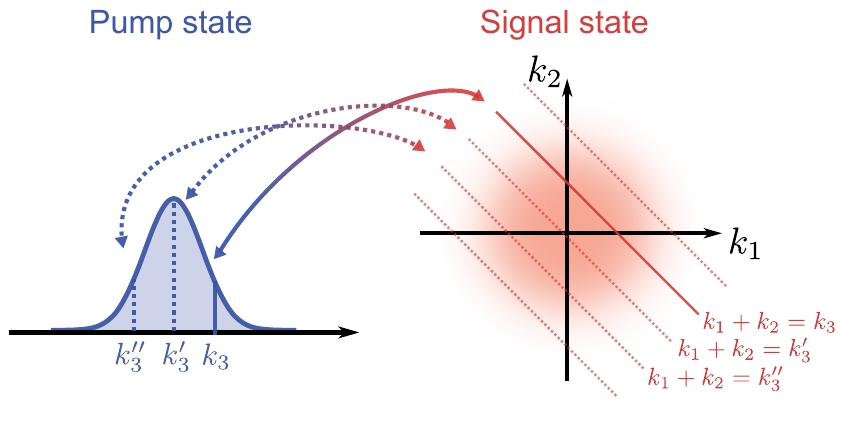}
    \caption{Illustration for the coupling structures of broadband single-photon-pumped PDC. Each spectral component of the pump pulse with wavenumber $k_3$ is coupled to a continuum of signal states with wavenumber $k_1$ and $k_2$ with $k_1+k_2=k_3$. There is no cross-talk between states with different total wavenumbers (\textit{i.e.}, momenta).}
    \label{fig:fano-pulse-shape-independence}
\end{figure}

We can also intuitively understand such behavior in the wavespace picture. A single-photon pump pulse is composed of a coherent superposition of monochromatic single-photon pump states $\ket{k_3}$, and each component $\ket{k_3}$ can downconvert to two-photon signal states $\ket{k_1,k_2}$ that satisfy the momentum conservation $k_1+k_2=k_3$ (see Fig.~\ref{fig:fano-pulse-shape-independence}). These downconverted signal photons, however, cannot interact with signal photons originated from a different pump photon because there is only one pump photon to begin with. Consequently, the state $\ket{k_1,k_2}$ always backconverts to $\ket{k_3}$, ensuring that the PDC dynamics are ``closed'' and independent within each of the Hilbert subspaces labeled by the pump wavenumber $k_3$. Because of the absence of cross-talk between the down-converted signal and idler pairs, the pulse-pumped PDC dynamics are insensitive to the distribution of spectral components, and therefore the overall pump pulse shape. We refer readers to Appendix.~\ref{sec:fano-pulse-pump-appendix} for more detailed discussions.

\paragraph{Further reading}
When the single-photon-pumped broadband PDC physics is extended from the single waveguide to a coupled-pair waveguide, the Hamiltonian accommodates two discrete states and a single continuum. When the energies of the two discrete states get close, a long-lived resonant state emerges due to Fano interference. The full discussions can be found in Ref.~\cite{Ryo2020Fano}.

\subsection{Model reduction with matrix-product states}\label{sec:MPS}
In the deep-quantum regime, multimode dynamics face a Hilbert space that grows exponentially, both in mode number and photon number. This section details a numerical approach relying on the general heuristic that entanglement in a one-dimensional quantum many-body system is limited~\cite{Vidal2004}. When entanglement between photons is well localized, and the total number of photons in the system is small, efficient simulation of quantum pulse propagation on a one-dimensional waveguide can be realized using a matrix-product state (MPS) formalism~\cite{Vidal2003, Yanagimoto2021}. A full treatment of this formalism is beyond the scope of this work, and we refer the readers to Refs.~\cite{Orus2014, Sornborger1999} for comprehensive reviews from the perspective of many-body physics. Also, Ref.~\cite{Quesada2022} provides a comprehensive tutorial on how one could use MPS to study general quantum photonic systems. Instead, the aim of this tutorial is to introduce the key concepts of MPS to readers with nonlinear-optics background, with applications focused on one-dimensional pulse propagation in nonlinear waveguides. While we consider one-dimensional $\chi^{(2)}$ waveguides as a case study, the formalism introduced in this subsection is general and can be applied to a broader class of systems, \textit{e.g.}, $\chi^{(3)}$ nonlinear waveguides. 

The Hamiltonian for $\chi^{(2)}$ nonlinear waveguide is given in \eqref{eq:Hamiltonian-multimode-finite-L} as
\begin{align}
\hat{H}=\hat{H}_\text{a}+\hat{H}_\text{b}+\hat{H}_\text{NL}
\end{align}
with the nonlinear part
\begin{align}
\hat{H}_\text{NL}/\hbar=\frac{r}{2}\int\mathrm{d}z\,\left(\hat{a}_z^2\hat{b}_z^\dagger+\hat{a}_z^{\dagger 2}\hat{b}_z\right),
\end{align}
and the linear part
\begin{align}
    \hat{H}_\text{u}/\hbar=\int\mathrm{d}z\,\hat{u}_z^\dagger \delta\omega_u(-\mathrm{i}\partial_z)\hat{u}_z
\end{align}
for the signal $(u=a)$ and the pump ($u=b$) modes, where the continuum field operators fulfill commutation relations $[\hat{u}_z,\hat{u}_{z'}]=\delta(z-z')$. Unlike previous sections, we specifically write down the Hamiltonian in the position basis instead of in the wavespace. 

\begin{figure}
    \centering
    \includegraphics[width=\textwidth]{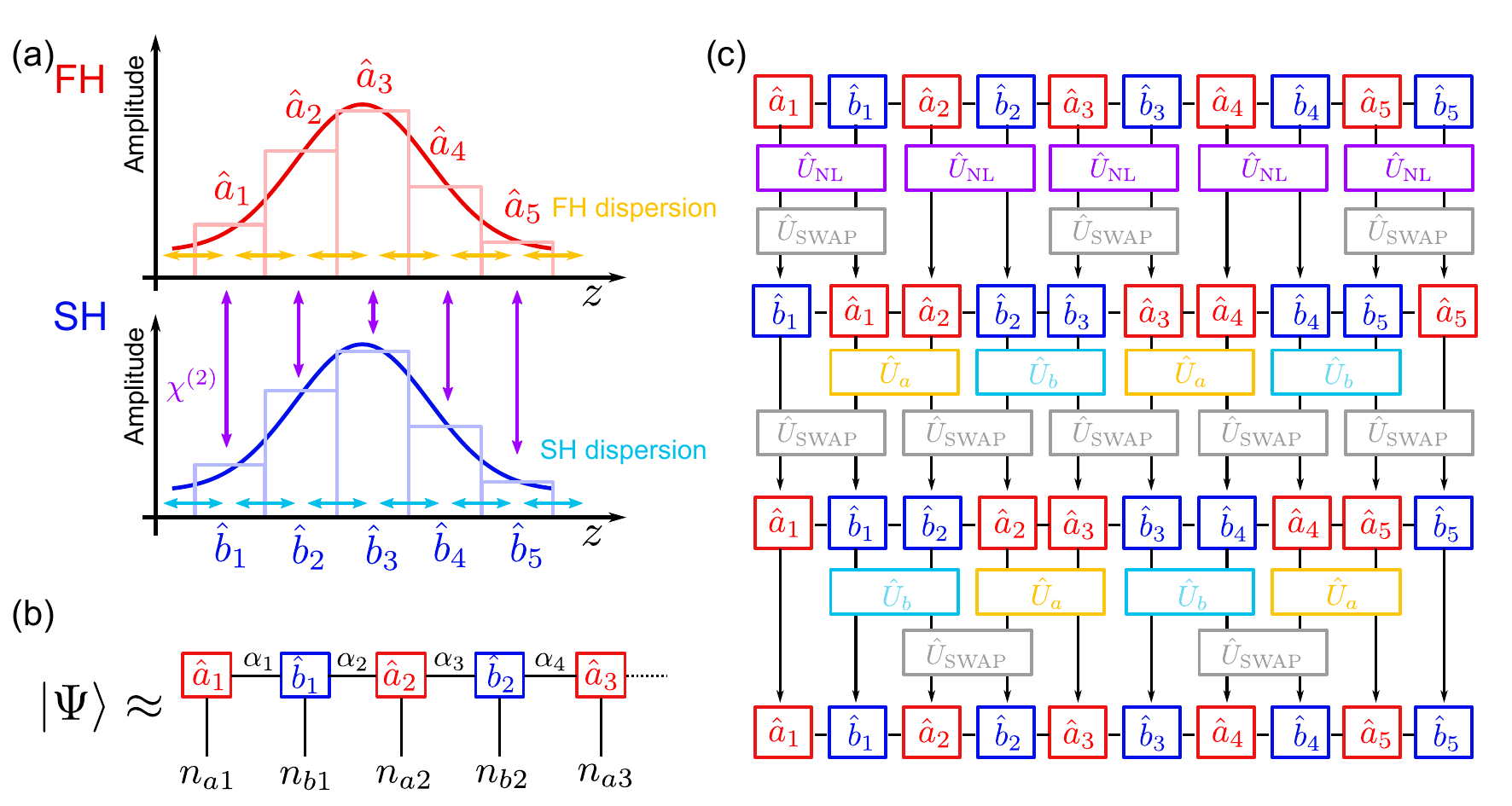}
    \caption{(a) Continuous spatial coordinate is discretized into spatial bins, and field annihilation operators are assigned to each bin. (b) A matrix product state representation of a full-quantum state as a product of low-rank tensors. (c) An example implementation of a one-time step of TEDB for quantum pulse propagation under $\chi^{(2)}$ nonlinear interaction. Figure is adapted from Ref.~\cite{Yanagimoto2023-thesis}.}
    \label{fig:mps}
\end{figure}

For numerical evaluations, we impose open boundary condition $-L/2 \leq z\leq L/2$ and discretize the space into $N$ spatial bins with size $\Delta z=L/N$, where the field operators are discretized to become
\begin{align}
    \hat{u}_{j\Delta z-L/2}\mapsto \frac{\hat{u}_j}{\Delta z},
\end{align}
with $j\in\{1,2,\dots,N\}$, and the commutation relations of the discretized operators are given by a Kronecker delta $[\hat{u}_j,\hat{u}_{j'}]=\delta_{j,j'}$. The Hamiltonian is rewritten in the discretized spatial basis as
\begin{align}
\label{eq:discretized-hamiltonian}
    \hat{H}=&\sum_{j}\hat{H}_{\text{NL},j}+\sum_{u\in\{a,b\}}\sum_j\hat{H}_{u,j},
\end{align}
with the nonlinear terms
\begin{subequations}
\begin{align}
    \hat{H}_{\text{NL},j}/\hbar=\frac{r}{2\sqrt{\Delta z}}\left(\hat{a}_j^2\hat{b}_j^\dagger+\hat{a}_j^{\dagger 2}\hat{b}_j\right)
\end{align}
and the linear terms
\begin{align}
\begin{split}
   \label{eq:mps-discretized-linear} \hat{H}_{u,j}/\hbar=&\underbrace{\delta\omega_u(0)\hat{u}_j^\dagger\hat{u}_j}_{\mathrm{relative~phase-velocity}}-\underbrace{\frac{\mathrm{i}\delta\omega_u'(0)}{\Delta z}\left(\hat{u}_j^\dagger\hat{u}_{j+1}-\hat{u}_j\hat{u}_{j+1}^\dagger\right)}_{\mathrm{relative~group-velocity}}\\
    &+\underbrace{\frac{\delta\omega_u''(0)}{2\Delta z^2}\left(\hat{u}_j^\dagger\hat{u}_{j+1}+\hat{u}_j\hat{u}_{j+1}^\dagger-2\hat{u}_j^\dagger\hat{b}_j\right)}_{\mathrm{group-velocity~dispersion}},
    \end{split}
\end{align}
\end{subequations}
where we have expanded the dispersion up to the second order. Physically, the first and second terms in \eqref{eq:mps-discretized-linear} represent phase and group velocity relative to the reference values, respectively, and the third term represents the group-velocity dispersion. The definitions of teh quantities involved and their connections to classical theories are provided in Sec.~\ref{sec:rosetta-waveguide-hamiltonian}. Notice that the Hamiltonian takes a highly localized form where only up to nearest-neighbor interactions are present (see Fig.~\ref{fig:mps}(a)). Such a local Hamiltonian structure provides an intuitive explanation for the spatially localized photon-photon correlation structure we observed in previous sections. This is to be contrasted to the Hamiltonian in the wavespace, where nonlinear interactions take a highly nonlocal form. 

A generic quantum state in the Hilbert space can be expressed in the Fock basis as
\begin{align}
\label{eq:full-ket}
    \ket{\psi}=\sum_\mathbf{n}c_\mathbf{n}\ket{\mathbf{n}}
\end{align}
where $\mathbf{n}=(n_{a1},n_{b1},\dots,n_{bN})^\intercal$, and $\ket{\mathbf{n}}=\ket{n_{a1}}\otimes \ket{n_{b1}}\cdots \ket{n_{bN}}$ is a tensor product of local Fock states. While the state representation \eqref{eq:full-ket} may appear simple in theory, it takes an exponentially large memory space to represent it numerically. For instance, assuming finite truncation to the Fock space as $n_{aj},n_{bj}<n_\text{max}$, the total dimension of the state space becomes $n_\mathrm{dim}=n_\mathrm{max}^{2N}$. Even for a moderate choice of parameters, \textit{e.g.}, $N=100$ and $n_\mathrm{max}=10$, we have $n_\mathrm{dim}=10^{200}$, which clearly exceeds the available memory on realistic computers. To enable full-quantum simulation of the photon dynamics, we thus need to employ a sophisticated model reduction technique to construct a tractable state representation. 

Here, we leverage the fact that the amount of entanglement in a one-dimensional quantum many-body system is heuristically limited, for which we employ Schmidt decomposition as a main theoretical tool. For the bipartite partition of the entire system to the left and right subsystems $\mathcal{L}$ and $\mathcal{R}$, the Schmidt decomposition is given as
\begin{align}
\label{eq:schmidt-decompostion}
    \ket{\psi}=\sum_{\alpha=1}^{\chi_\mathrm{SR}} \lambda_\alpha \ket{\phi_\alpha^{[\mathcal{L}]}}
    \ket{\phi_\alpha^{[\mathcal{R}]}},
\end{align}
where $\ket{\phi_\alpha^{[\mathcal{L}]}}$ and $\ket{\phi_\alpha^{[\mathcal{R}]}}$ are the eigenvectors of the reduced density matrices of the corresponding subsystems $\hat{\rho}^{[\mathcal{L}]}$ and $\hat{\rho}^{[\mathcal{R}]}$, respectively, with corresponding Schmidt weight $\lambda_\alpha$. The index $\alpha$ runs from $1$ to $\chi_\mathrm{SR}$, where the subscript ``SR'' stands for Schmidt rank. The distribution of $\lambda_\alpha$ is tightly related to the entanglement that is present in the system; The entanglement entropy between the subsystems is expressed as
\begin{align}
    \mathcal{E}=-\sum_{\alpha=1}^{\chi_\mathrm{SR}}|\lambda_\alpha|^2\log (|\lambda_\alpha|^2),
\end{align}
which implies that the entanglement is strong (weak) when $\lambda_\alpha$ decays slowy (quickly) as a function of $\alpha$ (Notice that $\ket{\psi}$ becomes separable if $\lambda_\alpha$ vanishes for all $\alpha$ but $\alpha=1$). When the Schmidt weight $\lambda_\alpha$ decays quickly, it is expected that \eqref{eq:schmidt-decompostion} could provide a good approximation of $\ket{\psi}$ even when the summation is truncated at $\alpha=\chi<\chi_\mathrm{SR}$. In other words, weakly entangled states can be efficiently approximated by truncating the sum decomposition at a minimum necessary ``bond dimension'' $\chi$.

To apply the intuition to the $2N$-mode state \eqref{eq:full-ket}, we first consider a Schmidt decomposition for the bipartition between the first mode (\textit{i.e.}, subsystem $[1]$) and the rest of the system (\textit{i.e.}, subsystem $[2\dots2N]$)
\begin{align}
\label{eq:first-schmidt-decomposition}
    \ket{\psi}=\sum_{\alpha_1}\lambda_{\alpha_1}^{[1]}\ket{\phi^{[1]}_{\alpha_1}}\ket{\phi^{[2\dots 2N]}_{\alpha_1}}=\sum_{n_{a1}}\sum_{\alpha_1}\Gamma_{1\alpha_1}^{[1]n_{a1}}\lambda_{\alpha_1}^{[1]}\ket{n_{a1}}\ket{\phi^{[2\dots 2N]}_{\alpha_1}},
\end{align}
where we have denoted $\ket{\phi^{[1]}_{\alpha_1}}=\sum_{n_{a1}}\Gamma_{1\alpha_1}^{[1]n_{a1}}\ket{n_{a1}}$. The vectors $\ket{\phi^{[2\dots 2N]}_{\alpha_1}}$ are eigenvectors of the reduced density matrix for the subsystem $[2\dots 2N]$ and can be decomposed as
\begin{align}
    \ket{\phi^{[2\dots 2N]}_{\alpha_1}}=\sum_{n_{b1}}\ket{n_{b1}}\ket{\tau^{[3\dots 2N]}_{\alpha_1,n_{b1}}}.
\end{align}
To factor out the dependence of $\ket{\tau^{[3\dots 2N]}_{\alpha_1,n_{b1}}}$ on $\alpha_1$, we write down
\begin{align}
    \ket{\tau^{[3\dots 2N]}_{\alpha_1,n_{b1}}}=\sum_{\alpha_2}\Gamma_{\alpha_1\alpha_2}^{[2]n_{b1}}\lambda_{\alpha_2}^{[2]}\ket{\phi_{\alpha_2}^{[3\dots 2N]}},
\end{align}
which is an expansion in terms of the eigenvectors of the reduced density matrix for the subsystem $[3\dots 2N]$. Finally, substituting these expressions into \eqref{eq:first-schmidt-decomposition} gives us
\begin{align}
\label{eq:second-schmidt-decomposition}
    \ket{\psi}=\sum_{n_{a1},n_{a2}}\sum_{\alpha_1,\alpha_2}\Gamma_{1\alpha_1}^{[1]n_{a1}}\lambda_{\alpha_1}^{[1]}\Gamma_{\alpha_1\alpha_2}^{[2]n_{b1}}\lambda_{\alpha_2}^{[2]}\ket{n_{a1}}\ket{n_{b1}}\ket{\phi^{[3\dots 2N]}_{\alpha_2}}.
\end{align}
We cascade this procedure to obtain an MPS representation of the form
\begin{align}
\label{eq:mps-ket}
    c_\mathbf{n}=\sum_{\alpha_1,\alpha_2,\dots,\alpha_{2N-1}}\Gamma^{[1]n_{a1}}_{1\alpha_1}\lambda_{\alpha_1}^{[1]}\Gamma^{[2]n_{b1}}_{\alpha_1\alpha_2}\lambda_{\alpha_2}^{[2]}\dots \lambda_{\alpha_{2N-1}}^{[2N-1]}\Gamma^{[2N]n_{bN}}_{\alpha_{2N-1}1},
\end{align}
whose structure is depicted in Fig.~\ref{fig:mps}(b). For an efficient state representation, we need to apply a finite truncation to the sum of the indices $\alpha_j\leq \chi$, where $\chi$ is referred to as the bond dimension of an MPS. As $\Gamma^{[j]}$ is a rank-3 tensor, we can interpret \eqref{eq:mps-ket} as the decomposition of the original rank-$2N$ tensor $c_\mathbf{n}$ to a product of low-rank tensors. As implied above, quantum states with strong and long-range entanglement require larger bond dimensions $\chi$ for their accurate description. Conversely, MPS is a particularly suitable representation for quantum states with limited entanglement. The total number of parameters required for the MPS representation is $n_\mathrm{dim}\approx 2N\chi^2n_\mathrm{max}$, which exhibits a favorable linear scaling with respect to the system size $N$ (c.f., the exponential scaling $n_\mathrm{dim}=n_\mathrm{max}^{2N}$ for the na\"ive state representation).

Now, our remaining task is to evolve an MPS in time to simulate the pulse dynamics. There are various means to update an MPS in time under a given Hamiltonian, and the readers can refer to Ref.~\cite{Paeckel2019} for a comprehensive review. In this tutorial, we introduce a scheme called ``time-evolving block decimation (TEDB)'', originally introduced by Vidal in his seminal work~\cite{Vidal2003}. In TEDB, unitary evolution under the system Hamiltonian is decomposed to local one-mode and two-mode unitary operations, which can be applied efficiently to an MPS.

A one-mode unitary operation acting on the $j$th mode of an MPS can be written as 
\begin{align}
    \hat{U}_1=\sum_{n,n'}U_1^{n,n'}\ket{n}\bra{n}.
\end{align}
To update the state under this unitary $\ket{\psi}\mapsto\hat{U}_1\ket{\psi}$, we simply need to update a single tensor $\Gamma^{[j]}$ as
\begin{align}
    \Gamma^{[j]n}_{\alpha_j\alpha_{j+1}}\mapsto \sum_{n'} U_1^{n,n'}\Gamma^{[j]n'}_{\alpha_j\alpha_{j+1}},
\end{align}
which can be done by tensor contractions.

Updating the state under a two-mode unitary operation is a bit more involved. A generic local two-mode unitary acting on $j$th and $(j+1)$th modes takes a form
\begin{align}
    \hat{U}_2=\sum_{n,\ell,n',\ell'}U_2^{(n,\ell),(n',\ell')}\ket{n}\ket{\ell}\bra{n'}\bra{\ell'}.
\end{align}
To update the state under the unitary as $\ket{\psi}\mapsto\hat{U}_2\ket{\psi}$, corresponding tensors are to be updated as
\begin{align}
    \sum_\beta\Gamma_{\alpha\beta}^{[j]n}\lambda^{[j]}_{\beta}\Gamma_{\beta\gamma}^{[j+1]\ell}\mapsto& \sum_\beta\sum_{n',\ell'}U_2^{(n,\ell),(n',\ell')}\Gamma_{\alpha\beta}^{[j+1]n'}\lambda^{[j]}_{\beta}\Gamma_{\beta\gamma}^{[j+1]\ell'}=\Theta^{n\ell}_{\alpha\gamma}.
\end{align}
Notice that the two-mode unitary has transformed a product of rank-3 tensors to a single rank-4 tensor $\Theta$. To recast the updated state to the original form, we need to decompose $\Theta$ back to a product of rank-3 tensors. To this, we reshape the tensor $\Theta_{\alpha\gamma}^{n\ell}$ to a matrix form $\Theta_{(n,\alpha),(\ell,\gamma)}$ and perform an SVD to get
\begin{align}
    \Theta_{(n,\alpha),(\ell,\gamma)}=\sum_{\beta=1}^{n_\mathrm{max}\chi}\lambda_{\beta}V_{(n,\alpha),\beta}W_{(\ell,\gamma),\beta}\approx\sum_{\beta=1}^{\chi}\lambda_{\beta}V_{(n,\alpha),\beta}W_{(\ell,\gamma),\beta},
\end{align}
where $V_{(n,\alpha),\beta}$ and $W_{(\ell,\gamma),\beta}$ are singular vectors of the matrix $\Theta$ with a singular value $\lambda_\beta$. To get the final expression, we sort $\lambda_\beta$ in ascending order and apply truncation to the singular vector component. As a result, we obtain closed-form update rules $\lambda^{[m]}_\beta=\lambda_\beta$, $\Gamma^{[m]n}_{\alpha\beta}=V_{(n,\alpha),\beta}$, and $\Gamma^{[m+1]\ell}_{\beta\gamma}=W_{(\ell,\gamma),\beta}$. Because this procedure involves an SVD of a square matrix with dimension $n_\mathrm{max}\chi$, the computational cost of each step scales as $\mathcal{O}(n_\mathrm{max}^3\chi^3)$, which usually becomes the computational bottleneck in TEDB.

Since unitary operations under the Hamiltonian generally induce nonlocal interactions, we need to decompose the total time evolution to a product of local one-mode and two-mode operations. To this, we consider implementing an infinitesimal time evolution $\hat{U}=e^{-\mathrm{i}\hat{H}\delta t}$. By virtue of the Torotter-Suzuki decomposition, $\hat{U}$ can be approximated as the product of infinitesimal two-mode unitaries as
\begin{align}
\label{eq:trotter-suzuki}
    \hat{U}\approx (\hat{U}_{\mathrm{NL},1}\hat{U}_{\mathrm{NL},2}\dots \hat{U}_{\mathrm{NL},m})(\hat{U}_{a,1}\hat{U}_{a,3}\dots)(\hat{U}_{b,1}\hat{U}_{b,3}\dots)(\hat{U}_{a,2}\hat{U}_{a,4}\dots)(\hat{U}_{b,2}\hat{U}_{b,4}\dots),
\end{align}
where $\hat{U}_\text{NL,m}=\exp(-\mathrm{i}\hat{H}_{\text{NL},m}\delta t/\hbar)$ represents $\chi^{(2)}$ interactions, and $\hat{U}_\text{u,m}=\exp(-\mathrm{i}\hat{H}_{u,m}\delta t/\hbar)$ represents the dispersion interactions. Because we encode FH and SH alternatively in an MPS, some unitary operations in \eqref{eq:trotter-suzuki} are not local. To bring the corresponding modes next to each other in an MPS representation, we can insert SWAP operations at appropriate locations. An example implementation of a single-TEDB step that implements $\hat{U}$ is shown in Fig.~\ref{fig:mps}(c). The figure shows the lowest-order decomposition, and one can further employ a higher-order Trotter-Suzuki decomposition to improve the numerical efficiency~\cite{Sornborger1999}. 

\begin{figure}[ht]
    \centering
    \includegraphics[width=\textwidth]{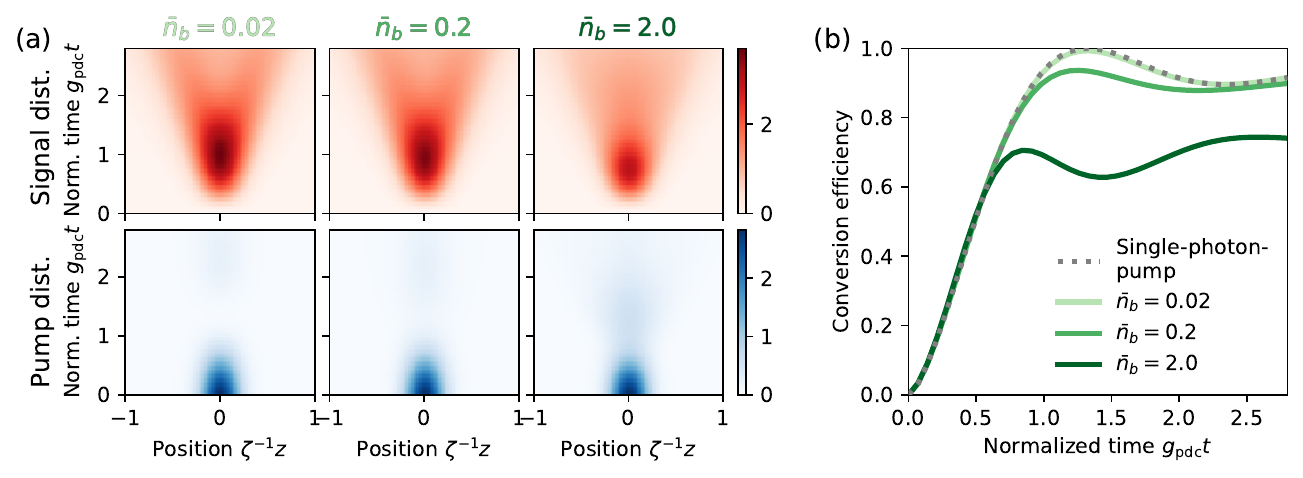}
    \caption{Dynamics of broadband PDC pumped by coherent pump states with various average photon number $\bar{n}_b$. (a) Spatial distribution of the signal and pump photons $\zeta^{-1}_\mathrm{pdc}\langle \hat{u}_z^\dagger\hat{u}_z\rangle$, where the characteristic spatial correlation length $\zeta_\mathrm{pdc}$ is defined in Appendix.~\ref{sec:fano-broadband-eigen}. (b) Overall conversion efficiency as a function of time. Grey dashed lines represent theoretical prediction for single-photon-pumped broadband PDC. For all the simulations, we assume normalized phase-mismatch of $\xi=(\delta\omega_b(0)-2\delta\omega_a(0))/g_\mathrm{pdc}$=1.9, matched group-velocity of $\delta\omega_a'(0)=\delta\omega_b'(0)$, and $\delta\omega_b''(0)=2\delta\omega_a''(0)$. Initial coherent pump state has spatial amplitide $\langle\hat{b}_z\rangle=\bar{n}_b^{1/2}\pi^{-1/4}\sigma^{-1/2}\exp(-z^2/2\sigma^2) $ with pulse width $\sigma=0.2\zeta^{-1}_\mathrm{pdc}$. Figure is adapted from Ref.~\cite{Yanagimoto2023-thesis}.}
    \label{fig:mps-pulsed-pdc}
\end{figure}

As an example case study to show the utility of the MPS framework in studying broadband photon dynamics in the deep-quantum regime, we simulate the dynamics of PDC pumped with a coherent pump state. As shown in Fig.~\ref{fig:mps-pulsed-pdc}, the PDC dynamics converge to that of single-photon-pumped PDC (see Sec.~\ref{sec:PDC}) in the weak-pump limit because a weak coherent pump state can be well-approximated as a superposition of a vacuum and a small amplitude of a single-photon pump state. As the pump photon number increases, however, multiphoton processes become more pronounced, which we observe limits the conversion efficiency.

\subsection{Multimode decoherence in nonlinear-optical quantum computation}\label{sec:decoherence}
\begin{figure}[bh]
    \centering
    \includegraphics[width=\textwidth]{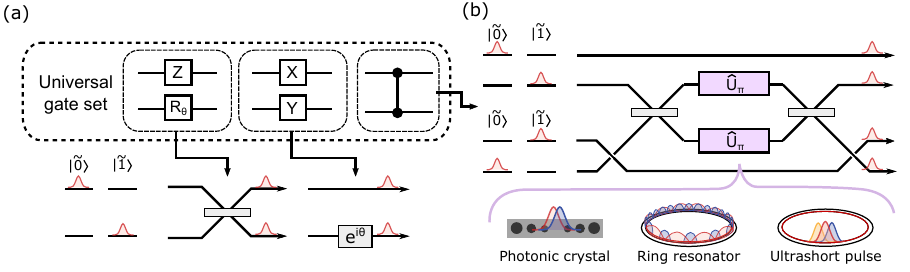}
    \caption{Implementation of the universal gate set for photonic qubits. (a) Single-qubit gates can be implemented using linear optics. (b) For a two-qubit entangling gate (\textit{e.g.}, CZ gate), one needs a strong nonlinearity. A CZ gate can be constructed by embedding NS gate $\hat{U}_\pi$ to both arms of an MZI. Various nonlinear-optical Implementation of an NS gate is possible. Figure is adapted from Ref.~\cite{Yanagimoto2023-thesis}.}
    \label{fig:gates-cz-gate}
\end{figure}

In previous sections, we noted that few-photon propagation in the deep-quantum regime displays complex dynamics that cannot be reduced to a single-mode description. While this complexity opens up possibilities for novel quantum devices, the lack of single-mode interaction presents challenges for conventional quantum applications. A seminal work by Shapiro (Ref.~\cite{Shapiro2006}) illustrated this point, demonstrating that high-fidelity quantum gates cannot be realized using traveling light waves on a nonlinear waveguide. This provides a ``no-go argument'' against using nonlinear optics for quantum computation.

Here we discuss multimode photon dynamics in the deep-quantum regime with a focus on realizing gates for optical quantum computation (OQC). In the context of multimode quantum optics, OQC is a particularly challenging task, as any deviation from single-mode physics leads to gate infidelities. Single-photon qubits are the leading candidate for optical quantum computation. In particular, the dual-rail basis, where quantum information is encoded to polarization, time-bin, or path, is an attractive approach since all single-qubit gates can be implemented using linear optics~\cite{Obrien2007}. To complete the universal gate set, however, we need one two-qubit entangling gate, such as a controlled-Z (CZ) gate. As illustrated in Fig.~\ref{fig:gates-cz-gate}, a popular implementation of a CZ gate is to embed nonlinear sign (NS) gates $\hat{U}_\pi$ to a Mach-Zehnder interferometer (MZI). An NS gate implements $\pi$-phase-shift only on the two-photon input as
\begin{align}
\hat{U}_\pi(c_0\ket{0}+c_1\ket{1}+c_2\ket{2})=c_0\ket{0}+c_1\ket{1}-c_2\ket{2},
\end{align}
which requires single-photon optical nonlinearity. For instance, unitary evolution under the single-mode Kerr Hamiltonian $\hat{H}/\hbar=\frac{1}{2}\hat{a}^{\dagger2}\hat{a}^2$ for time $t_\pi=\pi \chi^{-1}$ can realize an NS gate. Here, we consider implementing an NS gate using a single-mode $\chi^{(2)}$ interaction with Hamiltonian
\begin{align}
\label{eq:temporal-trap-single-mode-H}
    \hat{H}/\hbar=\frac{g}{2}(\hat{a}^{\dagger 2}\hat{b}+\hat{a}^2\hat{b}^\dagger).
\end{align}
For the initial input state of $(c_0\ket{0}+c_1\ket{1}+c_2\ket{2})\ket{0}$, coherent evolution under \eqref{eq:temporal-trap-single-mode-H} for time $t_\pi=\sqrt{2}\pi g^{-1}$ impinges a $\pi$-phase-shift only on the state $\ket{2\,0}$, realizing an NS gate. Here, we have denoted $\ket{n_a\,n_b}$ a product state of $n_a$-photon-FH state and $n_b$-photon-SH state. 

In the discussions above on quantum gate operations, it is implicitly assumed that the form of the modes in which relevant photons are present do not change as a function of time. However, in reality, we need to be aware that these ``computational modes'' generally consist of collective excitation of continuous quantum fields, and their structure may vary as a function of time. In other words, the complete specification of the quantum gate operations needs to account for both the Hamiltonian and the form of input/output modes $\hat{a}_\mathrm{in/out}=\int\mathrm{d}z\,\Psi^*_{\mathrm{in/out}}(z)\hat{a}_z$. With these formulations, an NS gate is generalized as
\begin{align}
\hat{U}_\pi\left(c_0\ket{0_\mathrm{in}}+c_1\ket{1_\mathrm{in}}+c_2\ket{2_\mathrm{in}}\right)=c_0\ket{0_\mathrm{out}}+c_1\ket{1_\mathrm{out}}-c_2\ket{2_\mathrm{out}},
\end{align}
where $\ket{n_\mathrm{in/out}}=\frac{1}{\sqrt{n!}}\hat{a}_\mathrm{in/out}^{\dagger n}\ket{0}$. As we have seen in the section for pulse-pumped squeezing, the structure of the computational modes can ``morph'' continuously during the gate evolution. Thus, we generalize the input/output waveforms at intermediate times as $\Psi(t,z)$, where $\Psi_\mathrm{in}(z)=\Psi(0,z)$ and $\Psi_\mathrm{out}(z)=\Psi(t_\pi,z)$.

For a given system Hamiltonian $\hat{H}$, we aim to choose the input/output waveforms so that the unitary evolution $e^{-\mathrm{i}\hat{H}t_\pi/\hbar}$ approximates the action of $\hat{U}_\pi$ as faithfully as possible. In this tutorial, we consider a $\chi^{(2)}$ nonlinear waveguide with Hamiltonian
\begin{subequations}
\label{eq:gates-waveguide-hamiltonian}
\begin{align}
\hat{H}=\hat{H}_\text{NL}+\sum_{u\in\{a,b\}}\hat{H}_u
\end{align}
with a nonlinear term
\begin{align}
\hat{H}_\text{NL}/\hbar=\frac{r}{2}\int\mathrm{d}z\,\left(\hat{a}_z^2\hat{b}_z^\dagger+\hat{a}_z^{\dagger 2}\hat{b}_z\right),
\end{align}
and linear terms
\begin{align}
\hat{H}_u/\hbar=\int\mathrm{d}z\,\hat{u}_z^\dagger \hat{G}_u(z)\hat{u}_z,
\end{align}
\end{subequations}
where we have introduced a generic function $G_u(z)$ to represent the linear dynamics of the system (see Sec.~\ref{sec:rosetta-waveguide-hamiltonian} for the derivation of the Hamiltonian). A nominal homogeneous waveguide is described as a special case $G_u(z)=\delta\omega_u(-\mathrm{i}\partial_z )$. Because the evolution of the vacuum input is trivial and exhibits no error, a reasonable strategy to choose $\Psi_{\mathrm{in}/\mathrm{out}}(z)$ is set such that the single-photon input has no error, \textit{i.e.},
\begin{align}
\ket{1_\mathrm{out}}=e^{-\mathrm{i}\hat{H}t_\pi/\hbar}\ket{1_\mathrm{in}},
\end{align}
which is fulfilled by taking $\Psi(t,z)$ as a solution of
\begin{align}
\label{eq:gates-mode-dynamics}
\mathrm{i}\partial_t\Psi(t,z)=G_u(z)\Psi(t,z).
\end{align}
Assuming that $\Psi(t,z)$ is defined as such, the only remaining source of error is the deviation of the two-photon-state output $e^{-\mathrm{i}\hat{H}t_\pi/\hbar}\ket{2_\mathrm{in}}$ from the expected output $-\ket{2_\mathrm{out}}$, which we can quantify using a distance measure
\begin{align}
\label{eq:gates-error-D}
    \mathcal{D}=\Vert e^{-\mathrm{i}\hat{H}t_\pi/\hbar}\ket{2_\mathrm{in}}+\ket{2_\mathrm{out}}\Vert.
\end{align}
Another intuitive measure of the gate performance is the time evolution of the signal/pump photon number as a function of time. If we observe a clean sinusoidal Rabi-oscillation in the photon conversion dynamics, it strongly indicates the presence of single-mode physics. Conversely, any decay in oscillation amplitudes can be attributed to multimode dynamics that would degrade the gate performance.

\subsubsection{Gate operations with monochromatic resonator modes}
\label{sec:microscopic-monochromatic}
A conventional approach to realize single-mode physics is to use single-mode micro-resonators~\cite{lu2020toward, zhao2022ingap}, which we revisit in this section to discuss how single-mode physics is enforced. The intuition obtained in this section can also highlight the core challenges toward realizing the same physics in broadband pulse propagation.

To model the physics of a nonlinear waveguide resonator, we assume a periodic boundary condition $-L/2\leq z\leq L/2$ to the waveguide Hamiltonian \eqref{eq:gates-waveguide-hamiltonian}. Then, a trivial solution to \eqref{eq:gates-mode-dynamics} would be the stationary monochromatic modes, \textit{i.e.}, $\Psi_{um}(z,t)\propto e^{2\pi\mathrm{i}mz/L}~u\in\{a,b\}$ with $m\in \mathbb{Z}$. Without a loss of generality, we assume the mode with $m=0$ is the computational mode. To see how single-mode physics emerges from such construction, let us define a basis spanned by monochromatic modes
\begin{align}
    \hat{u}_m(t)=\int\mathrm{d}z\,\Psi_{um}^*(t,z)\hat{u}_z,
\end{align}
with
\begin{align}
    \Psi_{um}(t,z)=\frac{1}{\sqrt{L}}e^{-\mathrm{i}\delta\omega_u(2\pi m/L)t}e^{2\pi\mathrm{i}mz/L}.
\end{align}
Then, the Hamiltonian can be rewritten as
\begin{align}
\label{eq:gates-temporal-cw-H}
\hat{H}/\hbar=\sum_{m,n}\frac{g}{2}e^{\mathrm{i}\delta_{m,n}t}\hat{a}_m^\dagger(t)\hat{a}_n^\dagger(t)\hat{b}_{m+n}(t)+\mathrm{h.c.}
\end{align}
with nonlinear coupling
\begin{align}
g=\frac{r}{\sqrt{L}},
\end{align}
and phase-mismatch 
\begin{align}
    \delta_{ mn}=\delta\omega_{b}(2\pi (m+n)/L)-\delta\omega_{a}(2\pi m/L)-\delta\omega_{a}(-2\pi m/L),
\end{align}
which we expand up to the second order.

\begin{figure}[bt]
    \centering
    \includegraphics[width=1.0\textwidth]{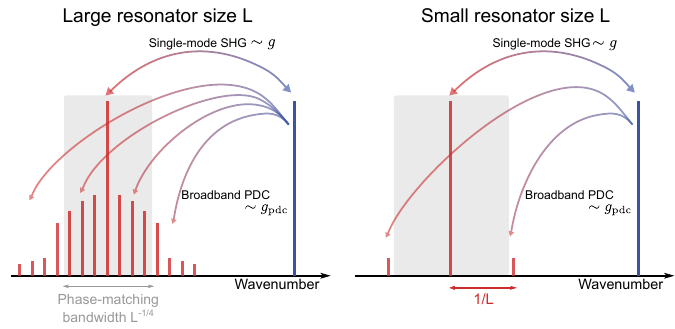}
    \caption{Physical process involved in the dynamics of initial two-photon state in the dc mode $\ket{2_\mathrm{in}}=\frac{1}{\sqrt{2}}\hat{a}_0^{\dagger2}\ket{0}$. When the resonator size $L$ is large (left figure), many FH modes are present in the phase-matching bandwidth (grey-shaded region), to which broadband PDC can occur, playing the role of an effective decay channel. For a small resonator size $L$ (right figure), mode spacing $1/L$ is large enough to push all but one FH mode out of the phase-matching bandwidth, realizing an effective single-mode interaction. Figure is adapted from Ref.~\cite{Yanagimoto2023-thesis}.}
    \label{fig:gates-resonator}
\end{figure}

Note that \eqref{eq:gates-temporal-cw-H} mediates multimode coupling among all the modes that fulfill momentum conservation, which can have an undesirable effect of leaking photons from the computational mode. As shown in Fig.~\ref{fig:gates-resonator}, while the initial two-photon FH state $\ket{a_0}=\frac{1}{\sqrt{2}}\hat{a}_0^{\dagger2}\ket{0}$ is only coupled to a monochromatic SH state $\ket{b_0}=\hat{b}_0^{\dagger}\ket{0}$ via SHG, a photon that has upconverted to $\ket{b_0}=\hat{b}_0^{\dagger}\ket{0}$ can leak to 
\begin{align}
\ket{a_m}=\hat{a}_m^\dagger\hat{a}_{-m}^\dagger\ket{0}
\end{align}
via the Hamiltonain terms $\hat{b}_{0}\hat{a}^\dagger_m\hat{a}^\dagger_{-m}$. Because $\ket{a_m}~(m>0)$ is a state outside of the computational basis, broadband PDC process to these modes should be seen as an effective decoherence channel. 

The nature of broadband PDC as multimode interaction can be more clearly seen from the structure of the Hamiltonian. For the relevant states, the diagonal matrix elements of the Hamiltonian are
\begin{subequations}
\label{gates:discrete-hamiltonian-elements}
\begin{align}
    &\frac{1}{\hbar}\langle b_0\vert\hat{H}\vert b_0\rangle=0 &\frac{1}{\hbar}\langle a_m\vert\hat{H}\vert a_m\rangle=-\delta_{m,-m}=\frac{4\pi^2\delta\omega_a''(0)}{L^2}m^2,
\end{align}
where we have chosen a reference phase so that $\delta\omega_b(0)=0$, assumed the phase-matching between carrier modes, i.e., $\delta\omega_b(0)-2\delta\omega_a(0)=0$, and approximated the FH dispersion up second order for concreteness. The off-diagonal elements are
\begin{align}
   \frac{1}{\hbar} \langle a_{m}\vert\hat{H}\vert b_0\rangle=\left\{\begin{array}{ll}
         & \frac{g}{\sqrt{2}}\quad(m=0)\\
         & g \quad(m\neq0)
    \end{array}\right..
\end{align}
\end{subequations}
The characteristic structure of the Hamiltonian, where a discrete SH state is coupled to multiple FH states, is reminiscent of what we saw in Sec.~\ref{sec:PDC} for single-photon-pumped broadband PDC. In fact, in the limit of large resonator size $L\rightarrow\infty$, the band formed by discrete FH states converges to a continuum, recovering the physics of discrete-continuum interaction. (Notice that the energy gap among neighboring FH states converge to zero in this limit in \eqref{gates:discrete-hamiltonian-elements}). As we have seen, dephasing caused by the continuum prohibits the realization of single-mode physics.


Notably, such multimode interactions, which are inevitable in a large resonator, can be suppressed when the size of resonator $L$ is reduced to a small value. To see this, we note that the magnitude of each off-diagonal element $\frac{1}{\hbar}\langle a_m\vert\hat{H}\vert b_0\rangle$ is $g$, implying that its effect is only significant when the magnitude of corresponding phase-mismatch $\delta_{m,-m}$ is smaller than $g$. More formally, we define the phase-matching bandwidth for single-photon-pumped PDC as the frequency window within which the magnitude of phase-mismatch
\begin{align}
    |\delta_{m,-m}|=\left|\frac{4\pi^2\delta\omega_a''(0)}{L^2}\right|m^2=g\left|\frac{g}{g_\mathrm{pdc}}\right|^3m^2
\end{align}
is smaller than $g$. Here, $g_\text{pdc}$ is the effective coupling rate for single-photon-pumped broadband PDC
\begin{equation}
    g_\mathrm{pdc}=(r^4/4\pi^2\delta\omega_a''(0))^{1/3},\label{eqn:g_pdc}
\end{equation}
which is the same as $g_\mathrm{pdc}$ we defined in Sec.~\ref{sec:fano-broadband-eigen}. The number of modes (other than $m=0$ mode) within the phase-matching bandwidth can be found as the largest $m$ that fulfills $|\delta_{m,-m}|<g$, given by
\begin{align}
\label{eq:gates-m-pdc}
    m_\mathrm{pdc}=(g_\mathrm{pdc}/g)^{3/2}.
\end{align}
This can be interpreted as an effective number of ``decay channels'' through which photons can leak out from the computational modes. 

\begin{figure}
    \centering
    \includegraphics[width=0.57\textwidth]{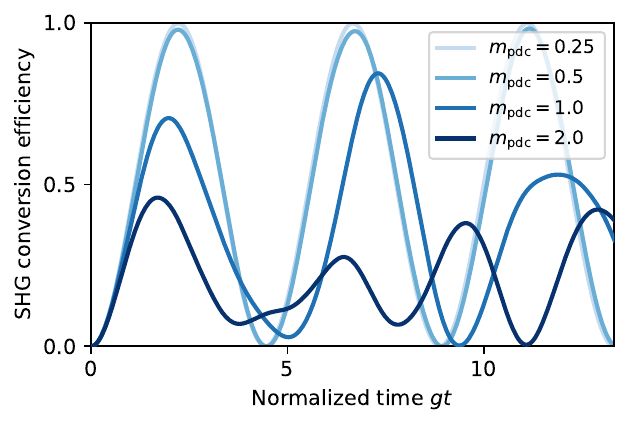}
    \caption{SHG conversion efficiency (\textit{i.e.}, pump photon number) for an initial two-photon FH state $\ket{a_0}$ shown as a function of time for various number of effective decay channels $m_\mathrm{pdc}\propto L^{3/4}$. Figure is adapted from Ref.~\cite{Yanagimoto2023-thesis}.}
    \label{fig:gates-transition}
\end{figure}

Here, we note that $g_\mathrm{pdc}$ is invariant to the resonator size $L$, while the coupling rate among the CW modes $g\sim L^{-1/2}$ decreases as a function of $L$, reflecting the decrease of the electric field per photon. The number of modes within the phase-matching bandwidth therefore scales as $m_\mathrm{pdc}\propto L^{3/4}$, indicating that we can reduce the number of ``decay'' channels by making the resonator size smaller, as illustrated in Fig.~\ref{fig:gates-resonator}.
When $L$ is small enough to ensure $m_\mathrm{pdc}\ll1$ is fulfilled, the multimode interactions are suppressed, and we can effectively realize a single-mode interaction with a Hamiltonian
\begin{align}
\hat{H}/\hbar=\frac{g}{2}\left(\hat{a}_0^{\dagger2}(t)\hat{b}_0(t)+\hat{a}_0^{2}(t)\hat{b}_0^\dagger(t)\right).
\end{align}
Another way to interpret how small resonators realize single-mode physics is to note that the limit of $m_\mathrm{pdc}\ll1$ implies $g\gg g_\mathrm{pdc}$. In this limit, SHG (with characteristic coupling rate $g$) can occur before any multimode effects occur due to broadband PDC process (with  rate $g_\mathrm{pdc}$). To see the transition from multimode to single-mode physics, in Fig.~\ref{fig:gates-transition}, we show the SHG dynamics for various $m_\mathrm{pdc}$, where we recover clear Rabi oscillation the limit of $m_\mathrm{pdc}\rightarrow0$.

These discussions, together with the fact that the nonlinear coupling $g\propto L^{-1/2}$ increases with smaller $L$, might seem to suggest that a design principle for nonlinear-optical implementation of quantum gates is always to make resonators smaller. In a realistic experiment, however, larger bending loss and surface roughness loss induce a critical tradeoff between the nonlinear coupling $g$ and the linear loss rate $\kappa$, which eventually limits the attainable figure of merit $g/\kappa$. The following sections discuss how short pulses can be used to mitigate these trade-offs.

\subsubsection{Pulsed operation}

To circumvent the trade-off between the coupling strength $g$ and loss rate $\kappa$ that occurs when using monochromatic resonator modes, we must find approaches that enhances the nonlinear coupling without introducing the large loss rates that occur in extremely small resonators. One possible resolution is the use of short optical pulses to localize the field associated with the in-coupled photons, where, intuitively, the pulse duration plays the role of the size of an effective ``flying cavity.'' While temporarily localized waveforms cannot be an eigenvalue of the dispersion operator $\delta\omega_a(-\mathrm{i}\partial_z)$, we can still choose the computational signal mode $\Psi(z,t)$ as a nonstationary solution of \eqref{eq:gates-mode-dynamics}, which eliminates the gate error for the single-photon input state $\ket{1_\mathrm{in}}$. Since the gate error on a vacuum input $\ket{0_\mathrm{in}}$ is always zero, only the source of gate error is the action of the Hamiltonian on the two-photon input state $\ket{2_\mathrm{in}}$. 

\begin{figure}[t]
    \centering
    \includegraphics[width=0.65\textwidth]{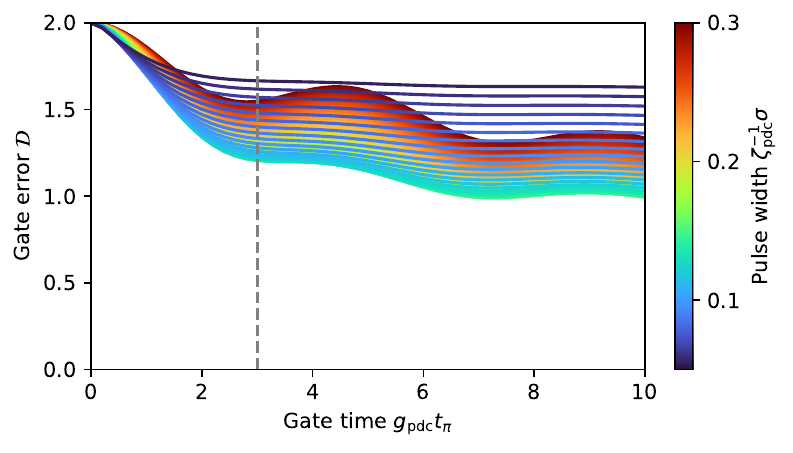}
    \caption{Gate error $\mathcal{D}$ for an NS gate implemented using the SHG process for various total gate time $t_\pi$ and pulse width. We assume $\delta\omega_b''(0)=2\delta\omega_a''(0)$, and the carriers are phase- and group-velocity matched. The grey dashed lines represent a gate time $g_\mathrm{pdc}t_\pi=3$. Figure is adapted from Ref.~\cite{Yanagimoto2023-thesis}.}
    \label{fig:gates-gate-error}
\end{figure}

To be concrete, let us assume zero phase and group velocity mismatch and expand the dispersion up to second order as $\delta\omega_a(k)=\frac{1}{2}\delta\omega_a''(0)k^2$. Then, \eqref{eq:gates-mode-dynamics} has a solution of chirped-Gaussian function 
\begin{align}
\label{eq:gates-gaussian-waveform}
\Psi(z,t)=\frac{\pi^{-1/4}\sigma_z^{-1/2}}{\sqrt{1+\mathrm{i}(\delta\omega_{a}''(0)/4\pi^2\sigma_z^2)(t-t_\mathrm{0})}}\exp\left(-\frac{1}{2}\frac{(z/\sigma_z)^2}{1+\mathrm{i}(\delta\omega_{a}''(0)/4\pi^2\sigma_z^2)(t-t_\mathrm{0})}\right),
\end{align}
where $t_0$ is the time at which there is no chirp, and $\sigma_z$ is the width of the pulse at $t=t_0$. To minimize the chirp during the gate operation, it is reasonable to set $t_0=t_\pi/2$ for the total gate time of $t_\pi$. The annihilation operator for the computational modes are defined as
\begin{align}
    \hat{a}_{\mathrm{in/out}}(t)=\int\mathrm{d}z\,\Psi^*_\mathrm{in/out}(z)\hat{a}_z
\end{align}
with $\Psi_\mathrm{in}(z)=\Psi(z,0)$ and $\Psi_\mathrm{out}(z)=\Psi(z,t_\pi)$.

In Fig.~\ref{fig:gates-gate-error}, we show the gate error $\mathcal{D}$ as defined in Eqn.~\eqref{eq:gates-error-D}, as a function of both gate time and the pulse duration. Long pulses (red curves) exhibit Rabi-like oscillations as the input biphoton undergoes cycles of up- and down-conversion. For these long pulses, we observe poor gate fidelities due to multimode decoherence. For short pulses (blue curves), we observe an enhanced rate for the Rabi oscillations that comes with faster decoherence and poorer gate fidelities. With an intermediate pulse duration (green curves), we obtain optimal gate performance, with both an enhanced gate fidelity and a faster effective coupling rate. However, even for this optimal configuration, the performance is far from perfect. These observations are to be seen as a reformulation of Shapiro's no-go argument for $\chi^{(2)}$ systems, showing that one cannot realize high-fidelity quantum gates using nonlinear propagation of optical pulses. 

\begin{figure}[b]
    \centering
    \includegraphics{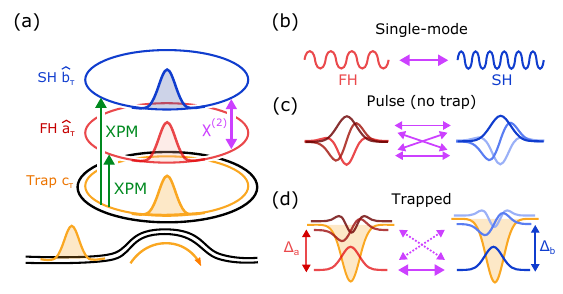}
    \caption{(a) Illustration for the temporal trapping scheme to achieve $\chi^{(2)}$ nonlinear interactions between ultrashort pulses. A trap pulse co-propagates with quantum signal pulses, creating a potential in the temporal domain via cross-phase modulation (XPM). (b) In a micro-resonator, dispersion can ensure that interactions among monochromatic cavity modes are phase-mismatched except for a single pair. This mechanism enables the realization of effective single-mode physics. (c) For traveling pulses without a trap, pulse waveforms are energetically degenerate. This degeneracy causes nonlinear coupling among them to be phase-matched, resulting in no single-mode subspace. (d) The temporal potentials created by the trap pulse lift the degeneracy among pulse waveforms, suppressing undesired multimode interactions. Through appropriate dispersion engineering, only the bound modes of the trap are phase-matched. Figure is adapted from Ref.~\cite{Yanagimoto2023-thesis}.}
    \label{fig:gates-pulse-supermode}
\end{figure}

As illustrated in Fig.~\ref{fig:gates-pulse-supermode}, the issue stems from the fact that the temporal supermodes have near-degenerate energies. As a result, nonlinear interactions generally induce all-to-all coupling, leading to a loss of photons from the computational basis. This observation highlights the need of to eliminate the degeneracy among the temporal supermodes, which is difficult with dispersion engineering alone. The following section on temporal trapping discusses techniques for introducing strong frequency shifts between the temporal modes; this approach introduces a strong phase-mismatch that suppresses coupling to any modes outside the computational basis while retaining the enhancements the coupling rate enabled by using ultrafast pulses.

\subsection{Temporal trapping}\label{sec:temporal_trapping}
In the previous section, we have observed that nonlinear optics is inherently multimode in the deep-quantum regime, and such multimode dynamics opens effective decoherence channels, \textit{e.g.}, when quantum gate operations are considered. The issue is particularly critical for traveling-pulse implementation, where single-mode dynamics cannot be realized even with optimized pulse waveforms. In this section, we introduce the prescription using ``temporal trapping'' to simultaneously resolve the issue of multimode decoherence and $g/\kappa$ tradeoff~\cite{Yanagimoto2022_temporal}. 

The core idea behind temporal trapping is depicted in Fig.~\ref{fig:gates-pulse-supermode}. In addition to the quantum signal pulses, we consider a co-propagating auxiliary ``trap pulse'' that induces variation in the index of refraction in the temporal (\textit{i.e.}, longitudinal) dimension. With an appropriate choice of waveguide dispersion, this refractive index variation can form bound pulsed modes that are energetically separated from the continuum modes, which confines the dynamics to an effective single-mode subspace spanned by the bound modes. This is to be contrasted to the normal pulse propagation without a trap, where all pulse waveforms are energetically degenerate, which makes multimode interactions inevitable.

For the generic Hamiltonian for a $\chi^{(2)}$ nonlinear waveguide \eqref{eq:gates-waveguide-hamiltonian}, we generalize the linear operator as
\begin{align}
    G_u(z)=\delta\omega_u(-\mathrm{i}\partial_z)+V_u(z),
\end{align}
where $V_u(z)$ represents position-dependent potential for signal ($u=a$) and pump ($u=b$). Such a potential can be generated, \textit{e.g.}, via cross-phase modulation induced by an auxiliary trap pulse that co-propagates with the quantum signal pulses. In this construction, the depth of the potential is given as
\begin{align}
    V_u(\tau)=-(n_2/n)\omega_{u,0}|c_z|^2/A,
\end{align}
where $\omega_{u,0}$ is the carrier frequency, $c_z$ is the amplitude of trap pulse, $n_2$ is the nonlinear index, and $A$ is the mode area. We note that the following discussions do not rely on the specific construction of the temporal trap. Also, while we assume a finite resonator size $-L/2\leq z\leq L/2$ to make the comparison to conventional resonator-based operation clear, our construction does not rely on a specific value of $L$ and is valid even in the limit of $L\rightarrow\infty$, corresponding to a traveling-wave implementation.

For concreteness, we assume group-velocity matching and include dispersion up to second order as $\delta\omega_u(k)=\delta\omega_{u}(0)+\frac{1}{2}\delta\omega_u''(0)k^2$, which defines an eigenproblem of the linear operator as
\begin{align}
\label{eq:temporal-trap-eigen}
    \left(\delta\omega_{u}(0)-\frac{1}{2}\delta\omega_u''(0)\partial_z^2+V_u(z)\right)\Psi_{um}(z)=\lambda_{um}\Psi_{um}(z),
\end{align}
where eigenvalues $\lambda_{um}$ are sorted in an ascending order. Notably, when the dispersion is anomalous, \textit{i.e.}, $\delta\omega_u''(0)>0$, \eqref{eq:temporal-trap-eigen} has at least one bound-state solution $\Psi_{u0}(z)$ with a finite energy gap $\Delta_u=\lambda_{u1}-\lambda_{u0}$ from the rest of the eigenmodes. Intuitively, when the trap is deep enough, the resultant energy $\Delta_u$ can be large enough to phase-mismatch all the undesired multimode interactions, effectively realizing single-mode interaction only between the bound signal and pump modes $\Psi_{a0}$ and $\Psi_{b0}$.

To see the emergence of effective single-mode interaction more clearly, we rewrite the Hamiltonian using temporal supermodes defined by the eigenfunctions of the linear operators as
\begin{align}
\label{eq:temporal-trap-supermode-H}
\hat{H}/\hbar=\sum_{\ell,m,n}\frac{g_{\ell mn}}{2}e^{\mathrm{i}\delta_{\ell mn}t}\hat{b}_\ell(t)\hat{a}_m^\dagger(t)\hat{a}_n^\dagger(t)+\mathrm{h.c.},
\end{align}
where
\begin{align}
\label{eq:temporal-trap-supermode}
\hat{u}_m(t)=\int\mathrm{d}z\,e^{\mathrm{i}\delta_{um}t}\Psi_{u,m}^*(z)\hat{u}_z
\end{align}
are temporal supermodes. The nonlinear coupling tensor
\begin{align}
    g_{\ell mn}=r\int\mathrm{d}z\,\Psi_{b\ell}^*(z)\Psi_{am}(z)\Psi_{an}(z)
\end{align}
represents the strength of parametric interaction among $\ell$th pump and $m$th and $n$th signal supermodes with corresponding phase-mismatch of
\begin{align}
    \delta_{\ell mn}=\delta_{b\ell}-\delta_{am}-\delta_{an}.
\end{align}
Notice that the specific choice of the mode basis \eqref{eq:temporal-trap-supermode} has eliminated linear coupling terms from the Hamiltonian \eqref{eq:temporal-trap-supermode-H}. 

The Hamiltonian \eqref{eq:temporal-trap-supermode-H} has a more complicated structure than that for monochromatic modes in a homogeneous resonator \eqref{eq:gates-temporal-cw-H}, but they share qualitative features; Both of them are composed of a sum of multimode parametric interactions with corresponding phase-mismatch. For the case of monochromatic mode, the small resonator size has allowed us to push undesired modes out of phase-matching bandwidth to realize an effective single-mode interaction. Analogously, for the case of an ultrashort pulse, we can utilize the tight temporal confinement provided by the temporal trap to make undesired multimode interactions phase-mismatched (\textit{i.e.}, off-resonant).

For concreteness, let us assume $\delta\omega_a''(0)=\delta\omega_b''(0)/2$ and trap with with $z_\mathrm{trap}$
\begin{align}
    V_a(z)=V_b(z)/2=-\delta\omega_a''(0)z_\mathrm{trap}^{-2}\sech^2(z/z_\mathrm{trap}),
\end{align}
for which we have bound-modes
\begin{align}
    \Psi_{a0}(z)=\Psi_{b0}(z)=\frac{1}{\sqrt{2z_\mathrm{trap}}}\sech(z/z_\mathrm{trap}).
\end{align}
The characteristic energy gap $\Delta$ is defined as
\begin{align}
\Delta=\Delta_a=\Delta_b/2=\frac{\delta\omega_a''(0)}{2z_\mathrm{trap}^2}.
\end{align}
We assume the phase-matching condition between the carriers $\delta\omega_a(0)=\delta\omega_b(0)/2$, which ensures that the interaction between the bound-modes is phase-matched, \textit{i.e.}, $\delta_{000}=0$. The corresponding coupling strength is
\begin{align}
    g_\mathrm{trap}=g_{000}=\frac{\pi r}{4\sqrt{2z_\mathrm{trap}}}.
\end{align}
We note here that the nonlinear coupling is now determined by the pulse width $z_\mathrm{trap}$, rather than the cavity length $L$. The use of few-cycle pulses therefore enables coupling rates comparable to wavelength-scale cavities, irrespective of the physical size of the resonator. This approach to realizing enhanced nonlinearities allows resonator loss to be engineered separately from the nonlinearity and the mode-structure of the cavity (\textit{e.g.} intermediate-scale ring cavities can be fabricated to avoiding bending losses).

Here, we note that one the terms in Eqn.\eqref{eq:temporal-trap-supermode-H} of the form $\hat{b}_0\hat{a}_m^\dagger\hat{a}_n^\dagger$ or $\hat{b}_\ell^\dagger\hat{a}_0\hat{a}_0$ can scatter photons out of the bound modes. For these terms, we can show that coupling is upper bounded as $g_{\ell mn}\leq g_\mathrm{trap}$, and the magnitude of phase-mismatch is lower-bounded as $\vert\delta_{\ell mn}\vert\geq \Delta$. As a result, if the condition
\begin{align}
    \frac{\Delta}{g_\mathrm{trap}}\gg 1
\end{align}
is met, all of the deleterious nonlinear interactions are suppressed, realizing an effective single-mode interaction
\begin{align}
    \hat{H}/\hbar=\frac{g_\mathrm{trap}}{2}(\hat{a}_0^{\dagger2}\hat{b}_0+\hat{a}^2\hat{b}_0^\dagger).
\end{align}

\begin{figure}
    \centering
\includegraphics[width=1.0\textwidth]{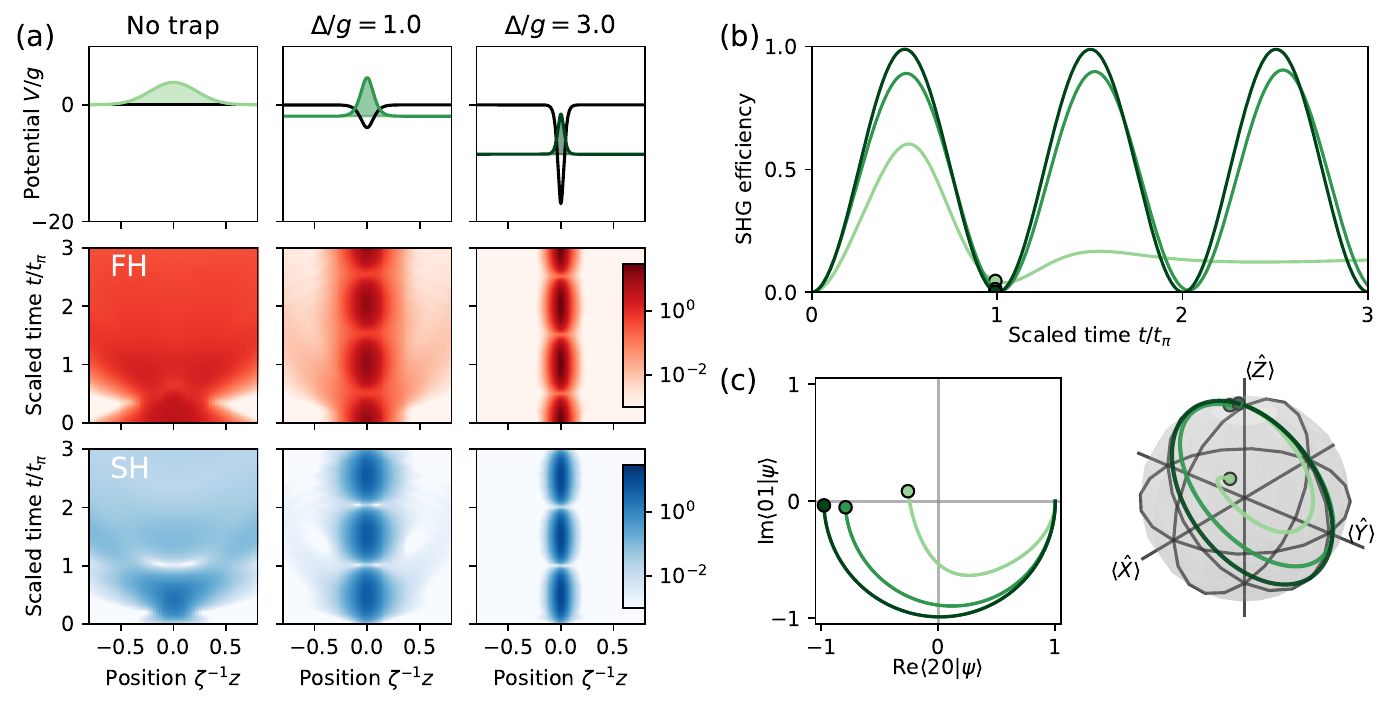}
    \caption{(a) Dynamics of the photon density distribution of FH (middle row) and SH (bottom row) photons with and without temporal trap (top row). (b) Overall SHG conversion efficiency of the nonlinear pulse propagation. (c) Left: System state is projected onto the computational subspace spanned by $\ket{20}$ and $\ket{01}$. Right: Projection of the dynamics on the pseudo-Bloch sphere characterized by the expectation values of pseudo-Pauli operators $\hat{X} =(\hat{a}^{\dagger2}\hat{b}+\hat{a}^2\hat{b}^\dagger)/\sqrt{2}$, $\hat{Y} =(\hat{a}^{\dagger 2}\hat{b}-\hat{a}^2\hat{b}^\dagger)/\sqrt{2}\mathrm{i}$, $\hat{Z}= \frac{1}{2} \hat{a}^{\dagger 2}\hat{a}^2-\hat{b}^\dagger\hat{b}$. Figure is adapted from Ref.~\cite{Yanagimoto2023-thesis}.}
    \label{fig:temporal-trap-gate-operations}
\end{figure}

In Fig.~\ref{fig:temporal-trap-gate-operations}(a), we show the dynamics of the photon distribution with and without a temporal trap. For the case without a trap, we choose the initial chirp and the width of an initial FH pulse to minimize the gate error $\mathcal{D}$. Even for such an optimized waveform, however, photons disperse quickly, and photon conversion dynamics exhibit highly damped Rabi-like oscillation (see Fig.~\ref{fig:temporal-trap-gate-operations}(b)). On the other hand, a temporal trap can critically suppress the dispersion of photons, and we can observe high-contrast Rabi-oscillation for multiple periods, highlighting that single-mode dynamics are enforced. We project the dynamics onto the subspace spanned by computational states in Fig.~\ref{fig:temporal-trap-gate-operations}(c), where we can clearly observe that the temporal trap enables the implementation of a high-fidelity NS gate.

We close this section by emphasizing that temporal trapping is one of many approaches that are currently being explored to realize ultrafast quantum gates. While the issue of multimode interactions was originally pointed out by Shapiro~\cite{Shapiro2006} regarding the use of XPM interactions for CZ gate~\cite{Chuang1995}, Ref.~\cite{He2011} later showed that the use of wavepackets with a large group-velocity mismatch could partially resolve this issue. In particular, the authors showed that single-mode XPM interactions can occur between a large coherent state and a few-photon state encoded in wavepacket with vastly different group velocities. At the same time, they showed this approach still does not enable high-fidelity CZ gates between two few-photon wavepackets. The use of wavepackets with large group-velocity mismatch was later revisited in Ref.~\cite{Xia2016}, where they showed an effective three-wave mixing interaction generated by coherent photon conversion (CPC) can realize a high-fidelity CZ-gate. Further theoretical studies are presented in Ref.~\cite{Viswanathan2018}. More recently, Ref.~\cite{Babushkin2022} showed that CZ gate based on CPC with a large group velocity mismatch could have a favorable property of being wave-shape tolerant, in the sense that gate operations can be realized with high fidelity regardless of the shape of the input wavepackets. These topics are sufficiently recent that there have not yet been any successful experimental demonstrations of ultrafast quantum gates, and we note that there are likely many approaches that have yet to be discovered.

\section{Conclusions and future work}

As platforms for ultrafast nonlinear photonics continue to mature, an extraordinary number of new possibilities emerge. At this time, the behaviors of classical devices are pushing both to new operating regimes enabled by dispersion engineering, and to new energy scales, enabled by the combination of tight confinement (in space and time) with long interaction lengths. Together, these features make ultrafast nonlinear optics an exciting platform for quantum optics, where the interplay between strong few-photon nonlinear interactions and extremely broadband multimode behaviors produces qualitatively new operating regimes at both the meso- and microscale. Beyond these exciting new device behaviors, quantum circuits based on optical nonlinearities may potentially resolve many of the outstanding challenges in scaling quantum technology by operating at room temperature and readily interfacing with telecom components. While few-photon nonlinearities have long been believed to be far beyond the reach of nonlinear optics, a more systematic study of the parameter space available in nonlinear nanophotonics suggests that these operating regimes will be accessed in the near future.


In anticipation of the considerable richness to be explored in this emerging field, we have attempted to provide a unified treatment of ultrafast quantum nonlinear optics that links classical- and quantum-nonlinear behaviors by using a mean-field approximation. This correspondence connects the parameters used in quantum models, such as the nonlinear coupling and the dispersion relations, to easily measured classical parameters such as the normalized efficiency of a waveguide or the bandwidth of an SHG transfer function. Furthermore, this correspondence allows us to extend the design tools used for classical nonlinear waveguides to the design of quantum devices, and enables the use of classical diagnostic techniques to verify the mean-field behaviors of fabricated devices before using them to explore exotic quantum features. We then extend this treatment to study the quantum physics of nonlinear optics at various energy scales, spanning the semi-classical, mesoscopic, and microscopic regimes. In particular, we argue that exotic non-Gaussian quantum features form even with the tens or hundreds of photons found in the mesoscopic regime. While such non-Gaussian physics is beyond the scope of conventional framework of Gaussian quantum optics, the intuition we acquire from classical and semi-classical nonlinear optics can be used to navigate this regime by working in the Gaussian interaction frame. Finally, we introduce when and how some of classical intuitions break down in the miscroscopic regime, discussing ``what it means for nonlinear optics to be quantum.'' We hope that the approach taken here enables the analysis of next-generation quantum nonlinear devices operating at and beyond mesoscale, and that this connection between classical and quantum nonlinear optics stimulates new directions for the field.

We conclude this tutorial by saying that much work is left to be done, and that the route forward likely involves innovations on many fronts: new materials, new approaches to photonic design, new model reduction strategies, and new operating regimes. In this regard, there are an extraordinary number of avenues for further study. On the theoretical front, there are likely many undiscovered operating regimes and model reduction strategies that may enable non-Gaussian quantum features to be more easily accessed using current technologies. As an example, temporal trapping was presented here as a proof of concept for how to realize few-photon nonlinearities with coupling rates enhanced by the bandwidths of the interacting pulses. In reality, many more undiscovered strategies likely exist that may be realized with greatly relaxed experimental requirements. Additional work is not only needed to better understand how to access novel quantum features, but also to better understand how these quantum features will be \emph{used}, \textit{e.g.} for metrology and computation.

Experimental studies of ultrafast quantum behavior in the saturated limit are in a similarly early stage. At this time, a relatively small number of photonics platforms have been able to access the combination of large nonlinearity and dispersion engineering. For the few platforms that have achieved dispersion-engineered interactions many of the dynamical regimes discussed here have yet to be demonstrated. Thus far, the few device demonstrations that have studied ultra-low power nonlinear interactions using femtosecond pulses are still several orders of magnitude away from the threshold of exotic quantum features, saturating with pulse energies of femtojoules and picojoules, rather than the attojoule and zeptojoule scales that are predicted to be possible. In addition, these devices tend to operate at wavelengths that are not convenient for efficiently detecting the generated quantum states. Attempts at accessing more favorable (\textit{i.e.} shorter) wavelength ranges have proved difficult, and the realization of short-wavelength devices is an outstanding challenge for the field. Promising routes forward include advanced approaches to dispersion engineering, such as inverse-designed photonic crystal waveguides or multilayer claddings, which enable greater design freedom than simple ridge structures alone.

We hope that this tutorial provides a helpful reference for researchers interested in working at the boundary of ultrafast and quantum nonlinear optics. While still quite young, this field is evolving rapidly, and there are many opportunities for further study. As the technical hurdles discussed above are resolved, we envision a new frontier both for exploring new phenomenology in quantum nonlinear optics, and for scalable quantum technologies, where densely integrated multi-functional photonics can be used to generate, manipulate, and detect non-classical light.

\begin{backmatter}

\bmsection{Funding}
The authors wish to thank NTT Research for their financial and technical support.

\bmsection{Acknowledgements}
The authors thank Noah Flemens, Evan Laksono, Heesoo Kim, Niharika Gunturu, Jean Wang, Huiting Liu, Chris Gustin, Daniel Wennberg, and Taewon Park for their valuable feedback on this article.

\bmsection{Disclosures}
The authors declare no conflicts of interest.

\bmsection{Data availability} No new data was generated or analyzed in this manuscript.

\end{backmatter}

\appendix

\section{The coupled-wave equations in nonlinear waveguides}\label{sec:CWEs}


To derive the coupled-wave equations for nonlinear waveguides, we must first treat the relevant aspects of waveguide modes with a particular focus on their complex reciprocity relations~\cite{SnyderLove}. These relations will be used to derive closed-form expressions for the phase- and group-velocity associated with each waveguide mode, which will be crucial for designing nonlinear waveguides. We then introduce a nonlinear polarization to Maxwell's equations, and repeat this analysis to derive the coupled-wave equations.

\subsection{Waveguide modes}

Waveguide modes are solutions to Maxwell’s equations that result when the dielectric tensor varies in only two spatial dimensions, $\mathbf{\epsilon}(x,y,z,\omega)=\mathbf{\epsilon}(x,y,\omega)$, in the absence of a driving current and a nonlinear polarization. We begin with Maxwell's equations for a dielectric medium,
\begin{subequations}
\begin{align}
\nabla\cdot \mathbf{D}(\mathbf{r},t) &= 0,\label{eqn:maxwell_1_t}\\
\nabla\cdot \mathbf{B}(\mathbf{r},t) &= 0,\label{eqn:maxwell_2_t}\\
\nabla\times \mathbf{H}(\mathbf{r},t) &= \partial_t \mathbf{D}(\mathbf{r},t),\label{eqn:maxwell_3_t}\\
\nabla\times \mathbf{E}(\mathbf{r},t) &= -\partial_t \mathbf{B}(\mathbf{r},t)\label{eqn:maxwell_4_t},
\end{align}
\end{subequations}
where the constitutive relations for $\mathbf{D}$ and $\mathbf{B}$ are typically given in the frequency domain. We Fourier transform Eqns.~\ref{eqn:maxwell_1_t}-\ref{eqn:maxwell_4_t}, using the convention
\begin{equation}
    \mathbf{E}(\mathbf{r},t)\equiv\int_{-\infty}^{\infty}\mathbf{E}(x,y,z,\omega)\exp(i\omega t)\frac{d\omega}{2\pi}=\frac12 \int_{-\infty}^{\infty}\mathbf{E}(x,y,z,\omega)\exp(i\omega t)\frac{d\omega}{2\pi} + c.c.\label{eqn:E-fourier},
\end{equation}
where the latter form follows from $\mathbf{E}(\mathbf{r},t)$ being real. The frequency-domain Maxwell equations are given by
\begin{subequations}
\begin{align}
\nabla\cdot \mathbf{D}(x,y,z,\omega) &= 0,\label{eqn:maxwell_1_f}\\
\nabla\cdot \mathbf{B}(x,y,z,\omega) &= 0,\label{eqn:maxwell_2_f}\\
\nabla\times \mathbf{H}(x,y,z,\omega) &= i\omega\mathbf{D}(x,y,z,\omega),\label{eqn:maxwell_3_f}\\
\nabla\times \mathbf{E}(x,y,z,\omega) &= -i\omega\mu_0\mathbf{H}(x,y,z,\omega),\label{eqn:maxwell_4_f}
\end{align}
\end{subequations}
where we have used the typical constitutive relations for a dielectric given by
\begin{subequations}
\begin{align}
\mathbf{D}(x,y,z,\omega) &= \mathbf{\epsilon}(x,y,\omega)\cdot \mathbf{E}(x,y,z,\omega),\\
\mathbf{B}(x,y,z,\omega) &= \mu_0\mathbf{H}(x,y,z,\omega).
\end{align}
\end{subequations}
$\mathbf{\epsilon}(x,y,\omega)$ is a second-rank tensor. In lossless media, $\mathbf{\epsilon}(x,y,\omega)$ is Hermitian, $\mathbf{\epsilon}_{ij}(x,y,\omega)=\mathbf{\epsilon}_{ji}^*(x,y,\omega)$, and for non-gyrotropic media $\mathbf{\epsilon}(x,y,\omega)$ is symmetric, $\mathbf{\epsilon}_{ij}(x,y,\omega)=\mathbf{\epsilon}_{ji}(x,y,\omega)$. Throughout this tutorial we will assume a lossless, non-gyrotropic medium, which renders $\mathbf{\epsilon}(x,y,\omega)$ both real and symmetric. In addition, for most cases of interest the crystal axes of the nonlinear medium are aligned to the waveguide coordinates, which renders $\mathbf{\epsilon}(x,y,\omega)$ a diagonal tensor
\begin{equation}
\mathbf{\epsilon}(x,y,\omega)=\epsilon_0\left(
\begin{array}{ccc}
     \epsilon_{xx}(x,y,\omega) & 0 & 0\\
     0 & \epsilon_{yy}(x,y,\omega) & 0\\
     0 & 0 & \epsilon_{zz}(x,y,\omega),
\end{array}{}
\right)
\end{equation}
where $\epsilon_0$ is the permittivity of free space, and $\epsilon_{xx}$ is the relative permittivity along the waveguide x-coordinate. In all further discussion, we will use capital letters $(X,Y,Z)$ to denote crystal axes, and lower-case letters $(x,y,z)$ to denote waveguide coordinates. For propagation along lab coordinate $z$ in a uniaxial medium, the direction of propagation is typically taken along the crystalline Y-axis so that one of the transverse fields is aligned along the crystalline Z-axis. For example, in X-cut lithium niobate thin films where the crystalline Z-axis aligned with the waveguide x-axis we have $\epsilon_{xx} = \epsilon_{ZZ}$, $\epsilon_{yy} = \epsilon_{XX}$, $\epsilon_{zz} = \epsilon_{XX}$. Similarly, in Z-cut lithium niobate thin films, which have their crystalline Z-axis aligned with the waveguide y-axis, we have $\epsilon_{xx} = \epsilon_{XX}$, $\epsilon_{yy} = \epsilon_{ZZ}$, $\epsilon_{zz} = \epsilon_{XX}$. Typical realizations of dispersion-engineered waveguides in thin-film lithium niobate have used TE$_{00}$ modes in X-cut films, which exhibit both large nonlinearities and allow for dispersion engineering at many wavelengths of interest. We note here that most materials with induced nonlinearities, such as silicon and silicon nitride, are sufficiently weakly perturbed by the DC Kerr effect that we may approximate them as having isotropic permittivities, $\epsilon_{ij}(x,y,\omega)=\epsilon(x,y,\omega) \delta_{ij}$, where $\delta_{ij}$ is the Kronecker delta function.

For waveguides $\mathbf{\epsilon}(x,y,\omega)$ is assumed to be translation invariant with respect to $z$. In this case, Maxwell's equations may be reduced to an eigenvalue problem by assuming solutions of the form $E(x,y,z,\omega) = \mathbf{E}_{\mu}(x,y,\omega)\exp(-ik_\mu(\omega) z)$, and $H(x,y,z,\omega) = \mathbf{H}_{\mu}(x,y,\omega)\exp(-ik_\mu(\omega) z)$. For the simple case of a diagonal $\epsilon$, which contains all of the essential features of a more general analysis, we first combine Maxwell's curl equations, to find
\begin{equation}
    \nabla\times\left(\epsilon^{(-1)}\cdot\nabla\times\mathbf{H}\right)-\omega^2\mu_0\mathbf{H}=0\label{eqn:waveeqn}
\end{equation}
This equation may be further simplified by eliminating $H_z$ using $\nabla\cdot B=0$ and using $\partial_z H(x,y,z,\omega) = -i k_\mu(\omega) H(x,y,z,\omega)$ for waveguide modes,
\begin{equation}
    H_{\mu,z} = \frac{\partial_x H_{\mu,x}+\partial_y H_{\mu,y}}{ik_\mu}\label{eqn:Hz}.
\end{equation}
Substituting Eqns.~\ref{eqn:modesE}-\ref{eqn:modesH} into Eqn.~\ref{eqn:waveeqn}, with Eqn.~\ref{eqn:Hz} yields
\begin{subequations}
\begin{align}
\left(\frac{\epsilon_{yy}}{\epsilon_{zz}}\partial_y^2+\partial_x^2+\epsilon_{yy}\omega^2\mu_0\right)H_{\mu,x}(x,y)+\left(1-\frac{\epsilon_{yy}}{\epsilon_{zz}}\right)\partial_x\partial_yH_{\mu,y}(x,y)=k_\mu^2 H_{\mu,x}(x,y),\label{eqn:eigenproblem1}\\
-\left(1-\frac{\epsilon_{xx}}{\epsilon_{zz}}\right)\partial_x\partial_yH_{\mu,x}(x,y)+\left(\frac{\epsilon_{xx}}{\epsilon_{zz}}\partial_x^2+\partial_y^2+\epsilon_{xx}\omega^2\mu_0\right)H_{\mu,y}(x,y)=k_\mu^2 H_{\mu,y}(x,y).\label{eqn:eigenproblem2}
\end{align}
\end{subequations}
When cast in this form, the transverse magnetic fields $[H_{\mu,x}, H_{\mu,y}]^\intercal$ are eigenfunctions of a linear operator with an eigenvalue given by $k_\mu^2$. Further generalizations of the presentation shown here to include off-diagonal components in $\mathbf{\epsilon}$ are given in~\cite{Fallahkhair2008}.

For a lossless, non-gyrotropic medium, where $\mathbf{\epsilon}$ is real, the fields associated with each waveguide mode exhibit a number of useful symmetry properties. Firstly, we note that $H_{\mu,x}$ and $H_{\mu,y}$ can both be taken to be real functions which, with Eqn.~\ref{eqn:Hz}, implies $H_{\mu,z}$ is imaginary. Similarly, using Eqn.~\ref{eqn:maxwell_3_f}, we may show that the transverse displacement fields $D_{\mu,x}$ and $D_{\mu,y}$ are both real and that the longitudinal displacement field $D_{\mu,z}$ is imaginary. Secondly, we note that the field distributions associated with backwards-propagating waves can be found by performing the coordinate transformation $(x,y,z)\rightarrow (x,y,-z)$, which implies that
\begin{equation}
    E_{\mu,x}^{-} = E_{\mu,x}^{+},\qquad E_{\mu,y}^{-} = E_{\mu,y}^{+},\qquad E_{\mu,z}^{-} = -E_{\mu,z}^{+},
\end{equation}
where $+$ and $-$ denote forward- and backwards-propagating propagating modes, respectively. Magnetic fields are pseudo-vectors, and therefore transform differently under coordinate inversion than electric fields. We may use Maxwell's equations to find
\begin{equation}
    H_{\mu,x}^{-} = -H_{\mu,x}^{+}, \qquad H_{\mu,y}^{-} = -H_{\mu,y}^{+}, \qquad H_{\mu,z}^{-} = H_{\mu,z}^{+}.   
\end{equation}
These symmetry relations will be useful in deriving closed-form expressions for $k_\mu$ and $\partial_\omega k_\mu$ in terms of the fields associated with a waveguide mode.

We now extend these solutions to the more general case, where the fields propagating in the waveguide can be expanded as a series of modes that only exhibit phase evolution along the z coordinate. In this case, for each frequency $\omega$ we have
\begin{subequations}
\begin{align}
\mathbf{E}(x,y,z,\omega) &= \sum_\mu a_{\mu}(\omega)\mathbf{E}_{\mu}(x,y,\omega)e^{-ik_\mu(\omega) z},\label{eqn:modesE}\\    
\mathbf{H}(x,y,z,\omega) &= \sum_\mu a_{\mu}(\omega)\mathbf{H}_{\mu}(x,y,\omega)e^{-ik_\mu(\omega) z},\label{eqn:modesH}\\
\mathbf{D}(x,y,z,\omega) &= \sum_\mu a_{\mu}(\omega)\epsilon(x,y)\cdot\mathbf{E}_{\mu}(x,y,\omega)e^{-ik_\mu(\omega) z},\label{eqn:modesD}\\
\mathbf{B}(x,y,z,\omega) &= \sum_\mu a_{\mu}(\omega)\mu_0\mathbf{H}_{\mu}(x,y,\omega)e^{-ik_\mu(\omega) z},\label{eqn:modesB}
\end{align}
\end{subequations}
where $a_{\mu}$ represents the component of $\mathbf{E}$ contained in mode $\mu$ around frequency $\omega$. Here each of the field distributions associated with each mode, $\mathbf{E}_\mu$ and $\mathbf{H}_\mu$, are eigenfunctions of Maxwell's equations at frequency $\omega$, with a corresponding eigenvalue $k_\mu^2(\omega)$. The complex reciprocity relations derived in the following sections will be used to calculate the coefficients $a_\mu$ needed to decompose $\mathbf{E}(x,y,z,\omega)$ into waveguide modes.

\subsection{Complex reciprocity relations}\label{sec:reciprocity}

We now derive the complex reciprocity relations for waveguide modes following the treatment in~\cite{SnyderLove}. The reciprocity relations will yield the orthogonality relations for the modes in linear waveguides and resonators, as well as closed-form solutions for the eigenvalues $k_\mu$ in terms of the field distributions. In following sections we will use complex reciprocity to derive the coupled-wave equations between modes in nonlinear devices. We begin by noting that throughout this section we will make frequent use of the two dimensional divergence theorem,
\begin{equation}
    \int_S \nabla\cdot\mathbf{v}(x,y) dx dy = \partial_z \int_S \mathbf{v}(x,y)\cdot\hat{\mathbf{z}} dx dy + \oint_\ell \mathbf{v}\cdot\hat{\textbf{n}} d\ell,
\end{equation}
where $S$ is a disk normal to $z$ and $\ell$ parameterizes a path around the boundary of $S$. For all cases of interest here $|\mathbf{v}(x,y)|$ vanishes for large $x$ and $y$, which yields
\begin{equation}
    \int_{A_\infty} \nabla\cdot\mathbf{v}(x,y) dx dy =\partial_z \int_{A_\infty} \mathbf{v}(x,y)\cdot\hat{\mathbf{z}} dx dy,
\end{equation}
where $\int_{A_\infty} \cdots dx dy$ denotes an integral over the full x-y plane.

\subsubsection{Complex reciprocity: conjugated form}

The conjugated form the of reciprocity relations establishes both orthogonality relations between modes and their normalization in lossless media. We begin by defining a vector $\mathbf{G}_c=\mathbf{E}_1\times\mathbf{H}^*_2$, where $\mathbf{E}_1$ and $\mathbf{H}_1$ satisfy Maxwell's equations, and $\mathbf{E}_2^*$ and $\mathbf{H}_2^*$ satisfy the conjugated form of Maxwell's equations,
\begin{align*}
\nabla\times\mathbf{H}_2^*(x,y,z,\omega) &= -i\omega\epsilon^*(x,y)\cdot\mathbf{E}_2^*(x,y,z,\omega),\\
\nabla\times\mathbf{E}_2^*(x,y,z,\omega) &= i\omega\mu_0\mathbf{H}_2^*(x,y,z,\omega).
\end{align*}
We evaluate the divergence
\begin{equation}
    \nabla\cdot\left(\mathbf{E}_1\times\mathbf{H}^*_2\right) = \mathbf{H}_2^*\cdot(\nabla\times\mathbf{E}_1) -\mathbf{E}_1\cdot(\nabla\times\mathbf{H}_2^*),
\end{equation}
using the vector identity $\nabla\cdot\left(\mathbf{A}\times\mathbf{B}\right)=\mathbf{B}\cdot\left(\nabla\times\mathbf{A}\right)-\mathbf{A}\cdot\left(\nabla\times\mathbf{B}\right)$. Substituting in Maxwell's equations, we find
\begin{equation}
    \nabla\cdot\left(\mathbf{E}_1\times\mathbf{H}^*_2\right) = -i\omega\mu_0\mathbf{H}_2^*\cdot \mathbf{H}_1+i\omega\mathbf{E}_1\cdot\epsilon^*(x,y)\cdot\mathbf{E}_2^*,\label{eqn:divGc}
\end{equation}
We may eliminate the magnetic fields from Eqn.~\ref{eqn:divGc} by evaluating the conjugate of this expression with the mode numbers interchanged ($\mathbf{G}_r=\mathbf{E}_2^*\times\mathbf{H}_1$),
\begin{equation}
    \nabla\cdot\left(\mathbf{E}_2^*\times\mathbf{H}_1\right) = i\omega\mu_0\mathbf{H}_1\cdot \mathbf{H}_2^*-i\omega\mathbf{E}_2^*\cdot\epsilon(x,y)\cdot\mathbf{E}_1.\label{eqn:divGr}
\end{equation}
Adding together Eqns.~\ref{eqn:divGc}-\ref{eqn:divGr}, we arrive at the conjugated  reciprocity equation
\begin{equation}
    \nabla\cdot\left(\mathbf{E}_2^*\times\mathbf{H}_1+\mathbf{E}_1\times\mathbf{H}^*_2\right)=-i\omega\left(\mathbf{E}_2^*\cdot\epsilon(x,y)\cdot\mathbf{E}_1-\mathbf{E}_1\cdot\epsilon^*(x,y)\cdot\mathbf{E}_2^*\right),
\end{equation}
which can be rewritten as
\begin{equation}
    \nabla\cdot\left(\mathbf{E}_2^*\times\mathbf{H}_1+\mathbf{E}_1\times\mathbf{H}^*_2\right)=-i\omega\mathbf{E}_2^*\cdot\left(\epsilon(x,y)-\epsilon^\dagger(x,y)\right)\cdot\mathbf{E}_1,\label{eqn:conjugatedreciprocity}
\end{equation}
where $\epsilon_{ij}^\dagger = \epsilon_{ji}^*$. For a nonabsorbing medium $\epsilon^\dagger(x,y)=\epsilon(x,y)$, and therefore 
\begin{equation}
\nabla\cdot\left(\mathbf{E}_1\times\mathbf{H}^*_2 + \mathbf{E}_2^*\times\mathbf{H}_1\right)=0.
\end{equation}

We now derive mode orthogonality by considering two different modes, labeled $\mu$ and $\nu$, with associated fields
\begin{subequations}
\begin{align}
    \mathbf{E}_1=\mathbf{E}_\mu(x,y)\exp(-ik_\mu z),&\qquad \mathbf{H}_1=\mathbf{H}_\mu(x,y)\exp(-ik_\mu z),\label{eqn:e_ccr}\\ \mathbf{E}_2=\mathbf{E}_\nu(x,y)\exp(-ik_\nu z),&\qquad \mathbf{H}_2=\mathbf{H}_\nu(x,y)\exp(-ik_\nu z).\label{eqn:ebar_ccr}
\end{align}
\end{subequations}
Substituting Eqns.~\ref{eqn:e_ccr}-\ref{eqn:ebar_ccr} into Eqn.~\ref{eqn:conjugatedreciprocity}, and applying the two dimensional divergence theorem, we find
\begin{equation}
    (k_\mu-k_\nu)\exp\left(-i(k_\mu-k_\nu)z\right)\int_{A_\infty}\left(\mathbf{E}_\mu\times\mathbf{H}^*_\nu+\mathbf{E}^*_\nu\times\mathbf{H}_\mu\right)\cdot\hat{z}dxdy=0.\label{eqn:recip1}
\end{equation}
The exponential term in Eqn.~\ref{eqn:recip1} does not contribute any meaningful insights, and will be dropped in all further discussion unless otherwise noted. When $k_\mu\neq k_\nu$, the integral must vanish. Conversely, when $k_\mu=k_\nu$ the integral can be evaluated using Poynting's theorem,
\begin{eqnarray}
\frac{1}{2}\int_{A}\mathrm{Re}\left(\left(\mathbf{E}_\mu\times \mathbf{H}_\mu^*\right) \cdot \hat{z}\right)dxdy =\mathrm{P},\label{eqn:poynting}
\end{eqnarray}
where, $\mathrm{P}$ is an arbitrary normalization constant. By convention, we choose to normalize the fields such that $\mathrm{P}=1$~W, and therefore the power contained in each mode of Eqn.~\ref{eqn:modesE} is given by $\mathrm{P}|a_\mu|^2$. It can be shown using the orthogonality relations derived below that the total power contained in the waveguide modes is given simply by $\sum_\mu P|a_\mu|^2$.

Repeating this derivation with a backwards-propagating mode ($k_{-\nu}=-k_\nu$, $\mathbf{E}_{t,-\mu}=\mathbf{E}_{t,\mu}$, $\mathbf{H}_{t,-\mu}=-\mathbf{H}_{t,\mu}$), we have
\begin{equation}
    (k_\mu+k_\nu)\exp\left(-i(k_\mu-k_\nu)z\right)\int_{A_\infty}\left(-\mathbf{E}_\mu\times\mathbf{H}^*_\nu+\mathbf{E}^*_\nu\times\mathbf{H}_\mu\right)\cdot\hat{z}dxdy=0.\label{eqn:recip2}
\end{equation}
Combining Eqn.~\ref{eqn:recip2} with Eqn.~\ref{eqn:recip1}, we find the orthogonality relations,
\begin{eqnarray}
\frac{1}{2}\int_{A}\mathrm{Re}\left(\left(\mathbf{E}_\mu\times \mathbf{H}_\nu^*\right) \cdot \hat{z}\right)dxdy =\mathrm{P}\delta_{\mu,\nu}.\label{eqn:orth}
\end{eqnarray}
Equation~\ref{eqn:orth} is one of the main results of this section. In addition to establishing the normalization of each waveguide mode, the orthogonality relations enable an efficient description of linear propagation within a waveguide. At any point $z$, the field in a waveguide can be decomposed the into the basis of waveguide modes using the orthogonality relations. The propagation of each mode $\mu$ from $z$ to any other position $z'$ is given by an overall phase, $\exp(-i k_\mu z)$, as in Eqn.~\ref{eqn:modesE}, and the field at point $z'$ can be reconstituted as a sum over all of the modes. In nonlinear waveguides we will continue to work in this basis of waveguide modes, and will use the orthogonality relations to derive the equations of motion describing the contribution of the nonlinear polarization to the evolution of each mode.

Having established the normalization conditions for each mode, it is convenient to express the mode profiles using dimensionless functions $\mathbf{e}(x,y)$ and $\mathbf{h}(x,y)$
\begin{subequations}
\begin{align}
\mathbf{E}_\mu(x,y) = \sqrt{\frac{2Z_0 \mathrm{P}}{n_\mu A_{\mathrm{mode},\mu}}}\mathbf{e}_\mu(x,y)\label{eqn:E_mode},\\
\mathbf{H}_\mu(x,y) = \sqrt{\frac{2n_\mu \mathrm{P}}{Z_0 A_{\mathrm{mode},\mu}}}\mathbf{h}_\mu(x,y)\label{eqn:H_mode},
\end{align}
\end{subequations}
where $Z_0 = 377$ Ohms is the impedance of free space and $n_\mu(\omega) = c k_\mu(\omega)/\omega$ is the effective refractive index of mode $\mu$ at frequency $\omega$. The choice of normalization for the dimensionless field distributions $\mathbf{e}(x,y)$ and $\mathbf{h}(x,y)$ is arbitrary; the convention presented here is only used to give clear physical intuition. We note here that as a consequence of (\ref{eqn:orth}), the area of mode $\mu$ is given by $A_{\mathrm{mode},\mu}=\int\mathrm{Re}(\mathbf{e}_\mu\times \mathbf{h}_\mu^*)\cdot \hat{z}dx dy$. If the field distributions are normalized such that the peak value of $\mathrm{Re}(\mathbf{e}_\mu\times \mathbf{h}_\mu^*) \cdot \hat{z}$ is unity, then $A_\mathrm{mode}$ is the ratio between the peak intensity of the waveguide mode and the average power contained the the mode. This definition of modal area is an intuitive measure of how tightly confined a mode is, with more tightly confined modes producing stronger nonlinear couplings. We note here that other choices of normalization for $\mathbf{e}_\mu$ only change the definition of the mode area, but ultimately yield the same result for any physically measurable value such as the nonlinear coupling. The definitions used here establish a correspondence between modes in nanowaveguides, which require a fully-vectorial description, and the transverse modes that occur in loosely-guiding waveguides (and in bulk media). As an example, for an x-polarized Gaussian beam propagating in free space, $\mathbf{e}(x,y)=\exp(-(x^2+y^2)/w^2)\hat{\mathbf{x}}$, $\mathbf{h}(x,y)=\exp(-(x^2+y^2)/w^2)\hat{\mathbf{y}}$, and $A_\mathrm{mode}=\pi w^2/2$. It should be clear from this example that $\mathbf{e}(x,y)$ and $\mathbf{h}(x,y)$ are simply dimensionless quantities that describe the shape of the field, and the resulting $A_\mathrm{mode}$ is the conventional mode area for a Gaussian beam.

We now use the complex reciprocity relations to derive closed-form expressions for the propagation constant of a mode in terms of $\epsilon(x,y)$ and the field distributions. Here, we again consider the vector $\mathbf{G}_c=\mathbf{E}_1\times\mathbf{H}_2^*$ and follow the same steps as above. The divergence of $\mathbf{G}_c$ is given by (Eqn.~\ref{eqn:divGc})
\begin{equation*}
    \nabla\cdot\left(\mathbf{E}_1\times\mathbf{H}^*_2\right) = -i\omega\mu_0\mathbf{H}_2^*\cdot \mathbf{H}_1+i\omega\mathbf{E}_1\cdot\epsilon^*(x,y)\cdot\mathbf{E}_2^*,
\end{equation*}
We now evaluate $\nabla\cdot\mathbf{G}_c$ for the case where the two fields are associated with a forward- and backward-propagating mode, respectively.
\begin{subequations}
\begin{align}
    \mathbf{E}_1=\mathbf{E}_\mu(x,y)\exp(-ik_\mu z),&\qquad \mathbf{H}_1=\mathbf{H}_\mu(x,y)\exp(-ik_\mu z),\label{eqn:e_beta}\\ \mathbf{E}_2=\mathbf{E}^-_\mu(x,y)\exp(ik_\mu z),&\qquad \mathbf{H}_2=\mathbf{H}^-_\mu(x,y)\exp(ik_\mu z).\label{eqn:ebar_beta}
\end{align}
\end{subequations}
Recalling that $E_\mu^{-} = E_{\mu,x}\hat{x} + E_{\mu,y}\hat{y} - E_{\mu,z}\hat{z}$, and that both $E_{\mu,x}$ and $E_{\mu,y}$ can be taken to be purely real, thereby causing $E_{\mu,z}$ to be purely imaginary, we have $\mathbf{E}_2^* = \mathbf{E}_\mu\exp(-i k_\mu z)$. Similarly, $\mathbf{H}_2^* = -\mathbf{H}_\mu\exp(-i k_\mu z)$. Substituting Eqns.~\ref{eqn:e_beta}-\ref{eqn:ebar_beta} into Eqn.~\ref{eqn:divGc} and applying the two dimensional divergence theorem, we find that the propagation constant of mode $\mu$ is given by
\begin{equation}
k_\mu(\omega) = \frac{\frac{1}{4}\int_{A_\infty} \left(\omega\mu_0\right) \mathbf{H}_\mu \cdot \mathbf{H}_\mu + \mathbf{E}_\mu\cdot \left(\omega\epsilon^*(x,y,\omega)\right)\cdot\mathbf{E}_\mu dxdy}{\frac{1}{2}\int_{A_\infty}\left(\mathbf{E}_\mu\times \mathbf{H}_\mu^*\right)\cdot \hat{z}dxdy}.\label{eqn:disp1}
\end{equation}

\subsubsection{Complex reciprocity: generalized form}\label{sec:generalized_reciprocity}

We now establish the generalized complex reciprocity relations, which allows for variations in the frequency between the two fields. This form of the complex reciprocity relations will be useful for developing orthogonality relations in resonators and for finding closed-form expressions for the inverse group velocity $\partial_\omega k_\mu$. In this case, the fields $\mathbf{E}_1$ and $\mathbf{H}_1$ correspond to solutions of Maxwell's equations for frequency $\omega_1$, and similarly, the fields $\mathbf{E}_2$ and $\mathbf{H}_2$ correspond to solutions of Maxwell's equations for frequency $\omega_2$,
\begin{subequations}
\begin{align}
    \mathbf{E}_1=\mathbf{E}_\mu(x,y,\omega_1)\exp(-ik_\mu(\omega_1)z),&\qquad \mathbf{H}_1=\mathbf{H}_\mu(x,y,\omega_1)\exp(-ik_\mu(\omega_1)z),\label{eqn:e_gccr}\\ \mathbf{E}_2=\mathbf{E}_\nu(x,y,\omega_2)\exp(-ik_\nu(\omega_2)z),&\qquad \mathbf{H}_2=\mathbf{H}_\nu(x,y,\omega_2)\exp(-ik_\nu(\omega_2)z).\label{eqn:ebar_gccr}
\end{align}
\end{subequations}
Repeating our analysis for $\mathbf{G}_c = \mathbf{E}_1\times\mathbf{H}_2^*$, we have
\begin{equation}
    \nabla\cdot\left(\mathbf{E}_1\times\mathbf{H}^*_2\right) = -i\omega_1\mu_0\mathbf{H}_2^*\cdot \mathbf{H}_1+i\omega_2\mathbf{E}_1\cdot\epsilon^*(x,y,\omega_2)\cdot\mathbf{E}_2^*.\label{eqn:divGc_generalized}
\end{equation}
Similarly, for $\mathbf{G}_r = \mathbf{E}_2^*\times\mathbf{H}_1$
\begin{equation}
    \nabla\cdot\left(\mathbf{E}_2^*\times\mathbf{H}_1\right) = i\omega_2\mu_0\mathbf{H}_1\cdot \mathbf{H}_2^*-i\omega_1\mathbf{E}_2^*\cdot\epsilon(x,y,\omega_1)\cdot\mathbf{E}_1.\label{eqn:divGr_generalized}
\end{equation}
Adding together Eqns.~\ref{eqn:divGc_generalized}-\ref{eqn:divGr_generalized} to calculate $\mathbf{F}_c=\mathbf{E}_1\times\mathbf{H}_2^*+\mathbf{E}_2^*\times\mathbf{H}_1$, we have
\begin{equation}
    \nabla\cdot\mathbf{F}_c=-i\mu_0(\omega_1-\omega_2)\mathbf{H}_1\cdot\mathbf{H}_2^*-i(\mathbf{E}_2^*\cdot\omega_1\epsilon(x,y,\omega_1)\cdot\mathbf{E}_1-\mathbf{E}_1\cdot\omega_2\epsilon^*(x,y,\omega_2)\cdot\mathbf{E}^*_2).
\end{equation}
Applying the two dimensional divergence theorem yields
\begin{align}
    (k_\mu(\omega_1)-k_\nu(\omega_2))\int_{A_\infty}\big(\mathbf{E}_\mu\times\mathbf{H}^*_\nu &+ \mathbf{E}^*_\nu\times\mathbf{H}_\mu\big)\cdot\mathbf{dA}=\label{eqn:generalizedreciprocity}\\
    \int_{A_\infty}\mu_0(\omega_1-\omega_2)\mathbf{H}_\mu\cdot\mathbf{H}_\nu^* + \mathbf{E}_\nu^*\cdot\big(\omega_1 \epsilon(x,y,\omega_1) &-\omega_2\epsilon^\dagger(x,y,\omega_2)\big)\cdot \mathbf{E}_\mu dxdy.\nonumber
\end{align}
There are two key insights from Eqn.~\ref{eqn:generalizedreciprocity}. We first consider the case where $\mu\neq\nu$ and $k_\mu(\omega_1)=k_\nu(\omega_2)$, which will be relevant in resonators. In this case, we find an alternative version of the orthogonality relations,
\begin{equation}
    \int_{A_\infty}\mu_0(\omega_1-\omega_2)\mathbf{H}_\mu\cdot\mathbf{H}_\nu^*+\mathbf{E}_\nu^*\cdot\big(\omega_1\epsilon(x,y,\omega_1)-\omega_2 \epsilon^\dagger(x,y,\omega_2))\cdot \mathbf{E}_\mu dxdy=0.\label{eqn:spatialmodeorthogonality}
\end{equation}
The second insight from Eqn.~\ref{eqn:generalizedreciprocity} is a closed-form expression for the inverse group velocity of mode $\mu$. Assuming $\mu=\nu$ and a lossless medium, dividing both sides by $\omega_1-\omega_2$ and taking the limit as $\omega_2\rightarrow\omega_1(=\omega)$, we find
\begin{equation*}
k'_\mu(\omega) = \frac{\int_{A_\infty}\mu_0\mathbf{H}_\mu\cdot\mathbf{H}_\mu^*+\mathbf{E}_\mu^*\cdot\partial_\omega(\omega\epsilon(x,y,\omega))\cdot\mathbf{E}_\mu dxdy}{\int_{A_\infty}\left(\mathbf{E}_\mu(x,y,\omega)\times\mathbf{H}^*_\mu(x,y,\omega)+\mathbf{E}^*_\mu(x,y,\omega)\times\mathbf{H}_\mu(x,y,\omega)\right)\cdot\hat{\mathbf{z}}dxdy}.
\end{equation*}
\begin{equation}
k'_\mu(\omega) = \frac{\frac{1}{4}\int_{A_\infty}\mu_0\mathbf{H}_\mu\cdot\mathbf{H}_\mu^*+\mathbf{E}_\mu^*\cdot\partial_\omega(\omega\epsilon(x,y,\omega))\cdot\mathbf{E}_\mu dxdy}{\frac{1}{2}\int_{A_\infty}\mathrm{Re}\left(\mathbf{E}_\mu(x,y,\omega)\times\mathbf{H}^*_\mu(x,y,\omega_2)\right)\cdot\hat{\mathbf{z}}dxdy}.\label{eqn:disp2}
\end{equation}
We note here that Eqn.~\ref{eqn:disp2} is the ratio of the energy density carried by an electromagnetic wave in a dispersive dielectric to the power flux~\cite{haus1984waves,Raymer2020}.

\subsection{Nonlinear coupling}\label{sec:kappa}

Having reviewed waveguide modes, their dispersion relations, and their normalization, we now consider nonlinear interactions between these modes. The treatment used in the following sections accounts for the fully-vectorial nature of the modes~\cite{Fejer1986,Kolesik2004}, with each field component of $\mathbf{E}_\mu$ coupled together by the full nonlinear tensor, $\chi^{(2)}_{ijk}$, of the media that comprise the waveguide. We note here that for SHG, $d_{ijk}=\chi^{(2)}_{ijk}/2$ is commonly used for tabulated values of the nonlinear susceptibility.

The derivation of the coupled-wave equations follows the treatment of the previous section, where complex reciprocity was used to establish the orthogonality relations between waveguide modes. In this case, we will consider reciprocity between two solutions to Maxwell's equations: $\mathbf{E}_1$ and $\mathbf{H}_1$, which evolve in the presence of a nonlinear polarization, $\mathbf{P}_\text{NL}$, and $\mathbf{E}_2$ and $\mathbf{H}_2$, which are solutions to Maxwell's equations in the absence of a nonlinear polarization. The nonlinear polarization is incorporated in the Amp\`{e}re-Maxwell curl equation,
\begin{equation}
\nabla\times \mathbf{H}(x,y,z,\omega)=i\omega\epsilon(x,y,\omega)\cdot\mathbf{E}(x,y,z,\omega)+i\omega \mathbf{P}_\mathrm{NL}(x,y,\omega).
\end{equation}
When $|\mathbf{P}_\mathrm{NL}(x,y,\omega)|$ is weak compared to $|\epsilon(x,y,\omega)\cdot\mathbf{E}(x,y,z,\omega)|$, $\mathbf{E}_1$ and $\mathbf{H}_1$ can be expanded as linear combination of waveguide modes, $\mathbf{E}_\mu$ and $\mathbf{H}_\mu$, where the mode expansion coefficients $a_\mu(z)$ now evolve during propagation due to the presence of $\mathbf{P}_\text{NL}$. We therefore generalize equations~\ref{eqn:modesE}-\ref{eqn:modesB} to include the evolution of $a_\mu(z)$ by assuming an ansatz of the form $\mathbf{E}_1 = \sum_\mu a_\mu(z) \mathbf{E}_\mu(x,y,\omega)\exp(-i k_\mu z)$, and $\mathbf{H}_1 = \sum_\mu a_\mu(z) \mathbf{H}_\mu(x,y,\omega)\exp(-i k_\mu z)$, with similar expressions for $\mathbf{D}$ and $\mathbf{B}$. We choose $\mathbf{E}_2 = \mathbf{E}_\nu(x,y,\omega)\exp(-i k_\nu z)$ to be a pure transverse mode, which allows projection of a single mode amplitude from the sum in $\mathbf{E}_1$.

As before, we evaluate the divergence of $\mathbf{E}_1\times\mathbf{H}_2^*$
\begin{equation}
    \nabla\cdot\left(\mathbf{E}_1\times\mathbf{H}_2^*\right) = \mathbf{H}_2^*\cdot(\nabla\times\mathbf{E}_1) -\mathbf{E}_1\cdot(\nabla\times\mathbf{H}_2^*).
\end{equation}
Substituting in Maxwell's equations, we have
\begin{equation}
    \nabla\cdot\left(\mathbf{E}_1\times\mathbf{H}_2^*\right) = -i\omega\mu_0\mathbf{H}_2^*\cdot \mathbf{H}_1+i\omega\mathbf{E}_1\cdot\left(\epsilon^*(x,y)\cdot\mathbf{E}_2^*\right).\label{eqn:divGcNL}
\end{equation}
Similarly, conjugating and interchanging these fields we find
\begin{equation}
    \nabla\cdot\left(\mathbf{E}_2^*\times\mathbf{H}_1\right) = i\omega\mu_0\mathbf{H}_1\cdot \mathbf{H}_2^*-i\omega\mathbf{E}_2^*\cdot\left(\epsilon(x,y)\cdot\mathbf{E}_1 + \mathbf{P}_\mathrm{NL}\right).\label{eqn:divGrNL}
\end{equation}
Adding together Eqns.~\ref{eqn:divGcNL}-\ref{eqn:divGrNL} yields
\begin{equation}
    \nabla\cdot\left(\mathbf{E}_1\times\mathbf{H}_2^*+\mathbf{E}_2^*\times\mathbf{H}_1\right) = -i\omega\mathbf{E}_2^*\cdot\left(\epsilon(x,y)-\epsilon^\dagger(x,y)\right)\cdot\mathbf{E}_1 -i\omega\left(\mathbf{E}_2^*\cdot\mathbf{P}_\mathrm{NL}\right).\label{eqn:recipNL}
\end{equation}
We now assume a lossless non-gyrotropic medium ($\epsilon(x,y) = \epsilon^\dagger(x,y)$), integrate both sides over the x-y plane, and apply the two-dimensional divergence theorem ($\int_S \nabla\cdot\mathbf{v}(x,y) dx dy = \partial_z \int_S \mathbf{v}(x,y)\cdot\hat{\mathbf{z}} dx dy$) to relate the evolution of the fields along $z$ to the nonlinear polarization,
\begin{equation}
    \partial_z\int_{A_\infty}\left(\mathbf{E}_1\times\mathbf{H}_2^*+\mathbf{E}_2^*\times\mathbf{H}_1\right)\cdot d\mathbf{A} = -i\omega\int_{A_\infty}\left(\mathbf{E}_2^*\cdot\mathbf{P}_\mathrm{NL}\right)dA.\label{eqn:recipNL_integrated}
\end{equation}
The left hand side of Eqn.~\ref{eqn:recipNL_integrated} can be evaluated using our ansatz for the fields. Starting with $\mathbf{E}_1\times\mathbf{H}_2^*=\sum_{\mu}a_\mu(z) \mathbf{E}_\mu\times\mathbf{H}_\nu^*\exp\left(-i(k_\mu-k_\nu)z\right)$, we find
\begin{align}
    \partial_z\left(\mathbf{E}_1\times\mathbf{H}_2^*\right)& = \label{eqn:NLrecip1}\\
    \sum_{\mu} & a_\mu(z) \left(\frac{\partial_z a_\mu(z)}{a_\mu(z)} - i(k_\mu - k_\nu) \right)\mathbf{E}_\mu\times\mathbf{H}^*_\nu\exp\left(-i(k_\mu-k_\nu)z\right)\nonumber.
\end{align}
Similarly, we evaluate $\mathbf{E}_2^*\times\mathbf{H}_1=\sum_{\mu} a_\mu(z)\mathbf{E}_\nu^*\times\mathbf{H}_\mu\exp\left(-i(k_\mu-k_\nu)z\right)$ to find
\begin{align}
    \partial_z\left(\mathbf{E}_2^*\times\mathbf{H}_1\right)& = \label{eqn:NLrecip2}\\
    \sum_{\mu} &  a_\mu(z)\left(\frac{\partial_z a_\mu(z)}{a_\mu(z)} - i(k_\mu - k_\nu) \right)\mathbf{E}^*_\nu\times\mathbf{H}_\mu\exp\left(-i(k_\mu-k_\nu)z\right).\nonumber
\end{align}
We add Eqns.~\ref{eqn:NLrecip1}-\ref{eqn:NLrecip2} and evaluate the integral in Eqn.~\ref{eqn:recipNL_integrated} using the orthogonality relations. This integral is only nonzero for terms where $\mu=\nu$, which yields
\begin{equation}
    \partial_z\int_{A_\infty}\left(\mathbf{E}_1\times\mathbf{H}_2^*+\mathbf{E}_2^*\times\mathbf{H}_1\right)\cdot d\mathbf{A} = 4\mathrm{P} \partial_z a_\nu(z),\label{eqn:modeevolution}
\end{equation}
where $\mathrm{P}= 1\,\mathrm{W} = \frac14\int_{A_\infty}\left(\mathbf{E}_\mu\times\mathbf{H}_\mu^* + \mathbf{E}^*_\mu\times\mathbf{H}_\mu\right)\cdot d\mathbf{A}$. Comparing the right-hand sides of Eqn.~\ref{eqn:modeevolution} and Eqn.~\ref{eqn:recipNL_integrated} we find that $a_\nu(z)$ evolves as
\begin{equation}
\partial_z a_{\nu}(z,\omega) = \frac{-i\omega}{4\mathrm{P}}e^{i k_{\nu}z}\int \mathbf{E}_{\nu}^*(x,y,\omega) \cdot \mathbf{P}_\mathrm{NL}(x,y,\omega) dx dy. \label{eqn:CWE}
\end{equation}
Eqn.~\ref{eqn:CWE} is the main result of this section, and can be used to derive the evolution of \emph{any} transverse mode, $a_\nu(z)$, in the presence of an arbitrary nonlinear polarization.



For second-harmonic generation in the limit where one pair of modes is close to phasematching, we calculate $P_\mathrm{NL,2}$ using one mode, $\mu$, for the fundamental at frequency $\omega$ and one mode, $\nu$, for the second harmonic at frequency $2\omega$. More general expressions are obtained simply by summing over all possible mode pairs. For the remainder of this section, the field amplitudes associated with the relevant mode of fundamental and second harmonic will be referred to as $a_{\omega}$ and $a_{2\omega}$ for the fundamental and second harmonic, respectively. In this case, the nonlinear polarization is given by
\begin{eqnarray}
\mathbf{P}_\mathrm{NL,\omega} = 2\epsilon_0 d_\mathrm{eff} a_{\nu,2\omega}(z)a_{\mu,\omega}^*(z)\sum_{jk}\bar{d}_{ijk}\mathrm{E}_{\nu,j}(x,y,2\omega)\mathrm{E}_{\mu,k}^*(x,y,\omega)e^{-i(k_{\nu,2\omega}-k_{\mu,\omega})z}\label{PNL1}\\
\mathbf{P}_\mathrm{NL,2\omega} = \epsilon_0  d_\mathrm{eff} a_{\mu,\omega}^2(z)\sum_{jk}\bar{d}_{ijk}\mathrm{E}_{\mu,j}(x,y,\omega)\mathrm{E}_{\mu,k}(x,y,\omega)e^{-2i k_{\omega} z}\label{PNL2}
\end{eqnarray}
where $i,j,k\in\lbrace x,y,z\rbrace$. For nonlinear interactions between modes polarized predominantly along the crystalline Z-axis in lithium niobate, $d_\mathrm{eff} = \frac{2}{\pi}d_{33}$ is the effective nonlinear coefficient for a 50\% duty cycle periodically poled waveguide, and $\bar{d}_{ijk}$ is the normalized $\chi^{(2)}$ tensor. This is expressed using contracted notation~\cite{Nye1985} in the coordinates of the crystal as
\begin{equation*}
\bar{d}_{iJ}=\frac{1}{d_{33}}\left(
\begin{array}{c c c c c c}
0 & 0 & 0 & 0 & d_{15} & d_{16}\\
d_{16} & -d_{16} & 0 & d_{15} & 0 & 0\\
d_{15} & d_{15} & d_{33} & 0 & 0 & 0
\end{array}
\right).
\end{equation*}

We arrive at the coupled-wave equations for SHG by inserting Eqns. (\ref{PNL1}-\ref{PNL2}) into (\ref{eqn:CWE}), substituting the normalizations found in Eqn.~\ref{eqn:E_mode}-\ref{eqn:H_mode}, and defining $A_{\omega}=\sqrt{\mathrm{P}}a_{\omega}$
\begin{eqnarray}
\partial_z A_{\omega} = -i\kappa A_{2\omega}A_{\omega}^* e^{-i\Delta k z},\label{eqn:A.CWE.1}\\
\partial_z A_{2\omega} = -i\kappa^* A_{\omega}^2 e^{i\Delta k z}.\label{eqn:A.CWE.2}
\end{eqnarray}
The nonlinear coupling, $\kappa$, and the associated effective area are given by 
\begin{align}
\kappa &= \frac{\sqrt{2 Z_0} \omega d_\mathrm{eff}}{c n_{\omega} \sqrt{A_{\mathrm{eff}}n_{2\omega}}}\exp(-i\phi_\kappa),\label{eqn:kappa_appendix}\\
A_\mathrm{eff} &=\frac{A_{\mathrm{mode,}\omega}^2 A_{\mathrm{mode,}2\omega}}{\left|\mathcal{O}_{\nu\mu\mu}\right|^2},\label{eqn:Aeff}\\
\mathcal{O}_{\nu\mu\mu} = \int_{A_\infty} \sum_{i,j,k} \bar{d}_{ijk}e^*_{\nu,i}&(x,y,2\omega)e_{\mu,j}(x,y,\omega)e_{\mu,k}(x,y,\omega) dx dy.
\end{align}
We emphasize here that all of the geometric contributions to $\kappa$ are contained in the effective area, $A_\text{eff}$, and that the numerator and denominator in Eqn.~\ref{eqn:Aeff} contain the same number of field envelopes. Therefore, any overall scale factor of the field distributions is cancelled, which greatly simplifies numerical calculations of $\kappa$. In practice, any electric field found using a numerical mode solver can be used for $\mathbf{e}_\mu$, as this only introduces an overall scale factor. Similarly, simply dividing the magnetic fields output from the same mode solver by $Z_0/n_\mu$ will produce a valid $\mathbf{h}_\mu$, with the same overall scale factor as $\mathbf{e}_\mu$. Using these fields in Eqn.~\ref{eqn:Aeff} will yield the correct nonlinear coupling, albeit with rescaled values for $A_\text{mode}$. 

The expression for the coupling coefficient $\kappa$ can be further simplified by noting that the phase of $\kappa$ is arbitrary. In general, $\kappa$ is complex in a nanophotonic waveguide due to coupling between the purely real transverse components of the fields with the purely imaginary z-component of the fields. The phase of the nonlinear coupling is determined by the field overlap integral, $\phi_\kappa=\arg(\mathcal{O}_{\nu\mu\mu})$, and can be neglected without loss of generality. When $\phi_\kappa$ is nonzero, the nonlinear coupling imparts a small phase shift between each of the interacting envelopes, but does not contribute any meaningful change in the resulting nonlinear dynamics. We can remove this phase from the coupled-wave equations by shifting phase reference of the second harmonic, $A_{2\omega}(z)\rightarrow A_{2\omega}(z)\exp(-i\phi_k)$.

The usual figure of merit for a nonlinear waveguide is the normalized efficiency, $\eta_0=\kappa^2$, which determines the power and device length needed to achieve efficient conversion; devices with larger $\eta_0$ can operate with either less power or shorter propagation length. The smallest possible effective area for a given wavelength is comparable to $A_\mathrm{eff}\sim (\lambda/n)^2$. Given the scale invariance of Maxwell's equations, the $A_\mathrm{eff}$ of any given device scales as $\lambda^2$, provided that all of the dimensions of the waveguide are rescaled. Therefore, we expect $\eta_0$ to exhibit a quartic scaling with frequency as given designs are rescaled to shorter wavelengths, with a factor of $\omega^2$ coming from the explicit $\omega$-dependence of $\kappa$, and another factor of $\omega^2$ coming from $A_\mathrm{eff}$. In practice, the scaling of $\eta_0$ for a given waveguide is slightly greater than $\omega^4$ due to the dispersion of $d_\mathrm{eff}$.

\subsection{Dispersive pulse propagation}\label{sec:pulse_prop}

Thus far, our treatment of nonlinear interactions has focused on quasi-continuous wave limits, where each interacting harmonic contains a single spectral mode. Our goal is to generalize this treatment to coupled-wave equations of the form $\partial_z A_\omega(z,t)$ that describe the evolution of broadband pulses in nonlinear waveguides. To facilitate this picture, we first review linear pulse propagation. The following section will combine both linear and nonlinear pulse propagation to derive the coupled-wave equations for ultrafast pulses.

We begin by Fourier transforming the time-domain electric field,
\begin{equation}
    \mathbf{E}(\mathbf{r},t)\equiv\int_{-\infty}^{\infty}\mathbf{E}(\mathbf{r},\omega)\exp(i\omega t)\frac{d\omega}{2\pi} = \int_{0}^{\infty}\mathbf{E}(\mathbf{r},\omega)\exp(i\omega t)\frac{d\omega}{2\pi} + c.c.\label{eqn:E-fourier_2},
\end{equation}
where the latter form of Eqn.~\ref{eqn:E-fourier_2} follows from $\mathbf{E}(\mathbf{r},t)$ being real. Since the integral in Eqn.~\ref{eqn:E-fourier_2} can be taken over positive frequencies, we can express $\mathbf{E}(\mathbf{r},t)$ as the Fourier transform of a single-sided distribution $\mathbf{E}^+(\mathbf{r},\omega)=2\mathbf{E}(\mathbf{r},\omega)H(0)$, where $H(\omega)$ is the Heaviside step function. Expanding $\mathbf{E}(\mathbf{r},\omega)$ in terms of waveguide modes we have 
\begin{align}
    \mathbf{E}(\mathbf{r},t) &= \frac12\int_{-\infty}^{\infty}\sum_\mu a_{\mu}(\omega)\mathbf{E}_{\mu}(x,y,\omega)\exp\left(i\omega t-ik_\mu(\omega) z\right)\left(\frac{d\omega}{2\pi}\right)+c.c.,\label{eqn:E-phasor}
\end{align}
where $a_\mu(\omega)$ is a single-sided distribution, $a_\mu(\omega) = 0$ for $\omega < 0$.

In principle, Eqn.~\ref{eqn:E-phasor} is sufficient for calculating the evolution of the time-domain electric field, provided $a_\mu(\omega)$ is known. In practice, since $a_\mu(\omega)$ is typically localized around a large frequency, $\omega_0$, $\mathbf{E}(\mathbf{r},t)$ has extremely rapid phase variations on the timescale of $\omega_0^{-1}$ and length scale of $k_\mu^{-1}(\omega_0)$. We may gain much clearer insights by removing these rapid phase variations and studying the behaviors of a residual pulse envelope, $\tilde{a}_\mu(z,t)$. First, we use the Fourier shift theorem to rewrite Eqn.~\ref{eqn:E-phasor} in terms of the offset frequency $\Omega = \omega - \omega_0$,
\begin{align*}
    \mathbf{E}(\mathbf{r},t) &= \frac{1}{2}\int_{-\infty}^{\infty}\sum_\mu a_{\mu}(\Omega)\mathbf{E}_{\mu}(x,y,\Omega)\exp\left(i\Omega t-ik_\mu(\Omega) z + i\omega_0 t\right)\left(\frac{d\Omega}{2\pi}\right)+c.c.
\end{align*}
We define the slowly-varying envelope $\tilde{a}_{\mu}(z,\Omega)$ by introducing a rotating frame,
\begin{align}
    \tilde{a}_{\mu}(z,\Omega) &= a_{\mu}(\Omega)\exp\left(-i\left[k_\mu(\Omega)-k_{\mu,0} - v_\text{g,ref}^{-1}\Omega\right] z\right),\label{eqn:envelope_def}\\
    &= \tilde{a}_{\mu}(0,\Omega)\exp\left(-i\left(k_\mu(\Omega)-k_{\mu,0} - v_\text{g,ref}^{-1}\Omega\right) z\right).\nonumber
\end{align}
where $k_{\mu,0} = k_\mu(\Omega = 0)$, and $v_\text{g,ref}$ is an arbitrary reference group velocity chosen to shift the time coordinate to be co-moving with the pulse envelope. With Eqn.~\ref{eqn:envelope_def} for the field envelope, Eqn.~\ref{eqn:E-phasor} becomes
\begin{align*}
    \mathbf{E}(\mathbf{r}, t') &= \frac12\int_{-\infty}^{\infty}\sum_\mu \tilde{a}_{\mu}(\Omega)\mathbf{E}_{\mu}(x,y,\omega)\exp\left(i\Omega t'+i\omega_0 t-ik_{\mu,0} z\right)\left(\frac{d\omega}{2\pi}\right)+c.c.,
\end{align*}
where $t' = t - v_\text{g,ref}^{-1}z$. In the context of linear propagation, choosing $v_\text{g,ref} = v_\text{g}(\omega_0)$ renders the field envelope stationary in the absence of second- and higher-order dispersion. Later, when studying nonlinear interactions between pulse envelopes, we will see that this reference velocity can be chosen to greatly simplify the equations of motion, and in practice our choice of reference velocity will depend on the particular problem being studied. We note that throughout the main text and this appendix, various choices of reference velocity will be used. Rather than explicitly defining many separate time coordinates, (\textit{e.g.} $t'$, $t''$, $t'''$, ...), we will instead suppress the prime on the shifted coordinate and simply describe our choice of reference velocity in each context. The terms ``lab frame'', and ``non-moving'' frame ($v_\text{g,ref} = 0$) will be used interchangeably.

The time domain pulse envelope is given by
\begin{equation}
    \tilde{a}_{\mu}(z,t) = \int_{-\infty}^{\infty}\tilde{a}_{\mu}(z,\Omega)\exp(i\Omega t)\frac{d\Omega}{2\pi},
\end{equation}
To find the propagation equation for $\tilde{a}_{\mu}(z,t)$, we first take the derivative of of Eqn.~\ref{eqn:envelope_def} with respect to $z$ to find a propagation equation for $\tilde{a}_{\mu}(z,\Omega),$
\begin{align}
\partial_z \tilde{a}_{\mu}(z,\Omega) &= -i\left(k_\mu(\omega_0 + \Omega) - k_{\mu,0} - v_\text{g,ref}^{-1}\Omega\right)\tilde{a}_{\mu}(z,\Omega).\label{eqn:a.prop1}
\end{align}
We convert Eqn.~\ref{eqn:a.prop1} to a propagation equation for the time-domain pulse envelope $\tilde{a}_{\mu}(z,t)$ with an inverse Fourier transform,
\begin{align}
\partial_z \tilde{a}_{\mu}(z,t) &= \int -i\left(k_\mu(\omega_0 + \Omega) - k_{\mu,0} - v_\text{g,ref}^{-1}\Omega\right)\tilde{a}_{\mu}(z,\Omega)\exp(i\Omega t)\frac{d\Omega}{2\pi}.\label{eqn:a.prop2}
\end{align}
To evaluate Eqn.~\ref{eqn:a.prop2} we will make use of the Fourier rule for derivatives, $\partial_t^n \leftrightarrow (i\Omega)^n$, by first series expanding the dispersion relations
\begin{align}
    k_\mu(\Omega) - k_{\mu,0} - v_\text{g,ref}^{-1}\Omega &= \left(\partial_\Omega k_\mu - v_\text{g,ref}^{-1}\right) \Omega + \frac12\partial_\Omega^2 k_\mu \Omega^2 + \frac16\partial_\Omega^3k_\mu\Omega^3 + ...\\
    &= \left(\partial_\Omega k_\mu - v_\text{g,ref}^{-1}\right) \Omega + D_{\text{int},\mu}(i\Omega),
\end{align}
where $D_{\text{int},\mu}(i\Omega) = \sum_{m=2}^{\infty} \frac{(-i)^m}{m!}
(i\Omega)^m k_\mu^{(m)}$ is the dispersion operator for mode $\mu$, and $k_\mu^{(m)} = \partial_\Omega k_\mu(\Omega)|_{\Omega=0}$ is the $m$th derivative of $k_\mu(\Omega)$, evaluated at $\Omega = 0$. With this form of the propagation equation, we may now evaluate Eqn.~\ref{eqn:a.prop2} to find
\begin{align}
\partial_z \tilde{a}_{\mu}(z,t) &= -(k_\mu'-v_\text{g,ref}^{-1})\partial_t\tilde{a}_{\mu}(z,t) -iD_{\text{int},\mu}(\partial_t)\tilde{a}_{\mu}(z,t).\label{eqn:a.prop3}
\end{align}
Equation~\ref{eqn:a.prop3} is the main result of this section and describes the evolution of a time-domain envelope with respect to propagation along $z$ in a linear waveguide.

Once Eqn.~\ref{eqn:envelope_def} or Eqn.~\ref{eqn:a.prop3} is solved for the slowly-varying envelopes, the steps described herein can be reversed to obtain $\mathbf{E}(\mathbf{r},t)$ using Eqn.~\ref{eqn:E-phasor}, with an identical expression holding for the magnetic fields. In practice, since the fields associated with each eigenmode are weak functions of frequency across the typical bandwidth of a pulse, Eqn.~\ref{eqn:E-phasor} can be approximated as
\begin{align}
    \mathbf{E}(\mathbf{r},t) &\approx \frac{1}{2}\sum_\mu \tilde{a}_{\mu}(z,t-v_\text{g,ref}^{-1}z)\mathbf{E}_{\mu}(x,y,\omega_0)\exp\left(i\omega_0 t-ik_{\mu,0} z\right)+c.c.\label{eqn:E-phasor2}
\end{align}
When Eqn.~\ref{eqn:E-phasor2} is valid $\tilde{a}_{\mu}(z,t)$ can be used to calculate the instantaneous power envelope in mode $\mu$, obtained by cycle-averaging the Poynting flux,
\begin{align}
    \frac{1}{T}\int_0^T\int_{A_\infty}\mathbf{E}(x,y,t)\times\mathbf{H}(x,y,t)\cdot d\mathbf{A}dt = \mathrm{P}\sum_\mu |\tilde{a}_{\mu}(z,t)|^2 ,
\end{align}
where $T={\frac{2\pi}{\omega_0}}$ and the normalization $\mathrm{P}=1$ Watt was introduced in Sec.~\ref{sec:reciprocity}. As with our previous analysis of nonlinear coupling, we introduce the instantaneous power envelope associated with the bandwidth contained in mode $\mu$ as $A_{\mu}(z,t)=\sqrt{P}\tilde{a}_{\mu}(z,t)$. In the context of nonlinear optics, for typical systems only one or two transverse modes will be relevant for each interacting wave, and therefore the mode index $\mu$ will often be dropped for compactness.

We close this section by briefly addressing practical considerations for how to calculate and to use $\tilde{a}_{\mu}$. First, we note that for pulses with a bandwidth much smaller than the carrier frequency, $D_{\text{int},\mu}$ is simply given by truncating the Taylor series of $k_\mu(\omega)$ at low order. Second order dispersion is sufficient for analysing most problems, though occasionally we will see throughout this paper that third- or fourth-order terms become relevant. We note, however, that the Taylor series of $k_\mu(\omega)$ truncated at a large but finite order should \emph{not} be used when simulating multi-octave effects. While this Taylor series can be evaluated, the radius of convergence is zero due to the underlying Sellmeier equations typically used to describe dispersive materials having real poles. For typical waveguides, the global behavior of $D_{\text{int},\mu}$ is usually best approximated with a Taylor series truncated around fourth or fifth order, and treating bandwidths beyond this approximation requires a different strategy altogether. For nonlinear propagation with multi-octave bandwidths, the best approach is split-step Fourier methods with the full $k_\mu(\omega)$ used to calculate linear propagation.

\subsection{The coupled-wave equations for short pulses}

In a nonlinear waveguide driven by short pulses we have multiple envelopes (one for each interacting wave), each of which comprise many frequencies. For each of these envelopes, the Fourier components $a_{\mu}(\omega)\mapsto a_{\mu}(z,\omega)$ now evolve with $z$ due to the nonlinear polarization $P_\text{NL}(\omega)$ according to Eqn.~\ref{eqn:CWE}, copied below for convenience,
\begin{equation}
\partial_z a_{\mu}(z,\omega) = \frac{-i\omega}{4\mathrm{P}}e^{i k_{\mu}(\omega) z}\int \mathbf{E}_{\mu}^*(x,y,\omega) \cdot \mathbf{P}_\mathrm{NL}(x,y,z,\omega) dx dy, \label{eqn:CWE_Omega}
\end{equation}
where $\mathbf{P}_\text{NL}(x,y,z,\omega)$ is calculated by integrating over all pairs of fields producing a contribution at frequency $\omega$. For broadband envelopes, coupled-wave equations are most easily derived using the constitutive relations for $\mathbf{P}_\text{NL}(x,y,z,\omega)$ in the frequency domain. Thus far, we have used $d_{ijk}$ for the constitutive relations of $\mathbf{P}_\text{NL}$, which is has been used historically for narrowband fields and weakly non-dispersive nonlinearities. In the context of broadband second harmonic generation, the nonlinear polarization around the fundamental is more conveniently given by the frequency-domain constitutive relation
\begin{align}
    \mathbf{P}_\text{NL}(x,y,z,\omega_1) = &\epsilon_0\int_{-\infty}^{\infty}\chi^{(2)}_{ijk}(\omega_1; \omega_2, \omega_2-\omega_1)E_{\nu,j}(x,y,\omega_2)E_{\mu,k}^*(x,y,\omega_2-\omega_1)\label{eqn:PNL_broadband}\\
    &\times a_{\nu}(z,\omega_2)a_{\mu}^*(z,\omega_2-\omega_1)\exp\left(-i(k_{\nu}(\omega_2)-k_{\mu}(\omega_2-\omega_1))z\right)\frac{d\omega_2}{2\pi}\nonumber,
\end{align}
where $\chi^{(2)}_{ijk} = 2 d_{ijk}$ for SHG. Our choice of $\chi^{(2)}_{ijk}$, rather than $d_{ijk}$, allows for easier accounting, since definitions of $d_{ijk}$ vary by factors of two when treating SHG and SFG, and the frequency integral in Eqn.~\ref{eqn:PNL_broadband} contains both. For notational consistence, we will later revert back to $d_{ijk}$ when the nonlinear coupling is assumed to be weakly dispersive. Defining $A_\mu = \sqrt{\mathrm{P}}a_\mu$, together with Eqn.~\ref{eqn:PNL_broadband}, Eqn.~\ref{eqn:CWE_Omega} becomes
\begin{align}
    \partial_z A_\mu(z,\omega_1) = -i\int_{-\infty}^{\infty}\kappa_{\mu\nu\mu}(\omega_1,\omega_2)A_{\nu}(z,\omega_2)A_{\mu}^*(z,\omega_2-\omega_1)\exp\left(-i\Delta k(\omega_1,\omega_2)z\right)\frac{d\omega_2}{2\pi},\label{eqn:broadband_CWE1}
\end{align}
where the phase-mismatch is given by $\Delta k(\omega_1,\omega_2) = k_\nu(\omega_2) - k_\mu(\omega_2-\omega_1) - k_\mu(\omega_1)$. The nonlinear coupling is given by identical expressions to the continuous-wave case, but now with the frequency dependence of the overlap integrals made explicit
\begin{align}
    \kappa_{\mu\nu\mu}(\omega_1,\omega_2) = &\frac{\sqrt{Z_0} \omega_1 \chi^{(2)}_\text{eff}(\omega_1; \omega_2, \omega_2-\omega_1)}{c \sqrt{2A_{\mathrm{eff}}(\omega_1,\omega_2)n_{\mu}(\omega_1)n_{\nu}(\omega_2)n_{\mu}(\omega_2-\omega_1)}},\label{eqn:kappa_broadband}\\
    A_\mathrm{eff}(\omega_1,\omega_2) = &\frac{A_\mathrm{mode,\mu}(\omega_1)A_\mathrm{mode,\nu}(\omega_2)A_\mathrm{mode,\mu}(\omega_2-\omega_1)}{\left|\mathcal{O}_{\mu\nu\mu}(\omega_1,\omega_2)\right|^2},\label{eqn:Aeff_broadband}\\
    \mathcal{O}_{\mu\nu\mu}(\omega_1,\omega_2) = &\int_{A_\infty} \sum_{i,j,k} \bigg(\bar{\chi}_{ijk}^{(2)}(\omega_1; \omega_2, \omega_2-\omega_1)e_{\mu,i}^*(x,y,\omega_1)\\
    &e_{\nu,j}(x,y,\omega_2)e_{\mu,k}(x,y,\omega_2-\omega_1)\bigg) dx dy,\nonumber
\end{align}
where $\chi^{(2)}_\text{eff}$ and $\bar{\chi}_{ijk}^{(2)}$ are defined analogously to $d_\text{eff}$ and $\bar{d}_{ijk}$. Equation~\ref{eqn:broadband_CWE1} is one of the main results of this section, and can generally be used to propagate the broadband coupled-wave equations for frequencies centered around the fundamental. A similar expression holds for the second-harmonic,
\begin{align}
    \partial_z A_\nu(z,\omega_2) = -i\int_{-\infty}^{\infty}\kappa_{\nu\mu\mu}(\omega_2,\omega_1)A_{\mu}(z,\omega_1)A_{\mu}(z,\omega_2-\omega_1)\exp\left(i\Delta k(\omega_1,\omega_2)z\right)\frac{d\omega_1}{2\pi},\label{eqn:broadband_CWE2}
\end{align}
where $\kappa_{\nu\mu\mu}(\omega_2,\omega_1)=\omega_2\kappa^*_{\mu\nu\mu}(\omega_1,\omega_2)/\omega_1$.

For many cases of interest, Eqns.~\ref{eqn:broadband_CWE1}-\ref{eqn:broadband_CWE2} can be greatly simplified by ignoring many of the dispersive terms contained in the integrand. The effective area, $A_\text{eff}$, is a weak function of $\omega_1$ and $\omega_2$ for frequencies near degeneracy, $\omega_2 \approx 2\omega_1$. This can be seen by the symmetry of factors such as $n_\mu(\omega_1)n_\mu(\omega_2-\omega_1)$ under the interchange $\omega_1 \leftrightarrow \omega_2-\omega_1$, which implies that these terms are maximized when $\omega_2 = 2\omega_1$. Since the derivative $\partial_{\omega_1}\left(n_\mu(\omega_1)n_\mu(\omega_2-\omega_1)\right)$ vanishes at the degenerate point, these functions vary slowly with increasing $\omega_1$ away from degeneracy. The same argument holds for $\mathcal{O_{\mu\nu\mu}}$ and $A_\mathrm{mode,\mu}(\omega_1)A_\mathrm{mode,\mu}(\omega_2-\omega_1)$. As a result, the dominant effect resulting from the dependence of $\kappa(\omega_1,\omega_2)$ on $\omega_2$ is a slight decrease of $A_\mathrm{mode,\nu}(\omega_2)A_\mathrm{mode,\mu}(\omega_2-\omega_1)$ with increasing $\omega_2$, which is partially cancelled by a corresponding decrease of $\mathcal{O}_{\mu\nu\mu}$. Together, these cancellations suggests that $\kappa(\omega_1,\omega_2)$ is well approximated by the value near degeneracy,
\begin{equation}
    \kappa_{\mu\nu\mu}(\omega_1=\omega + \Omega, \omega_2 = 2\omega)\approx \frac{\sqrt{2 Z_0} (\omega + \Omega) d_\mathrm{eff}}{c \sqrt{A_{\mathrm{eff}}(\omega,2\omega)n_{\nu}(2\omega)n_{\mu}^2(\omega)}}.
\end{equation}
The prefactor of $\omega + \Omega$ contributes a self-steepening term, and can often be neglected ($\omega+\Omega \approx \omega$) when considering pulses with duration greater than a few optical cycles. With the above approximations, namely ignoring dispersive contributions to $\kappa$ in the integrand, the coupled-wave equations become
\begin{align}
    \partial_z A_\mu(z,\omega_1) &\approx -i\kappa_{\mu\nu\mu}(\omega_1, 2\omega)\int_{-\infty}^{\infty}A_{\nu}(z,\omega_2)A_{\mu}^*(z,\omega_2 - \omega_1)\exp\left(-i\Delta k(\omega_1,\omega_2)z\right)d\omega_2,\label{eqn:broadband_CWE3}\\
    \partial_z A_\nu(z,\omega_2) &\approx -i\kappa_{\nu\mu\mu}(\omega_2, \omega)\int_{-\infty}^{\infty}A_{\mu}(z,\omega_1)A_{\mu}(z,\omega_2-\omega_1)\exp\left(i\Delta k(\omega_1,\omega_2)z\right)d\omega_1.\label{eqn:broadband_CWE4}
\end{align}

As with linear pulse propagation, we now define a rotating envelope $\tilde{A}_{\mu,\omega}(z,\Omega) = A_\mu(z,\omega+\Omega)\exp\left(-iD_{\text{int},\mu}(i\Omega)z\right)$ for the fundamental centered around $\omega$, and $\tilde{A}_{\nu,2\omega}(z,\Omega') = A_\nu(z,2\omega+\Omega')\exp\left(-i(\Delta k'\Omega'+D_{\text{int},\nu}(i\Omega')z)\right)$ for the second-harmonic centered around $2\omega$, where $\Delta k' = v_\mathrm{g,2\omega}^{-1}-v_\mathrm{g,\omega}^{-1}$ is the group-velocity mismatch between the interacting waves. This choice of rotating frame corresponds to setting $v_\text{g,ref} = v_{\text{g},\omega}$, which is useful for undepleted SHG, where the envelope of the fundamental in the co-moving frame is now constant. In some contexts, such as undepleted OPA, using the group velocity of the second-harmonic as the reference velocity of the fundamental yields clearer insights. We move between these rotating frames freely throughout the main text. With these definitions, Eqns.~\ref{eqn:broadband_CWE3}-\ref{eqn:broadband_CWE4} become convolution integrals,
\begin{align}
    \partial_z \tilde{A}_{\mu,\omega}(z,\Omega) \approx& -i\kappa(\omega+\Omega, 2\omega)\exp\left(-i\Delta k z\right)\int_{-\infty}^{\infty}\tilde{A}_{\nu,2\omega}(z,\Omega')\tilde{A}_{\mu,\omega}^*(z,\Omega' - \Omega)d\Omega'\nonumber\\
    &- iD_{\text{int},\mu}(i\Omega)\tilde{A}_{\mu,\omega}(z,\Omega),\label{eqn:broadband_CWE5}\\
    \partial_z \tilde{A}_{\nu,2\omega}(z,\Omega') \approx& -i\kappa(2\omega + \Omega', \omega)\exp\left(i\Delta k z\right)\int_{-\infty}^{\infty}\tilde{A}_{\mu,\omega}(z,\Omega)\tilde{A}_{\mu,\omega}(z,\Omega'-\Omega)d\Omega\nonumber\\
    &- (iD_{\text{int},\nu}(i\Omega')+i\Delta k'\Omega)\tilde{A}_{\nu,2\omega}(z,\Omega'),\label{eqn:broadband_CWE6}
\end{align}
where $\Delta k = k_\nu(2\omega) - 2k_\mu(\omega)$ is the phase-mismatch between the carrier frequencies of the envelopes. Finally, inverse Fourier-transforming Eqns.~\ref{eqn:broadband_CWE5}-\ref{eqn:broadband_CWE6}, we find the time-domain coupled-wave equations that describe the evolution of two interacting pulse envelopes,
\begin{subequations}
\begin{align}
\partial_z \tilde{A}_{\omega}(z,t) &= -i\kappa(1+\omega^{-1}\partial_t) \tilde{A}_{2\omega}\tilde{A}_{\omega}^* \exp(-i\Delta k z) - iD_{\text{int},\omega}(\partial_t) \tilde{A}_{\omega},\label{eqn:A.CWE.t.1}\\
\partial_z \tilde{A}_{2\omega}(z,t) &= -i\kappa(1+\omega^{-1}\partial_t) \tilde{A}_{\omega}^2 \exp(i\Delta k z) - \Delta k' \tilde{A}_{2\omega} - iD_{\text{int},2\omega}(\partial_t) \tilde{A}_{2\omega},\label{eqn:A.CWE.t.2}
\end{align}
\end{subequations}
where $\kappa$ is the continuous-wave coupling coefficient given by Eqn.~\ref{eqn:kappa_appendix}. Throughout the main text, we will drop the tildes that denote a rotating frame and drop the subscripts $\mu$ and $\nu$, since the relevant mode pairs will be clear by context. For all closed-form solutions, we neglect the self-steepening term, $\omega^{-1}\partial_t$. Equations~\ref{eqn:A.CWE.t.1} and \ref{eqn:A.CWE.t.2} are the main results of this Appendix, and are the equations of motion predominantly considered throughout the main text. This simplified time-domain version of the coupled-wave equations only relies on the assumption that the dispersion of $\kappa(\omega_1,\omega_2)$ within Eqns.~\ref{eqn:broadband_CWE1}-\ref{eqn:broadband_CWE2} can be neglected. For situations involving extremely broadband envelopes, the dispersion of $\kappa(\omega_1,\omega_2)$ can be greatly reduced by using frequency-dependent mode normalizations such as $A'_{\mu,\omega}(\Omega)=\sqrt{P/A_\mathrm{norm}(\Omega)}$~\cite{phillips2012broadband}. This approach yields the same equations of motion, now with a slightly different meaning attached to $A_{\mu}$.

\section{The coupled-wave equations for spatial modes}\label{sec:time-propagating-spatial-modes}

As discussed in Sec.~\ref{sec:Rosetta_stone}, the Heisenberg equations of motion in quantum optics take the form
\begin{align*}
    \partial_t\hat{a}=F_\mathrm{q}(\hat{a},\hat{a}^\dagger)
\end{align*}
with $F_\mathrm{q}(\hat{a},\hat{a}^\dagger)=\mathrm{i}[\hat{H},\hat{a}]$, whereas the classical coupled-wave equations take the form 
\begin{align*}
    \partial_z\alpha(z,t)=f_\mathrm{c}(\alpha(z,t),\alpha^*(z,t)).
\end{align*}
To establish the correspondence between classical and quantum nonlinear optics, we first need to reformulate classical nonlinear optics in terms of time-propagating spatial modes of the form,
\begin{align*}
    \partial_t\alpha=F_\mathrm{c}(\alpha,\alpha^*).
\end{align*}
rather than space-propagating temporal modes. We will see that this approach to nonlinear optics is formally equivalent to the conventional space-propagating approach, but with somewhat non-trivial boundary conditions. In practice this approach is rarely used for simulating classical nonlinear optics, but is an excellent stepping stone for linking the classical and quantum theories.

The approach taken here follows the treatment used throughout Sec.~\ref{sec:CWEs}, where the coupled-wave equations are derived using complex reciprocity relations. We note here that this reciprocity-based derivation is rather different than the usual Hamiltonian-based approach found throughout the literature. There are two common pitfalls of the Hamiltonian approach stemming from the use of constitutive relations $\mathbf{D}(\mathbf{E})$~\cite{Hillery1984,sipe2009photons,Quesada2017}. The most common mistake when using a Hamiltonian to derive the coupled-wave equations is the assumption that the components of $\mathbf{E}$, rather than $\mathbf{D}$, are the canonical momenta when the components of the vector potential $\mathbf{A}$ are taken to be canonical coordinates~\cite{Hillery1984}. Even when the correct Hamiltonian $\mathcal{H}$ is used, the equations of motion generated by this Hamiltonian, \textit{e.g.} $\partial_t B_i = \lbrace B_i, \mathcal{H}\rbrace$, will not yield Maxwell's equations when the wrong canonical momenta are used to calculate the Poisson bracket. The second common mistake is the assumption that the electromagnetic energy density contributed by the nonlinear polarization is given by $\mathbf{E}\cdot\mathbf{P}_\text{NL}$~\cite{Hillery1984,Quesada2017}. In reality, the contribution to the energy density by both the electric field and the material polarization is contained in the integral $\int \mathbf{E}\cdot d\mathbf{D}$. Both of these pitfalls lead to some subtleties when calculating the Hamiltonian for both dispersive and nonlinear dielectrics, where constitutive relations must now be given in terms of $\mathbf{E}(\mathbf{D})$ rather than the usual $\mathbf{D}(\mathbf{E})$~\cite{sipe2009photons, Quesada2017}. As a result, dispersion is typically treated using frequency derivatives of the inverse permittivity, and nonlinearity is now incorporated by expanding $\mathbf{E}$ as a power series in $\mathbf{D}$ with expansion coefficients corresponding to inverse susceptibilities, \textit{e.g.} $\eta^{(2)}_{ijk}$. By avoiding a Hamiltonian and working directly with Maxwell's equations, the reciprocity-based approach adopted here avoids these pitfalls, and can be used to derive coupled-wave wave equation for any constitutive relation, including the commonly encountered $\mathbf{P}_\text{NL}(\mathbf{E})$. Furthermore, by building upon the usual derivation of the orthogonality relations, this approach yields coupled-wave equations for $a_{\mu,m}(t)$ with relatively few algebraic manipulations when starting from the orthogonality relations established in linear media.

\subsection{Normal modes}

The time-propagating approach to nonlinear optics can be motivated by the observation that all physical devices have either periodic boundary conditions, as in a resonator, or a finite spatial extent, such as a waveguide. In either case, the spatial distribution of the fields is constrained to a finite domain $z\in[0,L]$, provided that $z$ is identified with the appropriate coordinate. In a waveguide, $z$ is the usual propagation coordinate, whereas in a ring resonator $z$ may be remapped to $s$, the angular coordinate around the ring. In any situation where the propagation coordinate is constrained to a finite or periodic domain, each field may be expressed using a Fourier series, $A_\omega(z,t)=\sum_m A_{\omega,m}(t) \exp(i k_m z)$ where $k_m = 2\pi m/L$. For a finite-domain signals, $A_\omega(z,t)$ can be obtained by windowing the periodic signal synthesized by a Fourier series with a rectangle function $\Pi_L(z)$. Here, the rectangle function is given by
\[
\Pi_L(z) = \begin{cases}
1 & z\in[0, L] \\
0 & \text{otherwise}
\end{cases}.
\]
Under these conditions, the solutions to Maxwell's equations for a uniform waveguide (or a ring resonator) now contain discrete $k_m$, corresponding to longitudinal modes of the resonator. The expansion of $\mathbf{E}(\mathbf{r},t)$ in terms of longitudinal and transverse modes takes a similar form as Eqn.~\ref{eqn:E-phasor},
\begin{subequations}
\begin{align}
\mathbf{E}(\mathbf{r},t) &= \frac12\left(\sum_{\mu,m}a_{\mu,m}\mathbf{E}_{\mu}(x,y,\omega_\mu(k_m))\exp(-i\omega_\mu(k_m)t + i k_m z) + c.c.\right),\label{eqn:spatial_mode_E}\\
\mathbf{D}(\mathbf{r},t) &= \frac12\left(\sum_{\mu,m}a_{\mu,m}\epsilon(x,y,\omega_\mu)\cdot\mathbf{E}_{\mu}(x,y,\omega_\mu(k_m))\exp(-i\omega_\mu(k_m)t + i k_m z) + c.c.\right),\label{eqn:spatial_mode_D}\\
\mathbf{H}(\mathbf{r},t) &= \frac12\left(\sum_{\mu,m}a_{\mu,m}\mathbf{H}_{\mu}(x,y,\omega_\mu(k_m))\exp(-i\omega_\mu(k_m)t + i k_m z)+ c.c.\right)\label{eqn:spatial_mode_H},\\
\mathbf{B}(\mathbf{r},t) &= \frac12\left(\sum_{\mu,m}a_{\mu,m}\mu_0\mathbf{H}_{\mu}(x,y,\omega_\mu(k_m))\exp(-i\omega_\mu(k_m)t + i k_m z)+ c.c.\right)\label{eqn:spatial_mode_B}.
\end{align}
\end{subequations}
These solutions can be generalized to more complicated structures, such as photonic crystal microcavities, by including longitudinal variations in $\mathbf{E}_\mu$ and $\mathbf{H}_\mu$. To better establish a correspondence with the traveling-wave nonlinear couplings studied in Sec.~\ref{sec:CWEs}, we restrict our focus here to the simple longitudinal modes discussed above. As with linearly-propagating pulses, the coefficients $a_{\mu,m}$ are constant in the absence of nonlinear coupling. When we include nonlinear couplings in the following sections, the Fourier components will evolve with propagation time, $a_{\mu,m}(t)$, rather than space. The complex fields given by 
\begin{align*}
	\mathbf{E}_{\mu,m} &= \exp(-i\omega_\mu(k_m)t+ik_m z)\mathbf{E}_{\mu}(x,y,\omega_\mu(k_m)),\\
	\mathbf{D}_{\mu,m} &= \exp(-i\omega_\mu(k_m)t+ik_m z)\epsilon(x,y,\omega_\mu(k_m))\cdot\mathbf{E}_{\mu}(x,y,\omega_\mu(k_m)),\\
    \mathbf{H}_{\mu,m} &= \exp(-i\omega_\mu(k_m)t+ik_m z)\mathbf{H}_{\mu}(x,y,\omega_\mu(k_m)),\\
	\mathbf{B}_{\mu,m} &= \exp(-i\omega_\mu(k_m)t+ik_m z)\mu_0\mathbf{H}_{\mu}(x,y,\omega_\mu(k_m)),
\end{align*}
independently satisfy Maxwell's equations, however $\omega_\mu(k_m)$ is now interpreted as the eigenvalue associated with propagation constant $k_m$, with corresponding eigenfunctions given by $\mathbf{E}_\mu$ and $\mathbf{H}_\mu$. $\mathbf{E}_{\mu,m}$ are best thought of as a decomposition of $\mathbf{E}(x,y,z,t)$ into complex time-harmonic phasors
\begin{equation*}
	\mathbf{E}(x,y,z,t) = \frac12\sum_{\mu,m}(\mathbf{E}_{\mu,m} + \mathbf{E}_{\mu,m}^*).
\end{equation*}
Throughout this appendix, we will refer to $\mathbf{E}_{\mu,m}$ as a spatial mode, or normal mode, of Maxwell's equations, with an associated longitudinal mode $\exp(i k_m z)$ and transverse mode $\mathbf{E}_{\mu}(x,y,\omega_\mu(k_m))$. In contrast with $z$-propagating waveguide modes, we will see that $|a_{\mu,m}|^2$ can be used to calculate the energy contained in mode $\mu$, rather than the power. In this approach to solving Maxwell's equations, each mode now evolves independently in time, $\exp(-i\omega_\mu(k_m) t)$, and the spatial distribution of the intracavity fields at a given time is obtained by summing the Fourier series over $m$. This behavior is a direct space-time analog of our approach to pulse propagation in Sec.~\ref{sec:pulse_prop}, where Fourier synthesis was used to describe a temporal envelope that evolves while propagating along the waveguide coordinate $z$, as opposed to a spatial envelope that evolves with time $t$. In accordance with $t$, as opposed to $z$, being the propagation coordinate our chosen phasor convention is now $\exp(-i\omega t + i k z)$.

\subsection{Orthogonality and normalization for spatial modes}\label{sec:resonator_orth}

The orthogonality relations for spatial modes can be derived by extending our previous analysis of the generalized complex reciprocity relations in Section.~\ref{sec:generalized_reciprocity}. In this case, we now use the spatial eigenmodes $\mathbf{E}_{\mu,m}$, $\mathbf{H}_{\mu,m}$, $\mathbf{E}_{\nu,n}$, and $\mathbf{H}_{\nu,n}$ as our ansatz
\begin{subequations}
\begin{align}
    \mathbf{E}_{\mu,m}=\mathbf{E}_\mu(x,y,\omega_\mu)\exp(-i\omega_\mu + i k_m z),&\quad \mathbf{H}_{\mu,m}=\mathbf{H}_\mu(x,y,\omega_\mu)\exp(-i\omega_\mu + ik_m z),\label{eqn:e_gccr_m}\\
    \mathbf{E}_{\nu,n}=\mathbf{E}_\nu(x,y,\omega_\nu)\exp(-i\omega_\nu + ik_n z),&\quad \mathbf{H}_{\nu,n}=\mathbf{H}_\nu(x,y,\omega_\nu)\exp(-i\omega_\nu + ik_nz),\label{eqn:ebar_gccr_n}
\end{align}
\end{subequations}
where we have suppressed the arguments of $\omega_\nu(k_n)$ and $\omega_\mu(k_m)$ for compactness. As with our previous treatments of reciprocity, we evaluate the divergence
\begin{equation}
    \nabla\cdot\left(\mathbf{E}_{\mu,m}\times\mathbf{H}_{\nu,n}^*\right) = \mathbf{H}_{\nu,n}^*\cdot(\nabla\times\mathbf{E}_{\mu,m}) -\mathbf{E}_{\mu,m}\cdot(\nabla\times\mathbf{H}_{\nu,n}^*).
\end{equation}
Substituting in Maxwell's curl equations, we have
\begin{equation}
    \nabla\cdot\left(\mathbf{E}_{\mu,m}\times\mathbf{H}_{\nu,n}^*\right) = -\mu_0\mathbf{H}_{\nu,n}^*\cdot\left(\partial_t\mathbf{H}_{\mu,m}\right)-\mathbf{E}_{\mu,m}\cdot\left(\partial_t\mathbf{D}^*_{\nu,n} + \partial_t\mathbf{P}_{\mathrm{NL},n}^*\right),\label{eqn:divGcNL_t}
\end{equation}
where $\mathbf{P}_{\mathrm{NL},n}$ now corresponds to a discrete spatial frequency $k_n$, rather than a frequency $\omega$. Similarly, conjugating these fields and interchanging the mode indices $(\mu, m)\leftrightarrow(\nu, n)$, we find
\begin{equation}
    \nabla\cdot\left(\mathbf{E}^*_{\nu,n}\times\mathbf{H}_{\mu,m}\right) = -\mu_0\mathbf{H}_{\mu,m}\cdot\left(\partial_t\mathbf{H}^*_{\nu,n}\right)-\mathbf{E}_{\nu,n}^*\cdot\left(\partial_t\mathbf{D}_{\mu,m} + \partial_t\mathbf{P}_{\mathrm{NL},m}\right).\label{eqn:divGrNL_t}
\end{equation}
Adding together Eqns.~\ref{eqn:divGcNL_t}-\ref{eqn:divGrNL_t} yields
\begin{align}
    \nabla\cdot \big(\mathbf{F}_c\big) = &-\mu_0\mathbf{H}_{\nu,n}^*\cdot\left(\partial_t\mathbf{H}_{\mu,m}\right) - \mu_0\mathbf{H}_{\mu,m}\cdot\left(\partial_t\mathbf{H}^*_{\nu,n}\right)\label{eqn:recipNL_t}\\
    &-\mathbf{E}_{\mu,m}\cdot\left(\partial_t\mathbf{D}^*_{\nu,n}
     + \partial_t\mathbf{P}_{\mathrm{NL},n}^* \right)-\mathbf{E}_{\nu,n}^*\cdot\left(\partial_t\mathbf{D}_{\mu,m} + \partial_t\mathbf{P}_{\mathrm{NL},m}\right),\nonumber
\end{align}
where, as before, $\mathbf{F}_c=\mathbf{E}_{\mu,m}\times\mathbf{H}_{\nu,n}^*+\mathbf{E}_{\nu,n}^*\times\mathbf{H}_{\mu,m}$.

In the absence of a nonlinear polarization, the time derivatives can be evaluated using Eqns.~\ref{eqn:e_gccr_m}-\ref{eqn:ebar_gccr_n} to find
\begin{align}
    \nabla\cdot \big(\mathbf{F}_c\big) = &i\mu_0\mathbf{H}_{\nu,n}^*\cdot\big(\omega_\mu-\omega_\nu\big)\mathbf{H}_{\mu,m}+i\mathbf{E}_{\nu,n}^*\cdot\left(\omega_\mu\epsilon_\mu -\omega_\nu \epsilon_\nu^\dagger\right)\cdot \mathbf{E}_{\mu,m}\label{eqn:div_setup},
\end{align}
where $\epsilon_\mu=\epsilon(x,y,\omega_\mu(k_m))$ and $\epsilon_\nu=\epsilon(x,y,\omega_\nu(k_n))$ have been introduced for compactness. We now integrate Eqn.~\ref{eqn:div_setup} over the volume defined by an infinite cross section $A_\infty$ and $z\in\left[0, L\right]$. Applying the divergence theorem to the left hand size of Eqn.~\ref{eqn:div_setup} and noting that $\int_{A_\infty}\mathbf{F}_c(x,y,z=0_+) \cdot\hat{z} dx dy = -\int_{A_\infty}\mathbf{F}_c(x,y,z=L) \cdot\hat{z} dx dy$, we find that the integral must vanish,
\begin{align}
    (\omega_\mu-\omega_\nu)\int_{A_\infty}\int_0^L \mu_0\mathbf{H}_{\mu,m}\cdot\mathbf{H}_{\nu,n}^*+\mathbf{E}_{\nu,n}^*\cdot\left(\frac{\omega_\mu\epsilon_\mu-\omega_\nu \epsilon_\nu^\dagger}{\omega_\mu-\omega_\nu}\right)\cdot \mathbf{E}_{\mu,m} dV=0.\label{eqn:volumeorthogonality}
\end{align}
Noting that the integral with respect to $z$ is zero unless $k_m = k_n \equiv k$, we find
\begin{equation}
    (\omega_\mu-\omega_\nu)L\int_{A_\infty}\mu_0\mathbf{H}_{\mu,m}\cdot\mathbf{H}_{\nu,m}^*+\mathbf{E}_{\nu,m}^*\cdot\left(\frac{\omega_\mu\epsilon_\mu-\omega_\nu \epsilon_\nu^\dagger}{\omega_\mu-\omega_\nu}\right)\cdot \mathbf{E}_{\mu,m} dxdy=0,\label{eqn:spatialmodeorthogonality_revisited}
\end{equation}
where $\omega_\mu(k)$ and $\omega_\nu(k)$ are now the eigenfrequencies of modes $\mu$ and $\nu$ associated with wavenumber $k$. When $\omega_\mu \neq \omega_\nu$, the integral in Eqn.~\ref{eqn:spatialmodeorthogonality_revisited} must vanish.

When $\mu = \nu$ the integral does not vanish and can be evaluated assuming a lossloss and non-gyrotropic medium ($\epsilon = \epsilon^\dagger$) by taking the limit $\omega_\nu\rightarrow\omega_\mu(\equiv\omega)$, which is equivalent to the energy density of a linear dispersive dielectric~\cite{Raymer2020,haus1984waves}
\begin{equation}
    \frac{1}{4}\int \mu_0\mathbf{H}_\mu\cdot\mathbf{H}_\mu^*+\mathbf{E}_\mu^*\cdot\left(\epsilon(x,y,\omega)+\omega\partial_\omega\epsilon(x,y,\omega)\right)\cdot \mathbf{E}_\mu dV=\text{U}_0.\label{eqn:spatialmode_energydensity}
\end{equation}
Here, $\text{U}_0$ is an arbitrary normalization constant with units of energy. For classical nonlinear optics (in MKS units), one can simply take $\text{U}_0$ to be $1$ Joule, in analogy to choosing $1$ Watt to normalize the power propagating in a waveguide. In anticipation of other choices for our field normalization, such as $\text{U}_0 = \hbar\omega$, we will leave this normalization arbitrary. Together with Eqn.~\ref{eqn:spatialmodeorthogonality_revisited}, the orthogonality relations for spatial modes in lossless dielectrics are given by
\begin{equation}
    \frac{1}{4}\int \mu_0\mathbf{H}_{\mu,m}\cdot\mathbf{H}_{\nu,m}^*+\mathbf{E}_{\nu,m}^*\cdot\left(\frac{\omega_\mu\epsilon_\mu-\omega_\nu \epsilon_\nu^\dagger}{\omega_\mu-\omega_\nu}\right)\cdot \mathbf{E}_{\mu,m} dV=\text{U}_0\delta_{\mu\nu},\label{eqn:spatialmode_orthogonality_H_final_1}
\end{equation}
where the $\mu=\nu$ case is evaluated using Eqn.~\ref{eqn:spatialmode_energydensity}. While the treatment presented here focused on longitudinal modes with a simple $\exp(i k_m z)$ dependence, these orthogonality relations are readily extended to more complicated systems where $\mathbf{E}_{\mu}$ and $\mathbf{H}_{\mu}$ have non-trivial $z$-dependence.

As with waveguide modes, the orthogonality relations derived here can be used to decompose electric and magnetic fields confined to a resonator at any time $t$ into a sum over spatial modes. The fields at any other time $t'$ can be found by evolving the phase of each mode with $\exp(-i \omega (t'-t))$. These orthogonality relations can also be used to express the total energy stored in the waveguide as the sum of the energy stored in each mode. The proof proceeds as follows: starting from the expression for the energy density in a dielectric medium,
\begin{align}
    U=\int \int^t_{-\infty}(\mathbf{H}\cdot \partial_{t'}\mathbf{B}+\mathbf{E}\cdot \partial_{t'}\mathbf{D})\mathrm{d}t'dV,\label{eqn:classical_energy}
\end{align}
and inserting Eqns.~\ref{eqn:spatial_mode_H}-\ref{eqn:spatial_mode_B} into Eqn.~\ref{eqn:classical_energy}, we find the contribution to the total energy from the magnetic fields associated with each mode
\begin{align}
	\mathbf{H}\cdot\partial_t\mathbf{B} =& \frac{\mu_0}{4}\left(\sum_{\mu,m}a_{\mu,m}\mathbf{H}_{\mu}(x,y,\omega_\mu(k_m))\exp(-i\omega_\mu(k_m)t+i k_m z)+ c.c.\right)\nonumber\\
&\cdot	\left(\sum_{\nu,n}i\omega_\nu(k_n)a_{\nu,n}^*\mathbf{H}_{\nu}^*(x,y,\omega_\nu(k_n))\exp(i\omega_\nu(k_n)t-i k_n z)+ c.c.\right).\nonumber
\end{align}
Similarly, inserting Eqns.~\ref{eqn:spatial_mode_E}-\ref{eqn:spatial_mode_D} into Eqn.~\ref{eqn:classical_energy}, we find the contribution to the total energy from the electric fields
\begin{align}
	\mathbf{E}\cdot \partial_{t}\mathbf{D} =& \frac{1}{4}\left(\sum_{\mu,m}a_{\mu,m}\mathbf{E}_{\mu}(x,y,\omega_\mu(k_m))\exp(-i\omega_\mu(k_m)t+i k_m z)+ c.c.\right)\nonumber\\
&\cdot	\left(\sum_{\nu,n}i\omega_\nu(k_n)a_{\nu,n}^*\epsilon_\nu^*\cdot\mathbf{E}_{\nu}^*(x,y,\omega_\nu(k_n))\exp(i\omega_\nu(k_n)t-i k_n z)+ c.c.\right)\nonumber.
\end{align}
We now calculate the total energy by integrating over the volume defined by $A_\infty$ and $z\in\left[0,L\right]$. For simplicity, we first evaluate the integral with respect to $z$ to eliminate the sum over longitudinal modes ($\sum_n$) using the orthogonality of the complex exponentials,
\begin{align}
	\int_0^L \mathbf{H}\cdot\partial_t\mathbf{B}dz = &\frac{\mu_0 L}{4}\sum_{m,\mu,\nu}-ia_{\mu,m}a_{\nu,m}^*\exp(-i(\omega_\mu(k_m)-\omega_\nu(k_m))t)\\
	&\cdot \mathbf{H}_{\nu}^*(x,y,\omega_\nu(k_m))\cdot(\omega_\mu(k_m)-\omega_\nu(k_m))\mathbf{H}_{\mu}(x,y,\omega_\mu(k_m))\nonumber,
\end{align}
The electric fields can be evaluated similarly,
\begin{align}
	\int_0^L \mathbf{E}\cdot\partial_t\mathbf{D}dz = &\frac{L}{4}\sum_{m,\mu,\nu} -ia_{\mu,m}a_{\nu,m}^*\exp(-i(\omega_\mu(k_m)-\omega_\nu(k_m))t)\\
	&\cdot \mathbf{E}_{\nu}^*(x,y,\omega_\nu(k_m))\cdot\big(\omega_\mu(k_m)\epsilon_\mu-\omega_\nu(k_m)\epsilon_\nu^\dagger\big)\cdot\mathbf{E}_{\mu}(x,y,\omega_\mu(k_m))\nonumber.
\end{align}
We now evaluate the integral $\int_{-\infty}^{t} dt'$ for each of these equations,
\begin{align}
	\int_{-\infty}^{t}\int_0^L \mathbf{H}\cdot\partial_{t'}\mathbf{B}dz dt' = &\frac{\mu_0 L}{4}\sum_{m,\mu,\nu}a_{\mu,m}a_{\nu,m}^*\exp(-i(\omega_\mu(k_m)-\omega_\nu(k_m))t)\\
	&\cdot \mathbf{H}_{\nu}^*(x,y,\omega_\nu(k_m))\cdot\mathbf{H}_{\mu}(x,y,\omega_\mu(k_m))\nonumber\\
	\int_{-\infty}^{t}\int_0^L \mathbf{E}\cdot\partial_t'\mathbf{D}dz dt' = &\frac{L}{4}\sum_{m,\mu,\nu} a_{\mu,m}a_{\nu,m}^*\exp(-i(\omega_\mu(k_m)-\omega_\nu(k_m))t')\\
	\cdot \mathbf{E}_{\nu}^*(x,y,& \omega_\nu(k_m))\cdot \left(\frac{\omega_\mu(k_m)\epsilon_\mu-\omega_\nu(k_m)\epsilon_\nu^\dagger}{\omega_\mu(k_m)-\omega_\nu(k_m)}\right)\cdot\mathbf{E}_{\mu}(x,y,\omega_\mu(k_m))\nonumber
\end{align}
Adding these equations together, evaluating the integral $\int_{A_\infty} dxdy$, and invoking Eqn.~\ref{eqn:spatialmode_orthogonality_H_final_1}, the energy contained in the waveguide is given simply by
\begin{align*}
    U=\sum_{\mu,m}|a_{\mu,m}|^2 \text{U}_0.
\end{align*}

Before we treat nonlinear interactions, we first present several equivalent forms of Eqn.~\ref{eqn:spatialmode_orthogonality_H_final_1} that are expressed only in terms of the electric fields, $E_{\mu,m}$ and $E_{\nu,n}$. These orthogonality relations are more convenient for developing the coupled-wave equations in the following sections. We begin with the general vector quantity,
\begin{equation}
	\mathbf{M} = \mathbf{A}\times(\nabla\times \mathbf{B}).
\end{equation}
The divergence of $\mathbf{M}$ can be written using well-known vector identities
\begin{equation}
	\nabla\cdot\mathbf{M} = (\nabla\times \mathbf{A})\cdot(\nabla\times \mathbf{B}) - \mathbf{A}\cdot(\nabla\times\nabla\times\mathbf{B}).
\end{equation}
We integrate both sides using the divergence theorem to find
\begin{equation}
	\int \nabla\cdot\mathbf{M} dV = \oint \mathbf{A}\times(\nabla\times\mathbf{B})\cdot d\mathbf{S} = \int(\nabla\times \mathbf{A})\cdot(\nabla\times \mathbf{B}) - \mathbf{A}\cdot(\nabla\times\nabla\times\mathbf{B})dV.\label{eqn:resonator_vector_theorem}
\end{equation}
Throughout this section we will substitute the fields associated with resonator normal modes, such as $\mathbf{E}_{\mu,m}$ for $\mathbf{A}$ and $\mathbf{B}$ in Eqn.~\ref{eqn:resonator_vector_theorem}. All of the main insights found here stem from the surface integral in Eqn.~\ref{eqn:resonator_vector_theorem} vanishing when the volume of integration is taken to be arbitrarily large.

We first consider the case where $\mathbf{A} = \mathbf{E}_{\mu,m}$ and $\mathbf{B} = \mathbf{E}_{\mu,m}^*$. This case can be used to establish a connection between $\int \mathbf{H}_{\mu,m}^*\cdot \mathbf{H}_{\mu,m} dV$ and $\int \mathbf{E}_{\mu,m}^*\cdot \epsilon(x,y,\omega_\mu(k_m))\cdot\mathbf{E}_{\mu,m} dV$, which will later allow us to simplify the orthogonality relations by eliminating terms that go as $\int \mathbf{H}_{\mu,m}^*\cdot \mathbf{H}_{\mu,m} dV$ in favor of expressions containing electric fields. With this choice of $\mathbf{A}$ and $\mathbf{B}$, and assuming the surface integral vanishes, Eqn.~\ref{eqn:resonator_vector_theorem} becomes
\begin{equation}
	\int(\nabla\times \mathbf{E}_{\mu,m})\cdot(\nabla\times \mathbf{E}_{\mu,m}^*) - \mathbf{E}_{\mu,m}\cdot(\nabla\times\nabla\times\mathbf{E}_{\mu,m}^*)dV=0.\label{eqn:orth_Emu_Emu}
\end{equation}
Equation~\ref{eqn:orth_Emu_Emu} can be evaluated using Maxwell's curl equations, $\nabla\times\mathbf{E}_{\mu,m} = -i\omega_\mu\mu_0 \mathbf{H}_{\mu,m}$ and $\nabla\times\mathbf{H}_{\mu,m}=i\omega_\mu\epsilon_\mu\cdot \mathbf{E}_{\mu,m}$, where we have defined the short-hand notation $\epsilon_\mu = \epsilon(x,y,\omega_\mu)$ and have suppressed the argument of $\omega_\mu(k_m)$ for compactness. Evaluating the double-curl in Eqn.~\ref{eqn:orth_Emu_Emu} yields a closed-form expression for the eigenvalue $\omega_\mu$ in terms of the electric fields
\begin{equation}
	\omega_\mu^2 = \frac{\int(\nabla\times \mathbf{E}_{\mu,m})\cdot(\nabla\times \mathbf{E}_{\mu,m}^*)dV}{\int \mathbf{E}_{\mu,m}\cdot(\mu_0\epsilon_\mu^*)\cdot\mathbf{E}_{\mu,m}^*dV}.\label{eqn:omega_dispersion}
\end{equation}
Equation~\ref{eqn:omega_dispersion} is a direct analog of Eqn.~\ref{eqn:disp1} for resonator normal modes. We note here that for typical nonlinear media, which are lossless and non-gyrotropic, we have $\epsilon^\dagger = \epsilon$. Throughout this presentation, we will leave $\epsilon^*$ and $\epsilon^\dagger$ in each expression since this allows one to more easily trace which curl equation the factors of $\epsilon$ originated from.  Evaluating the curl equations in the numerator of Eqn.~\ref{eqn:omega_dispersion} allows us to express the energy stored in the magnetic field in terms of electric fields
\begin{equation}
	\mu_0\frac14\int\mathbf{H}_{\mu,m}\cdot\mathbf{H}_{\mu,m}^* dV = \frac14\int \mathbf{E}_{\mu,m}\cdot\epsilon_\mu^*\cdot\mathbf{E}_{\mu,m}^*dV.\label{eqn:Emu_Hmu_correspondence}
\end{equation}
Equation~\ref{eqn:Emu_Hmu_correspondence} is typically interpreted as an equipartitioning of the electric and magnetic energy stored in a dielectric medium. While the left-hand side of Eqn.~\ref{eqn:Emu_Hmu_correspondence} is the energy stored in the magnetic field, the right-hand side of Eqn.~\ref{eqn:Emu_Hmu_correspondence} only corresponds to the energy stored in the electric field for a non-dispersive dielectric. Put simply, the electromagnetic energy is \emph{not} equally partitioned between electric and magnetic fields in dispersive media. We emphasize here that Eqn.~\ref{eqn:Emu_Hmu_correspondence} is still valid in dispersive dielectrics, and therefore Eqn.~\ref{eqn:spatialmode_energydensity} for the energy stored in a lossless, non-gyrotropic medium can be written in terms of electric fields as
\begin{equation}
    \frac{1}{4}\int \mathbf{E}_\mu^*\cdot\big(2\epsilon_\mu+\omega\partial_\omega\epsilon_\mu\big)\cdot \mathbf{E}_\mu dV=\text{U}_0.\label{eqn:spatialmode_energydensity_2}
\end{equation}

We now repeat the above analysis with $\mathbf{A} = \mathbf{E}_{\mu,m}$ and $\mathbf{B} = \mathbf{E}_{\nu,n}^*$, which will yield several orthogonality relations for resonator normal modes. Starting from Eqn.~\ref{eqn:orth_Emu_Emu}, now with $\mathbf{E}_{\mu,m}$ and $\mathbf{E}_{\nu,n}^*$, we have
\begin{equation}
	\int(\nabla\times \mathbf{E}_{\mu,m})\cdot(\nabla\times \mathbf{E}_{\nu,n}^*) - \mathbf{E}_{\mu,m}\cdot(\nabla\times\nabla\times\mathbf{E}_{\nu,n}^*)dV=0.\label{eqn:orth_Emu_Enu}
\end{equation}
Evaluating all of the curl equations in Eqn.~\ref{eqn:orth_Emu_Enu} using Maxwell's equations yields
\begin{equation}
	\int \mu_0\omega_\mu \mathbf{H}_{\mu,m}\cdot\mathbf{H}_{\nu,n}^* - \omega_\nu \mathbf{E}_{\mu,m}\cdot\epsilon_\nu^*\cdot\mathbf{E}_{\nu,n}^*dV=0.\label{eqn:orth_H_E_1}
\end{equation}
Repeating this analysis with the fields interchanged, and subtracting, recovers our first version of the orthogonality relations (Eqn.~\ref{eqn:volumeorthogonality}),
\begin{equation*}
	(\omega_\mu-\omega_\nu)\int \mu_0 \mathbf{H}_{\nu,n}^*\cdot\mathbf{H}_{\mu,m} + \mathbf{E}_{\nu,n}^*\cdot\left(\frac{\omega_\mu \epsilon_\mu-\omega_\nu \epsilon_\nu^\dagger}{\omega_\mu-\omega_\nu}\right)\cdot\mathbf{E}_{\mu,m}dV=0.
\end{equation*}
Having established that this approach is equivalent to the orthogonality relations derived above, we now extend this analysis to two equivalent expressions containing only electric fields. Starting from Eqn.~\ref{eqn:orth_Emu_Enu}, with only the double-curl evaluated, we find
\begin{equation}
	\int(\nabla\times \mathbf{E}_{\mu,m})\cdot(\nabla\times \mathbf{E}_{\nu,n}^*) - \mathbf{E}_{\mu,m}\cdot\big(\omega_\nu^2\mu_0\epsilon_\nu^*\big)\cdot\mathbf{E}_{\nu,n}^*dV=0.
\end{equation}
Interchanging $\mathbf{E}_{\mu,m}$ and $\mathbf{E}_{\nu,n}^*$ and subtracting yields a second version of the orthogonality relations found in~\cite{haus2000electromagnetic},
\begin{equation}
	\int \mathbf{E}_{\nu,n}^*\cdot\big(\omega_\mu^2\epsilon_\mu-\omega_\nu^2\epsilon_\nu^\dagger\big)\cdot\mathbf{E}_{\mu,m} dV=0.
\end{equation}
The third version of the orthogonality relations, which will be used in the following sections to derive nonlinear coupling, follows from re-expressing the magnetic field integral in terms of electric fields (from Eqn.~\ref{eqn:orth_H_E_1}),
\begin{align}
	\int \mu_0 \mathbf{H}_{\mu,m}\cdot\mathbf{H}_{\nu,n}^* dV = & \int\mathbf{E}_{\mu,m}\cdot\left(\frac{\omega_\nu\epsilon_\nu^*}{\omega_\mu}\right)\cdot\mathbf{E}_{\nu,n}^*dV,\label{eqn:H_in_terms_of_E}\\
	= & \int\mathbf{E}_{\nu,n}^*\cdot\left(\frac{\omega_\mu\epsilon_\mu}{\omega_\nu}\right)\cdot\mathbf{E}_{\mu,m}dV,\nonumber
\end{align}
where the latter form is found by interchanging $\mathbf{E}_{\mu,m}$ and $\mathbf{E}_{\nu,n}^*$ in Eqn.~\ref{eqn:orth_Emu_Enu}. Substituting Eqn.~\ref{eqn:H_in_terms_of_E} into Eqn.~\ref{eqn:volumeorthogonality} yields
\begin{equation}
	(\omega_\mu-\omega_\nu)\int \mathbf{E}_{\nu,n}^*\cdot\left(\frac{\omega_\mu\epsilon_\mu}{\omega_\nu}\right)\cdot\mathbf{E}_{\mu,m} + \mathbf{E}_{\nu,n}^*\cdot\left(\frac{\omega_\mu \epsilon_\mu-\omega_\nu \epsilon_\nu^\dagger}{\omega_\mu-\omega_\nu}\right)\cdot\mathbf{E}_{\mu,m}dV=0.\label{eqn:volumeorthogonality_2}
\end{equation}
When $\omega_\mu = \omega_\nu$, the integrand in Eqn.~\ref{eqn:volumeorthogonality_2} can be evaluated using Eqn.~\ref{eqn:spatialmode_energydensity_2}. Otherwise, the integrand is zero. Therefore, the final version of the orthogonality relations is given by,
\begin{equation}
	\frac14\int \mathbf{E}_{\nu,n}^*\cdot\left(\frac{\omega_\mu\epsilon_\mu}{\omega_\nu}\right)\cdot\mathbf{E}_{\mu,m} + \mathbf{E}_{\nu,n}^*\cdot\left(\frac{\omega_\mu \epsilon_\mu-\omega_\nu \epsilon_\nu^\dagger}{\omega_\mu-\omega_\nu}\right)\cdot\mathbf{E}_{\mu,m}dV=\text{U}_0\delta_{\mu\nu},\label{eqn:volumeorthogonality_3}
\end{equation}
with an equivalent expression holding for $(\omega_\nu\epsilon_\nu^\dagger/\omega_\mu)$ in the first parenthesis. For the case of nonlinear coupling, we will make use of Eqn.~\ref{eqn:volumeorthogonality_3} by assuming a weakly dispersive medium. Defining $\omega$ as the mean frequency between $\omega_\mu$ and $\omega_\nu$, we have $\omega_\mu = \omega \pm \delta\omega$, and $\omega_\nu = \omega \mp \delta\omega$. Similarly, for the permittivities, we have $\epsilon_\mu\approx\epsilon \pm \delta\omega \partial_\omega\epsilon$, and $\epsilon_\nu\approx\epsilon \mp \delta\omega \partial_\omega\epsilon$. Neglecting terms of order $(\delta\omega)^2$, and assuming $\epsilon_\nu^\dagger = \epsilon_\nu$, Eqn.~\ref{eqn:volumeorthogonality_3} becomes
\begin{equation}
	\frac14\int \mathbf{E}_{\nu,n}^*\cdot\big(\epsilon + \partial_\omega(\omega\epsilon)\big)\cdot\mathbf{E}_{\mu,m}dV\approx\text{U}_0\delta_{\mu\nu}.\label{eqn:nondispersive_orthogonality}
\end{equation}
In practice, since we have assumed the medium to be weakly dispersive, Eqn.~\ref{eqn:nondispersive_orthogonality} will be applied with the terms inside the parenthesis evaluated using the local frequencies, \textit{i.e.} $\epsilon_\mu + \partial_\omega(\omega\epsilon)|_{\omega_\mu}$, with an identical expression holding for $\mu\leftrightarrow\nu$. This approach to deriving the coupled-wave equations, where the fields are normalized using Eqn.~\ref{eqn:spatialmode_energydensity_2} but a weakly-dispersive projector is used to calculate the nonlinear coupling between the modes, is currently common practice in the field~\cite{Quesada2022}.

\subsection{Nonlinear coupling}

We now derive the nonlinear coupling between spatial modes induced by a nonlinear polarization, $\mathbf{P}_\text{NL}$ using the complex reciprocity relations established in the previous section. The following treatment relies on several approximations. First, we assume that the nonlinear coupling is sufficiently weak that the fields are still well described using an expansion in terms of the linear resonators modes, $\mathbf{E}_{\mu}$, that independently satisfy Maxwell's equations in the absence of a nonlinear polarization. Therefore, to leading order, the nonlinear polarization now induces time evolution in $a_{\mu,m}(t)$. This approximation is valid whenever the field evolution is slow compared to an optical cycle, $|\partial_t a_{\mu,m}(t)| \ll \omega_\mu(k_m) |a_{\mu,m}(t)|$. We note here that a common misconception is that this approximation breaks down in strongly-coupled (\textit{i.e.} few-photon) nonlinear devices. In reality, the fields in strongly-coupled devices evolve over timescales comparable to a cavity lifetime, on the order of nanoseconds. This is extremely slow compared to a typical optical cycle, which is on the order of femtoseconds. The use of modal expansions, with the slowly-varying envelope approximation made above, will also be used to evaluate the time derivative $\partial_t\mathbf{D}$. Finally, we note that the Eqn.~\ref{eqn:nondispersive_orthogonality} for the orthogonality relations in weakly-dispersive media will be used to isolate the contribution of the nonlinear polarization to the time evolution of each $a_{\mu,m}(t)$.

As with traveling-wave interactions, we assume solutions to Maxwell's equations in the presence of a nonlinear polarization are given by 
\begin{subequations}
\begin{align}
\mathbf{E}_1(x,y,z,t) = \frac12\sum_{\mu,m}\mathbf{E}_{\mu,m}(x,y,z,t)+c.c.,\\
\mathbf{H}_1(x,y,z,t) = \frac12\sum_{\mu,m}\mathbf{H}_{\mu,m}(x,y,z,t)+c.c.,
\end{align}
\end{subequations}
where $\mathbf{E}_{\mu,m}$ and $\mathbf{H}_{\mu,m}$ are nonzero when $m>0$, and are given by
\begin{subequations}
\begin{align}
    \mathbf{E}_{\mu,m}(x,y,z,t) = a_{\mu,m}(t)\mathbf{E}_\mu(x,y,\omega_\mu)\exp(-i\omega_\mu t + i k_m z),\label{eqn:e_gccr_NL_m}\\
    \mathbf{H}_{\mu,m}(x,y,z,t) = a_{\mu,m}(t)\mathbf{H}_\mu(x,y,\omega_\mu)\exp(-i\omega_\mu t + ik_m z)\label{eqn:h_gccr_NL_m}.
\end{align}
\end{subequations}
Our goal is to extract the time evolution of each $a_{\mu,m}(t)$ using the reciprocity relations established in the previous sections. Following the treatment used in Sec.~\ref{sec:kappa}, we choose $\mathbf{E}_2 =  \left(\mathbf{E}_{\nu,n} + \mathbf{E}_{\nu,n}^*\right)/2$ as a single-mode mode solution to Maxwell's equations in the absence of nonlinear polarization to project the evolution of the mode amplitudes from the sum contained in $\mathbf{E}_1$. Noting that the expansion coefficients $a_{\mu,m}$ evolve in time, we begin with Maxwell's curl equations in the time domain,
\begin{subequations}
\begin{align}
\nabla\times\mathbf{E}_1 = &-\mu_0\partial_t\mathbf{H}_1\label{eqn:curl_NL_1},\\
\nabla\times\mathbf{H}_1 = &\partial_t \mathbf{D}_1 + \partial_t\mathbf{P}_\text{NL},\label{eqn:curl_NL_2}
\end{align}
\end{subequations}
where the time derivative of $\mathbf{H}_1$ can be evaluated by expanding the fields in terms of normal modes,
\begin{align}
	\partial_t\mathbf{H}_{\mu,m} &= \left(\frac{\partial_t a_{\mu,m}(t)}{a_{\mu,m}(t)} - i\omega_\mu \right)\mathbf{H}_{\mu,m}.\label{eqn:dt_H}
\end{align}
The time derivative of $\mathbf{D}_1$ can be evaluated by accounting for the bandwidth of $a_{\mu,m}(t)$ in the constitutive relations. The constitutive relations for each mode are more clearly written in the frequency domain,
$$\mathbf{D}_{\mu,m}(t) = \int a_{\mu,m}(\Omega)\epsilon(x,y,\omega_\mu + \Omega)\cdot \mathbf{E}_{\mu}(x,y,\omega_\mu + \Omega)\exp\big(-i(\omega_\mu+\Omega) t + i k_m z\big)d\Omega.$$
We may evaluate $\partial_t \mathbf{D}_{\mu,m}$ by approximating $\epsilon(x,y,\omega_\mu + \Omega) \approx  \epsilon_\mu(x,y,\omega_\mu) + \Omega \epsilon_\mu'$, neglecting terms of order $\Omega^2$, and inverse transforming to find
\begin{align}
		\partial_t\mathbf{D}_{\mu,m} &\approx \left(\frac{\partial_t a_{\mu,m}(t)}{a_{\mu,n}(t)}\partial_\omega(\omega_\mu \epsilon_\mu) - i\omega_\mu \epsilon_\mu\right)\cdot\mathbf{E}_{\mu,m}.\label{eqn:dt_D}
\end{align}
Recalling that $\mathbf{E}_2$ corresponds to a solution of Maxwell's equations without the nonlinear polarization, the above relations can be evaluated for $\mathbf{E}_{\nu,n}$ and $\mathbf{H}_{\nu,n}$ by setting $\partial_t a_{\nu,n} = 0$.

With Eqns.~\ref{eqn:dt_H}-\ref{eqn:dt_D} for $\partial_t\mathbf{H}$ and $\partial_t\mathbf{D}$, we now repeat the steps taken in Sec.~\ref{sec:resonator_orth} to derive the generalized reciprocity relations, 
\begin{align}
    \nabla\cdot \big(\mathbf{E}_1\times\mathbf{H}_2 + \mathbf{E}_2\times\mathbf{H}_1\big) = &-\mathbf{E}_{\nu,n}^*\cdot \partial_t\boldsymbol{P}_\mathrm{NL}\label{eqn:recipNL2}\\
    & -\sum_{\mu,m}\mu_0\mathbf{H}_{\nu,n}^*\cdot\left(\partial_t\mathbf{H}_{\mu,m}\right) + \mu_0\mathbf{H}_{\mu,m}\cdot\left(\partial_t\mathbf{H}^*_{\nu,n}\right)\nonumber\\
    & -\sum_{\mu,m}\mathbf{E}_{\mu,m}\cdot\partial_t\mathbf{D}^*_{\nu,n}+\mathbf{E}_{\nu,n}^*\cdot\partial_t\mathbf{D}_{\mu,m},\nonumber
\end{align}
where the nonlinear polarization is written as a sum of complex phasors, $\mathbf{P}_\text{NL} = \frac12 \boldsymbol{P}_\text{NL} + c.c.$ For compactness, we have suppressed overall factors of $1/4$ and terms of the form $\mathbf{E}_{\nu,n}\cdot\epsilon_\mu\cdot\mathbf{E}_{\mu,m}$, $\mathbf{H}_{\nu,n}\cdot\mathbf{H}_{\mu,m}$, and $\mathbf{E}_{\nu,n}\cdot\partial_t \boldsymbol{P}_\text{NL}$ in Eqn.~\ref{eqn:recipNL2}. Since $a_{\mu,m}$ is only nonzero for positive $m$, these omitted terms vanish upon evaluating the volume integral of Eqn.~\ref{eqn:recipNL2}. We now insert Eqns.~\ref{eqn:dt_H}-\ref{eqn:dt_D} for the time derivatives in Eqn.~\ref{eqn:recipNL2} and evaluate $\int\nabla\cdot (\mathbf{E}_1\times\mathbf{H}_2 + \mathbf{E}_2\times\mathbf{H}_1)dV=0$ to find
\begin{align}
    \int \mathbf{E}_{\nu,n}^*\cdot \partial_t\boldsymbol{P}_\mathrm{NL}dV & = \int\sum_{\mu,m}\mu_0\mathbf{H}_{\nu,n}^*\cdot\mathbf{H}_{\mu,m}\left(i\omega_\mu-i\omega_\nu\right)dV\label{eqn:recipNL3}\\
    +&\int\sum_{\mu,m}\mathbf{E}_{\nu,n}^*\cdot\left(i\omega_\mu\epsilon_\mu-i\omega_\nu\epsilon_\nu^\dagger\right)\cdot\mathbf{E}_{\mu,m}dV\nonumber\\
    -\int\sum_{\mu,m} \bigg(\frac{\partial_t a_{\mu,m}(t)}{a_{\mu,m}(t)}\bigg)(\mu_0\mathbf{H}_{\nu,n}^*&\cdot\mathbf{H}_{\mu,m}+\mathbf{E}_{\nu,n}^*\cdot\partial_\omega(\omega_\mu\epsilon_\mu)\cdot\mathbf{E}_{\mu,m})dV.\nonumber
\end{align}
The first two terms on the right-hand side of Eqn.~\ref{eqn:recipNL3} cancel (Eqn.~\ref{eqn:spatialmodeorthogonality_revisited}), and the remaining terms are evaluated using Eqn.~\ref{eqn:nondispersive_orthogonality} for the orthogonality relations in weakly-dispersive media, resulting in
\begin{align}
	\partial_t a_{\nu,n}(t) = &-\frac{1}{4\text{U}_0}\int\mathbf{E}_{\nu,n}^*\cdot \partial_t\boldsymbol{P}_{\mathrm{NL}}dV\label{eqn:CWE_general_time}\\
 = &-\frac{\exp(i\omega_\nu t)}{4\text{U}_0} \partial_t\left(\int\mathbf{E}_\nu^*(x,y,\omega_\nu(k_n))\cdot\boldsymbol{P}_{\mathrm{NL}}\exp(-ik_n z)dV\right).\nonumber
\end{align}
Equation~\ref{eqn:CWE_general_time} is the time-propagating analogue to Eqn.~\ref{eqn:CWE}, here describing the evolution of each spatial mode as a function of time for an arbitrary nonlinear polarization, and is the main result of this section. We will obtain further insights in the following section by working through the case of SHG and OPA.

At this point, the approximations we have made in deriving Eqn.~\ref{eqn:CWE_general_time} are Eqn.~\ref{eqn:e_gccr_NL_m}-\ref{eqn:h_gccr_NL_m} for the field amplitudes, a slowly-varying approximation in Eqn.~\ref{eqn:dt_D}, and the use of weakly-dispersive orthogonality relations in evaluating Eqn.~\ref{eqn:recipNL3}. It is informative to verify the approximations \emph{implied} by this approach by evaluating Maxwell's wave equation,
\begin{equation}
	\nabla\times\nabla\times\mathbf{E} = -\mu_0\partial_t^2\mathbf{D}-\mu_0\partial_t^2\mathbf{P}_\text{NL}.\label{eqn:wave_eqn_NL}
\end{equation}
Expanding the left-hand side of Eqn.~\ref{eqn:wave_eqn_NL} using a modal expansion, we have
\begin{equation}
	\frac12 \sum_{\mu,m}\nabla\times\nabla\times\mathbf{E}_{\mu,m} + c.c = \frac{\mu_0}{2} \sum_{\mu,m}\omega_\mu^2 \epsilon_\mu \cdot \mathbf{E}_{\mu,m} + c.c.\label{eqn:wave_eqn_NL_LHS}
\end{equation}
The time derivative of $\mathbf{D}$ on right-hand side of Eqn.~\ref{eqn:wave_eqn_NL} is evaluated following the same procedure as Eqn.~\ref{eqn:dt_D}. We write $\mathbf{D}_{\mu,m}(t)$ as a Fourier integral, apply the time derivative operator twice to $\exp(-i(\omega_\mu + \Omega)t)$, and neglect terms of order $\Omega^2$ to find
\begin{equation}
	\partial_t^2\mathbf{D}_{\mu,m}(t) = \bigg(-\omega_\mu^2\epsilon_\mu -i\omega_\mu\big(\epsilon_\mu+\partial_\omega(\omega_\mu\epsilon_\mu)\big)\frac{\partial_t a_{\mu,m}(t)}{a_{\mu,m}(t)}\bigg)\cdot \mathbf{E}_{\mu,m}(t).\label{eqn:wave_eqn_NL_RHS}
\end{equation}
Inserting Eqns.~\ref{eqn:wave_eqn_NL_LHS}-\ref{eqn:wave_eqn_NL_LHS} into Eqn.~\ref{eqn:wave_eqn_NL}, we have
\begin{equation}
	\sum_{\mu,m} i\omega_\mu\frac{\partial_t a_{\mu,m}(t)}{a_{\mu,m}(t)}\big(\epsilon_\mu+\partial_\omega(\omega_\mu\epsilon_\mu)\big)\cdot \mathbf{E}_{\mu,m}(t) + c.c. = \partial_t^2\boldsymbol{P}_{\mathrm{NL}} + c.c.
\end{equation}
Multiplying both sides by $\mathbf{E}_{\nu,n}^*$, integrating with respect to volume, and applying Eqn.~\ref{eqn:nondispersive_orthogonality} for the weakly-dispersive orthogonality relations, we obtain
\begin{equation}
	-i\omega_\mu \partial_t a_{\mu,m}(t) = -\frac{1}{4\text{U}_0}\int\mathbf{E}_{\mu,m}^*\cdot \partial_t^2\boldsymbol{P}_{\mathrm{NL}}dV.\label{eqn:CWE_general_time_2}
\end{equation}
Comparing Eqn.~\ref{eqn:CWE_general_time_2} with Eqn.~\ref{eqn:CWE_general_time}, we find that the approximations made above imply that the nonlinear polarization can be decomposed into time-harmonic phasors that independently drive each resonator mode,
\begin{align*}
\int\mathbf{E}_{\mu,m}^*\cdot \partial_t\boldsymbol{P}_{\mathrm{NL}}dV \approx -i\omega_\mu(k_m)\int\mathbf{E}_{\mu,m}^*\cdot\boldsymbol{P}_{\mathrm{NL}}dV.
\end{align*}
We will see that this approximation is valid when the nonlinear coupling is constant across a typical phase-matching bandwidth. An equivalent approximation has been made in our treatment of traveling-wave interactions between ultrafast pulses, where the self-steepening due to the dispersion of the nonlinear coupling is neglected.

\subsection{The coupled-wave equations}

As with traveling-wave interactions, we restrict our focus to a single transverse mode of fundamental, $\mu$, and second-harmonic, $\nu$. For convenience, we introduce normalized field distributions,
\begin{subequations}
\begin{align}
\mathbf{E}_\mu(x,y,\omega_\mu(k_m)) = \sqrt{\frac{2\mathrm{U}_0}{\epsilon_0 n_\mu^2(k_m) V_{\mathrm{mode},\mu}(k_m)}}\mathbf{e}_\mu(x,y,\omega_\mu(k_m))\label{eqn:E_mode_V},\\
\mathbf{H}_\mu(x,y,\omega_\mu(k_m)) = \sqrt{\frac{2\mathrm{U}_0}{\mu_0 V_{\mathrm{mode},\mu}(k_m)}}\mathbf{h}_\mu(x,y,\omega_\mu(k_m))\label{eqn:H_mode_V},
\end{align}
\end{subequations}
where $\mathbf{e}_\mu(x,y,\omega_\mu(k_m))$ and $\mathbf{h}_\mu(x,y,\omega_\mu(k_m))$ are dimensionless vectors that capture the shape of the electric and magnetic fields, and $n_\mu(k_m) = c k_m/\omega_\mu(k_m)$ is the effective phase index of the mode. With Eqn.~\ref{eqn:spatialmode_energydensity} for the energy density in a dispersive dielectric, the effective mode volume is given by
\begin{equation}
	V_{\text{mode},\mu}(k_m) = \frac12 \int \mathbf{h}_{\mu}^*\cdot \mathbf{h}_{\mu} + \mathbf{e}_{\mu}^*\cdot\left(\frac{\epsilon_\mu + \omega_\mu\partial_\omega\epsilon_\mu}{n_\mu^2\epsilon_0}\right)\cdot \mathbf{e}_{\mu} dV.\label{eqn:V_mode}
\end{equation}
For translation-invariant structures, this mode volume can be linked to the previously defined mode area using Eqns.~\ref{eqn:E_mode}-\ref{eqn:H_mode} for the waveguide normalization and Eqn.~\ref{eqn:disp2} for the group velocity ($\text{P} = \mathrm{U}_0 v_{g,\mu}/L$). Together, we find for this choice of normalization $V_\mathrm{mode} = A_\mathrm{mode} L n_{g,\mu}/n_{\mu}$. We reiterate here that conventions for normalizing the fields are arbitrary and in many cases other choices can be advantageous, such as for simulating ultra-broadband interactions. The normalization chosen in equations~\ref{eqn:E_mode_V}-\ref{eqn:H_mode_V} is meant to establish intuitive connections between the ideas of mode volume and mode area in waveguides, and will be used to define an effective volume for the nonlinear coupling, $V_\text{eff}$, in analogy to the effective interaction area, $A_\text{eff}$. 

Equation~\ref{eqn:V_mode} for the mode volume greatly simplifies when the modal fields are transverse and the waveguide is weakly dispersive. For transverse fields, Eqn.~\ref{eqn:omega_dispersion} for the eigenvalue $\omega_\mu$ becomes $n_\mu^2 \int \epsilon_0|\mathbf{e}_\mu|^2 dV \approx \int \mathbf{e}_\mu^*\cdot\epsilon_\mu\cdot \mathbf{e}_\mu dV$. Similarly, with Eqn.~\ref{eqn:Emu_Hmu_correspondence} for the energy contained in the magnetic fields, we have
\begin{equation*}
	\int \mathbf{h}_{\mu}^*\cdot \mathbf{h}_{\mu} dV = \int \mathbf{e}_{\mu}^*\cdot\left(\frac{\epsilon_\mu}{n_\mu^2\epsilon_0}\right)\cdot\mathbf{e}_{\mu} dV \approx \int \mathbf{e}_{\mu}^*\cdot\mathbf{e}_{\mu} dV.
\end{equation*}
In a dispersionless medium, Eqn.~\ref{eqn:V_mode} simplifies to
\begin{equation*}
	V_{\text{mode},\mu} \approx \int \mathbf{h}_{\mu}^*\cdot \mathbf{h}_{\mu}  dV \approx \int \mathbf{e}_{\mu}^*\cdot\mathbf{e}_{\mu} dV.
\end{equation*}
When the peak value of the integrand in Eqn.~\ref{eqn:V_mode} (or the peak value of $\mathbf{h}_\mu^*\cdot\mathbf{h}_\mu$ for transverse fields) is chosen to be unity, the mode volume $V_{\text{mode},\mu}$ represents the the ratio of the total energy contained in mode $\mu$ to the peak energy density.

Having established our choice of mode normalization, we now calculate the coupled-wave equations for SHG and degenerate OPA, assuming a single relevant transverse mode, $\mu$ and $\nu$, for the fundamental and second-harmonic respectively. The nonlinear polarization can be expanded in terms of the interacting resonator normal modes. Noting that the nonlinear couplings will be evaluated using $\int \mathbf{E}_\nu^*\cdot\partial_t\boldsymbol{P}_{\mathrm{NL}}\exp(-ik_n z)dV$ for SHG and SFG of frequencies centered around the second harmonic, and $\int \mathbf{E}_\mu^*\cdot\partial_t\boldsymbol{P}_{\mathrm{NL}}\exp(-ik_m z)dV$ for OPA and DFG of frequencies centered around the fundamental, we only retain terms in $\boldsymbol{P}_\text{NL}$ that contain positive momentum in anticipation that the remaining terms vanish upon integration. For the fundamental, the relevant terms are
\begin{align*}
	\boldsymbol{P}_{\mathrm{NL}}^{(\text{OPA})}=&\epsilon_0\sum_{i,j,k}\sum_{\ell,n}a_{\mu,\ell}^*(t)a_{\nu,n}(t)\exp\big(i\omega_\mu(k_\ell)t-i\omega_\nu(k_n)t\big)\exp\left(i (k_n-k_\ell) z\right)\\
	\times\bigg( \hat{\boldsymbol{i}}
	\chi^{(2)}_{ijk}&\big(-\omega_\mu(k_m);\omega_\nu(k_n),-\omega_\mu(k_\ell)\big)E_{\mu,j}^*(x,y,\omega_\mu(k_\ell))E_{\nu,k}(x,y,\omega_\nu(k_n))\bigg),
\end{align*}
where the indices $i,j,k\in\left[x,y,z\right]$ refer to the Cartesian components of each eigenmode $\mathbf{E}_\mu$. The signs in the argument of $\chi^{(2)}_{ijk}(-\omega_\mu(k_m);\omega_\nu(k_n),-\omega_\mu(k_\ell))$ are chosen to correspond with those of the interacting k-vectors ($k_n - k_\ell - k_m$), which helps keep track of which fields are being coupled. By inspecting the argument of $\chi^{(2)}_{ijk}$, we can quickly ascertain that the nonlinear coupling contains a product of $\mathbf{E}_{\mu,m}^*$, $\mathbf{E}_{\nu,n}$, and $\mathbf{E}_{\mu,\ell}^*$. Proceeding similarly for the second harmonic, the relevant terms are
\begin{align*}
	\boldsymbol{P}_{\mathrm{NL}}^{(\text{SFG})}=&\epsilon_0\sum_{i,j,k}\sum_{\ell,m}a_{\mu,\ell}(t)a_{\mu,m}(t)\exp\big(i\omega_\mu(k_\ell)t+i\omega_\mu(k_m)t\big)\exp\left(i (k_m+k_\ell) z\right)\\
	\times\bigg( \hat{\boldsymbol{i}}
	\chi^{(2)}_{ijk}&\big(-\omega_\nu(k_n);\omega_\mu(k_m),\omega_\mu(k_\ell)\big)E_{\mu,j}(x,y,\omega_\mu(k_\ell))E_{\mu,k}(x,y,\omega_\mu(k_m))\bigg).
\end{align*}
We note here that the OPA and SFG terms contained in $\boldsymbol{P}_{\mathrm{NL}}=\boldsymbol{P}_{\mathrm{NL}}^{(\text{SFG})}+\boldsymbol{P}_{\mathrm{NL}}^{(\text{OPA})}+...$ contains all possible momentum conserving interactions with the exception of optical rectification terms generated by $\chi^{(2)}_{ijk}(0;\omega_\mu(k_m),-\omega_\mu(k_m))$. Conjugated terms obtained by interchanging the signs of each $\omega$ and $k$ in $\boldsymbol{P}_{\mathrm{NL}}^{(\text{SFG})}$ and $\boldsymbol{P}_{\mathrm{NL}}^{(\text{OPA})}$ drive the evolution of $a_{\nu,n}^*$ and $a_{\mu,m}^*$, respectively.

Following the presentation taken for traveling-wave interactions, we now evaluate the coupling coefficients by substituting Eqn.~\ref{eqn:E_mode_V} for the field distributions into $\boldsymbol{P}_{\mathrm{NL}}^{(\text{SFG})}$ and $\boldsymbol{P}_{\mathrm{NL}}^{(\text{OPA})}$. To eliminate any factors of $\text{U}_0$ from the coupled-wave equations, we define complex ``mode amplitudes'' $u_{\mu,m}(t)=a_{\mu,m}(t)\sqrt{\text{U}_0}$, where the energy contained in the $m^\text{th}$ longitudinal mode of transverse mode $\mu$ is given by $|u_{\mu,m}|^2$. With these substitutions, the coupled-wave equations take the form
\begin{subequations}
\begin{align}
	\partial_t u_{\mu,m}(t) = -i\sum_{n,\ell} \varsigma_{\mu\nu\mu}&(-m;n,-\ell) u_{\nu,n}(t)u_{\mu,\ell}^*(t)\delta_{\ell,n-m}\label{eqn:cwe_fh_time}\\
	\times &\exp\big(-i\big(\omega_\nu(k_n)-\omega_\mu(k_m)-\omega_\mu(k_\ell)\big) t\big),\nonumber\\
	\partial_t u_{\nu,n}(t) = -i\sum_{n,\ell} \varsigma_{\nu\mu\mu}&(-n;m,\ell) u_{\mu,m}(t)u_{\mu,\ell}(t)\delta_{\ell,n-m}\label{eqn:cwe_sh_time}\\
	\times &\exp\big(i\big(\omega_\nu(k_n)-\omega_\mu(k_m)-\omega_\mu(k_\ell)\big) t\big).\nonumber
\end{align}
\end{subequations}
The notation introduced here for the coupling coefficients $\varsigma_{\mu\nu\mu}(-m;n,-\ell)$ parallels that used for the $\chi^{(2)}$ tensor to account for which fields are interacting. Here, we continue our convention of using Greek subscripts for transverse modes, rather than Cartesian components. The argument $(-m;n,-\ell)$ follows the same sign convention used for the nonlinear tensor, $(-\omega_\mu(k_m); \omega_\nu(k_n), -\omega_\mu(k_\ell))$, to denote that longitudinal mode $m$ is being generated by difference-frequency mixing of modes $n$ and $\ell$. We note here that the Kronecker deltas originating from the orthogonality relations of the longitudinal modes enforce momentum conservation, and that $\omega_\nu(k_n)-\omega_\mu(k_m)-\omega_\mu(k_\ell)$ now plays the same role in this time-propagating model as phase mismatch does in space-propagating models. The double sum in Eqns.~\ref{eqn:cwe_fh_time}-\ref{eqn:cwe_sh_time} can be eliminated by replacing $\ell = n - m$. We have left $-\ell$ in the equations of motion, rather than $m - n$, to help clarify which fields are conjugated when calculating the overlap integrals for the coupling coefficients.

The coupling coefficients are obtained by evaluating the integrals in Eqn.~\ref{eqn:CWE_general_time}. As previously discussed, by comparing Eqn.~\ref{eqn:CWE_general_time} to Eqn.~\ref{eqn:CWE_general_time_2} the approximations used to derive the coupled-wave equations imply that time-derivatives of $\boldsymbol{P}_\text{NL}$ appearing inside the overlap integral are given simply by
\begin{align*}
	\int \mathbf{E}_{\mu,m}^*\cdot \partial_t\boldsymbol{P}_\text{NL}^\text{OPA} dV \approx -i\omega_\mu(k_m)\int \mathbf{E}_{\mu,m}^*\cdot\boldsymbol{P}_\text{NL}^\text{OPA} dV.
\end{align*}
We can verify this approximation by first noting that evaluating $\partial_t\boldsymbol{P}_\text{NL}^\text{OPA}$ generates pre-factors of the form $(-i\omega_\nu(k_n) + i\omega_\mu(k_\ell) + \partial_t a_{\mu,\ell}^*/a_{\mu,\ell}^* + \partial_t a_{\nu,n}/a_{\nu,n})$ inside the summation over $n$ and $\ell$. Secondly, we note that evaluating the integral over the propagation coordinate, $\int \exp(-i k_m z)\partial_t \boldsymbol{P}_\text{NL} dz$, eliminates the sum over $\ell$ and sets $\ell = n - m$. Therefore, the above approximation for $\partial_t \boldsymbol{P}_\text{NL}$ is equivalent to having applied a slowly-varying envelope approximation and having assumed $\omega_\nu(k_n) - \omega_\mu(k_{n-m})\approx \omega_\mu(k_{m})$ for each term inside the sum. With these approximations, the coupling coefficients are given by
\begin{subequations}
\begin{align}
	\varsigma_{\mu\nu\mu}(-m;n,-\ell) = \frac{\sqrt{Z_0 c}\omega_\mu\chi^{(2)}_\text{eff}\delta_{\ell,n-m}}{\sqrt{2}n_\mu(k_m)n_\nu(k_n)n_\mu(k_{\ell})}V_{\text{eff},\mu\nu\mu}^{-1/2}(-m;n,-\ell),\\
	\varsigma_{\nu\mu\mu}(-n;m,\ell) = \frac{\sqrt{Z_0 c}\omega_\nu\chi^{(2)}_\text{eff}\delta_{\ell,n-m}}{\sqrt{2}n_\mu(k_m)n_\nu(k_n)n_\mu(k_{\ell})}V_{\text{eff},\nu\mu\mu}^{-1/2}(-n;m,\ell),
\end{align}
\end{subequations}
with an effective interaction volume given by
\begin{subequations}
\begin{align}
	V_{\text{eff},\mu\nu\mu}^{-1/2}(-m;n,-\ell) &= \frac{V_{\text{overlap},\mu\nu\mu}(-m;n,-\ell)}{\sqrt{V_{\text{mode},\mu}(k_m)V_{\text{mode},\nu}(k_n)V_{\text{mode},\mu}(k_\ell)}},\\
	V_{\text{overlap},\mu\nu\mu}(-m;n,-\ell) &= \\
	\int\sum_{ijk}\bar{\chi}^{(2)}_{ijk}&\big(-\omega_\mu(k_m);\omega_\nu(k_n),\omega_\mu(k_\ell)\big)e_{\mu,i}^*(k_\ell)e_{\nu,k}(k_{n})e_{\mu,j}^*(k_m)dV.\nonumber
\end{align}
\end{subequations}
Here we have suppressed the arguments of $\mathbf{e}_\nu(x,y,\omega_\nu(k_n))\equiv \mathbf{e}_\nu(k_n)$ for compactness, and have normalized the nonlinear tensor by the largest component, $\chi^{(2)}_{ijk} = \chi^{(2)}_\text{eff}\bar{\chi}^{(2)}_{ijk}$. For quasi-phasematched nonlinear media $\chi^{(2)}(x,y,z)$ can be expanded in terms of Fourier components along the propagation direction $z$, and for the case of a 50\% duty-cycle square wave along $z$ with arbitrary extent in the x-y plane, the $\chi^{(2)}$ tensor is simply scaled by the relevant Fourier component of the grating, \textit{e.g.} for the first-order Fourier component of a square-wave $\chi^{(2)}_\text{eff} = 2\,\text{max}_{ijk}(|\chi^{(2)}_{ijk}|)/\pi$. The additional phase-factor of $\exp(i k_G z)$ contributed by the grating shifts which Fourier components of the fields are phase-matched to $k_n = k_m + k_\ell + k_G$, which modifies the phase-mismatch to $\Delta \omega = \omega_\nu(k_n) - \omega_\mu(k_m) - \omega_\mu(k_{n-m} - k_G)$.



As with $z$-propagating temporal modes, we may further simplify Eqns.~\ref{eqn:cwe_fh_time}-\ref{eqn:cwe_sh_time} by defining rotating waves $\tilde{u}$ that remove fast oscillations of these envelopes due to their respective carrier frequencies and their group velocities. Defining (spatial) reference frequencies $k_{\text{ref},\mu}$ and $k_{\text{ref},\nu}$, as well as reference group velocity $v_{g,\text{ref}}$, the slowly-varying envelopes are given by
\begin{subequations}
\begin{align}
	\tilde{u}_{\mu,m}(t) &=	u_{\mu,m}(t)\exp\left(-i\omega_\mu(k_m)t + i\omega_\mu(k_{\text{ref},\mu})t + iv_{g,\text{ref}}\delta k t\right),\label{eqn:u_envelope_1}\\
	\tilde{u}_{\nu,n}(t) &=	u_{\nu,n}(t)\exp\left(-i\omega_\nu(k_n)t + i\omega_\nu(k_{\text{ref},\nu})t + iv_{g,\text{ref}}\delta k' t\right),\label{eqn:u_envelope_2}
\end{align}
\end{subequations}
where $\delta k = k_m - k_{\text{ref},\mu}$ and $\delta k' = k_n - k_{\text{ref},\nu}$ represent the frequency detuning from $k_{\text{ref},\mu}$ and $k_{\text{ref},\nu}$, respectively. Again extending techniques established for $z$-propagating modes, we series expand the dispersion relations for the fundamental, $\omega_\mu(k_m)$ and $\omega_\nu(k_n)$, around the reference frequencies $k_{\text{ref},\mu}$ and $k_{\text{ref},\nu}$, respectively. With these definitions, the dispersion relations can be written as
\begin{align*}
	\omega_\mu(k_m) &= \omega_\mu(k_{\text{ref},\mu}) + \partial_k\omega_\mu \delta k + \frac12 \partial_k^2\omega_\mu \delta k^2 + ...\\
	&=\omega_\mu(k_{\text{ref},\mu}) + v_{g,\mu}\delta k + \hat{D}_{\mu}(i\delta k)\\
	\omega_\nu(k_n) &= \omega_\nu(k_{\text{ref},\nu}) + \partial_k\omega_\nu \delta k' + \frac12 \partial_k^2\omega_\nu (\delta k')^2 + ...\\
	&=\omega_\nu(k_{\text{ref},\nu}) + v_{g,\nu}\delta k' + \hat{D}_{\nu}(i\delta k').
\end{align*}
The coupled-wave equations can now be written several equivalent ways,
\begin{subequations}
\begin{align}
	\partial_t \tilde{u}_{\mu,m}(t) = &-i\sum_n \varsigma_{\mu\nu\mu}(-m;n,m-n) \tilde{u}_{\mu,n-m}^*(t)\tilde{u}_{\nu,n}(t)\exp\left(-i\Delta\omega t\right)\label{eqn:cwe_fh_env_time}\\
	 &-i(\omega_\mu(k_m)-\omega_\mu(k_{\text{ref},\mu}) - v_{g,\text{ref}}\delta k)\tilde{u}_{\mu,m}(t),\nonumber\\
	\partial_t \tilde{u}_{\nu,n}(t) = &-i\sum_m \varsigma_{\nu\mu\mu}(-n;m,n-m) \tilde{u}_{\mu,n-m}(t)\tilde{u}_{\mu,m}(t)\exp\left(i\Delta\omega t\right)\label{eqn:cwe_sh_env_time}\\
	&-i(\omega_\nu(k_m)-\omega_\nu(k_{\text{ref},\nu}) - v_{g,\text{ref}}\delta k')\tilde{u}_{\nu,n}(t),\nonumber
\end{align}
\end{subequations}
or in terms of dispersion operators,
\begin{subequations}
\begin{align*}
	\partial_t \tilde{u}_{\mu,m}(t) = &-i\sum_n \varsigma_{\mu\nu\mu}(-m;n,m-n) \tilde{u}_{\mu,n-m}^*(t)\tilde{u}_{\nu,n}(t)\exp\left(-i\Delta\omega t\right)\\
	 &-(v_{g,\mu}-v_{g,\text{ref}})(i \delta k)\tilde{u}_{\mu,m}(t) - iD_{\text{int},\mu}(i \delta k)\tilde{u}_{\mu,m}(t),\nonumber\\
	\partial_t \tilde{u}_{\nu,n}(t) = &-i\sum_m \varsigma_{\nu\mu\mu}(-n;m,n-m) \tilde{u}_{\mu,n-m}(t)\tilde{u}_{\mu,m}(t)\exp\left(i\Delta\omega t\right)\\
	&-(v_{g,\nu}-v_{g,\text{ref}})(i \delta k')\tilde{u}_{\nu,n}(t) - iD_{\text{int},\nu}(i \delta k')\tilde{u}_{\nu,n}(t),\nonumber
\end{align*}
\end{subequations}
where $\Delta \omega = \omega_\nu(k_{\text{ref},\nu}) - \omega_\mu(k_{\text{ref},\mu}) - \omega_\mu(k_{\text{ref},\nu}-k_{\text{ref},\mu})$ is the phase-mismatch between the reference frequencies. For second-harmonic generation, the reference frequencies are typically chosen to satisfy $\omega_\mu(k_{\text{ref},\mu})\equiv \omega$, and $k_{\text{ref},\nu} = 2 k_{\text{ref},\mu}$.

Following the procedures established for traveling-wave interactions, we construct time-evolving (spatial) pulse envelopes, $u_\mu(z,t) = \sum_m \tilde{u}_{\mu,m}(t)/\sqrt{L}$, hereafter referred to as the complex amplitude density of the electromagnetic fields. In analogy to the complex field amplitudes describing the instantaneous power flowing through a point $z$, $U_\omega(z) = \int|A_\omega(z,t')|^2 dt'$, these envelopes satisfy
\begin{equation}
	U_\mu(t) = \int_0^L |u_\mu(z,t)|^2 dz = \sum_m |u_{\mu,m}(t)|^2,
\end{equation}
and therefore the square-magnitude, $|u_\mu(z,t)|^2$, is the energy per unit length stored along the longitudinal coordinate, $z$, at a given time $t$. We may derive coupled-wave equations for the complex amplitude densities by assuming the coupling coefficients are weakly dispersive ($\varsigma_{\mu\nu\mu}(-m;n,m-n)\approx\varsigma_{\mu\nu\mu}(-m_\text{ref};2m_\text{ref},-m_\text{ref})\equiv\varsigma$). In this case, noting that the sums over all longitudinal modes correspond to a convolution between the interacting fields, the coupled-wave equations become
\begin{subequations}
\begin{align}
	\partial_t u_{\mu}(z,t) = &-i\sigma u_{\mu}^*(z,t)u_{\nu}(z,t)\exp\left(-i\Delta\omega t\right)\label{eqn:cwe_fh_density_time}\\
	 &-(v_{g,\mu}-v_{g,\text{ref}})\partial_zu_{\mu,n}(z,t)- iD_{\text{int},\mu}(\partial_z)u_{\mu}(z,t),\nonumber\\
	\partial_t u_{\nu}(z,t) = &-i\sigma u_{\mu}^2(z,t)\exp\left(i\Delta\omega t\right)\label{eqn:cwe_sh_density_time}\\
	&-(v_{g,\nu}-v_{g,\text{ref}})\partial_z u_{\nu,n}(z,t) - iD_{\text{int},\nu}(\partial_z)u_{\nu}(z,t),\nonumber
\end{align}
\end{subequations}
where $\sigma = \varsigma\sqrt{L}$ is the coupling coefficient for the amplitude densities, and was separately introduced in Sec.~\ref{sec:time-prop}. Equations~\ref{eqn:cwe_fh_density_time}-\ref{eqn:cwe_sh_density_time} are equivalent to the coupled-wave equations derived for amplitude densities by heuristic means in the main text, and are the main results of this section. Noting that for a waveguide the effective volume can be rewritten as $V_\text{eff}=A_\text{eff}L n_{g,\mu}^2n_{g,\nu}/n_{\mu}^2 n_{\nu}$, we see that the coupling coefficients $\sigma$ are invariant with respect to the length of the waveguide,
\begin{align*}
	\sigma=&\frac{\sqrt{2 Z_0 c}\omega_\mu d_\mathrm{eff}}{\sqrt{A_\mathrm{eff}n_\mu^2 n_\nu n_{g,\mu}^2 n_{p,\nu}}}=\kappa\sqrt{v_{g,\mu}^2v_{g,\nu}},
\end{align*}
where $n_\mu$, $n_\nu$, $n_{g,\mu}$, and $n_{g,\mu}$ are evaluated at the reference frequencies of the fundamental and second harmonic, and $2 d_\text{eff}=\chi^{(2)}_\text{eff}$ in accordance with our assumption of a weakly-dispersive nonlinearity.

We close this section by revisiting the correspondence with the $z$-propagating coupled-wave equations by converting the line density to a power envelope using $\sqrt{v_{g,\mu}}u_\mu(z,t) = A_\mu(z,t)$ in a non-moving frame ($v_{g,\text{ref}} = 0$),
\begin{subequations}
\begin{align}
	\partial_t A_{\mu}(z,t) = &-i\frac{\sigma}{\sqrt{v_{g,\nu}}} A_{\mu}^*(z,t)A_{\nu}(z,t)\exp\left(-i\Delta\omega t\right)\label{eqn:cwe_fh_A_time}\\
	 &-v_{g,\mu}\partial_z A_{\mu}(z,t) - iD_{\text{int},\mu}(\partial_z)A_{\mu}(z,t),\nonumber\\
	\partial_t A_{\nu}(z,t) = &-i\sigma\frac{\sqrt{v_{g,\nu}}}{v_{g,\mu}} A_{\mu}^2(z,t)\exp\left(i\Delta\omega t\right)\label{eqn:cwe_sh_A_time}\\
	&-v_{g,\nu}\partial_z A_{\nu}(z,t) - iD_{\text{int},\nu}(\partial_z)A_{\nu}(z,t),\nonumber
\end{align}
\end{subequations}
As with treatment in Sec.~\ref{sec:time-prop}, there is some subtlety involved when finishing the conversion between time- and space-propagating models. Simply subtracting $\partial_t A_\mu - v_{g,\mu}A_\mu$ from both sides of Eqn.~\ref{eqn:cwe_fh_A_time} and dividing by $v_{g,\mu}$ recovers the correct form of the nonlinear coupling and temporal walk-off terms. However, as discussed in the main text, $v_{g,\mu}^{-1}D_{\text{int},\mu}(\partial_z)\approx D_{\text{int},\mu}(\partial_t)$ only approximates the higher-order dispersion, and in reality $D_{\text{int},\mu}(\partial_t)$ must be calculated using the series expansion of $k_\mu(\omega)$. Similarly, the conversion $\Delta \omega t \rightarrow \Delta k z$ is achieved by examining the phase-characteristics for each wave separately, rather than moving between space and time using a single group velocity. In other words, while the heuristic expressions $\Delta k \approx \Delta\omega v_g^{-1}$ and $v_{g,\mu}^{-1} D_{\text{int},\mu}(\partial_z)\approx D_{\text{int},\mu}(\partial_t)$ are helpful for remembering the conversion between space- and time-propagating models, the formal conversion is obtained by moving back to the Fourier domain, and using temporal frequency modes, $\omega_\mu$, rather than the spatial frequency modes, $k_m$, to synthesize the $z$-propagating pulse envelope.

\section{Quasi-static solutions to the coupled-wave equations}\label{sec:QS_theory}

In this appendix we describe in more detail the solution to the quasi-static equations of motion. This discussion follows the treatment of~\cite{Armstrong1962,Eckardt1984}. Our goal is to solve for the pulsed envelopes in the absence of dispersion,
\begin{subequations}
    \begin{align*}
        \partial_z A_\omega(z,t) &=-i\kappa A_{2\omega}(z,t)A_\omega^*(z,t) \exp(-i\Delta k z),\\
        \partial_z A_{2\omega}(z,t) &= -i\kappa A_\omega^2(z,t)(z,t) \exp(i\Delta k z).
    \end{align*}
\end{subequations}
We solve for the dynamics of each time-slice separately following the treatment of\cite{Armstrong1962,Eckardt1984}. We begin by putting the fields in phase-amplitude form, $A_\omega(z,t) = \rho_\omega(z,t) \exp(i \phi_\omega(z,t))$, and $A_{2\omega}(z,t)=\rho_{2\omega}(z,t)\exp(i\phi_{2\omega}(z,t))$, which converts the CWEs to \begin{subequations}
\begin{align*}
\partial_z\rho_\omega(z,t) + i\rho_\omega(z,t) \partial_z\phi_\omega(z,t) = -i\kappa \rho_\omega(z,t) \rho_{2\omega}(z,t)\exp(-i\theta(z,t)),\\
\partial_z\rho_{2\omega}(z,t) + i\rho_{2\omega}(z,t)\partial_z\phi_{2\omega}(z,t) = -i\kappa \rho_\omega^2(z,t) \exp(i\theta(z,t)),
\end{align*}
\end{subequations}
where $\theta(z,t)=2\phi_\omega(z,t) - \phi_{2\omega}(z,t) + \Delta k z$. The real and imaginary parts of these equations determine the amplitude and phase evolution
\begin{subequations}
\begin{align*}
\partial_z \rho_\omega(z,t) &= -\kappa \rho_\omega(z,t) \rho_{2\omega}(z,t)\sin(\theta(z,t)),\\
\partial_z\phi_\omega(z,t) &= -\kappa  \rho_{2\omega}(z,t)\cos(\theta(z,t)),\\
\partial_z\rho_{2\omega}(z,t) &= \kappa \rho_\omega^2(z,t)\sin(\theta(z,t)),\\
\partial_z\phi_{2\omega}(z,t) &= -\kappa  \frac{\rho_\omega^2(z,t)}{\rho_{2\omega}(z,t)}\cos(\theta(z,t)),\\
\partial_z \theta(z,t) = 2\partial_z\phi_\omega(z,t)-\partial_z\phi_{2\omega}(z,t)+\Delta k &= -\kappa \left(2\rho_{2\omega}(z,t)-\frac{\rho_\omega^2(z,t)}{\rho_{2\omega}(z,t)}\right)\cos(\theta(z,t))+\Delta k.
\end{align*}
\end{subequations}
We can identify the conserved quantities of these equations by rewriting them as
\begin{subequations}
\begin{align*}
\partial_z \mathrm{ln}\left(\rho_\omega(z,t)\right) &= -\kappa \rho_{2\omega}(z,t)\sin(\theta(z,t)),\\
\partial_z \mathrm{ln}\left(\rho_{2\omega}(z,t)\right) &= \kappa \frac{\rho_\omega^2(z,t)}{\rho_{2\omega}(z,t)}\sin(\theta(z,t)),\\
\partial_z \theta(z,t) &= \partial_z \mathrm{ln} \left(\rho_{2\omega}^2(z,t)\rho_\omega(z,t)\right)\frac{\cos\left(\theta(z,t)\right)}{\sin\left(\theta(z,t)\right)}+\Delta k
\end{align*}
\end{subequations}
Each time bin locally conserves power, which allows us to normalize the fields into the notation used by Bloembergen, $\sqrt{P(t)}u(z,t) = \rho_\omega(z,t)$, $\sqrt{P(t)}v(z,t) = \rho_{2\omega}(z,t)$, where $P(t) = |A_\omega(z,t)|^2 + |A_{2\omega}(z,t)|^2$. We also define the characteristic nonlinear length for each time bin as $\zeta(t)=\kappa \sqrt{P(t)}z$, and normalized phase mismatch as $\Delta s(t) = \Delta k/\left(\kappa\sqrt{P(t)}\right)$. With this normalization, we have
\begin{subequations}
\begin{align}
\partial_\zeta u(\zeta,t) &= -u(\zeta,t)v(\zeta,t)\sin(\theta(\zeta,t)),\label{du}\\
\partial_\zeta v(\zeta,t) &= u^2(\zeta,t)\sin(\theta(\zeta,t)),\label{dv}\\
\partial_\zeta \phi_\omega(\zeta,t) &= -v(\zeta,t) \cos(\theta(\zeta,t)),\label{dphaseu}\\
\partial_\zeta \phi_{2\omega}(\zeta,t) &= -\frac{u^2(\zeta,t)}{v(\zeta,t)} \cos(\theta(\zeta,t)),\label{dphasev}\\
\partial_\zeta \theta(\zeta,t) &= \Delta s + \partial_\zeta \mathrm{ln} (u^2(\zeta,t) v(\zeta,t))\frac{\cos\theta(\zeta,t)}{\sin\theta(\zeta,t)}\label{dtheta}
\end{align}
\end{subequations}
Throughout this section, we will solve these equations for $u(\zeta,t)$, $v(\zeta,t)$, and $\theta(\zeta,t)$ for different boundary conditions. Then, given the solutions for $u$ and $v$, we may directly integrate equations \ref{dphaseu} and \ref{dphasev} to get the phase evolution of each time bin. Almost all of the cases we will consider here have $\cos(\theta(0,t))=0$, and in many cases we take $\phi_\omega(0,t) = 0$, which then results in $\theta(0,t)=-\phi_{2\omega}(0,t)$.

\subsection{Phase-matched case}

We begin with phase-matched SHG, which has no input second harmonic, $v(0,t)=0$, and no phase-mismatch, $\Delta s(t)=0$. The coupled wave equations in this limit are
\begin{subequations}
\begin{align}
\partial_\zeta u(\zeta,t) &= -u(\zeta,t)v(\zeta,t)\sin(\theta(\zeta,t)),\\
\partial_\zeta v(\zeta,t) &= u^2(\zeta,t)\sin(\theta(\zeta,t)),\\
\partial_\zeta \theta(\zeta,t) &= \partial_\zeta \mathrm{ln} (u^2(\zeta,t) v(\zeta,t))\frac{\cos\theta(\zeta,t)}{\sin\theta(\zeta,t)}
\end{align}
\end{subequations}
The third of these equations is integrated to find the first constant of motion,
$$\Gamma(t) = u(\zeta,t)^2 v(\zeta,t) \cos(\theta(\zeta,t)).$$
Since $v(0,t)=0$, we have $\Gamma(t)=0$, which renders $\cos(\theta(\zeta,t))=0$ and $\sin(\theta(\zeta,t))=1$ for all $\zeta$. These conditions, with $\phi_\omega(\zeta,t)=0$ and $\Delta s = 0$, give $\phi_{2\omega}(0,t)=-\pi/2$ and the equation for the second harmonic becomes
$$\partial_\zeta v = (1-v^2),$$
which is integrated to find $v(\zeta,t) = \tanh(\zeta(t))$. Converting back to our original units, we find
\begin{subequations}
\begin{align*}
A_{2\omega}(z,t) &= -i A_\omega(0,t)\tanh(\kappa A_\omega(0,t) z),\\
A_{\omega}(z,t) &= A_\omega(0,t)\mathrm{sech}(\kappa A_\omega(0,t) z).
\end{align*}
\end{subequations}
As expected, each time bin separately evolves from undepleted SHG ($A_{2\omega}(z,t) \approx -i\kappa A_\omega^2(0,t)z$) to full conversion ($A_{2\omega} = -iA_\omega(0,t)$), with a characteristic conversion length given by the local amplitude of the input pulse envelope.

\subsection{Phase-mismatched case}

In the phase-mismatched case, the easiest starting point is the coupled-wave equations in normalized units (Eqns.~\ref{du}-\ref{dtheta}),
\begin{subequations}
\begin{align*}
\partial_\zeta u(\zeta,t) &= -u(\zeta,t)v(\zeta,t)\sin(\theta(\zeta,t)),\\
\partial_\zeta v(\zeta,t) &= u^2(\zeta,t)\sin(\theta(\zeta,t)),\\
\partial_\zeta \theta(\zeta,t) &= \Delta s(t) + \partial_\zeta \mathrm{ln} (u^2(\zeta,t) v(\zeta,t))\frac{\cos(\theta(\zeta,t))}{\sin(\theta(\zeta,t))}.
\end{align*}
\end{subequations}
The last of these equations may be integrated to find a constant of motion
\begin{equation}
\Gamma_\Delta(t) = v(\zeta,t) u^2(\zeta,t) \cos(\theta(\zeta,t)) + \frac{1}{2}\Delta s(t) v^2(\zeta,t),
\end{equation}
or, in terms of $\Gamma(t) = u_0^2(t) v_0(t)\cos(\theta_0(t))\equiv u^2(0,t) v(0,t)\cos(\theta(0,t))$,
\begin{equation}
v(\zeta,t) u^2(\zeta,t)\cos(\theta(\zeta,t)) = \Gamma(t) + \frac{1}{2}\Delta s (v_0^2(t)-v^2(\zeta,t)).\label{eqn:JE_step}
\end{equation}
Squaring both sides and re-arranging terms, recasts Eqn.~\ref{eqn:JE_step} in terms of $\sin(\theta(\zeta,t))$,
\begin{equation}
v u^2\sin(\theta) = \pm \sqrt{(v u^2)^2 - \left(\Gamma+\frac{1}{2}\Delta s\big(v_0^2-v^2\big)\right)^2}\label{eqn:JE_step_2},
\end{equation}
which allows the left-hand side of Eqn.~\ref{eqn:JE_step_2} to be rewritten using the coupled wave equation for the second harmonic, Eqn.~\ref{dv},
\begin{equation}
\frac{1}{2} \partial_\zeta v^2 = \pm \sqrt{v^2 \big(1-v^2\big)^2 - \left(\Gamma+\frac{1}{2}\Delta s(v_0^2-v^2)\right)^2}\label{JE_CWE}.
\end{equation}
Noting that the spatial variation of each time bin occurs independently, we have suppressed the $(\zeta,t)$ and $(t)$ arguments of each term in Eqns.~\ref{eqn:JE_step_2}-\ref{JE_CWE}.

In general, we may rewrite the right hand side of Eqn.~\ref{JE_CWE} as the square root of a polynomial in $v^2$ with ordered roots defined as $v_a^2<v_b^2<v_c^2$ for each $t$,
\begin{equation}
\frac{1}{2} \partial_\zeta v^2 = \pm \sqrt{(v^2-v_a^2)(v^2-v_b^2)(v^2-v_c^2)},
\end{equation}
where $v_a(t)$, $v_b(t)$, and $v_c(t)$ are determined by $\Gamma(t)$, $\Delta s(t)$, and $v_0(t)$. The general solution to this differential equation is found using the following substitutions
\begin{equation}
d\zeta=\frac{\pm\frac{1}{2} dv^2}{\sqrt{(v^2-v_a^2)(v^2-v_b^2)(v^2-v_c^2)}} = \frac{\pm dy}{\sqrt{(v_c^2-v_a^2)(1-y^2)(1-\gamma^2 y^2)}},
\end{equation}
where $y = (v^2 -v_a^2)/(v_b^2-v_a^2)$, and $\gamma^2 = (v_b^2-v_a^2)/(v_c^2-v_a^2)$. Integrating both sides yields
\begin{equation}
v^2 = v_a^2 + (v_b^2-v_a^2) \mathrm{sn}^2\left(\pm \sqrt{v_c^2-v_a^2}(\zeta+\zeta_0)\big|\gamma^2 \right),
\end{equation}
where $\mathrm{sn}$ is the Jacobi-elliptic sine.

For SHG, $v_0 = 0$ and $\Gamma = 0$, which simplifies the polynomial on the RHS of Eqn \ref{JE_CWE},
\begin{equation}
\frac{1}{2} \partial_\zeta v^2 = \pm \sqrt{v^2 \big(1-v^2\big)^2 - \left(\frac{1}{2}\Delta s v^2\right)^2}.
\end{equation}
In this case, the roots are given by $v_a(t)=0$, and
\begin{equation}
v_\pm(t) = \frac{1}{4}\Delta s(t) \pm \sqrt{\left(\frac{1}{4}\Delta s(t)\right)^2+1},
\end{equation}
where $v_+(t) = v_c(t) = v_b^{-1}(t)$, and $v_-(t) = v_b(t)$. The resulting fundamental and second harmonic amplitudes are given by~\cite{Armstrong1962}
\begin{subequations}
\begin{align}
v(\zeta,t) &= v_b(t)\mathrm{sn}(\zeta v_c(t)|\gamma^2(t)),\\
u^2(\zeta,t) &= 1 - v^2(\zeta,t).
\end{align}
\end{subequations}

Having obtained the field evolution, we now calculate the phase evolution of the two harmonics with
\begin{subequations}
\begin{align}
\partial_\zeta \phi_1(\zeta,t) &= -v(\zeta,t) \cos(\theta(\zeta,t)),\label{dphaseu2}\\
\partial_\zeta \phi_2(\zeta,t) &= -\frac{u^2(\zeta,t)}{v(\zeta,t)} \cos(\theta(\zeta,t)).\label{dphasev2}
\end{align}
\end{subequations}
Integrating Eqn.~\ref{dphaseu2} with respect to $\zeta$, we find
$$\phi_1(\zeta,t) = \phi_1(0,t) - \int \frac{\Gamma - \frac{1}{2}\Delta s (v^2-v_0^2)}{1-v^2}d\zeta.$$
Similarly, integrating Eqn.~\ref{dphasev2} results in
$$\phi_2(\zeta,t) = \phi_2(0,t) - \int \frac{\Gamma - \frac{1}{2}\Delta s (v^2-v_0^2)}{v^2}d\zeta.$$
Noting that, $v_0(t)=0$ and $\Gamma(t)=0$ are both zero for SHG, and assuming an unchirped input pulse at the fundamental, $\phi_1(0,t)=0$, and $\phi_2(0,t)=-\theta_0(t)=-\pi/2$, we have
\begin{subequations}
\begin{align}
\phi_1(\zeta,t) &= \frac{\Delta s}{2} \int_0^\zeta \frac{v(\zeta',t)^2}{1-v(\zeta',t)^2}d\zeta'\\
\phi_2(\zeta,t) &= -\frac{\pi}{2} + \frac{\Delta s}{2}\zeta.
\end{align}
\end{subequations}
For a fundamental input pulse linearly chirped in frequency, $\phi_1(0,t)=b t^2$, we have $\phi_2(0,t)=2\phi_1(0,t)-\theta_0(t)$. In this case, the phases of the fundamental and second harmonic are given by
\begin{subequations}
\begin{align}
\phi_1(\zeta,t) &= bt^2 + \frac{\Delta s}{2} \int_0^\zeta \frac{v(\zeta',t)^2}{1-v(\zeta',t)^2}d\zeta'\\
\phi_2(\zeta,t) &= 2bt^2 - \frac{\pi}{2} + \frac{\Delta s}{2}\zeta.
\end{align}
\end{subequations}

\section{Single-photon-pumped PDC}
\label{sec:single-photon-pumped-PDC-appendix}
\subsection{Single-mode PDC}
\label{sec:single-mode-pdc-appendix}
Here, we introduce the physics of single-mode single-photon-pumped PDC. The Hamiltonian takes the form
\begin{align}
    \hat{H}/\hbar=\frac{g}{2}(\hat{a}^{\dagger 2}\hat{b}+\hat{a}^2\hat{b}^\dagger)+\delta\hat{b}^\dagger\hat{b}
\end{align}
with phase-mismatch $\delta$ (see \eqref{eq:single-mode-small-L-Hamiltonian} for the derivation of the Hamiltonian). Note that the Hamiltonian commutes with the operator representing the Manley-Rowe invariant
\begin{align}
\hat{n}_\mathrm{MR}=\frac{1}{2}\hat{n}_a+\hat{n}_b,
\end{align}
where $\hat{n}_a=\hat{a}^\dagger\hat{a}$ and $\hat{n}_b=\hat{b}^\dagger\hat{b}$ are signal and pump number operators, respectively. This implies that the Hilbert space is partitioned into subspaces $\mathcal{H}_{n_\mathrm{MR}}$ spanned by eigenstates of $\hat{n}_\mathrm{MR}$ with different eigenvalues, and quantum states in different subspace evolve independently without talking to other subspaces. 

For instance, for an initial weak coherent pump state $\hat{D}(\beta)\ket{0\,0}\approx \ket{0\,0}+\beta\ket{0\,1}$, where $\ket{n_a\,n_b}$ is the $n_a$-signal and $n_b$-pump photon state, $\ket{0\,0}$ and $\ket{0\,1}$ belong to $\mathcal{H}_{n_\mathrm{MR}=0}$ and $\mathcal{H}_{n_\mathrm{MR}=1}$, respectively. Because the dynamics inside the subspace $\mathcal{H}_{n_\mathrm{MR}=0}$ is trivial (\textit{i.e.}, constant), the simplest nontrivial dynamics occur in $\mathcal{H}_{n_\mathrm{MR}=1}$, which is the single-photon-pumped PDC with initial state $\ket{0\,1}$. In other words, by solving the dynamics of single-photon-pumped PDC, we can also understand the PDC pumped with a very weak coherent light. In the following, we focus on the subspace $\mathcal{H}_{n_\mathrm{MR}=1}$.

The Hilbert subspace $\mathcal{H}_{n_\mathrm{MR}=1}$ is spanned by the single-photon pump state $\ket{0\,1}$ and the two-photon signal state $\ket{2\,0}$. The diagonal elements of the Hamiltonian are
\begin{align}
    &\frac{1}{\hbar}\langle 0\,1\vert\hat{H}\vert 0\,1\rangle=\delta,&\frac{1}{\hbar}\langle 2\,0\vert\hat{H}\vert 2\,0\rangle=0 
\end{align}
while the off-diagonal elements are
\begin{align}
    &\frac{1}{\hbar}\langle 2\,0\vert\hat{H}\vert 0\,1\rangle=\frac{g}{\sqrt{2}}.
\end{align}
The Hamiltonian structure is analogous to a driven two-level atom, where $\ket{0\,1}$ and $\ket{2\,0}$ correspond to atomic ground state $\ket{g}$ and excited state $\ket{e}$, respectively. We note here that since there are only two basis states inside the subspace $\mathcal{H}_{n_\mathrm{MR}=1}$, the Hamiltonian is represented as a $2\times 2$ matrix.

To solve for eigenstates in $\mathcal{H}_{n_\mathrm{MR}=1}$, we posit the form of an eigenstate as
\begin{align}
    \ket{\phi(\omega)}=(c\hat{b}^\dagger+f\hat{a}^{\dagger2})\ket{0},
\end{align}
where $\hbar\omega$ is an eigenenergy, and the state must fulfill the normalization condition 
\begin{align}
    |c|^2+2|f|^2=1.
\end{align}
Note that this $\omega$ is a dummy parameter we introduce to represent the eigenenergy of a state and is not to be seen, e.g., as a carrier frequency of light. Using
\begin{align}
\frac{1}{\hbar}\hat{H}\ket{\phi(\omega)}=\left(\frac{g}{2}c\hat{a}^{\dagger2}+\delta c\hat{b}^\dagger+gf\hat{b}^\dagger\right)\ket{0}
\end{align}
and imposing the condition $\hat{H}\ket{\phi(\omega)}=\hbar\omega\ket{\phi(\omega)}$, we obtain equations defining eigenstates
\begin{align}
\label{eq:fano-single-mode-eigen}
    &\omega c=\delta c+gf &\omega f=\frac{g}{2}c.
\end{align}
Here, we convert variables to dimensionless form via
\begin{align}
\label{eq:xi-definition-single-mode}
    &\lambda=\omega/g &\xi=\delta/g,
\end{align}
with which \eqref{eq:fano-single-mode-eigen} is converted to a dimensionless universal form
\begin{align}
\label{eq:fano-single-mode-universal}
    &\lambda c_\xi=\xi c_\xi+f &\lambda f_\xi=\frac{1}{2}c_\xi,
\end{align}
where we have explicitly labeled the variables with the single-nontrivial parameter $\xi$. Multiple physical systems with different parameters but the same $\xi=\delta/g$ can be mapped to this universal form via trivial time scaling. The series equation \eqref{eq:fano-single-mode-universal} can be combined to give
\begin{align}
\label{eq:fano-single-mode-eigenvals}
    \lambda^2-\lambda\xi-\frac{1}{2}=0,
\end{align}
which we solve to obtain ``plus'' and ``minus'' eigenvalues
\begin{align}
\lambda^\pm_\xi=\frac{1}{2}\left(\xi\pm\sqrt{\xi^2+2}\right).
\end{align}
The corresponding eigenstates are characterized by
\begin{align}
    &c^\pm_\xi=\left(1+\frac{1}{2\lambda_\pm^2}\right)^{-1/2}&f^\pm_\xi=\frac{c_\pm}{2\lambda_\pm},
\end{align}
and we label the eigenstates as $\ket{\phi(\lambda_\pm)}=\ket{\phi_\pm}$ for notational simplicity. 

\begin{figure}
    \centering
    \includegraphics[width=0.57\textwidth]{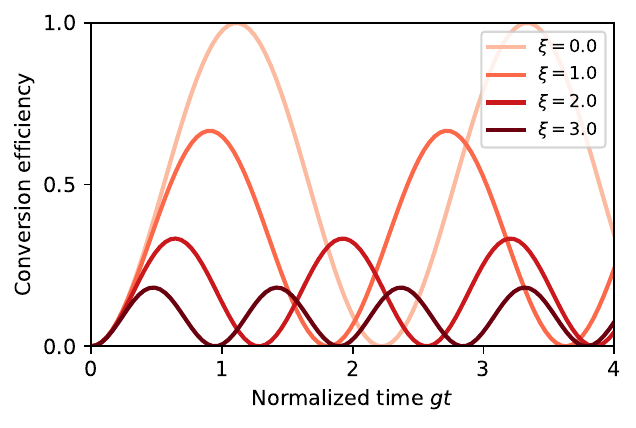}
    \caption{Conversion efficiency of single-photon-pumped single-mode PDC for various normalized phase-mismatch $\xi$. Figure is adapted from Ref.~\cite{Yanagimoto2023-thesis}.}
    \label{fig:fano-rabi-pump-number}
\end{figure}

With the analytic expressions for the eigenstates, we can, in principle, solve for any dynamics in the Hilbert subspace. Ro convert the solutions in the universal form to the original form, we can simply rescale the time. More concretely, for the initial state $\ket{\psi(0)}=\ket{0\,1}$, the PDC dynamics can be solved as
\begin{align}
    \ket{\psi(t)}=&e^{-\mathrm{i}\lambda_+\tau}c_\xi^{+*}\ket{\phi_+}+e^{-\mathrm{i}\lambda_-\tau}c^{-*}_\xi\ket{\phi_-}\nonumber\\
    =&e^{-\mathrm{i}\xi \tau}\left\{\cos\left(\frac{1}{2}\sqrt{\xi^2+2}\,\tau\right)-\frac{\mathrm{i}\xi}{\sqrt{\xi^2+2}}\sin\left(\frac{1}{2}\sqrt{\xi^2+2}\,\tau\right)\right\}\ket{0\,1}\nonumber\\
    &\quad-e^{-\mathrm{i}\xi \tau}\frac{\sqrt{2}\,\mathrm{i}}{\sqrt{\xi^2+2}}\sin\left(\frac{1}{2}\sqrt{\xi^2+2}\,\tau\right)\ket{2\,0}
\end{align}
with a normalized time $\tau=gt$. In analogy to the atomic two-level system, single-photon PDC can be seen as the dynamics of an initial ground state atom undergoing Rabi oscillations with a generalized Rabi frequency $g\sqrt{\xi^2+2}/2$. The signal photon population
\begin{align}
\label{eq:conversion-single-mode-PDC}
\bar{n}_a(t)=\langle\psi(t)\vert\hat{n}_a\vert\psi(t)\rangle=\frac{4}{\xi^2+2}\sin^2\left(\frac{1}{2}\sqrt{\xi^2+2}\,\tau\right)
\end{align}
exhibits clear sinusoidal oscillation with amplitude $4/(\xi^2+2)$, which is a manifestation of the ``single-mode'' nature of the interaction (see Fig.~\ref{fig:fano-rabi-pump-number}). When the $\chi^{(2)}$ process is phase-matched, \textit{i.e.}, $\xi=0$, the oscillation amplitude reaches $2$, enabling unit-efficiency parametric downconversion.

\subsection{Broadband PDC: Hamiltonian structure}
\label{sec:Hamiltonian-structure-appendix}
We now introduce the physics of broadband single-photon-pumped PDC. To clarify the parallelism with the single-mode discussions, we intentionally override notations for some variables introduced in Appendix~\ref{sec:single-mode-pdc-appendix}. In doing so, we make their definitions clear.

We consider the broadband Hamiltonian for a $\chi^{(2)}$-nonlinear waveguide in the wavespace
\begin{align}
\label{eq:fano-chi2-waveguide}
    \hat{H}/\hbar=\frac{r}{2}\iint\mathrm{d}s_1\mathrm{d}s_2\,\left(\hat{a}_{s_1}\hat{a}_{s_2}\hat{b}_{s_1+s_2}^\dagger+\hat{a}_{s_1}^\dagger\hat{a}_{s_2}^\dagger\hat{b}_{s_1+s_2}\right)+\sum_{u\in\{a,b\}}\int\mathrm{d}s\,\hat{u}_s^\dagger \delta\omega_u(2\pi s)\hat{u}_s,
\end{align}
which we derived in Sec.~\ref{sec:rosetta-waveguide-hamiltonian}. Note that the Hamiltonian has two explicit conserved quantities; The first is the multimode extension of the Manley-Rowe invariant
\begin{align}
&\hat{n}_\mathrm{MR}=\frac{1}{2}\int\mathrm{d}s\,\hat{a}^\dagger_s\hat{a}_s+\int\mathrm{d}s\,\hat{b}^\dagger_s\hat{b}_s, 
\end{align}
which represents the conservation of the generalized particle number. The second is the total wavenumber (i.e., momentum)
\begin{align}
\hat{s}=\int\mathrm{d}s'\,s'(\hat{a}_{s'}^\dagger\hat{a}_{s'}+\hat{b}_{s'}^\dagger\hat{b}_{s'}).
\end{align}
It is straightforward to check that these operators commute with the Hamiltonian $[\hat{H},\hat{n}_\mathrm{MR}]=[\hat{H},\hat{s}]=0$, meaning that they are conserved quantities of the dynamics under $\hat{H}$. As a result, the entire Hilbert space can be separated into Hilbert subspaces $\mathcal{H}_{n_\mathrm{MR},s}$ depending on the values of these conserved quantities, in which quantum states evolve independently. In particular, we are interested in the single-photon-pumped PDC, which occurs inside $\mathcal{H}_{n_\mathrm{MR}=1,s}$. 

We note that the Hilbert subspace $\mathcal{H}_{n_\mathrm{MR}=1,s}$ is spanned by states
\begin{align}
 &\ket{b_s}=\hat{b}^\dagger_s\ket{0}&\ket{a_{sp}}=\hat{a}^\dagger_{s/2-p}\hat{a}^\dagger_{s/2+p}\ket{0}
\end{align}
with wavenumber nondegeneracy $p\geq0$. The diagonal Hamiltonian elements depend on dispersion
\begin{subequations}
\label{eq:fano-hamiltonian-structure}
\begin{align}
\label{eq:fano-diagonal}
    &\frac{1}{\hbar}\langle a_{sp}\vert\hat{H}\vert a_{sp}\rangle=\delta_{as}(p) &\frac{1}{\hbar}\langle b_s\vert\hat{H}\vert b_s\rangle=\delta_{bs},
\end{align}
with $\delta_{bs}=\delta\omega_b(2\pi s)$ and $\delta_{as}(p)=\delta\omega_a(2\pi (s/2-p))+\delta\omega_a(2\pi (s/2+p))$, while off-diagonal elements take a constant value
\begin{align}
\label{eq:fano-off-diagonal}
    &\frac{1}{\hbar}\langle a_{sp}\vert\hat{H}\vert b_s\rangle=r.
\end{align}
\end{subequations}
The Hamiltonian structure \eqref{eq:fano-hamiltonian-structure} takes a distinctive form of discrete-continuum interaction, where a discrete pump state $\ket{b_{s}}$ is coupled to a continuum of signal states $\ket{a_{sp}}$.\footnote{Because a continuum of states exist in the Hilbert subspace $\mathcal{H}_{n_\mathrm{MR}=1,s}$, the Hamiltonian takes the form of an infinite-dimensional matrix. In practice, we can discretize the Fourier modes and truncate them within a finite bandwidth to approximate the Hamiltonian as a finite-dimensional matrix.}

Notably, such a Hamiltonian structure is analogous to atomic autoionization, where a discrete excited state $\ket{e}$ is coupled to a continuum of ionized states $\ket{\phi_\mathrm{ion}(E)}$. When the energy of the excited state lies within the energy continuum of the ionized states, an initial excited-state atom ionizes unitarily via autoionization. A theoretical framework to study the physics of such discrete-continuum interactions was initially established by Ugo Fano in his seminal work on atomic autoionization Ref.~\cite{Fano1961}, and subsequently, various physical systems have been shown to exhibit analogous physics. Here, by leveraging the analogy to atomic autoionization, we can adapt the theoretical machinery developed in atomic physics and elsewhere to analyze the dynamics of broadband parametric downconversion in the deep-quantum regime. 

To apply Fano's theory for discrete-continuum interaction to broadband PDC, we posit the form of discrete/continuum energy eigenstate in the subspace $\mathcal{H}_{n_\mathrm{MR}=1,s}$ as
\begin{align}
\ket{\phi_s^\mathrm{d/c}(\omega)}=c_{s}^\mathrm{d/c}(\omega)\ket{b_s}+\int_0^\infty\mathrm{d}p\,f_s^\mathrm{d/c}(\omega,p)\ket{a_{sp}}
\end{align}
with an eigenenergy $\omega$, where the labels ``d'' and ``c'' denote ``discrete'' and ``continuum'', respectively. Intuitively, the discrete pump state $\ket{b_s}$ gets hybridized with the signal states to form the discrete eigenstate $\ket{\phi_s^\mathrm{d}(\omega)}$, while each signal state turns into the continuum eigenstates $\ket{\phi_s^\mathrm{c}(\omega)}$. The discrete states are normalized as $\langle\phi_s^\mathrm{d}(\omega)\vert \phi_{s'}^\mathrm{d}(\omega)\rangle=\delta(s-s')$, while the states in the continuum energy spectrum fulfill
$\langle\phi_s^\mathrm{c}(\omega)\vert \phi_{s'}^\mathrm{c}(\omega')\rangle=\delta(s-s')\delta(\omega-\omega')$ , and thus, discrete and continuum states have different units $[\mathrm{length}^{1/2}\cdot \mathrm{frequency}^{1/2}]$ and $[\mathrm{length}^{1/2}]$, respectively. These orthonormalization conditions are equivalently written as
\begin{subequations}
\label{eq:fano-normalization-condition}
\begin{align}
    \label{eq:normalization-discrete}
    |c_s^\mathrm{d}(\omega)|^2+\int_0^\infty \mathrm{d}p\,|f_s^\mathrm{d}(\omega,p)|^2&=1&\mathrm{(for~discrete~state)}\\
    \label{eq:normalization-continuum}
    c_s^{\mathrm{c}*}(\omega)c_s^\mathrm{c}(\omega')+\int_0^\infty \mathrm{d}p\,f_s^\mathrm{c*}(\omega,p)f_s^\mathrm{c}(\omega',p)&=\delta(\omega-\omega')&\mathrm{(for~continuum~state)}
\end{align}
\end{subequations}

Because $\ket{\phi_s(\omega)}$ is an energy eigenstate with an eigenvalue $\hbar\omega$, it must fulfill
\begin{align}
\label{eq:fano-eigenstate-condition}
\frac{1}{\hbar}\hat{H}\ket{\phi_s(\omega)}=\omega\ket{\phi_s(\omega)}
\end{align}
whose right-hand side takes the form
\begin{align}
\label{eq:fano-hamiltonian-action}
\begin{split}
\frac{1}{\hbar}\hat{H}\ket{\phi_s(\omega)}=&\int_0^\infty\mathrm{d}p\,\Bigl(rc_s(\omega)+
\delta_{as}(p)f_s(\omega,p)\Bigr)\ket{a_{sp}}\\
&\quad+\left(\delta_{bs} c_s(\omega)\hat{b}_s^\dagger+r\int_0^\infty\mathrm{d}p\,f_{s}(\omega,p)\right)\ket{b_s}.
\end{split}
\end{align}
We have omitted the labels for discrete/continuum states because these equations hold for both cases. Inserting \eqref{eq:fano-hamiltonian-action} to \eqref{eq:fano-eigenstate-condition} and comparing coefficients for $\ket{b_s}$ and $\ket{a_{sp}}$, we get
\begin{subequations}
\label{eq:fano-lab-frame}
\begin{align}
    \omega c_s(\omega)&=\delta_{bs}c_s(\omega)+r\int_0^\infty\mathrm{d}p\,f_s(\omega,p)\\
    \omega f_s(\omega,p)&=rc_s(\omega)+\delta_{as}(p)f_s(\omega,p),
\end{align}
\end{subequations}
which, together with the normalization conditions \eqref{eq:fano-normalization-condition}, specify all the eienstates in $\mathcal{H}_{n_\mathrm{MR}=1,s}$. Notice that \eqref{eq:fano-lab-frame} can be seen as a multimode extension of the single-mode equations \eqref{eq:fano-single-mode-eigen}.

\subsection{Broadband PDC: eigenstates}
\label{sec:fano-broadband-eigen}
While we can, in principle, directly solve \eqref{eq:fano-lab-frame}, we first normalize the variables to dimensionless forms. This normalization allows us to simplify the equations to a universal form that involves only parameters that nontrivially alter the system behavior. Also, by factoring out trivial scaling factors, we can establish isomorphism among systems with different parameters that can be mapped to the same universal form.

\paragraph{Conversion to universal form} For concreteness, we approximate $\delta\omega_a$ as a quadratic function near the carrier as
\begin{align}
    \delta\omega_a(2\pi s)=\delta\omega_a(0)+\delta\omega_a'(0)(2\pi s)+\frac{1}{2}\delta\omega_a''(0)(2\pi s)^2,
\end{align}
which lets us define
\begin{align}
\delta_{as}(p)=\delta_{as}(0)+\sigma p^2
\end{align}
with $\delta_{as}(0)=2\delta\omega_a(0)+\delta\omega_a'(0)(2\pi s)+\frac{1}{4}\delta\omega_a''(0)(2\pi s)^2$ and $\sigma=4\pi^2\delta\omega_a''(0)>0$. Using the conversion formulas given in Table~\ref{tab:dispersion}, we can express $\sigma$ with classical experimental parameters as $\sigma=-4\pi^2k''_av_{g,a}^2$, where $v_{g,a}$ and $k''_a$ are the group velocity and GVD of signal mode, respectively. The curvature term $\sigma$, which physically represents an effective mass of a signal photon, becomes independent of $s$. 

Using the short-hand notation $\delta_{bs}=\delta\omega_b(2\pi s)$, we can also define
\begin{align}
    \delta_s=\delta_{bs}-\delta_{as}(0).
\end{align}
Intuitively, $\delta_s$ represents the energy deference between a pump photon with momentum $s$ and a two signal photons with momentum $s/2$. At the carrier wavenumner, $\delta_{s=0}$ reduces to the phase-mismatch $\delta_{s=0}=\delta\omega_b(0)-2\delta\omega_a(0)$. 

For the single-mode case discussed in Sec.~\ref{sec:single-mode-pdc-appendix}, we only scaled the time (\textit{i.e.}, energy) to obtain the universal form. Because multiple spectral components are involved in broadband PDC, we must also scale the wavenumber (\textit{i.e.}, space) to identify minimum set of nontrivial parameters. To find the scaling factors, we identify characteristic nonlinear coupling $g_\mathrm{pdc}$ and length scale $\zeta_\mathrm{pdc}$ via dimensional analysis. We note that $r$ has a unit of $[\mathrm{time}^{-1}\cdot \mathrm{length}^{1/2}]$, while $\sigma$ has a unit of $[\mathrm{time}^{-1}\cdot \mathrm{length}^{2}]$. As a result, it is motivated to define
\begin{align}
&g_\mathrm{pdc}=(r^4/\sigma)^{1/3} &\zeta_\mathrm{pdc}=(\sigma/r)^{2/3}.
\end{align}
It is worth noting the scaling of $g_\mathrm{pdc}\propto \sigma^{-1/3}\propto\mathrm{GVD}^{-1/3}$, implying that a waveguide with smaller dispersion exhibits stronger nonlinear coupling even if the pump state is CW. With these variables, we scale energy and wavenumber to dimensionless variables
\begin{align}
&\lambda=(\omega-\delta_{as}(0))/g_\mathrm{pdc}&\mu=\zeta_\mathrm{pdc} p,
\end{align}
which leaves us with a single-nontrivial parameter: Normalized phase-mismatch
\begin{align}
\label{eq:definition-xi}
\xi=\delta_s/g_\mathrm{pdc}.
\end{align}
The wavefunctions are also normalized to dimensionless forms via
\begin{subequations}
\label{eq:fano-conversion-amplitudes}
\begin{align}
&f^\mathrm{d}_\xi(\lambda,\mu)=\zeta^{-1/2}_\mathrm{pdc}\,f_s^\mathrm{d}(\omega,p)&c^\mathrm{d}_\xi(\lambda)=c_s^\mathrm{d}(\omega),
\end{align}
for discrete state and
\begin{align}
&f^\mathrm{c}_\xi(\lambda,\mu)=g^{1/2}_\mathrm{pdc}\zeta^{-1/2}_\mathrm{pdc}\,f_s^\mathrm{c}(\omega,p)&c^\mathrm{c}_\xi(\lambda)=g^{1/2}_\mathrm{pdc}\,c_s^\mathrm{c}(\omega)
\end{align}
\end{subequations}
for continuum states, which are defined to fulfill normalization conditions
\begin{subequations}
\label{eq:fano-normalization-dimensionless}
\begin{align}
    \label{eq:normalization-discrete-dimensionless}
    |c^\mathrm{d}_\xi(\omega)|^2+\int_0^\infty \mathrm{d}\mu\,|f^\mathrm{d}_\xi(\omega,\mu)|^2&=1&\mathrm{(for~discrete~state)}\\
    \label{eq:normalization-continuum-dimensionless}
    c^{\mathrm{c}*}_\xi(\lambda)c^\mathrm{c}_\xi(\lambda')+\int_0^\infty \mathrm{d}\mu\,f^{\mathrm{c}*}_\xi(\omega,\mu)f^\mathrm{c}_\xi(\omega',\mu)&=\delta(\lambda-\lambda')&\mathrm{(for~continuum~state)}
\end{align}
\end{subequations}
With these dimensionless variables, we can convert \eqref{eq:fano-lab-frame} to a universal form
\begin{subequations}
\label{eq:fano-universal}
\begin{align}
\label{eq:fano-pump}
    \lambda c_\xi(\lambda)&=\xi c_\xi(\lambda)+\int_0^\infty\mathrm{d}\mu\,f_\xi(\lambda,\mu)\\
\label{eq:fano-signal}
    \lambda f_\xi(\lambda,\mu)&=c_\xi(\lambda)+\mu^2f_\xi(\lambda,\mu),
\end{align}
\end{subequations}
which has a single nontrivial parameter $\xi=\delta_s/g_\mathrm{pdc}$. Again, we omitted the labels for discrete/continuum states because these relations apply to both cases. The universal form \eqref{eq:fano-universal} can be seen as a broadband extension of the single-mode universal form \eqref{eq:fano-single-mode-universal}. Because only $\xi$ depends nontrivially on $s$, once we find solutions to \eqref{eq:fano-universal} for any given $\xi$, we can use the results to solve the eigenproblem for any $s$.

\paragraph{Discrete eigenstate}
For $\lambda<0$, it turns out that we have a discrete bound-state solution called ``optical mesons''~\cite{Drummond1997} with binding energy $\lambda=-\bar{\lambda}_\xi<0$. Because there is only one solution, we explicitly denote $c^\mathrm{d}_\xi=c_\xi(-\bar{\lambda}_\xi)$ and $f^\mathrm{d}_\xi(\mu)=f_\xi(-\bar{\lambda}_\xi,\mu)$, and we solve \eqref{eq:fano-signal} for $f^\mathrm{d}_\xi$ to get
\begin{subequations}
\label{eq:meson-solutions}
\begin{align}
\label{eq:meson-signal}
f^\mathrm{d}_\xi(\mu)=-\frac{c^\mathrm{d}_\xi}{\bar{\lambda}_\xi+\mu^2}.
\end{align}
Then, inserting to \eqref{eq:meson-signal} to \eqref{eq:fano-pump} gives us
\begin{align}
\label{eq:meson-energy}
\bar{\lambda}_\xi=-\xi+\frac{\pi}{2\sqrt{\bar{\lambda}_\xi}},
\end{align}
which is an analogous equation to \eqref{eq:fano-single-mode-eigen} that determines a discrete eigenenergy. The eigenenergy has an asymptotic scaling $\bar{\lambda}_\xi\approx(\pi/2\xi)^2$ for $\xi\rightarrow +\infty$ and $\bar{\lambda}_\xi\approx-\xi$ for $\xi\rightarrow -\infty$. Using the normalization condition for discrete states \eqref{eq:normalization-discrete}, we can determine the value of $c^\mathrm{d}_\xi$ as
\begin{align}
c^\mathrm{d}_\xi=\left(1+\frac{\pi}{4\bar{\lambda}^{3/2}_\xi}\right)^{-1/2}.
\end{align}
\end{subequations}
The equations \eqref{eq:meson-solutions} provide analytic solutions for the discrete bound state of the system.

\paragraph{Continuum eigenstates}
We now focus on the solution $\lambda\geq0$, for which we have a continuum solution, as shown shortly. To explicitly denote the continuum nature of the eigenstates, we label the variables with superscript ``c'', \textit{e.g.}, $c^\mathrm{c}_\xi(\lambda)$ and $f^\mathrm{c}_\xi(\lambda,\mu)$. Upon solving \eqref{eq:fano-signal} for $f^\mathrm{c}_\xi(\lambda,\mu)$, we need to introduce a principal part of $f^\mathrm{c}_\xi(\lambda,\mu)$ to account for the singularity at $\lambda=\mu^2$ as
\begin{align}
\label{eq:fano-continuum-f}
f^\mathrm{c}_\xi(\lambda,\mu)=c^\mathrm{c}_\xi(\lambda)\left(\frac{1}{\lambda- \mu^2}+w_\xi(\lambda)\delta\left(\lambda-\mu^2\right)\right).
\end{align}
The value of the principal component $w_\xi(\lambda)$ is to be determined afterward to fulfill the normalization condition. Inserting \eqref{eq:fano-continuum-f} to \eqref{eq:fano-pump} yields an equality
\begin{align}
\label{eq:fano-continuum-energy}
\lambda=\xi+\frac{w_\xi(\lambda)}{2\sqrt{\lambda}}
\end{align}
where we have used $P\int\mathrm{d}\mu\,(\lambda-\mu^2)^{-1}=0$ with $P$ denoting the Cauchy principal value integration. While \eqref{eq:fano-continuum-energy} might seem similar to \eqref{eq:meson-energy} (and the single-mode case \eqref{eq:fano-single-mode-eigen}), this equation has qualitatively different characteristics and cannot be used to obtain a specific eigenenergy. This is because $w_\xi(\lambda)$ is a free variable that has not been set yet, making it possible for any $\lambda\geq 0$ to be a solution of \eqref{eq:fano-continuum-energy}. Instead, \eqref{eq:fano-continuum-energy} should be seen as an equation that determines the functional form of $w_\xi(\lambda)$ as
\begin{align}
\label{eq:fano-continuum-w}
    w_\xi(\lambda)=2\sqrt{\lambda}\,(\lambda-\xi)
\end{align}
for every $\lambda\geq 0$. 

Finally, we use the continuum normalization condition \eqref{eq:normalization-continuum} to determine the value of $c^\mathrm{c}_\xi(\lambda)$. For this, we note that
\begin{align}
    \frac{1}{(x-z)(y-z)}=\frac{1}{x-y}\left(\frac{1}{y-z}-\frac{1}{x-z}\right)+\pi^2\delta(x-y)\delta\left(z-(x+y)/2\right)
\end{align}
holds true~\cite{Fano1961}, which we use to derive
\begin{align}
    P\int_0^\infty\mathrm{d}\mu f^\mathrm{c*}_\xi(\lambda,\mu)&f^\mathrm{c}(\lambda',\mu)=c^\mathrm{c*}_\xi(\lambda)c^\mathrm{c}_\xi(\lambda')\left(-1+\frac{\pi^2+w^2_\xi(\lambda)}{2\sqrt{\lambda}}\delta(\lambda-\lambda')\right).
\end{align}
As a result, the normalization condition determines the value of $c^\mathrm{c}_\xi(\lambda)$ as
\begin{align}
\label{eq:fano-continuum-c}
c^\mathrm{c}_\xi(\lambda)=\left(\frac{2\sqrt{\lambda}}{\pi^2+w^2_\xi(\lambda)}\right)^{1/2}.
\end{align}
In summary \eqref{eq:fano-continuum-f}, \eqref{eq:fano-continuum-w}, and \eqref{eq:fano-continuum-c} provide analytic expressions for the continuum eigenstates with energy $\lambda\geq 0$.

\paragraph{Conversion to the orginal frame}
To obtain eigenstates for a given $\hat{s}=s$, we set $\xi=\delta_s/g_\mathrm{pdc}$ and solve \eqref{eq:fano-universal} to obtain $c^{\mathrm{d}/\mathrm{c}}_\xi(\lambda)$ and $f^{\mathrm{d}/\mathrm{c}}_\xi(\lambda,\mu)$. Using these solutions, we can denote the discrete and continuum eigenstates as
\begin{subequations}
\begin{align}
\ket{\phi_s^\mathrm{d}}&=c^\mathrm{d}_\xi\ket{b_s}+\zeta^{-1/2}_\mathrm{pdc}\int_0^\infty\mathrm{d}\mu\,f^\mathrm{d}_\xi(\mu)\ket{a_{s,\zeta^{-1}_\mathrm{pdc} \mu}}\\
\ket{\phi_s^\mathrm{c}(\omega)}&=g^{1/2}_\mathrm{pdc}\,c^\mathrm{c}_\xi(\lambda)\ket{b_s}+g^{1/2}_\mathrm{pdc}\zeta^{-1/2}_\mathrm{pdc}\int_0^\infty\mathrm{d}\mu\,f^\mathrm{c}_\xi(\lambda,\mu)\ket{a_{s,\zeta^{-1}_\mathrm{pdc}\mu}},
\end{align}
\end{subequations}
where $\lambda=(\omega-\delta_{as}(0))/g_\mathrm{pdc}$. With these eigenstates, we can analytically solve any pulse propagation dynamics in the subspace $\mathcal{H}_{n_\mathrm{MR}=1,s}$. For instance, the dynamics of an initial single-pump photon state $\ket{\psi_s(0)}=\ket{b_s}$ are
\begin{subequations}
\label{eq:fano-cw-propagation}
\begin{align}
\ket{\psi_s(t)}&=\ket{\psi_{as}(t)}+\ket{\psi_{bs}(t)},
\end{align}
where signal and pump wavefunctions are
\begin{align}
\ket{\psi_{as}(t)}&=e^{-\mathrm{i}\delta_{as}(0)t}\zeta^{-1/2}_\mathrm{pdc}\int_0^\infty\mathrm{d}\mu\left(e^{\mathrm{i}\bar{\lambda}\tau}c^\mathrm{d*}_\xi f^\mathrm{d}_\xi(\mu)+\int_{0}^\infty\mathrm{d}\lambda\,e^{-\mathrm{i}\lambda \tau}c^\mathrm{c*}_\xi(\lambda)f^\mathrm{c}_\xi(\lambda,\mu)\right)\ket{a_{s,\zeta^{-1}_\mathrm{pdc}\mu}}\nonumber\\
\ket{\psi_{bs}(t)}&=e^{-\mathrm{i}\delta_{as}(0)t}\left(e^{\mathrm{i}\bar{\lambda}\tau}|c^\mathrm{d}_\xi|^2+\int_{0}^\infty\mathrm{d}\lambda\,e^{-\mathrm{i}\lambda \tau}|c^\mathrm{c}_\xi(\lambda)|^2\right)\ket{b_s}
\end{align}
\end{subequations}
with $\tau=g_\mathrm{pdc}t$. We can now use \eqref{eq:fano-cw-propagation} to describe the propagation of a generic initial single-photon pump pulse $\ket{\psi(0)}=\int\mathrm{d}s\,h_s\ket{\psi_s(0)}=\int\mathrm{d}s\,h_s\ket{b_s}$;
\begin{align}
\label{eq:fano-generic-pulse-propagation}
\ket{\psi(t)}=\int\mathrm{d}s\,h_s\ket{\psi_s(t)}.
\end{align}
It is worth mentioning that \eqref{eq:fano-generic-pulse-propagation} shows each momentum component $\ket{\psi_s(t)}$ evolves independently under the parametric interactions. When we are interested in an observable that commutes with $\hat{s}$, \textit{e.g.}, signal/pump photon number $\hat{n}_c=\int\mathrm{d}s\,\hat{c}_s^\dagger\hat{c}_s~(c\in\{a,b\})$, the expectation value of the operator can also be written as a sum of independent contributions from every $\ket{\psi_s(t)}$ as
\begin{align}
\label{eq:fano-photon-number}
\bar{n}_{c}(t)=\langle\psi(t)\vert\hat{n}_{c}\vert\psi(t)\rangle=\int\mathrm{d}s\,|h_s|^2\bar{n}_{cs}(t)
\end{align}
with $\bar{n}_{cs}(t)=\langle\psi_s(t)\vert \hat{c}_s^\dagger\hat{c}_s\vert\psi_s(t)\rangle$. Therefore, once we understand the dynamics of each spectral component, the overall behavior of the system can be understood as a simple sum of these dynamics. In the following Sec.~\ref{sec:fano-pump-photon-dynamics} and Sec.~\ref{sec:fano-signal-photon-dynamics}, we study pump-photon dynamics and signal dynamics of each spectral component, respectively. Then, Sec.~\ref{sec:fano-pulse-pump-appendix} discusses pulse-pumped broadband PDC, studying the collective behavior of multiple spectral components. 

\subsection{Broadband PDC: dynamics of pump photon population}
\label{sec:fano-pump-photon-dynamics}
\begin{figure}[bt]
    \centering
    \includegraphics[width=0.57\textwidth]{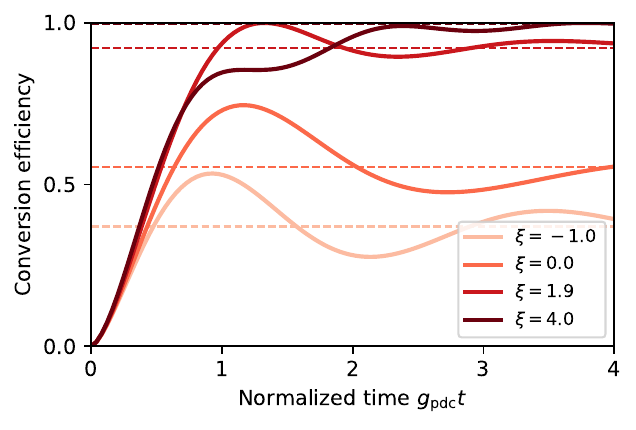}
    \caption{Conversion efficiency of single-photon-pumped broadband PDC as a function of interaction time for various values of phase-mismatch $\xi$. Dashed lines represent the asymptotic values to which the trajectories converge in the limit $t\rightarrow\infty$. For $\xi\approx 1.9$, the unit-efficiency conversion is realized at a finite time $\tau=g_\mathrm{pdc}t\approx 1.32$. Figure is adapted from Ref.~\cite{Yanagimoto2023-thesis}.}
    \label{fig:fano-pump-photon-number}
\end{figure}
Here, we focus on a single-spectral component and study the dynamics of the pump photon number
\begin{align}
\label{eq:fano-pump-photon-dynamics}
    \bar{n}_{bs}(t)=\Biggl\vert \underbrace{\left(1+\frac{\pi}{4\bar{\lambda}^{3/2}_\xi}\right)^{-1}}_{\mathrm{optical~meson~contribution}}+\int_0^\infty\mathrm{d}\lambda\,\underbrace{\frac{2\sqrt{\lambda}\,e^{-\mathrm{i}(\lambda+\bar{\lambda}_\xi)\tau}}{\pi^2+4\lambda(\lambda-\xi)^2}}_{\mathrm{continuum~contribution}}\Biggr\vert^2,
\end{align}
where the first and the second terms in the absolute value represent the contribution from the optical meson and the continuum, respectively. 

In Fig.~\ref{fig:fano-pump-photon-number}, we show the numerically calculated evolution of conversion efficiency $\mathcal{E}(t)=1-\bar{n}_{bs}(t)$ for various phase-mismatch $\xi$. Qualitatively, the behavior of $\mathcal{E}$ (thus that of $\bar{n}_{bs}(t)$) becomes more ``oscillatory'' for smaller $\xi$, exhibiting damped Rabi-like oscillations. Physically, the oscillation is caused by the interference between the optical meson and continuum contributions. Thus, its characteristic frequency is commensurate to the meson binding energy $\bar{\lambda}g_\mathrm{pdc}$. As $\xi$ increases, the oscillation is damped more strongly to exhibit more ``relaxational'' dynamics. For any values of $\xi$, the trajectories take features qualitatively distinctive from the single-mode physics shown in Fig.~\ref{fig:fano-rabi-pump-number}.

In the limit of $t\rightarrow\infty$, the continuum contribution in \eqref{eq:fano-pump-photon-dynamics} dephases and vanishes, leaving only the optical meson contribution as
\begin{align}
\lim_{t\rightarrow\infty}\bar{n}_{bs}(t)=\left(1+\frac{\pi}{4\bar{\lambda}^{3/2}}\right)^{-2},
\end{align}
and the asymptotic value $\bar{n}_{bs}(\infty)$ becomes smaller (larger) for larger (smaller) $\xi$. While $\bar{n}_{bs}(\infty)$ can only be zero asymptotically, due to nontrivial interference effects, $\bar{n}_{bs}(t)$ can transiently vanish, \textit{e.g.}, at $\xi\approx 1.9$ and $\tau=g_\mathrm{pdc}t\approx 1.32$, to enable unit-efficiency PDC.

Below, to obtain a more qualitative understanding, we consider the behavior of the pump photon number in two extreme regimes; (i) non-degenerate limit $\xi\gg1$, and (ii) degenerate limit $\xi\ll -1$. The phenomenology of the system for an intermediate phase-mismatch $\xi$ can be intuitively understood by interpolation between these two regimes.

\paragraph{Non-degenerate coupling limit}

For $\xi\gg1$, we are in a regime that we refer to as the ``non-degenerate coupling regime.'' As illustrated in Fig.~\ref{fig:fano-dissipative-dispersive}(a), in this regime, the pump-state energy lies deep within the energy band formed by the signal continuum. The PDC process excites signal frequency components with similar energy to the pump photon, leading to a Lorentzian lineshape in the signal continuum excitation, and the interaction is non-degenerate in the sense that the signal excitation is composed of two separate wavenumber components. In the spatial domain, such spectral features translate to two dispersive waves moving away from each other. This walkoff-like process suppresses backconversion, resulting in the strong damping of the oscillation in $\bar{n}_{bs}(t)$. To see this more quantitatively, we first note that the meson binding energy is well-approximated as $\bar{\lambda}_\xi\approx(\pi/2\xi)^2\approx 0$ for $\xi\gg 1$, implying that the meson contribution in \eqref{eq:fano-pump-photon-dynamics} is negligible. As to the continuum contribution, the integrand in \eqref{eq:fano-pump-photon-dynamics} can be well-approximated as a Lorentzian lineshape around $\lambda=\xi$ with full-width-at-half-maximum $\sqrt{\xi}/\pi$ as
\begin{align}
\frac{2\sqrt{\lambda}}{\pi^2+4\lambda(\lambda-\xi)^2}\approx \frac{2\sqrt{\xi}}{\pi^2+4\xi(\lambda-\xi)^2}.
\end{align}
As a result, the continuum contribution, which is given as a Fourier transform of a Lorentizan lineshape, exponentially decays to zero with characteristic decay time $\tau_\mathrm{decay}=\sqrt{\xi}/\pi$. In fact, we can show that the pump photon number exponentially decays as
\begin{align}
\label{eq:fano-pump-exponential}
    \bar{n}_{bs}(t)\approx e^{-\mathrm{\tau}/\tau_\mathrm{decay}}
\end{align}
in this regime. As shown in Fig.~\ref{fig:fano-dissipative-dispersive}(a), the approximate expression \eqref{eq:fano-pump-exponential} agrees well with the numerically calculated pump photon population.

\paragraph{Degenerate coupling limit}
For $\xi\ll-1$, we are in a regime that we refer to as the ``degenerate coupling regime.'' As shown in Fig.~\ref{fig:fano-dissipative-dispersive}, in this regime, the pump energy lies well below the energy band formed by the continuum. Because of the energy gap, the coupling is not highly energy-selective, and the pump photon can excite a broad spectrum of signal modes, which translates to a localized structure in the spatial domain. Since downconverted signal photon pairs can be physically close to each other for a longer time, backconversion is enhanced, leading to more oscillatory behaviors. To see this more quantitatively, we perform an approximation for the integrand in \eqref{eq:fano-pump-photon-dynamics} as
\begin{align}
\frac{2\sqrt{\lambda}}{\pi^2+4\lambda(\lambda-\xi)^2}\approx \frac{2\sqrt{\lambda}}{\pi^2+4\xi^2\lambda}.
\end{align}
The approximation holds well for $\lambda\ll |\xi|$, outside of which the integrant has negligible amplitude. After some math, we can show
\begin{align}
\label{eq:fano-pump-subpolynomial}
    \bar{n}_{bs}(t)=\left\vert 1-\frac{\pi}{4(-\xi^{3/2})}+\frac{\sqrt{\pi}}{2\xi^2\sqrt{\tau}}e^{\mathrm{i}(\xi\tau-\pi/4)}\right\vert,
\end{align}
where we have used $\bar{\lambda}\approx -\xi$ for $\xi\ll -1$. As shown in Fig.~\ref{fig:fano-dissipative-dispersive}(b), the approximate expression \eqref{eq:fano-pump-subpolynomial} agrees well with the numerically calculated pump photon population. Notably, $\bar{n}_{bs}(t)$ exhibits oscillations whose amplitude decays \emph{sub-polynomially} $\sim \tau^{-1/2}$. Due to the fast initial decay of the sub-polynomial decay, the oscillation $\bar{n}_{bs}(t)$ never converges to a canonical Rabi-oscillation in any limit of $\xi$, showing that the phenomenology of broadband PDC is qualitatively distinct from the single-mode PDC.

\subsection{Broadband PDC: dynamics of the signal wavefunction}
\label{sec:fano-signal-photon-dynamics}
\begin{figure}[bt]
    \centering
\includegraphics[width=\textwidth]{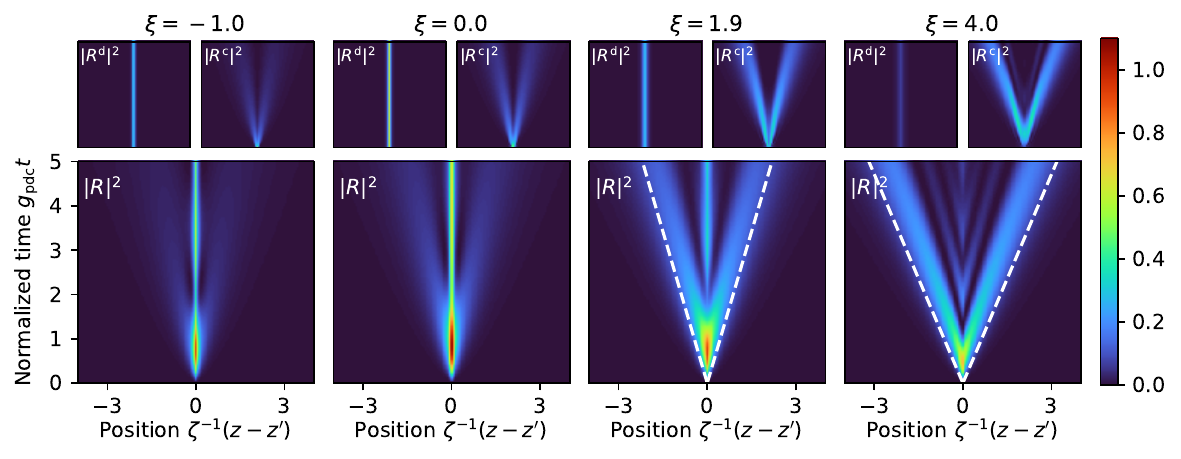}
    \caption{Dynamics of the spatial biphoton wavefunction for various values of the phase-mismatch $\xi$. Figures on the upper row show optical meson contribution $|R^\mathrm{d}|^2$ and continuum contribution $|R^\mathrm{c}|^2$. The white dashed lines represent the theoretically expected group velocity of the dispersive waves $\sqrt{\xi}\,g_\mathrm{pdc}\zeta_\mathrm{pdc}/\pi$. Figure is adapted from Ref.~\cite{Yanagimoto2023-thesis}.}
    \label{fig:fano-signal-spatial}
\end{figure}
We now turn our attention to the dynamics of the signal part of state $\ket{\psi_{as}(t)}$. In particular, we are interested in the spatial features of the signal state. To this end, we consider the signal biphoton wavefunction function $R(z,z')$, which is defined via
\begin{align}
\begin{split}
    \ket{\psi_{as}(t)}&=e^{-\mathrm{i\delta_{as}(0)}t}\zeta^{-1/2}_\mathrm{pdc}\int_0^\infty\mathrm{d}\mu \,Q(\mu)\ket{a_{s,\zeta^{-1}_\mathrm{pdc}\mu}}\\
    &=e^{-\mathrm{i\delta_{as}(0)}t}\int\mathrm{d}z\mathrm{d}z'e^{-2\pi\mathrm{i}s(z+z')}R(z,z')\hat{a}_z^\dagger\hat{a}_{z'}^\dagger\ket{0},
    \end{split}
\end{align}
where
\begin{align}
    Q(\mu)=-\frac{e^{\mathrm{i}\bar{\lambda}_\xi\tau}}{\bar{\lambda}_\xi+\mu^2}\left(1+\frac{\pi}{4\bar{\lambda}^{3/2}_\xi}\right)^{-1}+\int_0^\infty\mathrm{d}\lambda\,\frac{2\sqrt{\lambda}\,e^{-\mathrm{i}\lambda \tau}}{\pi^2+w^2_\xi(\lambda)}\left(\frac{1}{\lambda-\mu^2}+w_\xi(\lambda)\delta(\lambda-\mu^2)\right)
\end{align}
Physically, $R(z,z')$ is the wavefunction of the biphoton state written in the spatial domain, and $|R(z,z')|^2$ gives the joint probability distribution of finding two photons at position $z$ and $z'$. After some math, we can calculate the spatial biphoton wavefunction as
\begin{align}
&R(z,z')=\zeta^{-1/2}_\mathrm{pdc}\int_0^\infty\mathrm{d}\mu\,e^{2\pi\mathrm{i}\mu \zeta^{-1}_\mathrm{pdc} (z-z')}Q(\mu)\\
&=\underbrace{-\frac{2\pi\bar{\lambda}_\xi\exp\left(-2\pi \bar{\lambda}_\xi^{1/2}\zeta^{-1}_\mathrm{pdc}|z-z'|\right)}{\zeta^{1/2}_\mathrm{pdc}\left(\pi+4\bar{\lambda}^{3/2}_\xi\right)}e^\mathrm{i\bar{\lambda}_\xi\tau}}_{=R^\mathrm{d}~\mathrm{(optical~meson~contribution)}}+\underbrace{\int_0^\infty\mathrm{d}\lambda\,\frac{\cos\left(2\pi\sqrt{\lambda}\zeta^{-1}_\mathrm{pdc}|z-z'|+\Delta(\lambda)\right)e^{-\mathrm{i}\lambda \tau}}{\zeta^{1/2}_\mathrm{pdc}\sqrt{\pi^2+w^2_\xi(\lambda)}}}_{=R^\mathrm{c}~\mathrm{(continuum~contribution)}}\nonumber,
\end{align} 
where we denote the first and the second terms as $R^\mathrm{d}(z,z')$ and $R^\mathrm{c}(z,z')$, and $\Delta(\lambda)=-\arctan(\pi/w_\xi(\lambda))$ is the Fano parameter. In Fig.~\ref{fig:fano-signal-spatial}, we show the dynamics of the intensity of $R(z,z')$ alongside the optical meson contribution $R^\mathrm{d}(z,z')$ and the continuum contribution $R^\mathrm{c}(z,z')$. Due to the translational invariance of the system, $R(z,z')$ depends only on the distance between the photons $|z-z'|$.

For large $\xi$ (\textit{i.e.}, non-degenerate coupling regime), most of the signal contribution comes from the continuum excitation, which disperses in space as a function of time, where the group velocity of the dispersive waves is $\sqrt{\xi}\,g_\mathrm{pdc}\zeta_\mathrm{pdc}/\pi$. On the other hand, for small $\xi$ (\textit{i.e.}, degenerate coupling regime), the meson contribution dominates, and due to the nature of the optical meson as a bound state, the signal wavefunction remains exponentially localized. Finally, for an intermediate $\xi$, both continuum and meson contribution coexist, forming a triplet structure with interference patterns. These observations highlight that broadband PDC exhibits rich but localized spatial features. Note that all spatial coordinates are localized by the characteristic length scale $1/\xi\propto \mathrm{GVD}^{-2/3}$, meaning waveguides with smaller dispersion can realize more localized photon-photon correlation structures.

\subsection{Broadband PDC: pulsed-pump}
\label{sec:fano-pulse-pump-appendix}
So far, our analysis has focused on the behavior of a single spectral component. In this section, we extend the analysis to the collective behavior of multiple spectral components, studying PDC dynamics for an initial pulsed-pump state
\begin{align}
\label{eq:fano-pulse-initial}
\ket{\psi(0)}=\int\mathrm{d}s\,h_s\ket{b_s}.
\end{align}
As we saw in the previous sections, one of the main advantages of using such broadband pump pulse for semiclassical PDC (\textit{i.e.}, vacuum squeezing) was to increase the peak pump intensity, which significantly increases the conversion efficiency for the same average pump power. However, as we see shortly as a main takeaway message of this section, we show that the rate of single-photon-pumped PDC is almost independent of the pump-pulse shape, and thus, there exists no enhancement of conversion efficiency by broader pump bandwidth. Instead, the PDC rate is commensurate to $g$, which can be enhanced by broadening the \emph{signal fluorescence bandwidth}.

\begin{figure}
    \centering
    \includegraphics[width=0.6\textwidth]{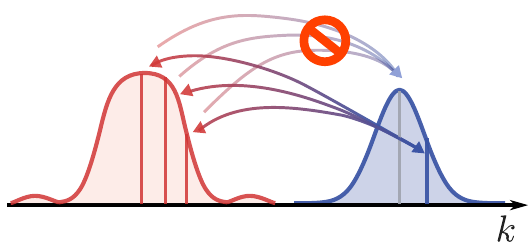}
    \caption{Illustration for the single-photon-pumped PDC with pulsed pump. Signal photons produced from a pump photon with momentum $s$ can only upconvert to the same pump mode. Figure is adapted from Ref.~\cite{Yanagimoto2023-thesis}.}
    \label{fig:fano-pulse-pump-schematics}
\end{figure}

A major difference between semiclassical PDC and single-photon-pumped PDC is the absence of cross-talk between different spectral components of the pump. As illustrated in Fig.~\ref{fig:fano-pulse-pump-schematics}, in single-photon-pumped PDC, signal photons that are produced by the downconversion of a pump photon with wavenumber (\textit{i.e.}, momentum) $s$ can only upconvert to the same pump mode, because the momentum of the signal photons $s/2\pm p$ always sums up to $s$. Here, it might seem possible for the pulsed nature of the pump to break this constraint. For instance, pulse waveform \eqref{eq:fano-pulse-initial} can contain non-zero amplitudes for 
$\ket{b_s}$ and $\ket{b_s'}$ with $s\neq s'$, which can downconvert to signal photons with momenta $s/2\pm p$ and $s'/2\pm p'$, respectively. Then, when signal photons with, say, momenta $s/2$ and $s'/2$ merge, the momentum of the resultant pump photon becomes $(s+s')/2$, which can be different from either $s$ or $s'$. However, such scattering processes are prohibited in single-photon-pumped PDC because the presence of $\ket{b_s}$ and $\ket{b_s'}$ are \emph{mutually exclusive}, \textit{i.e.}, a single pump-photon cannot simultaneously have momenta $s$ and $s'$. Consequently, the production of a signal-photon pair with momentum $s/2\pm p$ excludes the presence of another photon pair, forcing them to upconvert to the original pump mode with momentum $s$. For different pump spectral components to talk with each other, there must be at least two pump photons so that they can exchange momentum.

Because of the absence of cross-talk, the behavior of each spectral component is independent of the overall pulse shape $h_s$. This is reflected in the form of the system state
\begin{align}
\ket{\psi(t)}=\int\mathrm{d}s\,h_s\ket{\psi_s(t)},
\end{align}
where the explicit form of $\ket{\psi_s(t)}=e^{-\mathrm{i}\hat{H}t/\hbar}\ket{b_s}$ is provided in \eqref{eq:fano-cw-propagation} and is independent of $h_s$. As shown in the equation \eqref{eq:fano-cw-propagation}, the behavior of $\ket{\psi_s(t)}$ is solely determined by the normalized phase-mismatch $\xi=\delta_s/g_\mathrm{pdc}$ up to an overall phase. As a result, we can upper bound the overall conversion efficiency as
\begin{align}
\label{eq:fano-efficiency-inequality}
    \mathcal{E}(t)=\frac{1}{2}\int\mathrm{d}s\,|h_s|^2\langle\psi_s(t)\vert \hat{a}_s^\dagger\hat{a}_s\vert\psi_s(t)\rangle\leq \max_{s\in\mathbb{R}}\frac{1}{2}\langle\psi_s(t)\vert \hat{a}_s^\dagger\hat{a}_s\vert\psi_s(t)\rangle,
\end{align}
where we have used the independence of $h_s$ and $\ket{\psi_s(t)}$ to derive the inequality. The equation \eqref{eq:fano-efficiency-inequality} indicates that the photon conversion rate of pulse-pumped PDC cannot exceed that of an appropriately phase-matched CW-pumped PDC, which is rather counterintuitive from the viewpoint of classical NLO.

The explicit $s$-dependence of $\delta_s$ is given by
\begin{align}
\delta_s=\delta\omega_{b}(2\pi s)-\left(2\delta\omega_a(0)+\delta\omega_a'(0)(2\pi s)+\frac{1}{2}\delta\omega_a''(0)(2\pi s)^2\right).
\end{align}
Notably, when perfectly phase-matched conditions
\begin{align}
\label{eq:fano-extended-phase-matching}
&\delta\omega_b'(0)=\delta\omega_a'(0) &\delta\omega_b''(0)=\delta\omega_b''(0)/2
\end{align}
hold true, the $s$-dependence of $\delta_s$ vanishes. When this condition is met, $\ket{\psi_s(t)}$ becomes homogeneous up to an overall phase factor, and the inequality \eqref{eq:fano-efficiency-inequality} is always saturated. In other words, at least in terms of conversion efficiency, the PDC dynamics become completely independent of the pump pulse shape. To see this more clearly, note that the pump part evolves as
\begin{subequations}
\label{eq:fano-pulse-overall-pump-dynamics}
\begin{align}
    \ket{\psi_{b}(t)}&=\mathcal{C}_\xi(t)\int\mathrm{d}s\,e^{-\mathrm{i}\delta_{as}(0)t}h_s\ket{b_s}
\end{align}
where the overall amplitude is
\begin{align}
\mathcal{C}_\xi(t)=\left(e^{\mathrm{i}\bar{\lambda}\tau}|c^\mathrm{d}_\xi|^2+\int_{0}^\infty\mathrm{d}\lambda\,e^{-\mathrm{i}\lambda \tau}|c^\mathrm{c}_\xi(\lambda)|^2\right).
\end{align}
\end{subequations}
The form of the equation \eqref{eq:fano-pulse-overall-pump-dynamics} suggests the pump field distribution evolves as if there were only the linear part of the Hamiltonian, and the nonlinear dynamics affect the pump evolution only via the variation of the overall amplitude $C_\xi(t)$.

In Fig.~\ref{fig:fano-pulse-pumped}, we show the results of numerical simulations for single-photon-pumped PDC assuming \eqref{eq:fano-extended-phase-matching}. We consider various initial pump-pulse shapes, \textit{i.e.}, $m$th Hermite-Gaussian function for $m\in\{0,1,2\}$ (see Fig.~\ref{fig:fano-pulse-pumped}(a)), which leads to highly complicated dynamics as shown in Fig.~\ref{fig:fano-pulse-pumped}(c). However, for any pump pulse shape, the overall conversion efficiency follows an identical trajectory (see Fig.~\ref{fig:fano-pulse-pumped}(b)). For the phase-mismatch $\xi\approx 1.9$ considered for the simulation, unit conversion is achieved at time $\tau=g_\mathrm{pdc}t\approx 1.32$, represented by grey dashed lines in the figures.

In summary, we have seen that the nonlinear coupling for broadband single-photon-pumped PDC is \emph{independent} of the shape (\textit{i.e.,}, bandwidth) of the pump pulse. To enhance the PDC rate, one really needs to increase the characteristic coupling rate $g_\mathrm{pdc}=(r^4/\sigma)^{1/3}$. Aside from directly increasing waveguide nonlinearity $r$, this can be achieved by reducing the signal dispersion $\sigma$, implying that what determines the broadband PDC rate is the signal fluorescence bandwidth, not the pump bandwidth.

\bibliography{bibliography}

\begin{thebibliography}{100}
\newcommand{\enquote}[1]{``#1''}

\bibitem{lu2020toward}
J.~Lu, M.~Li, C.-L. Zou, A.~{Al Sayem}, and H.~X. Tang, \enquote{{Toward 1\%
  single photon nonlinearity with periodically-poled lithium niobate microring
  resonators},} {\protect\JournalTitle{Optica}} \textbf{7}, 1654--1659 (2020).

\bibitem{zhao2022ingap}
M.~Zhao and K.~Fang, \enquote{{InGaP quantum nanophotonic integrated circuits
  with 1.5\% nonlinearity-to-loss ratio},} {\protect\JournalTitle{Optica}}
  \textbf{9}, 258--263 (2022).

\bibitem{Yanagimoto2022_temporal}
R.~Yanagimoto, E.~Ng, M.~Jankowski, H.~Mabuchi, and R.~Hamerly,
  \enquote{{Temporal trapping: a route to strong coupling and deterministic
  optical quantum computation},} {\protect\JournalTitle{Optica}} \textbf{9},
  1289 (2022).

\bibitem{Fink2008}
J.~M. Fink, M.~G{\"o}ppl, M.~Baur, R.~Bianchetti, P.~J. Leek, A.~Blais, and
  A.~Wallraff, \enquote{{Climbing the Jaynes–Cummings ladder and observing
  its nonlinearity in a cavity QED system},} {\protect\JournalTitle{Nature}}
  \textbf{454}, 315--318 (2008).

\bibitem{Wallraff2004}
A.~Wallraff, D.~I. Schuster, A.~Blais, L.~Frunzio, R.-S. Huang, J.~Majer,
  S.~Kumar, S.~M. Girvin, and R.~J. Schoelkopf, \enquote{{Strong coupling of a
  single photon to a superconducting qubit using circuit quantum
  electrodynamics},} {\protect\JournalTitle{Nature}} \textbf{431}, 162--167
  (2004).

\bibitem{Nogues1999}
G.~Nogues, A.~Rauschenbeutel, S.~Osnaghi, M.~Brune, J.~M. Raimond, and
  S.~Haroche, \enquote{{Seeing a single photon without destroying it},}
  {\protect\JournalTitle{Nature}} \textbf{400}, 239 (1999).

\bibitem{Brune1996quantum}
M.~Brune, F.~Schmidt-Kaler, A.~Maali, J.~Dreyer, E.~Hagley, J.~Raimond, and
  S.~Haroche, \enquote{Quantum rabi oscillation: A direct test of field
  quantization in a cavity,} {\protect\JournalTitle{Physical review letters}}
  \textbf{76}, 1800 (1996).

\bibitem{Gleyzes2007}
S.~Gleyzes, S.~Kuhr, C.~Guerlin, J.~Bernu, S.~Del\'eglise, U.~B. Hoff,
  M.~Brune, J.-M. Raimond, and S.~Haroche, \enquote{{Quantum jumps of light
  recording the birth and death of a photon in a cavity},}
  {\protect\JournalTitle{Nature}} \textbf{446}, 297 (2007).

\bibitem{Thompson1992}
R.~J. Thompson, G.~Rempe, and H.~J. Kimble, \enquote{{Observation of
  normal-mode splitting for an atom in an optical cavity},}
  {\protect\JournalTitle{Phys. Rev. Lett.}} \textbf{68}, 1132 (1992).

\bibitem{Kimble1998}
H.~J. Kimble, \enquote{{Strong interactions of single atoms and photons in
  cavity QED},} {\protect\JournalTitle{Phys. Scr.}} \textbf{1998}, 127 (1998).

\bibitem{Lugiato1987}
L.~A. Lugiato and R.~Lefever, \enquote{Spatial dissipative structures in
  passive optical systems,} {\protect\JournalTitle{Phys. Rev. Lett.}}
  \textbf{58}, 2209--2211 (1987).

\bibitem{werner1993simulton}
M.~Werner and P.~Drummond, \enquote{Simulton solutions for the parametric
  amplifier,} {\protect\JournalTitle{JOSA B}} \textbf{10}, 2390--2393 (1993).

\bibitem{Grelu2012}
P.~Grelu and N.~Akhmediev, \enquote{{Dissipative solitons for mode-locked
  lasers},} {\protect\JournalTitle{Nat. Photon}} \textbf{6}, 84 (2012).

\bibitem{trillo1996bright}
S.~Trillo, \enquote{Bright and dark simultons in second-harmonic generation,}
  {\protect\JournalTitle{Optics letters}} \textbf{21}, 1111--1113 (1996).

\bibitem{mollenauer1984soliton}
L.~F. Mollenauer and R.~H. Stolen, \enquote{The soliton laser,}
  {\protect\JournalTitle{Optics letters}} \textbf{9}, 13--15 (1984).

\bibitem{siegman1986lasers}
A.~E. Siegman, \emph{Lasers} (University science books, 1986).

\bibitem{Haus1975}
H.~Haus, \enquote{Theory of mode locking with a slow saturable absorber,}
  {\protect\JournalTitle{IEEE Journal of Quantum Electronics}} \textbf{11},
  736--746 (1975).

\bibitem{Haus2000}
H.~Haus, \enquote{Mode-locking of lasers,} {\protect\JournalTitle{IEEE Journal
  of Selected Topics in Quantum Electronics}} \textbf{6}, 1173--1185 (2000).

\bibitem{Dudley2006}
J.~M. Dudley, G.~Genty, and S.~Coen, \enquote{Supercontinuum generation in
  photonic crystal fiber,} {\protect\JournalTitle{Rev. Mod. Phys.}}
  \textbf{78}, 1135--1184 (2006).

\bibitem{luo2017chip}
R.~Luo, H.~Jiang, S.~Rogers, H.~Liang, Y.~He, and Q.~Lin, \enquote{On-chip
  second-harmonic generation and broadband parametric down-conversion in a
  lithium niobate microresonator,} {\protect\JournalTitle{Optics express}}
  \textbf{25}, 24531--24539 (2017).

\bibitem{bruch2019opo}
A.~W. Bruch, X.~Liu, J.~B. Surya, C.-L. Zou, and H.~X. Tang, \enquote{On-chip
  $\chi$ (2) microring optical parametric oscillator,}
  {\protect\JournalTitle{Optica}} \textbf{6}, 1361--1366 (2019).

\bibitem{lu2019pplnring}
J.~Lu, J.~B. Surya, X.~Liu, A.~W. Bruch, Z.~Gong, Y.~Xu, and H.~X. Tang,
  \enquote{Periodically poled thin-film lithium niobate microring resonators
  with a second-harmonic generation efficiency of 250,000\%/w,}
  {\protect\JournalTitle{Optica}} \textbf{6}, 1455--1460 (2019).

\bibitem{lu2021ultralow}
J.~Lu, A.~Al~Sayem, Z.~Gong, J.~B. Surya, C.-L. Zou, and H.~X. Tang,
  \enquote{Ultralow-threshold thin-film lithium niobate optical parametric
  oscillator,} {\protect\JournalTitle{Optica}} \textbf{8}, 539--544 (2021).

\bibitem{McKenna2022}
T.~P. McKenna, H.~S. Stokowski, V.~Ansari, J.~Mishra, M.~Jankowski, C.~J.
  Sarabalis, J.~F. Herrmann, C.~Langrock, M.~M. Fejer, and A.~H. Safavi-Naeini,
  \enquote{Ultra-low-power second-order nonlinear optics on a chip,}
  {\protect\JournalTitle{Nature Communications}} \textbf{13} (2022).

\bibitem{rivoire2011multiply}
K.~Rivoire, S.~Buckley, and J.~Vu{\v{c}}kovi{\'c}, \enquote{Multiply resonant
  photonic crystal nanocavities for nonlinear frequency conversion,}
  {\protect\JournalTitle{Optics express}} \textbf{19}, 22198--22207 (2011).

\bibitem{Hickstein2019}
D.~D. Hickstein, D.~R. Carlson, H.~Mundoor, J.~B. Khurgin, K.~Srinivasan,
  D.~Westly, A.~Kowligy, I.~I. Smalyukh, S.~A. Diddams, and S.~B. Papp,
  \enquote{Self-organized nonlinear gratings for ultrafast nanophotonics,}
  {\protect\JournalTitle{Nature Photonics}} \textbf{13}, 494--499 (2019).

\bibitem{zhao2020high}
J.~Zhao, C.~Ma, M.~Rüsing, and S.~Mookherjea, \enquote{High quality entangled
  photon pair generation in periodically poled thin-film lithium niobate
  waveguides,} {\protect\JournalTitle{Physical Review Letters}} \textbf{124}
  (2020).

\bibitem{Jankowski2020}
M.~Jankowski, C.~Langrock, B.~Desiatov, A.~Marandi, C.~Wang, M.~Zhang, C.~R.
  Phillips, M.~Lon{\v{c}}ar, and M.~M. Fejer, \enquote{Ultrabroadband nonlinear
  optics in nanophotonic periodically poled lithium niobate waveguides,}
  {\protect\JournalTitle{Optica}} \textbf{7}, 40 (2020).

\bibitem{Jankowski2021SCG}
M.~Jankowski, C.~Langrock, B.~Desiatov, M.~Lon{\v{c}}ar, and M.~M. Fejer,
  \enquote{Supercontinuum generation by saturated second-order nonlinear
  interactions,} {\protect\JournalTitle{{APL} Photonics}} \textbf{8} (2023).

\bibitem{Singh2020}
N.~Singh, M.~Raval, A.~Ruocco, and M.~R. Watts, \enquote{Broadband 200-nm
  second-harmonic generation in silicon in the telecom band,}
  {\protect\JournalTitle{Light: Science {\&} Applications}} \textbf{9} (2020).

\bibitem{Jankowski2022}
M.~Jankowski, N.~Jornod, C.~Langrock, B.~Desiatov, A.~Marandi, M.~Lon{\v{c}}ar,
  and M.~M. Fejer, \enquote{Quasi-static optical parametric amplification,}
  {\protect\JournalTitle{Optica}} \textbf{9}, 273 (2022).

\bibitem{Jankowski2021-review}
M.~Jankowski, J.~Mishra, and M.~M. Fejer, \enquote{{Dispersion-engineered
  $\chi^{(2)}$ nanophotonics: a flexible tool for nonclassical light},}
  {\protect\JournalTitle{J. Phys. Photon.}} \textbf{3}, 042005 (2021).

\bibitem{Armen2006}
M.~A. Armen and H.~Mabuchi, \enquote{Low-lying bifurcations in cavity quantum
  electrodynamics,} {\protect\JournalTitle{Phys. Rev. A}} \textbf{73}, 063801
  (2006).

\bibitem{Olivares2012}
S.~Olivares, \enquote{{Quantum optics in the phase space - A tutorial on
  Gaussian states},} {\protect\JournalTitle{Eur. Phys. J. Special Topics}}
  \textbf{203}, 3--24 (2012).

\bibitem{Weedbrook2012}
C.~Weedbrook, S.~Pirandola, R.~García-Patrón, N.~J. Cerf, T.~C. Ralph, J.~H.
  Shapiro, and S.~Lloyd, \enquote{Gaussian quantum information,}
  {\protect\JournalTitle{Rev. Mod. Phys.}} \textbf{84}, 621 (2012).

\bibitem{Quesada2022}
N.~Quesada, L.~G. Helt, M.~Menotti, M.~Liscidini, and J.~E. Sipe,
  \enquote{Beyond photon pairs{\textemdash}nonlinear quantum photonics in the
  high-gain regime: a tutorial,} {\protect\JournalTitle{Advances in Optics and
  Photonics}} \textbf{14}, 291 (2022).

\bibitem{Triginer2020}
G.~Triginer, M.~D. Vidrighin, N.~Quesada, A.~Eckstein, M.~Moore, W.~S.
  Kolthammer, J.~Sipe, and I.~A. Walmsley, \enquote{Understanding high-gain
  twin-beam sources using cascaded stimulated emission,}
  {\protect\JournalTitle{Physical Review X}} \textbf{10}, 031063 (2020).

\bibitem{Guidry2022}
M.~A. Guidry, D.~M. Lukin, K.~Y. Yang, R.~Trivedi, and J.~Vu{\v c}kovi\'c,
  \enquote{{Quantum optics of soliton microcombs},} {\protect\JournalTitle{Nat.
  Photon.}} \textbf{16}, 52 (2022).

\bibitem{Kashiwazaki2020}
T.~Kashiwazaki, N.~Takanashi, T.~Yamashima, T.~Kazama, K.~Enbutsu, R.~Kasahara,
  T.~Umeki, and A.~Furusawa, \enquote{Continuous-wave 6-db-squeezed light with
  2.5-thz-bandwidth from single-mode ppln waveguide,}
  {\protect\JournalTitle{APL Photonics}} \textbf{5} (2020).

\bibitem{Vahlbruch2016}
H.~Vahlbruch, M.~Mehmet, K.~Danzmann, and R.~Schnabel, \enquote{Detection of 15
  db squeezed states of light and their application for the absolute
  calibration of photoelectric quantum efficiency,}
  {\protect\JournalTitle{Physical review letters}} \textbf{117}, 110801 (2016).

\bibitem{Bao2021}
C.~Bao, M.-G. Suh, B.~Shen, K.~{\c S}afak, A.~Dai, H.~Wang, L.~Wu, Z.~Yuan,
  F.~X. Yang, K{\"a}rtner, and K.~J. Vahala, \enquote{{Quantum diffusion of
  microcavity solitons},} {\protect\JournalTitle{Nat. Phys.}} \textbf{17},
  462--466 (2021).

\bibitem{Walschaers2021}
M.~Walschaers, \enquote{{Non-Gaussian Quantum States and Where to Find Them},}
  {\protect\JournalTitle{PRX Quantum}} \textbf{2}, 030204 (2021).

\bibitem{Ryo2023Mesoscopic}
R.~Yanagimoto, E.~Ng, M.~P. Jankowski, R.~Nehra, M.~P. Timothy, T.~Onodera,
  L.~G. Wright, R.~Hamerly, A.~Marandi, M.~M. Fejer, and H.~Mabuchi,
  \enquote{Mesoscopic ultrafast nonlinear optics -- the emergence of multimode
  quantum non-gaussian physics,} {\protect\JournalTitle{arXiv preprint
  arXiv:2311.13775}}  (2023).

\bibitem{Cantu2020}
S.~H. Cantu, A.~V. Venkatramani, W.~Xu, L.~Zhou, B.~Jelenkovi{\'c}, M.~D.
  Lukin, and V.~Vuleti{\'c}, \enquote{Repulsive photons in a quantum nonlinear
  medium,} {\protect\JournalTitle{Nature Physics}} \textbf{16}, 921--925
  (2020).

\bibitem{Firstenberg2013}
O.~Firstenberg, T.~Peyronel, Q.-Y. Liang, A.~V. Gorshkov, M.~D. Lukin, and
  V.~Vuleti{\'c}, \enquote{Attractive photons in a quantum nonlinear medium,}
  {\protect\JournalTitle{Nature}} \textbf{502}, 71--75 (2013).

\bibitem{Chang2014}
D.~Chang, C.~Langrock, Y.-W. Lin, C.~Phillips, C.~Bennett, and M.~M. Fejer,
  \enquote{Complex-transfer-function analysis of optical-frequency converters,}
  {\protect\JournalTitle{Optics Letters}} \textbf{39}, 5106--5109 (2014).

\bibitem{Shapiro2006}
J.~H. Shapiro, \enquote{{Single-photon Kerr nonlinearities do not help quantum
  computation},} {\protect\JournalTitle{Phys. Rev. A}} \textbf{73}, 062305
  (2006).

\bibitem{Hillery1984}
M.~Hillery and L.~D. Mlodinow, \enquote{Quantization of electrodynamics in
  nonlinear dielectric media,} {\protect\JournalTitle{Physical Review A}}
  \textbf{30}, 1860--1865 (1984).

\bibitem{Drummond1990}
P.~D. Drummond, \enquote{Electromagnetic quantization in dispersive
  inhomogeneous nonlinear dielectrics,} {\protect\JournalTitle{Physical Review
  A}} \textbf{42}, 6845--6857 (1990).

\bibitem{sipe2009photons}
J.~Sipe, \enquote{Photons in dispersive dielectrics,}
  {\protect\JournalTitle{Journal of Optics A: Pure and Applied Optics}}
  \textbf{11}, 114006 (2009).

\bibitem{Drummond2014}
P.~D. Drummond and M.~Hillery, \emph{The Quantum Theory of Nonlinear Optics}
  (Cambridge University Press, 2014).

\bibitem{Huttner1992}
B.~Huttner and S.~M. Barnett, \enquote{Quantization of the electromagnetic
  field in dielectrics,} {\protect\JournalTitle{Physical Review A}}
  \textbf{46}, 4306--4322 (1992).

\bibitem{Raymer2020}
M.~G. Raymer, \enquote{Quantum theory of light in a dispersive structured
  linear dielectric: a macroscopic hamiltonian tutorial treatment,}
  {\protect\JournalTitle{Journal of Modern Optics}} \textbf{67}, 196--212
  (2020).

\bibitem{Quesada2017}
N.~Quesada and J.~E. Sipe, \enquote{{Why you should not use the electric field
  to quantize in nonlinear optics},} {\protect\JournalTitle{Opt. Lett.}}
  \textbf{42}, 3443--3446 (2017).

\bibitem{Yanagimoto2022-non-Gaussian}
R.~Yanagimoto, E.~Ng, A.~Yamamura, T.~Onodera, L.~G. Wright, M.~Jankowski,
  M.~Fejer, P.~L. McMahon, and H.~Mabuchi, \enquote{Onset of non-gaussian
  quantum physics in pulsed squeezing with mesoscopic fields,}
  {\protect\JournalTitle{Optica}} \textbf{9}, 379--390 (2022).

\bibitem{Tezak2017}
N.~Tezak, N.~H. Amini, and H.~Mabuchi, \enquote{Low-dimensional manifolds for
  exact representation of open quantum systems,}
  {\protect\JournalTitle{Physical Review A}} \textbf{96}, 062113 (2017).

\bibitem{Manzoni2017}
M.~T. Manzoni, D.~E. Chang, and J.~S. Douglas, \enquote{{Simulating quantum
  light propagation through atomic ensembles using matrix product states},}
  {\protect\JournalTitle{Nat. Commun.}} \textbf{8}, 1743 (2017).

\bibitem{Yanagimoto2021_mps}
R.~Yanagimoto, E.~Ng, L.~G. Wright, T.~Onodera, and H.~Mabuchi,
  \enquote{{Efficient simulation of ultrafast quantum nonlinear optics with
  matrix product states},} {\protect\JournalTitle{Optica}} \textbf{8},
  1306--1315 (2021).

\bibitem{Lubasch2018}
M.~Lubasch, A.~A. Valido, J.~J. Renema, W.~S. Kolthammer, D.~Jaksch, M.~S. Kim,
  I.~Walmsley, and R.~Garc{\'i}a-Patr{\'o}n, \enquote{{Tensor network states in
  time-bin quantum optics},} {\protect\JournalTitle{Phys. Rev. A}} \textbf{97},
  062304 (2018).

\bibitem{Yanagimoto2023-thesis}
R.~Yanagimoto, \enquote{Quantum dynamics of broadband nonlinear photonics :
  from phenomenology to function,} Ph.D. thesis, Stanford University (2023).

\bibitem{Armstrong1962}
J.~A. Armstrong, N.~Bloembergen, J.~Ducuing, and P.~S. Pershan,
  \enquote{Interactions between light waves in a nonlinear dielectric,}
  {\protect\JournalTitle{Phys. Rev.}} \textbf{127}, 1918--1939 (1962).

\bibitem{Franken1963}
P.~A. Franken and J.~F. Ward, \enquote{Optical harmonics and nonlinear
  phenomena,} {\protect\JournalTitle{Rev. Mod. Phys.}} \textbf{35}, 23--39
  (1963).

\bibitem{Fejer1992}
M.~Fejer, G.~Magel, D.~Jundt, and R.~Byer, \enquote{Quasi-phase-matched second
  harmonic generation: tuning and tolerances,} {\protect\JournalTitle{{IEEE}
  Journal of Quantum Electronics}} \textbf{28}, 2631--2654 (1992).

\bibitem{Hum2007}
D.~S. Hum and M.~M. Fejer, \enquote{Quasi-phasematching,}
  {\protect\JournalTitle{Comptes Rendus Physique}} \textbf{8}, 180--198 (2007).

\bibitem{rao2016second}
A.~Rao, M.~Malinowski, A.~Honardoost, J.~R. Talukder, P.~Rabiei, P.~Delfyett,
  and S.~Fathpour, \enquote{Second-harmonic generation in periodically-poled
  thin film lithium niobate wafer-bonded on silicon,}
  {\protect\JournalTitle{Optics express}} \textbf{24}, 29941--29947 (2016).

\bibitem{wang2018ultrahigh}
C.~Wang, C.~Langrock, A.~Marandi, M.~Jankowski, M.~Zhang, B.~Desiatov, M.~M.
  Fejer, and M.~Lon{\v{c}}ar, \enquote{Ultrahigh-efficiency wavelength
  conversion in nanophotonic periodically poled lithium niobate waveguides,}
  {\protect\JournalTitle{Optica}} \textbf{5}, 1438 (2018).

\bibitem{rao2019actively}
A.~Rao, K.~Abdelsalam, T.~Sjaardema, A.~Honardoost, G.~F. Camacho-Gonzalez, and
  S.~Fathpour, \enquote{Actively-monitored periodic-poling in thin-film lithium
  niobate photonic waveguides with ultrahigh nonlinear conversion efficiency of
  4600\% w- 1 cm- 2,} {\protect\JournalTitle{Optics express}} \textbf{27},
  25920--25930 (2019).

\bibitem{Timurdogan2017}
E.~Timurdogan, C.~V. Poulton, M.~J. Byrd, and M.~R. Watts, \enquote{Electric
  field-induced second-order nonlinear optical effects in silicon waveguides,}
  {\protect\JournalTitle{Nature Photonics}} \textbf{11}, 200--206 (2017).

\bibitem{heydari2023degenerate}
D.~Heydari, M.~C{\u{a}}tuneanu, E.~Ng, D.~J. Gray, R.~Hamerly, J.~Mishra,
  M.~Jankowski, M.~Fejer, K.~Jamshidi, and H.~Mabuchi, \enquote{Degenerate
  optical parametric amplification in cmos silicon,}
  {\protect\JournalTitle{Optica}} \textbf{10}, 430--437 (2023).

\bibitem{Billat2017}
A.~Billat, D.~Grassani, M.~H.~P. Pfeiffer, S.~Kharitonov, T.~J. Kippenberg, and
  C.-S. Br{\`{e}}s, \enquote{Large second harmonic generation enhancement in
  si3n4 waveguides by all-optically induced quasi-phase-matching,}
  {\protect\JournalTitle{Nature Communications}} \textbf{8} (2017).

\bibitem{Porcel2017}
M.~A. Porcel, J.~Mak, C.~Taballione, V.~K. Schermerhorn, J.~P. Epping, P.~J.
  van~der Slot, and K.-J. Boller, \enquote{Photo-induced second-order
  nonlinearity in stoichiometric silicon nitride waveguides,}
  {\protect\JournalTitle{Optics Express}} \textbf{25}, 33143 (2017).

\bibitem{XLu2020}
X.~Lu, G.~Moille, A.~Rao, D.~A. Westly, and K.~Srinivasan, \enquote{Efficient
  photoinduced second-harmonic generation in silicon nitride photonics,}
  {\protect\JournalTitle{Nature Photonics}}  (2020).

\bibitem{chang2018heterogeneously}
L.~Chang, A.~Boes, X.~Guo, D.~T. Spencer, M.~Kennedy, J.~D. Peters, N.~Volet,
  J.~Chiles, A.~Kowligy, N.~Nader \emph{et~al.}, \enquote{Heterogeneously
  integrated gaas waveguides on insulator for efficient frequency conversion,}
  {\protect\JournalTitle{Laser \& Photonics Reviews}} \textbf{12}, 1800149
  (2018).

\bibitem{may2019second}
S.~May, M.~Kues, M.~Clerici, and M.~Sorel, \enquote{Second-harmonic generation
  in algaas-on-insulator waveguides,} {\protect\JournalTitle{Optics Letters}}
  \textbf{44}, 1339--1342 (2019).

\bibitem{chang2019low}
L.~Chang, A.~Boes, P.~Pintus, W.~Xie, J.~D. Peters, M.~Kennedy, W.~Jin, X.-W.
  Guo, S.-P. Yu, S.~B. Papp \emph{et~al.}, \enquote{Low loss (al) gaas on an
  insulator waveguide platform,} {\protect\JournalTitle{Optics Letters}}
  \textbf{44}, 4075--4078 (2019).

\bibitem{chiles2019multifunctional}
J.~Chiles, N.~Nader, E.~J. Stanton, D.~Herman, G.~Moody, J.~Zhu, J.~C. Skehan,
  B.~Guha, A.~Kowligy, J.~T. Gopinath \emph{et~al.}, \enquote{Multifunctional
  integrated photonics in the mid-infrared with suspended algaas on silicon,}
  {\protect\JournalTitle{Optica}} \textbf{6}, 1246--1254 (2019).

\bibitem{chang2019strong}
L.~Chang, A.~Boes, P.~Pintus, J.~D. Peters, M.~Kennedy, X.-W. Guo, N.~Volet,
  S.-P. Yu, S.~B. Papp, and J.~E. Bowers, \enquote{Strong frequency conversion
  in heterogeneously integrated gaas resonators,} {\protect\JournalTitle{APL
  Photonics}} \textbf{4} (2019).

\bibitem{stanton2020efficient}
E.~J. Stanton, J.~Chiles, N.~Nader, G.~Moody, N.~Volet, L.~Chang, J.~E. Bowers,
  S.~W. Nam, and R.~P. Mirin, \enquote{Efficient second harmonic generation in
  nanophotonic gaas-on-insulator waveguides,} {\protect\JournalTitle{Optics
  express}} \textbf{28}, 9521--9532 (2020).

\bibitem{chang2020ultra}
L.~Chang, W.~Xie, H.~Shu, Q.-F. Yang, B.~Shen, A.~Boes, J.~D. Peters, W.~Jin,
  C.~Xiang, S.~Liu \emph{et~al.}, \enquote{Ultra-efficient frequency comb
  generation in algaas-on-insulator microresonators,}
  {\protect\JournalTitle{Nature communications}} \textbf{11}, 1331 (2020).

\bibitem{xie2020ultrahigh}
W.~Xie, L.~Chang, H.~Shu, J.~C. Norman, J.~D. Peters, X.~Wang, and J.~E.
  Bowers, \enquote{Ultrahigh-q algaas-on-insulator microresonators for
  integrated nonlinear photonics,} {\protect\JournalTitle{Optics Express}}
  \textbf{28}, 32894--32906 (2020).

\bibitem{mahmudlu2021algaas}
H.~Mahmudlu, S.~May, A.~Angulo, M.~Sorel, and M.~Kues,
  \enquote{Algaas-on-insulator waveguide for highly efficient photon-pair
  generation via spontaneous four-wave mixing,} {\protect\JournalTitle{Optics
  Letters}} \textbf{46}, 1061--1064 (2021).

\bibitem{may2021supercontinuum}
S.~May, M.~Clerici, and M.~Sorel, \enquote{Supercontinuum generation in
  dispersion engineered algaas-on-insulator waveguides,}
  {\protect\JournalTitle{Scientific Reports}} \textbf{11}, 2052 (2021).

\bibitem{castro2022expanding}
J.~E. Castro, T.~J. Steiner, L.~Thiel, A.~Dinkelacker, C.~McDonald, P.~Pintus,
  L.~Chang, J.~E. Bowers, and G.~Moody, \enquote{Expanding the quantum photonic
  toolbox in algaasoi,} {\protect\JournalTitle{APL Photonics}} \textbf{7}
  (2022).

\bibitem{wu2023algaas}
L.~Wu, W.~Xie, H.-J. Chen, K.~Colburn, C.~Xiang, L.~Chang, W.~Jin, J.-Y. Liu,
  Y.~Yu, Y.~Yamamoto \emph{et~al.}, \enquote{Algaas soliton microcombs at room
  temperature,} {\protect\JournalTitle{Optics Letters}} \textbf{48}, 3853--3856
  (2023).

\bibitem{Rivoire2011second}
K.~Rivoire, S.~Buckley, F.~Hatami, and J.~Vu{\v{c}}kovi{\'{c}}, \enquote{Second
  harmonic generation in {GaP} photonic crystal waveguides,}
  {\protect\JournalTitle{Applied Physics Letters}} \textbf{98}, 263113 (2011).

\bibitem{Wilson2019}
D.~J. Wilson, K.~Schneider, S.~H\"{o}nl, M.~Anderson, Y.~Baumgartner,
  L.~Czornomaz, T.~J. Kippenberg, and P.~Seidler, \enquote{Integrated gallium
  phosphide nonlinear photonics,} {\protect\JournalTitle{Nature Photonics}}
  \textbf{14}, 57--62 (2019).

\bibitem{pantzas2022continuous}
K.~Pantzas, S.~Combri{\'e}, M.~Bailly, R.~Mandouze, F.~R. Talenti, A.~Harouri,
  B.~G{\'e}rard, G.~Beaudoin, L.~Le~Gratiet, G.~Patriarche \emph{et~al.},
  \enquote{Continuous-wave second-harmonic generation in orientation-patterned
  gallium phosphide waveguides at telecom wavelengths,}
  {\protect\JournalTitle{ACS photonics}} \textbf{9}, 2032--2039 (2022).

\bibitem{ueno1997second}
Y.~Ueno, V.~Ricci, and G.~I. Stegeman, \enquote{Second-order susceptibility of
  ga$_{0.5}$ in$_{0.5}$ p crystals at 1.5 $\mu$m and their feasibility for
  waveguide quasi-phase matching,} {\protect\JournalTitle{JOSA B}} \textbf{14},
  1428--1436 (1997).

\bibitem{poulvellarie2021efficient}
N.~Poulvellarie, C.~M. Arabi, C.~Ciret, S.~Combri{\'e}, A.~De~Rossi,
  M.~Haelterman, F.~Raineri, B.~Kuyken, S.-P. Gorza, and F.~Leo,
  \enquote{Efficient type ii second harmonic generation in an indium gallium
  phosphide on insulator wire waveguide aligned with a crystallographic axis,}
  {\protect\JournalTitle{Optics Letters}} \textbf{46}, 1490--1493 (2021).

\bibitem{Skauli2002}
T.~Skauli, K.~L. Vodopyanov, T.~J. Pinguet, A.~Schober, O.~Levi, L.~A. Eyres,
  M.~M. Fejer, J.~S. Harris, B.~Gerard, L.~Becouarn, E.~Lallier, and
  G.~Arisholm, \enquote{Measurement of the nonlinear coefficient of
  orientation-patterned {GaAs} and demonstration of highly efficient
  second-harmonic generation,} {\protect\JournalTitle{Optics Letters}}
  \textbf{27}, 628 (2002).

\bibitem{yu2007growth}
X.~Yu, L.~Scaccabarozzi, A.~C. Lin, M.~M. Fejer, and J.~S. Harris,
  \enquote{Growth of gaas with orientation-patterned structures for nonlinear
  optics,} {\protect\JournalTitle{Journal of crystal growth}} \textbf{301},
  163--167 (2007).

\bibitem{Lukin2019}
D.~M. Lukin, C.~Dory, M.~A. Guidry, K.~Y. Yang, S.~D. Mishra, R.~Trivedi,
  M.~Radulaski, S.~Sun, D.~Vercruysse, G.~H. Ahn, and J.~Vu{\v{c}}kovi{\'{c}},
  \enquote{4h-silicon-carbide-on-insulator for integrated quantum and nonlinear
  photonics,} {\protect\JournalTitle{Nature Photonics}} \textbf{14}, 330--334
  (2019).

\bibitem{Song:19}
B.-S. Song, T.~Asano, S.~Jeon, H.~Kim, C.~Chen, D.~D. Kang, and S.~Noda,
  \enquote{Ultrahigh-q photonic crystal nanocavities based on 4h silicon
  carbide,} {\protect\JournalTitle{Optica}} \textbf{6}, 991--995 (2019).

\bibitem{Guidry:20}
M.~A. Guidry, K.~Y. Yang, D.~M. Lukin, A.~Markosyan, J.~Yang, M.~M. Fejer, and
  J.~Vu\v{c}kovi\'{c}, \enquote{Optical parametric oscillation in silicon
  carbide nanophotonics,} {\protect\JournalTitle{Optica}} \textbf{7},
  1139--1142 (2020).

\bibitem{Lukin2020}
D.~M. Lukin, M.~A. Guidry, and J.~Vu{\v{c}}kovi{\'{c}}, \enquote{Integrated
  quantum photonics with silicon carbide: Challenges and prospects,}
  {\protect\JournalTitle{{PRX} Quantum}} \textbf{1} (2020).

\bibitem{xiong2011integrated}
C.~Xiong, W.~Pernice, K.~K. Ryu, C.~Schuck, K.~Y. Fong, T.~Palacios, and H.~X.
  Tang, \enquote{Integrated gan photonic circuits on silicon (100) for second
  harmonic generation,} {\protect\JournalTitle{Optics express}} \textbf{19},
  10462--10470 (2011).

\bibitem{hite2012development}
J.~Hite, M.~Twigg, M.~Mastro, J.~Freitas, J.~Meyer, I.~Vurgaftman,
  S.~O’Connor, N.~Condon, F.~Kub, S.~Bowman \emph{et~al.},
  \enquote{Development of periodically oriented gallium nitride for non-linear
  optics,} {\protect\JournalTitle{Optical Materials Express}} \textbf{2},
  1203--1208 (2012).

\bibitem{stassen2019high}
E.~Stassen, M.~Pu, E.~Semenova, E.~Zavarin, W.~Lundin, and K.~Yvind,
  \enquote{High-confinement gallium nitride-on-sapphire waveguides for
  integrated nonlinear photonics,} {\protect\JournalTitle{Optics letters}}
  \textbf{44}, 1064--1067 (2019).

\bibitem{zheng2022integrated}
Y.~Zheng, C.~Sun, B.~Xiong, L.~Wang, Z.~Hao, J.~Wang, Y.~Han, H.~Li, J.~Yu, and
  Y.~Luo, \enquote{Integrated gallium nitride nonlinear photonics,}
  {\protect\JournalTitle{Laser \& Photonics Reviews}} \textbf{16}, 2100071
  (2022).

\bibitem{hickstein2017ultrabroadband}
D.~D. Hickstein, H.~Jung, D.~R. Carlson, A.~Lind, I.~Coddington, K.~Srinivasan,
  G.~G. Ycas, D.~C. Cole, A.~Kowligy, C.~Fredrick \emph{et~al.},
  \enquote{Ultrabroadband supercontinuum generation and frequency-comb
  stabilization using on-chip waveguides with both cubic and quadratic
  nonlinearities,} {\protect\JournalTitle{Physical Review Applied}} \textbf{8},
  014025 (2017).

\bibitem{bruch2019chip}
A.~W. Bruch, X.~Liu, J.~B. Surya, C.-L. Zou, and H.~X. Tang, \enquote{On-chip
  $\chi$ (2) microring optical parametric oscillator,}
  {\protect\JournalTitle{Optica}} \textbf{6}, 1361--1366 (2019).

\bibitem{lu2020ultraviolet}
J.~Lu, X.~Liu, A.~W. Bruch, L.~Zhang, J.~Wang, J.~Yan, and H.~X. Tang,
  \enquote{Ultraviolet to mid-infrared supercontinuum generation in
  single-crystalline aluminum nitride waveguides,}
  {\protect\JournalTitle{Optics Letters}} \textbf{45}, 4499--4502 (2020).

\bibitem{liu2021aluminum}
X.~Liu, Z.~Gong, A.~W. Bruch, J.~B. Surya, J.~Lu, and H.~X. Tang,
  \enquote{Aluminum nitride nanophotonics for beyond-octave soliton microcomb
  generation and self-referencing,} {\protect\JournalTitle{Nature
  communications}} \textbf{12}, 5428 (2021).

\bibitem{liu2023aluminum}
X.~Liu, A.~W. Bruch, and H.~X. Tang, \enquote{Aluminum nitride photonic
  integrated circuits: from piezo-optomechanics to nonlinear optics,}
  {\protect\JournalTitle{Advances in Optics and Photonics}} \textbf{15},
  236--317 (2023).

\bibitem{guo2023ultrathin}
Q.~Guo, X.-Z. Qi, L.~Zhang, M.~Gao, S.~Hu, W.~Zhou, W.~Zang, X.~Zhao, J.~Wang,
  B.~Yan \emph{et~al.}, \enquote{Ultrathin quantum light source with van der
  waals nbocl2 crystal,} {\protect\JournalTitle{Nature}} \textbf{613}, 53--59
  (2023).

\bibitem{abdelwahab2022giant}
I.~Abdelwahab, B.~Tilmann, Y.~Wu, D.~Giovanni, I.~Verzhbitskiy, M.~Zhu,
  R.~Bert{\'e}, F.~Xuan, L.~d.~S. Menezes, G.~Eda \emph{et~al.}, \enquote{Giant
  second-harmonic generation in ferroelectric nboi2,}
  {\protect\JournalTitle{Nature Photonics}} \textbf{16}, 644--650 (2022).

\bibitem{fichtner2019alscn}
S.~Fichtner, N.~Wolff, F.~Lofink, L.~Kienle, and B.~Wagner, \enquote{Alscn: A
  iii-v semiconductor based ferroelectric,} {\protect\JournalTitle{Journal of
  Applied Physics}} \textbf{125} (2019).

\bibitem{wang2021fully}
P.~Wang, D.~Wang, N.~M. Vu, T.~Chiang, J.~T. Heron, and Z.~Mi, \enquote{Fully
  epitaxial ferroelectric scaln grown by molecular beam epitaxy,}
  {\protect\JournalTitle{Applied Physics Letters}} \textbf{118} (2021).

\bibitem{zhu2021strongly}
W.~Zhu, J.~Hayden, F.~He, J.-I. Yang, P.~Tipsawat, M.~D. Hossain, J.-P. Maria,
  and S.~Trolier-McKinstry, \enquote{Strongly temperature dependent
  ferroelectric switching in aln, al1-xscxn, and al1-xbxn thin films,}
  {\protect\JournalTitle{Applied Physics Letters}} \textbf{119} (2021).

\bibitem{yoshioka2021strongly}
V.~Yoshioka, J.~Lu, Z.~Tang, J.~Jin, R.~H. Olsson, and B.~Zhen,
  \enquote{Strongly enhanced second-order optical nonlinearity in
  cmos-compatible al1- xscxn thin films,} {\protect\JournalTitle{APL
  Materials}} \textbf{9} (2021).

\bibitem{suceava2023enhancement}
A.~Suceava, J.~Hayden, K.~P. Kelley, Y.~Xiong, B.~Fazlioglu-Yalcin, I.~Dabo,
  S.~Trolier-McKinstry, J.-P. Maria, and V.~Gopalan, \enquote{Enhancement of
  second-order optical nonlinearities and nanoscale periodic domain patterning
  in ferroelectric boron-substituted aluminum nitride thin films,}
  {\protect\JournalTitle{Optical Materials Express}} \textbf{13}, 1522--1534
  (2023).

\bibitem{yang2023domain}
F.~Yang, G.~Yang, D.~Wang, P.~Wang, J.~Lu, Z.~Mi, and H.~X. Tang,
  \enquote{Domain control and periodic poling of epitaxial scaln,}
  {\protect\JournalTitle{Applied Physics Letters}} \textbf{123} (2023).

\bibitem{Tassev:19}
V.~L. Tassev and S.~R. Vangala, \enquote{New heteroepitaxially grown materials
  for frequency conversion in the mid and longwave infrared,} in
  \emph{Nonlinear Optics (NLO),}  (Optical Society of America, 2019), p.
  NTu4A.33.

\bibitem{Vangala:19}
S.~Vangala, V.~Tassev, and M.~Snure, \enquote{Thick heteroepitaxial growth of
  znse on gaas substrates for frequency conversion in the mlwir,} in
  \emph{Nonlinear Optics (NLO),}  (Optical Society of America, 2019), p.
  NTu4A.40.

\bibitem{Schunemann2019}
P.~G. Schunemann and K.~T. Zawilski, \enquote{{Vapor transport growth of single
  crystal zinc selenide (Conference Presentation)},} in \emph{Nonlinear
  Frequency Generation and Conversion: Materials and Devices XVIII,}  vol.
  10902 P.~G. Schunemann and K.~L. Schepler, eds., International Society for
  Optics and Photonics (SPIE, 2019).

\bibitem{schwesyg2010light}
J.~Schwesyg, M.~Kajiyama, M.~Falk, D.~Jundt, K.~Buse, and M.~Fejer,
  \enquote{Light absorption in undoped congruent and magnesium-doped lithium
  niobate crystals in the visible wavelength range,}
  {\protect\JournalTitle{Applied Physics B}} \textbf{100}, 109--115 (2010).

\bibitem{Leidinger2015}
M.~Leidinger, S.~Fieberg, N.~Waasem, F.~K\"{u}hnemann, K.~Buse, and I.~Breunig,
  \enquote{Comparative study on three highly sensitive absorption measurement
  techniques characterizing lithium niobate over its entire transparent
  spectral range,} {\protect\JournalTitle{Opt. Express}} \textbf{23},
  21690--21705 (2015).

\bibitem{shams2022reduced}
A.~Shams-Ansari, G.~Huang, L.~He, Z.~Li, J.~Holzgrafe, M.~Jankowski,
  M.~Churaev, P.~Kharel, R.~Cheng, D.~Zhu \emph{et~al.}, \enquote{Reduced
  material loss in thin-film lithium niobate waveguides,}
  {\protect\JournalTitle{Apl Photonics}} \textbf{7} (2022).

\bibitem{li2023low}
Z.~Li, Z.~Qiu, R.~N. Wang, M.~Divall, and T.~J. Kippenberg,
  \enquote{Low-temperature and hydrogen-free silicon dioxide cladding for
  next-generation integrated photonics,} in \emph{2023 Conference on Lasers and
  Electro-Optics Europe \& European Quantum Electronics Conference
  (CLEO/Europe-EQEC),}  (IEEE, 2023), pp. 1--1.

\bibitem{gruenke2023surface}
R.~G. Gruenke, O.~A. Hitchcock, E.~A. Wollack, C.~J. Sarabalis, M.~Jankowski,
  T.~P. McKenna, N.~R. Lee, and A.~H. Safavi-Naeini, \enquote{Surface
  modification and coherence in lithium niobate saw resonators,}  (2023).

\bibitem{Zhao2023unveiling}
J.~Zhao, X.~Li, T.-C. Hu, A.~A. Sayem, H.~Li, A.~Tate, K.~Kim, R.~Kopf,
  P.~Sanjari, M.~Earnshaw, N.~K. Fontaine, C.~Wang, and A.~Blanco-Redondo,
  \enquote{Unveiling the origins of quasi-phase matching spectral imperfections
  in thin-film lithium niobate frequency doublers,}  (2023).

\bibitem{zhang2017monolithic}
M.~Zhang, C.~Wang, R.~Cheng, A.~Shams-Ansari, and M.~Lon{\v{c}}ar,
  \enquote{Monolithic ultra-high-q lithium niobate microring resonator,}
  {\protect\JournalTitle{Optica}} \textbf{4}, 1536--1537 (2017).

\bibitem{Imeshev2000a}
G.~Imeshev, M.~A. Arbore, M.~M. Fejer, A.~Galvanauskas, M.~Fermann, and
  D.~Harter, \enquote{Ultrashort-pulse second-harmonic generation with
  longitudinally nonuniform quasi-phase-matching gratings: pulse compression
  and shaping,} {\protect\JournalTitle{Journal of the Optical Society of
  America B}} \textbf{17}, 304 (2000).

\bibitem{Imeshev2000b}
G.~Imeshev, M.~A. Arbore, S.~Kasriel, and M.~M. Fejer, \enquote{Pulse shaping
  and compression by second-harmonic generation with quasi-phase-matching
  gratings in the presence of arbitrary dispersion,}
  {\protect\JournalTitle{Journal of the Optical Society of America B}}
  \textbf{17}, 1420 (2000).

\bibitem{phillips2012broadband}
C.~R. Phillips, \emph{Broadband optical sources based on highly nonlinear
  quasi-phasematched interactions} (Stanford University, 2012).

\bibitem{Crouch1988}
D.~D. Crouch, \enquote{Broadband squeezing via degenerate parametric
  amplification,} {\protect\JournalTitle{Physical Review A}} \textbf{38},
  508--511 (1988).

\bibitem{Sukhorukov1971}
A.~Sukhorukov and A.~Shchednova, \enquote{Parametric amplification of light in
  the field of a modulated laser wave,} {\protect\JournalTitle{SOVIET PHYSICS
  JETP}} \textbf{33}, 677--682 (1971).

\bibitem{Danielius1993}
R.~Danielius, G.~P. Banfi, P.~D. Trapani, R.~Righini, A.~Piskarskas, and
  A.~Stabinis, \enquote{Traveling-wave parametric generation of widely tunable,
  highly coherent femtosecond light pulses,} {\protect\JournalTitle{Journal of
  the Optical Society of America B}} \textbf{10}, 2222 (1993).

\bibitem{Manzoni2016}
C.~Manzoni and G.~Cerullo, \enquote{Design criteria for ultrafast optical
  parametric amplifiers,} {\protect\JournalTitle{Journal of Optics}}
  \textbf{18}, 103501 (2016).

\bibitem{Charbonneau_Lefort_2010}
M.~Charbonneau-Lefort, B.~Afeyan, and M.~M. Fejer, \enquote{Theory and
  simulation of gain-guided noncollinear modes in chirped quasi-phase-matched
  optical parametric amplifiers,} {\protect\JournalTitle{Journal of the Optical
  Society of America B}} \textbf{27}, 824 (2010).

\bibitem{Eckardt1984}
R.~Eckardt and J.~Reintjes, \enquote{Phase matching limitations of high
  efficiency second harmonic generation,} {\protect\JournalTitle{IEEE Journal
  of Quantum Electronics}} \textbf{20}, 1178--1187 (1984).

\bibitem{Ledezma2022}
L.~Ledezma, R.~Sekine, Q.~Guo, R.~Nehra, S.~Jahani, and A.~Marandi,
  \enquote{Intense optical parametric amplification in dispersion-engineered
  nanophotonic lithium niobate waveguides,} {\protect\JournalTitle{Optica}}
  \textbf{9}, 303 (2022).

\bibitem{nehra2022few}
R.~Nehra, R.~Sekine, L.~Ledezma, Q.~Guo, R.~M. Gray, A.~Roy, and A.~Marandi,
  \enquote{Few-cycle vacuum squeezing in nanophotonics,}
  {\protect\JournalTitle{Science}} \textbf{377}, 1333--1337 (2022).

\bibitem{gouy1890phase}
C.~R. Gouy, \enquote{Sur une propri{\'e}t{\'e} nouvelle des ondes lumineuses,}
  {\protect\JournalTitle{C. R. Acad. Sci. Paris}} \textbf{110} (1890).

\bibitem{akhmanov1969nonstationary}
S.~Akhmanov, A.~Sukhorukov, and A.~Chirkin, \enquote{Nonstationary phenomena
  and space-time analogy in nonlinear optics,} {\protect\JournalTitle{Sov.
  Phys. JETP}} \textbf{28}, 748--757 (1969).

\bibitem{Major2008}
H.~E. Major, C.~B. Gawith, and P.~G. Smith, \enquote{Gouy phase compensation in
  quasi-phase matching,} {\protect\JournalTitle{Optics Communications}}
  \textbf{281}, 5036--5040 (2008).

\bibitem{Babushkin2022_temporal}
I.~Babushkin, O.~Melchert, U.~Morgner, and A.~Demircan, \enquote{Photon
  trapping in a time cavity flying with a speed of light,}
  {\protect\JournalTitle{Preprint}}  (2022).

\bibitem{Guo2020}
H.~Guo, W.~Weng, J.~Liu, F.~Yang, W.~H\"ansel, C.~S. Br\`es, L.~Th\'evenaz,
  R.~Holzwarth, and T.~J. Kippenberg, \enquote{{Nanophotonic
  supercontinuum-based mid-infrared dual-comb spectroscopy},}
  {\protect\JournalTitle{Optica}} \textbf{7}, 1181--1188 (2020).

\bibitem{lukashchuk2019advanced}
A.~Lukashchuk, F.~Gremion, M.~Karpov, J.~Liu, and T.~J. Kippenberg,
  \enquote{Advanced dispersion engineering of dispersive waves in si3n4
  microresonators,} in \emph{CLEO: QELS\_Fundamental Science,}  (Optica
  Publishing Group, 2019), pp. FF2D--1.

\bibitem{Lucas2023}
E.~Lucas, S.-P. Yu, T.~C. Briles, D.~R. Carlson, and S.~B. Papp,
  \enquote{{Tailoring microcombs with inverse-designed, meta-dispersion
  microresonators},} {\protect\JournalTitle{Nat. Photon.}}  (2023).

\bibitem{moille2023fourier}
G.~Moille, X.~Lu, J.~Stone, D.~Westly, and K.~Srinivasan, \enquote{Fourier
  synthesis dispersion engineering of photonic crystal microrings for broadband
  frequency combs,} {\protect\JournalTitle{Communications Physics}} \textbf{6},
  144 (2023).

\bibitem{Vercruysse2020}
D.~Vercruysse, N.~V. Sapra, L.~Su, and J.~Vuckovic, \enquote{{Dispersion
  Engineering With Photonic Inverse Design},} {\protect\JournalTitle{IEEE J.
  Sel. Top. Quantum Electron.}} \textbf{26}, 8301706 (2020).

\bibitem{Ikeda1979}
K.~Ikeda, \enquote{Multiple-valued stationary state and its instability of the
  transmitted light by a ring cavity system,} {\protect\JournalTitle{Optics
  Communications}} \textbf{30}, 257--261 (1979).

\bibitem{Hamerly2016}
R.~Hamerly, A.~Marandi, M.~Jankowski, M.~M. Fejer, Y.~Yamamoto, and H.~Mabuchi,
  \enquote{Reduced models and design principles for half-harmonic generation in
  synchronously pumped optical parametric oscillators,}
  {\protect\JournalTitle{Phys. Rev. A}} \textbf{94}, 063809 (2016).

\bibitem{Ng2022sampling}
E.~Ng, T.~Onodera, S.~Kako, P.~L. McMahon, H.~Mabuchi, and Y.~Yamamoto,
  \enquote{Efficient sampling of ground and low-energy ising spin
  configurations with a coherent ising machine,}
  {\protect\JournalTitle{Physical Review Research}} \textbf{4} (2022).

\bibitem{Onodera2022}
T.~Onodera, E.~Ng, C.~Gustin, N.~Lörch, A.~Yamamura, R.~Hamerly, P.~L.
  McMahon, A.~Marandi, and H.~Mabuchi, \enquote{Nonlinear quantum behavior of
  ultrashort-pulse optical parametric oscillators,}
  {\protect\JournalTitle{Physical Review A}} \textbf{105} (2022).

\bibitem{haus1984waves}
H.~Haus, \enquote{Waves and fields in optoelectronics.}
  {\protect\JournalTitle{PRENTICE-HALL, INC., ENGLEWOOD CLIFFS, NJ 07632, USA,
  1984, 402}}  (1984).

\bibitem{panuski2020fundamental}
C.~Panuski, D.~Englund, and R.~Hamerly, \enquote{Fundamental thermal noise
  limits for optical microcavities,} {\protect\JournalTitle{Physical Review X}}
  \textbf{10}, 041046 (2020).

\bibitem{Wasilewski2006}
W.~Wasilewski, A.~I. Lvovsky, K.~Banaszek, and C.~Radzewicz, \enquote{Pulsed
  squeezed light: Simultaneous squeezing of multiple modes,}
  {\protect\JournalTitle{Physical Review A}} \textbf{73}, 063819 (2006).

\bibitem{Braunstein2005}
S.~Braunstein, \enquote{Squeezing as an irreducible resource,}
  {\protect\JournalTitle{Phys. Rev. A}} \textbf{71}, 055801 (2005).

\bibitem{Harris1967}
S.~E. Harris, M.~K. Oshman, and R.~L. Byer, \enquote{{Observation of Tunable
  Optical Parametric Fluorescence},} {\protect\JournalTitle{Phys. Rev. Lett.}}
  \textbf{18}, 265 (1967).

\bibitem{Xing2023}
W.~Xing and T.~C. Ralph, \enquote{Pump depletion in optical parametric
  amplification,} {\protect\JournalTitle{Phys. Rev. A}} \textbf{107}, 023712
  (2023).

\bibitem{Yanagimoto2023-qnd}
R.~Yanagimoto, R.~Nehra, R.~Hamerly, E.~Ng, A.~Marandi, and H.~Mabuchi,
  \enquote{{Quantum Nondemolition Measurements with Optical Parametric
  Amplifiers for Ultrafast Universal Quantum Information Processing},}
  {\protect\JournalTitle{PRX Quantum}} \textbf{4}, 010333 (2023).

\bibitem{Qin2022}
W.~Qin, A.~Miranowicz, and F.~Nori, \enquote{{Beating the 3 dB Limit for
  Intracavity Squeezing and Its Application to Nondemolition Qubit Readout},}
  {\protect\JournalTitle{Phys. Rev. Lett.}} \textbf{129}, 123602 (2022).

\bibitem{Florez2020}
J.~Fl\'orez, J.~S. Lundeen, and M.~V. Chekhova, \enquote{{Pump depletion in
  parametric down-conversion with low pump energies},}
  {\protect\JournalTitle{Opt. Lett.}} \textbf{45}, 4264--4267 (2020).

\bibitem{Langford2011}
N.~K. Langford, S.~Ramelow, R.~Prevedel, W.~J. Munro, G.~J. Milburn, and
  A.~Zeilinger, \enquote{Efficient quantum computing using coherent photon
  conversion,} {\protect\JournalTitle{Nature}} \textbf{478}, 360--363 (2011).

\bibitem{Chuang1995}
I.~L. Chuang and Y.~Yamamoto, \enquote{Simple quantum computer,}
  {\protect\JournalTitle{Physical Review A}} \textbf{52}, 3489 (1995).

\bibitem{Milburn1989}
G.~J. Milburn, \enquote{{Quantum optical Fredkin gate},}
  {\protect\JournalTitle{Phys. Rev. Lett.}} \textbf{62}, 2124 (1989).

\bibitem{Nielsen2000}
M.~A. Nielsen and I.~L. Chuang, \emph{Quantum Computation and Quantum
  Information} (Cambridge University Press, 2000).

\bibitem{Kenfack2004}
A.~Kenfack and K.~{\.Z}yczkowski, \enquote{{Negativity of the Wigner function
  as an indicator of non-classicality},} {\protect\JournalTitle{J. Opt. B:
  Quantum Semiclass. Opt.}} \textbf{6}, 396--404 (2004).

\bibitem{Wiseman2010}
H.~Wiseman and G.~Milburn, \emph{Quantum Measurement and Control} (Cambridge
  University Press, 2010).

\bibitem{Breuer2002}
H.~P. Breuer and F.~Petruccione, \emph{The Theory of Open Quantum Systems}
  (Oxford University Press, 2002).

\bibitem{Roberts2020}
D.~Roberts and A.~A. Clerk, \enquote{{Driven-Dissipative Quantum Kerr
  Resonators: New Exact Solutions, Photon Blockade and Quantum Bistability},}
  {\protect\JournalTitle{Phys. Rev. X}} \textbf{10}, 021022 (2020).

\bibitem{Rivera2023}
N.~Riveraa, J.~Sloanc, Y.~Salaminc, J.~D. Joannopoulosb, and M.~Solja{\v c}ic,
  \enquote{{Creating large Fock states and massively squeezed states in optics
  using systems with nonlinear bound states in the continuum},}
  {\protect\JournalTitle{PNAS}} \textbf{120}, e2219208120 (2023).

\bibitem{Seibold2022}
K.~Seibold, R.~Rota, F.~Minganti, and V.~Savona, \enquote{{Quantum dynamics of
  dissipative Kerr solitons},} {\protect\JournalTitle{Phys. Rev. A}}
  \textbf{105}, 053530 (2022).

\bibitem{deVega2017}
I.~{de Vega} and D.~Alonso, \enquote{{Dynamics of non-Markovian open quantum
  systems},} {\protect\JournalTitle{Rev. Mod. Phys.}} \textbf{80}, 015001
  (2017).

\bibitem{Walls1983}
D.~F. Walls, \enquote{{Squeezed states of light},}
  {\protect\JournalTitle{Nature}} \textbf{306}, 141--146 (1983).

\bibitem{Wu1986}
L.-A. Wu, H.~J. Kimble, J.~L. Hall, and H.~Wu, \enquote{{Generation of Squeezed
  States by Parametric Down Conversion},} {\protect\JournalTitle{Phys. Rev.
  Lett.}} \textbf{57}, 2520 (1986).

\bibitem{LIGO2013}
{The LIGO Scientific Collaboration}, \enquote{{Enhanced sensitivity of the LIGO
  gravitational wave detector by using squeezed states of light},}
  {\protect\JournalTitle{Nat. Photon.}} \textbf{7}, 613--619 (2013).

\bibitem{Furusawa1998}
A.~Furusawa, J.~L. S\o{}rensen, S.~L. Braunstein, C.~A. Fuchs, H.~J. Kimble,
  and E.~S. Polzik, \enquote{{Unconditional Quantum Teleportation},}
  {\protect\JournalTitle{Science}} \textbf{282}, 706 (1998).

\bibitem{Takeda2019}
S.~Takeda and A.~Furusawa, \enquote{{Toward large-scale fault-tolerant
  universal photonic quantum computing},} {\protect\JournalTitle{APL
  Photonics}} \textbf{4}, 060902 (2019).

\bibitem{Conteau2018}
C.~Couteau, \enquote{{Spontaneous parametric down-conversion},}
  {\protect\JournalTitle{Contemp. Phys.}} \textbf{59}, 291--304 (2018).

\bibitem{El-Ganainy2018}
R.~El-Ganainy, K.~G. Makris, M.~Khajavikhan, Z.~H. Musslimani, S.~Rotter, and
  D.~N. Christodoulides, \enquote{{Non-Hermitian physics and PT symmetry},}
  {\protect\JournalTitle{Nat. Phys.}} \textbf{14}, 11--19 (2018).

\bibitem{Pysher2009}
M.~Pysher, R.~Bloomer, C.~M. Kaleva, T.~D. Roberts, P.~Battle, and O.~Pfister,
  \enquote{Broadband amplitude squeezing in a periodically poled ktiopo 4
  waveguide,} {\protect\JournalTitle{Optics letters}} \textbf{34}, 256--258
  (2009).

\bibitem{Mondain2019}
F.~Mondain, T.~Lunghi, A.~Zavatta, E.~Gouzien, F.~Doutre, M.~De~Micheli,
  S.~Tanzilli, and V.~D’Auria, \enquote{Chip-based squeezing at a telecom
  wavelength,} {\protect\JournalTitle{Photonics Research}} \textbf{7}, A36--A39
  (2019).

\bibitem{Riemensberger2022}
J.~Riemensberger, N.~Kuznetsov, J.~Liu, J.~He, R.~N. Wang, and T.~J.
  Kippenberg, \enquote{{A photonic integrated continuous-travelling-wave
  parametric amplifier},} {\protect\JournalTitle{Nature}} \textbf{612}, 56--61
  (2022).

\bibitem{Hansryd2001}
J.~Hansryd and P.~A. Andrekson, \enquote{Broad-band continuous-wave-pumped
  fiber optical parametric amplifier with 49-db gain and wavelength-conversion
  efficiency,} {\protect\JournalTitle{IEEE Photonics Technology Letters}}
  \textbf{13}, 194--196 (2001).

\bibitem{Gouzien2020}
E.~Gouzien, S.~Tanzilli, V.~d’Auria, and G.~Patera, \enquote{Morphing
  supermodes: a full characterization for enabling multimode quantum optics,}
  {\protect\JournalTitle{Physical Review Letters}} \textbf{125}, 103601 (2020).

\bibitem{Brecht2015}
B.~Brecht, D.~V. Reddy, C.~Silberhorn, and M.~G. Raymer, \enquote{{Photon
  Temporal Modes: A Complete Framework for Quantum Information Science},}
  {\protect\JournalTitle{Phys. Rev. X}} \textbf{5}, 041017 (2015).

\bibitem{Grice1997}
W.~P. Grice and I.~A. Walmsley, \enquote{{Spectral information and
  distinguishability in type-II down-conversion with a broadband pump},}
  {\protect\JournalTitle{Phys. Rev. A}} \textbf{56}, 1627 (1997).

\bibitem{Law2000}
C.~K. Law, I.~A. Walmsley, and J.~H. Eberly, \enquote{{Continuous Frequency
  Entanglement: Effective Finite Hilbert Space and Entropy Control},}
  {\protect\JournalTitle{Phys. Rev. A}} \textbf{84}, 5304 (2000).

\bibitem{Uren2005}
A.~B. U'Ren, C.~Silberhorn, K.~Banaszek, I.~Walmsley, R.~Erdmann, W.~P. Grice,
  and M.~G. Raymer, \enquote{Generation of pure-state single-photon wavepackets
  by conditional preparation based on spontaneous parametric downconversion,}
  {\protect\JournalTitle{LASER PHYSICS}} \textbf{15} (2005).

\bibitem{Keller1997}
T.~E. Keller and M.~H. Rubin, \enquote{{Theory of two-photon entanglement for
  spontaneous parametric down-conversion driven by a narrow pump pulse},}
  {\protect\JournalTitle{Phys. Rev. A}} \textbf{56}, 1534 (1997).

\bibitem{Ansari2018}
V.~Ansari, J.~M. Donohue, B.~Brecht, and C.~Silberhorn, \enquote{Tailoring
  nonlinear processes for quantum optics with pulsed temporal-mode encodings,}
  {\protect\JournalTitle{Optica}} \textbf{5}, 534--550 (2018).

\bibitem{Quesada2014}
N.~Quesada and J.~E. Sipe, \enquote{{Effects of time ordering in quantum
  nonlinear optics},} {\protect\JournalTitle{Phys. Rev. A}} \textbf{90}, 063840
  (2014).

\bibitem{Kraemer2018}
S.~Kr\"amer, D.~Plankensteiner, L.~Ostermann, and H.~Ritsch,
  \enquote{{QuantumOptics.jl: A Julia framework for simulating open quantum
  systems},} {\protect\JournalTitle{Comput. Phys. Commun}} \textbf{227},
  109--116 (2018).

\bibitem{Schack1990}
R.~Schack and A.~Schenzle, \enquote{Moment hierarchies and cumulants in quantum
  optics,} {\protect\JournalTitle{Phys. Rev. A}} \textbf{41}, 3847 (1990).

\bibitem{Ng2023}
E.~Ng, R.~Yanagimoto, M.~Jankowski, M.~Fejer, and H.~Mabuchi, \enquote{Quantum
  noise dynamics in nonlinear pulse propagation,} {\protect\JournalTitle{arXiv
  preprint arXiv:2307.05464}}  (2023).

\bibitem{Hult2007}
J.~Hult, \enquote{A fourth-order runge–kutta in the interaction picture
  method for simulating supercontinuum generation in optical fibers,}
  {\protect\JournalTitle{J. Light. Technol.}} \textbf{25}, 3770--3775 (2007).

\bibitem{CUDAjl}
\enquote{{JuliaGPU/CUDA.jl},} \url{https://github.com/JuliaGPU/CUDA.jl}.

\bibitem{GaussianSSGjl}
\enquote{{ngedwin98/GaussianSSF.jl},}
  \url{https://github.com/ngedwin98/GaussianSSF.jl}.

\bibitem{Huang2022}
Y.-X. Huang, M.~Li, K.~Lin, Y.-L. Zhang, G.-C. Guo, and C.-L. Zou,
  \enquote{{Classical-to-quantum transition in multimode nonlinear systems with
  strong photon-photon coupling},} {\protect\JournalTitle{Phys. Rev. A}}
  \textbf{105}, 043707 (2022).

\bibitem{Plankensteiner2021}
D.~Plankensteiner and H.~Ritsch, \enquote{{QuantumCumulants.jl: A Julia
  framework for generalized mean-field equations in open quantum systems},}
  {\protect\JournalTitle{Quantum}} \textbf{6}, 617 (2021).

\bibitem{Birnbaum2005}
K.~M. Birnbaum, A.~Boca, R.~Miller, A.~D. Boozer, T.~E. Northup, and H.~J.
  Kimble, \enquote{Photon blockade in an optical cavity with one trapped atom,}
  {\protect\JournalTitle{Nature}} \textbf{436}, 87--90 (2005).

\bibitem{Darrick2014}
D.~E. Chang, V.~Vuleti\'c, and M.~D. Lukin, \enquote{{Quantum nonlinear optics
  — photon by photon},} {\protect\JournalTitle{Nat. Photonics}} \textbf{8},
  685--694 (2014).

\bibitem{Fano1961}
U.~Fano, \enquote{{Effects of Configuration Interaction on Intensities and
  Phase Shifts},} {\protect\JournalTitle{Phys. Rev.}} \textbf{124}, 1866
  (1961).

\bibitem{Ryo2020Fano}
R.~Yanagimoto, E.~Ng, M.~P. Jankowski, T.~Onodera, M.~M. Fejer, and H.~Mabuchi,
  \enquote{Broadband parametric downconversion as a discrete-continuum fano
  interaction,} {\protect\JournalTitle{arXiv preprint arXiv:2009.01457}}
  (2020).

\bibitem{Antonosyan2014}
D.~A. Antonosyan, A.~S. Solntsev, and A.~A. Sukhorukov, \enquote{{Single-photon
  spontaneous parametric down-conversion in quadratic nonlinear waveguide
  arrays},} {\protect\JournalTitle{Opt. Commun.}} \textbf{327}, 22--26 (2014).

\bibitem{Solntsev2022}
A.~S. Solntsev, S.~V. Batalov, N.~K. Langford, and A.~A. Sukhorukov,
  \enquote{{Complete conversion between one and two photons in nonlinear
  waveguides: theory of dispersion engineering},} {\protect\JournalTitle{New J.
  Phys.}} \textbf{24}, 065002 (2022).

\bibitem{Drummond1997}
P.~D. Drummond and H.~He, \enquote{{Optical mesons},}
  {\protect\JournalTitle{Phys. Rev. A}} \textbf{56}, R1107 (1997).

\bibitem{Vidal2004}
G.~Vidal, \enquote{{Efficient Simulation of One-Dimensional Quantum Many-Body
  Systems},} {\protect\JournalTitle{Phys. Rev. Lett.}} \textbf{93}, 040502
  (2004).

\bibitem{Vidal2003}
G.~Vidal, \enquote{{Efficient Classical Simulation of Slightly Entangled
  Quantum Computations},} {\protect\JournalTitle{Phys. Rev. Lett.}}
  \textbf{91}, 147902 (2003).

\bibitem{Yanagimoto2021}
R.~Yanagimoto, E.~Ng, L.~G. Wright, T.~Onodera, and H.~Mabuchi,
  \enquote{Efficient simulation of ultrafast quantum nonlinear optics with
  matrix product states,} {\protect\JournalTitle{Optica}} \textbf{8},
  1306--1315 (2021).

\bibitem{Orus2014}
R.~Or'us, \enquote{{A practical introduction to tensor networks: Matrix product
  states and projected entangled pair states},} {\protect\JournalTitle{Ann.
  Phys.}} \textbf{349}, 117--158 (2014).

\bibitem{Sornborger1999}
A.~T. Sornborger and E.~D. Stewart, \enquote{{Higher-order methods for
  simulations on quantum computers},} {\protect\JournalTitle{Phys. Rev. A}}
  \textbf{60}, 1956 (1999).

\bibitem{Paeckel2019}
S.~Paeckel, T.~K\"ohler, A.~Swoboda, S.~R. Manmana, U.~Schollw\"ock, and
  C.~Hubig, \enquote{{Time-evolution methods for matrix-product states},}
  {\protect\JournalTitle{Ann. Phys.}} \textbf{411}, 167998 (2019).

\bibitem{Obrien2007}
J.~L. O'Brien, \enquote{{Optical Quantum Computing},}
  {\protect\JournalTitle{Science}} \textbf{318}, 1567--1570 (2007).

\bibitem{He2011}
B.~He, Q.~Lin, and C.~Simon, \enquote{{Cross-Kerr nonlinearity between
  continuous-mode coherent states and single photons},}
  {\protect\JournalTitle{Phys. Rev. A}} \textbf{83}, 053826 (2011).

\bibitem{Xia2016}
K.~Xia, M.~Johnsson, P.~L. Knight, and J.~Twamley, \enquote{{Cavity-Free Scheme
  for Nondestructive Detection of a Single Optical Photon},}
  {\protect\JournalTitle{Phys. Rev. Lett.}} \textbf{116}, 023601 (2016).

\bibitem{Viswanathan2018}
B.~Viswanathan and J.~Gea-Banacloche, \enquote{{Analytical results for a
  conditional phase shift between single-photon pulses in a nonlocal nonlinear
  medium},} {\protect\JournalTitle{Phys. Rev. A}} \textbf{97}, 032314 (2018).

\bibitem{Babushkin2022}
I.~Babushkin, A.~Demircan, M.~Kues, and U.~Morgner,
  \enquote{{Wave-Shape-Tolerant Photonic Quantum Gates},}
  {\protect\JournalTitle{Phys. Rev. Lett.}} \textbf{128}, 090502 (2022).

\bibitem{SnyderLove}
A.~W. Snyder, J.~D. Love \emph{et~al.}, \emph{Optical waveguide theory}, vol.
  175 (Chapman and hall London, 1983).

\bibitem{Fallahkhair2008}
A.~B. Fallahkhair, K.~S. Li, and T.~E. Murphy, \enquote{Vector finite
  difference modesolver for anisotropic dielectric waveguides,}
  {\protect\JournalTitle{Journal of lightwave technology}} \textbf{26},
  1423--1431 (2008).

\bibitem{Fejer1986}
M.~M. Fejer, \emph{Single crystal fibers: Growth dynamics and nonlinear optical
  interactions} (Stanford University, 1986).

\bibitem{Kolesik2004}
M.~Kolesik and J.~V. Moloney, \enquote{Nonlinear optical pulse propagation
  simulation: From maxwell’s to unidirectional equations,}
  {\protect\JournalTitle{Physical Review E}} \textbf{70}, 036604 (2004).

\bibitem{Nye1985}
J.~F. Nye \emph{et~al.}, \emph{Physical properties of crystals: their
  representation by tensors and matrices} (Oxford university press, 1985).

\bibitem{haus2000electromagnetic}
H.~A. Haus, \emph{Electromagnetic noise and quantum optical measurements}
  (Springer Science \& Business Media, 2000).

\end{thebibliography}

\end{document}